%% file: arXiv_version.tex
\documentclass[conference]{IEEEtran}
\usepackage{cite}
\newcommand{\subparagraph}{}
\usepackage{titlesec}

\usepackage{graphicx}

\graphicspath{{./figures}}
\DeclareGraphicsExtensions{.pdf}
\usepackage[cmex10]{amsmath}
\usepackage{amssymb}
\usepackage{amsthm}
\usepackage{hyperref}
\usepackage{cleveref}
\usepackage{multirow}
\usepackage{bm}
\usepackage{algorithmic}
\usepackage{tikz}
\usetikzlibrary{shapes,arrows,positioning,plotmarks,shadows,calc,matrix,fit,backgrounds}
\usepackage{pgfplots}
\pgfdeclarelayer{foreground}
\pgfdeclarelayer{background}
\pgfsetlayers{background,main,foreground}
\usepackage{todonotes}
\usepackage{morefloats}
\usepackage{afterpage}

\usepackage[caption=false,font=footnotesize]{subfig}
\usepackage{fixltx2e}

\input{input/preamble}

\titlespacing*{\subsubsection}{0pt}{0.5em}{0pt}
\begin{document}
%
\title{The Passive Eavesdropper Affects my Channel: Secret-Key Rates under Real-World Conditions\\
--- Extended Version ---}


\author{Christan Zenger, Hendrik Vogt, Jan Zimmer, Aydin Sezgin, Christof Paar\\
Ruhr-Universit\"at Bochum, Germany\\
\{christian.zenger, hendrik.vogt, jan.zimmer, aydin.sezgin, christof.paar\}@rub.de }

\maketitle

\IEEEpeerreviewmaketitle

\input{input/contentICC}


\bibliographystyle{IEEEtran}
\bibliography{IEEEabrv,../references/Conf_abrv_new,../references/references,../references/PROPHYLAXE}


\input{input/appendix}

\end{document}

%% file: input/preamble.tex
\newcommand{\mybold}[1]{\bm{#1}}

\DeclareMathOperator{\mui}{I}

\newcommand{\rsk}{R_{\text{sk}}}
\newcommand{\csk}{C_{\text{sk}}}

\titlespacing*{\subsubsection}{0pt}{1em}{0pt}

\newtheoremstyle{remarkmod}
  {\topsep}   
  {\topsep}   
  {\normalfont}  
  {0pt}       
  {\itshape} 
  {.}         
  {5pt plus 1pt minus 1pt} 
  {}          
\theoremstyle{remarkmod}

\makeatletter
\newcommand*{\textoverline}[1]{$\overline{\hbox{#1}}\m@th$}
\makeatother

\tikzstyle{antenna} = [regular polygon,regular polygon sides=3, draw,shape
border rotate=180,minimum size=0.2pt,scale=1]
\tikzstyle{sum} = [draw, circle, node distance=1cm]

%% file: input/contentICC.tex
\begin{abstract}
%
Channel-reciprocity based key generation (CRKG) has gained significant importance as it has recently been proposed as a potential lightweight security solution for IoT devices. However, the impact of the attacker's position in close range has only rarely been evaluated in practice, posing an open research problem about the security of real-world realizations. Furthermore, this would further bridge the gap between theoretical channel models and their practice-oriented realizations.
%
%
For security metrics, we utilize cross-correlation, mutual information, and a lower bound on secret-key capacity. We design a practical setup of three parties such that the channel statistics, although based on joint randomness, are always \emph{reproducible}. We run experiments to obtain channel states and evaluate the aforementioned metrics for the impact of an attacker depending on his position.
%
%
It turns out the attacker himself affects the outcome, which has not been adequately regarded yet in standard channel models.
\end{abstract}

\section{Introduction}

The inherent randomness of the wireless medium can be utilized for extracting a shared secret, since wireless channels exhibit the feature of \emph{reciprocity}. This approach is referred to as channel-reciprocity based key generation (CRKG). The underlying assumption is that an eavesdropper (Eve) cannot obtain the same channel state, and thus cannot compute the key. The general feasibility of the approach has been reported by several early works in the literature~\cite{src:li2006securing,src:wilson2007channel},
which have been extended by subsequent studies related to practical key agreement~\cite{src:liu2012exploiting,src:pierrot2013}. In particular, there have been some works that deal with the removal of temporal correlation, by methods like principal component analysis (PCA)~\cite{src:chen2011secret}, beamforming~\cite{src:madiseh2012applying} or linear prediction~\cite{src:mcguire2014bounds}.   

Throughout the paper, we use \textit{cross-correlation}, \textit{mutual information}, and \emph{secret-key rates} as performance metric. The theoretical foundation of secret-key rates has been established by Maurer~\cite{src:maurer1993} and Ahlswede et al.~\cite{src:ahlswede1993}. They coined the information-theoretic \emph{source-type model}, where Alice, Bob and Eve have access to a jointly random source, and derived bounds on the secret-key \emph{capacity}. Their result is used in a large body of research, especially for Gaussian channels, e.g., reference~\cite{src:wallace2009key} for a multi-observation model or \cite{src:wilson2007channel} for the application to UWB channels. 

However, some of the popular beliefs regarding the capabilities of the eavesdropper have to be challenged. Many previous works, e.g.,~\cite{DBLP:conf/mobicom/MathurTMYR08sh,DBLP:conf/mobicom/JanaPCKPK09sh}, have relied on the assumption that the channel of Alice-to-Bob gets uncorrelated to that of Eve, as long as Eve is positioned more than half a wavelength away from Alice and Bob, commonly referred to as Jake's model~\cite[Chapter 3.2.1]{src:goldsmith2005wireless}. In the literature, this is usually referred to as \emph{spatial decorrelation}~\cite{src:zhang2016key}.  A study~\cite{src:pierrot2013} has questioned this assumption by practical evaluation. Recently, a comprehensive study~\cite{src:he2016toward} has shown that for many popular correlation models of scattering environments, the eavesdropper might obtain largely correlated observations, especially if Eve is located within the line-of-sight beam of Alice and Bob. 

In this work, we intend  and shed more light on the threats for CRGK from passive eavesdropping. 
As a consequence, we extend the work of~\cite{src:he2016toward} by providing more elaborated practical measurements. We quantify the leakage of Alice and Bob in relation to Eve with respect to the distance, especially for low ranges that introduce near-field effects. The measurement setup is designed with the objective to generate \emph{reproducible} results, such that we can justify
\emph{stationary} random processes. This is a fundamental necessity in order to obtain meaningful results, which has sometimes been overlooked in previous work. The cross-correlation and achievable secret-key rate serve as the performance metrics that indicate the common randomness available to Alice and Bob, and likewise, the information loss to Eve. We evaluate the metrics for the original data and the processed versions after down-sampling or decorrelation. The results demonstrate that the close physical presence of Eve in the communication setting significantly changes the channel statistics. This phenomenon is so far not covered by conventional channel models for CRKG.

Section~\ref{sec:systemmodel} introduces the system model and elaborates on both the processing of the measured data and the performance metrics on security. The measurement setup is described in section~\ref{sec:measurements}. The evaluation and results of the measurement campaign are presented in section~\ref{sec:evaluation}. Finally, section~\ref{sec:conclusion} concludes the paper. 

\section{System model} 
\label{sec:systemmodel}
\begin{figure}
\centering 
\begin{tikzpicture}[
block/.style={draw, drop shadow, fill=white, rectangle, minimum height=0.75cm, minimum width=3.5em},
publicch/.style={draw, drop shadow, fill=white, rectangle, rounded corners,minimum height=1.5em, minimum width=2.5em},
]
\node[block] (alice) at (0,0) {Alice};	
\node[block] (eve) at ($(alice)+(2.5,-1.25)$) {Eve};
\node[block] (bob) at ($(alice)+(5,0)$) {Bob};

\draw[thick,->] ($(alice)+(1,0.2)$) -- node[above] {$h_{ab,k}$} ($(bob)+(-1,0.2)$);
\draw[thick,->] ($(bob)+(-1,-0.1)$) -- node[below] {$h_{ba,k}$} ($(alice)+(1,-0.1)$);
\draw[thick,->,dashed] (alice) -- node[below,pos=0.5] {$h_{ae,k}$} (eve);
\draw[thick,->,dashed] (bob) -- node[below,pos=0.3] {$h_{be,k}$} (eve);
\end{tikzpicture}
\caption{Overview of the system model.} \label{fig:systemmodel}
\end{figure}
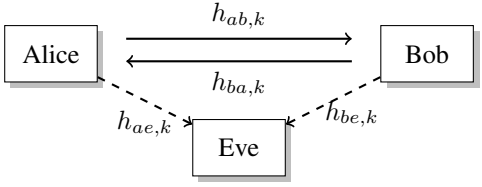
As depicted in Figure~\ref{fig:systemmodel}, we consider Alice, Bob and Eve measuring the channel $h_{ab,k}\in\mathbb{R}$, $h_{ba,k}\in\mathbb{R}$, $h_{ae,k}\in\mathbb{R}$ and $h_{be,k}\in\mathbb{R}$, which represent the state of Alice-to-Bob, Bob-to-Alice, Alice-to-Eve and Bob-to-Eve channels, respectively, and $k$ denotes a discrete time instant. We model these variables as joint stationary and ergodic random processes. In general, Eve gets two channel states $\left(h_{ae,k},h_{be,k}\right)$, however, in this study we focus on $h_{ae,k}$ only. In the following, we use the labels $x_k:=h_{ba,k}$ for Alice, $y_k:=h_{ab,k}$ for Bob, and $z_k:=h_{ae,k}$ for Eve. Furthermore, we define the vector process
$\mybold{v}_k:=\left(x_{k}, y_{k}, z_{k}\right)^T$. 

\subsection{Processing}
For different $k$, the random vectors $\mybold{v}_k$ are likely to exhibit correlation in time, since the wireless channel is varying only slowly in indoor environments. In order to remove the temporal dependencies, we perform two alternative options of processing, namely either downsampling or decorrelation. We show both options for $x_{k}$ only, since we have the same processing for $y_{k}$ and $z_{k}$. 

\subsubsection{Downsampling}
If we keep only every $N_m$th variable of the process $x_{k}$, we effectively
downsample by factor $N_m$ and obtain
\begin{align}
\label{eq:donwsampled}
x_{k}^{\text{ds}}=x_{kN_m}.
\end{align} 
The generated $x^{\text{ds}}_k$ can be assumed independent under the condition that the process does not exhibit any dependence after an interval of $N_m$ variables. Subsequently, we assume that the $\mybold{v}^{\text{ds}}_k=\left(x_{k}^{\text{ds}},y_{k}^{\text{ds}},z_{k}^{\text{ds}}\right)^T$ are identically and independently distributed (i.i.d.) for different $k$. 

\subsubsection{Decorrelation}
 We need to provide an estimator for the autocorrelation function 
\begin{align}
\label{eq:autocorrest}
\hat{r}_{xx}[l] = \frac{1}{N-l} \sum_{i=0}^{N-l-1}x_{i}x_{i+l}.
\end{align}
This estimator is unbiased if the process is correlation-ergodic. The linear forward predictor for $x_{k}$ of order $N_m$ is given by
\begin{align}
\label{eq:predictor}
\hat{x}_{k} = \sum_{i=1}^{N_m} a_i x_{k-i},
\end{align}
where $a_i\in\mathbb{R}$ are parameter coefficients, which can be computed by Levinson-Durbin recursion based on Yule-Walker equations~\cite{src:vaidyanathan2007the}. We define 
\begin{align}
\label{eq:decorr}
x_{k}^{\text{de}} = x_{k}-\hat{x}_{k}
\end{align}
as \emph{innovation sequence}, which is orthogonal to past $h_{ab,k-i}$ for $i>0$. However, orthogonal (or uncorrelated if zero-mean) variables do not necessarily imply independence, especially not \emph{joint} independence of $\mybold{v}^{\text{de}}_k=\left(x_{k}^{\text{de}},y_{k}^{\text{de}},z_{k}^{\text{de}}\right)^T$ for different $k$. Decorrelation is practically more relevant than downsampling (even if no i.i.d. can be achieved), since the information loss is significantly lower.

\subsection{Performance metrics}
Throughout the paper, we use (1) the Pearson correlation and (2) secret-key rates as performance metrics for security.

\subsubsection{Pearson correlation}
The Pearson correlation provides a measure of linear dependence between two data series. The values span between $-1$ and $1$, where $1$ refers to absolute correlation, $0$ to no correlation, and $-1$ to perfect inverse correlation. It is a wide-used metric for secrecy of practical secret-key generation~\cite{src:he2016toward}. Given a finite collection of $N$ pairs $\left(x_{k},y_{k}\right)$ from the process, we use the estimator
\begin{align}
\label{eq:pearson}
\rho_{xy}=\frac{\sum\limits_{i=0}^{N-1}\left(x_{i}-\bar{x}\right)\left(y_{i}-\bar{y}\right)}{\sqrt{\sum\limits_{i=0}^{N-1}\left(x_{i}-\bar{x}\right)^2}\sqrt{\sum\limits_{i=0}^{N-1}\left(y_{i}-\bar{y}\right)^2}},
\end{align}
where $\bar{x}=\frac{1}{N}\sum_{j=0}^{N-1}x_{j}$ and $\bar{y}=\frac{1}{N}\sum_{j=0}^{N-1}y_{j}$ are the sample means.

\subsubsection{Secret-key rate} 
\label{sec:sk}
We introduce the information-theoretic secret-key rate and use the downsampled process~\eqref{eq:donwsampled}.
Recall that the $\mybold{v}^{\text{ds}}_k$ are i.i.d. We characterize $\mybold{v}^{\text{ds}}_k$ by the joint probability density function $f_{\mybold{v}^{\text{ds}}_k}$. We apply a lower bound on secret-key capacity based on the source-type model, under the following conditions:
\begin{enumerate}
\item The joint probability density function $f_{\mybold{v}^{\text{ds}}_k}$ is known a priori at all terminals.
\item Alice and Bob exchange messages over an authenticated, public channel with unlimited communication capacity.
\item Eve remains passive at all times.
\end{enumerate}
Subsequently, the asymptotic bound is given by~\cite{src:ahlswede1993}
\begin{align}
\label{eq:sklower}
\csk &\geq \mui\left(x_{k}^{\text{ds}};y_{k}^{\text{ds}}\right) \notag\\
& \qquad-\min\left[ \mui\left(x_{k}^{\text{ds}};z_{k}^{\text{ds}}\right), \mui\left(y_{k}^{\text{ds}};z_{k}^{\text{ds}}\right) \right]=:\rsk
\end{align}
for each $k$, since the process is stationary. Since the actual probability distributions are unknown in practice, we evaluate the lower bound~\eqref{eq:sklower} by estimations, based on a finite number of measured samples. We utilize a $k$-nearest neighbor estimator (NNE) for the mutual information, which is based on the idea and implementation of~\cite{src:kraskov2004}. Mutual information is a function of joint and marginal probability densities. For  a measure of the joint density, the estimator computes the distance between a tuple of samples and its $k$th-next neighbor. A similar approach is provided for the marginal densities. To best of our knowledge, the reliability of the NNE has not been studied systematically. However, results in~\cite{src:kraskov2004} indicate that at least for multivariate Gaussian variables, the estimation error is very low if $N>10^4$ samples are used for the estimation.

Note that the bound~\eqref{eq:sklower} could have been defined with the original $\mybold{v}_k$ or the decorrelated processes~\eqref{eq:decorr}, such that less information is discarded than in case of downsampling. However, in order to obtain an accurate estimation of~\eqref{eq:sklower}, we require i.i.d. samples for the two following reasons: 
\begin{enumerate}
\item The bound~\eqref{eq:sklower} has been derived under the assumption of an unlimited number of i.i.d. observations from a random source. Therefore, a value of $\rsk$ measured in bits per observation, is meaningful only if the time series is i.i.d. as well.
\item The NNE of~\cite{src:kraskov2004} requires i.i.d. samples, since it relies on Khinchin's theorem~\cite[p. 277]{src:papoulis2002probability}. If the time series of samples exhibits some dependence in time, the estimator might induce an undesired bias.
\end{enumerate}
 
Therefore, if we apply the process $\mybold{v}_k$ or its decorrelated modification~\eqref{eq:decorr}, we have an approximation of the lower bound $\rsk$~\eqref{eq:sklower} only. While approximating the common information of Alice and Bob is a rather "safe" option, we need to be cautious regarding Eve. In order to minimize the risk of underestimating Eve, we verify our results obtained from $\mybold{v}_k$ or the decorrelated version~\eqref{eq:decorr} by comparing them with the downsampling approach, since it provides a more accurate description of the information leakage to Eve. Unfortunately, by removing samples from the estimation, the NNE gets more biased.

\begin{figure}[h]
		\centering
\includegraphics[width=0.5\textwidth]{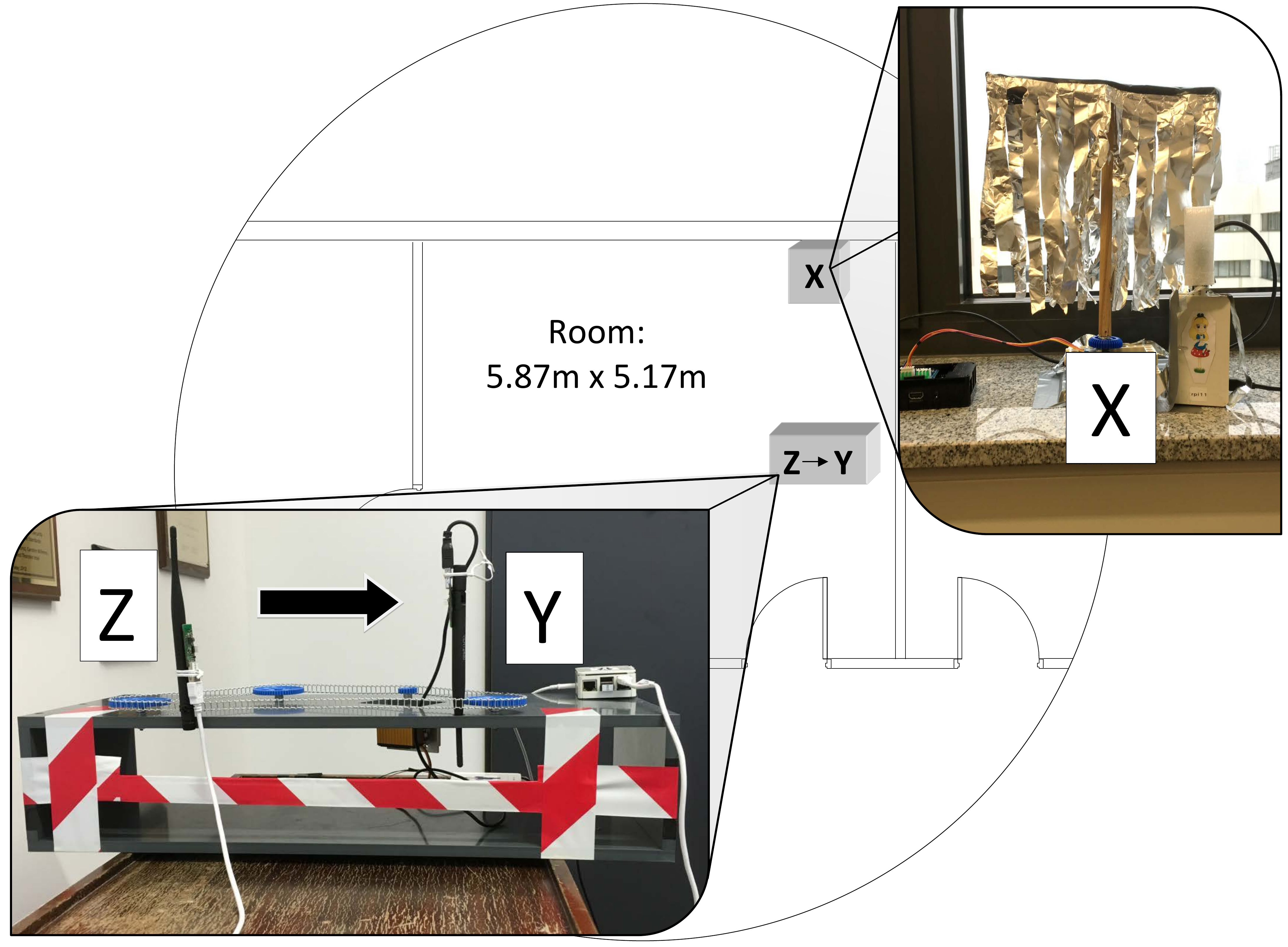}
  		\caption{The testbed includes several experimental setups for performance evaluation as well as for security analysis. 
  		Alice (X), Bob (Y) and Eve (Z) are mounted on a automated antenna positioning system.}
        \label{fig:setup}
\end{figure}

\section{Measurements}
\label{sec:measurements}
The testbed is applied at the premises of our research group, which is an office area in a university building. 
Alice is positioned at a predestined access point position. Bob and Eve are mounted on an automated antenna positioning setup, which is located at several predestined "end-device" positions (cf.\ Figure~\ref{fig:setup}). For this, we choose positions which are representative for security-related IoT devices, such as doorknobs (keyless entry systems), window frames (perimeter fence intrusion sensor), and wall (motion detectors) positions. Due to a lack of space, in this version of the paper we restrict  ourselves to a description of one representative realization of all experiments. We will also provide a full version of the paper with results of $23$ further positionings in the building.

\begin{table}
\caption{Parameters of the measurement setup}
\begin{center}
\begin{tabular}{|l | c | l| }
	\hline
	Parameter &  Variable & Value \\
	\hline\hline
	Sampling interval & $T_s$    & $100$ msec  \\\hline
	Probing duration & $T_p$    & $<5$ msec  \\\hline
	Step size & $\Delta_d$    & $5$ mm  \\\hline
	Accuracy of step size & $\hat{\Delta}_d$    & $\pm 0.05$ mm  \\\hline
	Geometrical distance Bob-Eve & $\Delta_{BE}$    & $[0,30]$ cm  \\\hline	
	Geometrical distance Alice-Bob & $\Delta_{AB}$    & $5$ m  \\\hline	
	Samples per step & $N$    & $3\cdot 10^5$   \\\hline
\end{tabular}
\end{center}

\label{tab:parameters}
\end{table}

We perform mobile, long-time narrow-band channel measurements on $2.4$~GHz (wavelength $12.5$~cm). The data exchange protocol is implemented on three Raspberry Pi $2$ platforms (credit-card sized computer). 
All devices are equipped with a CC$2531$ USB enabled IEEE $802.15.4$ communication interface\footnote{http://www.ti.com/tool/cc2531emk}. The CC$2531$ is a true SoC solution for IEEE $802.15.4$ applications, that is compatible to network layer standards for resource-constrained devices: ZigBee, WirelessHART, and 6LoWPAN. The platform is equipped with proprietary PCB antennas, i.e., \textit{Meandered Inverted-F antenna} (MIFA), with the size of $5\times 12$~mm. These antennas provide good performance with a small form factor. The platform and antenna design are widely used in commercial products and suited for systems where ultra-low-power consumption is required. 

In order to establish common channel probing, Alice periodically sends data frames to Bob and waits for acknowledgments. Eve also receives these request-response pairs. When receiving a probe, all three devices extract Received Signal Strength Indicators (RSSI) values and, thus, can measure a channel-dependent sequence over time. For evaluation of the channel measurements, we store and process the realizations of  $\mybold{v}_k:=\left(x_k, y_k, z_k\right)^T$, locally on a monitoring laptop.

Table~\ref{tab:parameters} lists the relevant parameters of our measurement setup. We obtain a complete realization of $\mybold{v}_k$ on every sampling interval $T_s=100$~msec. The protocol ensures that Alice, Bob, and Eve can probe the channel within a probing duration $T_p<5$~msec. We want to analyze the joint statistical properties of the samples with respect to the position of Eve in the scene. As a consequence, we apply an automated antenna positioning system, which is constructed from a low-reflective material, cf. Figure~\ref{fig:setup}. 
It moves the antenna of Eve on a linear guide towards the fixed antenna of Bob in step size $\Delta_d=5$~mm with accuracy $\hat{\Delta}_d=\pm 0.05$~mm. The variable distance $\Delta_{BE}$ ranges from $0$ to $30$~cm in order to provide $60$ different locations. Alice's antenna is placed orthogonal to the linear guiding at a fixed distance $\Delta_{AB}=5$~m. For each position of Eve's antenna, we record at least $N$ samples.  

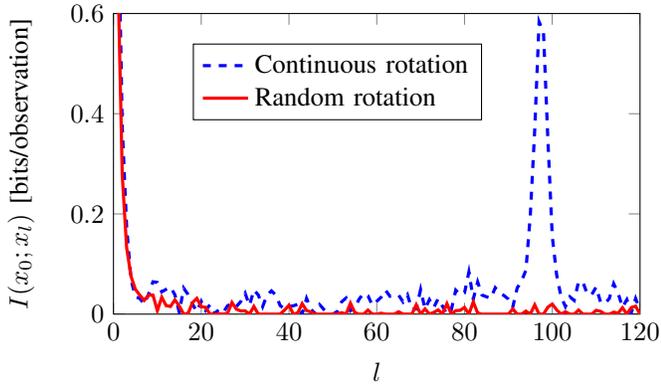
\begin{figure}
\centering
\begin{tikzpicture}
\begin{axis}[%
width=7cm,
height=4cm,
scale only axis,
xmin=0,
xmax=120,
xlabel={$l$},
ymin=0,
ymax=0.6,
ylabel={$I(x_0;x_l)$ [bits/observation]},
legend style={draw=black,fill=white,legend cell align=left,at={(0.7,0.9)}}
]
\addplot [color=blue,solid,very thick,dashed]
  table[]{figures/muiVsDelayAliceOld-1.tsv};
\addlegendentry{Continuous rotation}; 
\addplot [color=red,solid,very thick]
  table[]{figures/muiVsDelayAliceNew-1.tsv};
\addlegendentry{Random rotation}; 
\end{axis}
\end{tikzpicture}%
	\caption{Self-dependence of channel gains with respect to time delay. Setup is equipped with aluminum strips of either continuous or random rotation.}
	\label{fig:MI_cyclic}
\end{figure}

Alice and Bob extract the common randomness $x_k$ and $y_k$ from a time-varying channel. Since we aim for meaningful and reproducible results, we have to create an environment which provides the joint stationarity to the random process. 
Therefore, with a distance of $10$~cm to Alice's antenna, we deploy  a curtain of $30 \times 30$~cm aluminum strips that continuously rotates at $ \approx 0.1$~rotations per second, cf. Figure~\ref{fig:setup}. However, the rotation itself inserts a deterministic component into the channel. The evolution of the self-dependence of channel gains  --- we show exemplary $x^{\text{ds}}_k$ --- is illustrated in Figure~\ref{fig:MI_cyclic}. It shows that the mutual information decays rapidly and vanishes after four samples, corresponding to approximately $400$~ms. However, due to the continuously rotating curtain of aluminum strips, we discover strong stochastical dependencies after $96$ samples, corresponding to approximately $9.6$~s. Therefore, we adapt a random source (Unix file \texttt{/dev/urandom}) to the motor controller and program the instrument to rotate with random speed between $0.240$ rad/s and $1$~rad/s in random direction and with random interval lengths $0^\circ, 1^\circ, \ldots 60^\circ$ (uniformly distributed).
 Figure~\ref{fig:MI_cyclic} shows that no strong stochastical dependencies are given anymore.

\section{Evaluation and Results}
\label{sec:evaluation}
\begin{figure*}[htp!]
	\centering
	\subfloat[]{\includegraphics[trim=1.4cm 0.1cm 3.5cm 1.6cm, clip=true, height=0.224\textwidth]{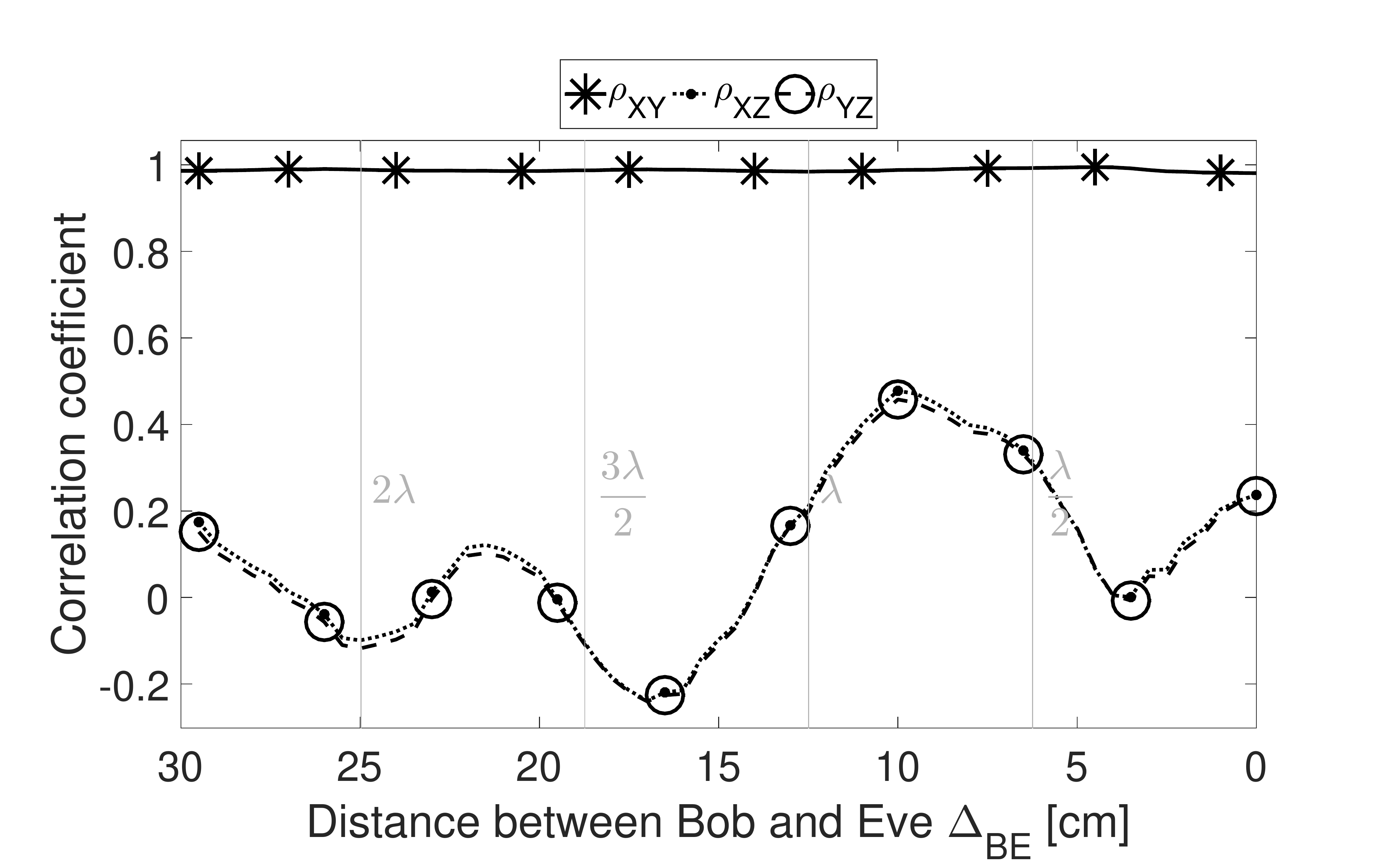}}
	\subfloat[]{\includegraphics[trim=0.5cm 0.1cm 3.5cm 1.6cm, clip=true, height=0.224\textwidth]{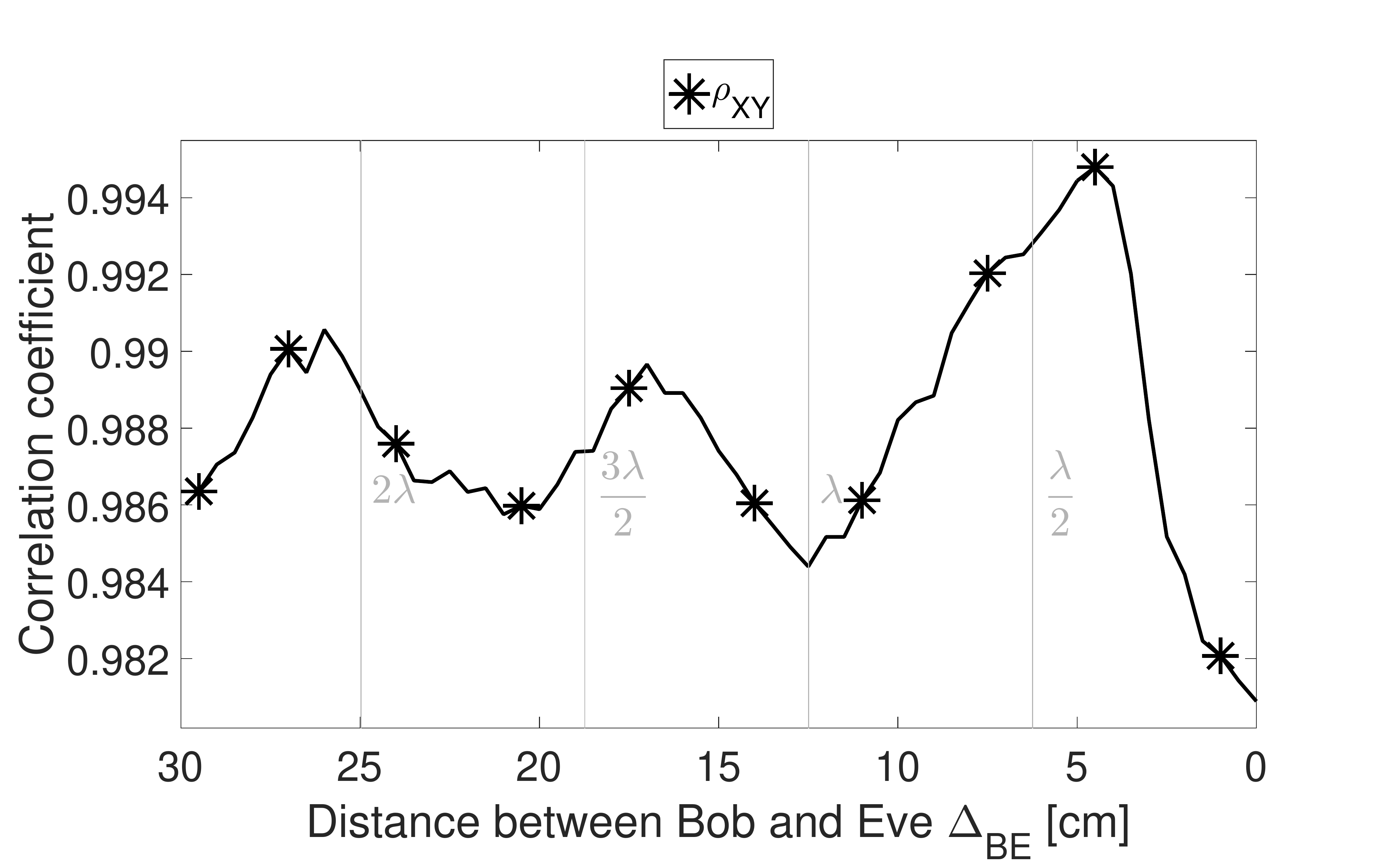}}
	\subfloat[]{\includegraphics[trim=2.2cm 0.1cm 3.5cm 1.6cm, clip=true, height=0.224\textwidth]{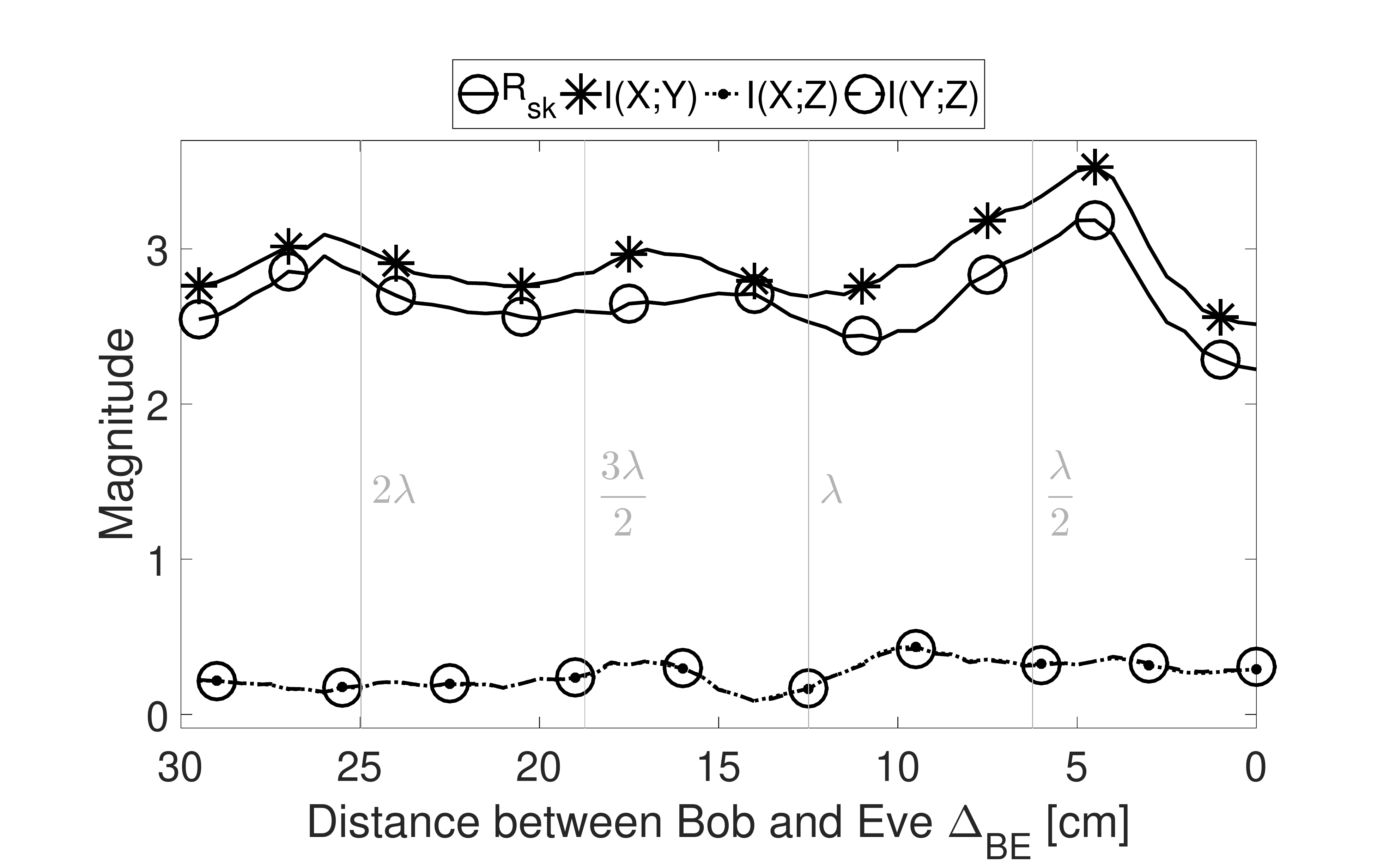}}
	\caption{Evaluation results of $\mybold{v}_k$. In (a) and (b) the cross-correlations is given; in (c) the mutual information as well as $\rsk$ is given.}
	\label{fig:original}
\end{figure*}

\begin{figure*}[htp!]
	\centering
	\subfloat[]{\includegraphics[trim=1.4cm 0.1cm 3.5cm 1.6cm, clip=true, height=0.224\textwidth]{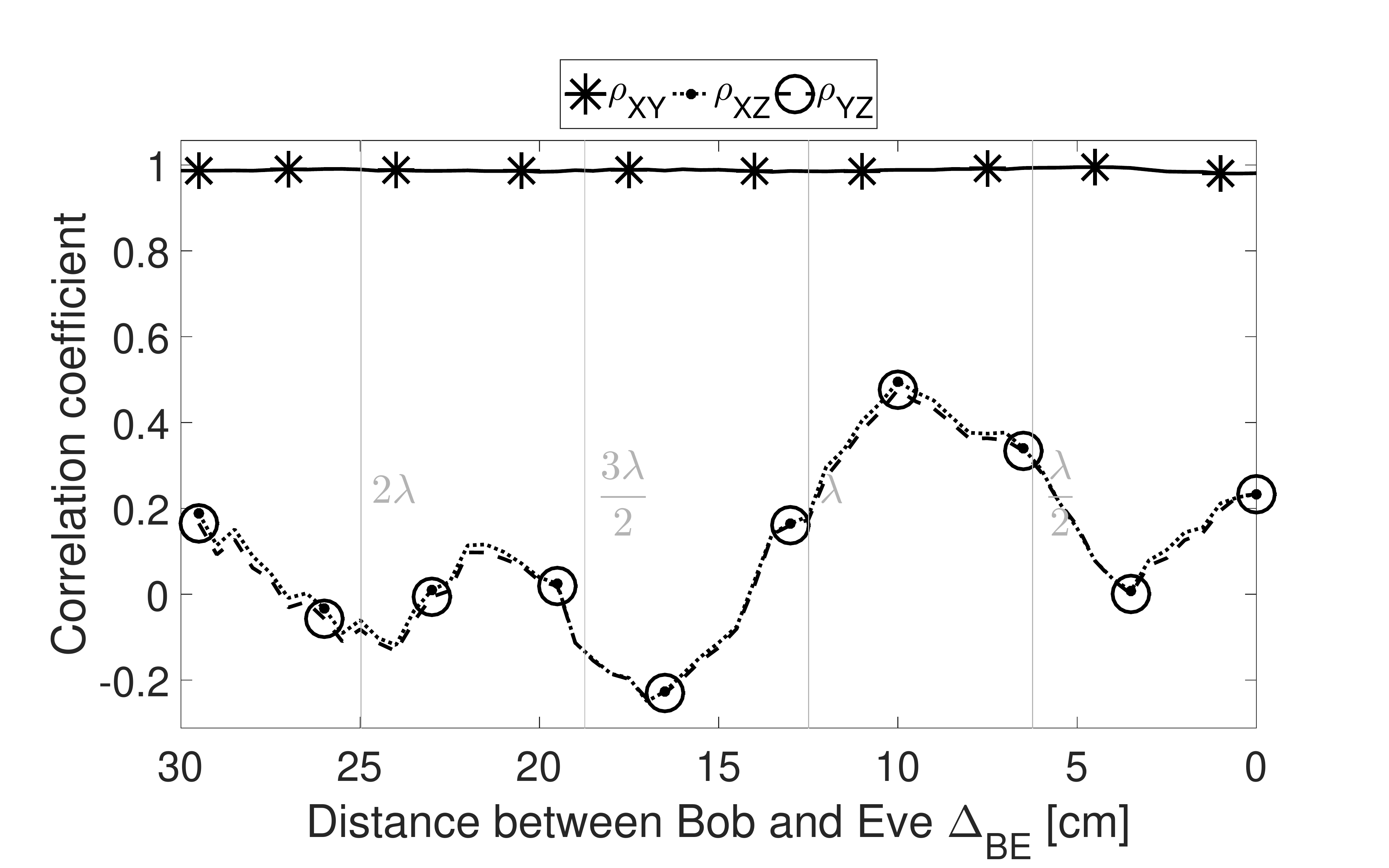}}
	\subfloat[]{\includegraphics[trim=0.5cm 0.1cm 3.5cm 1.6cm, clip=true, height=0.224\textwidth]{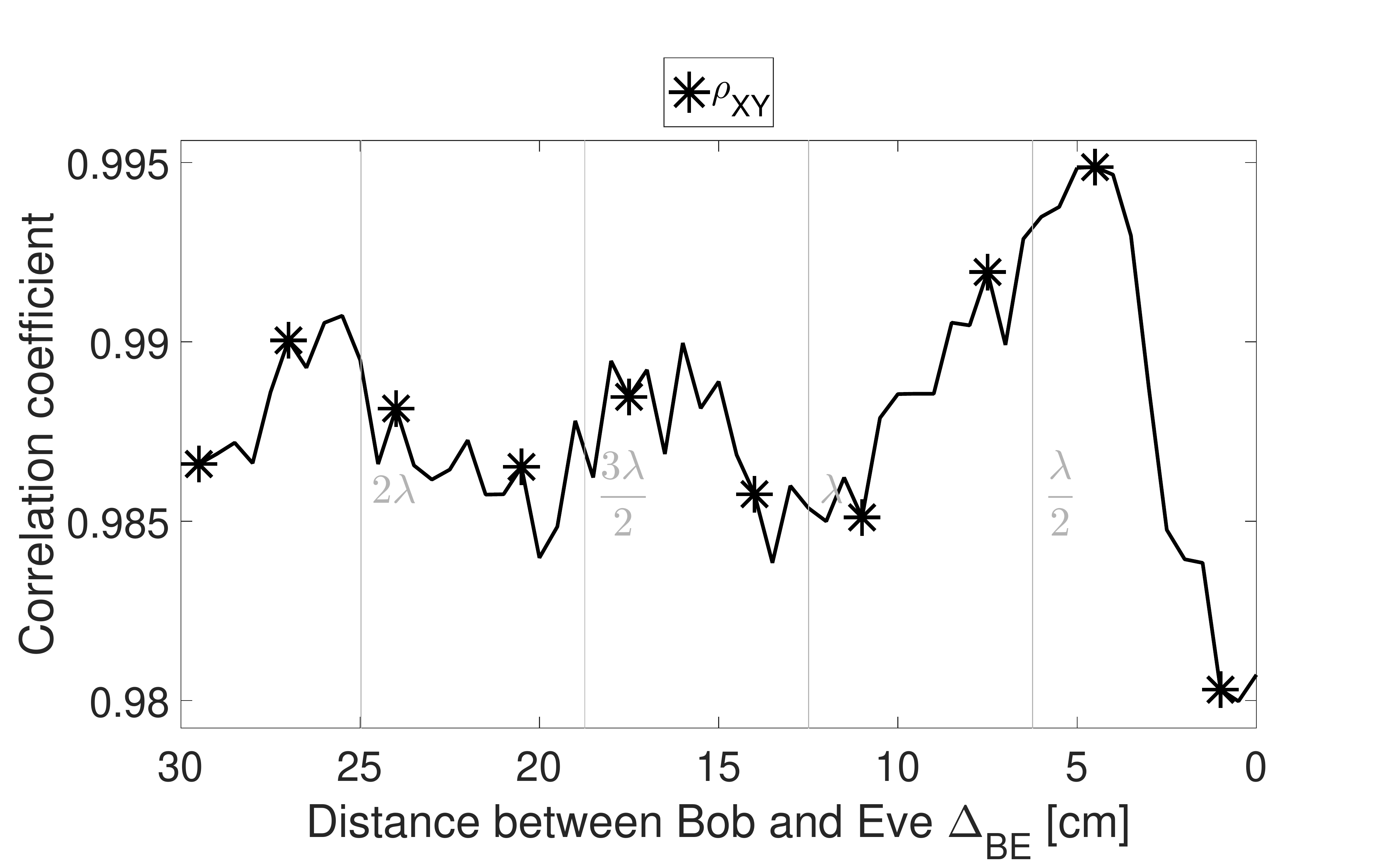}}
	\subfloat[]{\includegraphics[trim=2.2cm 0.1cm 3.5cm 1.6cm, clip=true, height=0.224\textwidth]{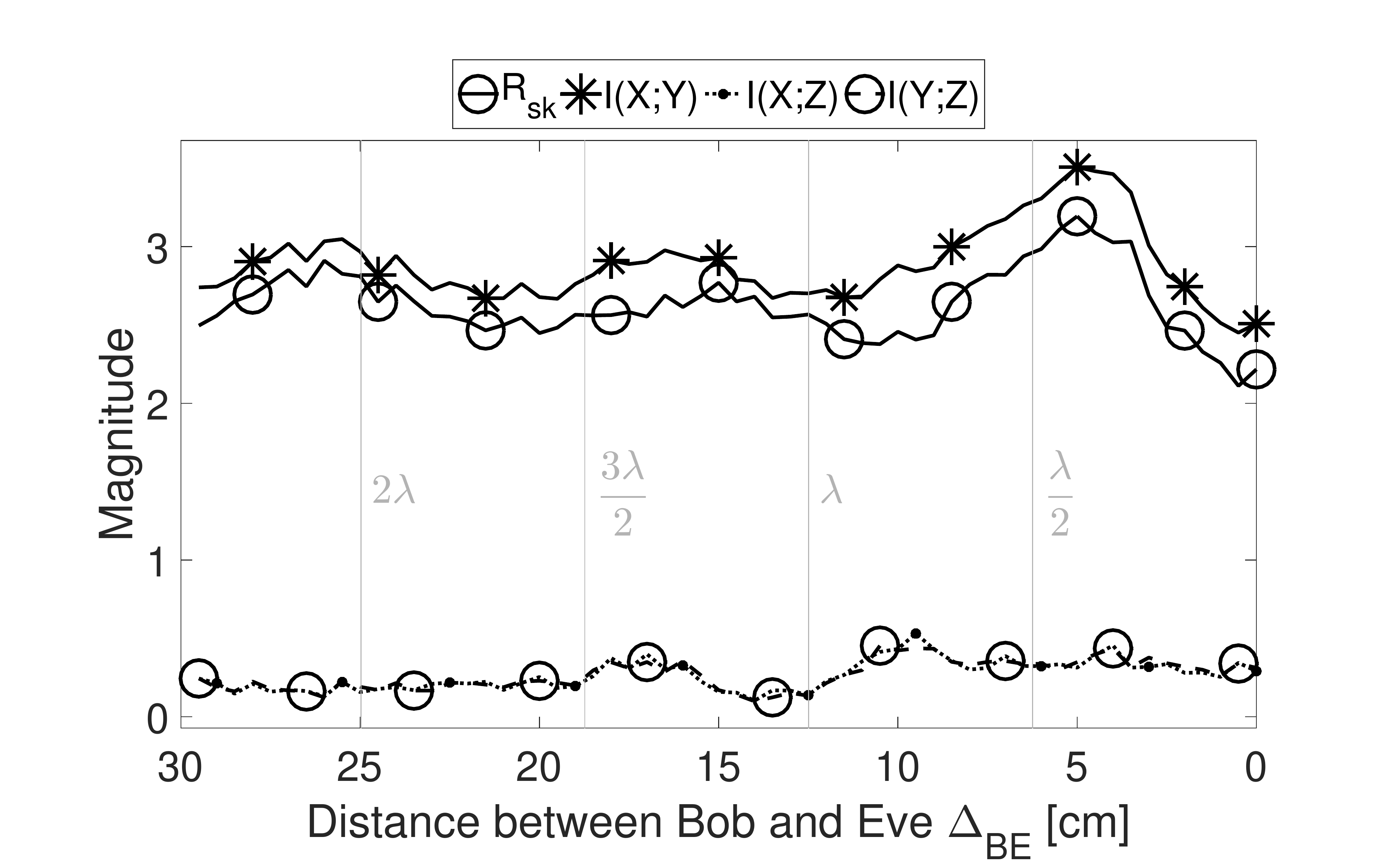}}
	\caption{Evaluation results of $\mybold{v}^{\text{ds}}_k$. In (a) and (b) the cross-correlations is given; in (c) the mutual information as well as $\rsk$ is given.}
	\label{fig:DS}
\end{figure*}

\begin{figure*}[t!]
	\centering
	\subfloat[]{\includegraphics[trim=1.4cm 0.1cm 3.5cm 1.6cm, clip=true, height=0.224\textwidth]{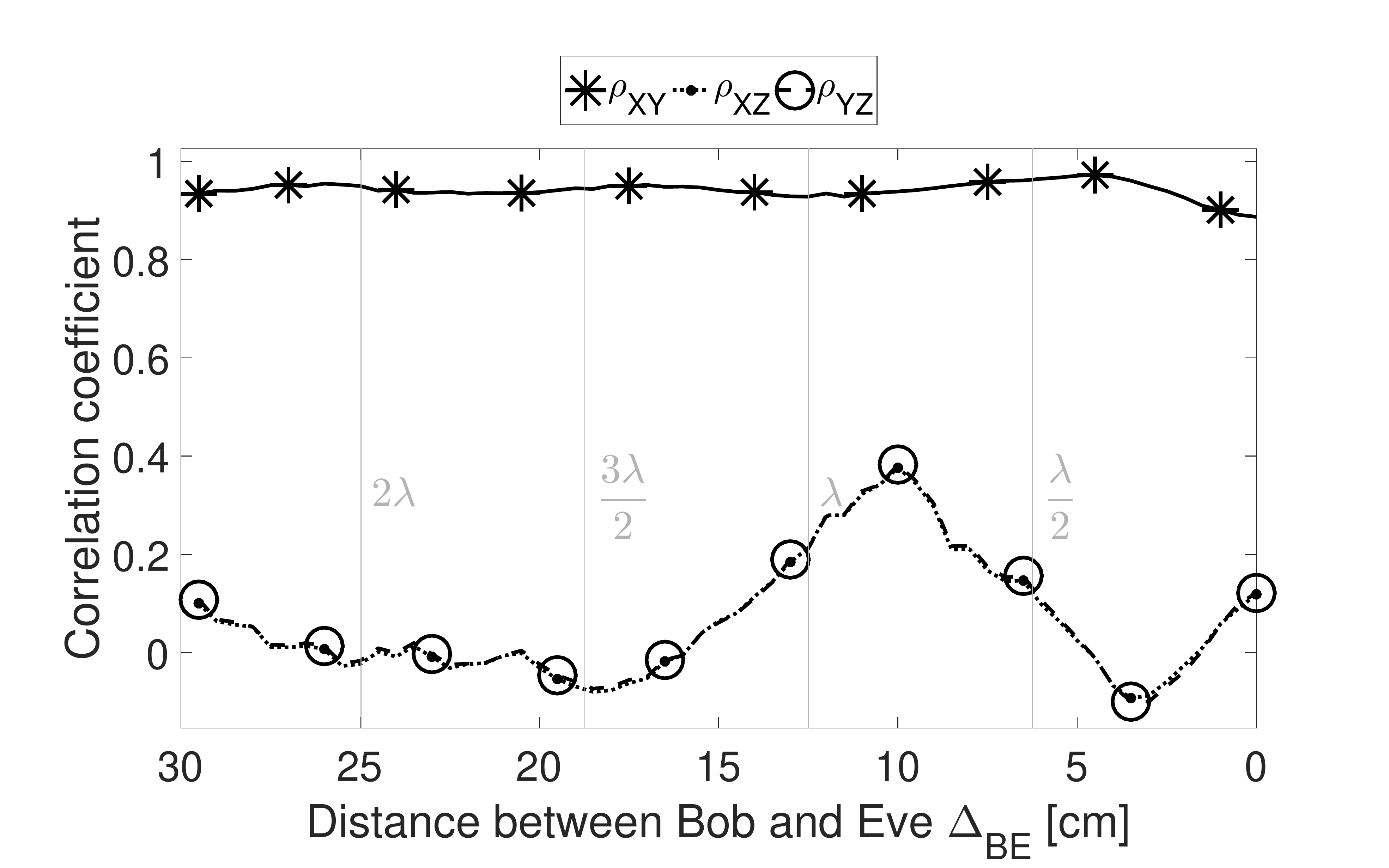}}
	\subfloat[]{\includegraphics[trim=1cm 0.1cm 3.5cm 1.6cm, clip=true, height=0.224\textwidth]{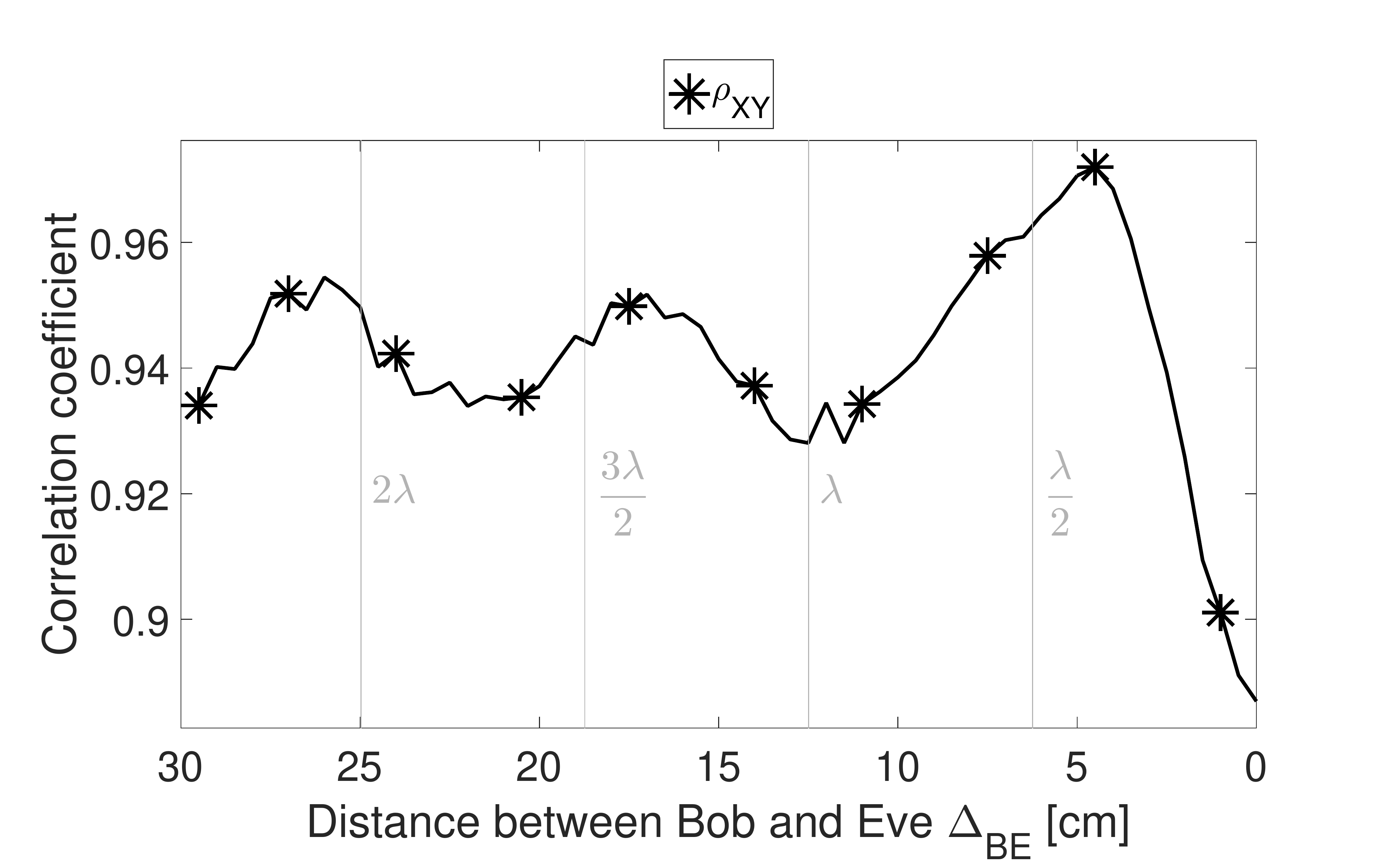}}
	\subfloat[]{\includegraphics[trim=1.8cm 0.1cm 3.5cm 1.6cm, clip=true, height=0.224\textwidth]{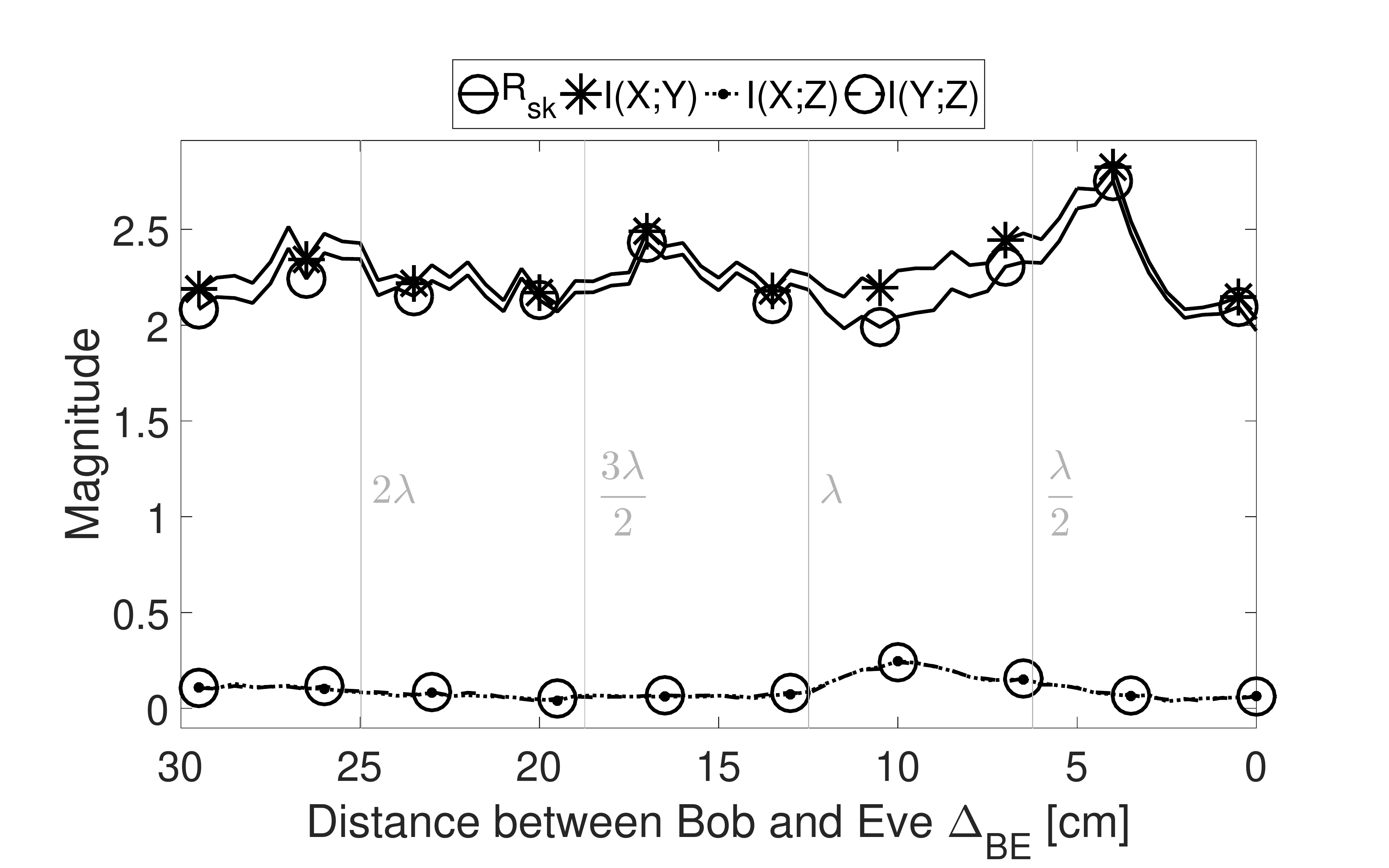}}
	\caption{Evaluation results of $\mybold{v}^{\text{de}}_k$. In (a) and (b) the cross-correlations is given; in (c) the mutual information as well as $\rsk$ is given.}
	\label{fig:decorr}
\end{figure*}

We now use the experimental measurements to evaluate and compare the results of the Pearson correlation~\eqref{eq:pearson}, mutual information, as well as the achievable bound of the secret-key capacity~\eqref{eq:sklower}, as a function of attacker's distance $\Delta_{BE}$ to Bob. We interpret the original measurements as realizations of $\mybold{v}_k$. In addition, we have the decorrelated and downsampled outcomes, denoted by the processes $\mybold{v}^{\text{de}}_k$ and $\mybold{v}^{\text{ds}}_k$, respectively. The decorrelated samples are obtained by a linear prediction of order $N_m = 30$. To generate the i.i.d. random vectors $\mybold{v}^{\text{ds}}_k$ we downsample $\mybold{v}_k$ by the factor $N_m = 30$. In subsubsection~\ref{sec:sk}, we have already outlined the necessity of i.i.d. random vectors to obtain accurate estimations. This is not given for $\mybold{v}_k$ and $\mybold{v}^{\text{de}}_k$, however, they provide valid approximations, as the results indicate later on.
We present three~Figures~\ref{fig:original}, \ref{fig:DS}, \ref{fig:decorr} with three Subfigures a)-c) each, which are arranged in a 3x3 matrix on the next page. The \emph{rows} denote the Figures as follows.
\begin{enumerate}
\item Fig.~\ref{fig:original} illustrates the results for the \emph{original} process $\mybold{v}_k$.
\item Fig.~\ref{fig:DS} shows the results for the \emph{downsampled} process $\mybold{v}^{\text{ds}}_k$ of~\eqref{eq:donwsampled}.
\item Fig.~\ref{fig:decorr} depicts the results for the \emph{decorrelated} process $\mybold{v}^{\text{de}}_k$ of~\eqref{eq:decorr}.
\end{enumerate}
The \emph{columns} constitute Subfigures as follows. For convenience, we introduce generic labels $X\in\left\lbrace x_k,x^{\text{de}}_k,x^{\text{ds}}_k \right\rbrace$ for Alice, $Y\in\left\lbrace y_k,y^{\text{de}}_k,y^{\text{ds}}_k \right\rbrace$ for Bob and $Z\in\left\lbrace z_k,z^{\text{de}}_k,z^{\text{ds}}_k \right\rbrace$ for Eve.
\begin{enumerate}
\item Subfigures a) show the Pearson correlation~\eqref{eq:pearson} vs. geometrical distance $\Delta_{BE}$ between the three pairs (Alice$\leftrightarrow$Bob $\rho_{XZ}$, Alice$\leftrightarrow$Eve $\rho_{XY}$, Bob$\leftrightarrow$Eve $\rho_{YZ}$).
\item Subfigures b) zoom into the correlation $\rho_{XY}$ of Alice$\leftrightarrow$Bob.
\item Subfigures c) depict the three mutual information results ($I(X;Y)$, $I(X;Z)$, $I(Y;Z)$) and the secret-key rate $\rsk$ of ~\eqref{eq:sklower} vs. geometrical distance $\Delta_{BE}$.
\end{enumerate}

Most of the practical key generation schemes use downsampling or decorrelation on the original observations $\mybold{v}_k$. We introduce the Figs. \ref{fig:original}, \ref{fig:DS} and \ref{fig:decorr} in order to analyze whether downsampling and decorrelation obscure certain features of the channel that are important for the security evaluation of the system. We start with a comparison of the cross-correlation behavior between Alice and Bob, as well as to a potential attacker. By comparing Figure~\ref{fig:original} (a-b) and Figure~\ref{fig:DS} (a-b) we see that no significant differences in $\rho_{XY}$ and $\rho_{XZ}$ occur after downsampling. (Further, $\rho_{XZ}$ and $\rho_{YZ}$ are almost identical due to channel reciprocity between Alice and Bob.) The high similarity is due to the fact that even the process $\mybold{v}_k$ does not exhibit much dependency in time, as already hinted in Figure~\ref{fig:MI_cyclic}. As a consequence, the results obtained for $\mybold{v}_k$ expose a valid approximation of the cross-correlation. As it can be seen from Figure~\ref{fig:DS}, in case of downsampling the results are more noisy, since much fewer samples are available for the estimations.


%
%
After decorrelation, the results (see Figure~\ref{fig:decorr}) show that (unlike in case of downsampling) the correlation decreases on average by $\approx 0.05$, which can have a significant negative impact on the performance of a potential quantization scheme, cf.~\cite[Figure 3]{WiComSec-Phy-QuantAna}. 
%
%
Furthermore, the difference between the minimum and maximum value significantly decreases. Whereas in the original (and downsampled) signal the difference is $0.995-0.98=0.015$, the difference is $0.97-0.89=0.08$ for the decorrelated signal. 
This probably stems from errors of the autocorrelation estimate~\eqref{eq:autocorrest}, which is necessary for the linear forward prediction. Another reason might be the Pearson correlation where single outliers (e.g., strong peaks) significantly influence the result. Analyzing the impact of decorrelation techniques on the reciprocity and security in detail is left for future work.

\begin{figure}[htp]
		\centering
\includegraphics[width=0.375\textwidth]{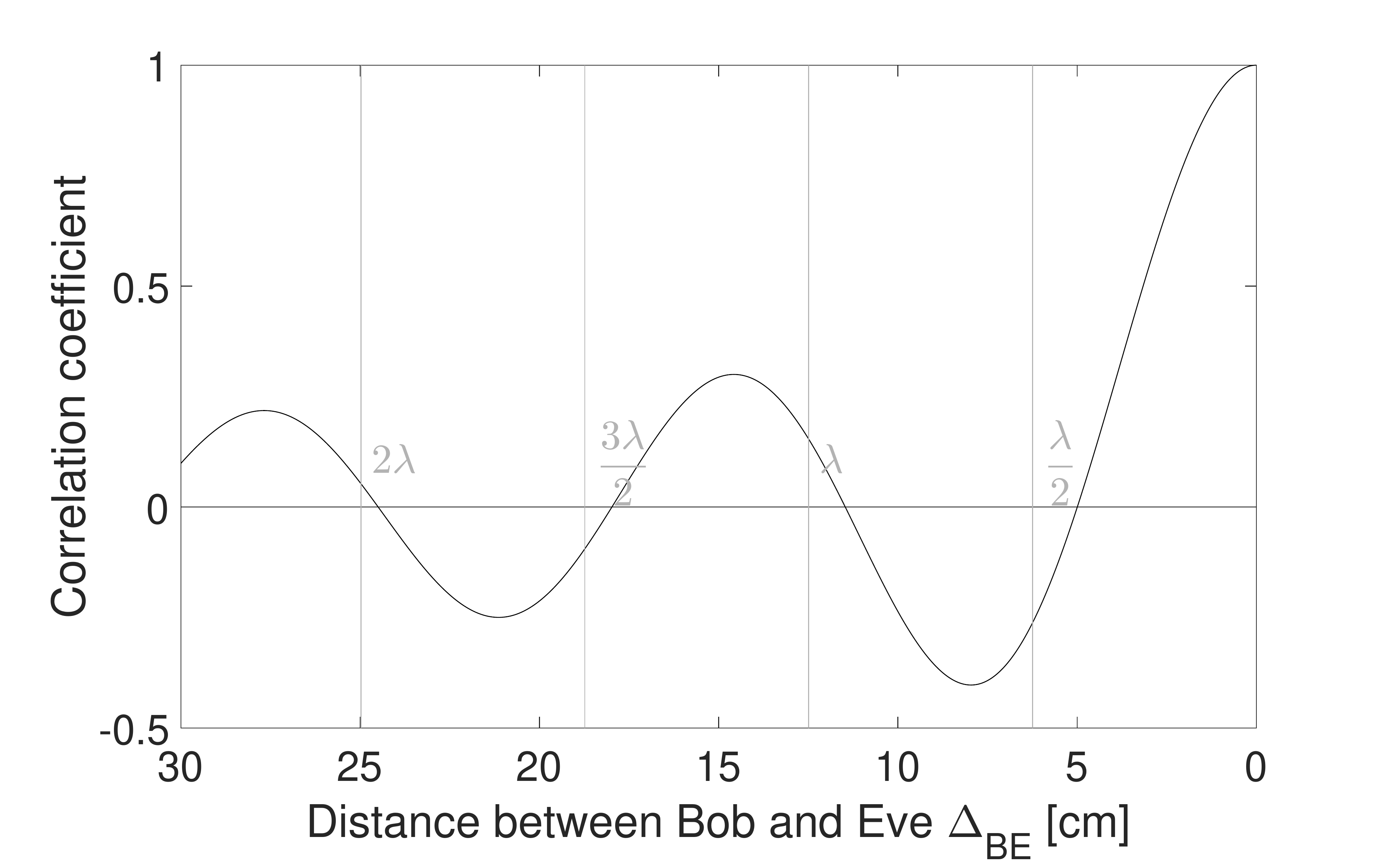}
  		\caption{Bessel function versus distance.}
        \label{fig:bessel}
\end{figure}

By analyzing the attacker's opportunity, we observe a wavelength dependent behavior of the correlation between $z_k$ and $x_k$ (or $y_k$), as illustrated in Subfigures a). The following findings hold for all three processes: $\mybold{v}_k$, $\mybold{v}^{\text{ds}}_k$, $\mybold{v}^{\text{de}}_k$. The correlation vs. distance function $\rho_{XZ}$ (and $\rho_{YZ}$) looks similar to the channel diversity function known from Jake's model~\cite{src:goldsmith2005wireless}, which is a zero-order Bessel function\footnote{A zero-order Bessel function is expected for the cross-correlation behavior of two receivers if uniformly distributed scatterers are given. According to Jake's model the first zero correlation is given after $\approx 0.4\lambda$, where $\lambda$ is the wavelength of the carrier~\cite{src:goldsmith2005wireless,DBLP:books/daglib/0025266}.} (cf. Figure~\ref{fig:bessel}). However, the highest correlation is not at distance $\Delta_{BE} =0$, where the correlation is only $0.2$. The highest cross-correlation is given at a distance of $\Delta_{BE} \approx 12.5$~cm, which is the wavelength of the $2.4$~GHz carrier. The first correlation of zero is given at a distance of $4$~cm. 

Note that the cross-correlation behavior of $x_k$ to $y_k$ is not independent of Eve's antenna position. Figure~\ref{fig:original}(b) illustrates the correlation behavior in detail. The correlation has an "oscillating" behavior with a wavelength of approximately $11$~cm, whereby at a distance of $5$~cm the curve decreases rapidly to the lowest level of $\approx 0.98$. The reason for that might be the non-perfect uniformly distributed scatterers in the environment, which are the basis of Jake's model. 
The oscillating behavior in Alice's and Bob's original observation is also given in the downsampled and decorrelated versions, cf. Figure~\ref{fig:DS}(b) and Figure~\ref{fig:decorr}(b).
This behavior is contradictory to theoretical approaches based on Jake's Doppler spectrum~\cite{DBLP:books/daglib/0025266}. The reason might be because the narrow band fading models do not include coupling and near field effects between both antennas for the spatial evaluation of autocorrelation, cross-correlation, and power spectral density (cf. \cite[Chapter 3.2]{src:goldsmith2005wireless}).

The boundary $B$ between the near field zone and the far field zone can usually be determined by the following relationship: $B\geq \frac{2D^2}{\lambda}$, where $D$ is the largest antenna size~\cite{DBLP:journals/comnet/DlugoszT10}. We estimated the size of our antenna to be $6$~cm. Therefore, the boundary is $\approx 5.7$~cm. 
Analyzing near field boundaries in detail is left for future work. 



Compared to the cross-correlation behavior between the i.i.d. samples   $x^{\text{ds}}_k$ and $y^{\text{ds}}_k$ (after downsampling), both mutual information $I(X;Y)$ and $R_{sk}$ have very similar oscillating behavior, shown in Subfigures c). The (minimum, maximum) values of the correlation are ($0.980$, $0.995$) and the ones of the mutual information are ($2.1$, $2.75$). 
By analyzing Eve's observation, we see only a slight similarity between the mutual information $I(X;Z)$ (and $I(Y;Z)$) to the correlation behavior of her observation $\rho_{XZ}$ (and $\rho_{YZ}$). The similarity can be found by comparing the maximum absolute values. For instance, the highest correlation occurs at $10$~cm with a value of $0.5$, and corresponds to the highest mutual information of $0.5$~bits per sample. 
%
%
However, the Bessel-like behavior is not evident. Notably is the fact that the attackers observation $z_k$ does not significantly impact $R_{sk}$. Our results show that $R_{sk}$ is mainly dependent on $x_k$ and $y_k$. 
However, Eve's antenna affects Alice's and Bob's observation and, therefore, affects $R_{sk}$. Table~\ref{tab:results} summarizes our results.

\begin{table}
\caption{Averaged results of our experiment.}
\begin{center}
\begin{tabular}{|l | c | c | c | }
    \hline
    & $\mybold{v}_k$ &  $\mybold{v}^{\text{ds}}_k$ &
$\mybold{v}^{\text{de}}_k$ \\
    \hline\hline
    $\rho_{x_k,y_k}$ & $\approx 0.99$  & $\approx 0.99$  & $0.94$  \\\hline
    $\rho_{y_k,z_k}$ & $\approx 0.09$  & $\approx 0.09$  & $ \approx
0.07$  \\\hline
    $I(X;Y)$         & $\approx 2.92$  & $\approx 2.89$  & $\approx 
2.31$  \\\hline
    $I(Y;Z)$         & $\approx 0.26$  & $\approx 0.27$  & $0.10$  \\\hline
    $\rsk$ & $\approx 2.67$  & $\approx 2.63$  & $\approx 2.22$  \\\hline

\end{tabular}
\end{center}

\label{tab:results}
\end{table}





\section{Conclusion}
\label{sec:conclusion}
In this work, we have provided an important pillar to bridge the gap between theory and practice-oriented approaches for CRKG. Our experimental study helps to provide a better understanding of channel statistics in wireless environments for security applications. 
We present reproducible results based on a relevant environment which justifies the joint stationarity of a random process.  
We show results of cross-correlation, mutual information and secret-key rates, which are dependent on attacker's (or third device's) position. 
As a result, we discovered that the \textit{observer effect} occurs, which most probably originates from near field distortions. 
%
%
We believe the effect needs to be considered in the future. Common channel models like Jake's model for channel diversity need to be extended in order to be valid for key generation setups. Furthermore, it might be pertinent, for instance, to detect the proximity of Eve. Basing on our results two bidirectionally communicating nodes might recognize a third device, its relative position, and its motion in the proximity. Further studies might use complex-valued channel profiles to analyze  third party positioning based and motion based influences.


%% file: input/appendix.tex
\begin{appendices}
\section{Full Measurement}
\begin{figure*}
		\centering
\includegraphics[width=0.75\textwidth]{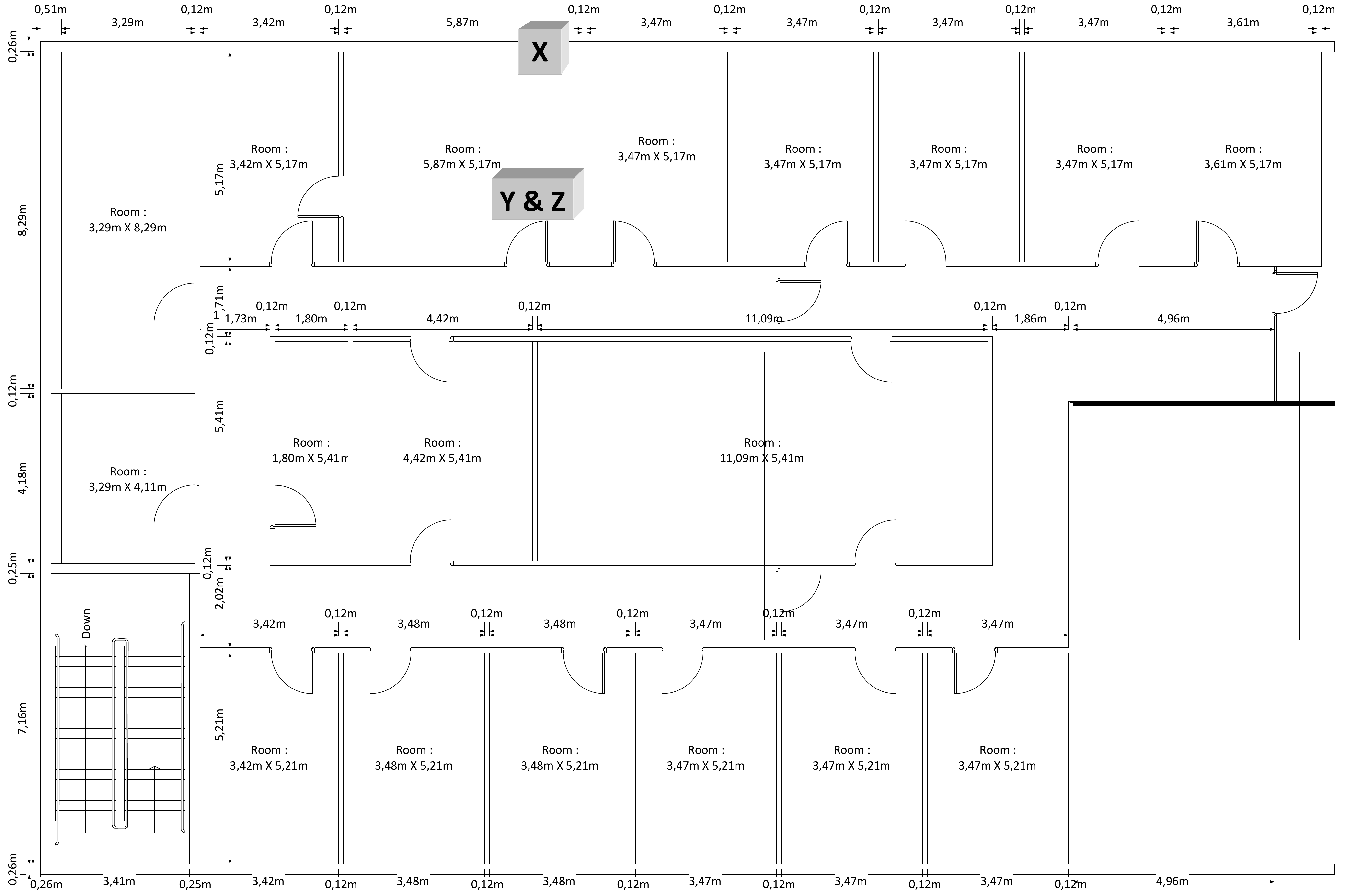}
  		\caption{The testbed includes several experimental setups for performance evaluation as well as for security analysis. 	Alice (X), Bob (Y) and Eve (Z) are mounted on a automated antenna positioning system.}
        \label{fig:setup_ur}
\end{figure*}

\begin{figure*}
	\centering
	\subfloat[]{\includegraphics[trim=1.4cm 0.1cm 3.5cm 1.6cm, clip=true, height=0.224\textwidth]{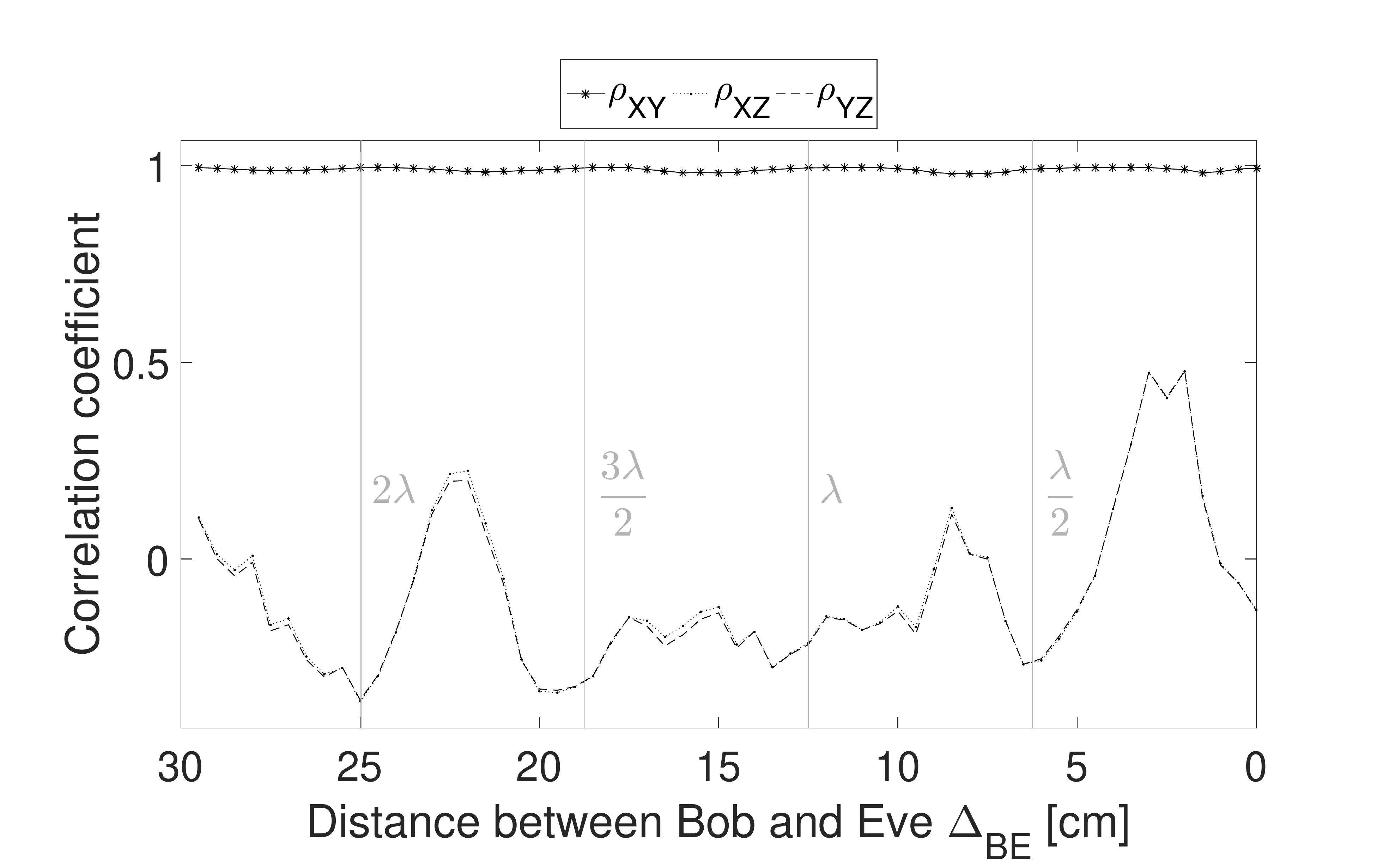}}
	\subfloat[]{\includegraphics[trim=0.5cm 0.1cm 3.5cm 1.6cm, clip=true, height=0.224\textwidth]{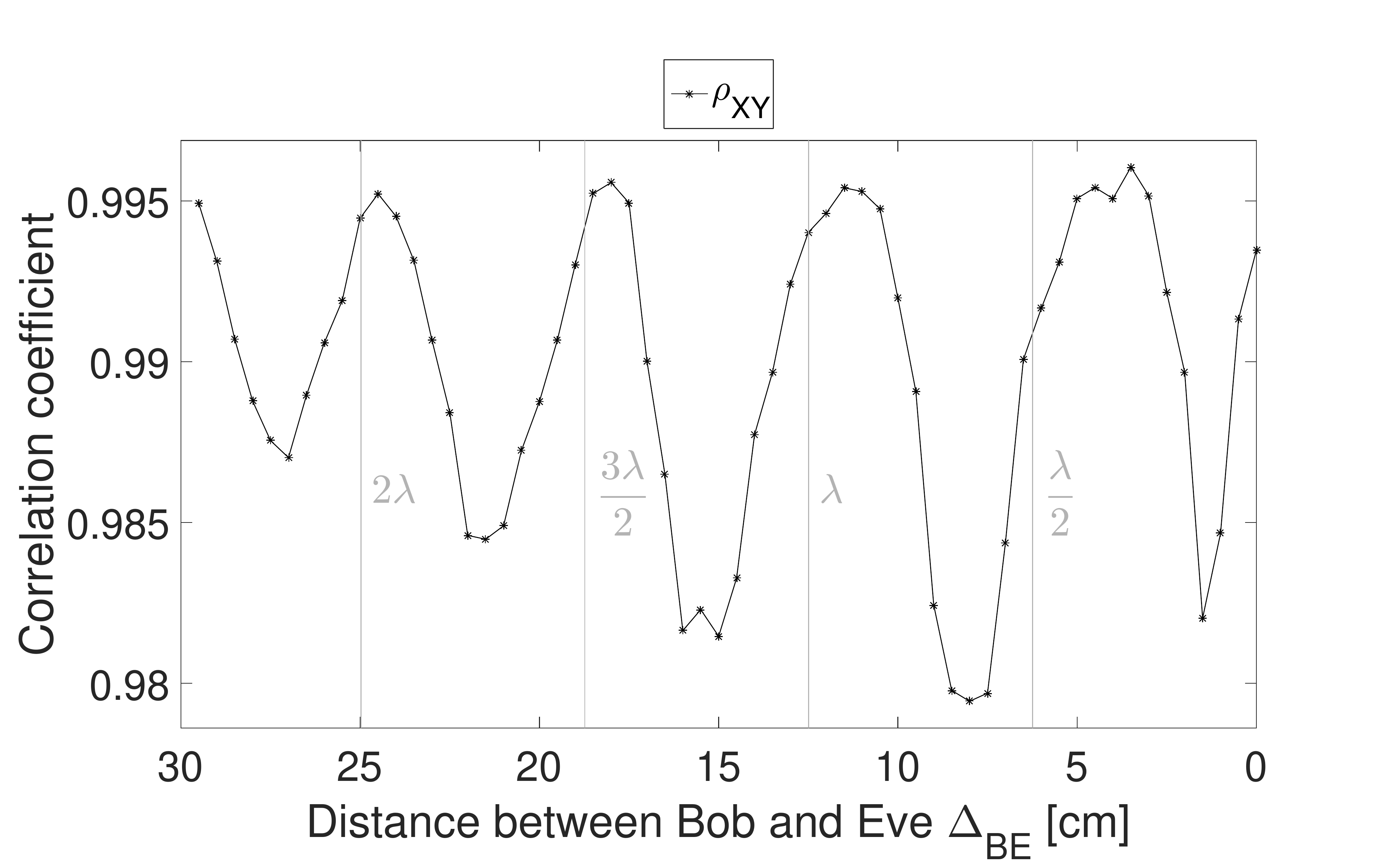}}
	\subfloat[]{\includegraphics[trim=2.2cm 0.1cm 3.5cm 1.6cm, clip=true, height=0.224\textwidth]{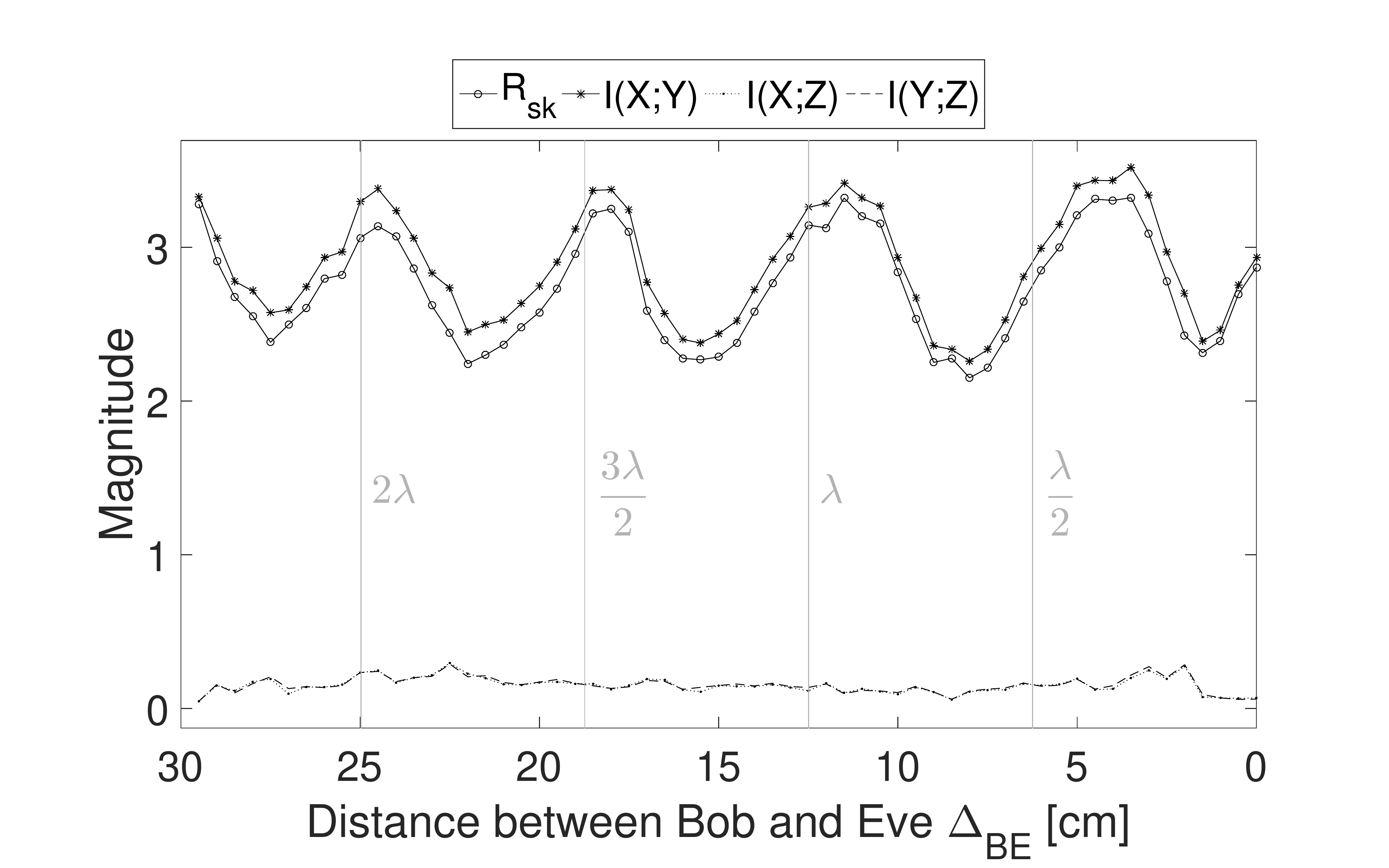}}
	\caption{Evaluation results of $\mybold{v}_k$. In (a) and (b) the cross-correlations is given; in (c) the mutual information as well as $\rsk$ is given. Position 0.}
	\label{fig:app_original_0}
\end{figure*}

\begin{figure*}
	\centering
	\subfloat[]{\includegraphics[trim=1.4cm 0.1cm 3.5cm 1.6cm, clip=true, height=0.224\textwidth]{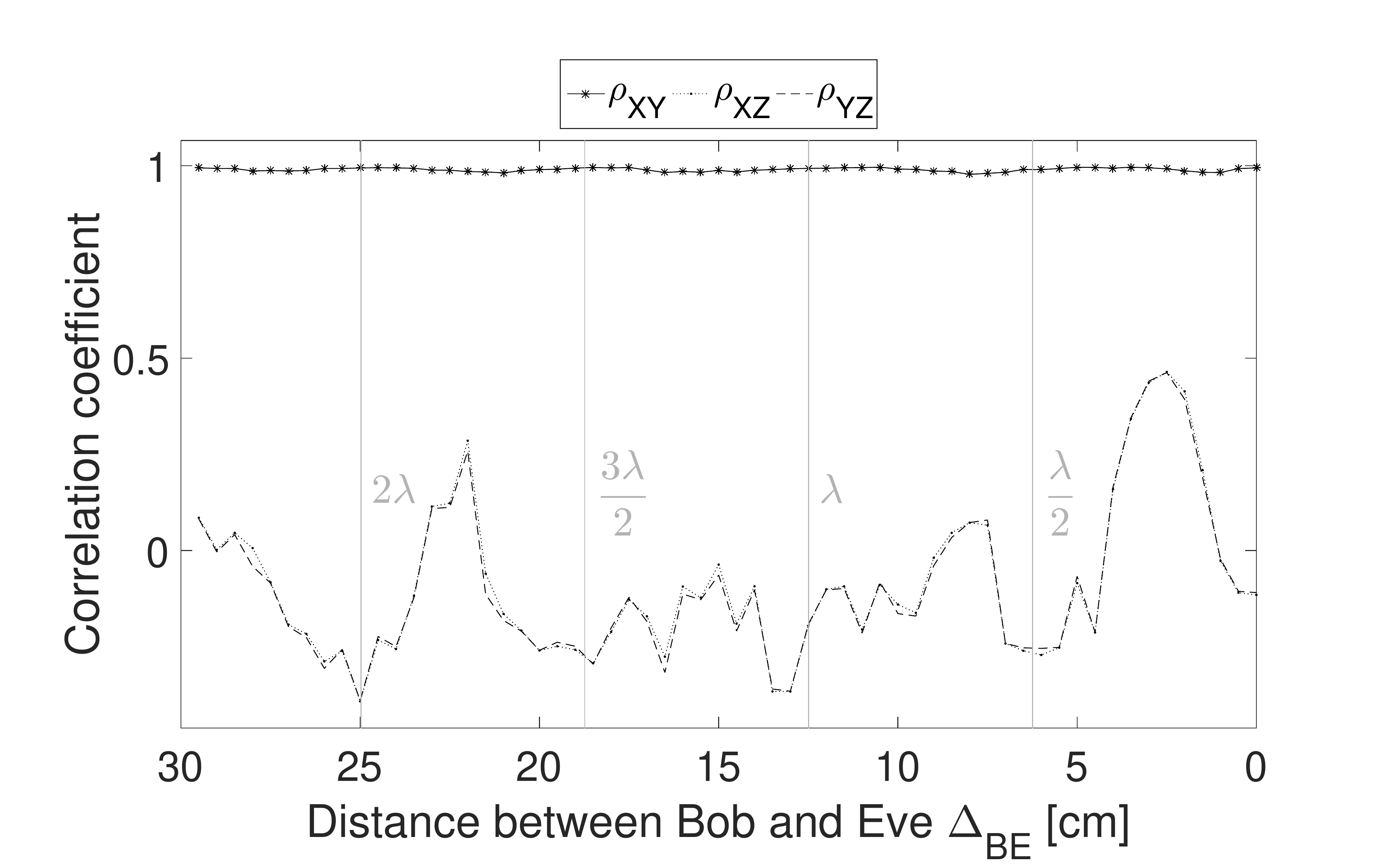}}
	\subfloat[]{\includegraphics[trim=0.5cm 0.1cm 3.5cm 1.6cm, clip=true, height=0.224\textwidth]{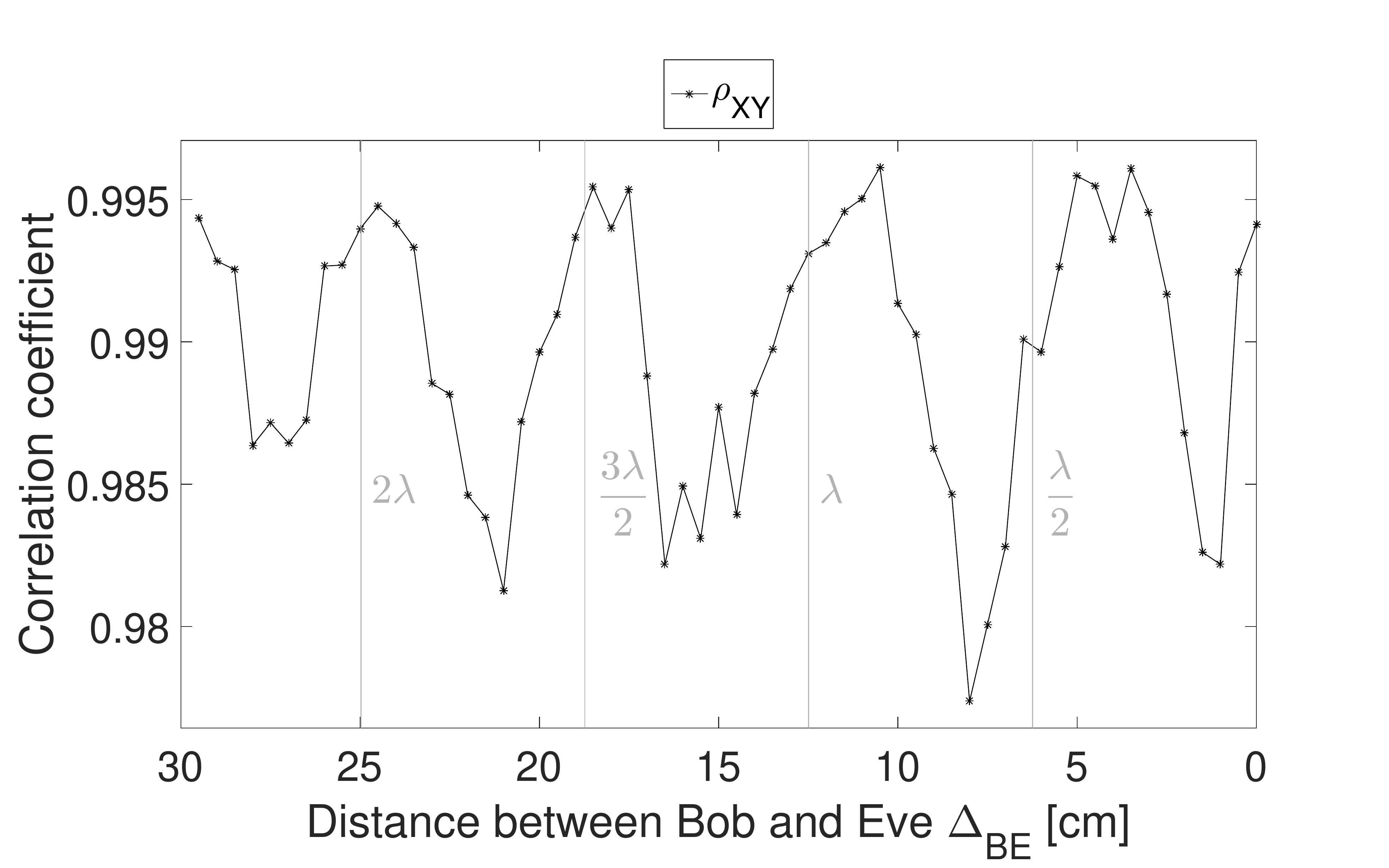}}
	\subfloat[]{\includegraphics[trim=2.2cm 0.1cm 3.5cm 1.6cm, clip=true, height=0.224\textwidth]{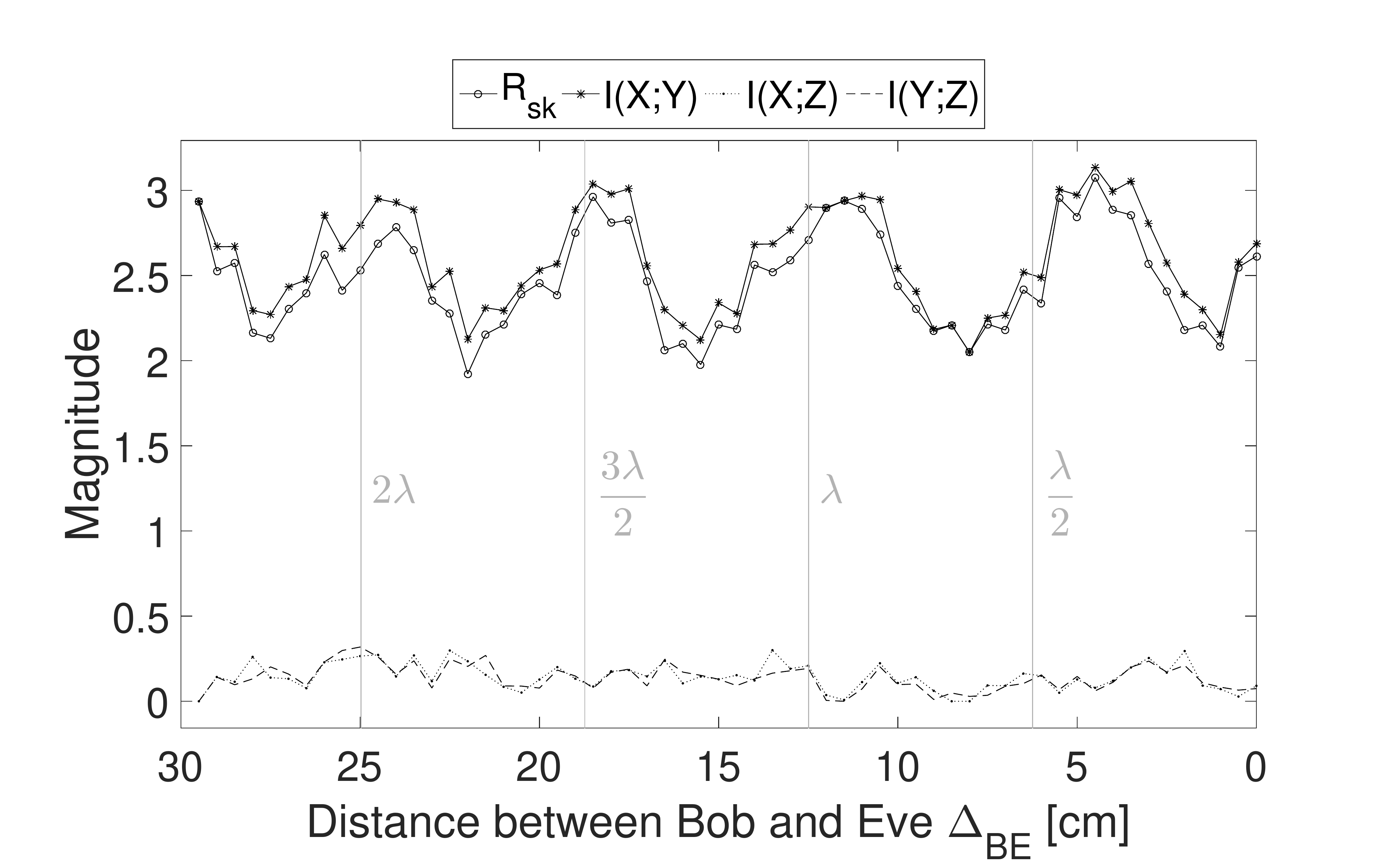}}
	\caption{Evaluation results of $\mybold{v}^{\text{ds}}_k$. In (a) and (b) the cross-correlations is given; in (c) the mutual information as well as $\rsk$ is given. Position 0.}
	\label{fig:app_ds_0}
\end{figure*}

\begin{figure*}
	\centering
	\subfloat[]{\includegraphics[trim=1.4cm 0.1cm 3.5cm 1.6cm, clip=true, height=0.224\textwidth]{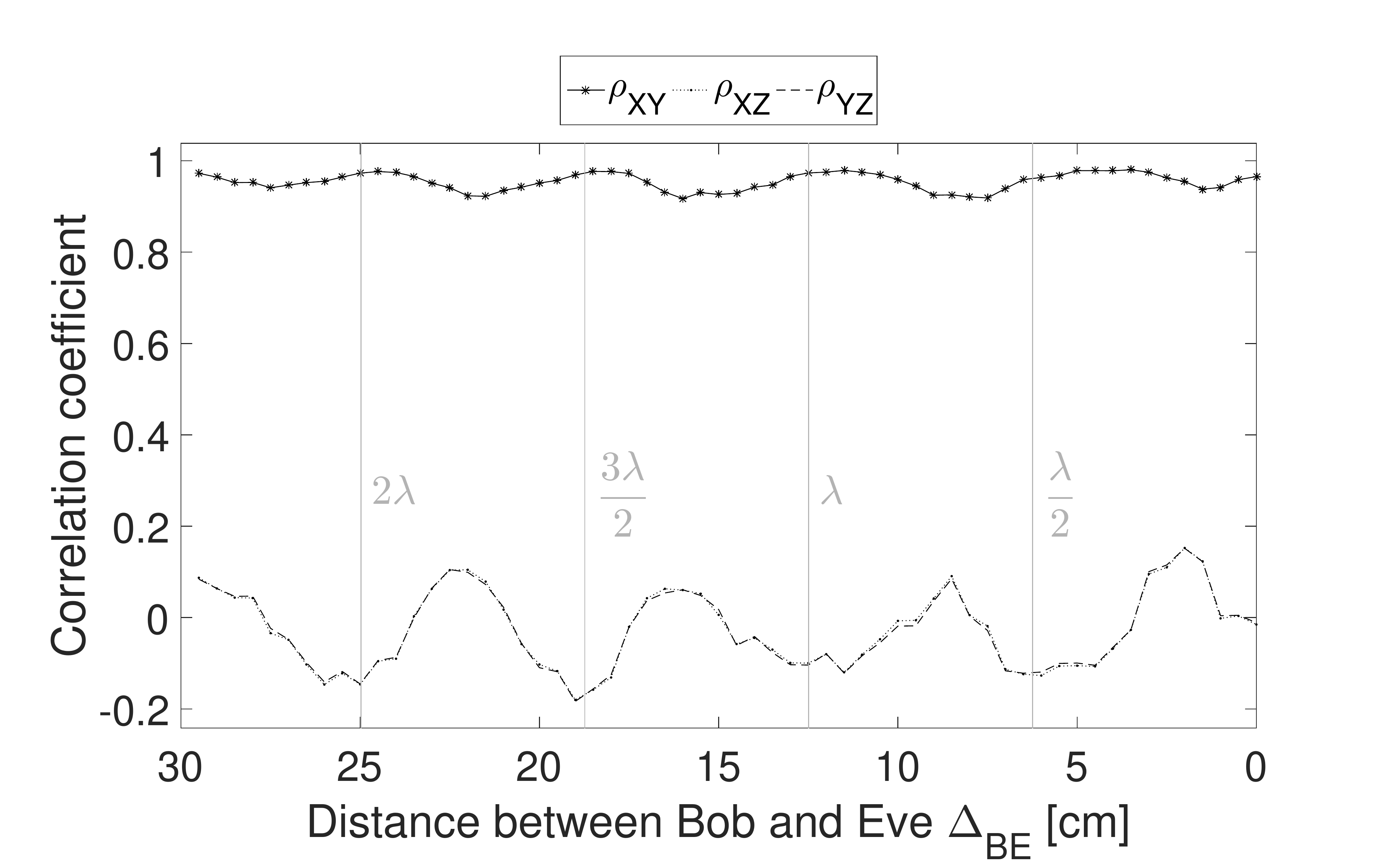}}
	\subfloat[]{\includegraphics[trim=1cm 0.1cm 3.5cm 1.6cm, clip=true, height=0.224\textwidth]{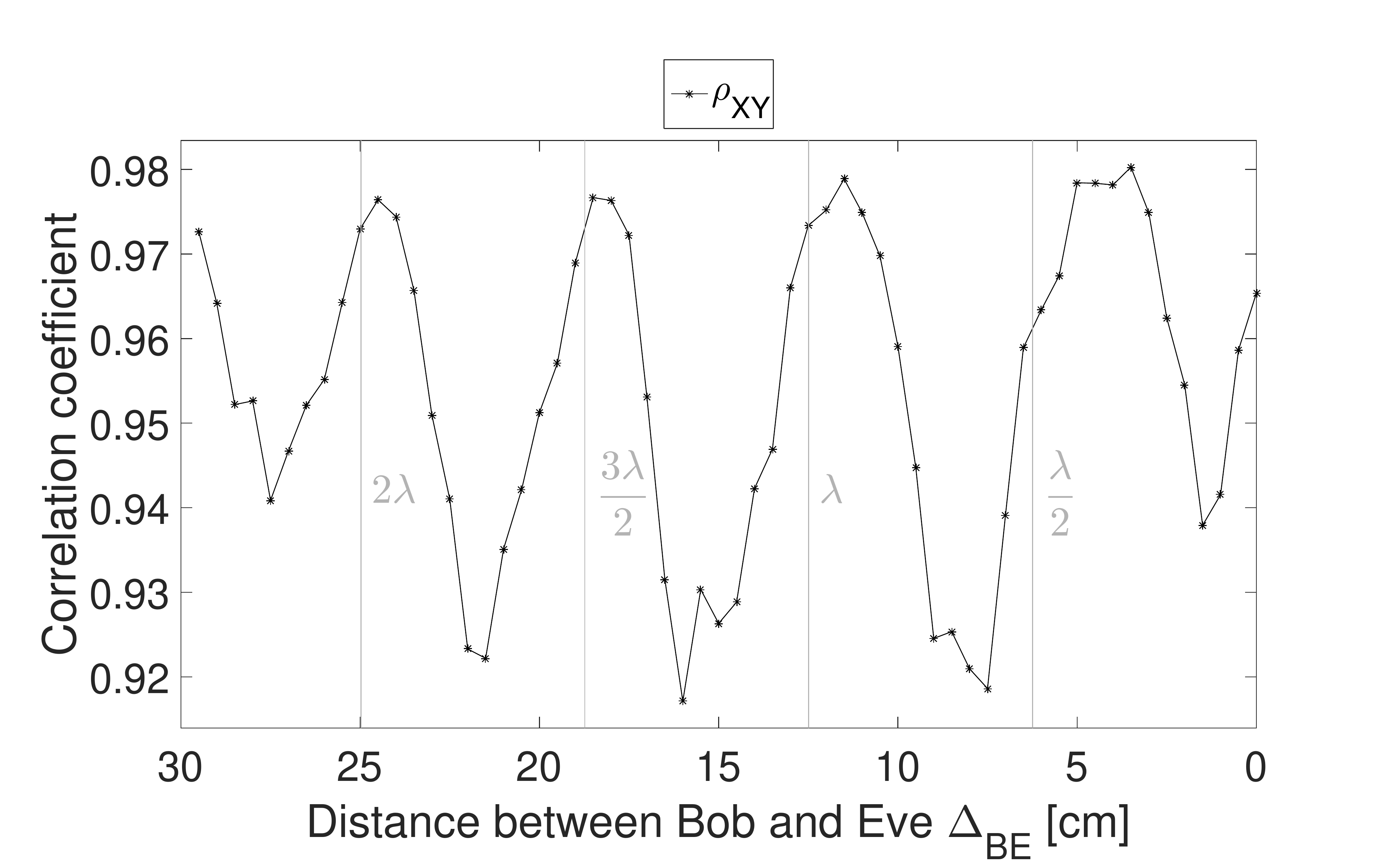}}
	\subfloat[]{\includegraphics[trim=1.8cm 0.1cm 3.5cm 1.6cm, clip=true, height=0.224\textwidth]{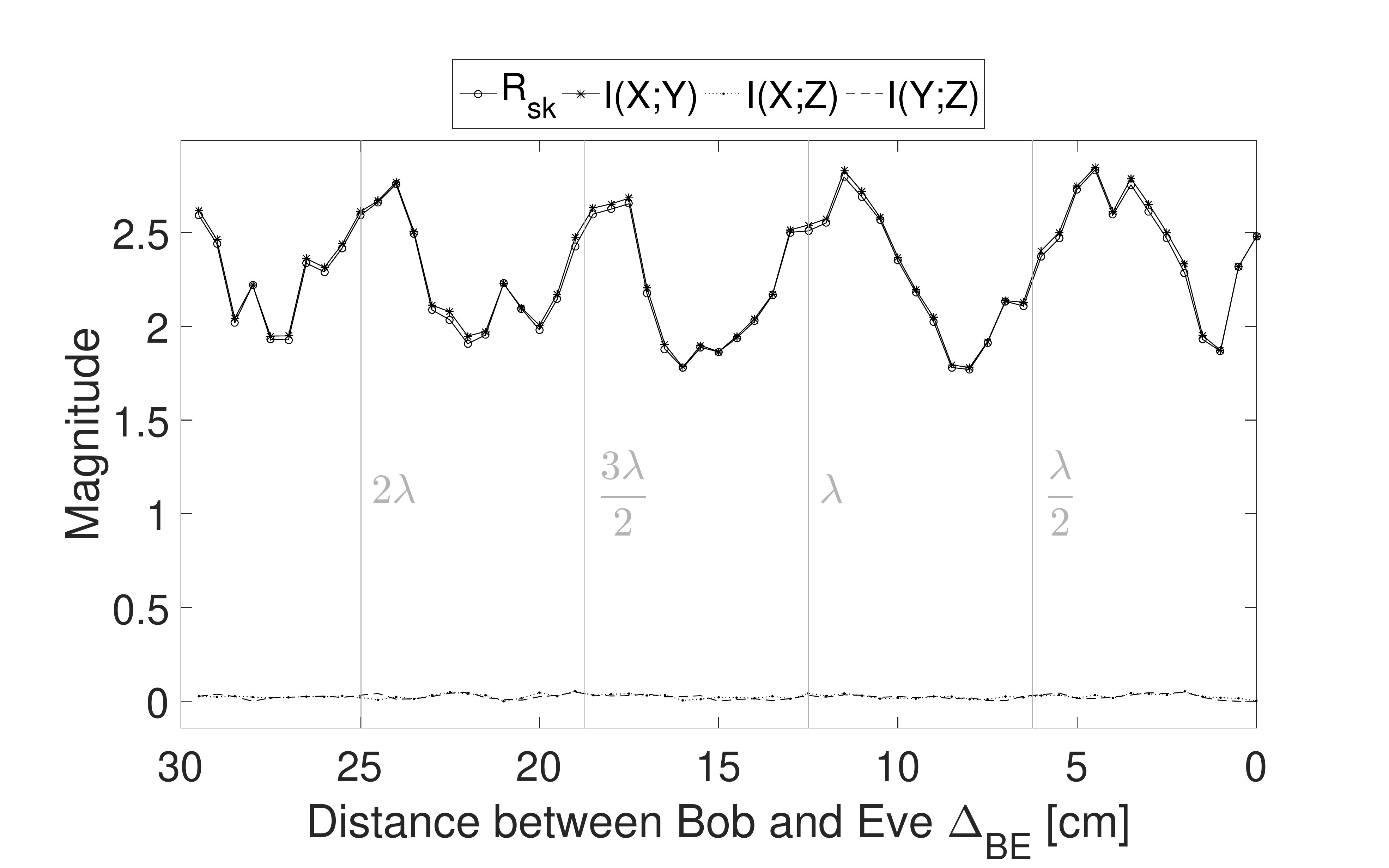}}
	\caption{Evaluation results of $\mybold{v}^{\text{de}}_k$. In (a) and (b) the cross-correlations is given; in (c) the mutual information as well as $\rsk$ is given. Position 0.}
	\label{fig:app_decorr_0}
\end{figure*}


\begin{figure*}
	\centering
	\subfloat[]{\includegraphics[trim=1.4cm 0.1cm 3.5cm 1.6cm, clip=true, height=0.224\textwidth]{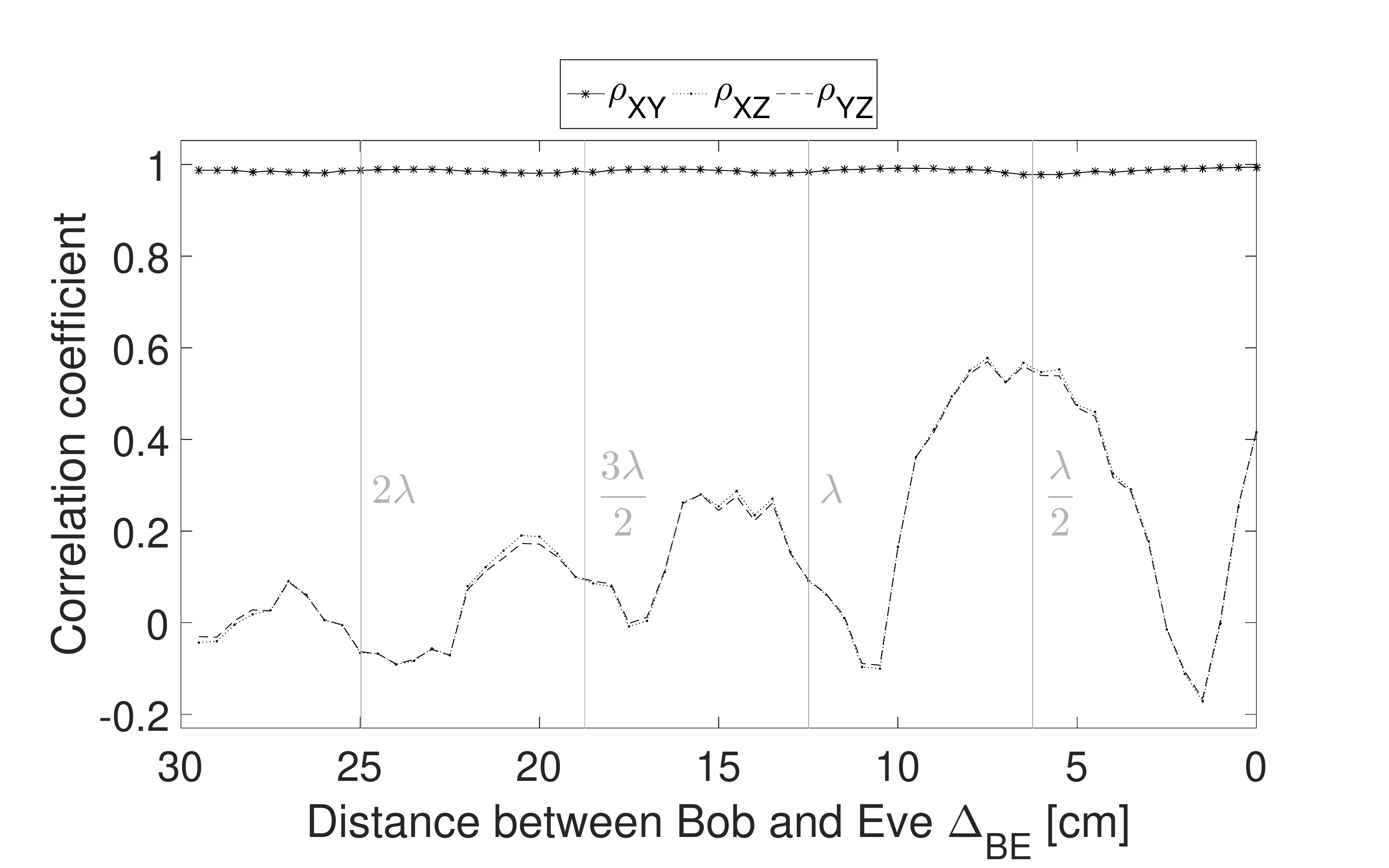}}
	\subfloat[]{\includegraphics[trim=0.5cm 0.1cm 3.5cm 1.6cm, clip=true, height=0.224\textwidth]{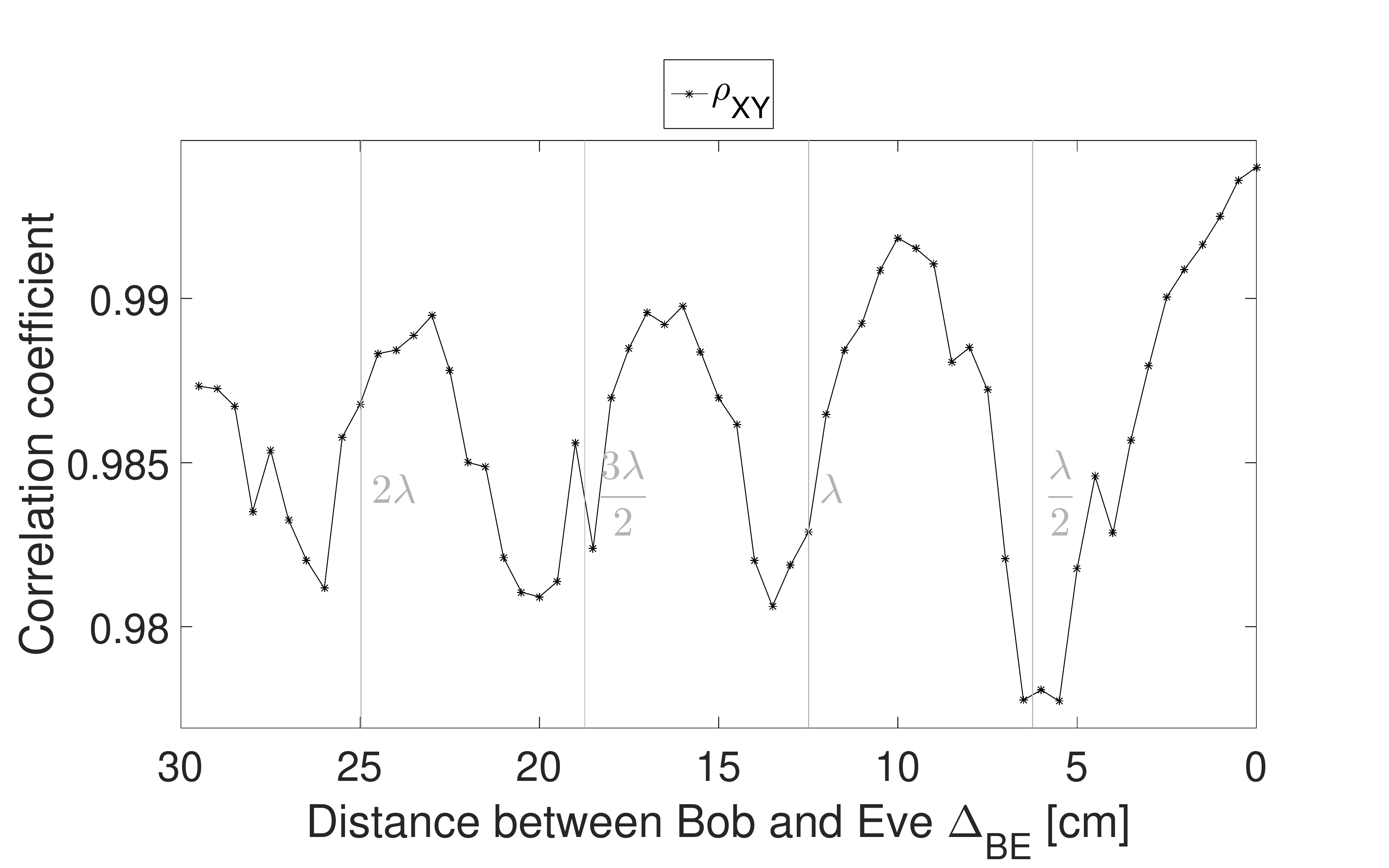}}
	\subfloat[]{\includegraphics[trim=2.2cm 0.1cm 3.5cm 1.6cm, clip=true, height=0.224\textwidth]{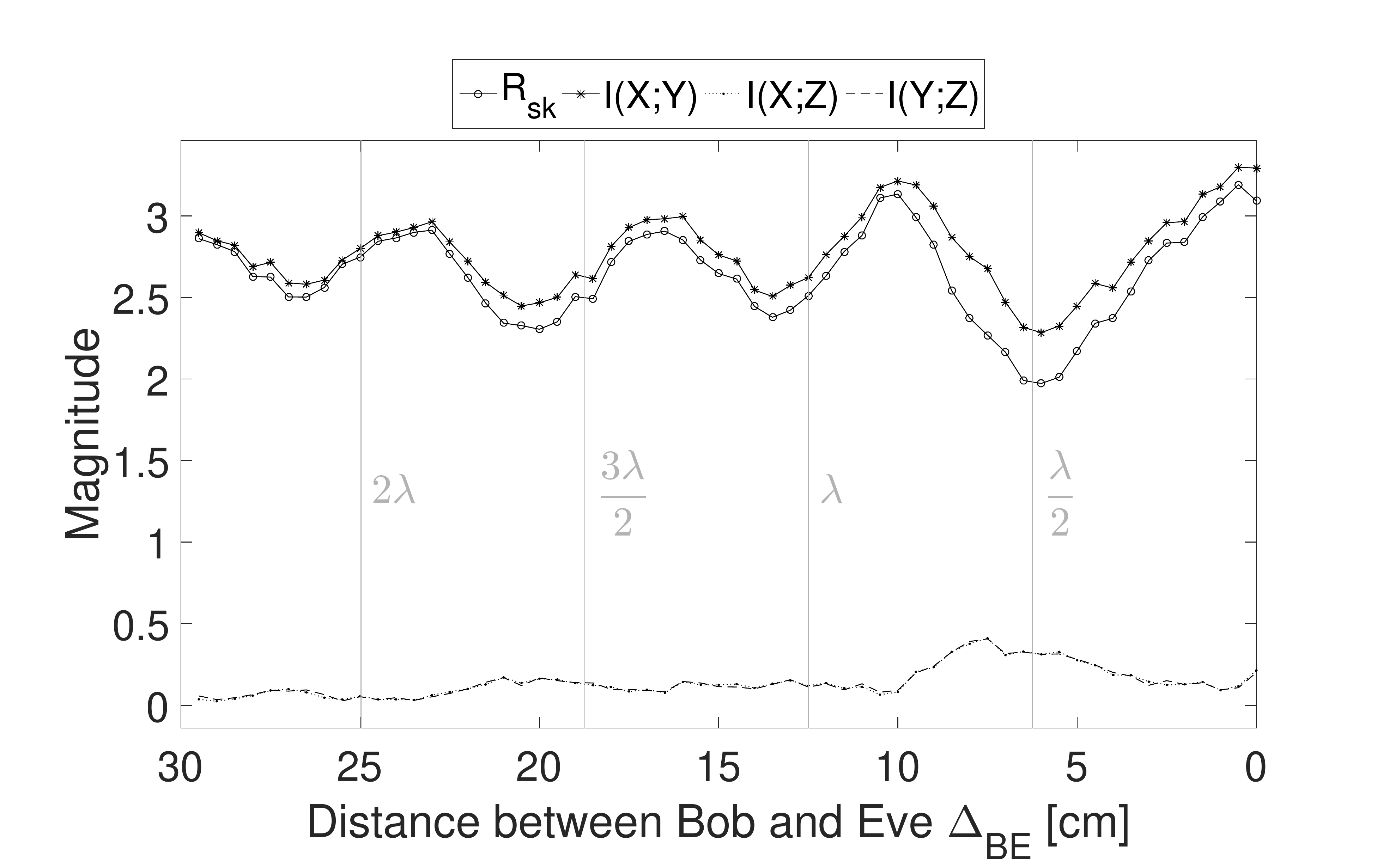}}
	\caption{Evaluation results of $\mybold{v}_k$. In (a) and (b) the cross-correlations is given; in (c) the mutual information as well as $\rsk$ is given. Position 1.}
	\label{fig:app_original_1}
\end{figure*}

\begin{figure*}
	\centering
	\subfloat[]{\includegraphics[trim=1.4cm 0.1cm 3.5cm 1.6cm, clip=true, height=0.224\textwidth]{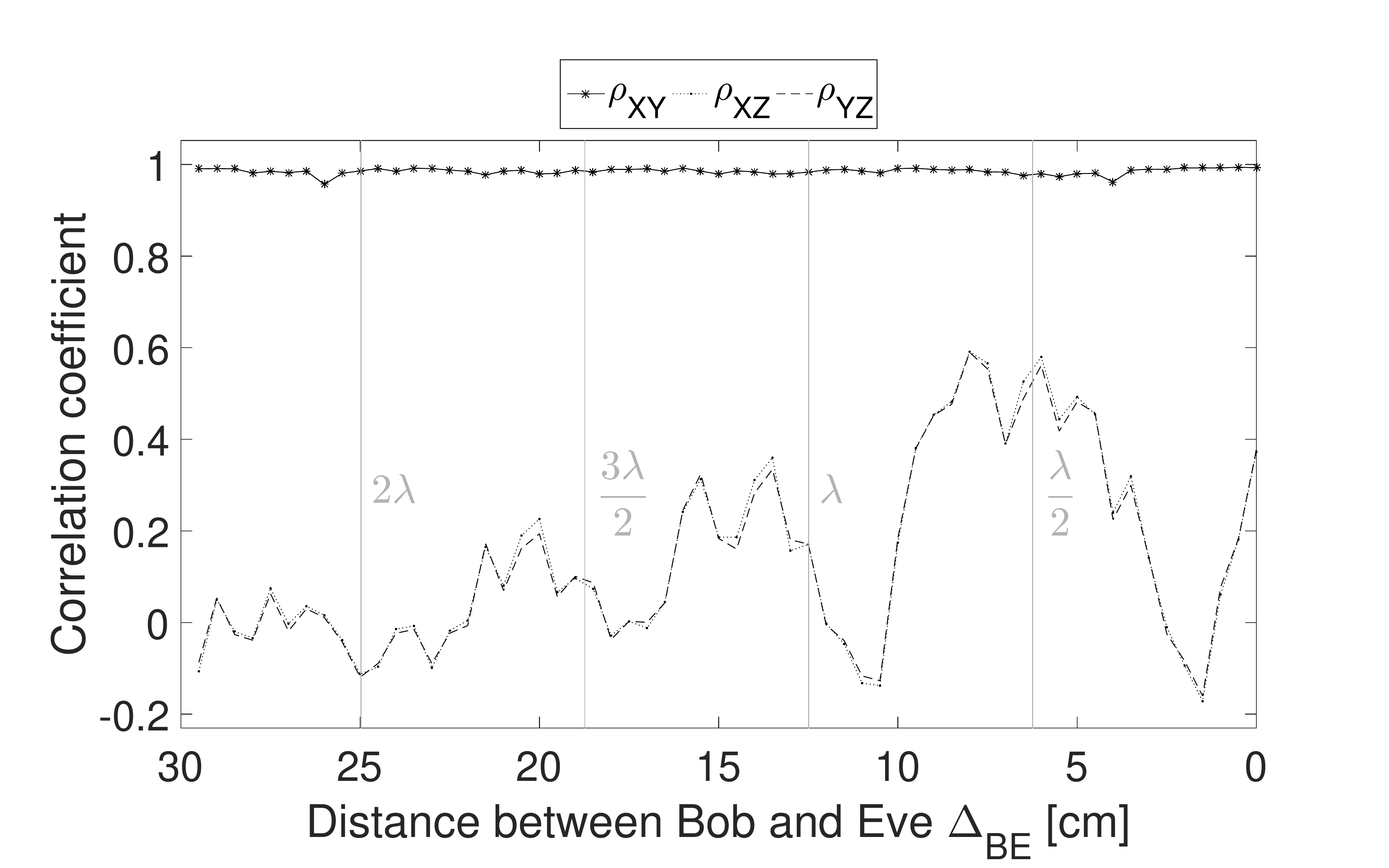}}
	\subfloat[]{\includegraphics[trim=0.5cm 0.1cm 3.5cm 1.6cm, clip=true, height=0.224\textwidth]{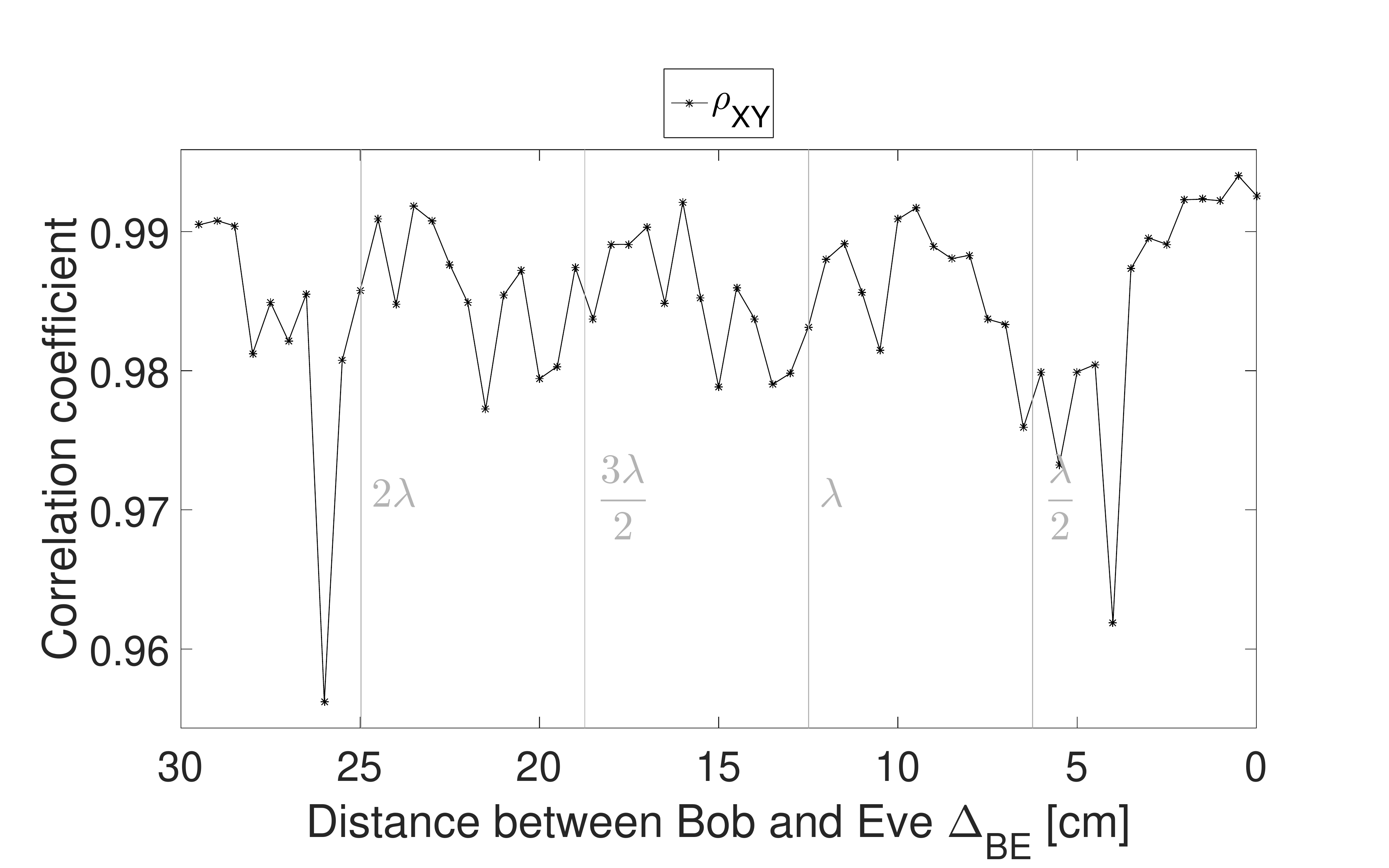}}
	\subfloat[]{\includegraphics[trim=2.2cm 0.1cm 3.5cm 1.6cm, clip=true, height=0.224\textwidth]{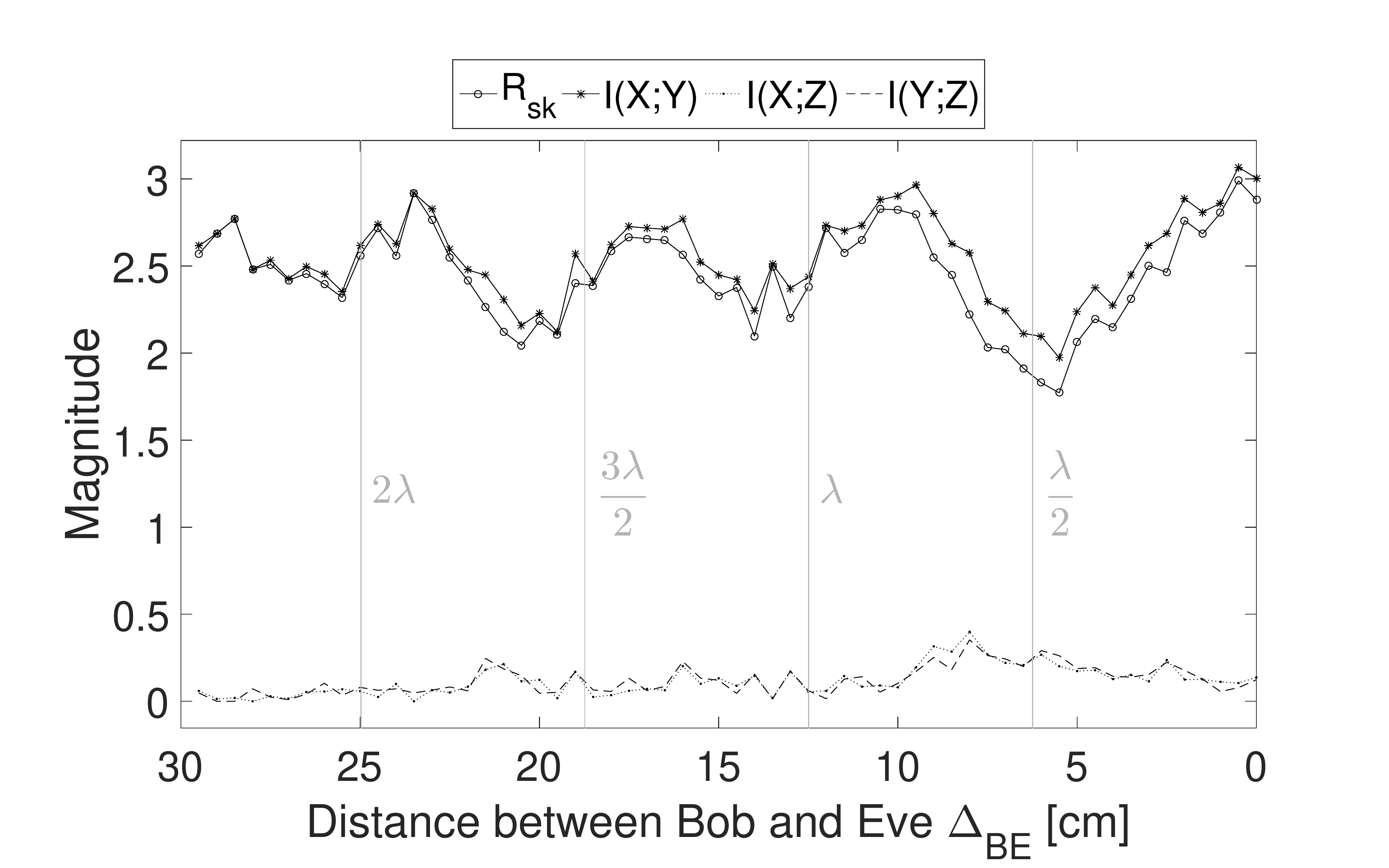}}
	\caption{Evaluation results of $\mybold{v}^{\text{ds}}_k$. In (a) and (b) the cross-correlations is given; in (c) the mutual information as well as $\rsk$ is given. Position 1.}
	\label{fig:app_ds_1}
\end{figure*}

\begin{figure*}
	\centering
	\subfloat[]{\includegraphics[trim=1.4cm 0.1cm 3.5cm 1.6cm, clip=true, height=0.224\textwidth]{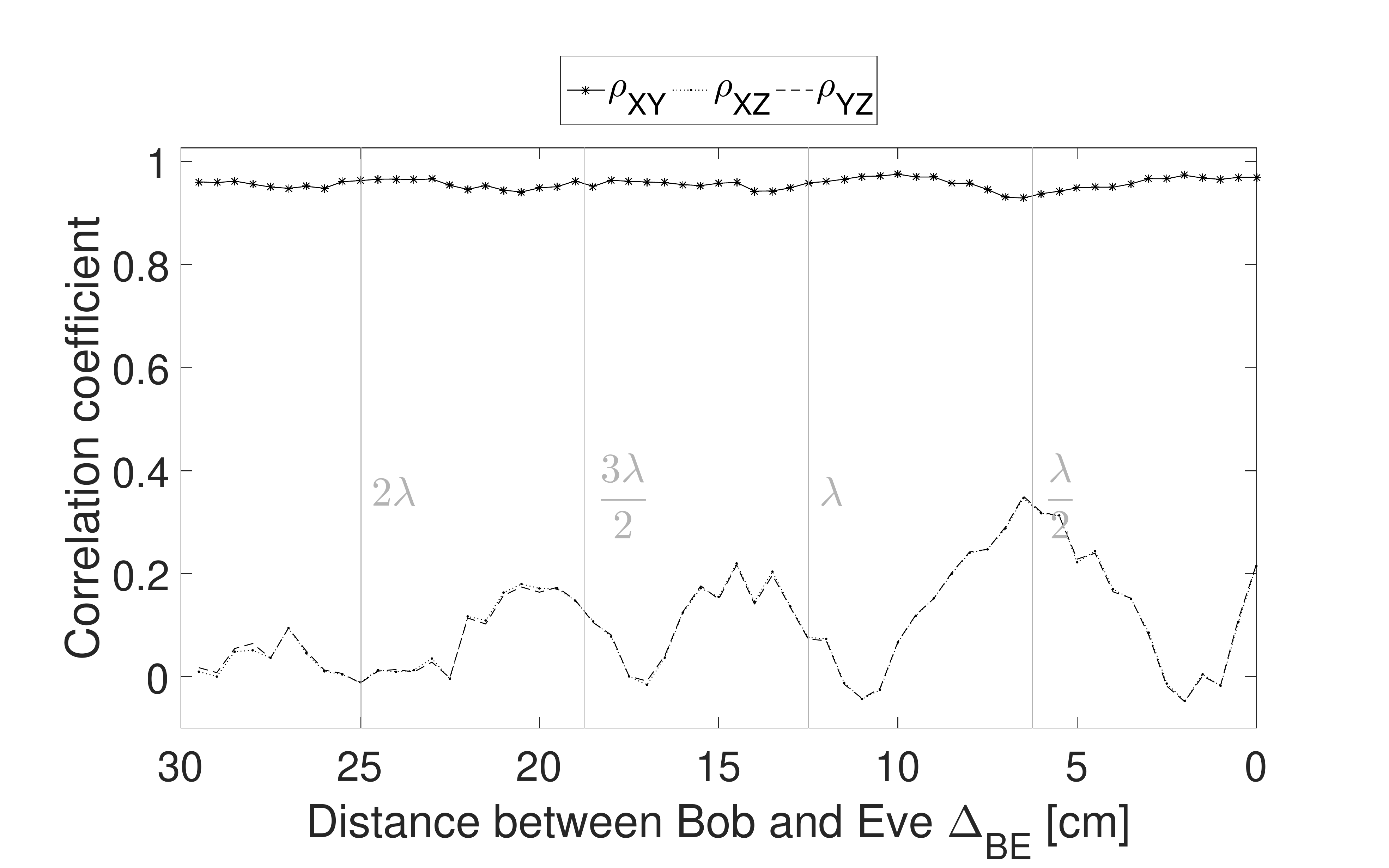}}
	\subfloat[]{\includegraphics[trim=1cm 0.1cm 3.5cm 1.6cm, clip=true, height=0.224\textwidth]{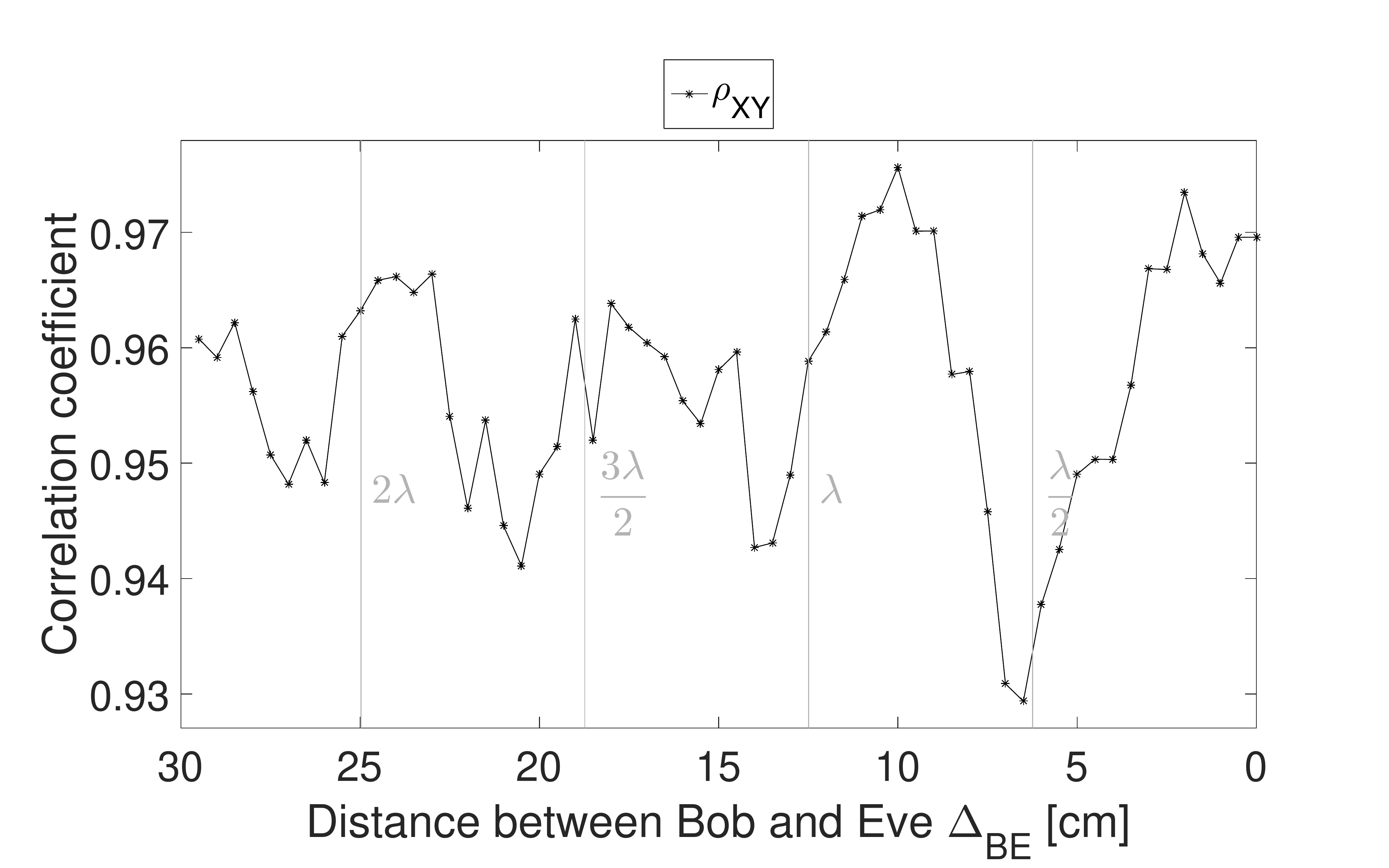}}
	\subfloat[]{\includegraphics[trim=1.8cm 0.1cm 3.5cm 1.6cm, clip=true, height=0.224\textwidth]{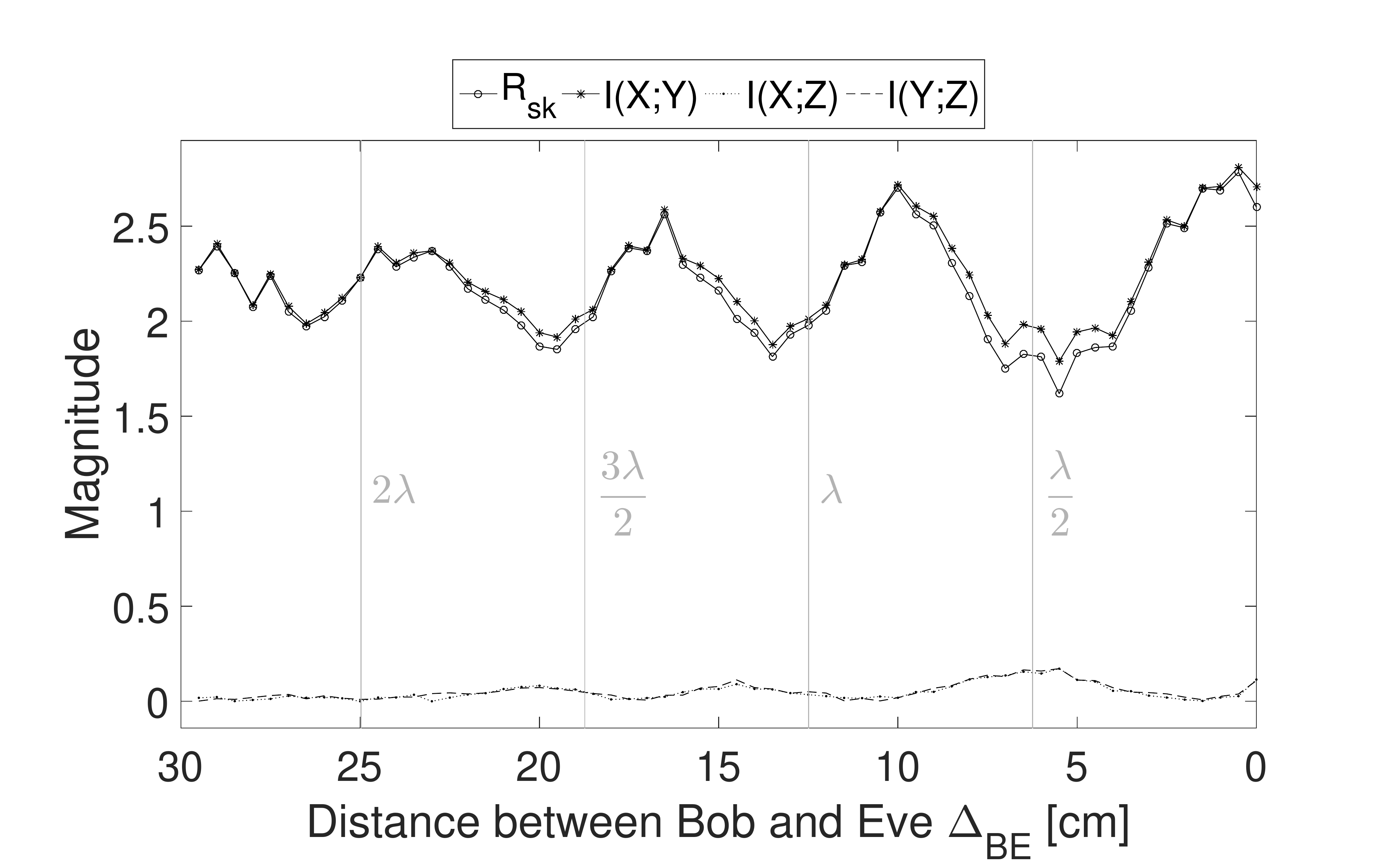}}
	\caption{Evaluation results of $\mybold{v}^{\text{de}}_k$. In (a) and (b) the cross-correlations is given; in (c) the mutual information as well as $\rsk$ is given. Position 1.}
	\label{fig:app_decorr_1}
\end{figure*}


\begin{figure*}
	\centering
	\subfloat[]{\includegraphics[trim=1.4cm 0.1cm 3.5cm 1.6cm, clip=true, height=0.224\textwidth]{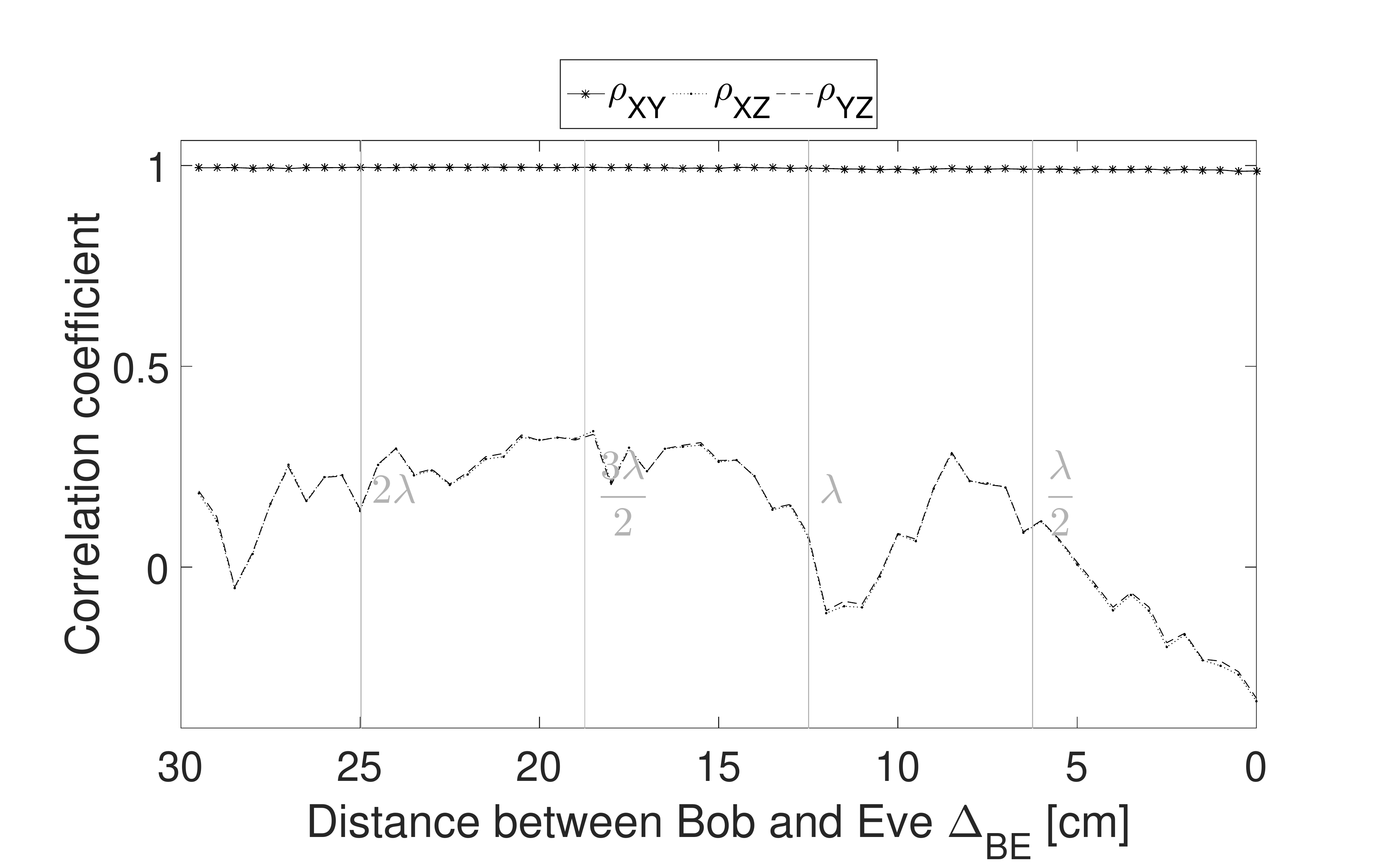}}
	\subfloat[]{\includegraphics[trim=0.5cm 0.1cm 3.5cm 1.6cm, clip=true, height=0.224\textwidth]{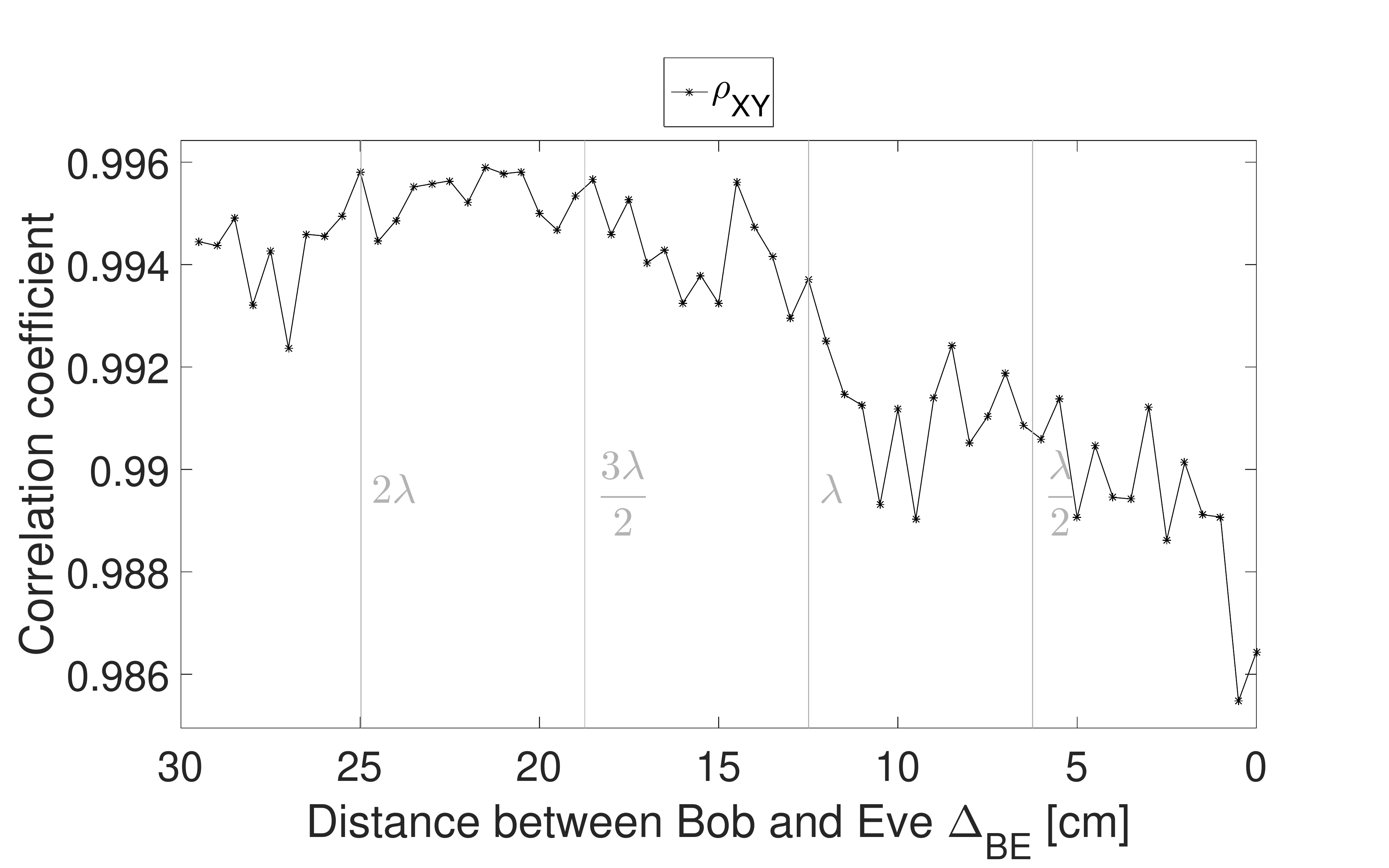}}
	\subfloat[]{\includegraphics[trim=2.2cm 0.1cm 3.5cm 1.6cm, clip=true, height=0.224\textwidth]{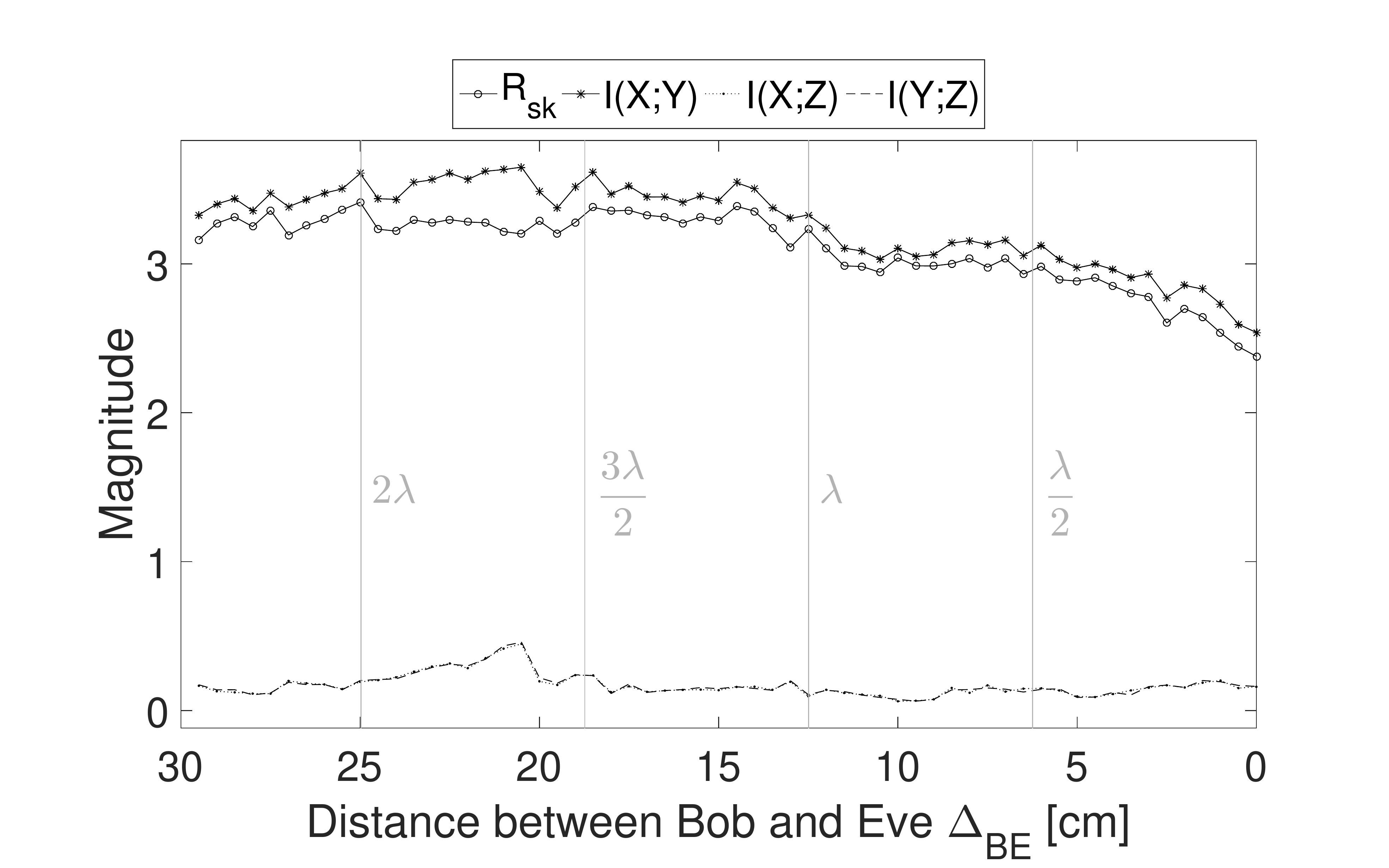}}
	\caption{Evaluation results of $\mybold{v}_k$. In (a) and (b) the cross-correlations is given; in (c) the mutual information as well as $\rsk$ is given. Position 2.}
	\label{fig:app_original_2}
\end{figure*}

\begin{figure*}
	\centering
	\subfloat[]{\includegraphics[trim=1.4cm 0.1cm 3.5cm 1.6cm, clip=true, height=0.224\textwidth]{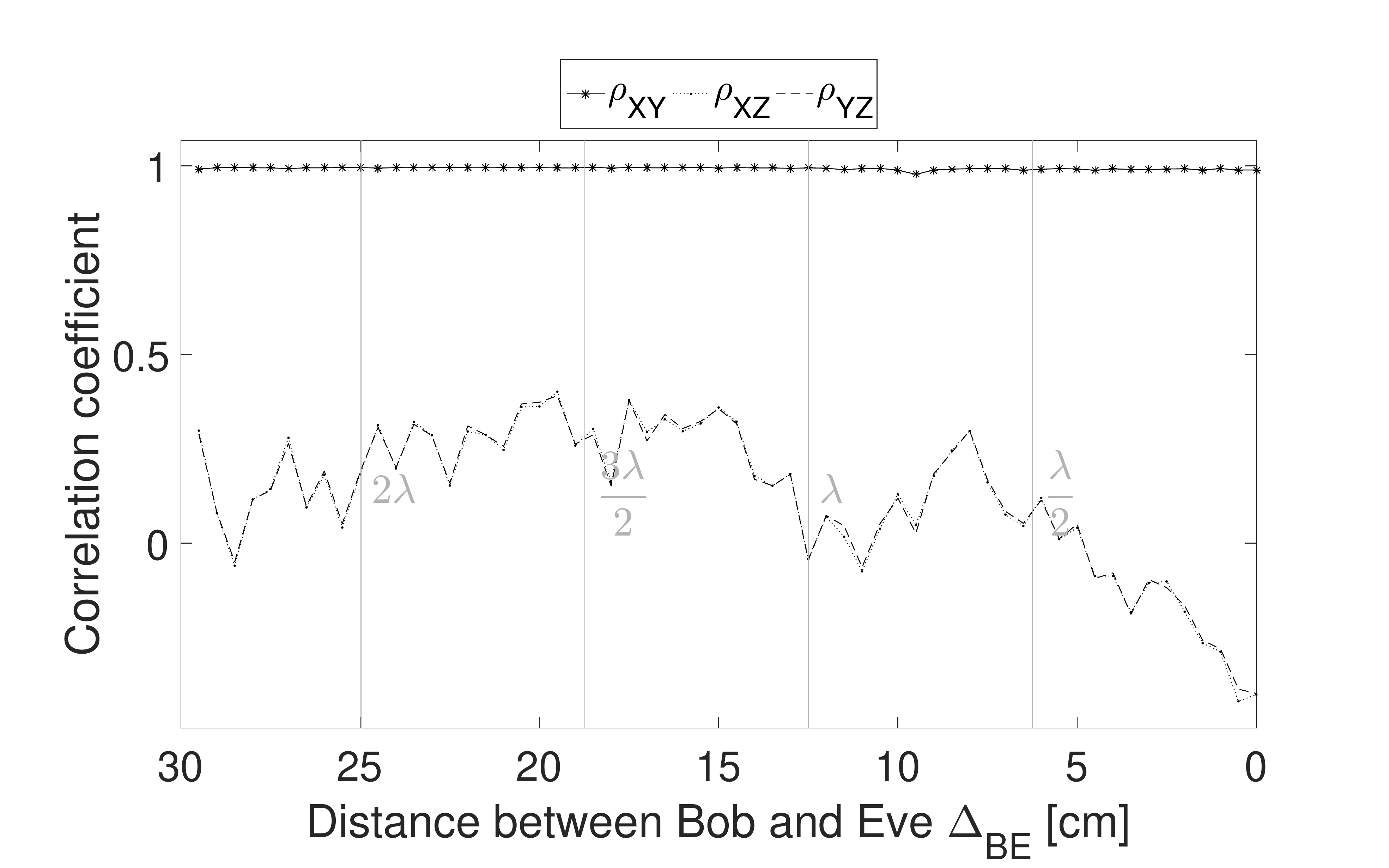}}
	\subfloat[]{\includegraphics[trim=0.5cm 0.1cm 3.5cm 1.6cm, clip=true, height=0.224\textwidth]{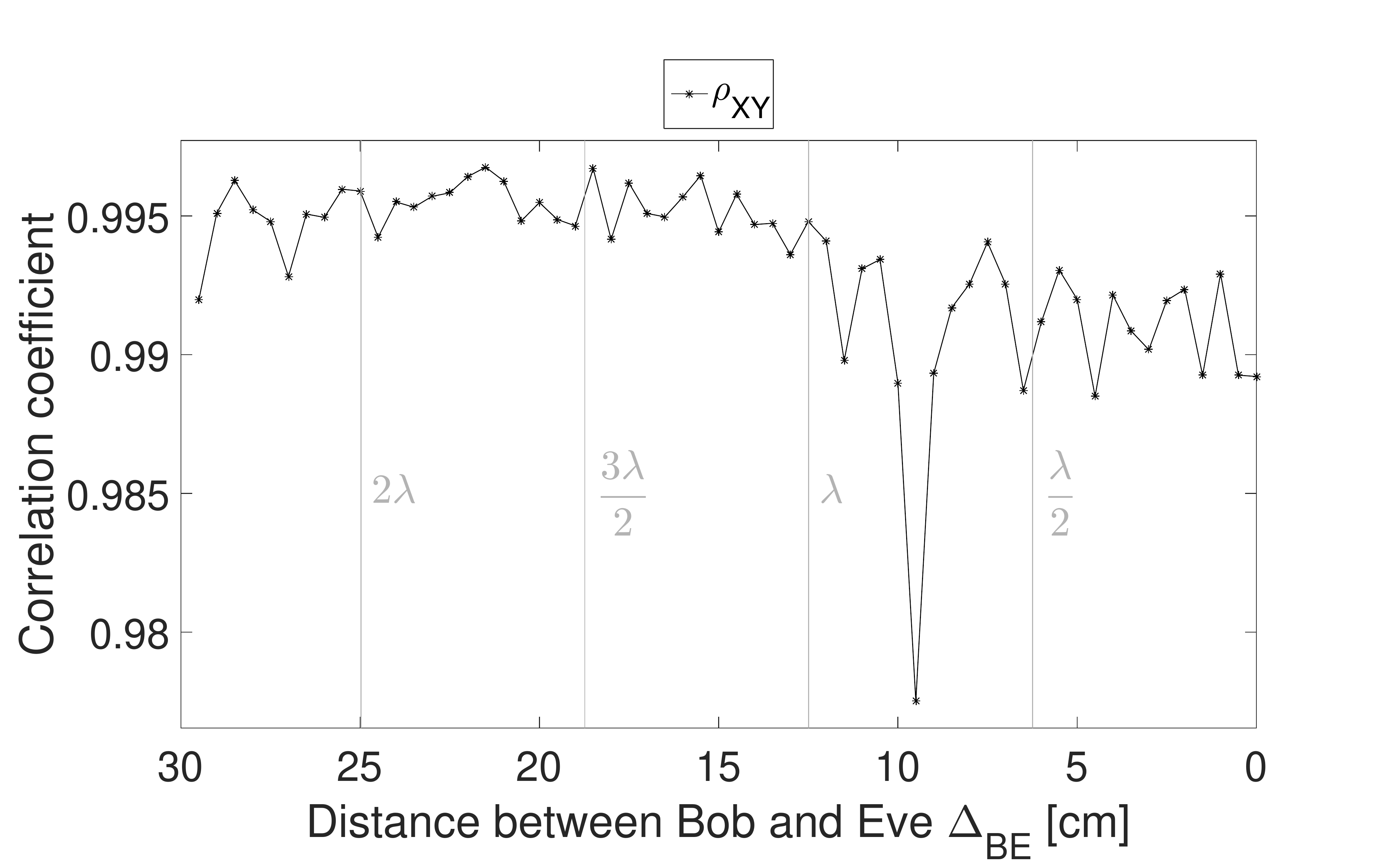}}
	\subfloat[]{\includegraphics[trim=2.2cm 0.1cm 3.5cm 1.6cm, clip=true, height=0.224\textwidth]{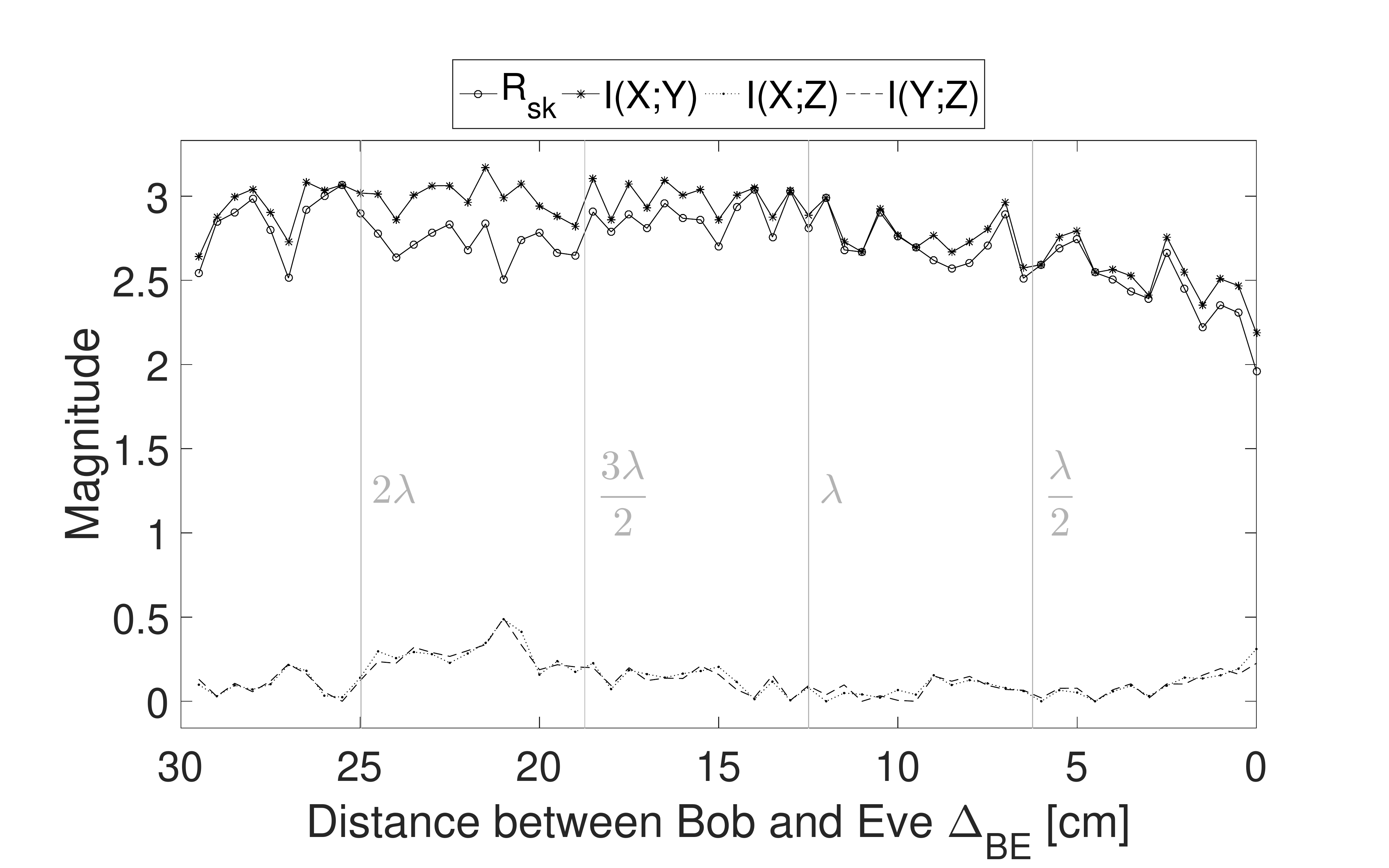}}
	\caption{Evaluation results of $\mybold{v}^{\text{ds}}_k$. In (a) and (b) the cross-correlations is given; in (c) the mutual information as well as $\rsk$ is given. Position 2.}
	\label{fig:app_ds_2}
\end{figure*}

\begin{figure*}
	\centering
	\subfloat[]{\includegraphics[trim=1.4cm 0.1cm 3.5cm 1.6cm, clip=true, height=0.224\textwidth]{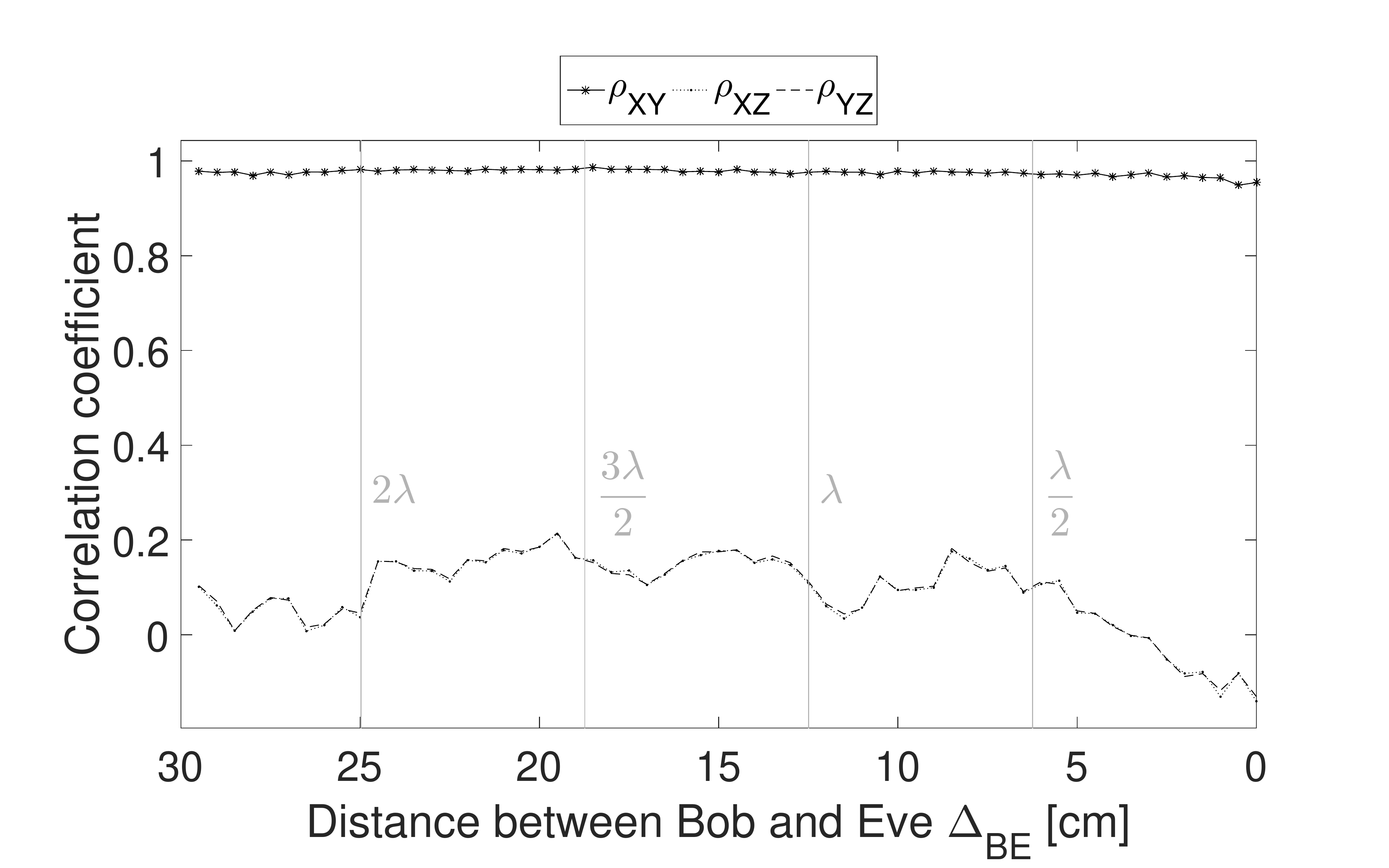}}
	\subfloat[]{\includegraphics[trim=1cm 0.1cm 3.5cm 1.6cm, clip=true, height=0.224\textwidth]{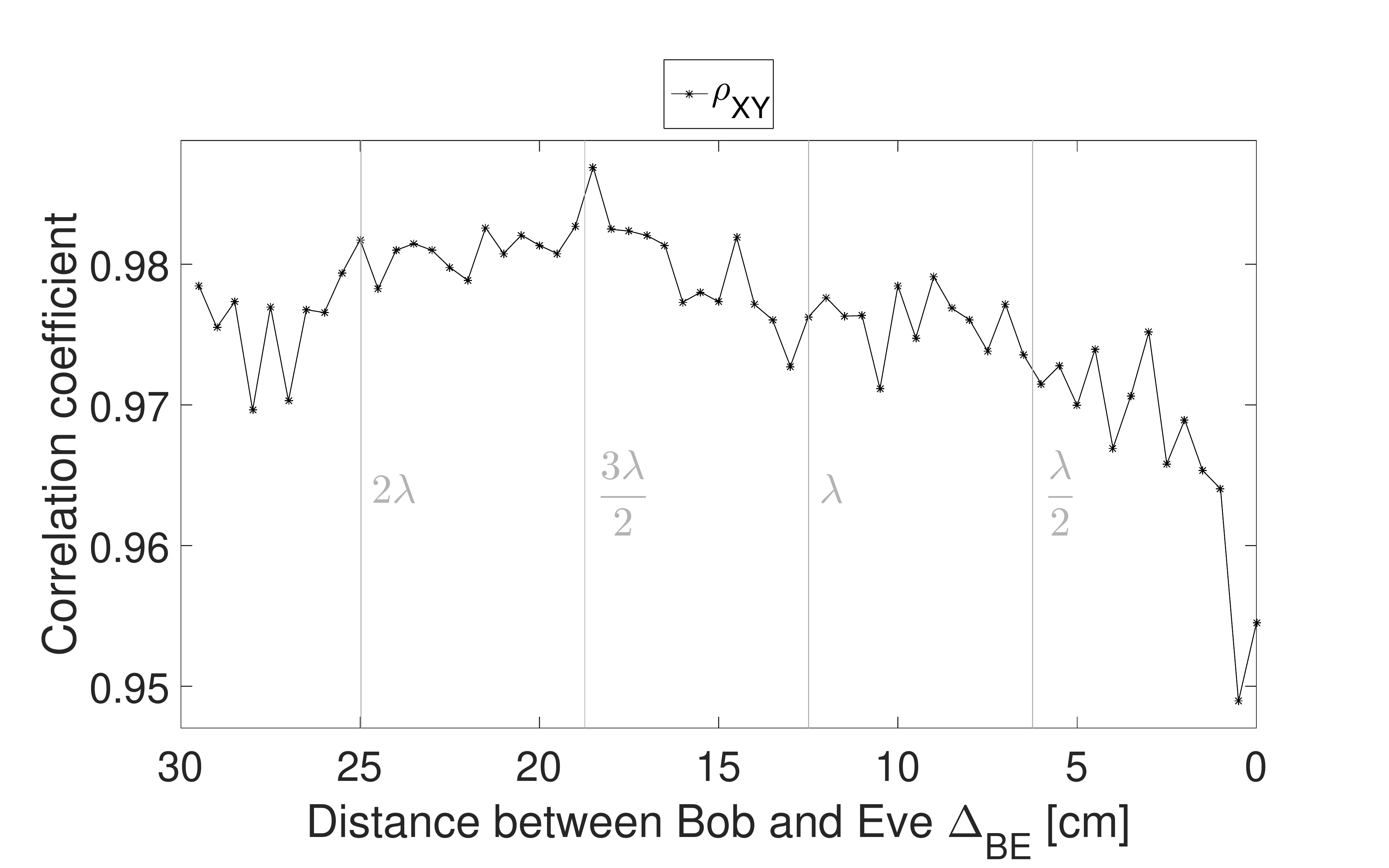}}
	\subfloat[]{\includegraphics[trim=1.8cm 0.1cm 3.5cm 1.6cm, clip=true, height=0.224\textwidth]{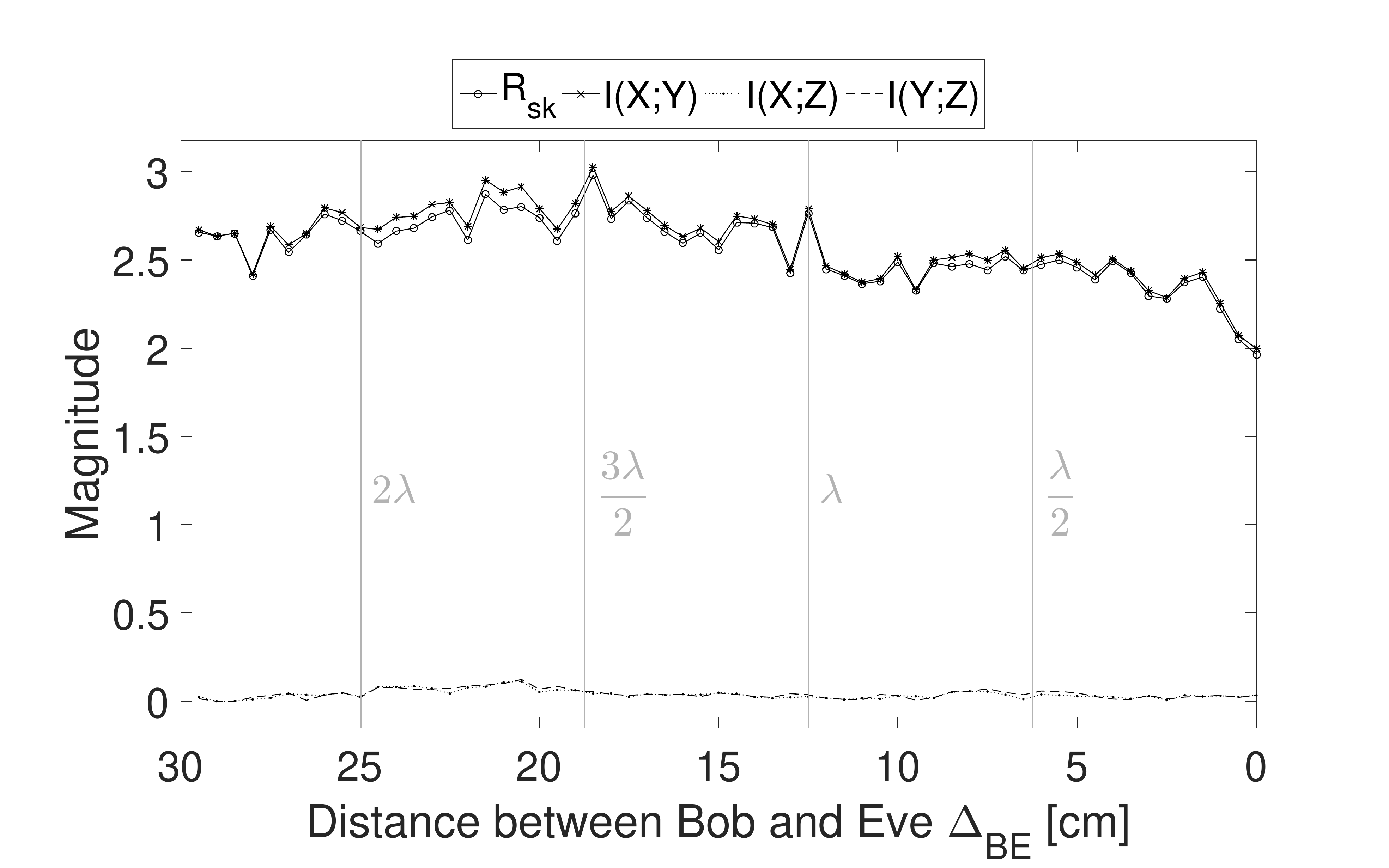}}
	\caption{Evaluation results of $\mybold{v}^{\text{de}}_k$. In (a) and (b) the cross-correlations is given; in (c) the mutual information as well as $\rsk$ is given. Position 2.}
	\label{fig:app_decorr_2}
\end{figure*}


\begin{figure*}
	\centering
	\subfloat[]{\includegraphics[trim=1.4cm 0.1cm 3.5cm 1.6cm, clip=true, height=0.224\textwidth]{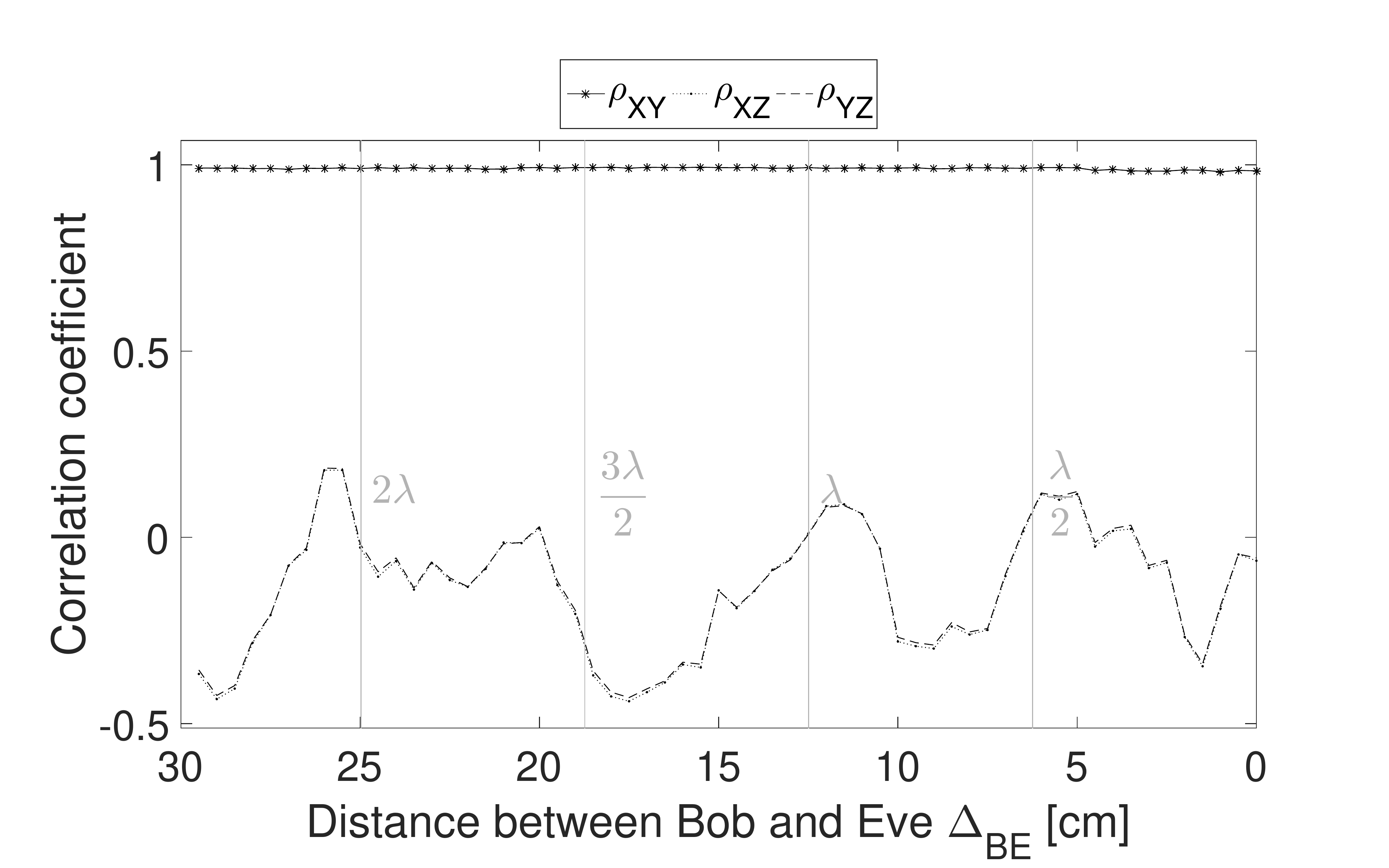}}
	\subfloat[]{\includegraphics[trim=0.5cm 0.1cm 3.5cm 1.6cm, clip=true, height=0.224\textwidth]{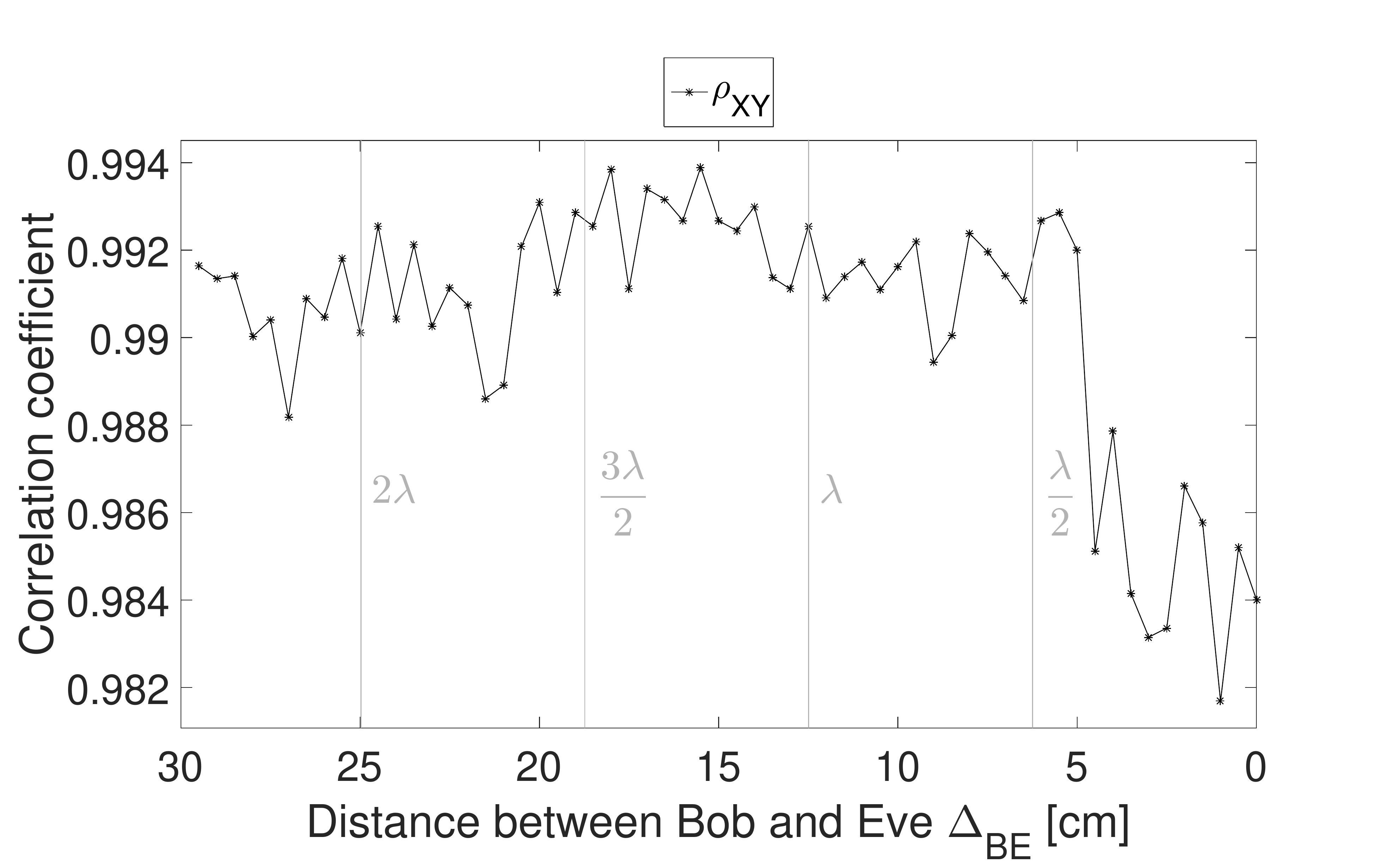}}
	\subfloat[]{\includegraphics[trim=2.2cm 0.1cm 3.5cm 1.6cm, clip=true, height=0.224\textwidth]{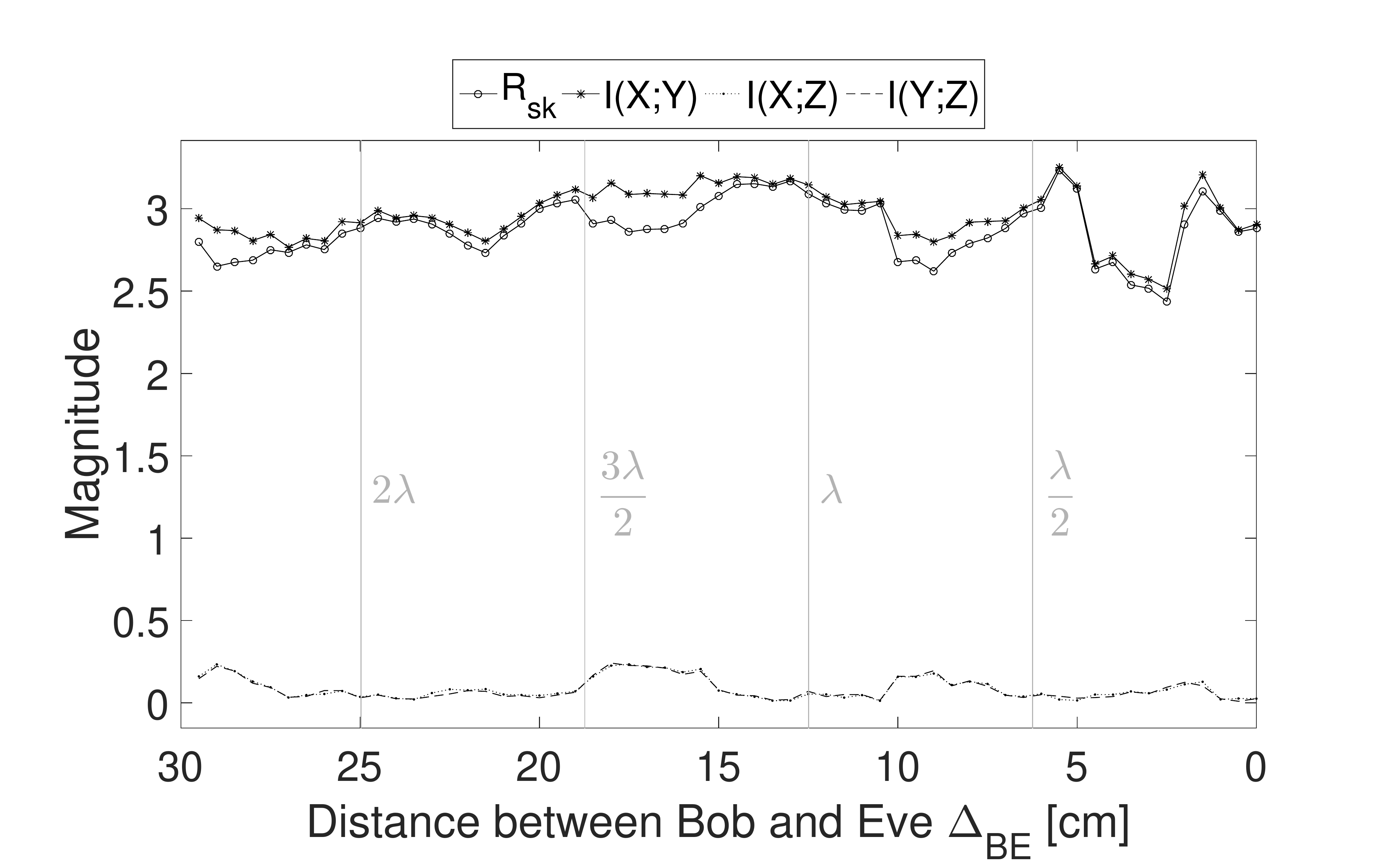}}
	\caption{Evaluation results of $\mybold{v}_k$. In (a) and (b) the cross-correlations is given; in (c) the mutual information as well as $\rsk$ is given. Position 3.}
	\label{fig:app_original_3}
\end{figure*}

\begin{figure*}
	\centering
	\subfloat[]{\includegraphics[trim=1.4cm 0.1cm 3.5cm 1.6cm, clip=true, height=0.224\textwidth]{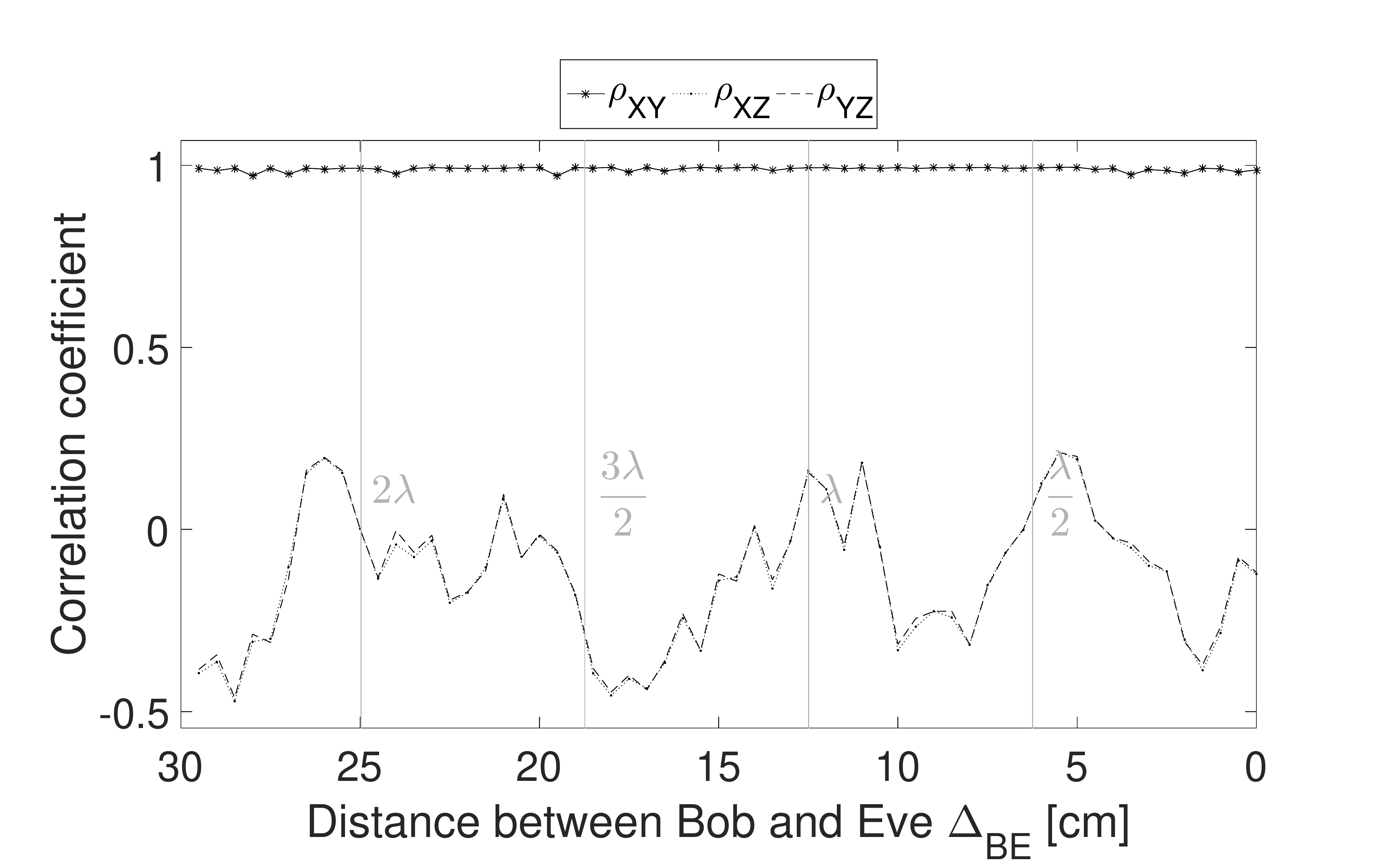}}
	\subfloat[]{\includegraphics[trim=0.5cm 0.1cm 3.5cm 1.6cm, clip=true, height=0.224\textwidth]{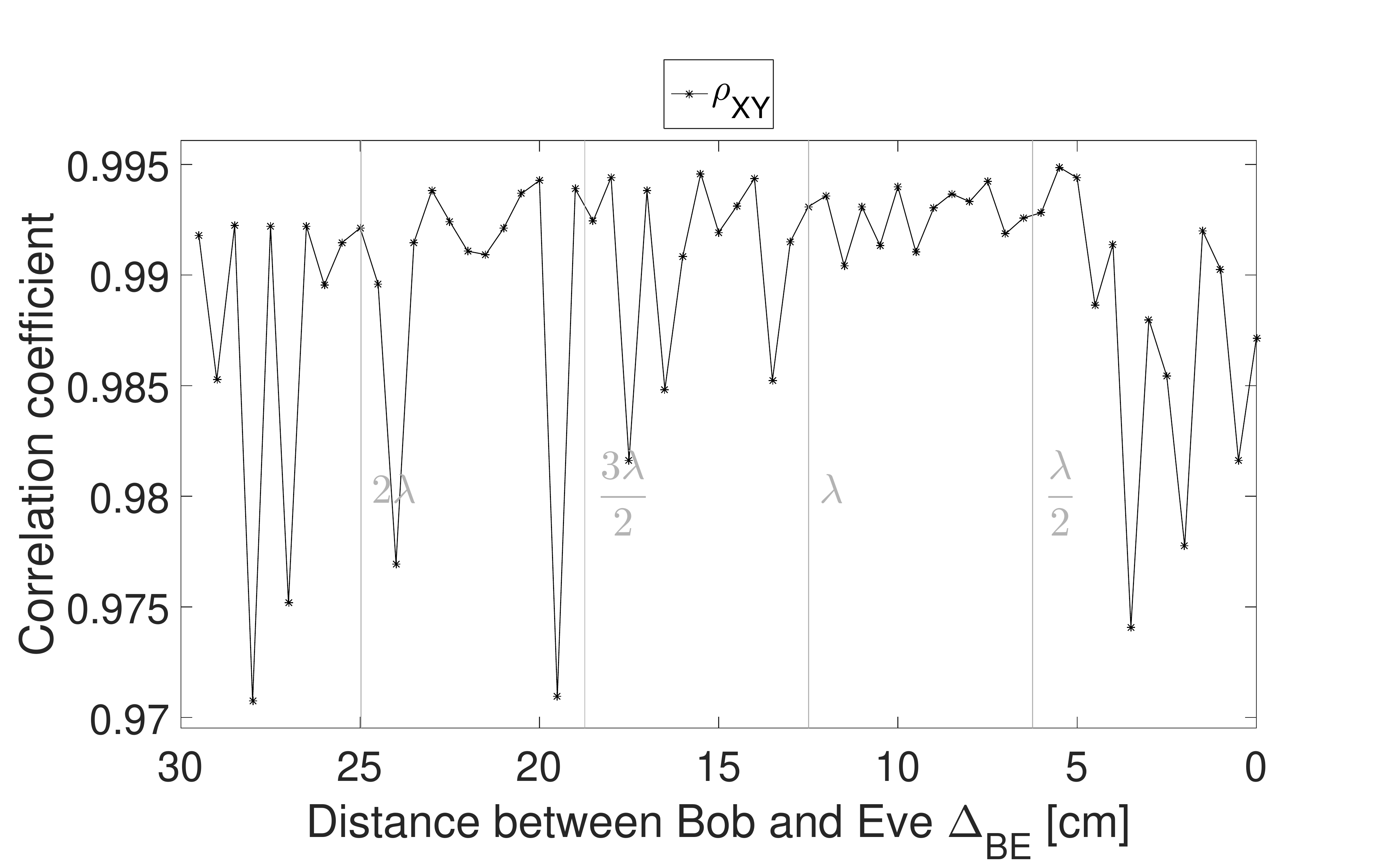}}
	\subfloat[]{\includegraphics[trim=2.2cm 0.1cm 3.5cm 1.6cm, clip=true, height=0.224\textwidth]{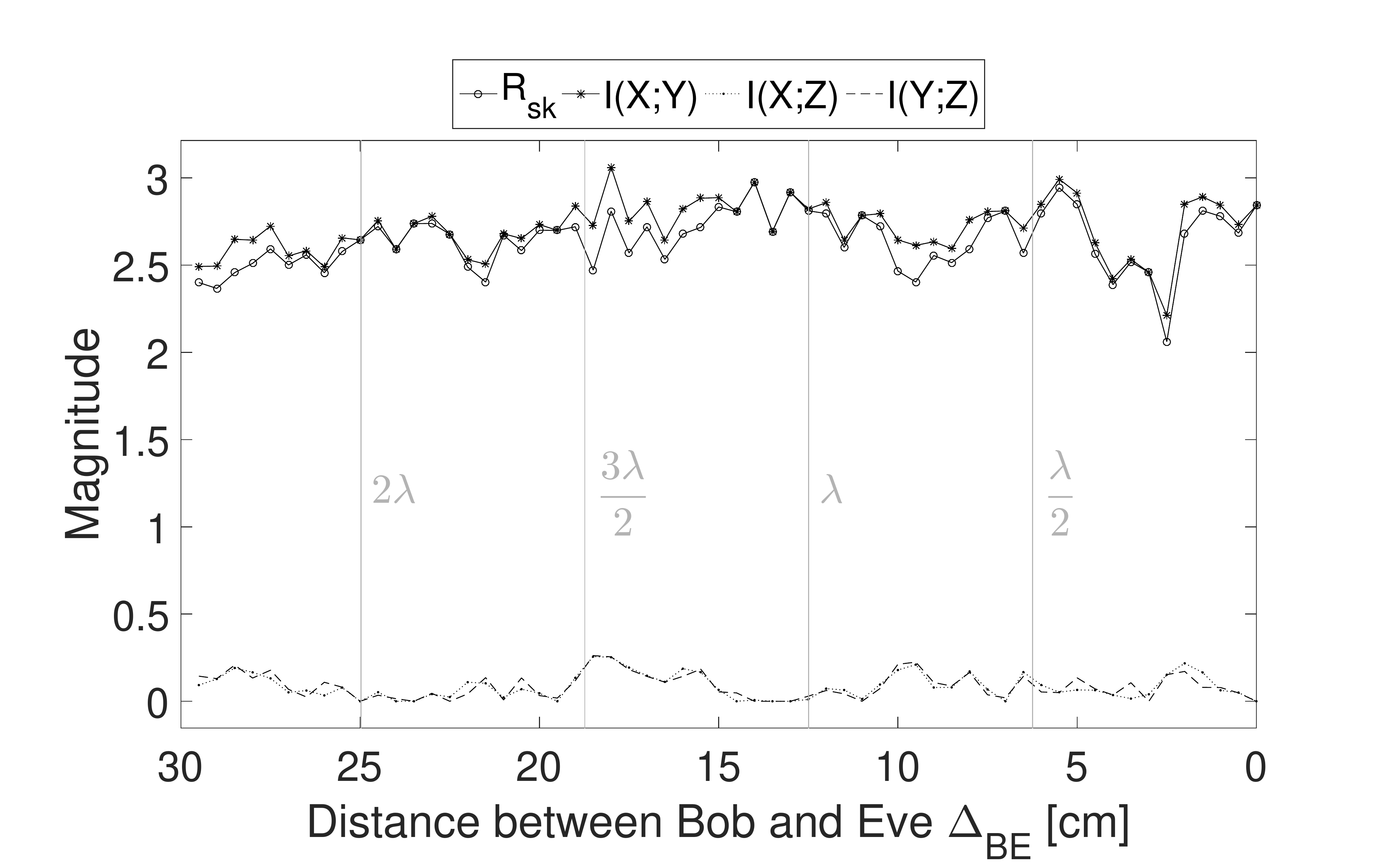}}
	\caption{Evaluation results of $\mybold{v}^{\text{ds}}_k$. In (a) and (b) the cross-correlations is given; in (c) the mutual information as well as $\rsk$ is given. Position 3.}
	\label{fig:app_ds_3}
\end{figure*}

\begin{figure*}
	\centering
	\subfloat[]{\includegraphics[trim=1.4cm 0.1cm 3.5cm 1.6cm, clip=true, height=0.224\textwidth]{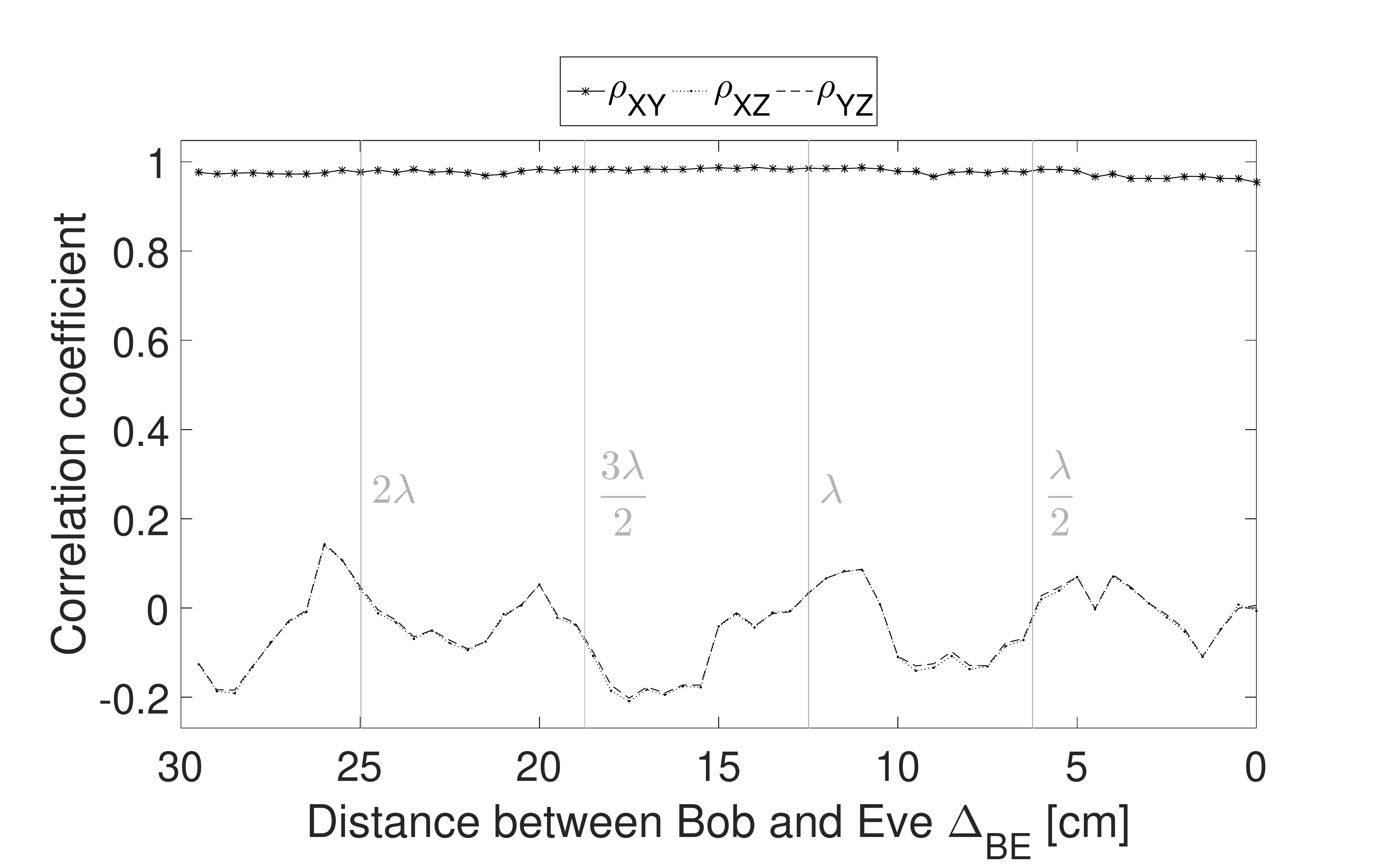}}
	\subfloat[]{\includegraphics[trim=1cm 0.1cm 3.5cm 1.6cm, clip=true, height=0.224\textwidth]{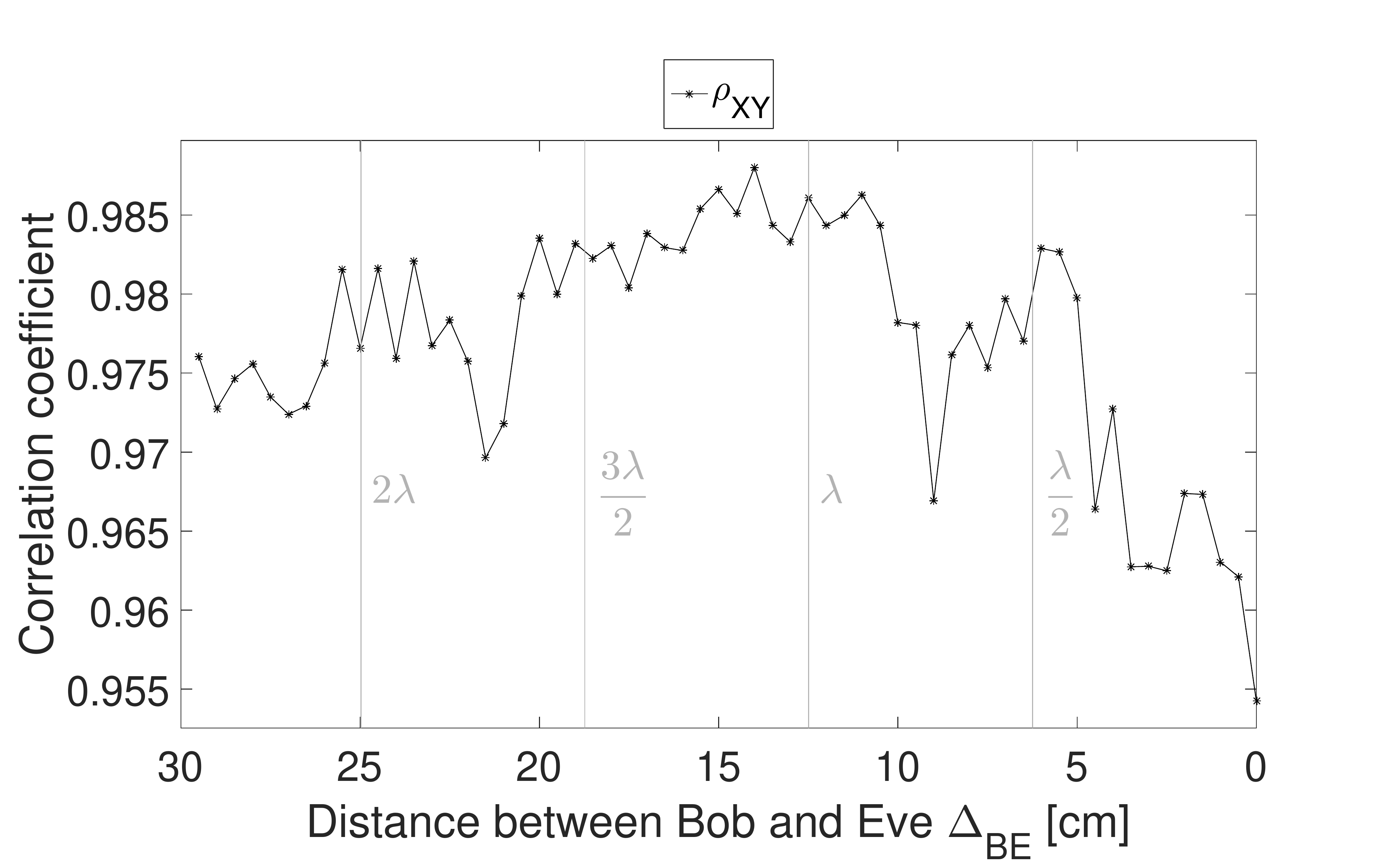}}
	\subfloat[]{\includegraphics[trim=1.8cm 0.1cm 3.5cm 1.6cm, clip=true, height=0.224\textwidth]{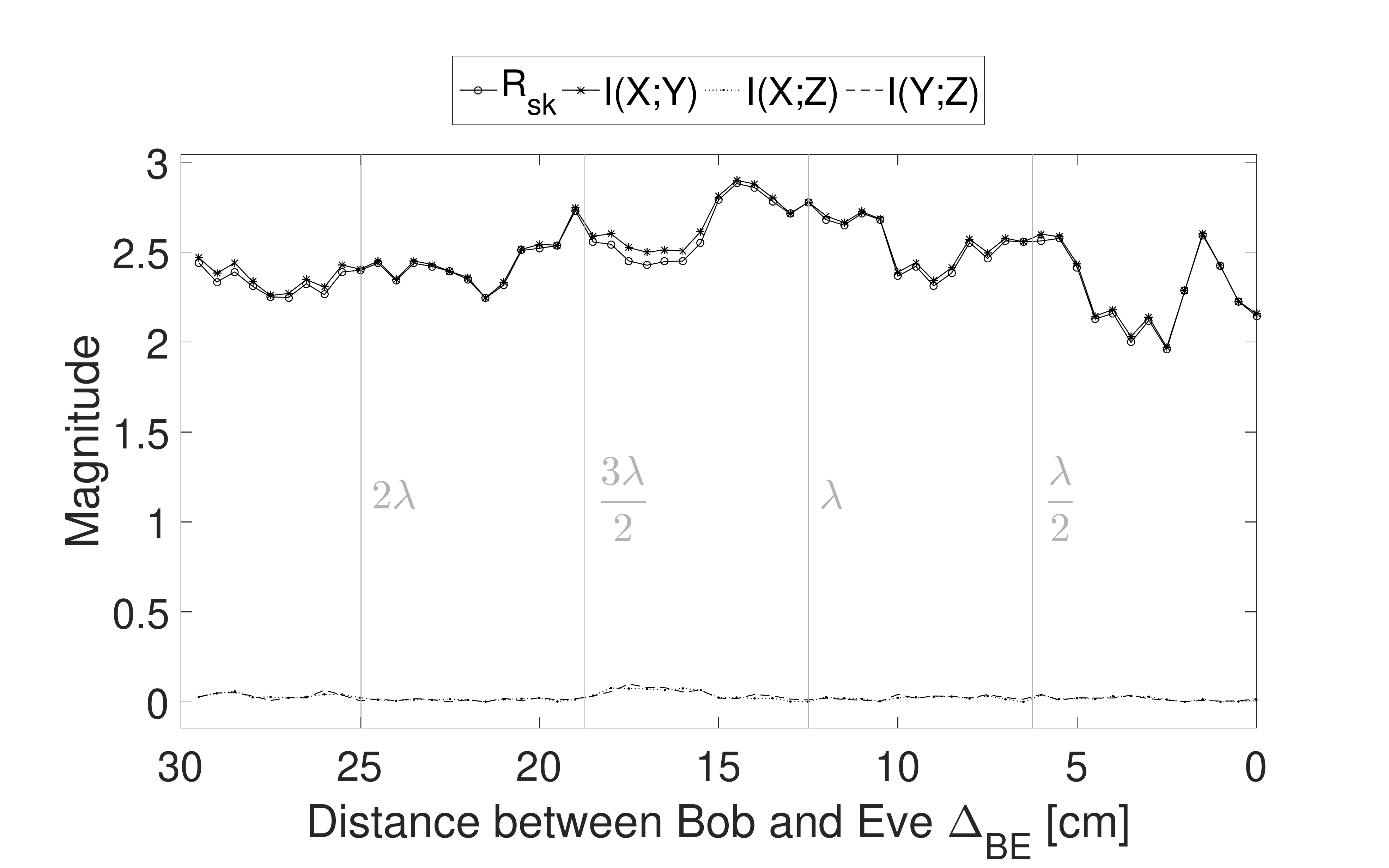}}
	\caption{Evaluation results of $\mybold{v}^{\text{de}}_k$. In (a) and (b) the cross-correlations is given; in (c) the mutual information as well as $\rsk$ is given. Position 3.}
	\label{fig:app_decorr_3}
\end{figure*}


\begin{figure*}
	\centering
	\subfloat[]{\includegraphics[trim=1.4cm 0.1cm 3.5cm 1.6cm, clip=true, height=0.224\textwidth]{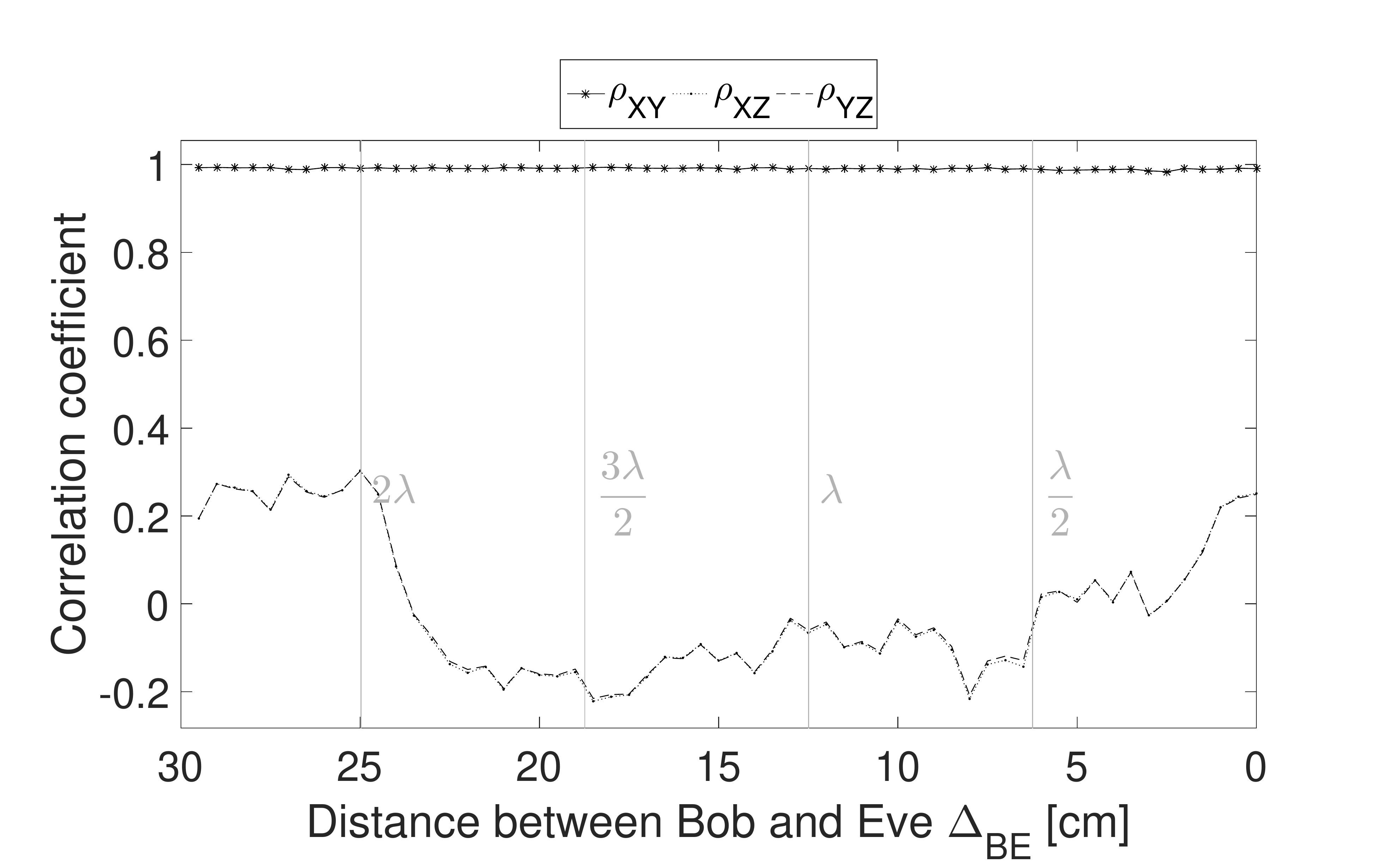}}
	\subfloat[]{\includegraphics[trim=0.5cm 0.1cm 3.5cm 1.6cm, clip=true, height=0.224\textwidth]{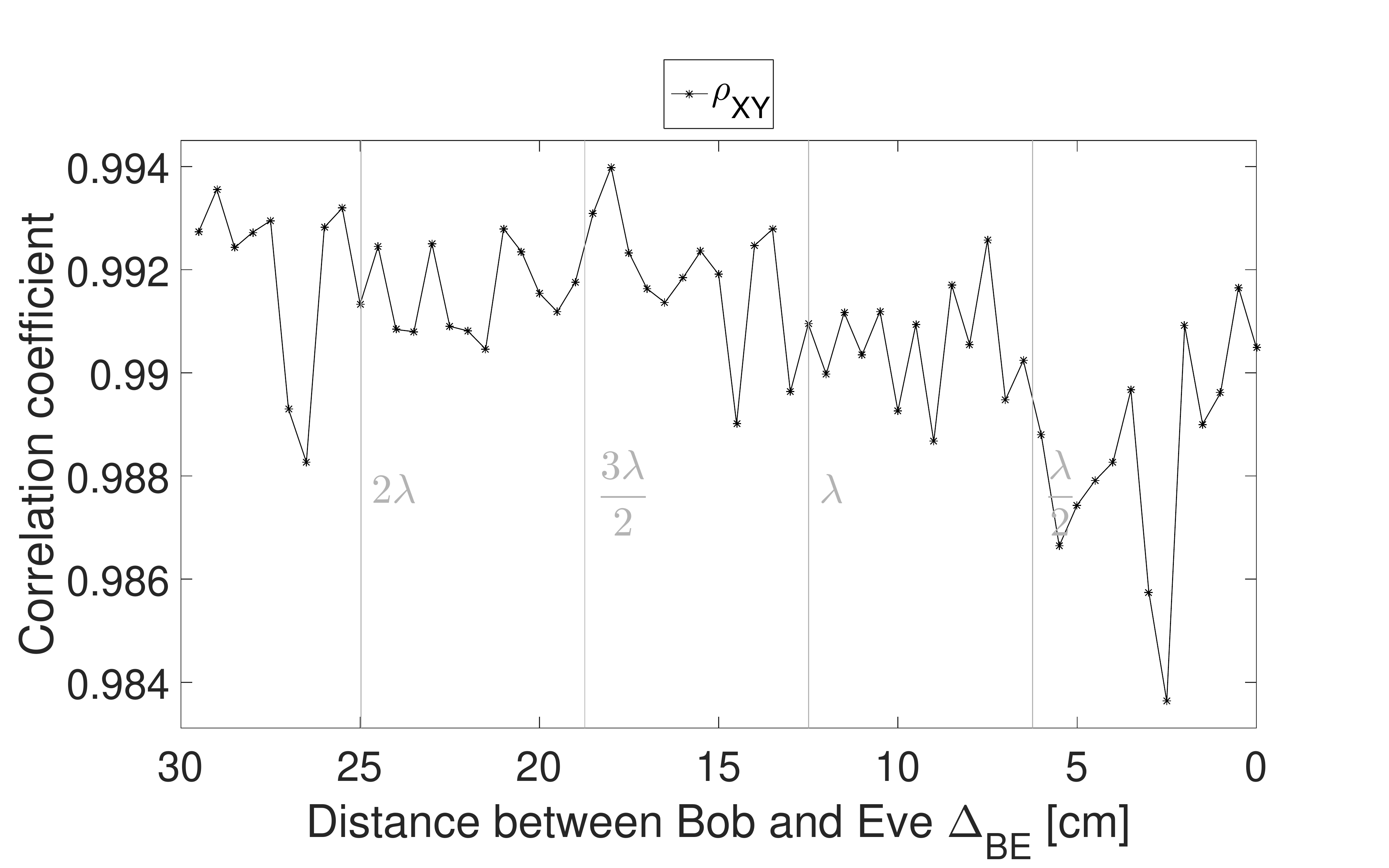}}
	\subfloat[]{\includegraphics[trim=2.2cm 0.1cm 3.5cm 1.6cm, clip=true, height=0.224\textwidth]{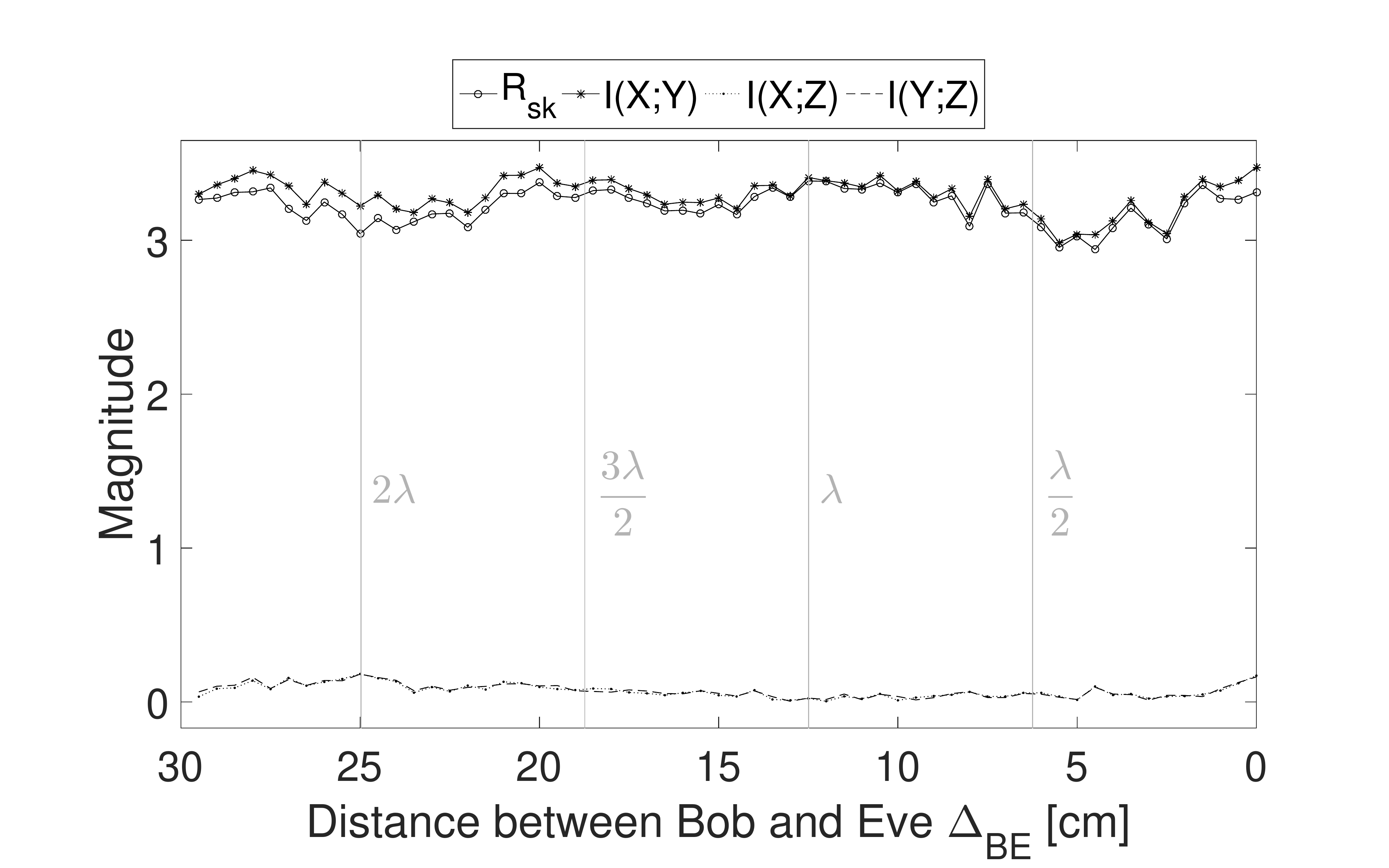}}
	\caption{Evaluation results of $\mybold{v}_k$. In (a) and (b) the cross-correlations is given; in (c) the mutual information as well as $\rsk$ is given. Position 4.}
	\label{fig:app_original_4}
\end{figure*}

\begin{figure*}
	\centering
	\subfloat[]{\includegraphics[trim=1.4cm 0.1cm 3.5cm 1.6cm, clip=true, height=0.224\textwidth]{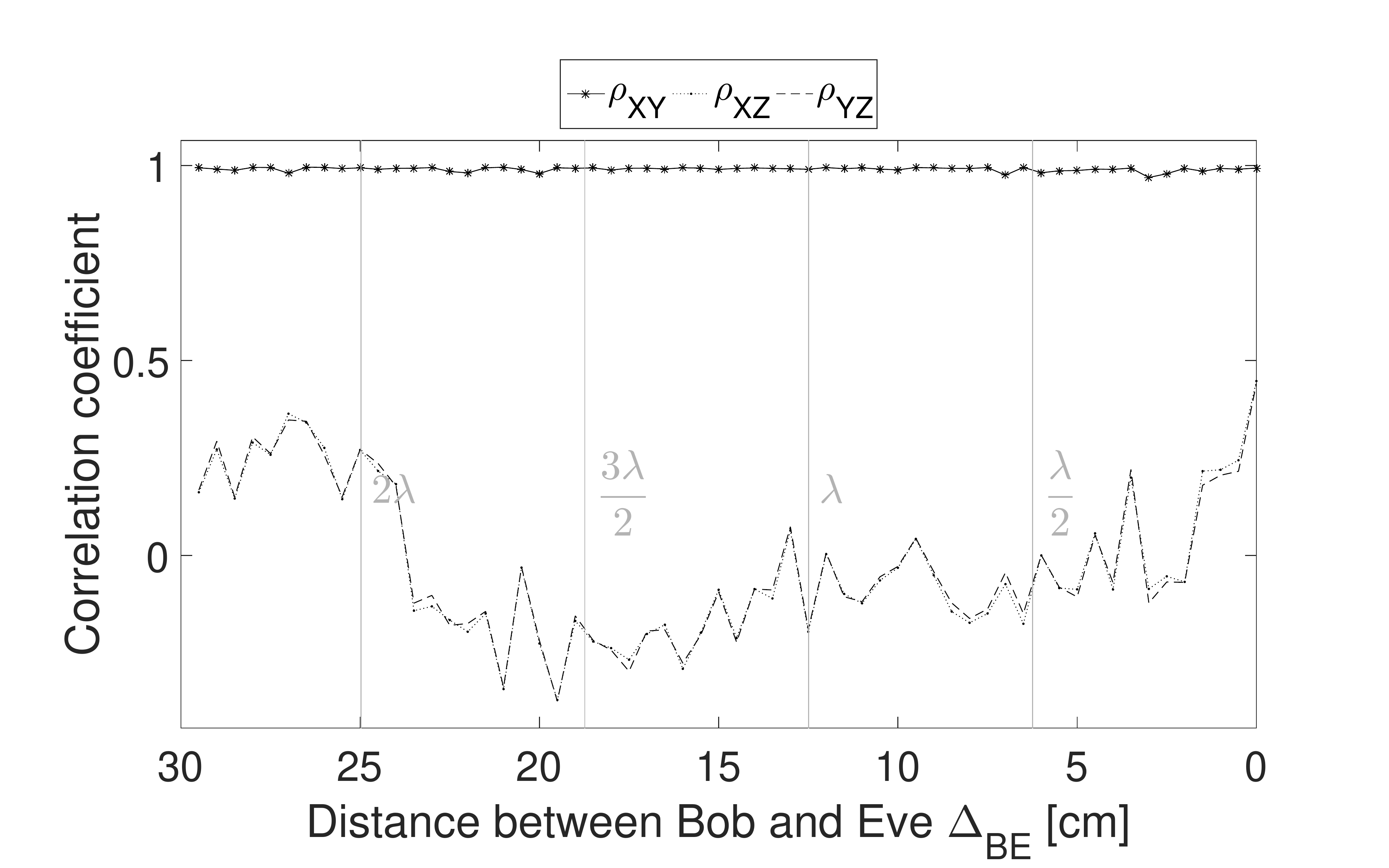}}
	\subfloat[]{\includegraphics[trim=0.5cm 0.1cm 3.5cm 1.6cm, clip=true, height=0.224\textwidth]{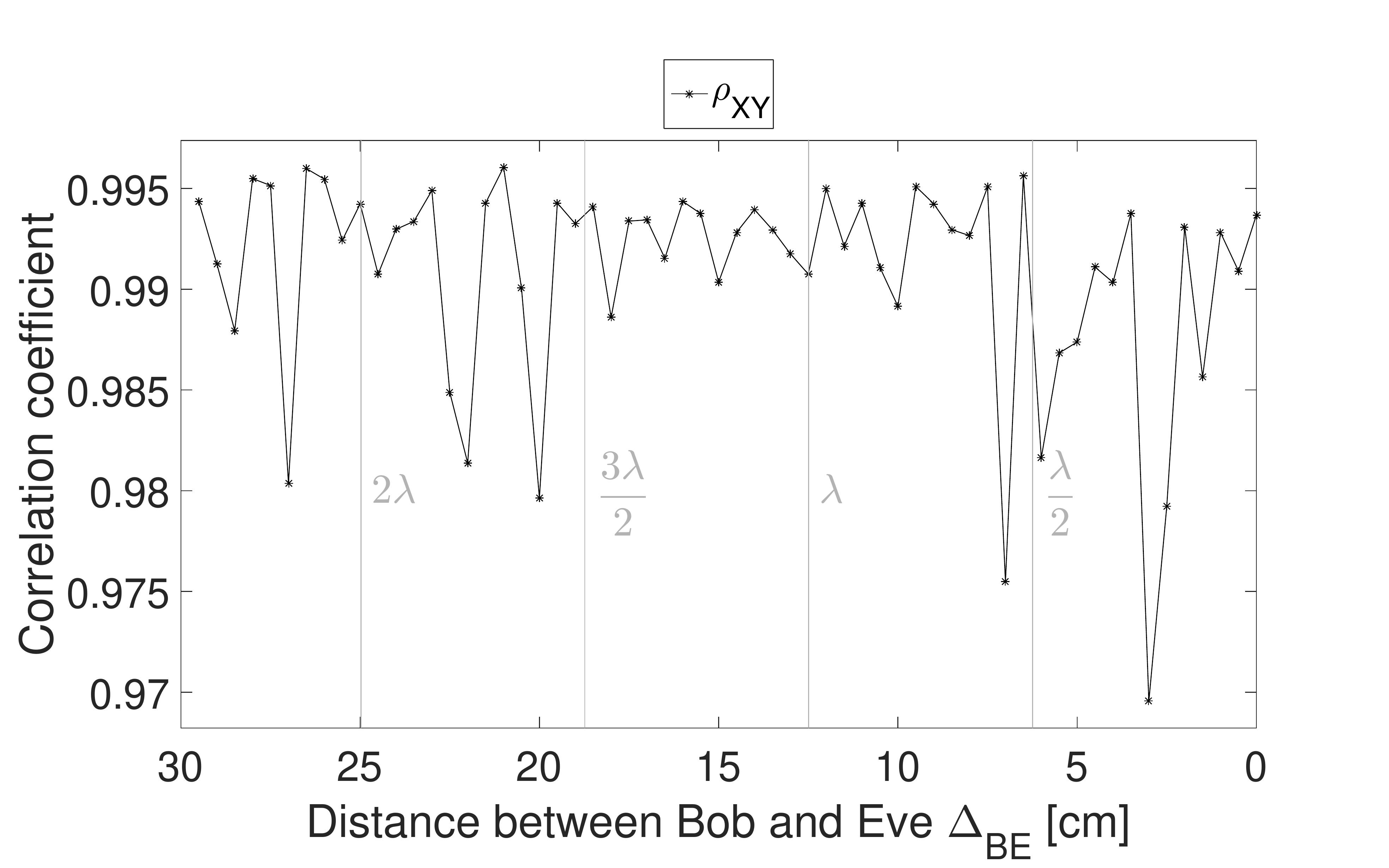}}
	\subfloat[]{\includegraphics[trim=2.2cm 0.1cm 3.5cm 1.6cm, clip=true, height=0.224\textwidth]{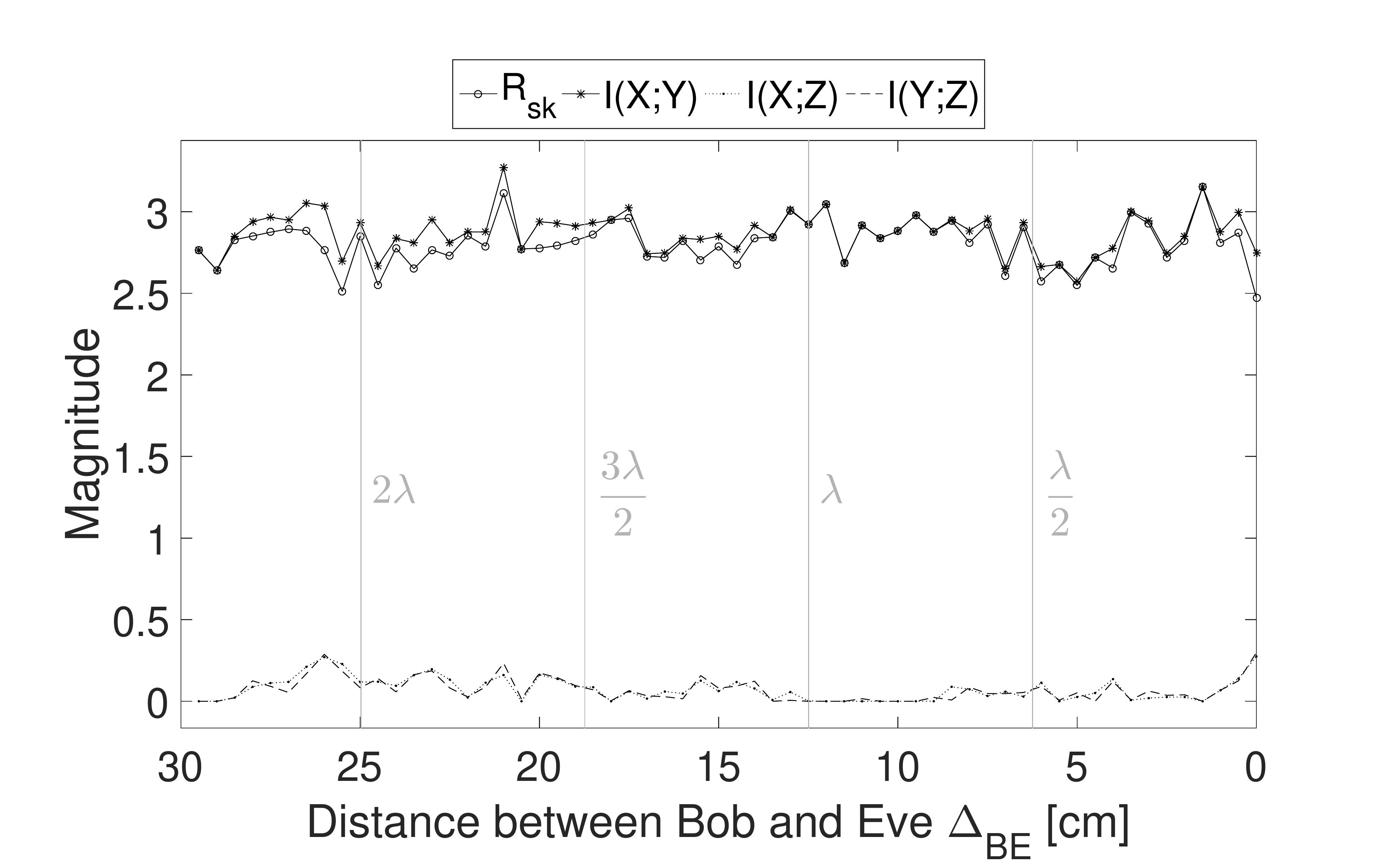}}
	\caption{Evaluation results of $\mybold{v}^{\text{ds}}_k$. In (a) and (b) the cross-correlations is given; in (c) the mutual information as well as $\rsk$ is given. Position 4.}
	\label{fig:app_ds_4}
\end{figure*}

\begin{figure*}
	\centering
	\subfloat[]{\includegraphics[trim=1.4cm 0.1cm 3.5cm 1.6cm, clip=true, height=0.224\textwidth]{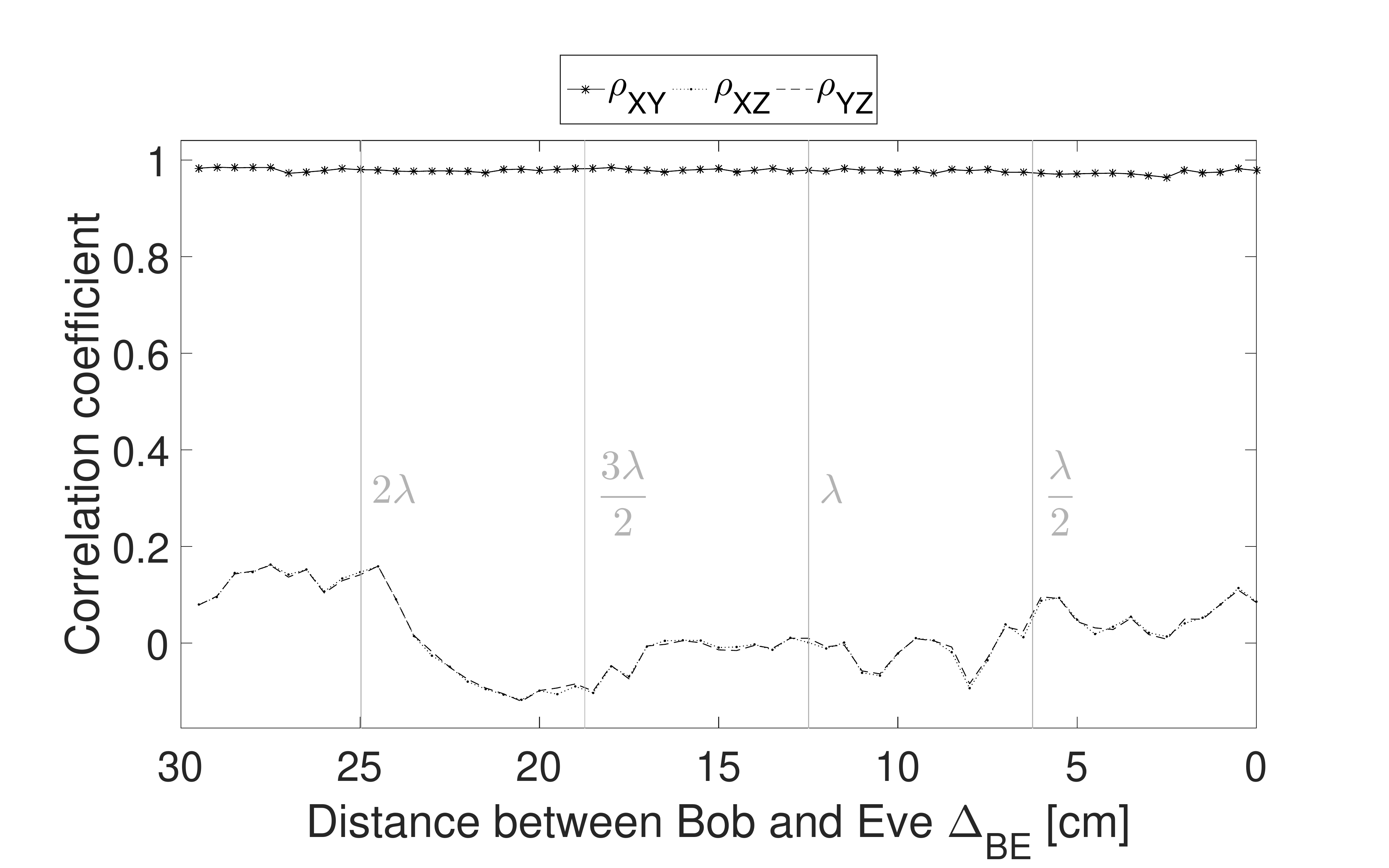}}
	\subfloat[]{\includegraphics[trim=1cm 0.1cm 3.5cm 1.6cm, clip=true, height=0.224\textwidth]{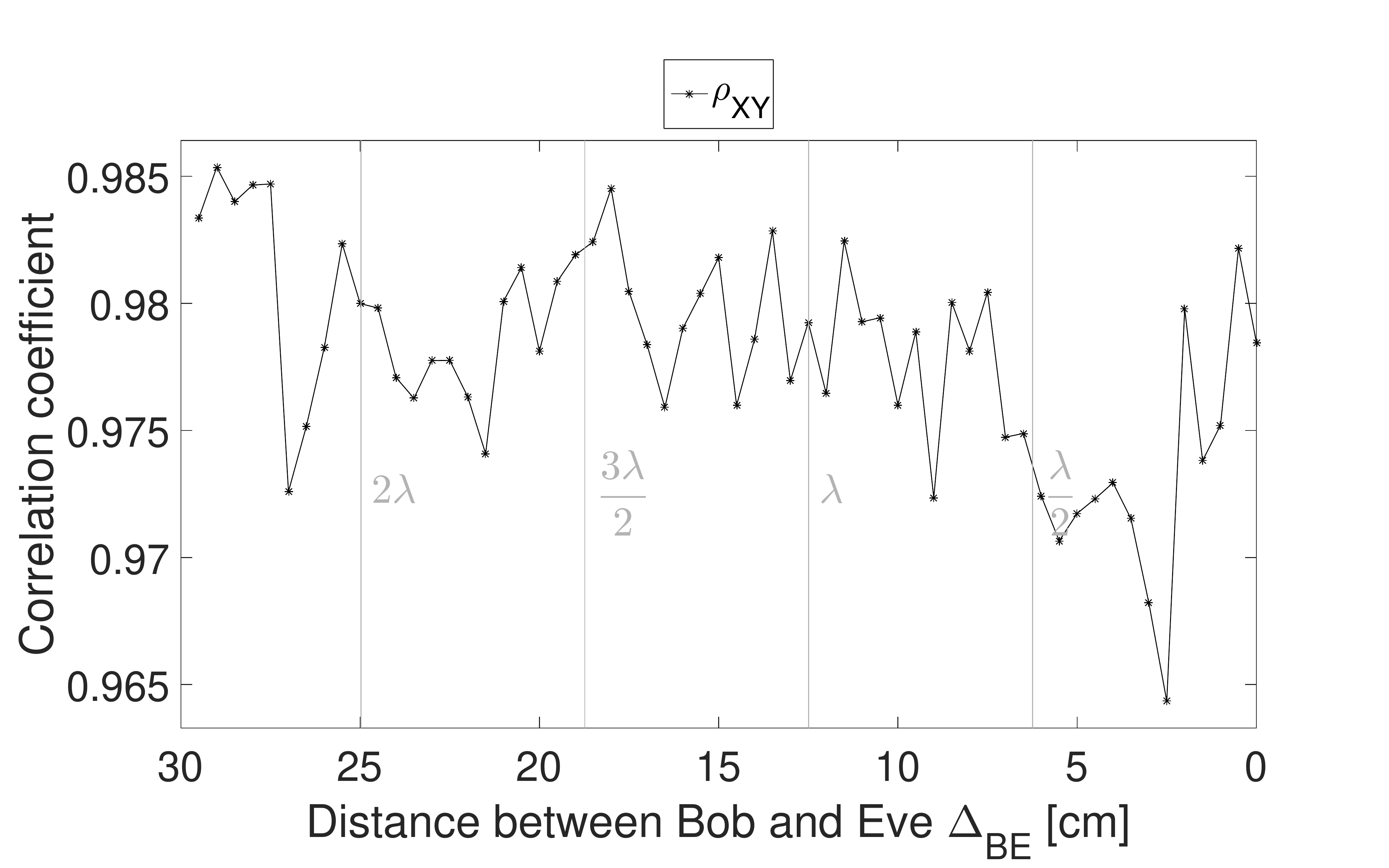}}
	\subfloat[]{\includegraphics[trim=1.8cm 0.1cm 3.5cm 1.6cm, clip=true, height=0.224\textwidth]{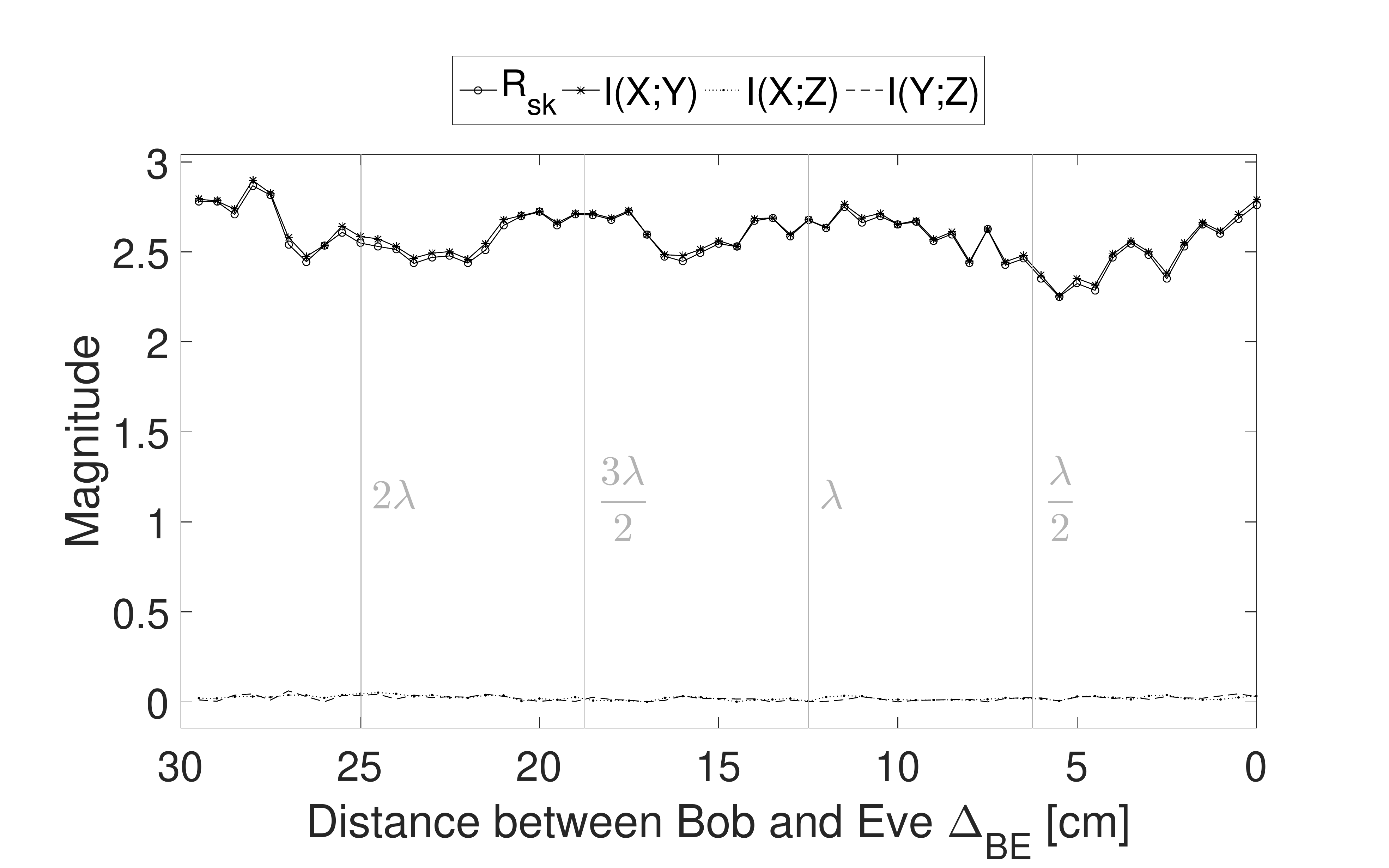}}
	\caption{Evaluation results of $\mybold{v}^{\text{de}}_k$. In (a) and (b) the cross-correlations is given; in (c) the mutual information as well as $\rsk$ is given. Position 4.}
	\label{fig:app_decorr_4}
\end{figure*}


\begin{figure*}
	\centering
	\subfloat[]{\includegraphics[trim=1.4cm 0.1cm 3.5cm 1.6cm, clip=true, height=0.224\textwidth]{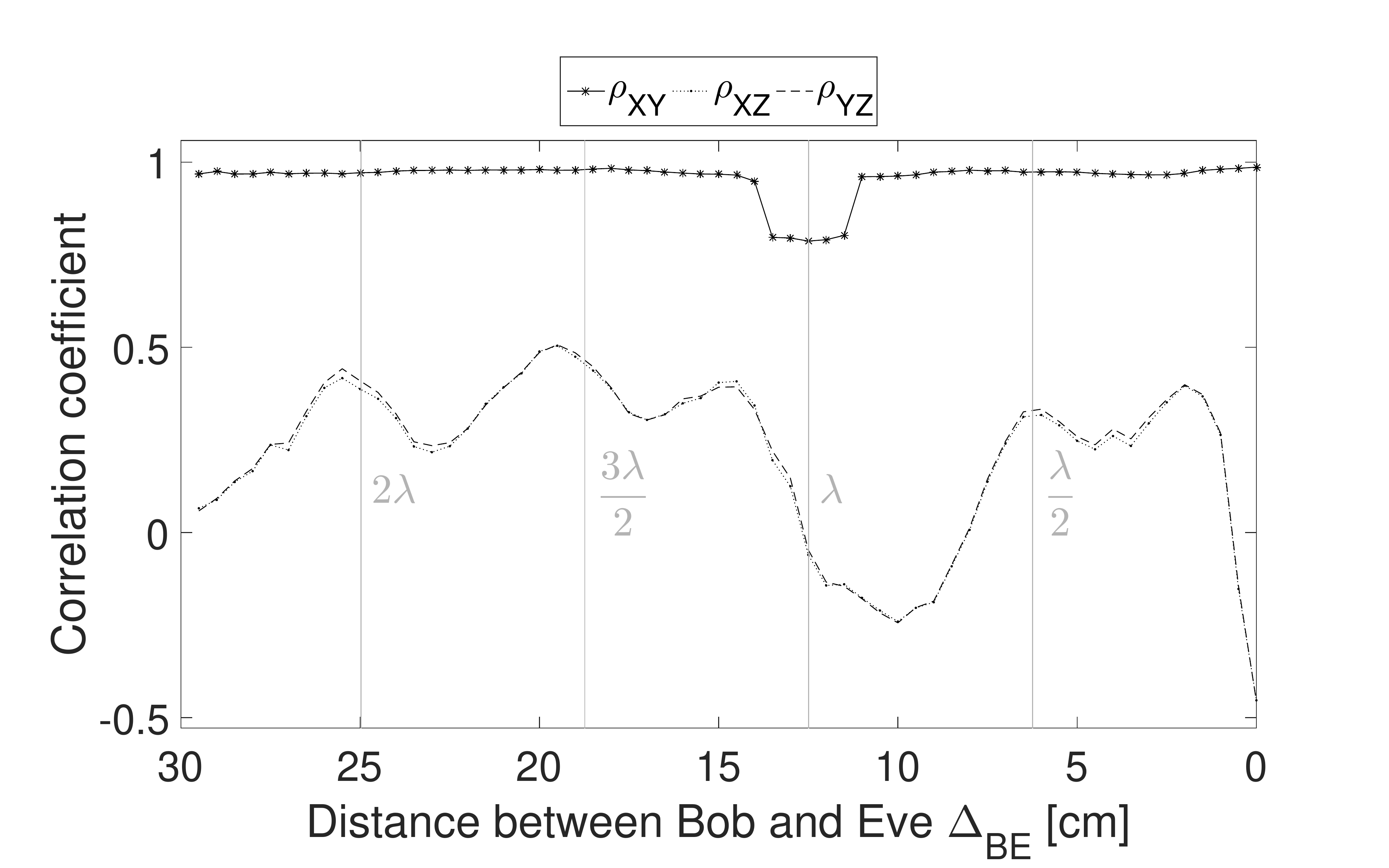}}
	\subfloat[]{\includegraphics[trim=0.5cm 0.1cm 3.5cm 1.6cm, clip=true, height=0.224\textwidth]{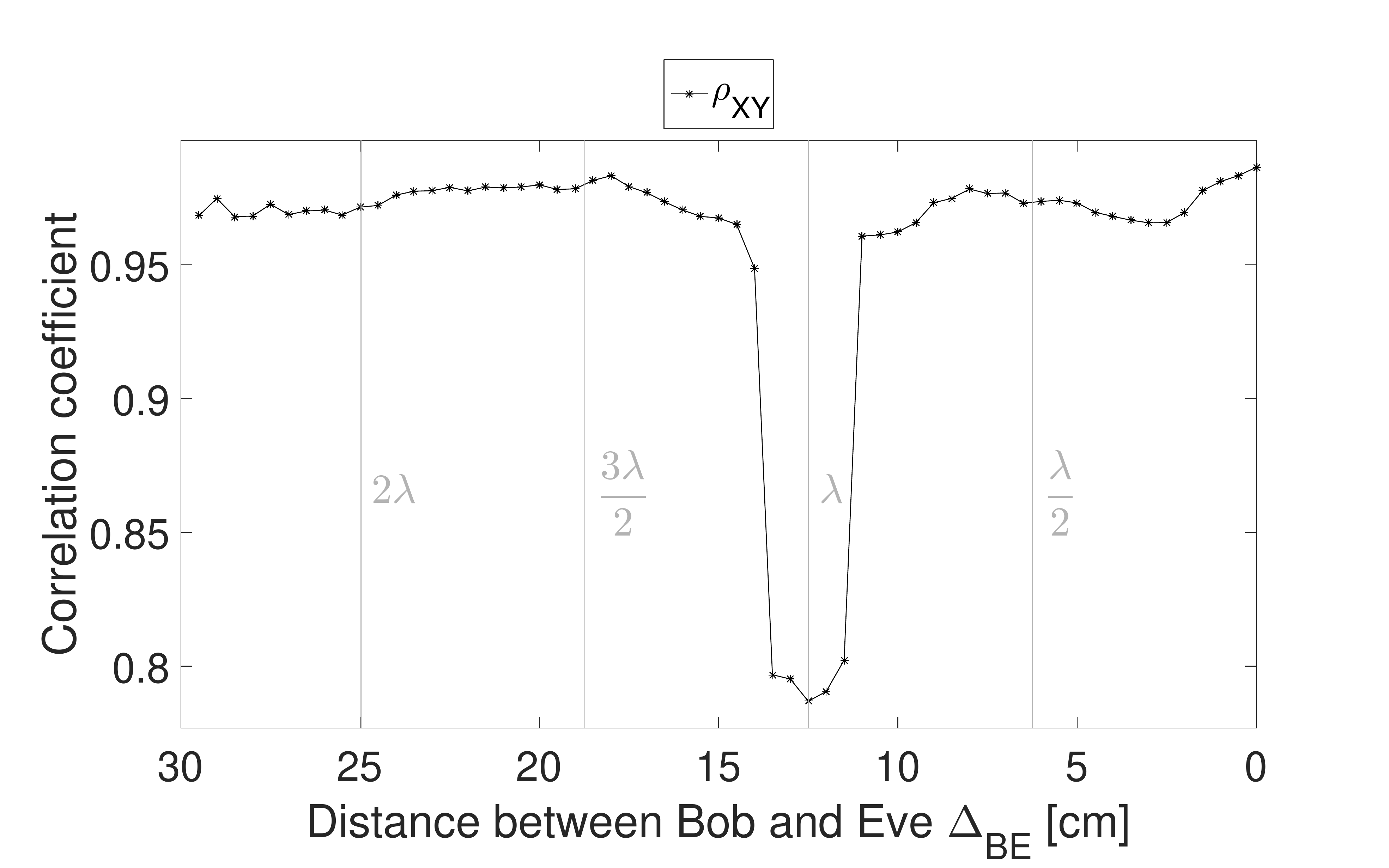}}
	\subfloat[]{\includegraphics[trim=2.2cm 0.1cm 3.5cm 1.6cm, clip=true, height=0.224\textwidth]{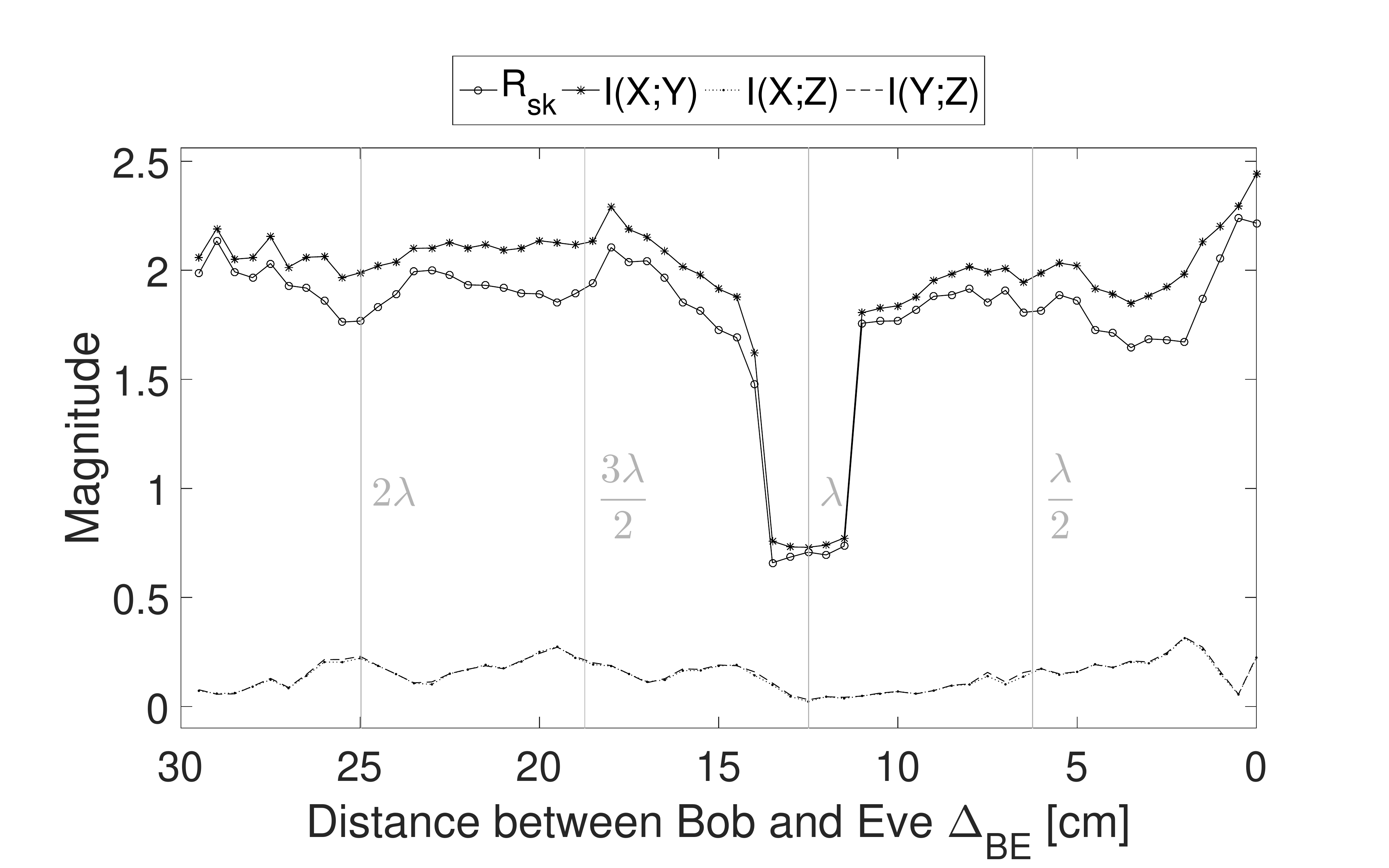}}
	\caption{Evaluation results of $\mybold{v}_k$. In (a) and (b) the cross-correlations is given; in (c) the mutual information as well as $\rsk$ is given. Position 5.}
	\label{fig:app_original_5}
\end{figure*}

\begin{figure*}
	\centering
	\subfloat[]{\includegraphics[trim=1.4cm 0.1cm 3.5cm 1.6cm, clip=true, height=0.224\textwidth]{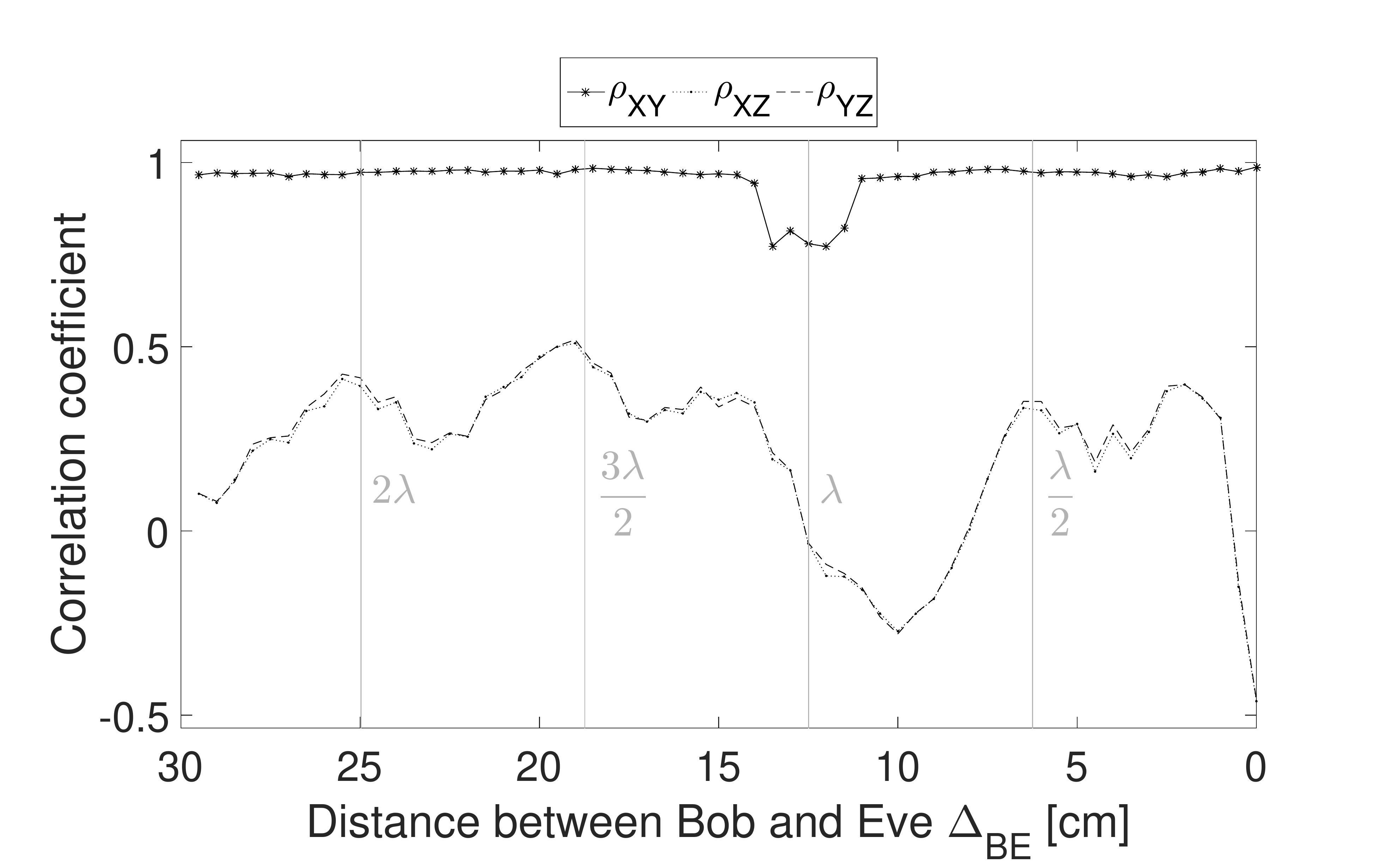}}
	\subfloat[]{\includegraphics[trim=0.5cm 0.1cm 3.5cm 1.6cm, clip=true, height=0.224\textwidth]{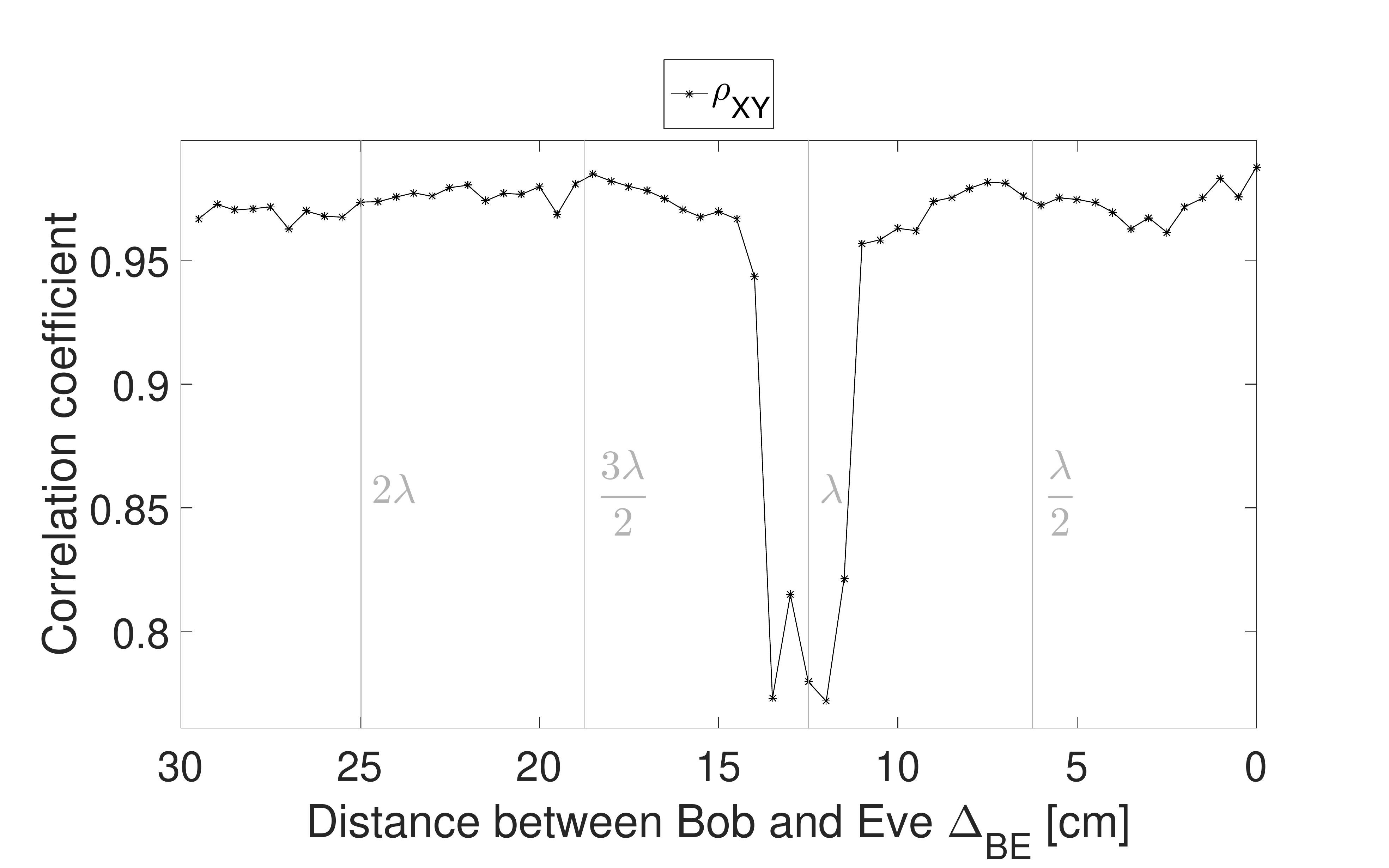}}
	\subfloat[]{\includegraphics[trim=2.2cm 0.1cm 3.5cm 1.6cm, clip=true, height=0.224\textwidth]{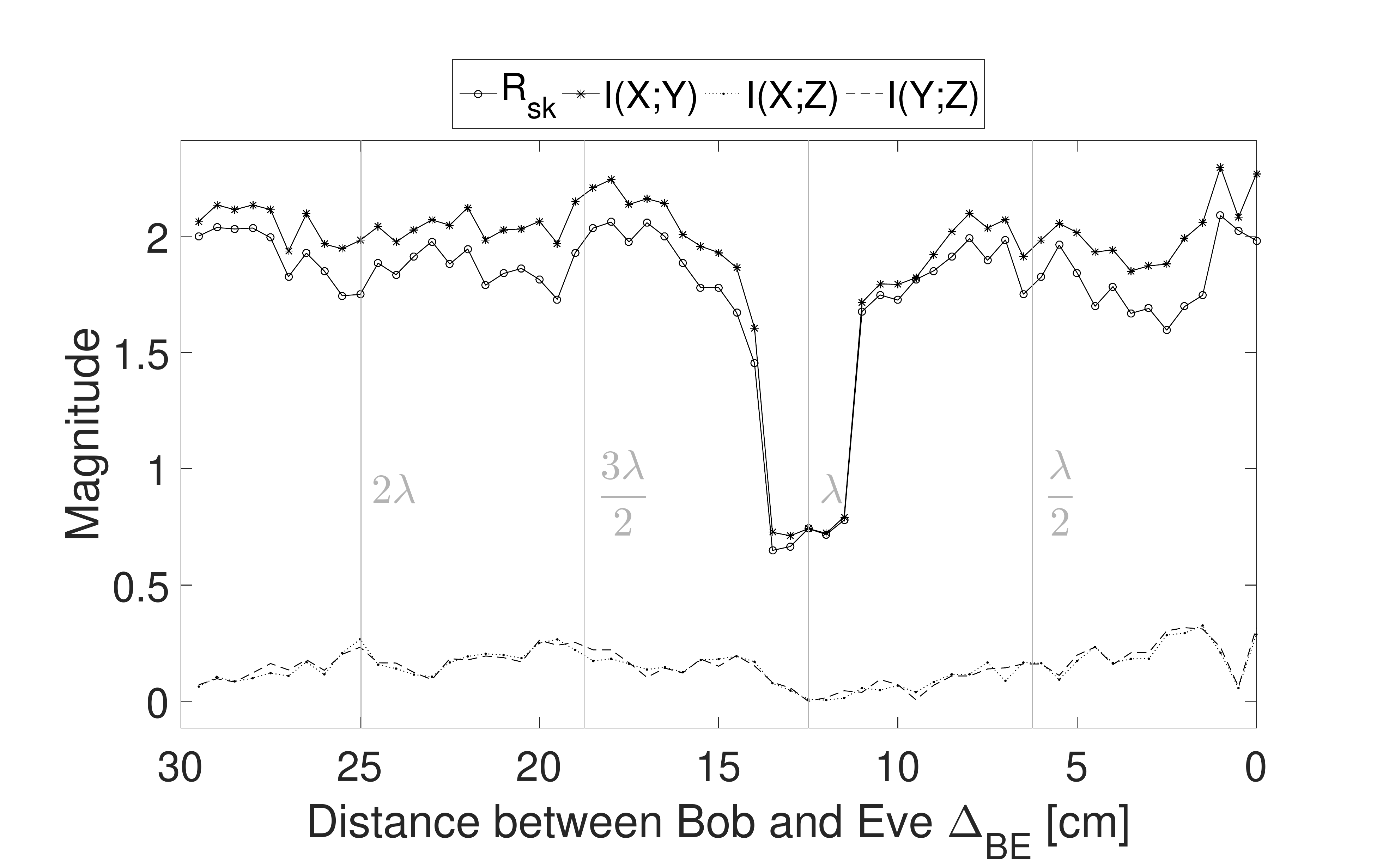}}
	\caption{Evaluation results of $\mybold{v}^{\text{ds}}_k$. In (a) and (b) the cross-correlations is given; in (c) the mutual information as well as $\rsk$ is given. Position 5.}
	\label{fig:app_ds_5}
\end{figure*}

\begin{figure*}
	\centering
	\subfloat[]{\includegraphics[trim=1.4cm 0.1cm 3.5cm 1.6cm, clip=true, height=0.224\textwidth]{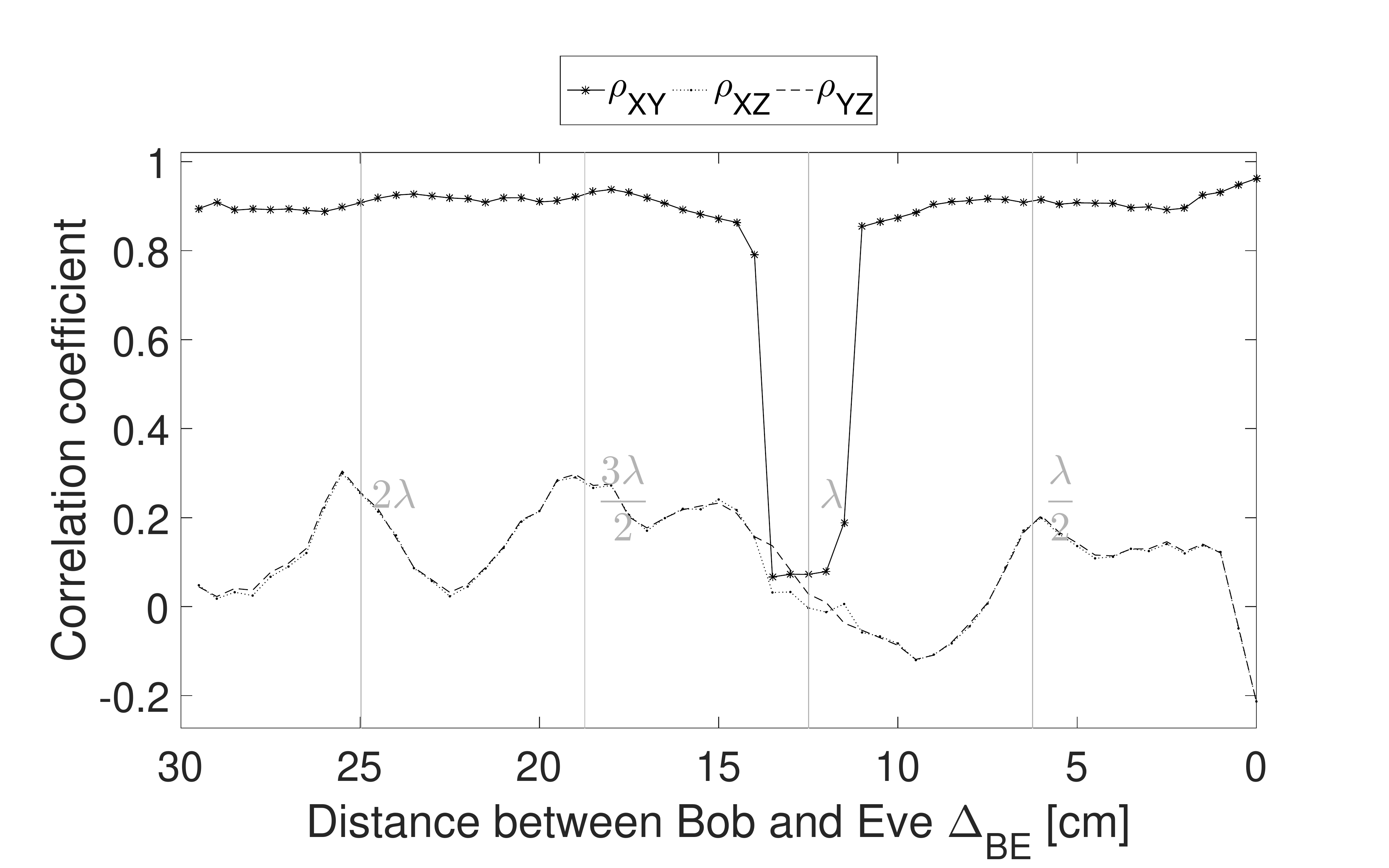}}
	\subfloat[]{\includegraphics[trim=1cm 0.1cm 3.5cm 1.6cm, clip=true, height=0.224\textwidth]{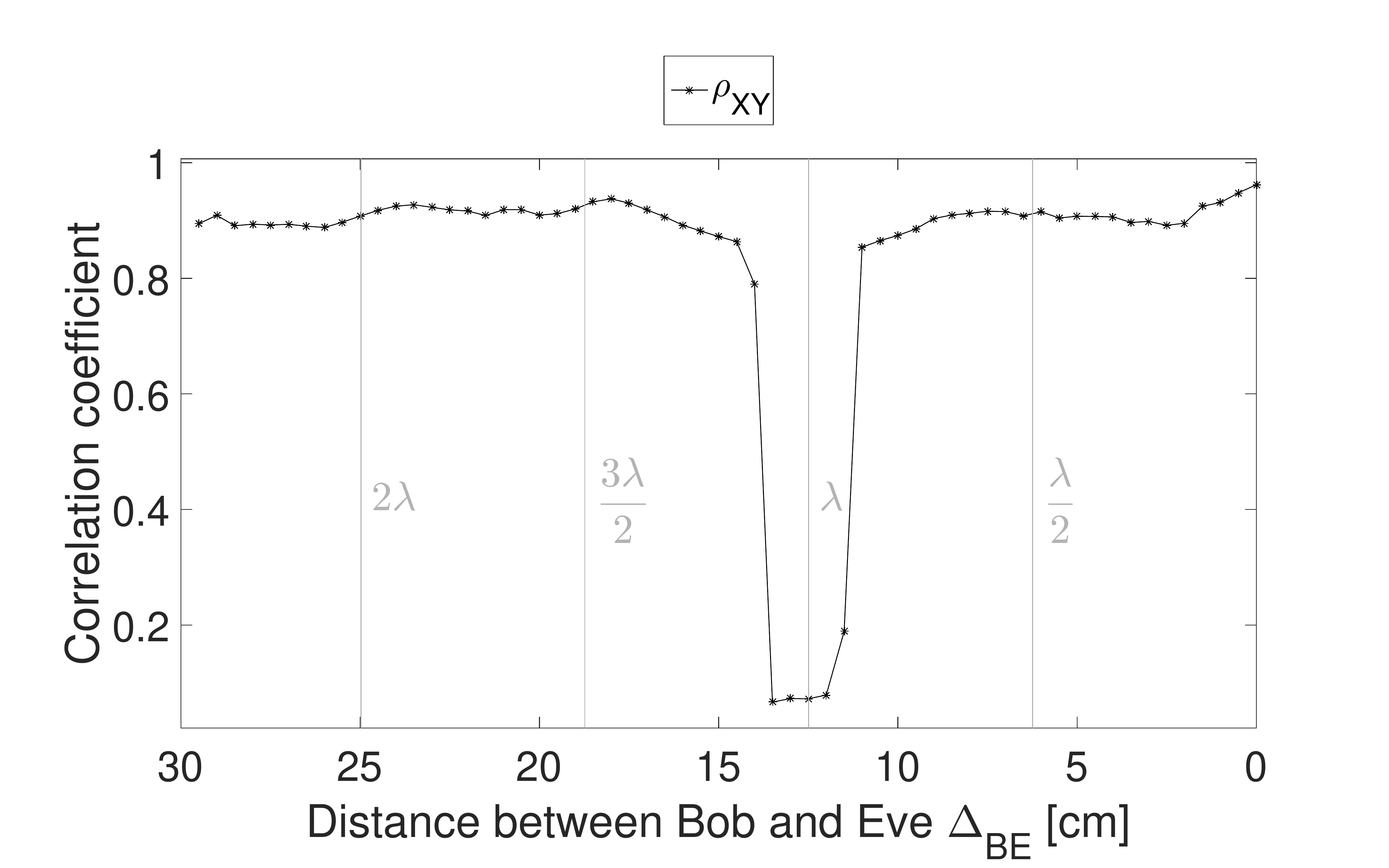}}
	\subfloat[]{\includegraphics[trim=1.8cm 0.1cm 3.5cm 1.6cm, clip=true, height=0.224\textwidth]{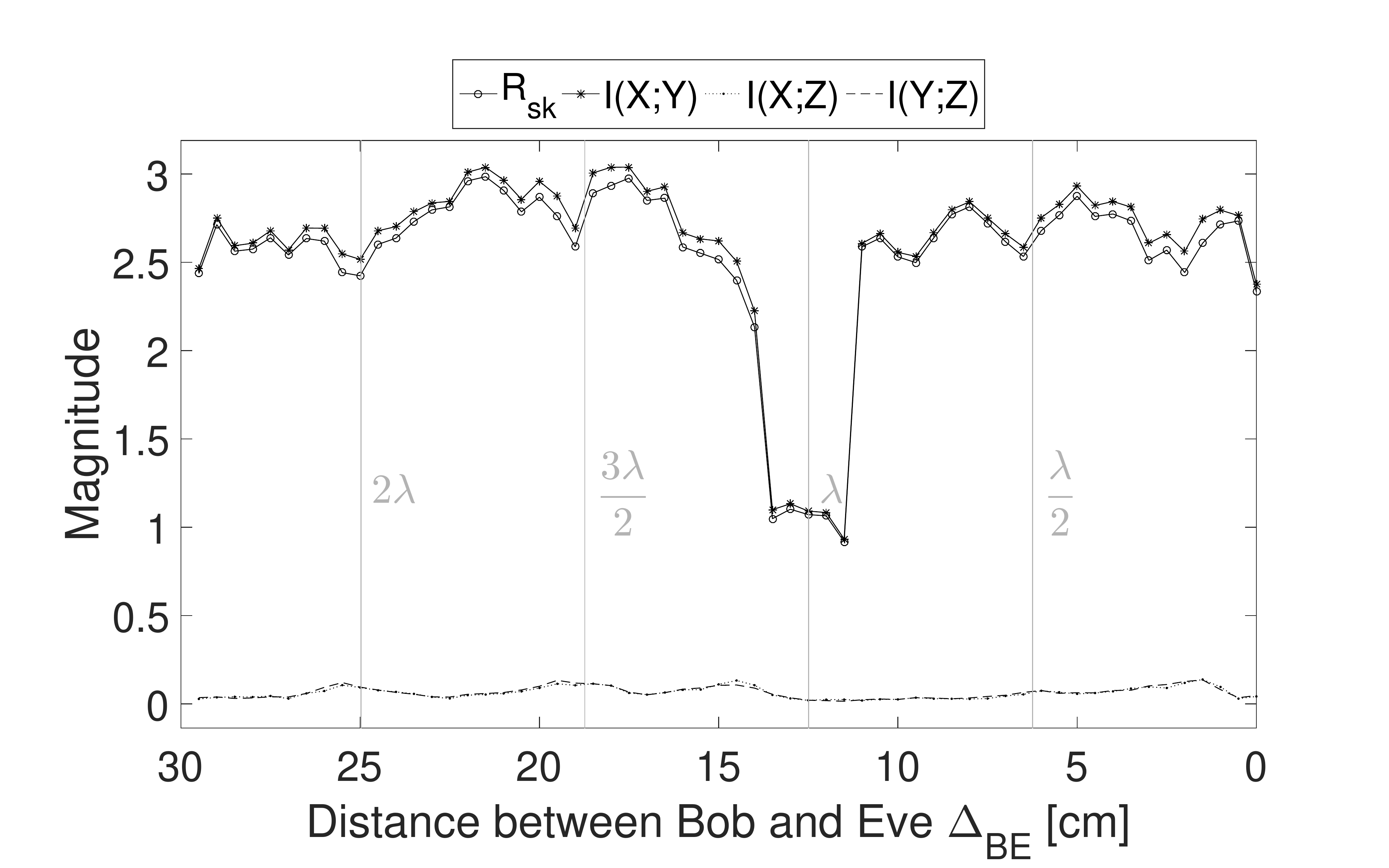}}
	\caption{Evaluation results of $\mybold{v}^{\text{de}}_k$. In (a) and (b) the cross-correlations is given; in (c) the mutual information as well as $\rsk$ is given. Position 5.}
	\label{fig:app_decorr_5}
\end{figure*}


\begin{figure*}
	\centering
	\subfloat[]{\includegraphics[trim=1.4cm 0.1cm 3.5cm 1.6cm, clip=true, height=0.224\textwidth]{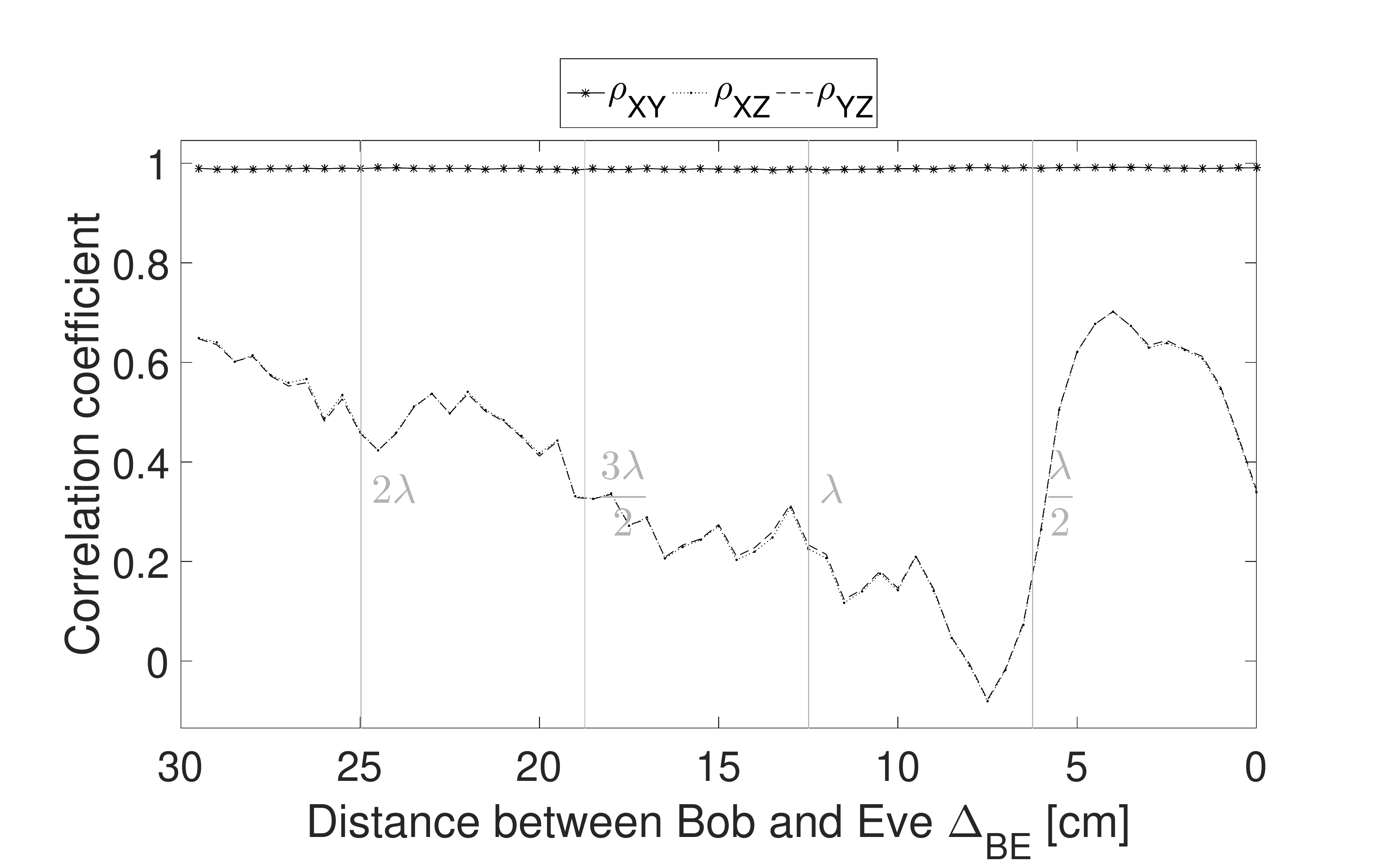}}
	\subfloat[]{\includegraphics[trim=0.5cm 0.1cm 3.5cm 1.6cm, clip=true, height=0.224\textwidth]{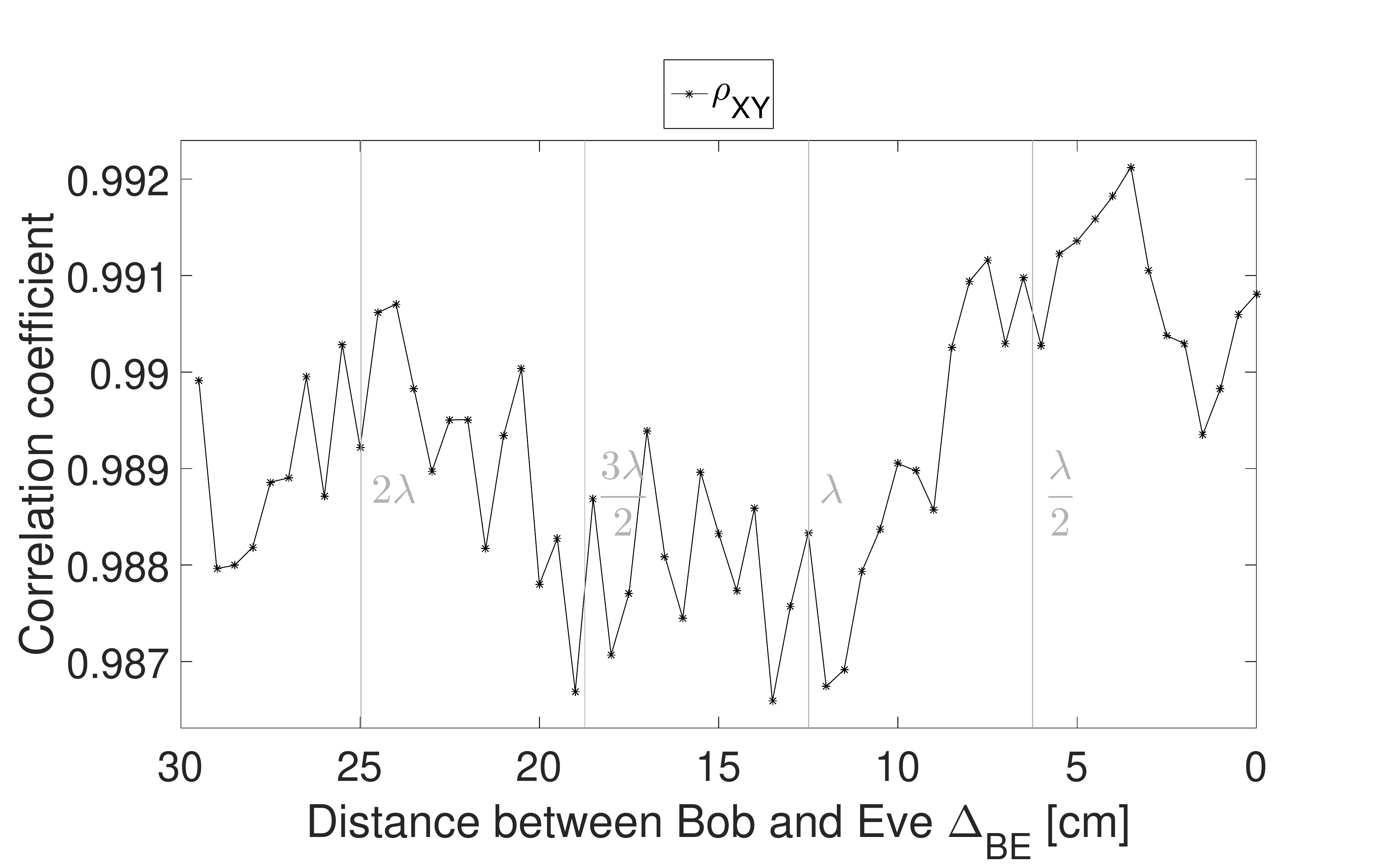}}
	\subfloat[]{\includegraphics[trim=2.2cm 0.1cm 3.5cm 1.6cm, clip=true, height=0.224\textwidth]{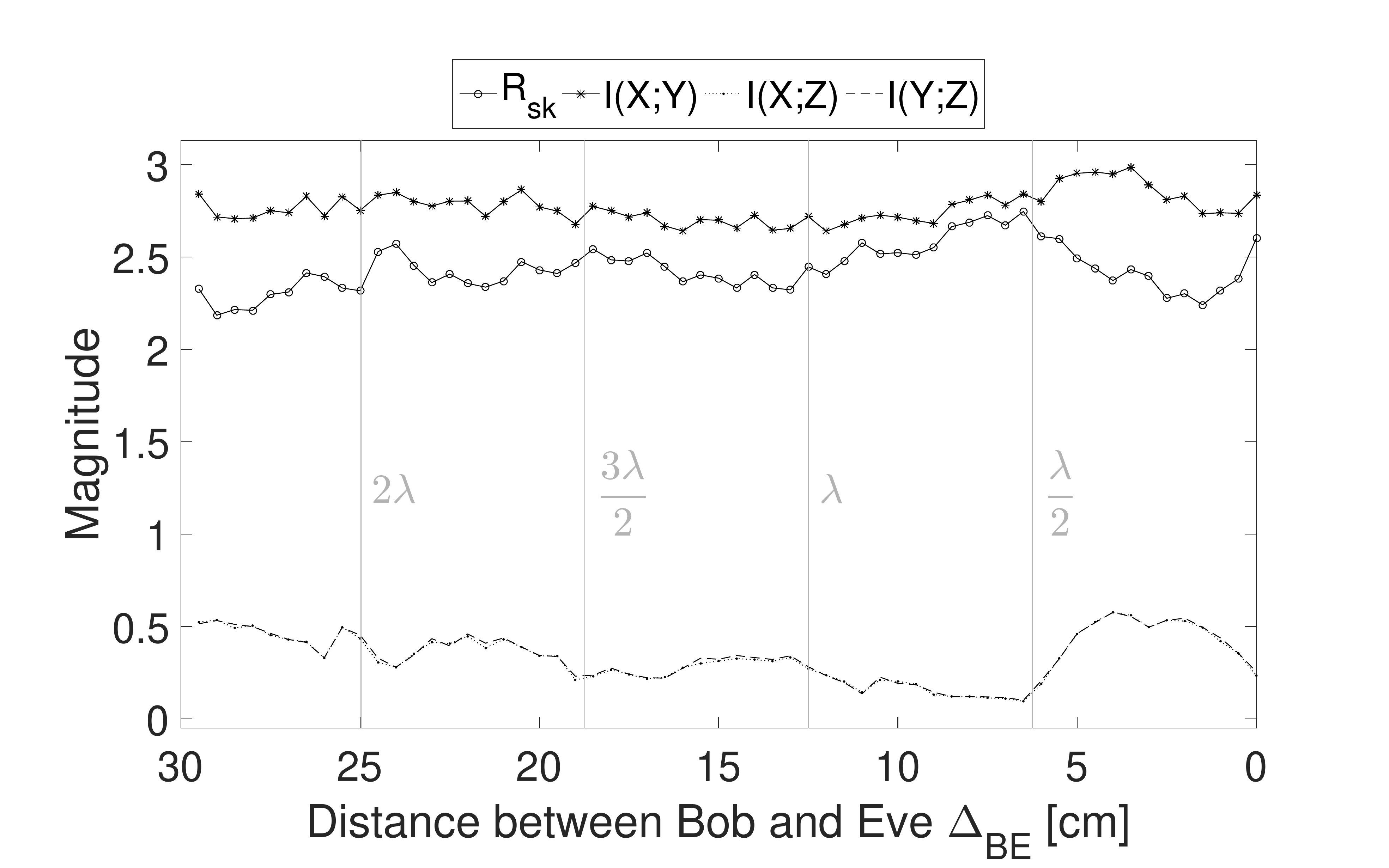}}
	\caption{Evaluation results of $\mybold{v}_k$. In (a) and (b) the cross-correlations is given; in (c) the mutual information as well as $\rsk$ is given. Position 6.}
	\label{fig:app_original_6}
\end{figure*}

\begin{figure*}
	\centering
	\subfloat[]{\includegraphics[trim=1.4cm 0.1cm 3.5cm 1.6cm, clip=true, height=0.224\textwidth]{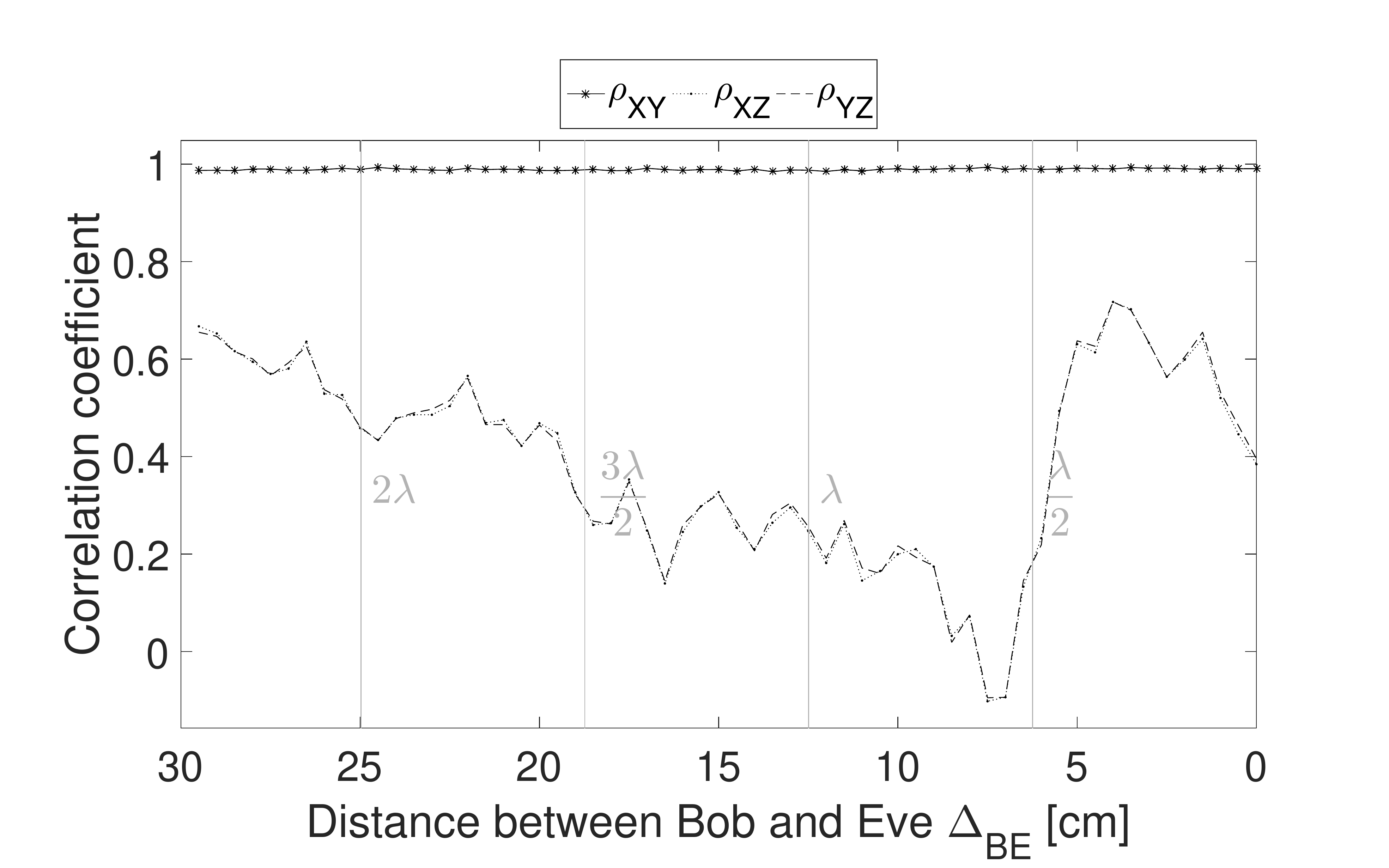}}
	\subfloat[]{\includegraphics[trim=0.5cm 0.1cm 3.5cm 1.6cm, clip=true, height=0.224\textwidth]{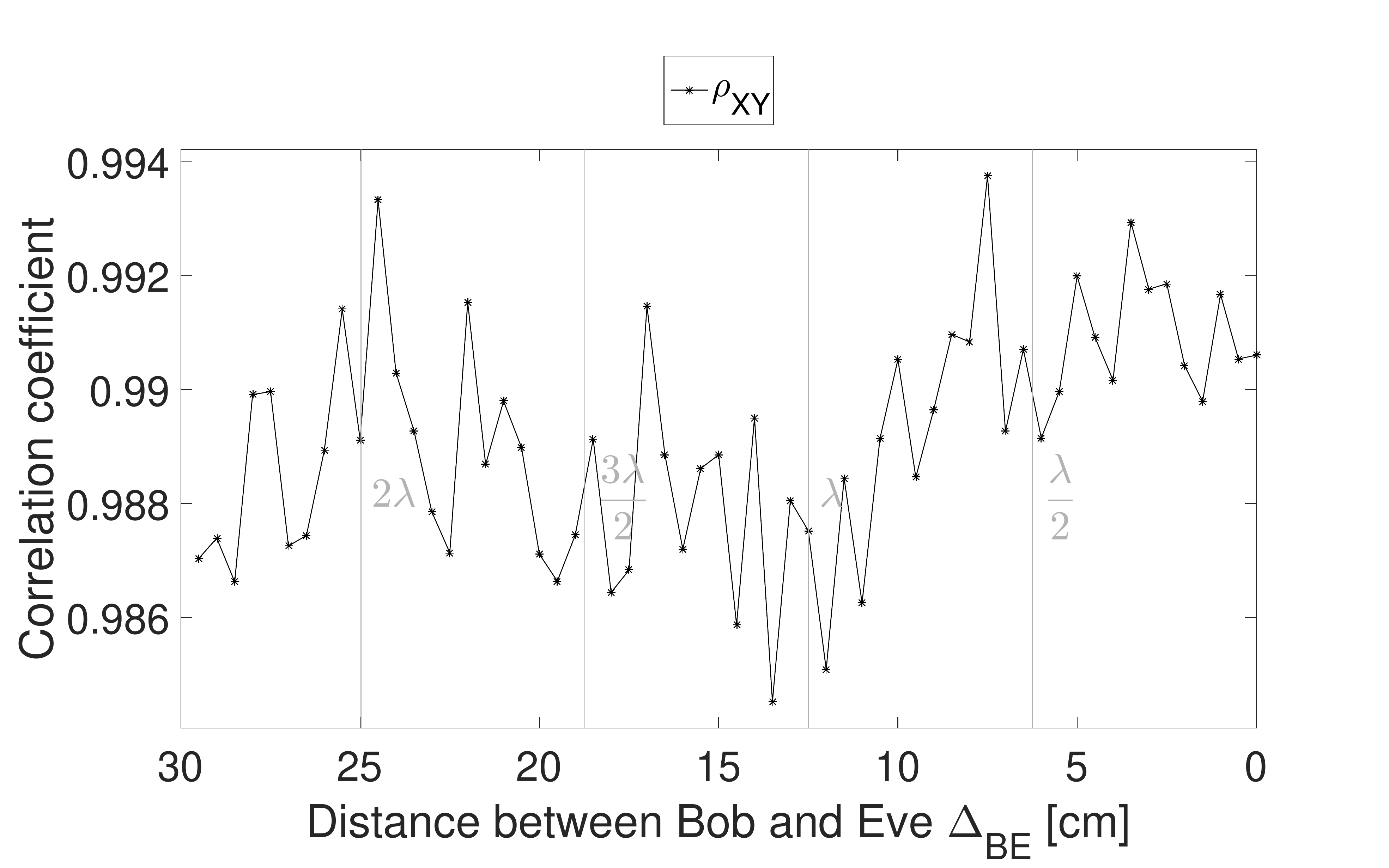}}
	\subfloat[]{\includegraphics[trim=2.2cm 0.1cm 3.5cm 1.6cm, clip=true, height=0.224\textwidth]{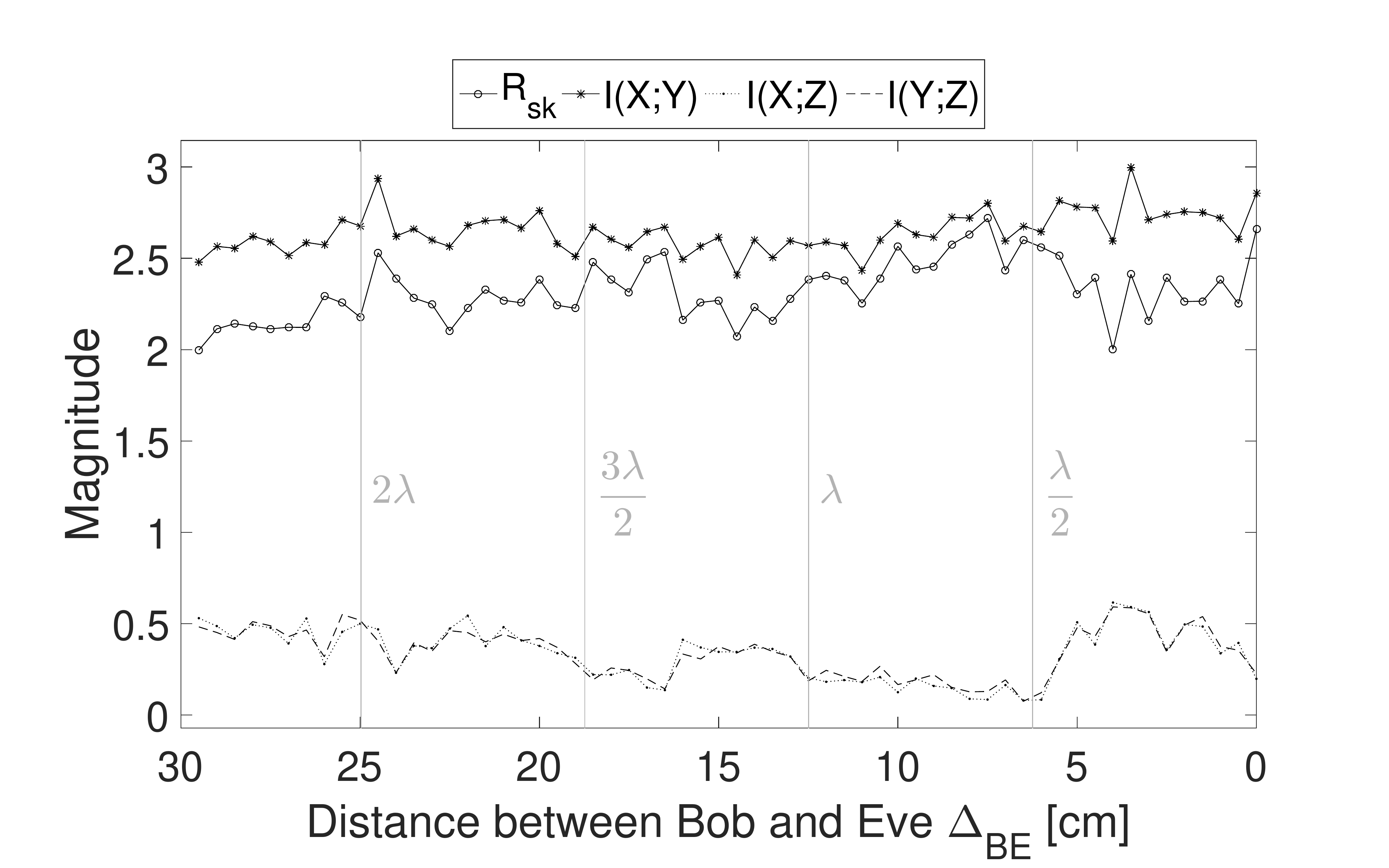}}
	\caption{Evaluation results of $\mybold{v}^{\text{ds}}_k$. In (a) and (b) the cross-correlations is given; in (c) the mutual information as well as $\rsk$ is given. Position 6.}
	\label{fig:app_ds_6}
\end{figure*}

\begin{figure*}
	\centering
	\subfloat[]{\includegraphics[trim=1.4cm 0.1cm 3.5cm 1.6cm, clip=true, height=0.224\textwidth]{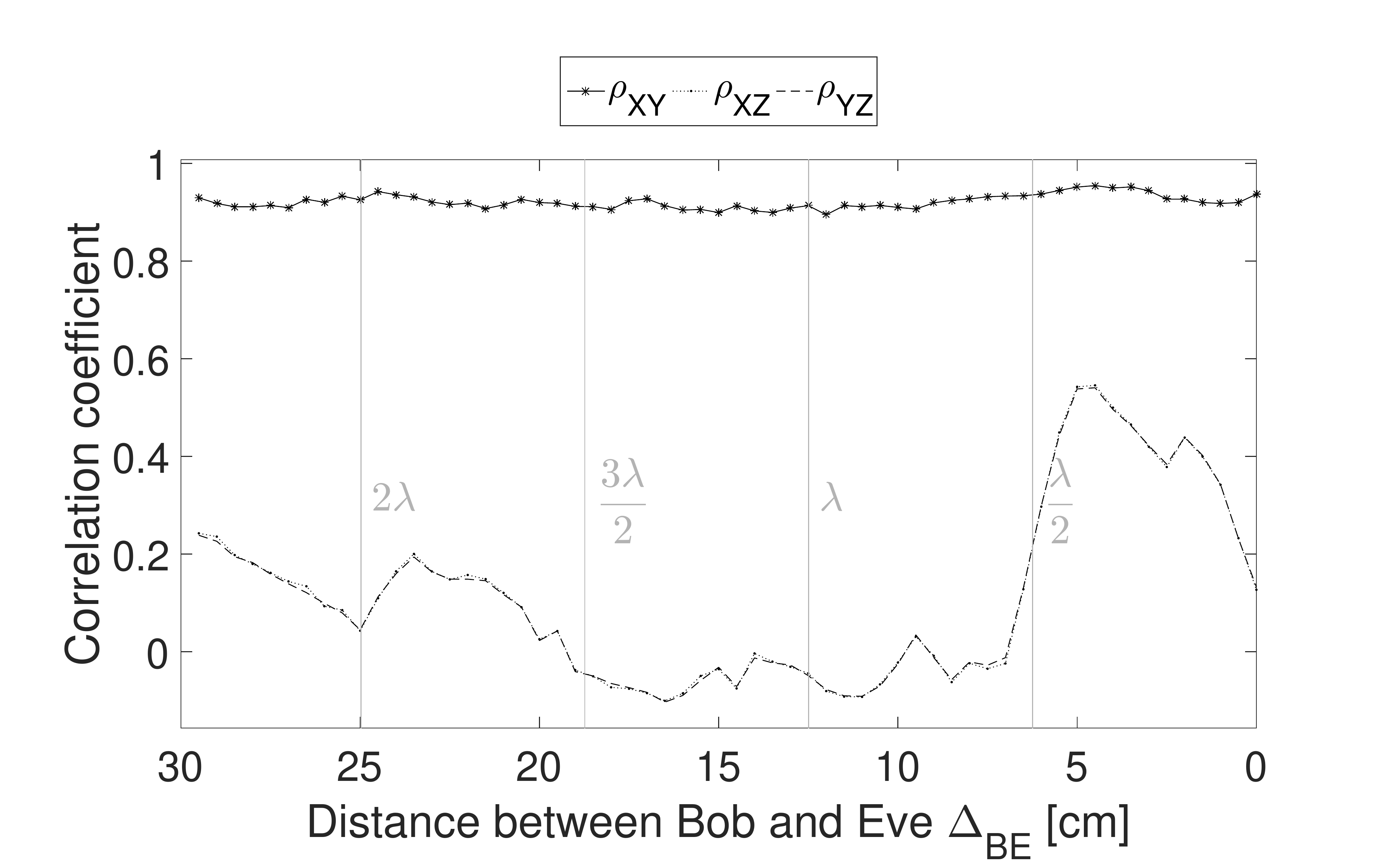}}
	\subfloat[]{\includegraphics[trim=1cm 0.1cm 3.5cm 1.6cm, clip=true, height=0.224\textwidth]{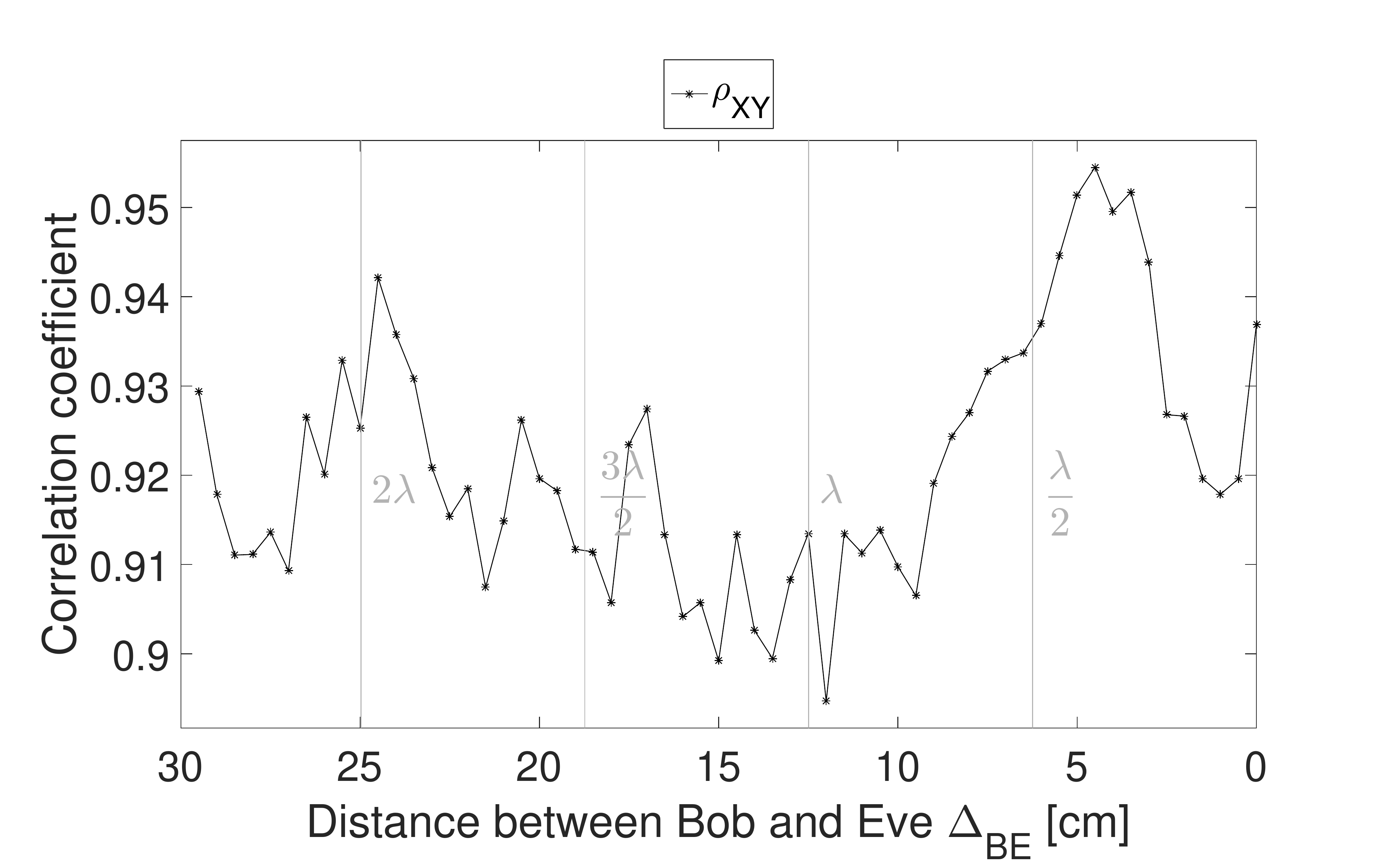}}
	\subfloat[]{\includegraphics[trim=1.8cm 0.1cm 3.5cm 1.6cm, clip=true, height=0.224\textwidth]{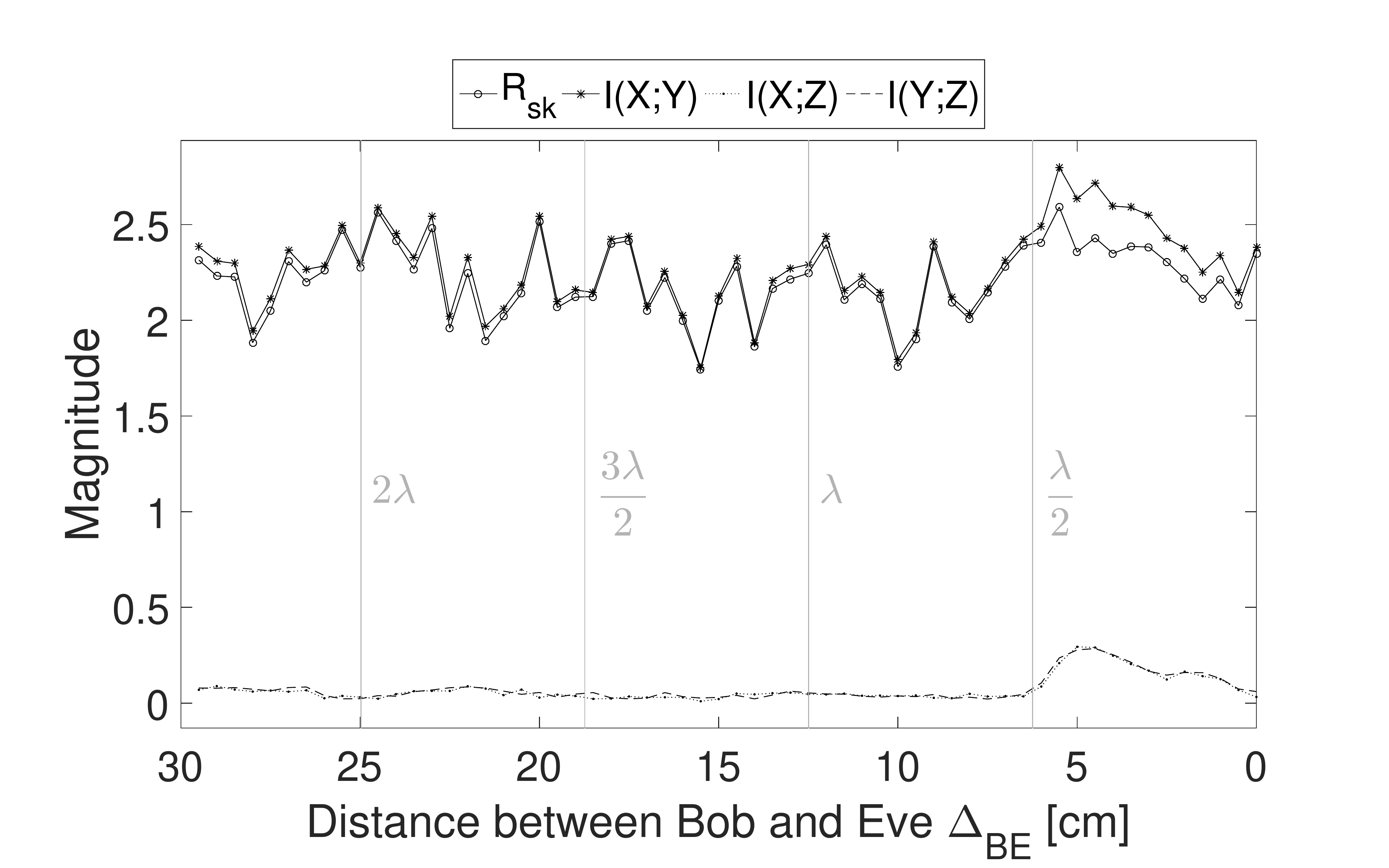}}
	\caption{Evaluation results of $\mybold{v}^{\text{de}}_k$. In (a) and (b) the cross-correlations is given; in (c) the mutual information as well as $\rsk$ is given. Position 6.}
	\label{fig:app_decorr_6}
\end{figure*}


\begin{figure*}
	\centering
	\subfloat[]{\includegraphics[trim=1.4cm 0.1cm 3.5cm 1.6cm, clip=true, height=0.224\textwidth]{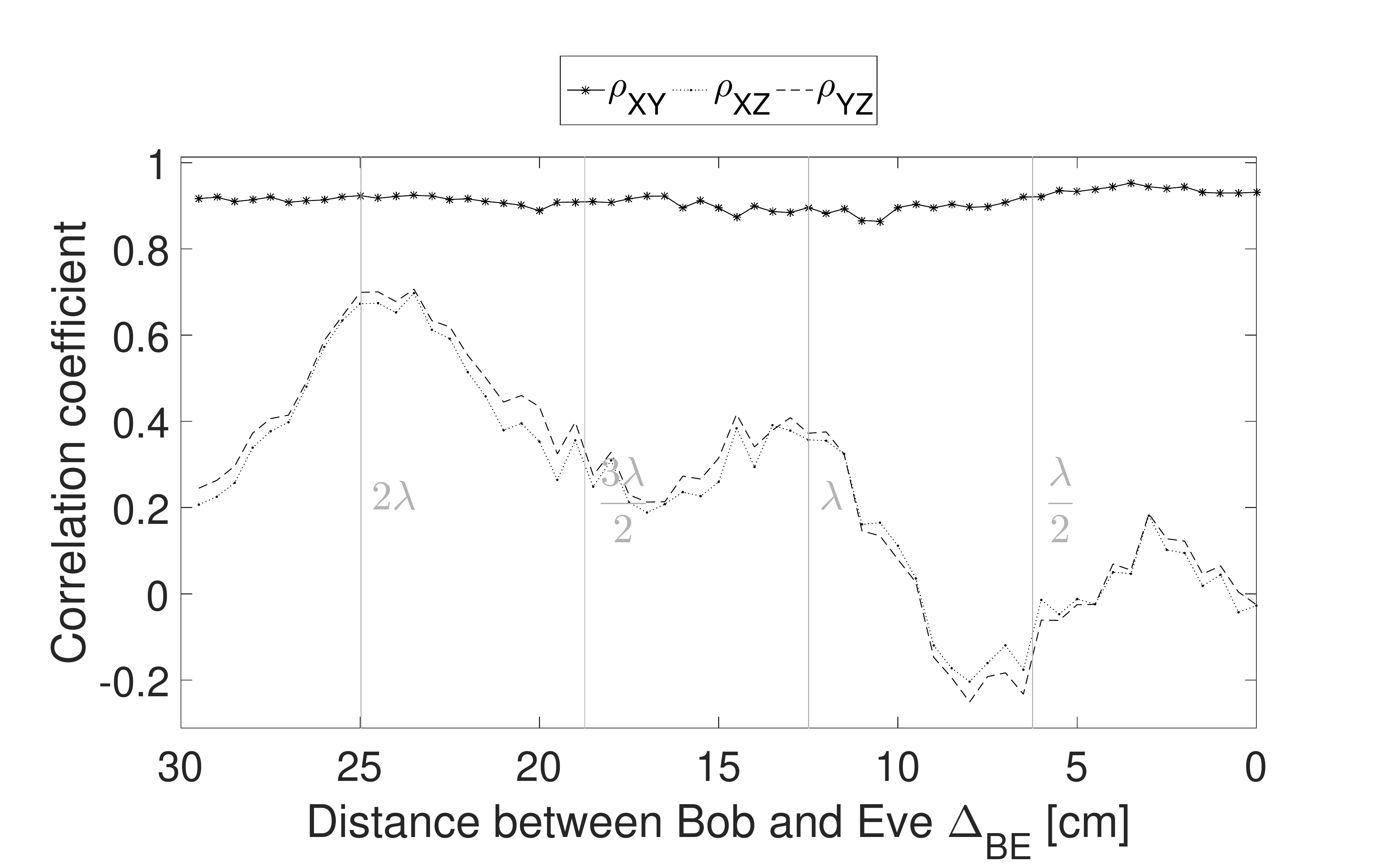}}
	\subfloat[]{\includegraphics[trim=0.5cm 0.1cm 3.5cm 1.6cm, clip=true, height=0.224\textwidth]{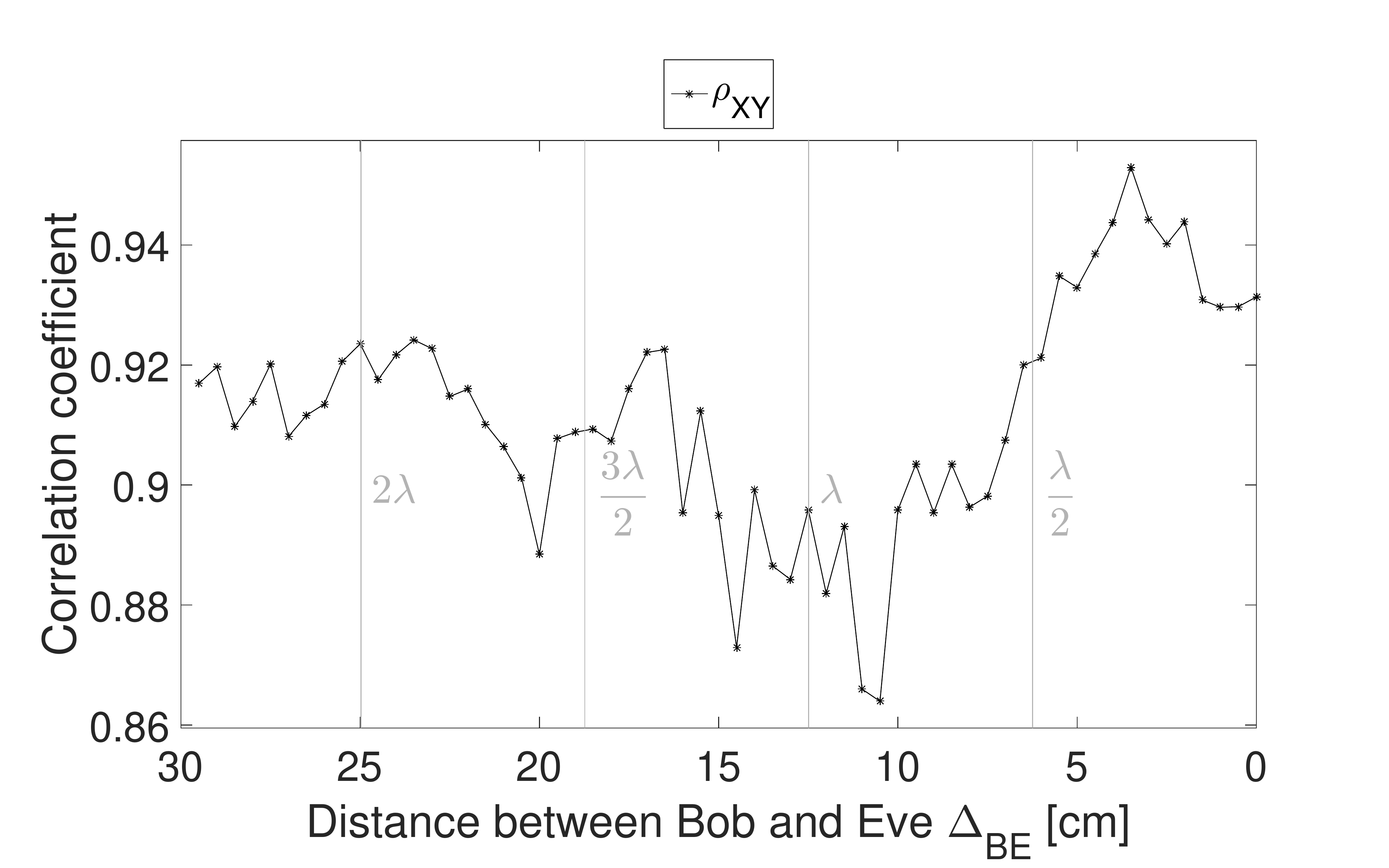}}
	\subfloat[]{\includegraphics[trim=2.2cm 0.1cm 3.5cm 1.6cm, clip=true, height=0.224\textwidth]{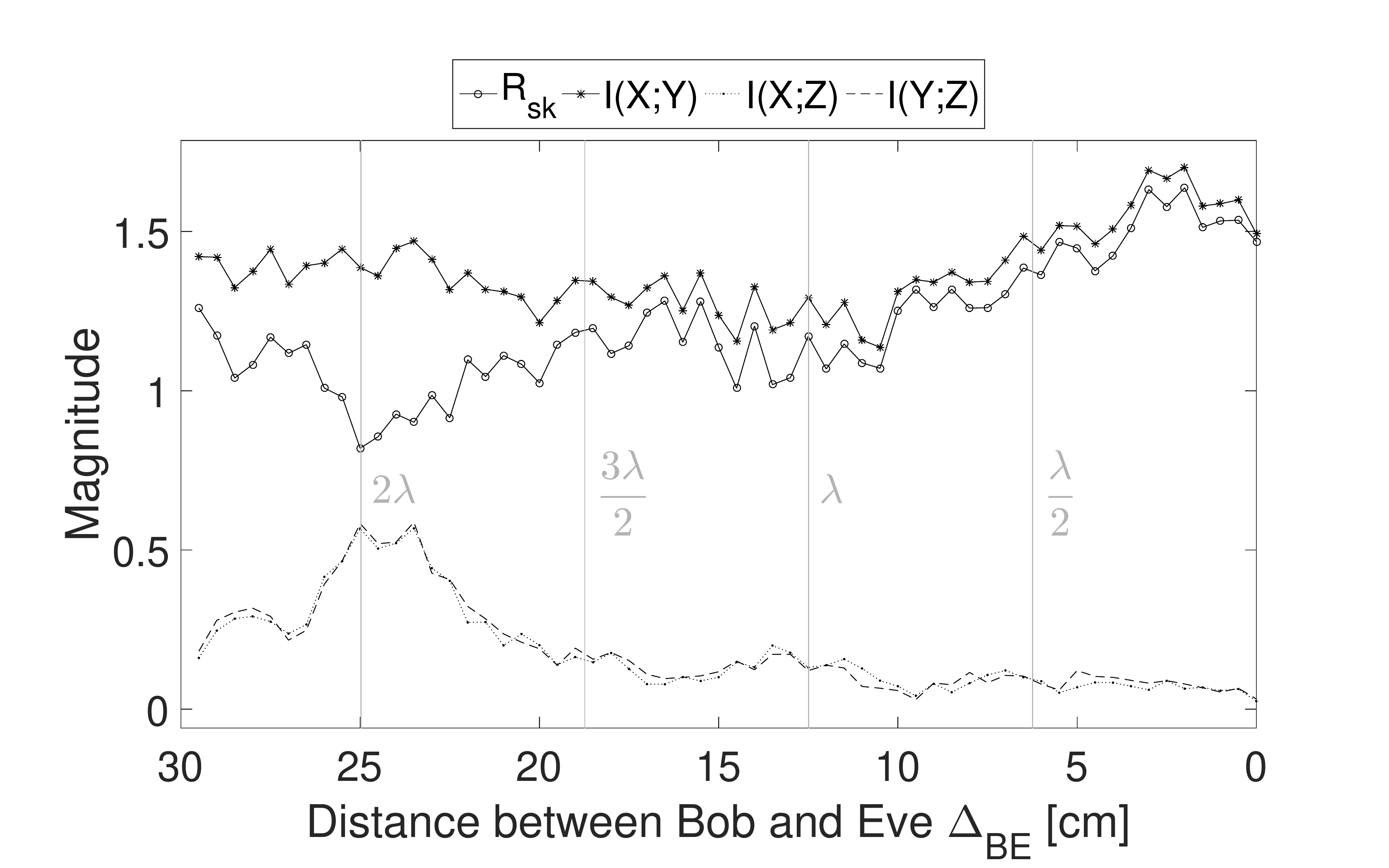}}
	\caption{Evaluation results of $\mybold{v}_k$. In (a) and (b) the cross-correlations is given; in (c) the mutual information as well as $\rsk$ is given. Position 7.}
	\label{fig:app_original_7}
\end{figure*}

\begin{figure*}
	\centering
	\subfloat[]{\includegraphics[trim=1.4cm 0.1cm 3.5cm 1.6cm, clip=true, height=0.224\textwidth]{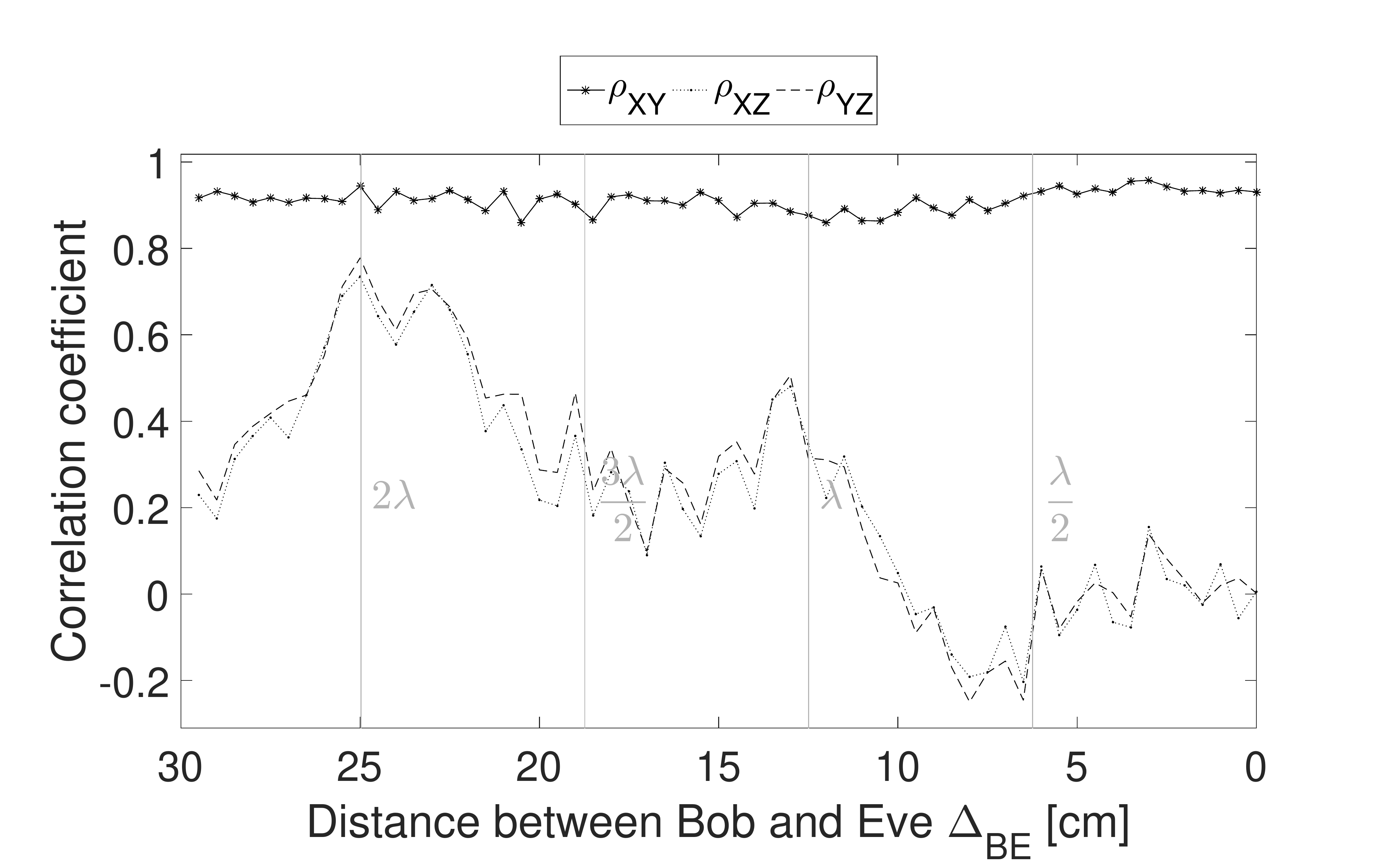}}
	\subfloat[]{\includegraphics[trim=0.5cm 0.1cm 3.5cm 1.6cm, clip=true, height=0.224\textwidth]{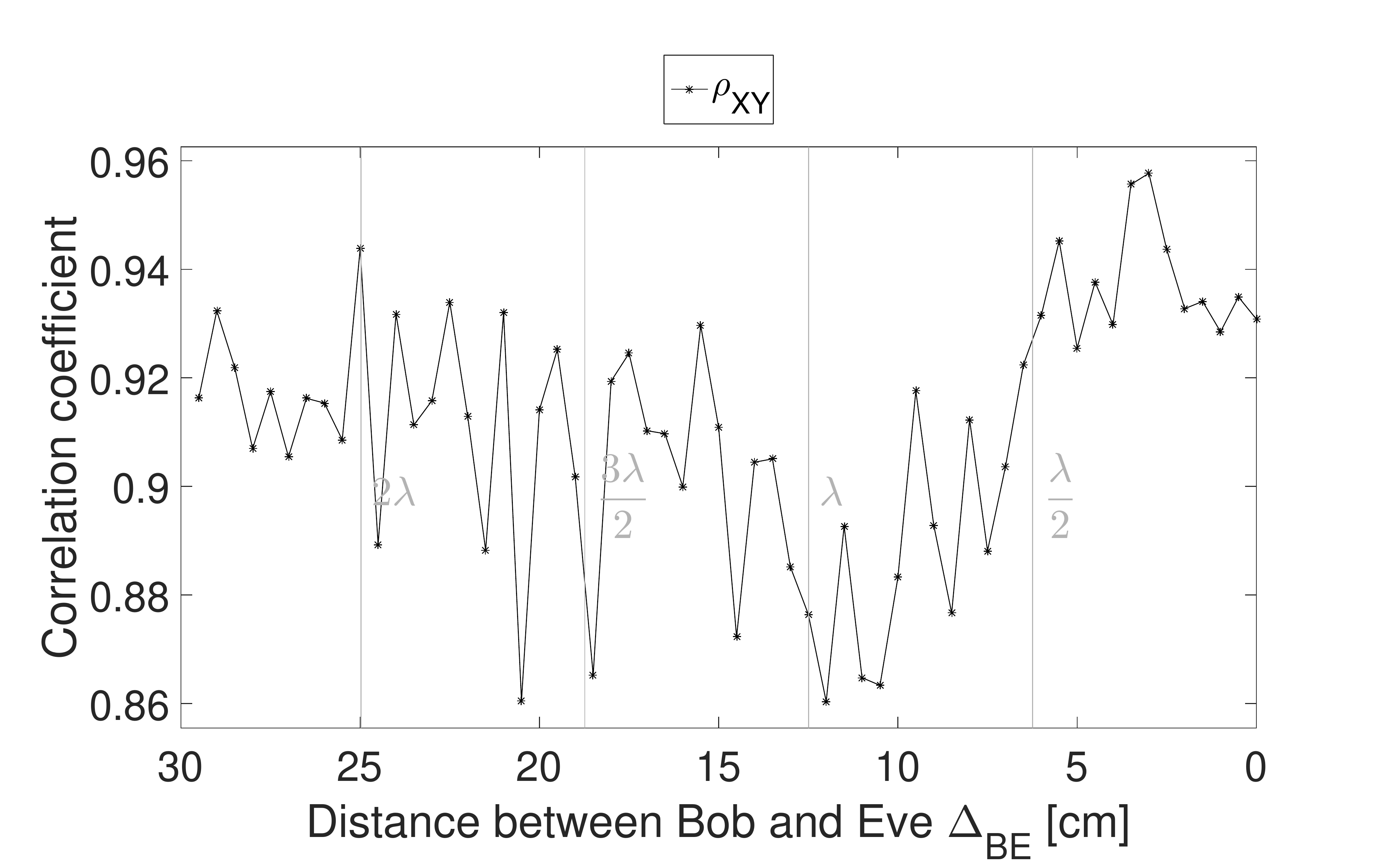}}
	\subfloat[]{\includegraphics[trim=2.2cm 0.1cm 3.5cm 1.6cm, clip=true, height=0.224\textwidth]{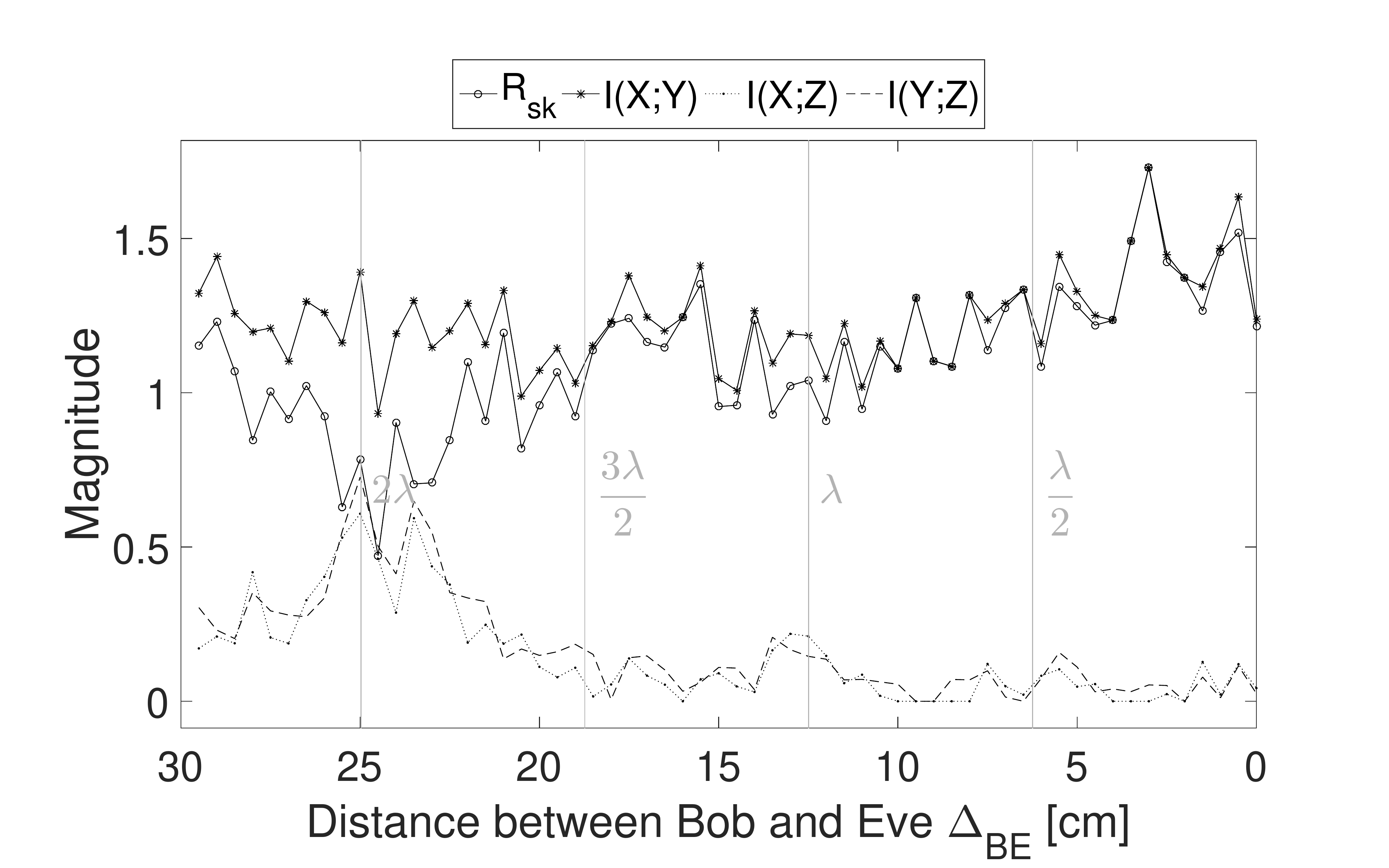}}
	\caption{Evaluation results of $\mybold{v}^{\text{ds}}_k$. In (a) and (b) the cross-correlations is given; in (c) the mutual information as well as $\rsk$ is given. Position 7.}
	\label{fig:app_ds_7}
\end{figure*}

\begin{figure*}
	\centering
	\subfloat[]{\includegraphics[trim=1.4cm 0.1cm 3.5cm 1.6cm, clip=true, height=0.224\textwidth]{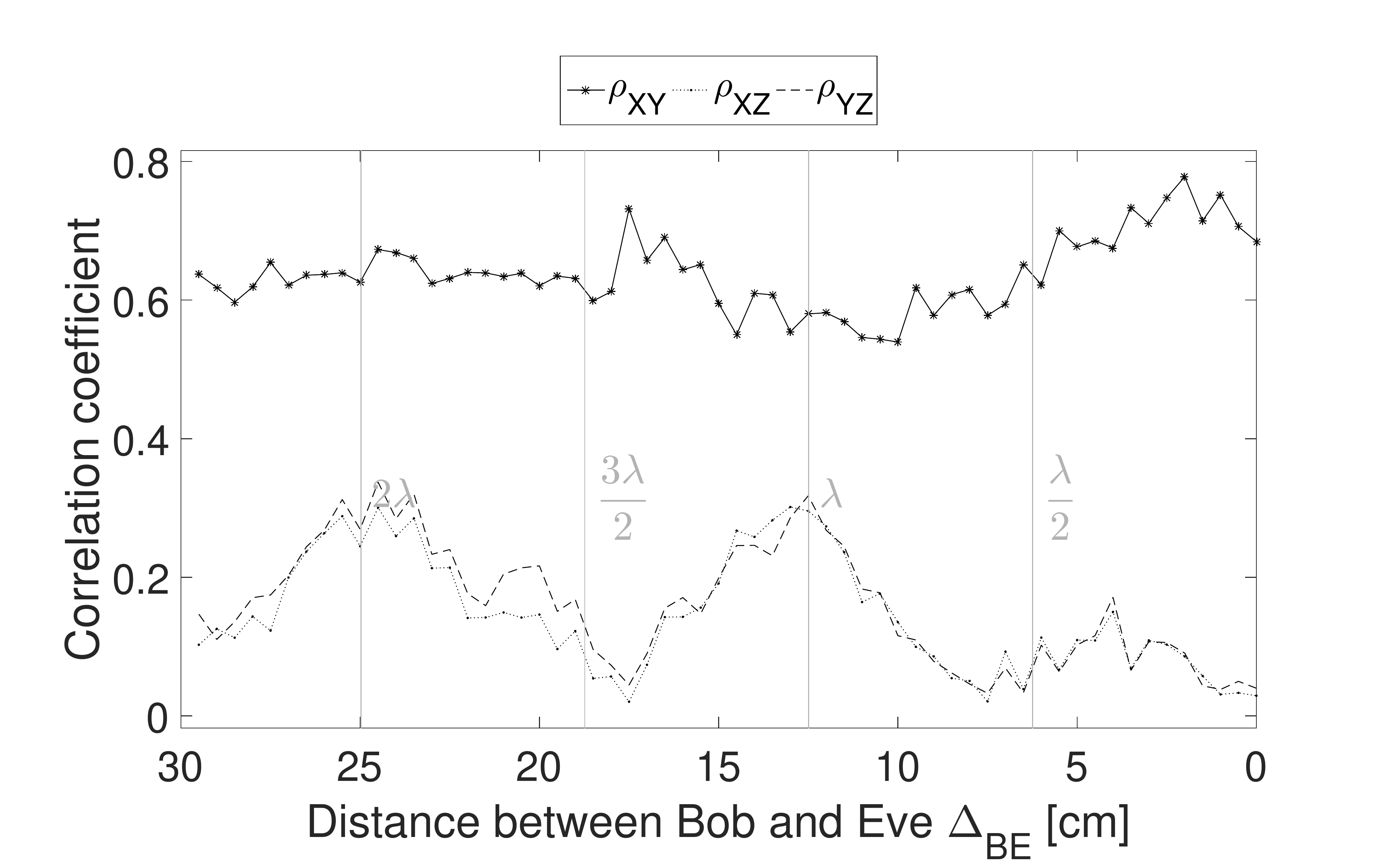}}
	\subfloat[]{\includegraphics[trim=1cm 0.1cm 3.5cm 1.6cm, clip=true, height=0.224\textwidth]{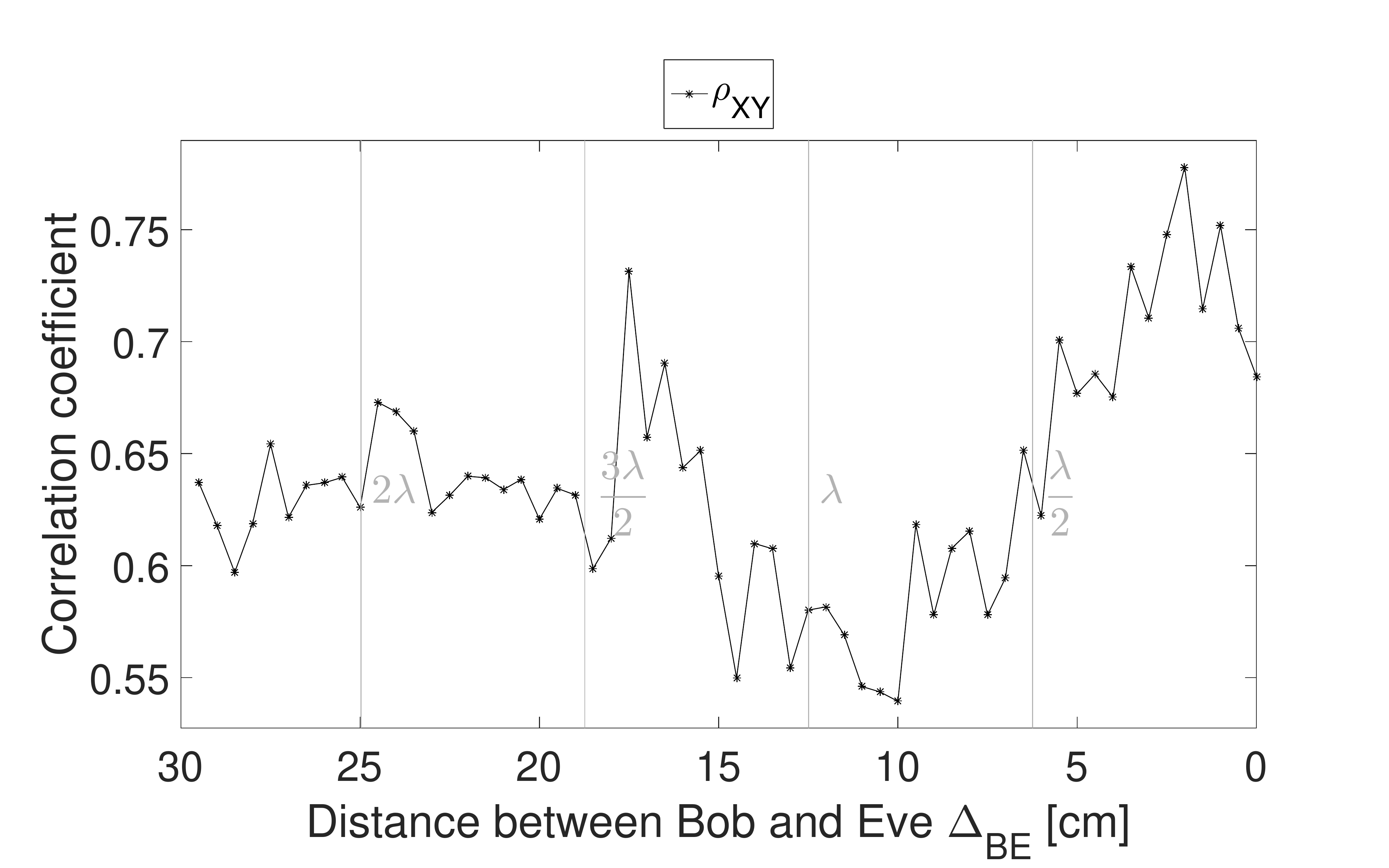}}
	\subfloat[]{\includegraphics[trim=1.8cm 0.1cm 3.5cm 1.6cm, clip=true, height=0.224\textwidth]{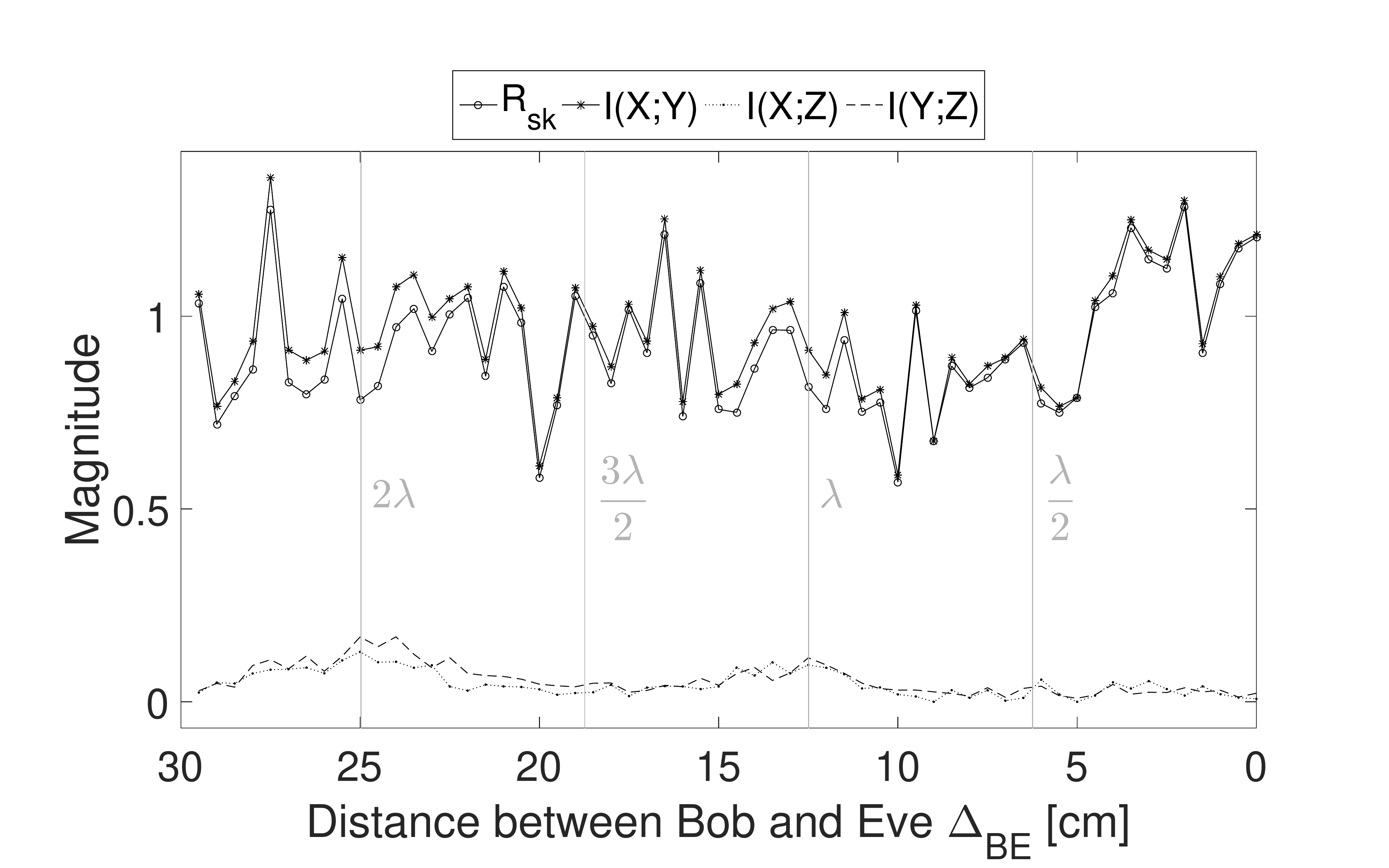}}
	\caption{Evaluation results of $\mybold{v}^{\text{de}}_k$. In (a) and (b) the cross-correlations is given; in (c) the mutual information as well as $\rsk$ is given. Position 7.}
	\label{fig:app_decorr_7}
\end{figure*}


\begin{figure*}
	\centering
	\subfloat[]{\includegraphics[trim=1.4cm 0.1cm 3.5cm 1.6cm, clip=true, height=0.224\textwidth]{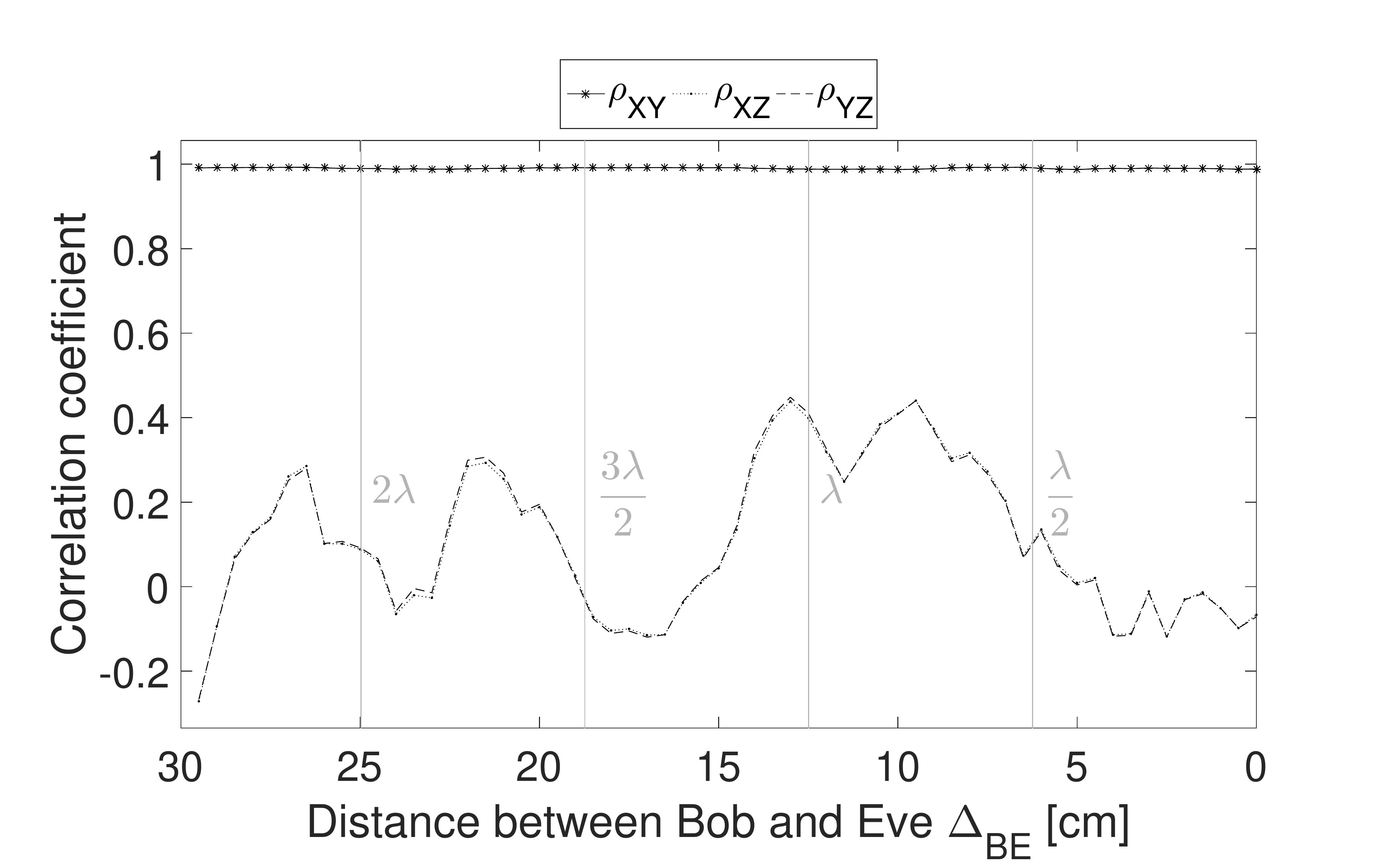}}
	\subfloat[]{\includegraphics[trim=0.5cm 0.1cm 3.5cm 1.6cm, clip=true, height=0.224\textwidth]{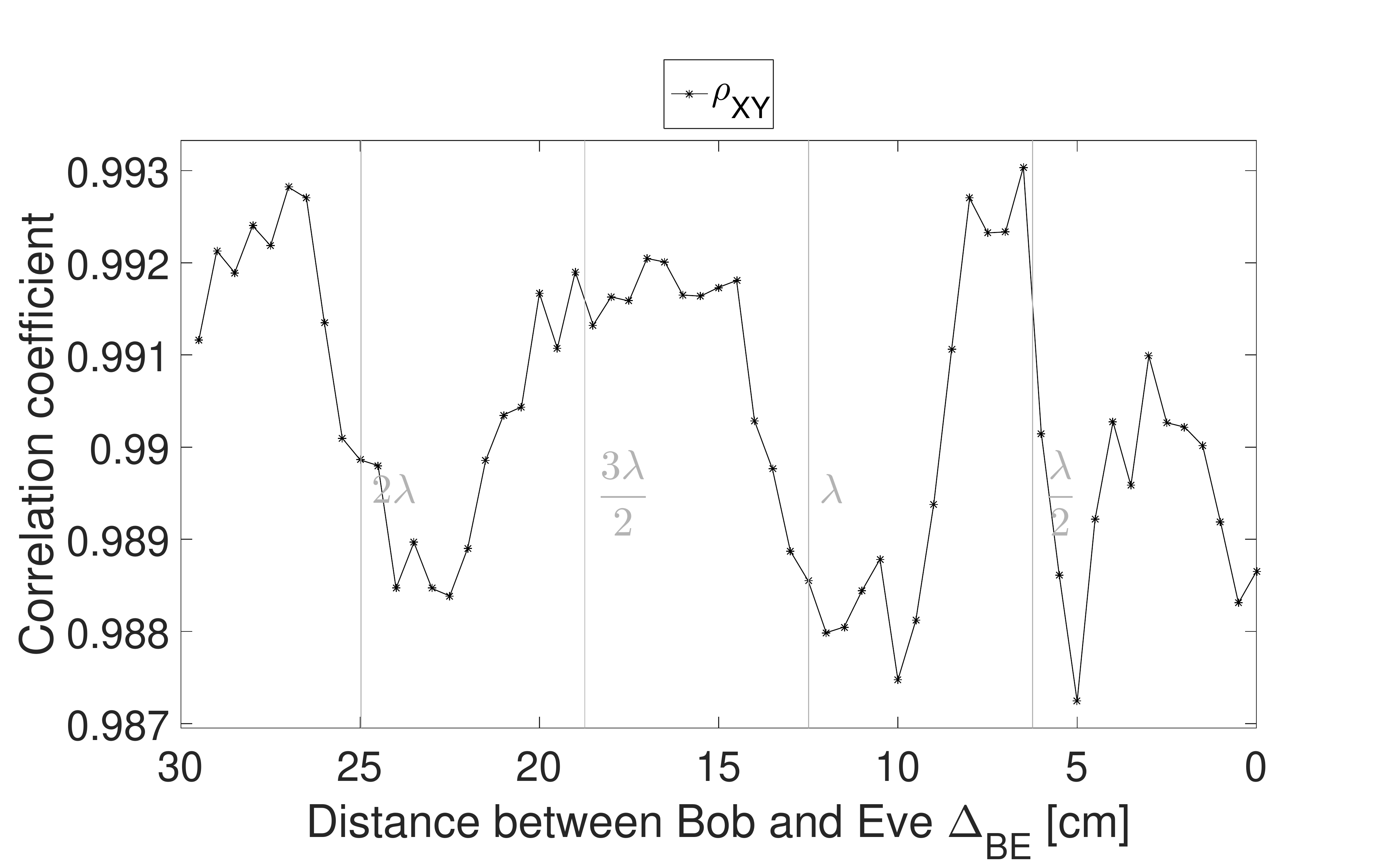}}
	\subfloat[]{\includegraphics[trim=2.2cm 0.1cm 3.5cm 1.6cm, clip=true, height=0.224\textwidth]{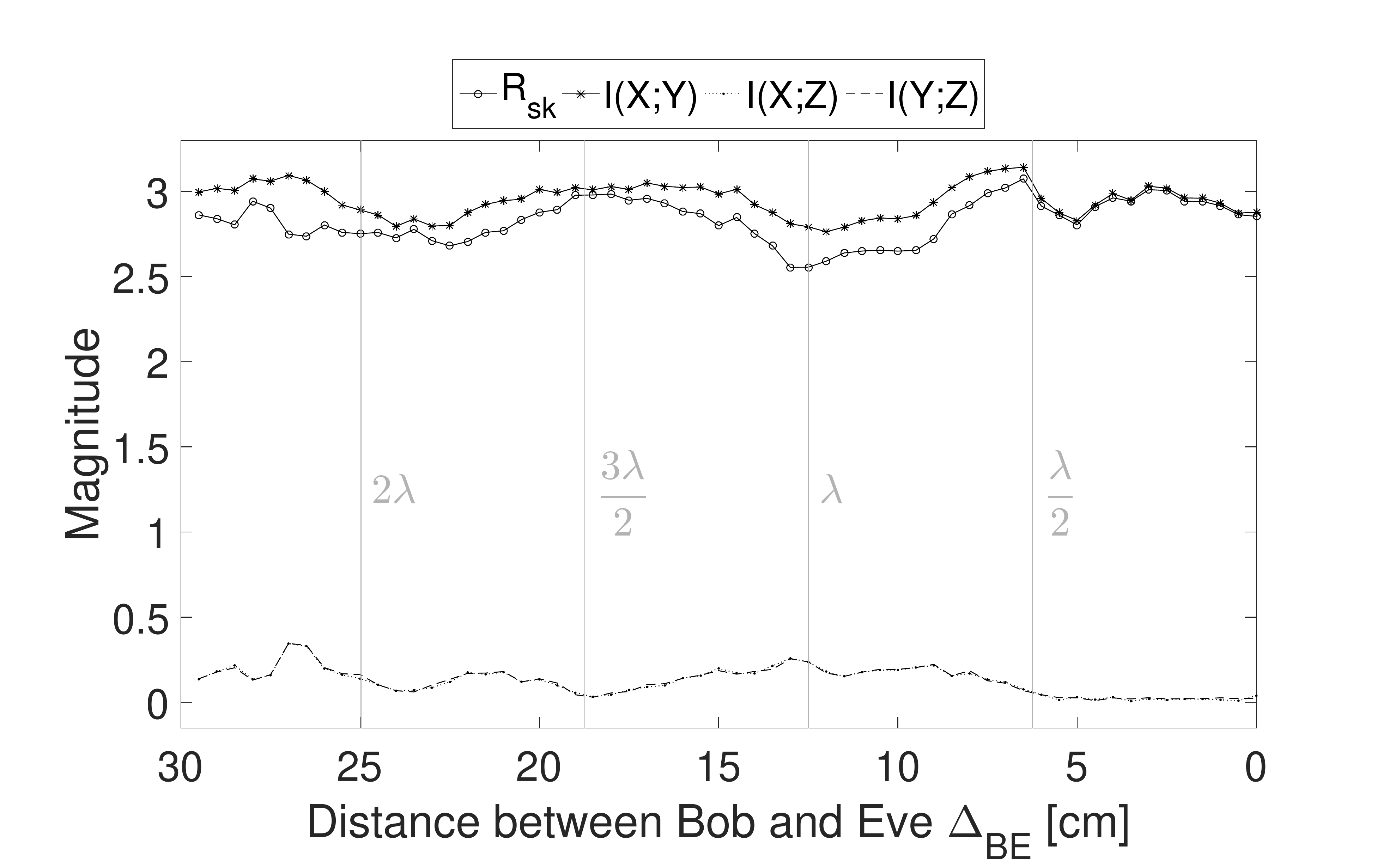}}
	\caption{Evaluation results of $\mybold{v}_k$. In (a) and (b) the cross-correlations is given; in (c) the mutual information as well as $\rsk$ is given. Position 8.}
	\label{fig:app_original_8}
\end{figure*}

\begin{figure*}
	\centering
	\subfloat[]{\includegraphics[trim=1.4cm 0.1cm 3.5cm 1.6cm, clip=true, height=0.224\textwidth]{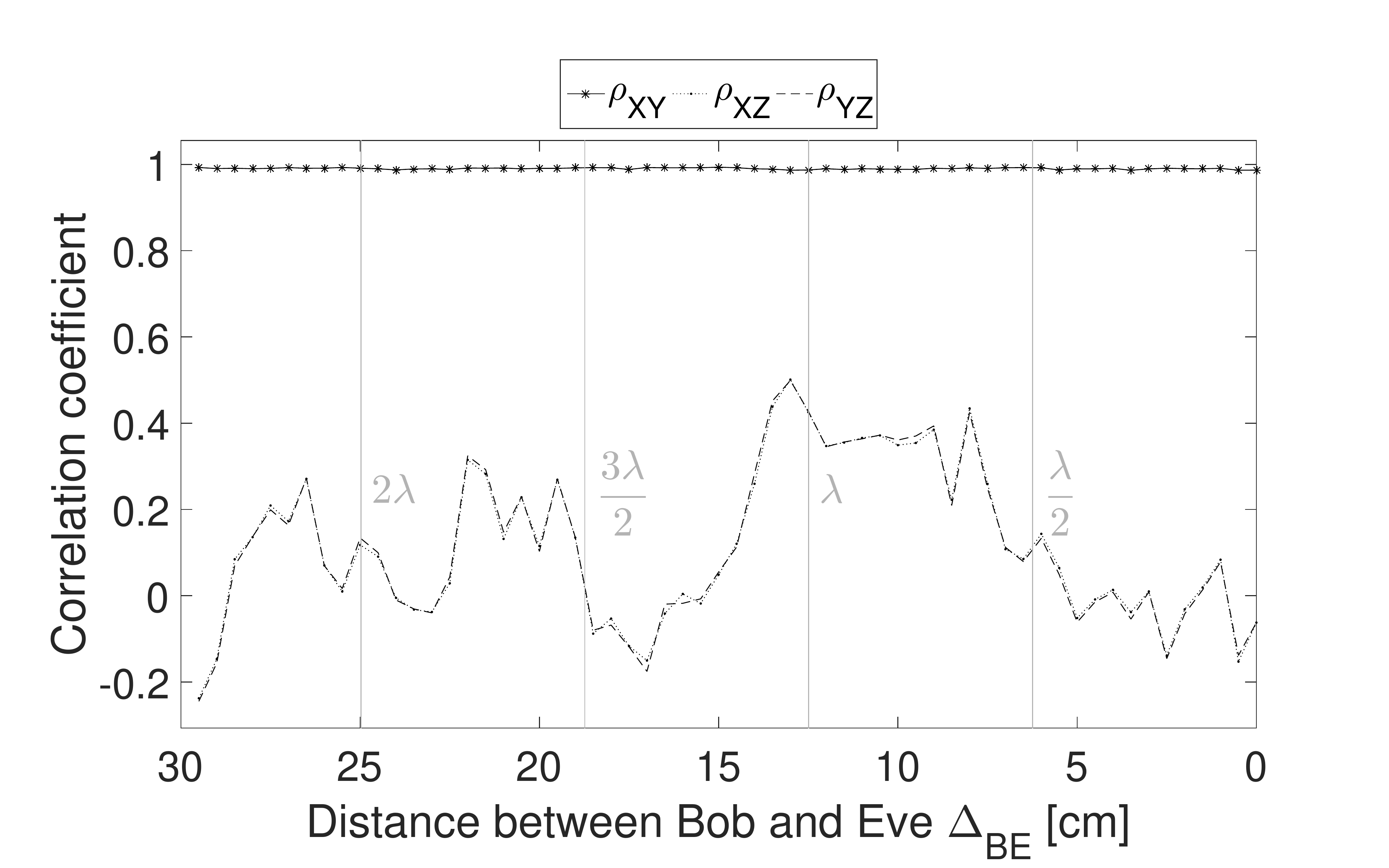}}
	\subfloat[]{\includegraphics[trim=0.5cm 0.1cm 3.5cm 1.6cm, clip=true, height=0.224\textwidth]{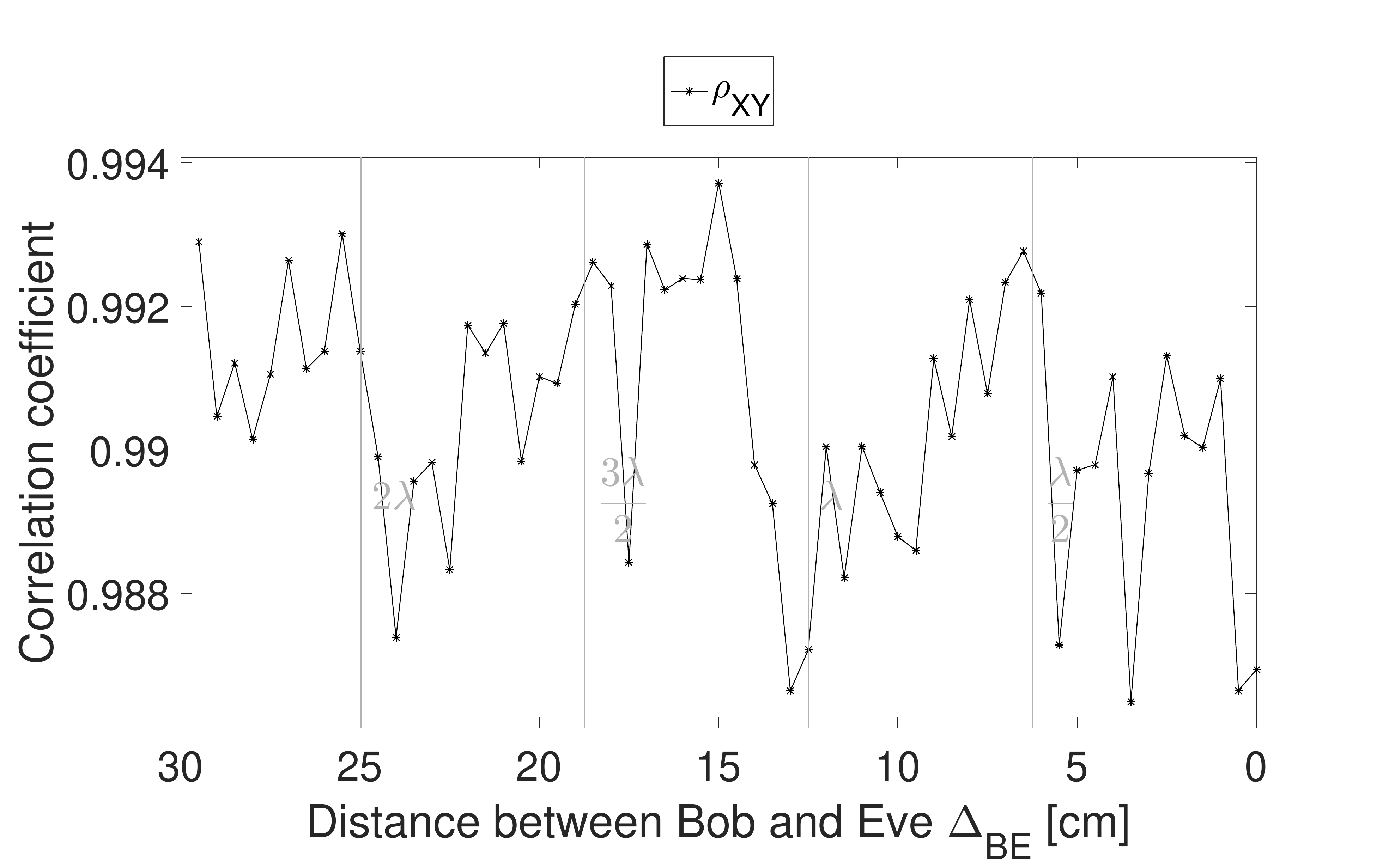}}
	\subfloat[]{\includegraphics[trim=2.2cm 0.1cm 3.5cm 1.6cm, clip=true, height=0.224\textwidth]{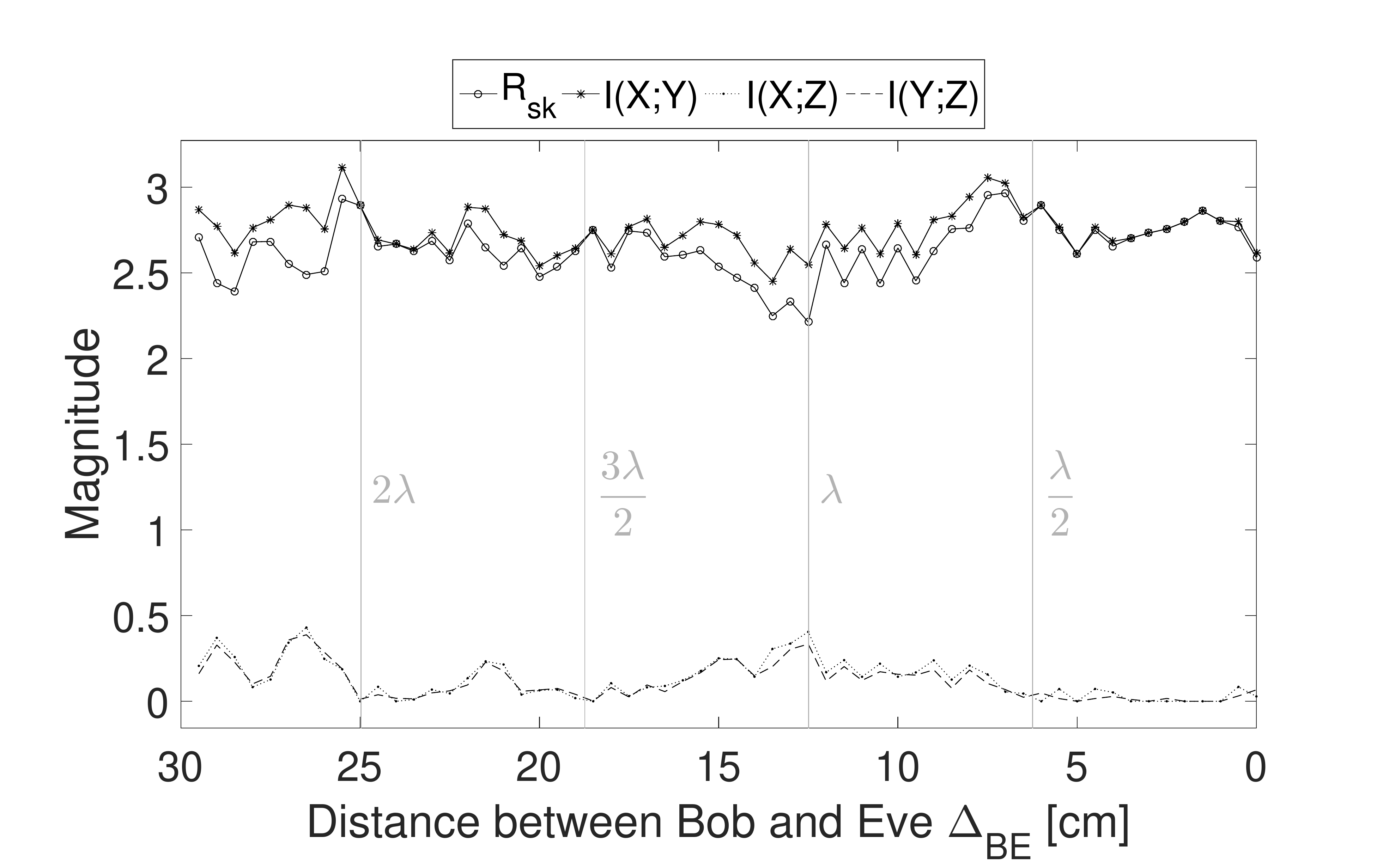}}
	\caption{Evaluation results of $\mybold{v}^{\text{ds}}_k$. In (a) and (b) the cross-correlations is given; in (c) the mutual information as well as $\rsk$ is given. Position 8.}
	\label{fig:app_ds_8}
\end{figure*}

\begin{figure*}
	\centering
	\subfloat[]{\includegraphics[trim=1.4cm 0.1cm 3.5cm 1.6cm, clip=true, height=0.224\textwidth]{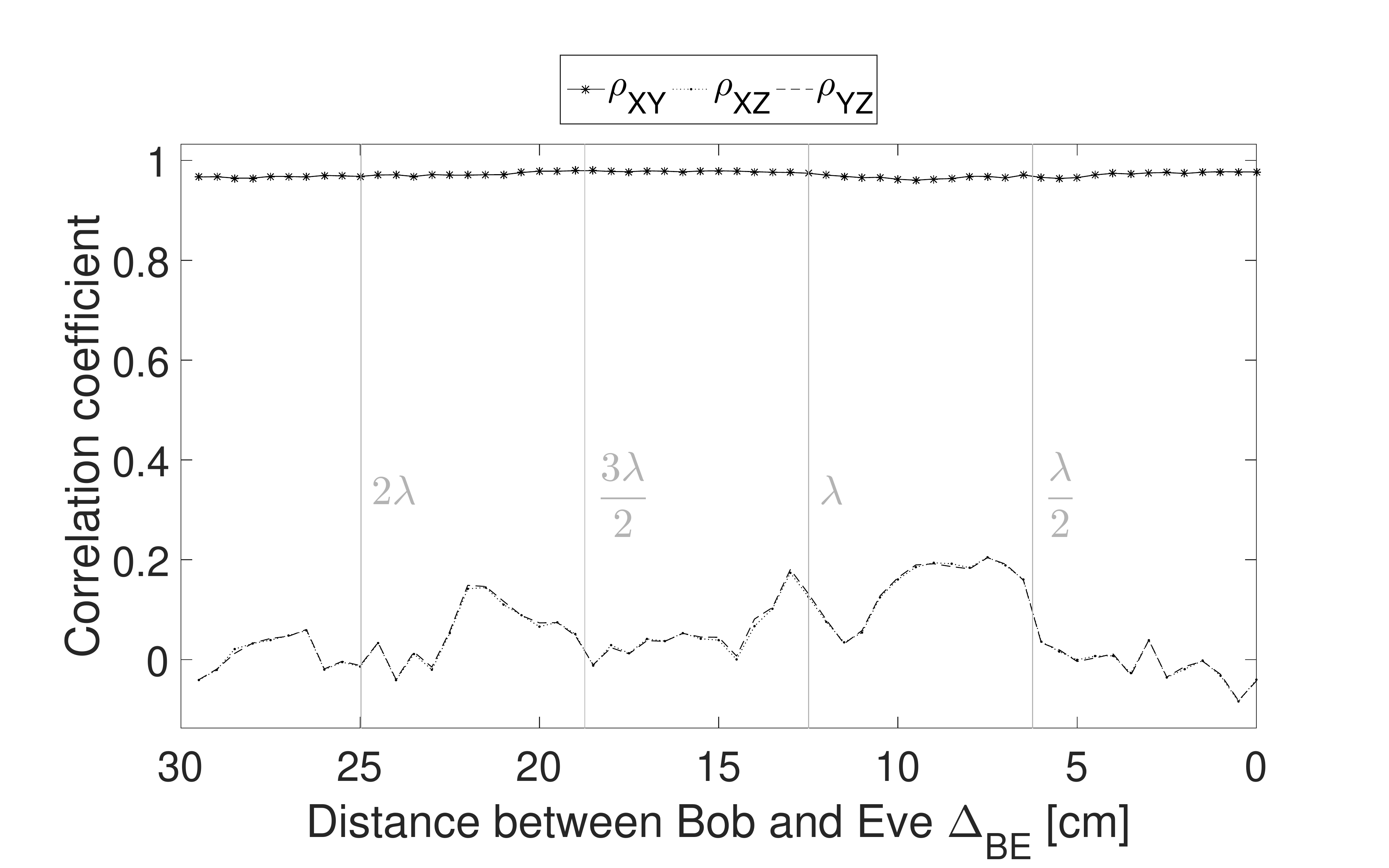}}
	\subfloat[]{\includegraphics[trim=1cm 0.1cm 3.5cm 1.6cm, clip=true, height=0.224\textwidth]{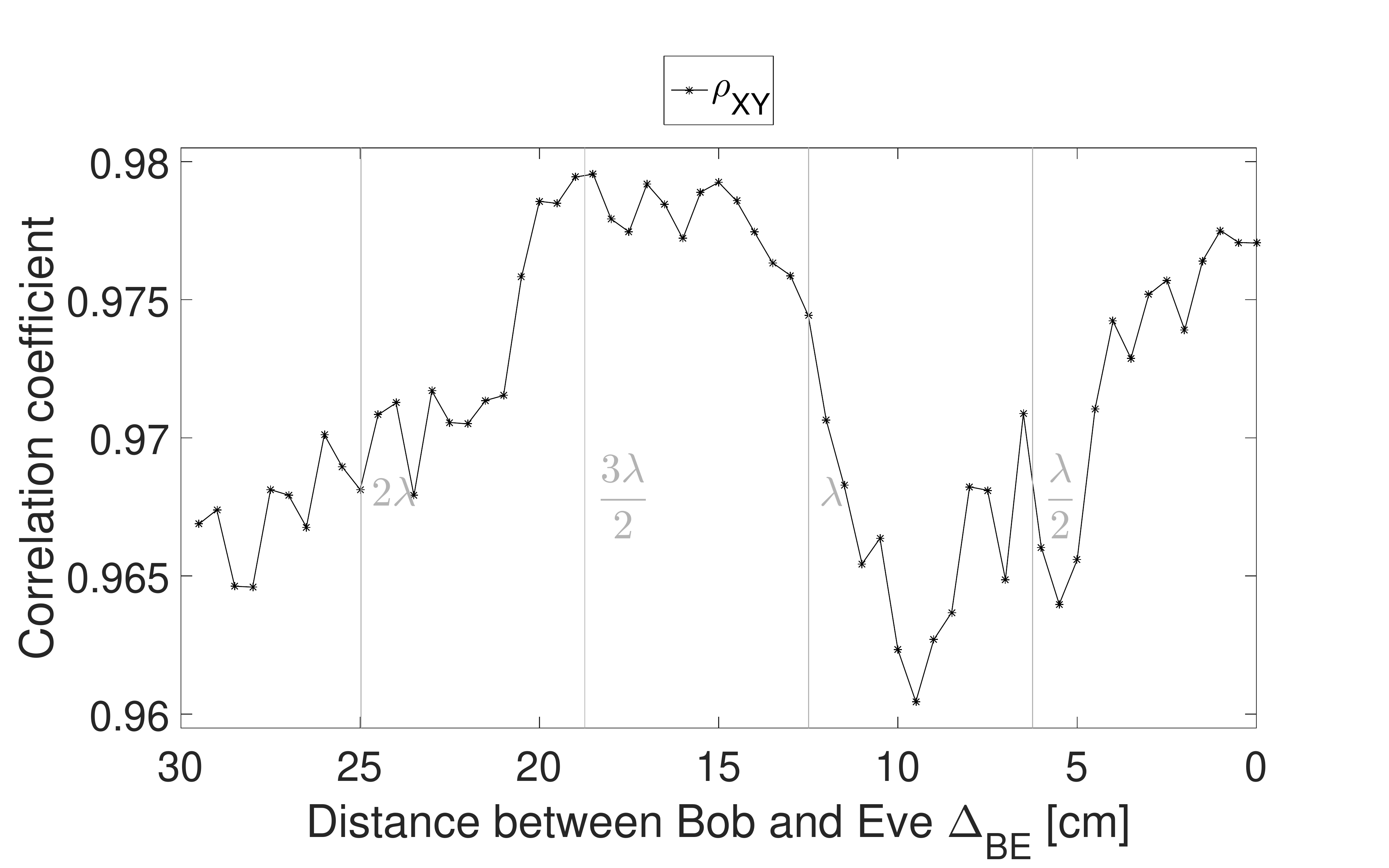}}
	\subfloat[]{\includegraphics[trim=1.8cm 0.1cm 3.5cm 1.6cm, clip=true, height=0.224\textwidth]{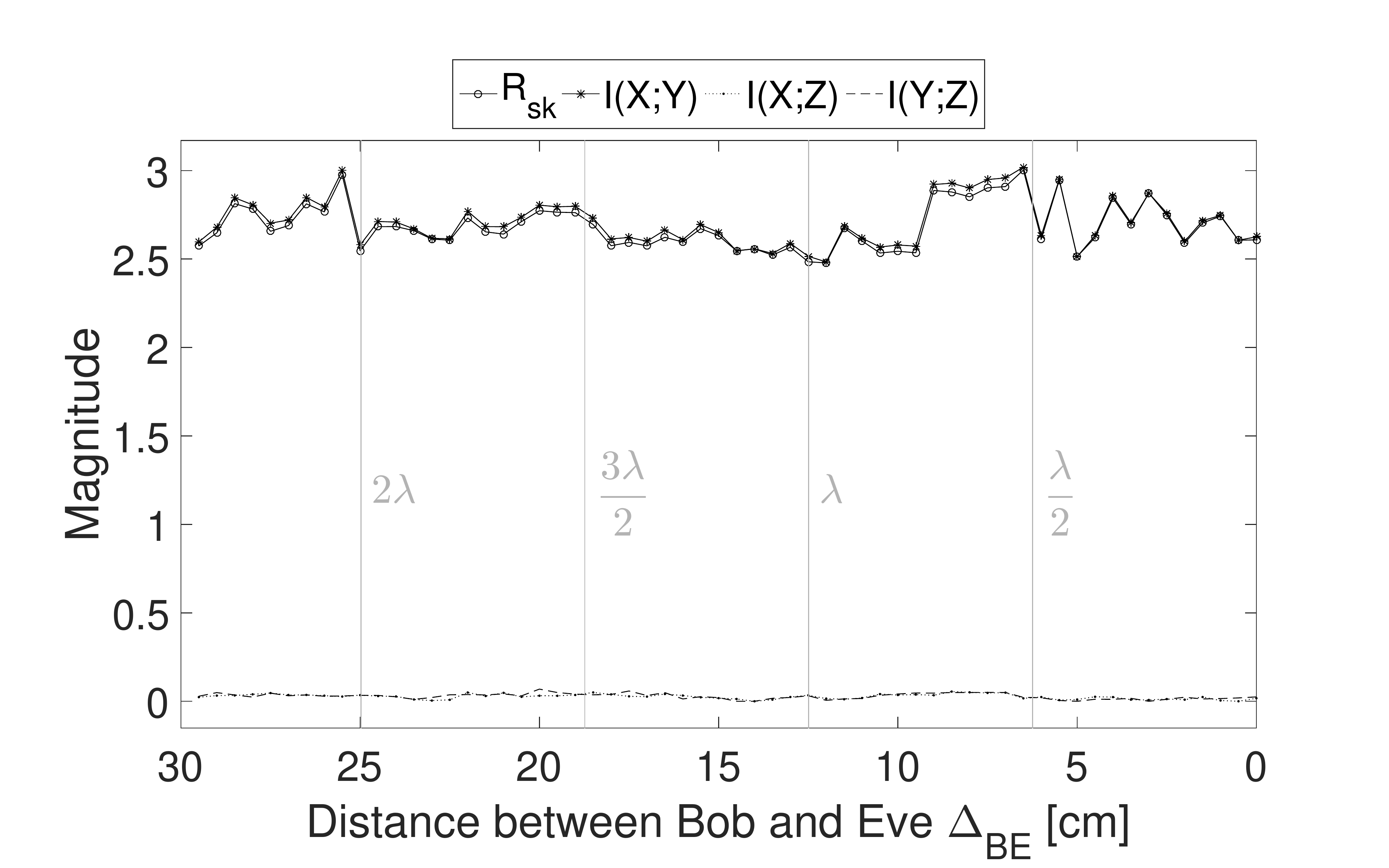}}
	\caption{Evaluation results of $\mybold{v}^{\text{de}}_k$. In (a) and (b) the cross-correlations is given; in (c) the mutual information as well as $\rsk$ is given. Position 8.}
	\label{fig:app_decorr_8}
\end{figure*}


\begin{figure*}
	\centering
	\subfloat[]{\includegraphics[trim=1.4cm 0.1cm 3.5cm 1.6cm, clip=true, height=0.224\textwidth]{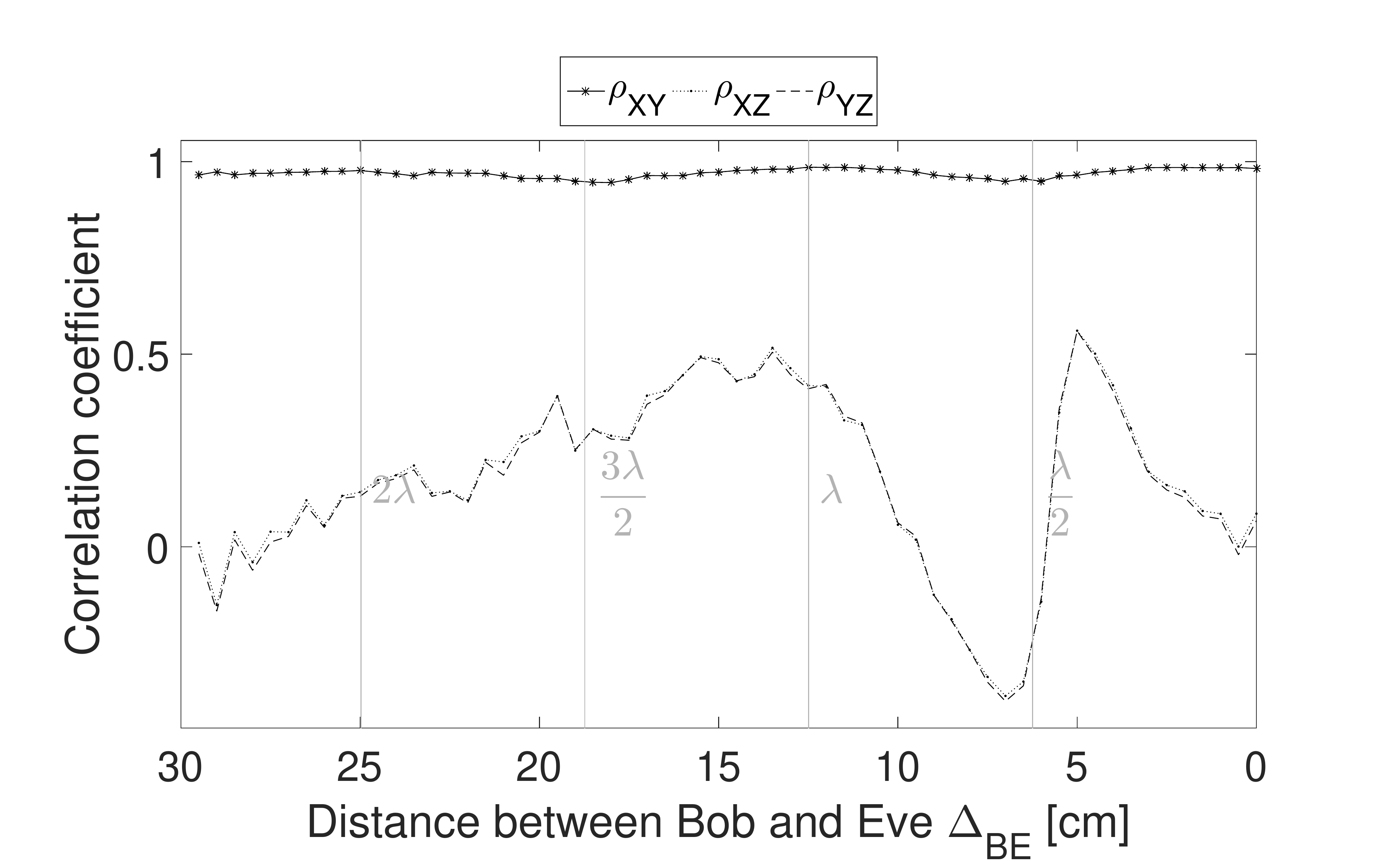}}
	\subfloat[]{\includegraphics[trim=0.5cm 0.1cm 3.5cm 1.6cm, clip=true, height=0.224\textwidth]{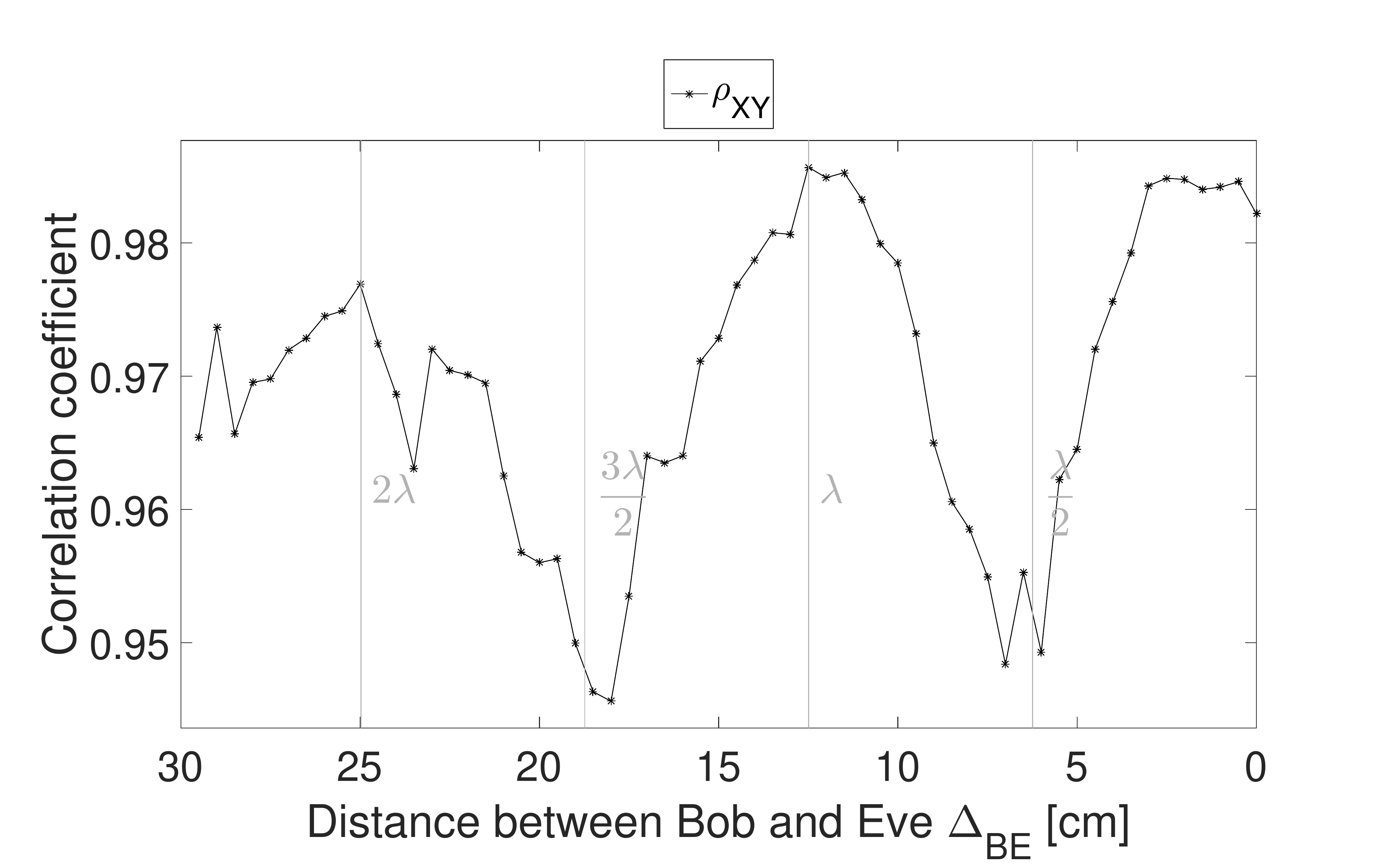}}
	\subfloat[]{\includegraphics[trim=2.2cm 0.1cm 3.5cm 1.6cm, clip=true, height=0.224\textwidth]{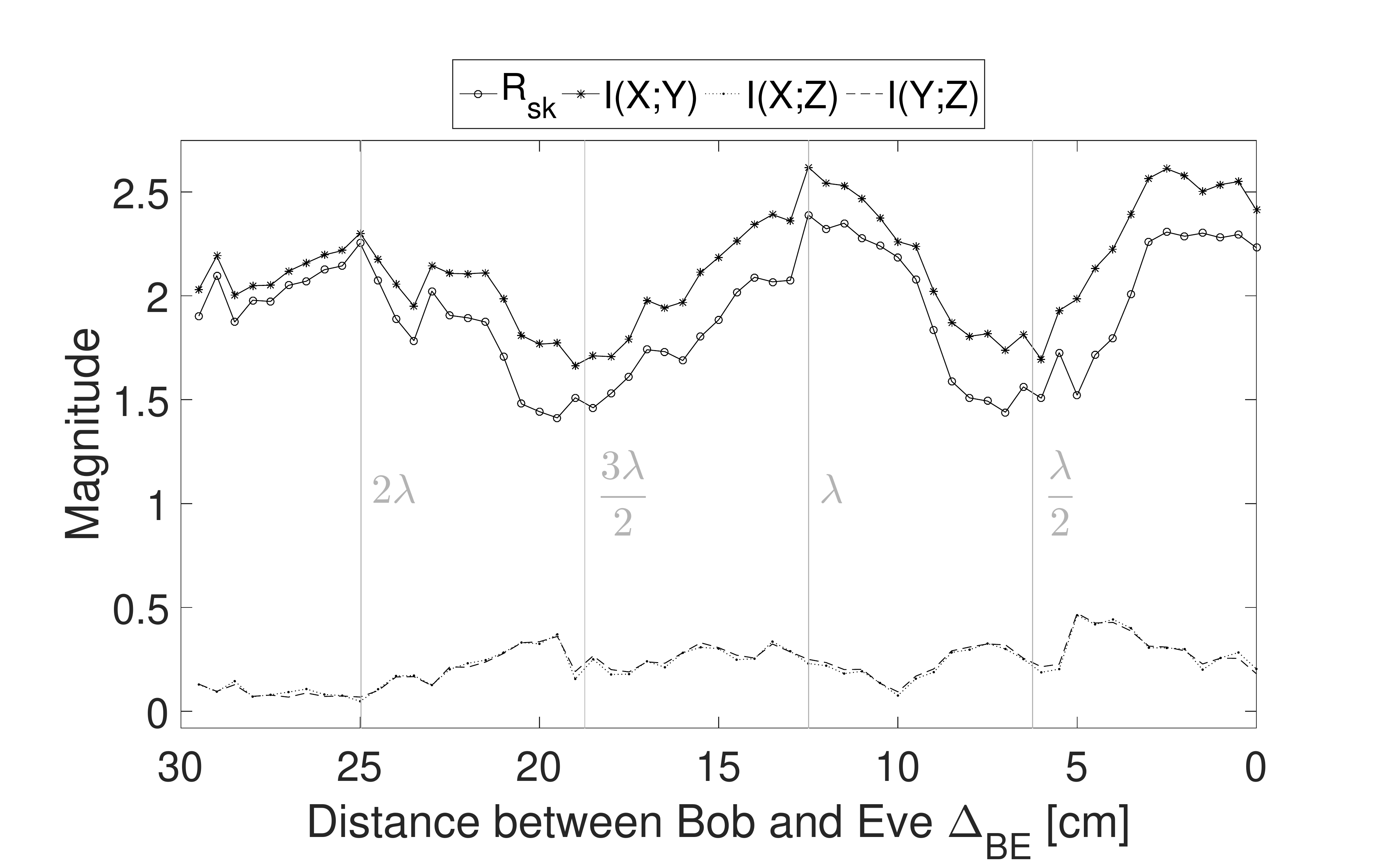}}
	\caption{Evaluation results of $\mybold{v}_k$. In (a) and (b) the cross-correlations is given; in (c) the mutual information as well as $\rsk$ is given. Position 9.}
	\label{fig:app_original_9}
\end{figure*}

\begin{figure*}
	\centering
	\subfloat[]{\includegraphics[trim=1.4cm 0.1cm 3.5cm 1.6cm, clip=true, height=0.224\textwidth]{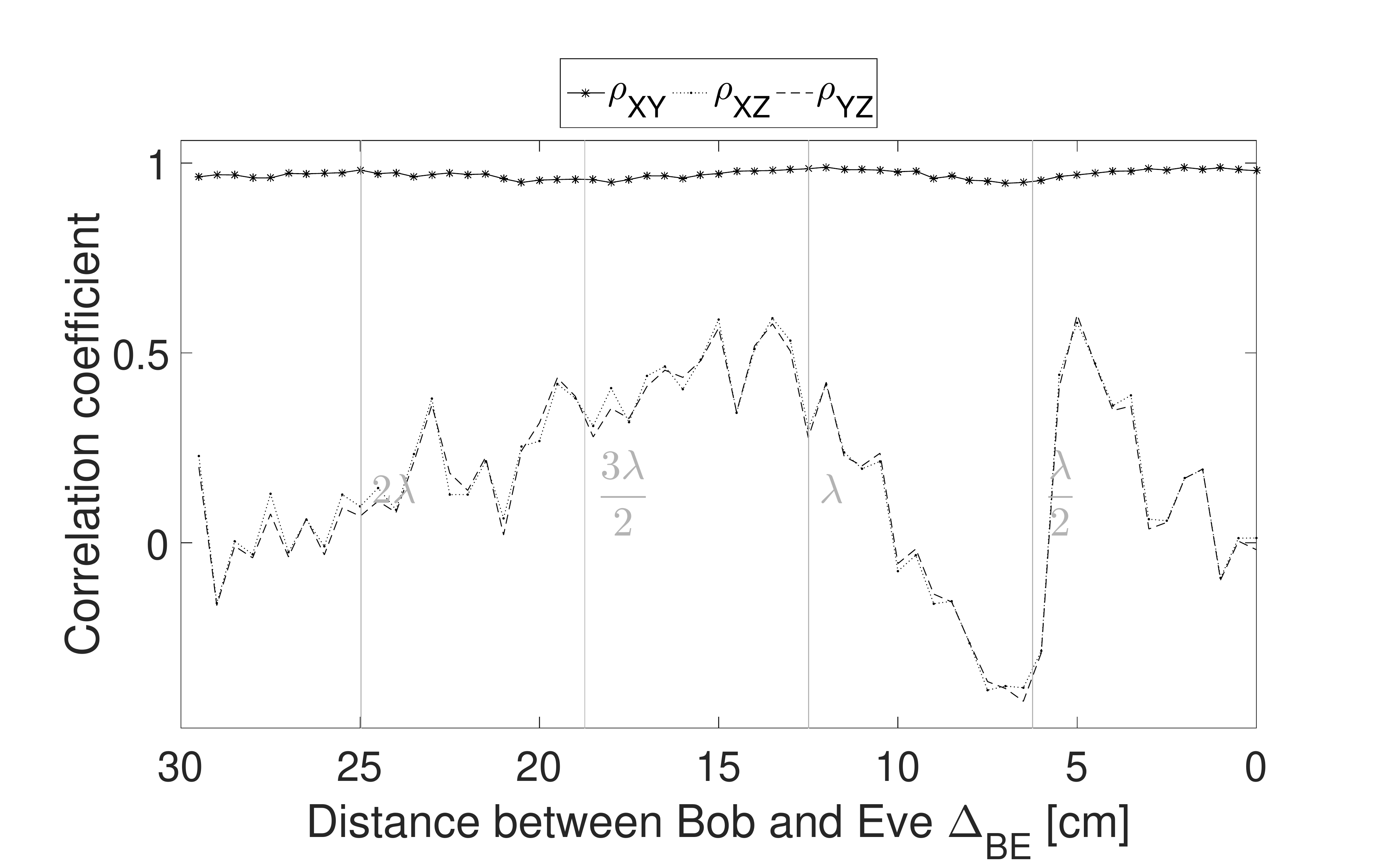}}
	\subfloat[]{\includegraphics[trim=0.5cm 0.1cm 3.5cm 1.6cm, clip=true, height=0.224\textwidth]{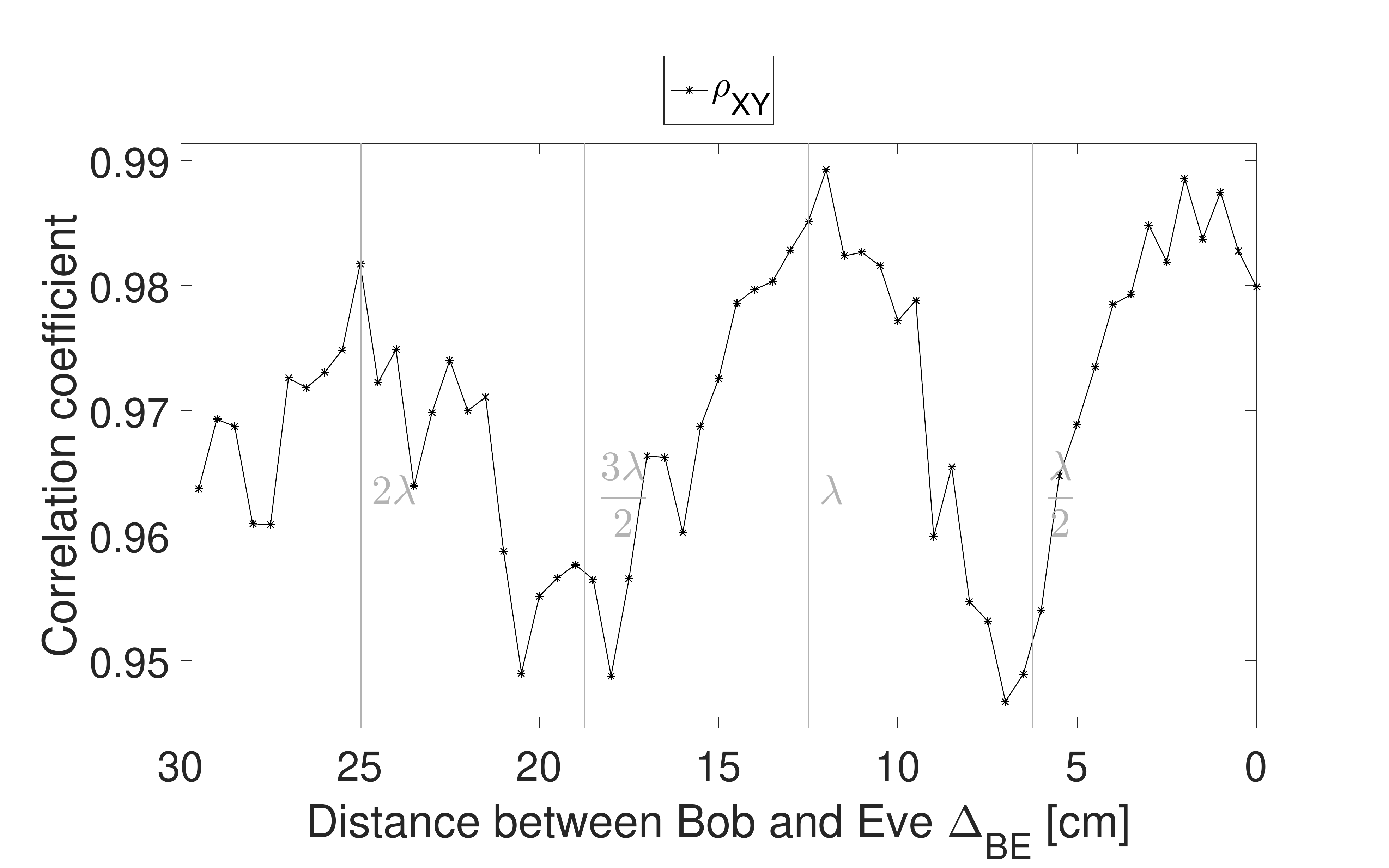}}
	\subfloat[]{\includegraphics[trim=2.2cm 0.1cm 3.5cm 1.6cm, clip=true, height=0.224\textwidth]{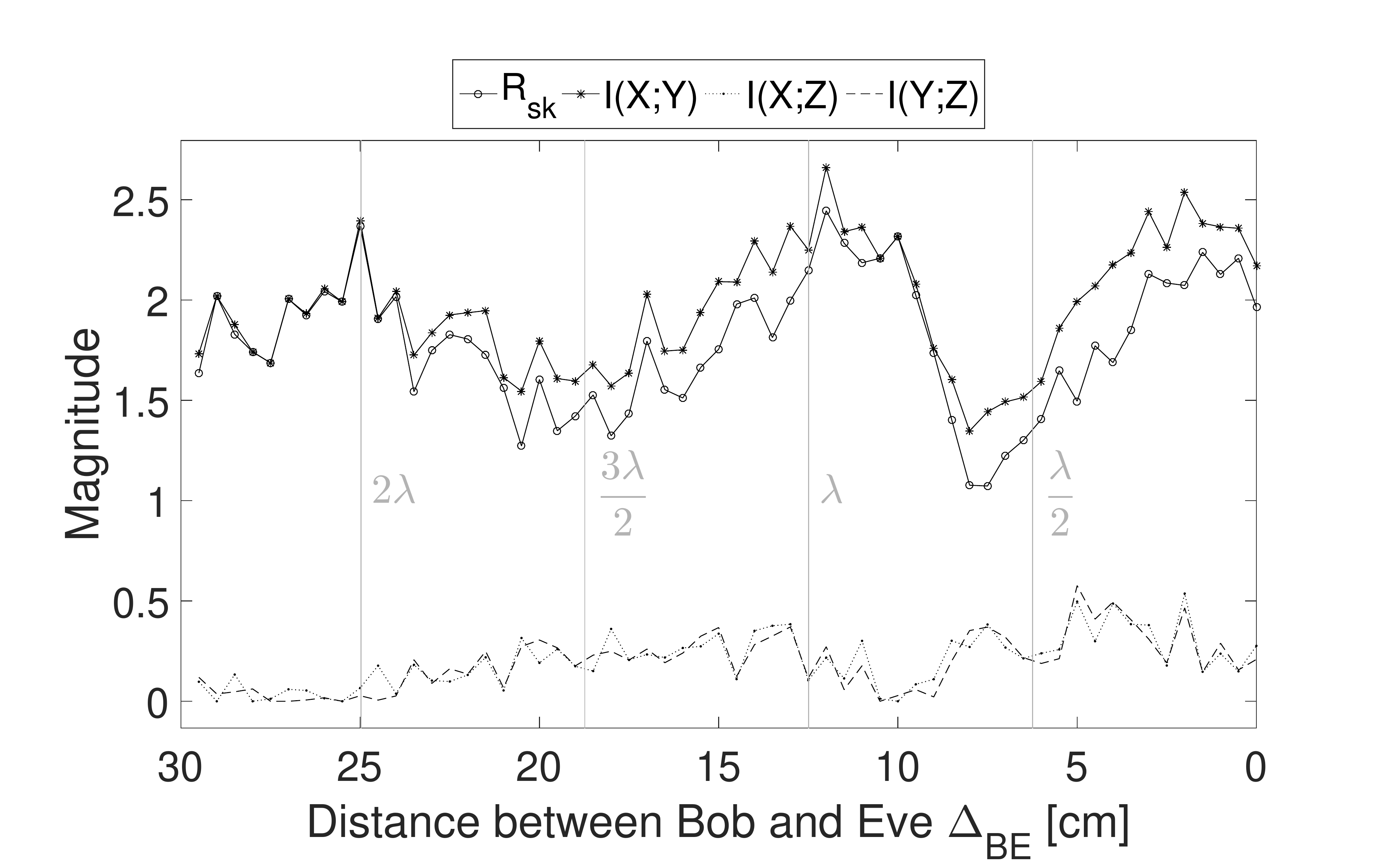}}
	\caption{Evaluation results of $\mybold{v}^{\text{ds}}_k$. In (a) and (b) the cross-correlations is given; in (c) the mutual information as well as $\rsk$ is given. Position 9.}
	\label{fig:app_ds_9}
\end{figure*}

\begin{figure*}
	\centering
	\subfloat[]{\includegraphics[trim=1.4cm 0.1cm 3.5cm 1.6cm, clip=true, height=0.224\textwidth]{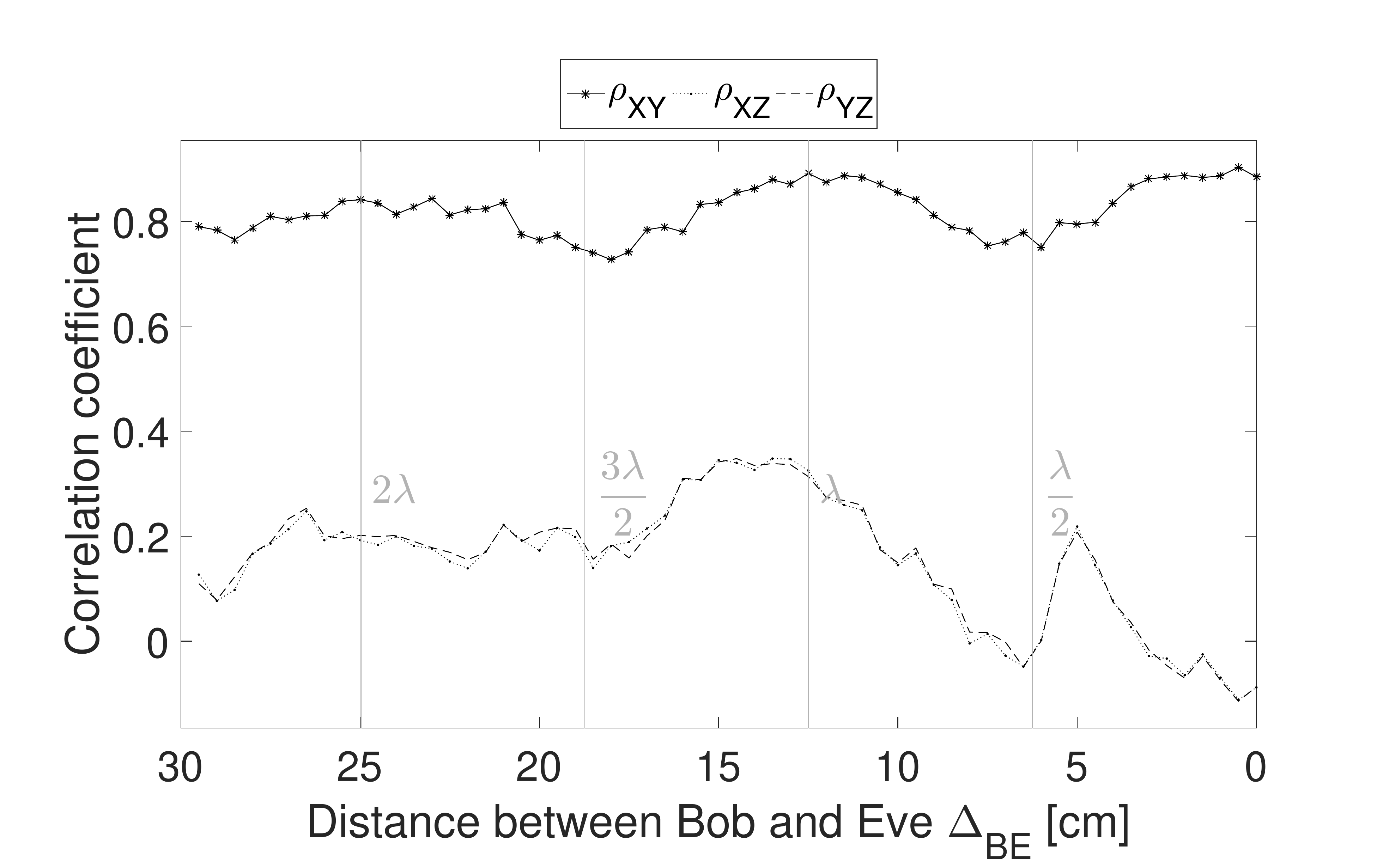}}
	\subfloat[]{\includegraphics[trim=1cm 0.1cm 3.5cm 1.6cm, clip=true, height=0.224\textwidth]{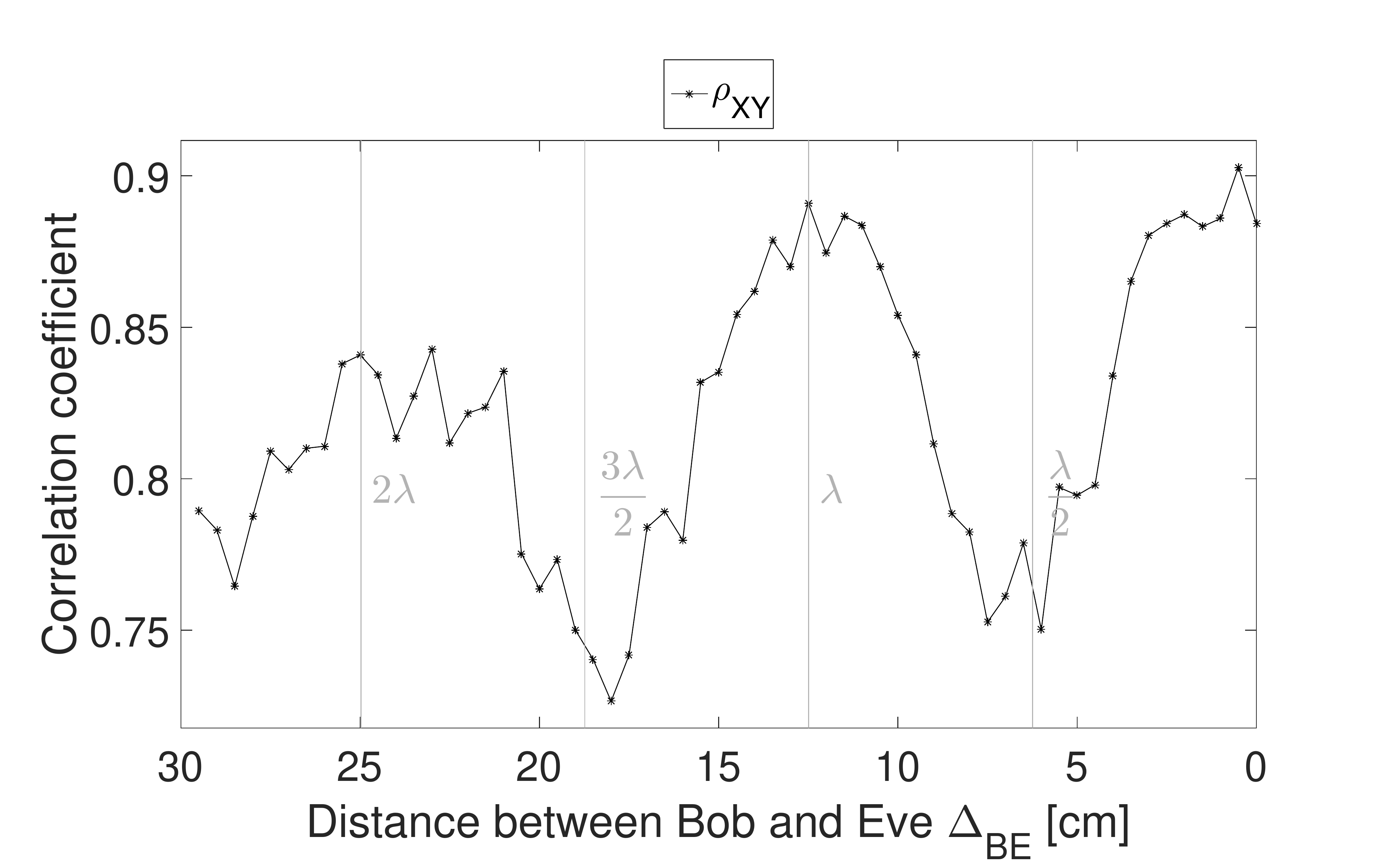}}
	\subfloat[]{\includegraphics[trim=1.8cm 0.1cm 3.5cm 1.6cm, clip=true, height=0.224\textwidth]{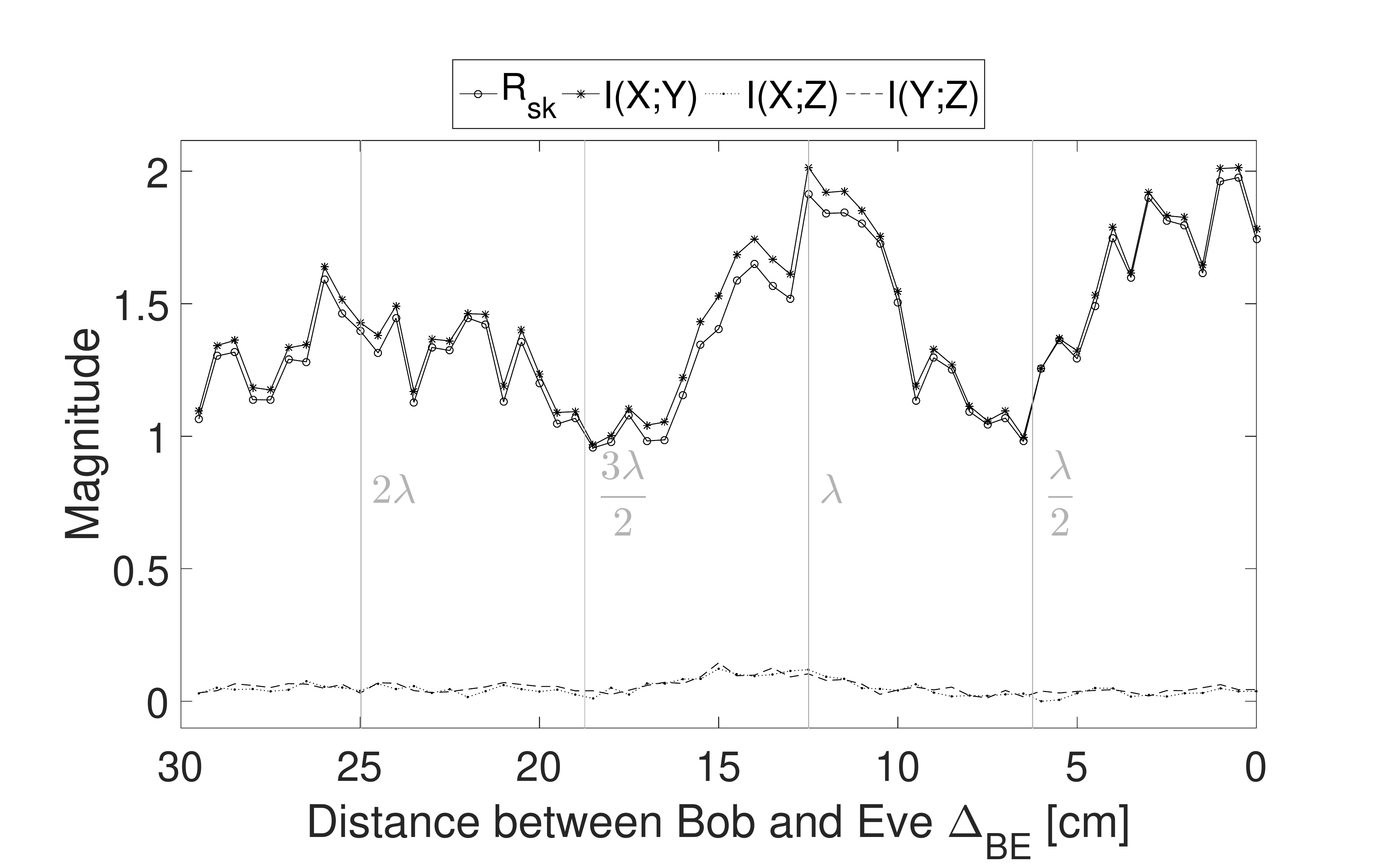}}
	\caption{Evaluation results of $\mybold{v}^{\text{de}}_k$. In (a) and (b) the cross-correlations is given; in (c) the mutual information as well as $\rsk$ is given. Position 9.}
	\label{fig:app_decorr_9}
\end{figure*}


\begin{figure*}
	\centering
	\subfloat[]{\includegraphics[trim=1.4cm 0.1cm 3.5cm 1.6cm, clip=true, height=0.224\textwidth]{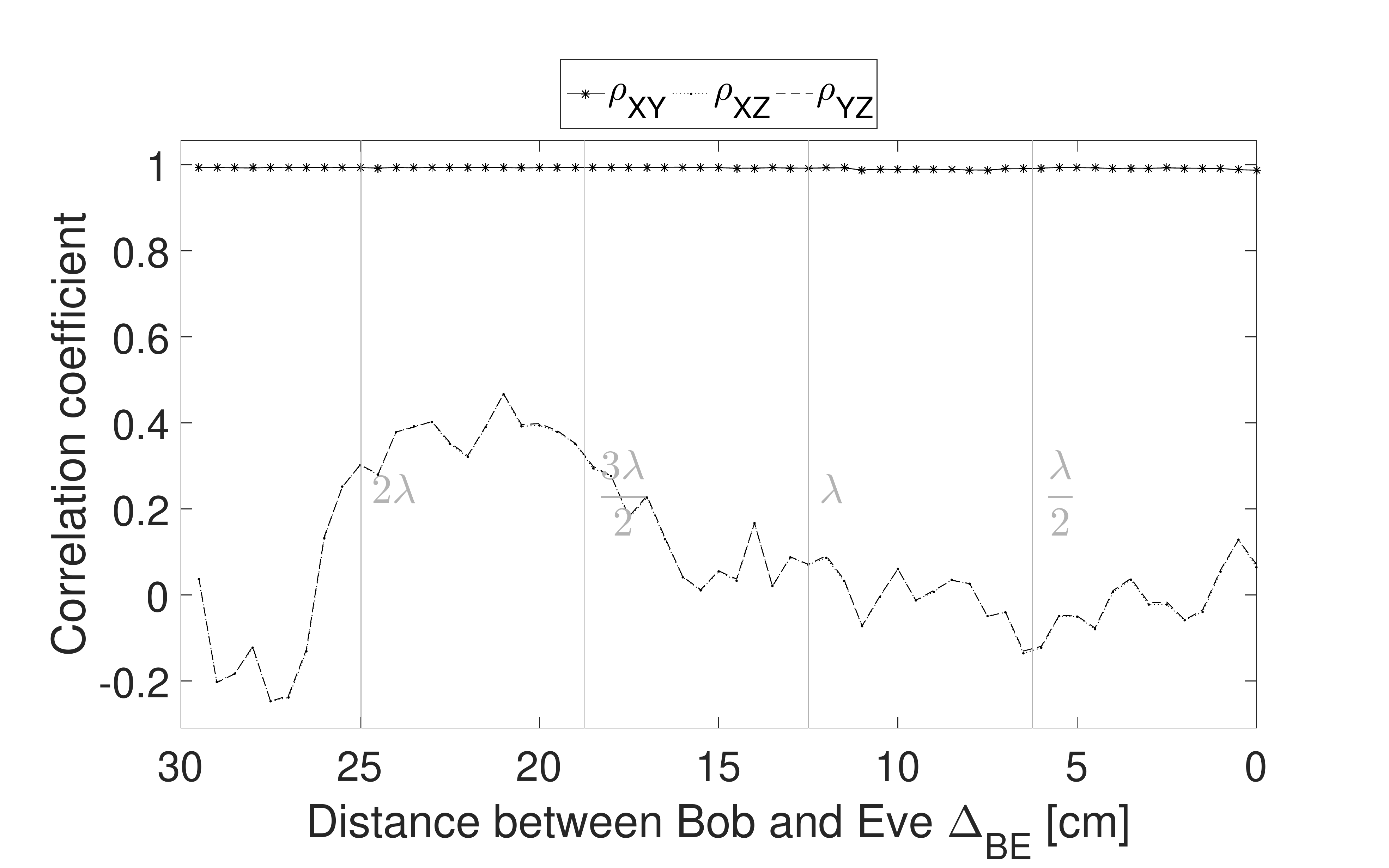}}
	\subfloat[]{\includegraphics[trim=0.5cm 0.1cm 3.5cm 1.6cm, clip=true, height=0.224\textwidth]{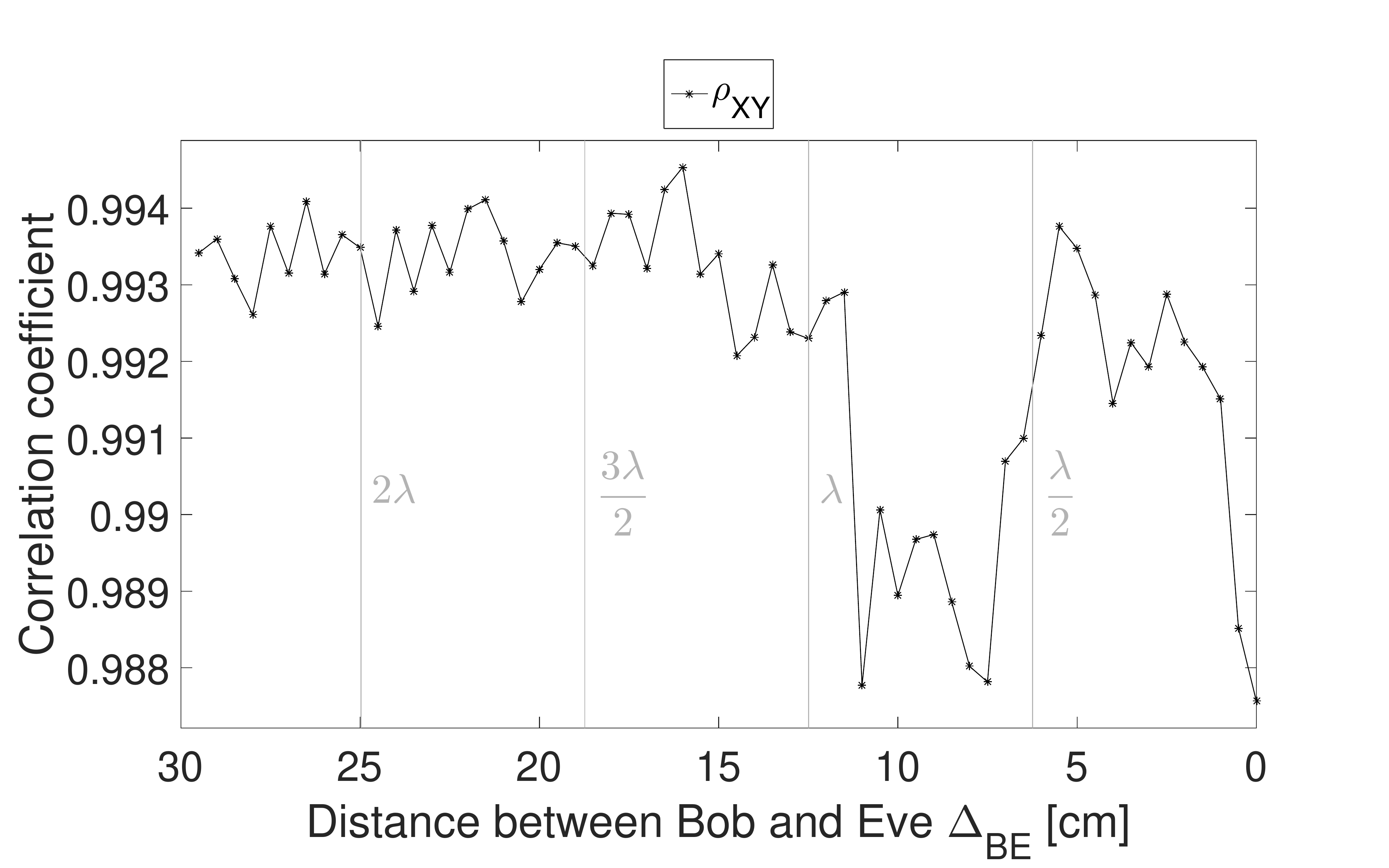}}
	\subfloat[]{\includegraphics[trim=2.2cm 0.1cm 3.5cm 1.6cm, clip=true, height=0.224\textwidth]{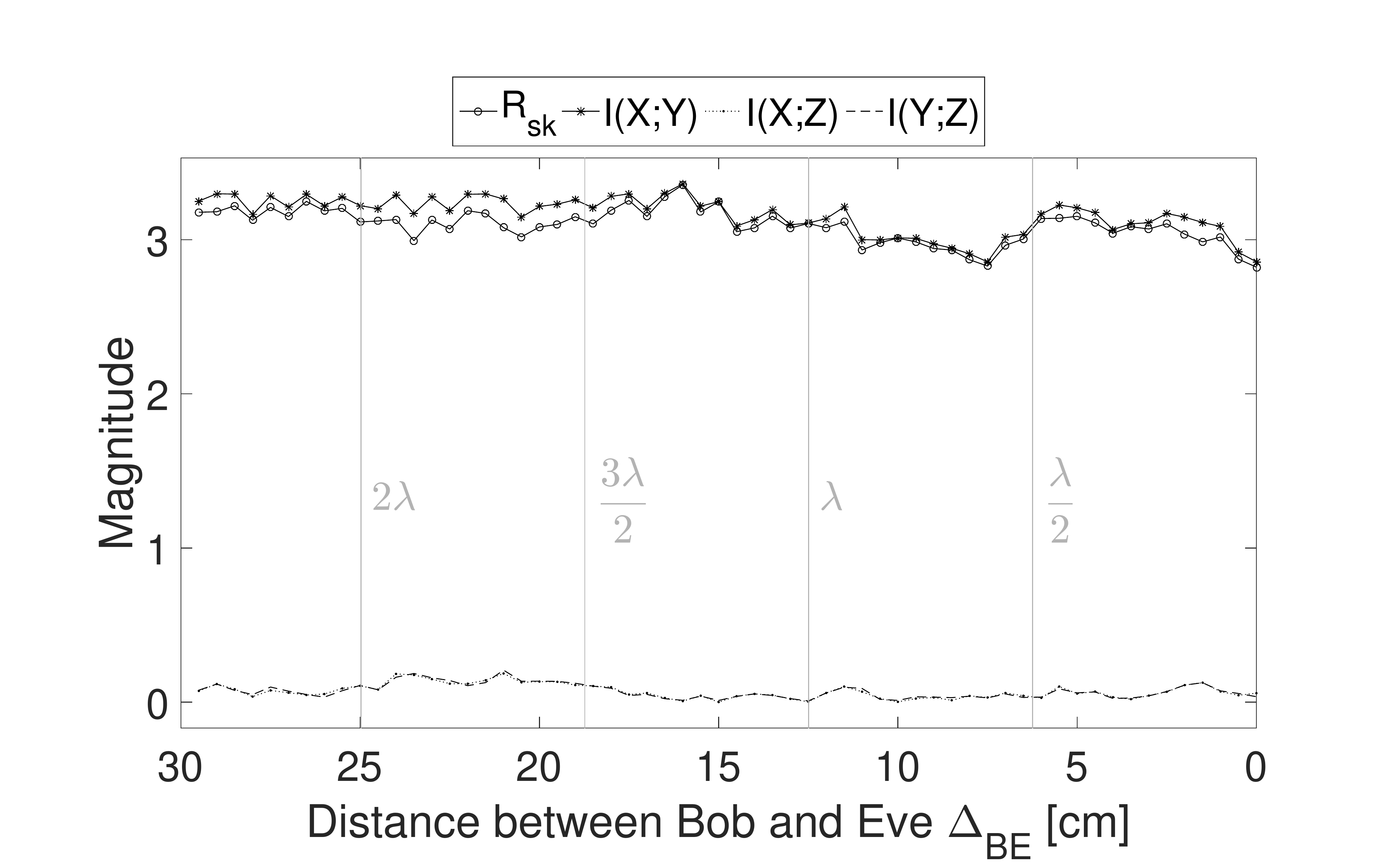}}
	\caption{Evaluation results of $\mybold{v}_k$. In (a) and (b) the cross-correlations is given; in (c) the mutual information as well as $\rsk$ is given. Position 10.}
	\label{fig:app_original_10}
\end{figure*}

\begin{figure*}
	\centering
	\subfloat[]{\includegraphics[trim=1.4cm 0.1cm 3.5cm 1.6cm, clip=true, height=0.224\textwidth]{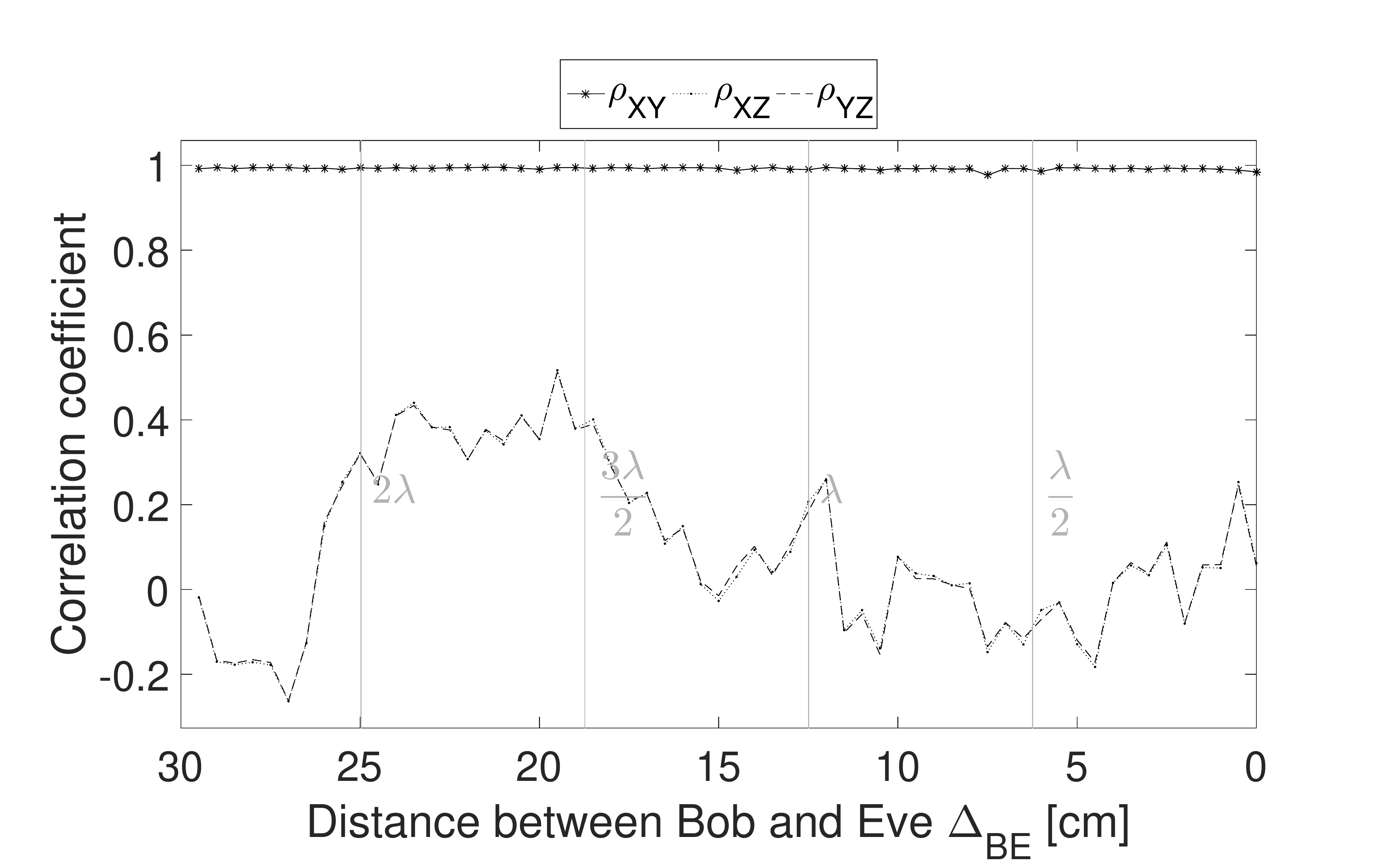}}
	\subfloat[]{\includegraphics[trim=0.5cm 0.1cm 3.5cm 1.6cm, clip=true, height=0.224\textwidth]{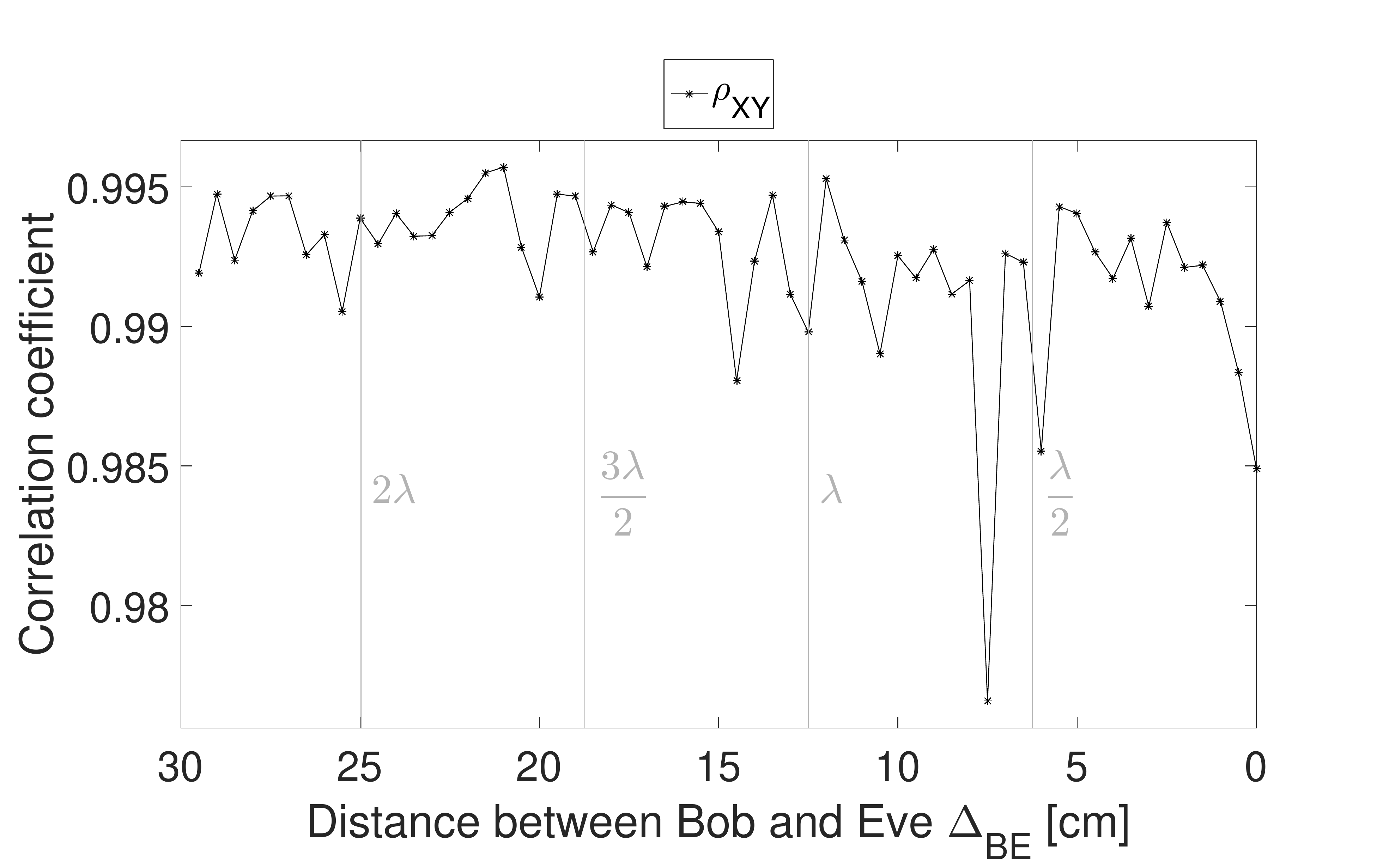}}
	\subfloat[]{\includegraphics[trim=2.2cm 0.1cm 3.5cm 1.6cm, clip=true, height=0.224\textwidth]{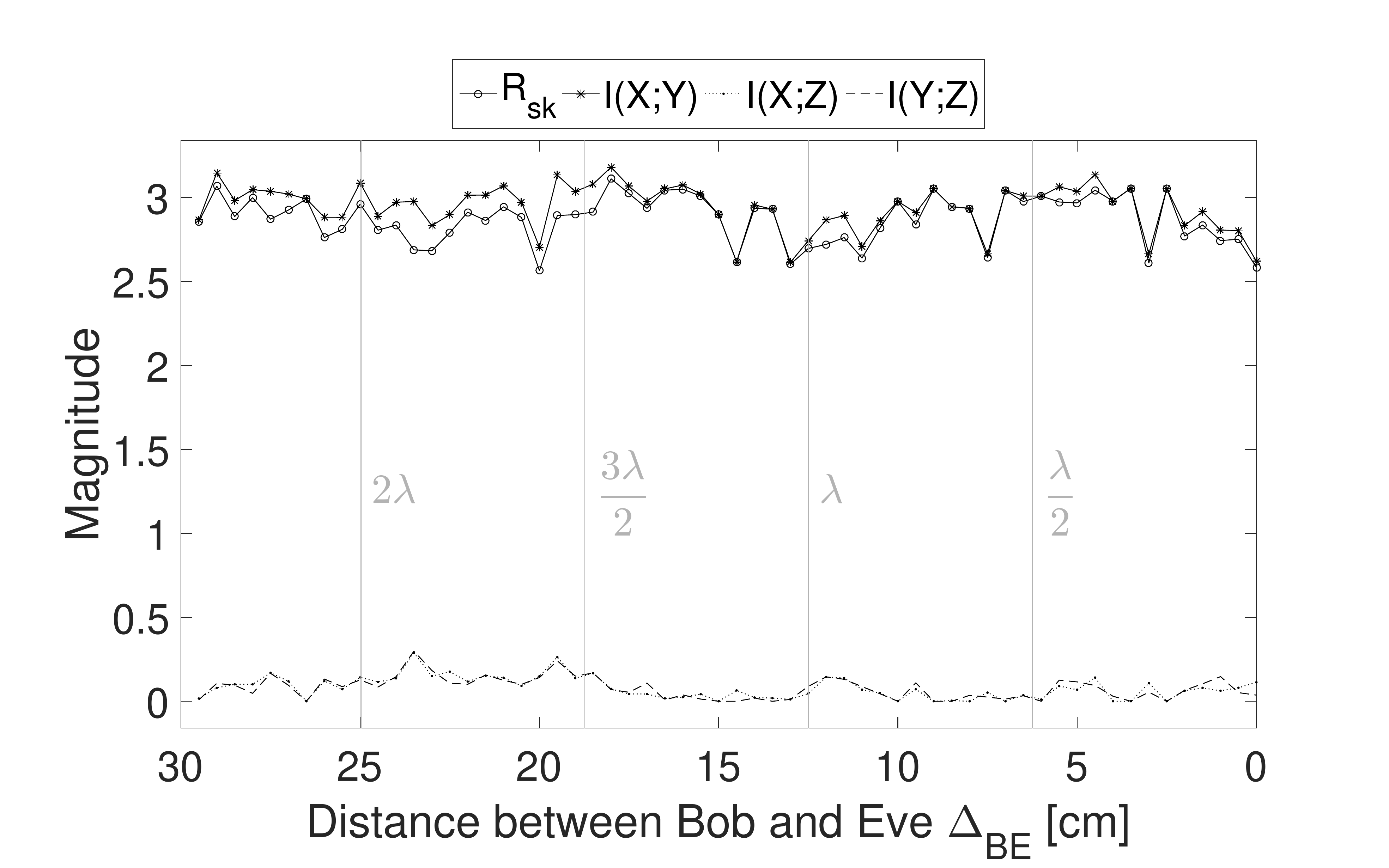}}
	\caption{Evaluation results of $\mybold{v}^{\text{ds}}_k$. In (a) and (b) the cross-correlations is given; in (c) the mutual information as well as $\rsk$ is given. Position 10.}
	\label{fig:app_ds_10}
\end{figure*}

\begin{figure*}
	\centering
	\subfloat[]{\includegraphics[trim=1.4cm 0.1cm 3.5cm 1.6cm, clip=true, height=0.224\textwidth]{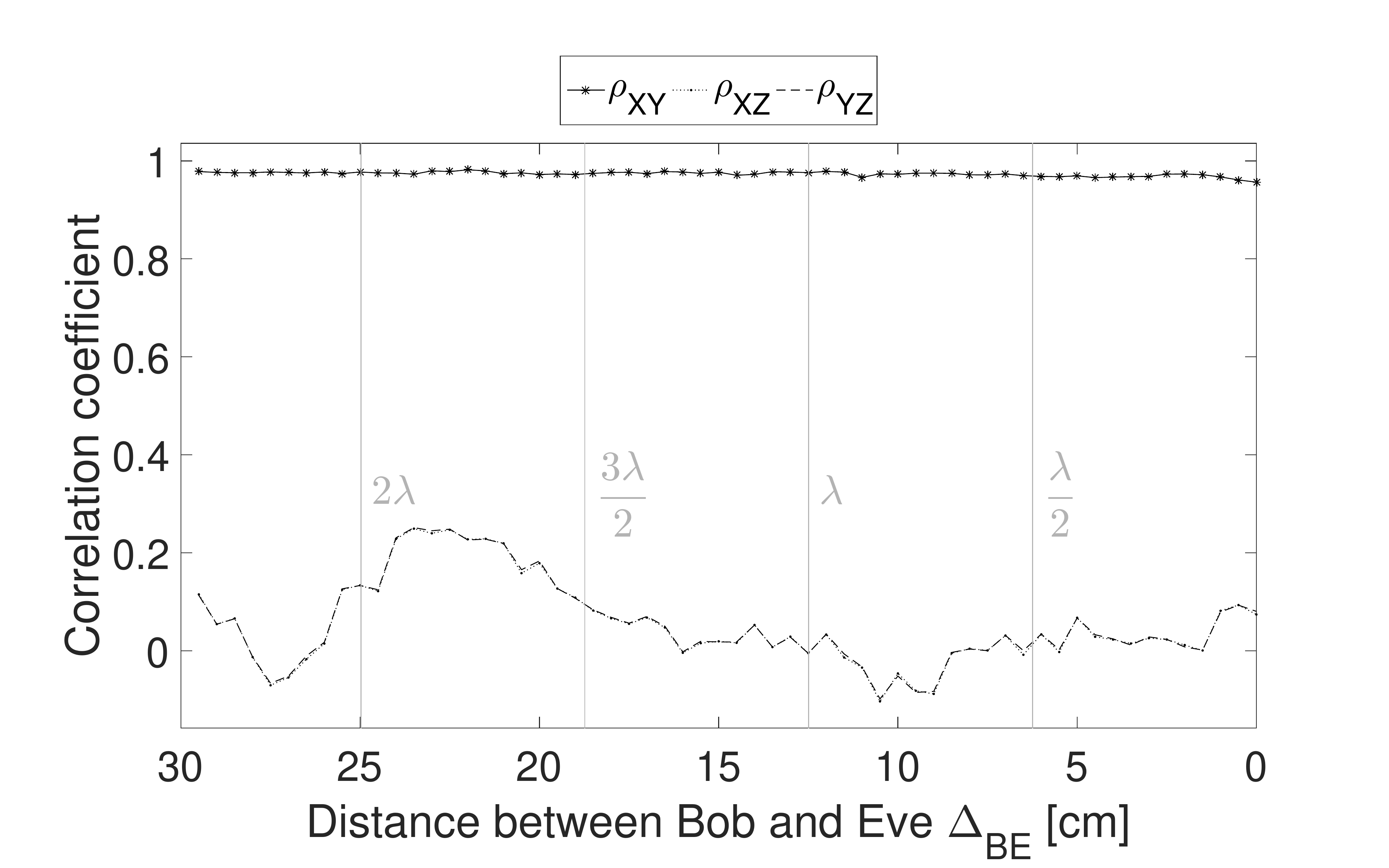}}
	\subfloat[]{\includegraphics[trim=1cm 0.1cm 3.5cm 1.6cm, clip=true, height=0.224\textwidth]{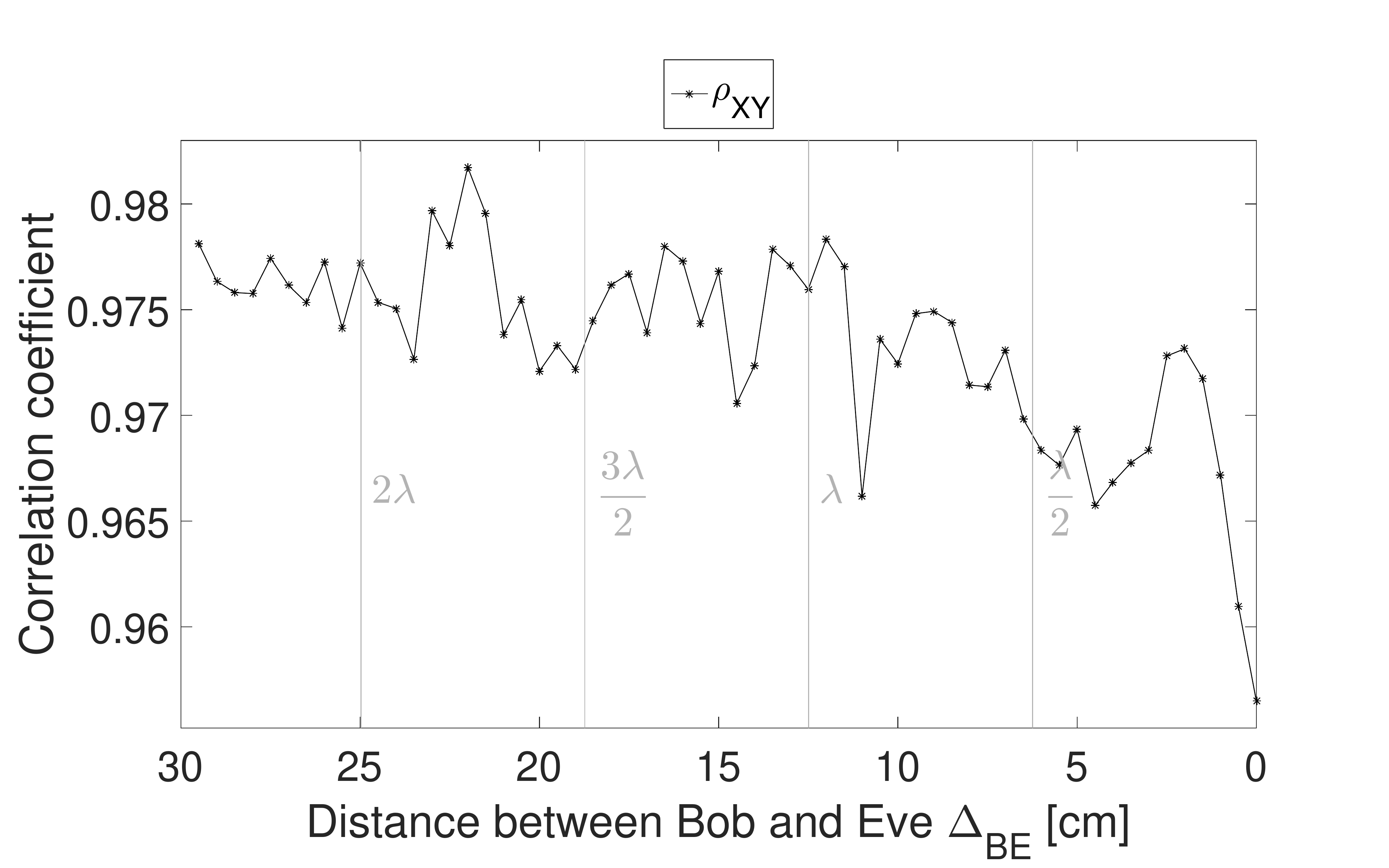}}
	\subfloat[]{\includegraphics[trim=1.8cm 0.1cm 3.5cm 1.6cm, clip=true, height=0.224\textwidth]{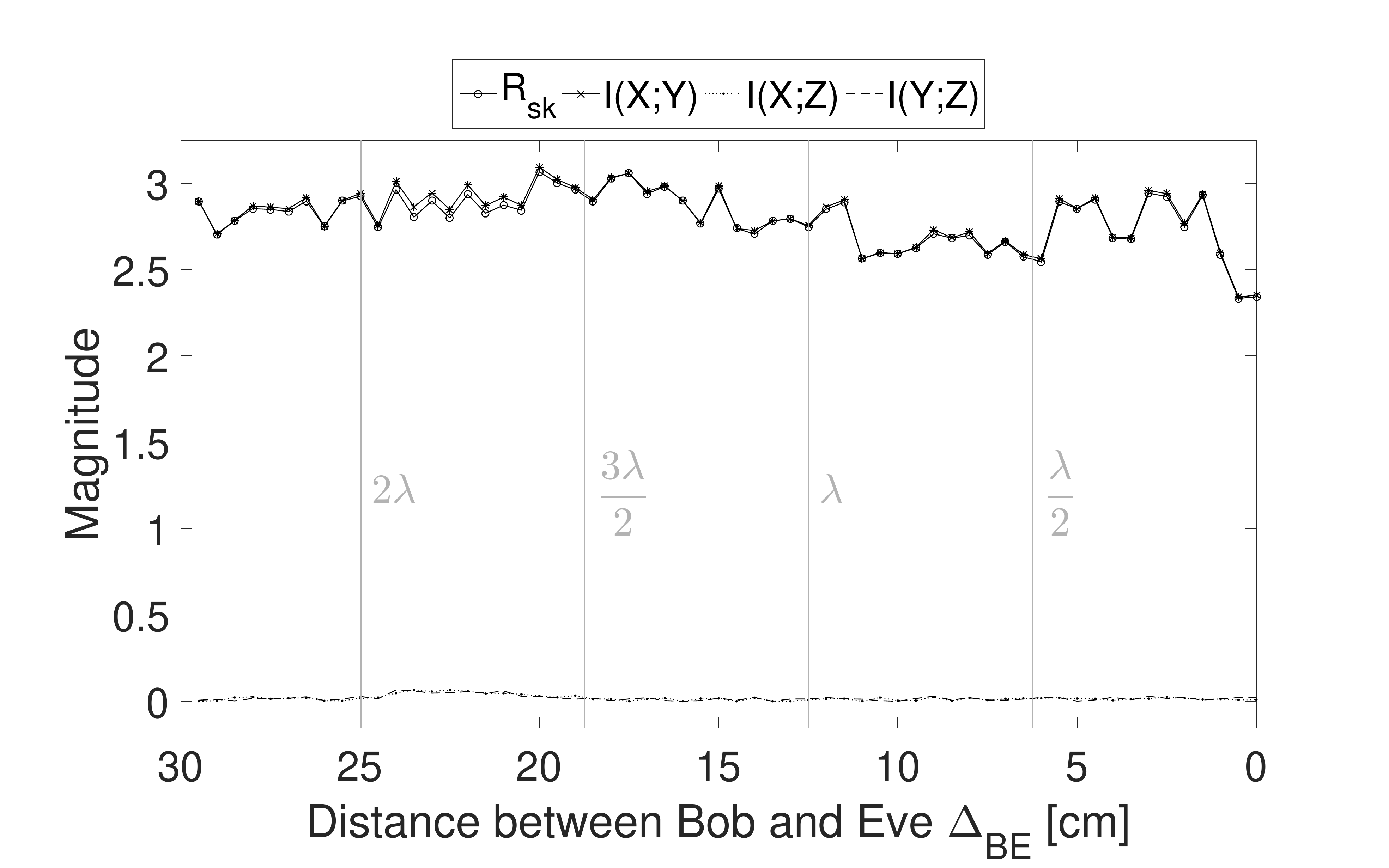}}
	\caption{Evaluation results of $\mybold{v}^{\text{de}}_k$. In (a) and (b) the cross-correlations is given; in (c) the mutual information as well as $\rsk$ is given. Position 10.}
	\label{fig:app_decorr_10}
\end{figure*}


\begin{figure*}
	\centering
	\subfloat[]{\includegraphics[trim=1.4cm 0.1cm 3.5cm 1.6cm, clip=true, height=0.224\textwidth]{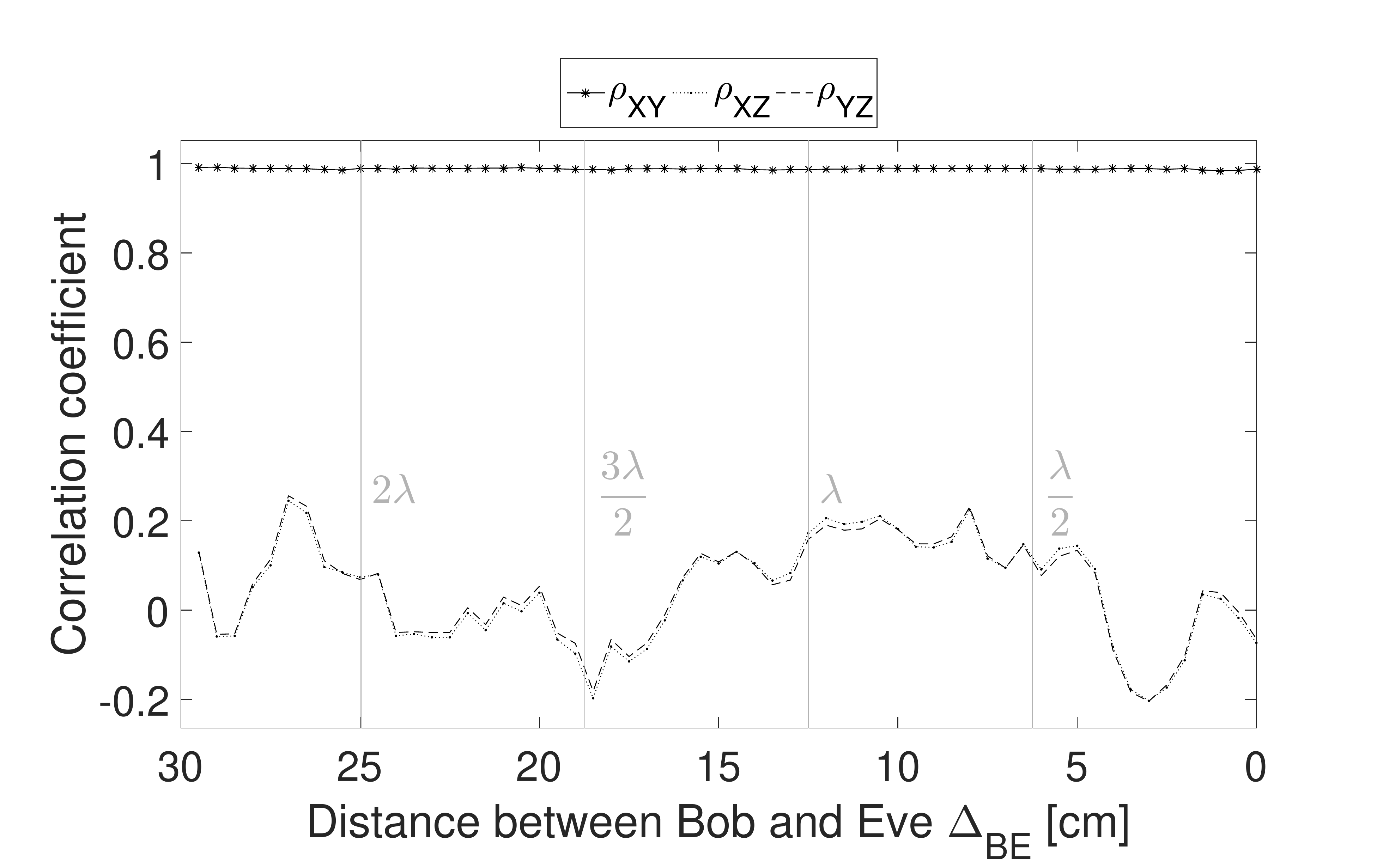}}
	\subfloat[]{\includegraphics[trim=0.5cm 0.1cm 3.5cm 1.6cm, clip=true, height=0.224\textwidth]{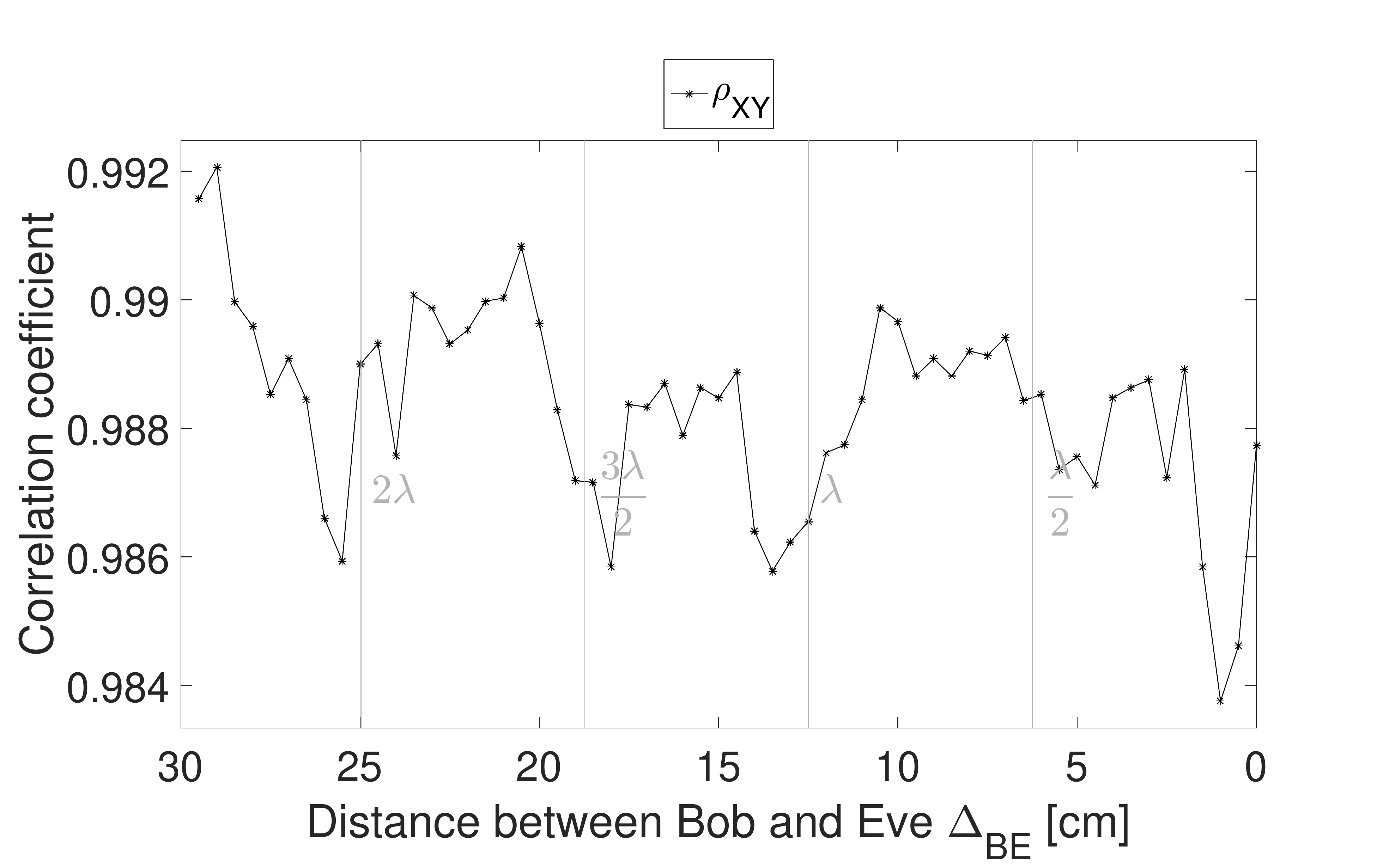}}
	\subfloat[]{\includegraphics[trim=2.2cm 0.1cm 3.5cm 1.6cm, clip=true, height=0.224\textwidth]{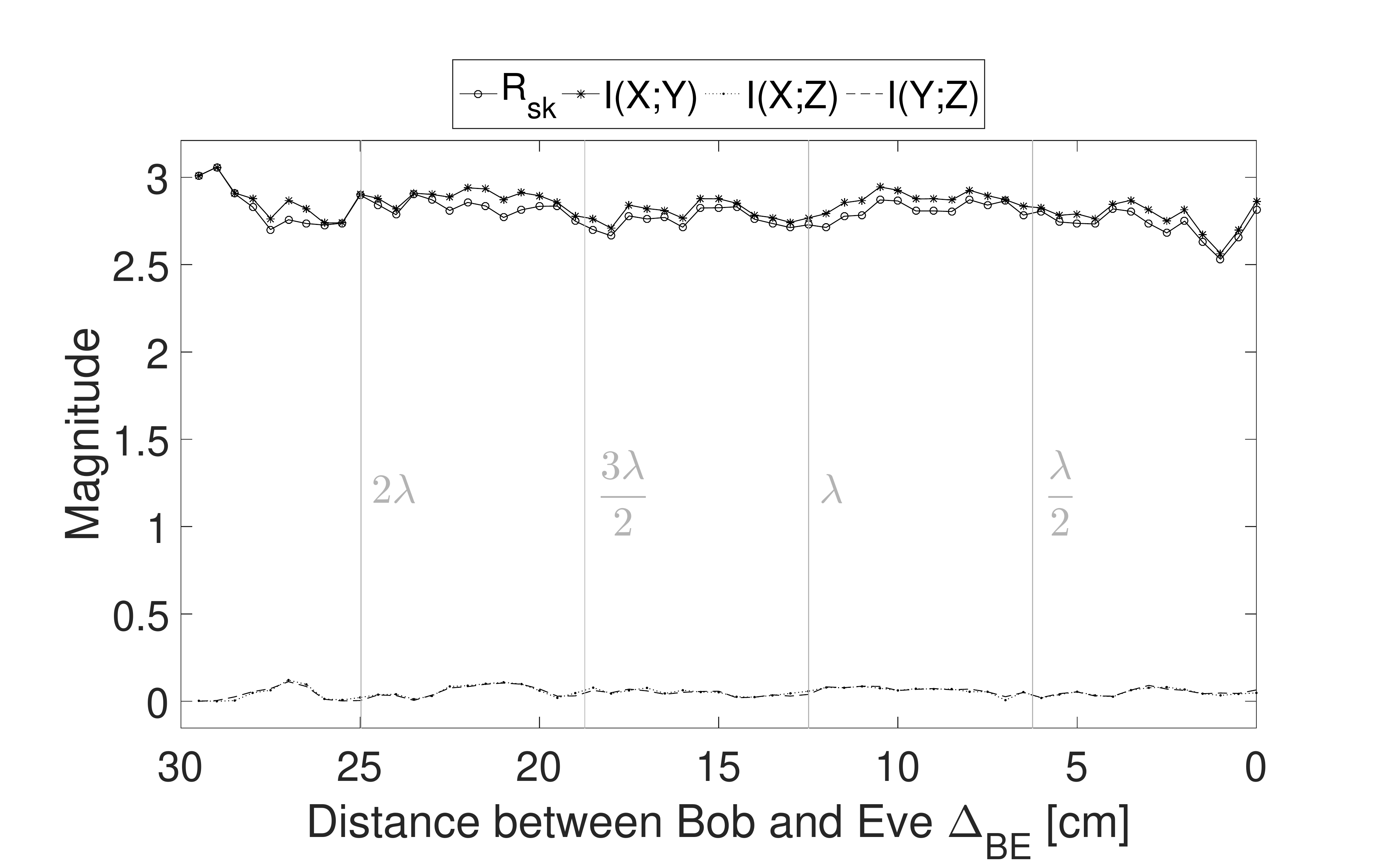}}
	\caption{Evaluation results of $\mybold{v}_k$. In (a) and (b) the cross-correlations is given; in (c) the mutual information as well as $\rsk$ is given. Position 11.}
	\label{fig:app_original_11}
\end{figure*}

\clearpage

\begin{figure*}
	\centering
	\subfloat[]{\includegraphics[trim=1.4cm 0.1cm 3.5cm 1.6cm, clip=true, height=0.224\textwidth]{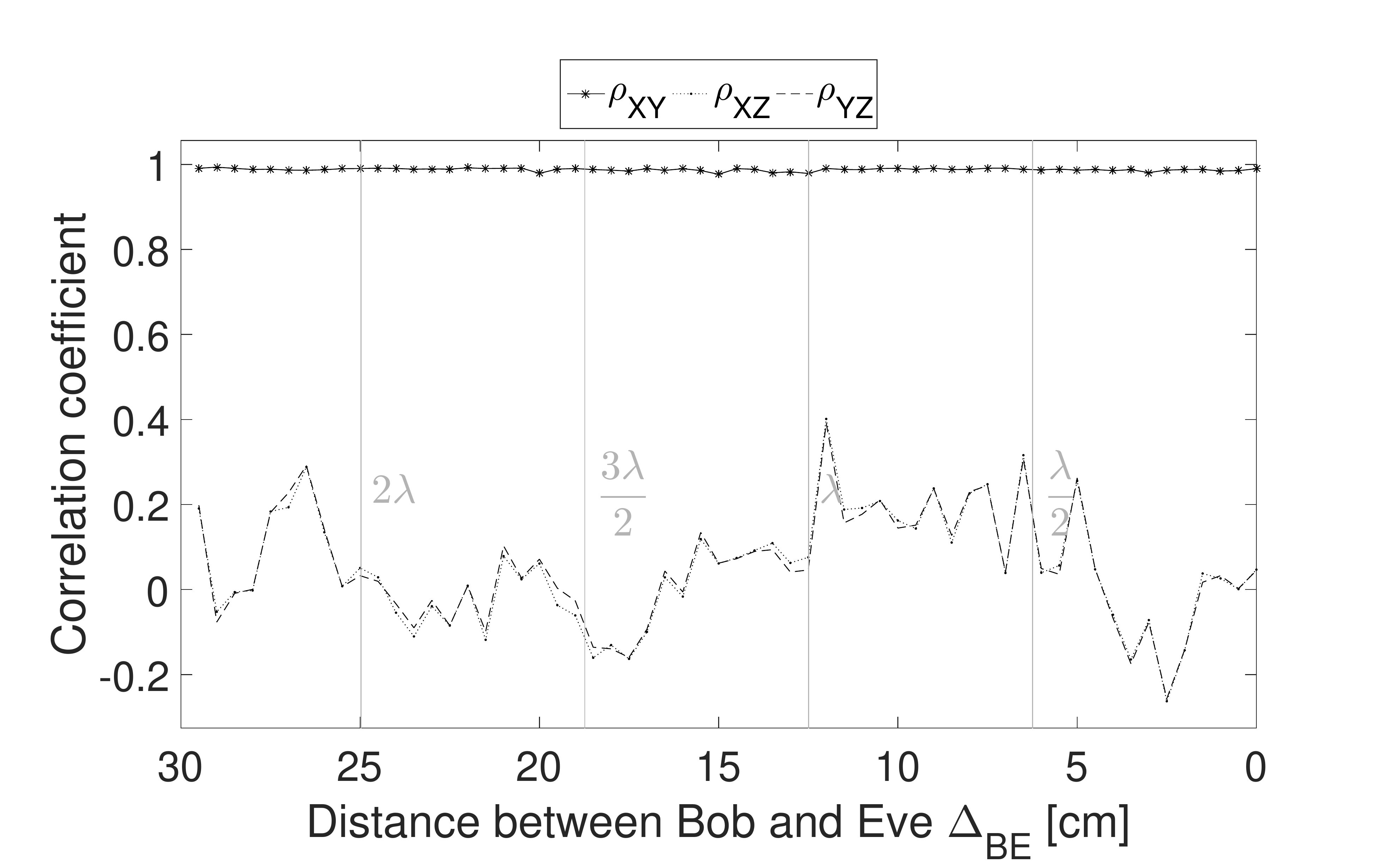}}
	\subfloat[]{\includegraphics[trim=0.5cm 0.1cm 3.5cm 1.6cm, clip=true, height=0.224\textwidth]{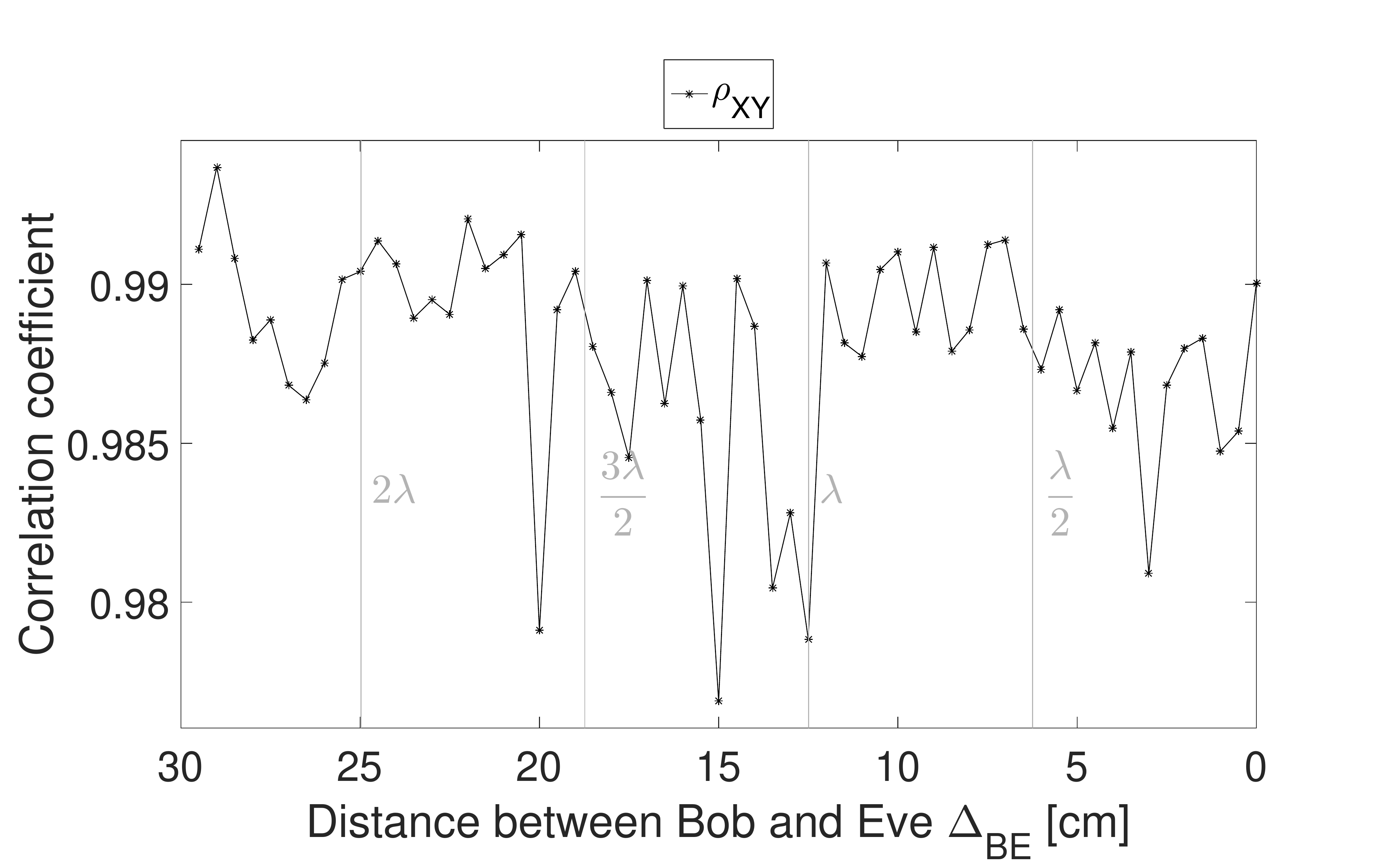}}
	\subfloat[]{\includegraphics[trim=2.2cm 0.1cm 3.5cm 1.6cm, clip=true, height=0.224\textwidth]{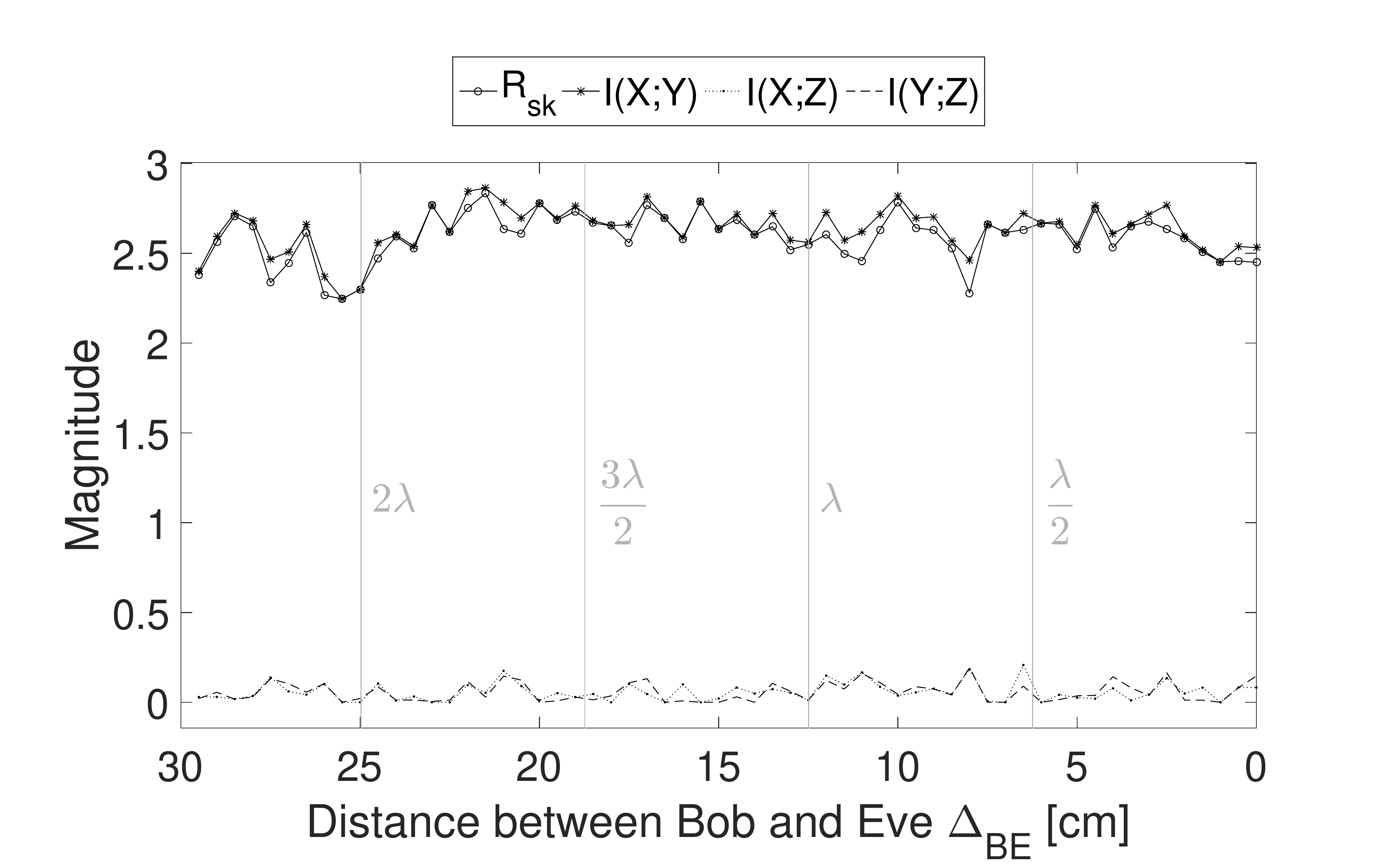}}
	\caption{Evaluation results of $\mybold{v}^{\text{ds}}_k$. In (a) and (b) the cross-correlations is given; in (c) the mutual information as well as $\rsk$ is given. Position 11.}
	\label{fig:app_ds_11}
\end{figure*}

\begin{figure*}
	\centering
	\subfloat[]{\includegraphics[trim=1.4cm 0.1cm 3.5cm 1.6cm, clip=true, height=0.224\textwidth]{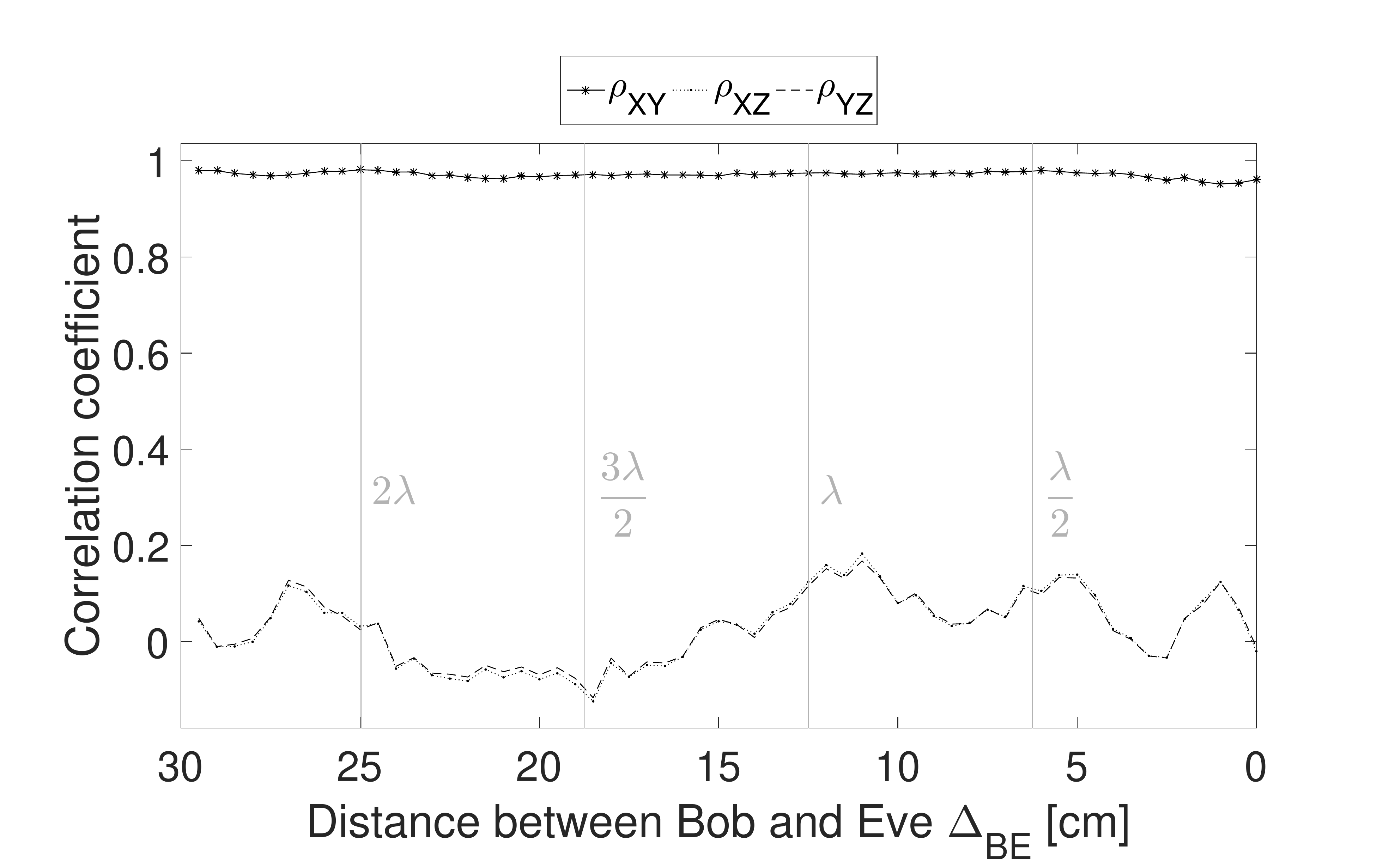}}
	\subfloat[]{\includegraphics[trim=1cm 0.1cm 3.5cm 1.6cm, clip=true, height=0.224\textwidth]{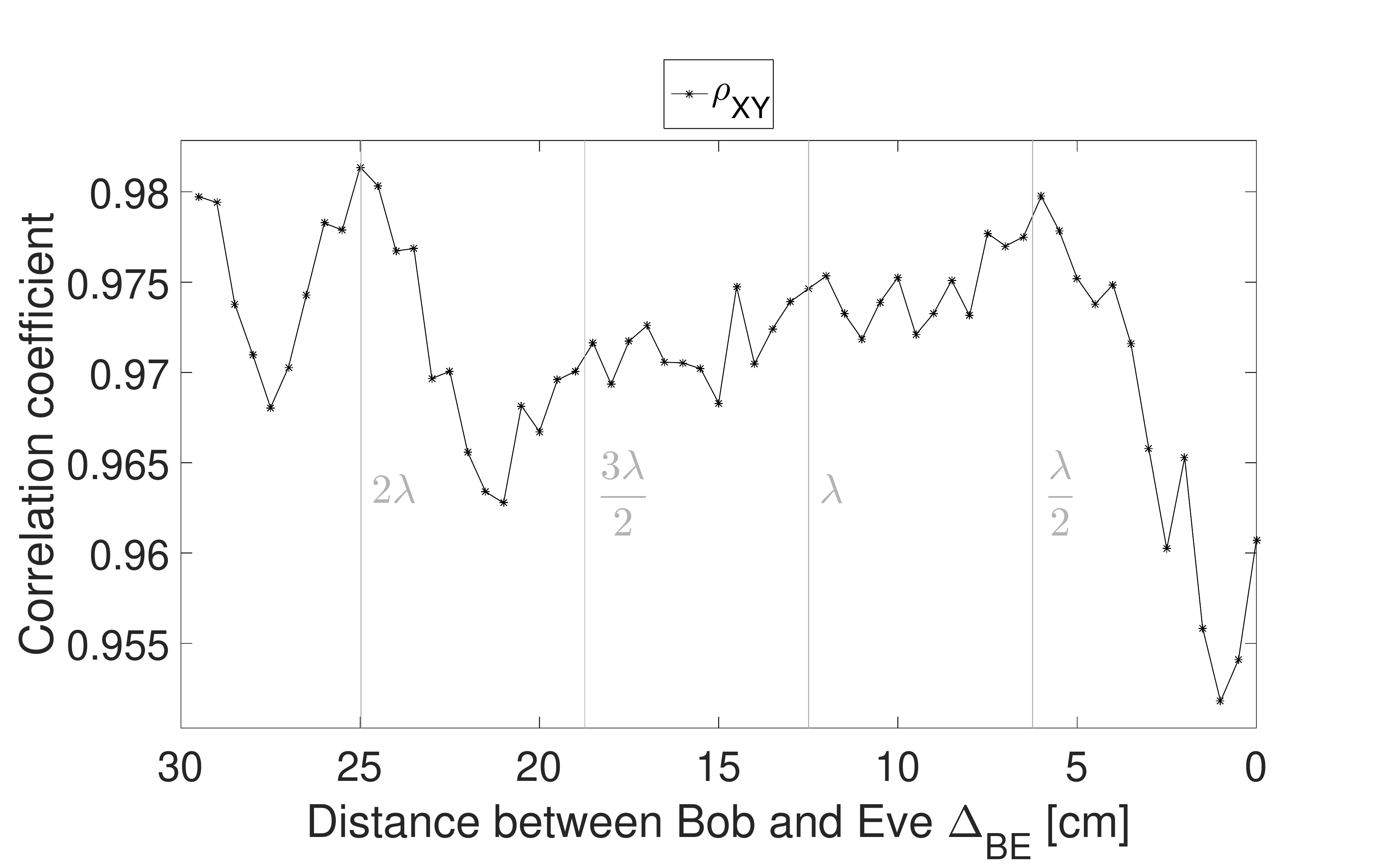}}
	\subfloat[]{\includegraphics[trim=1.8cm 0.1cm 3.5cm 1.6cm, clip=true, height=0.224\textwidth]{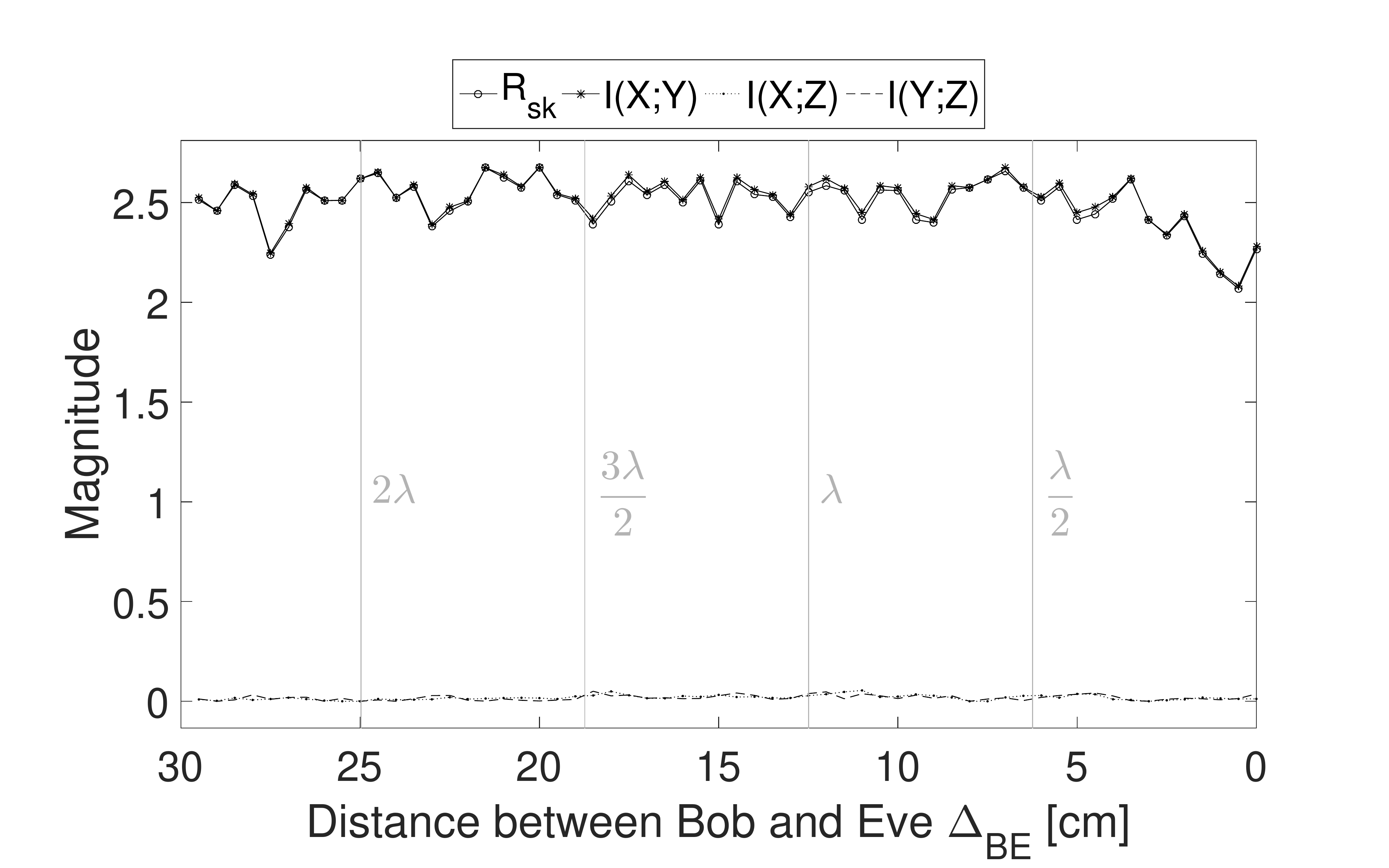}}
	\caption{Evaluation results of $\mybold{v}^{\text{de}}_k$. In (a) and (b) the cross-correlations is given; in (c) the mutual information as well as $\rsk$ is given. Position 11.}
	\label{fig:app_decorr_11}
\end{figure*}


\begin{figure*}
	\centering
	\subfloat[]{\includegraphics[trim=1.4cm 0.1cm 3.5cm 1.6cm, clip=true, height=0.224\textwidth]{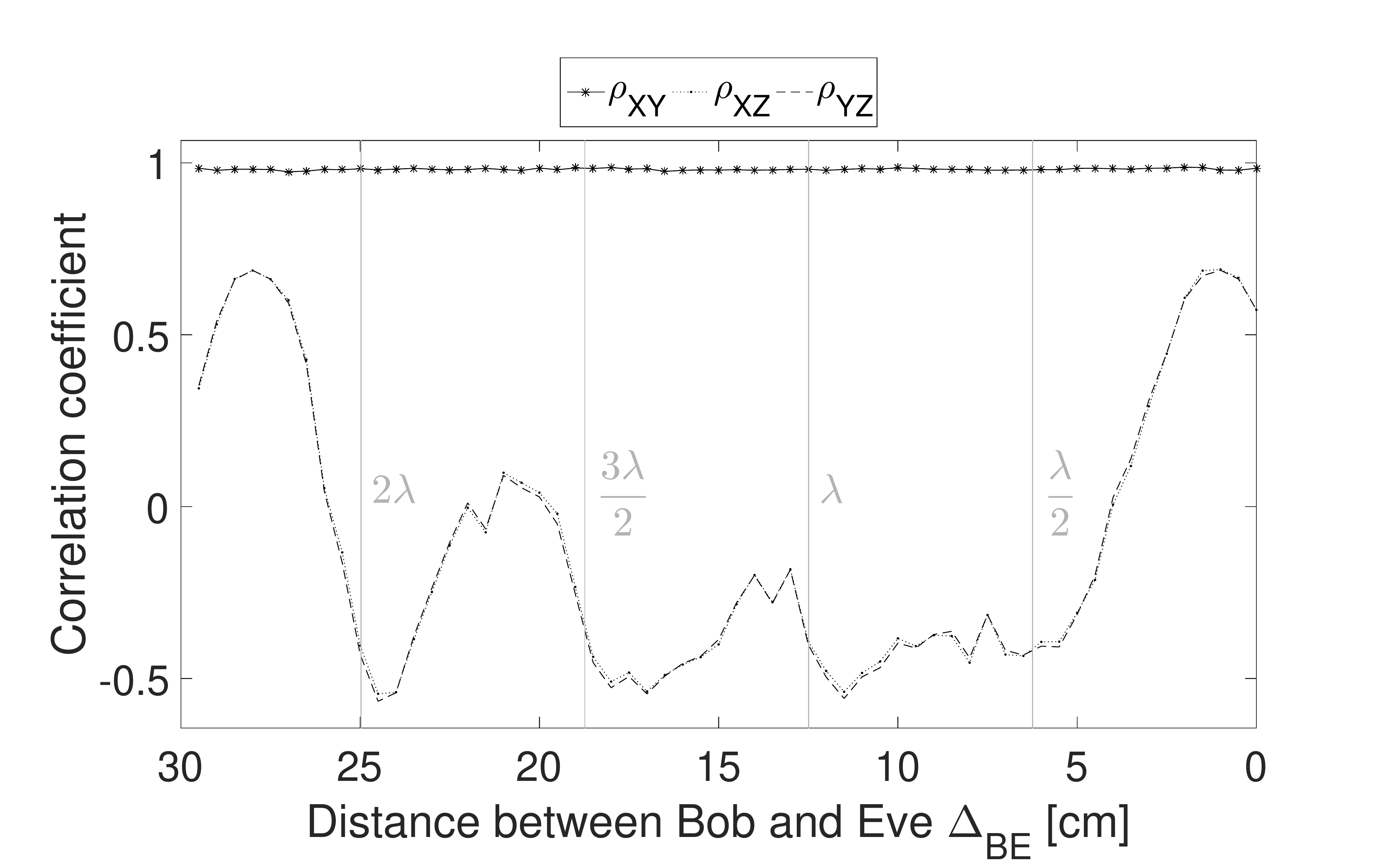}}
	\subfloat[]{\includegraphics[trim=0.5cm 0.1cm 3.5cm 1.6cm, clip=true, height=0.224\textwidth]{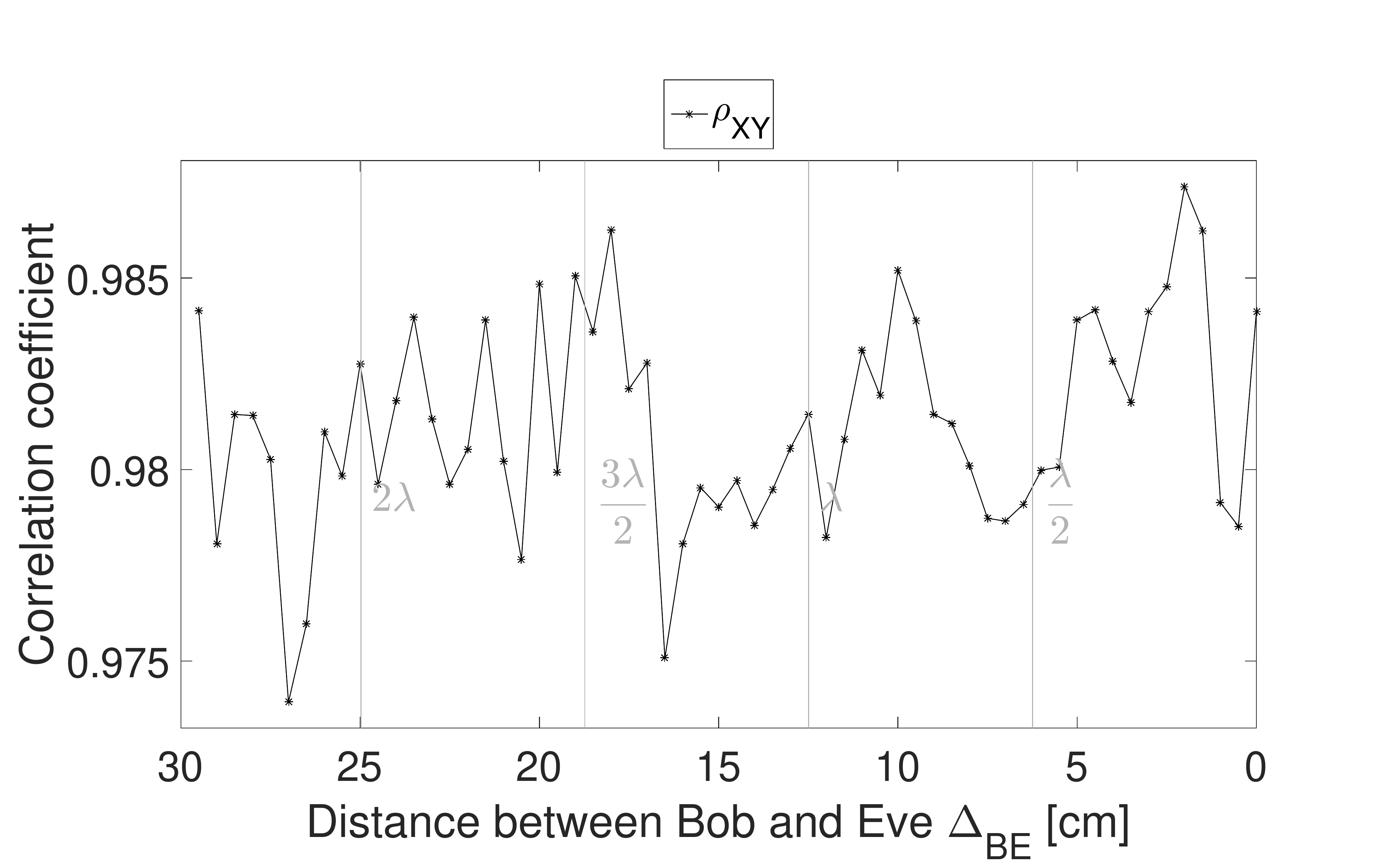}}
	\subfloat[]{\includegraphics[trim=2.2cm 0.1cm 3.5cm 1.6cm, clip=true, height=0.224\textwidth]{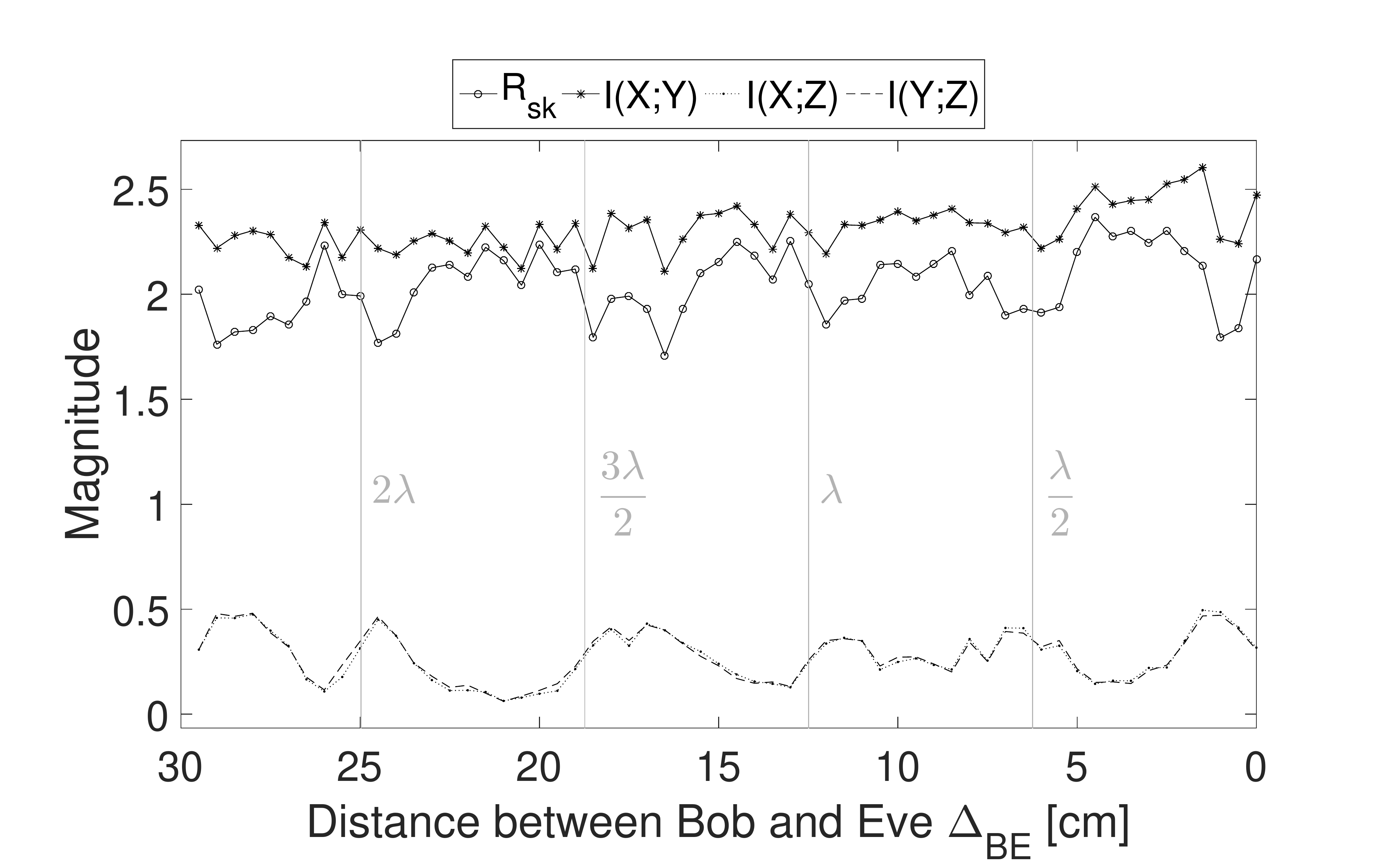}}
	\caption{Evaluation results of $\mybold{v}_k$. In (a) and (b) the cross-correlations is given; in (c) the mutual information as well as $\rsk$ is given. Position 12.}
	\label{fig:app_original_12}
\end{figure*}

\begin{figure*}
	\centering
	\subfloat[]{\includegraphics[trim=1.4cm 0.1cm 3.5cm 1.6cm, clip=true, height=0.224\textwidth]{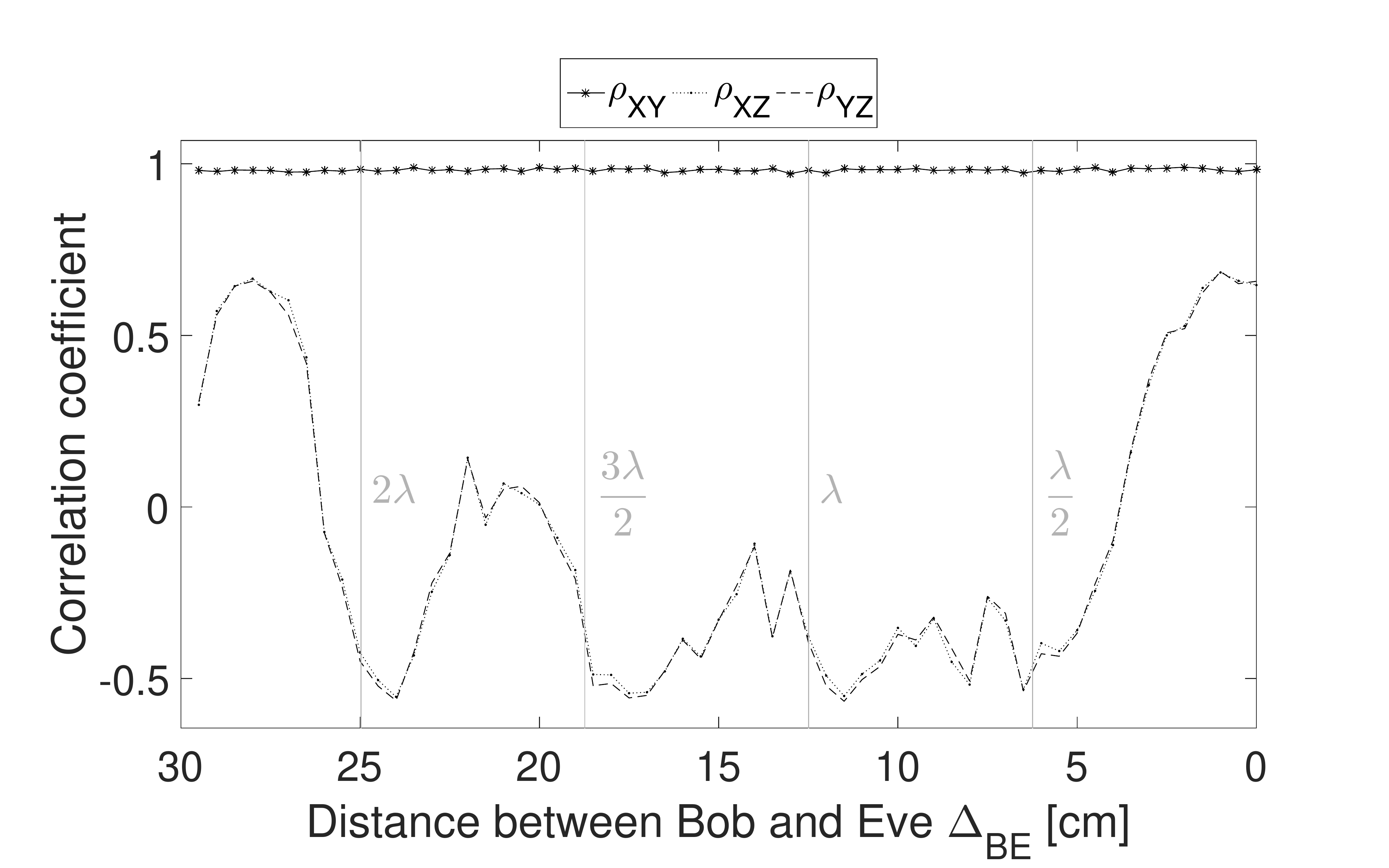}}
	\subfloat[]{\includegraphics[trim=0.5cm 0.1cm 3.5cm 1.6cm, clip=true, height=0.224\textwidth]{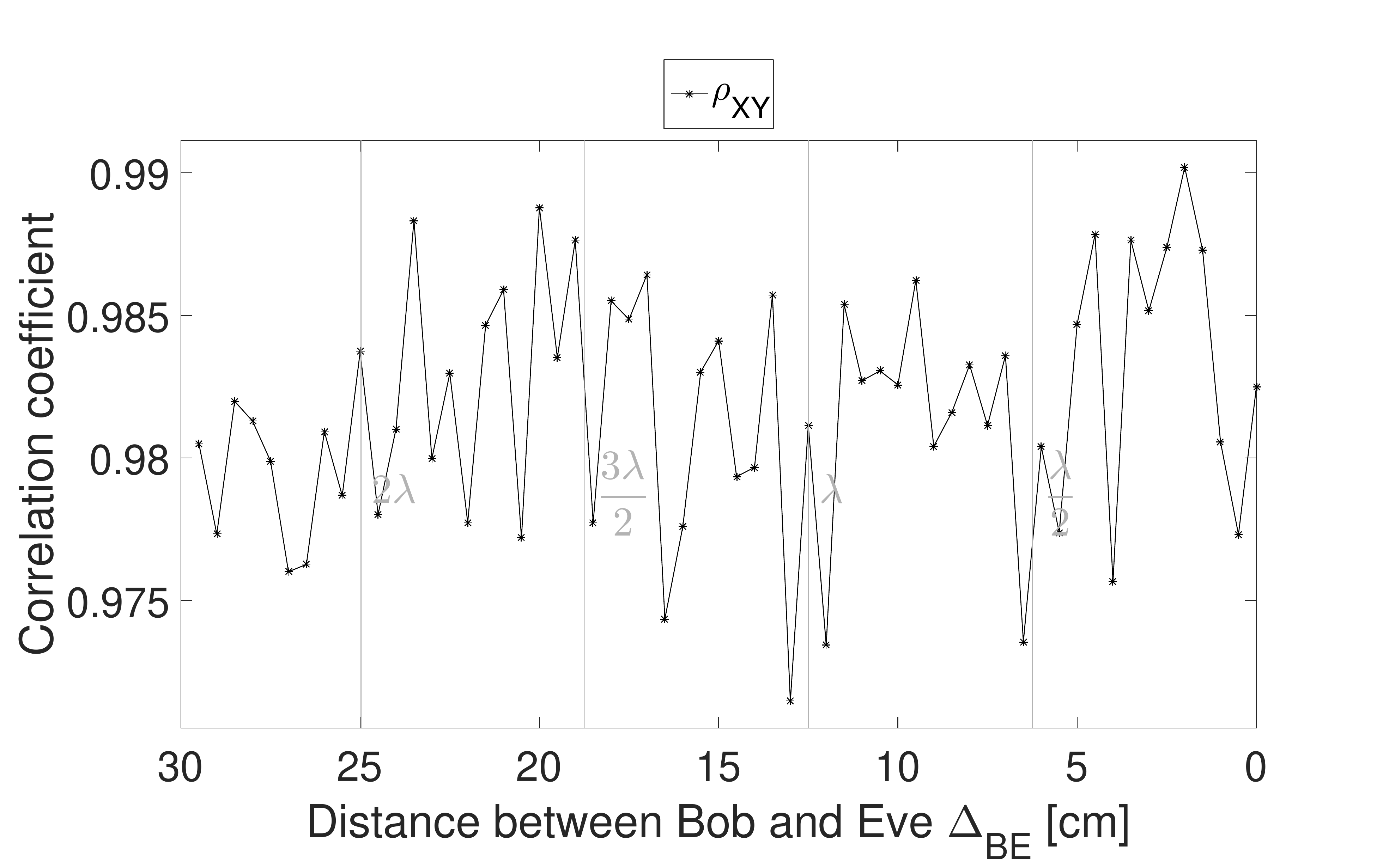}}
	\subfloat[]{\includegraphics[trim=2.2cm 0.1cm 3.5cm 1.6cm, clip=true, height=0.224\textwidth]{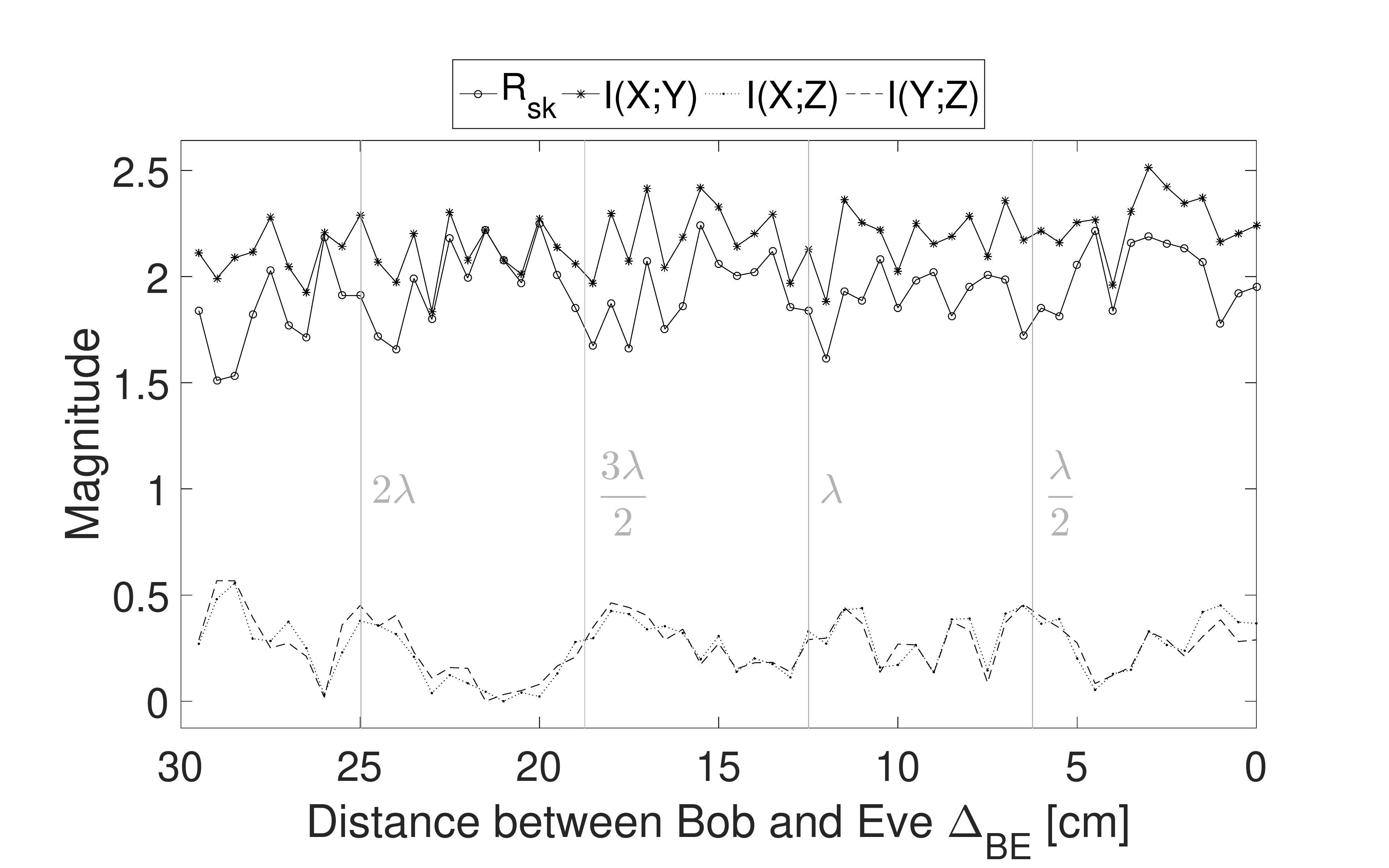}}
	\caption{Evaluation results of $\mybold{v}^{\text{ds}}_k$. In (a) and (b) the cross-correlations is given; in (c) the mutual information as well as $\rsk$ is given. Position 12.}
	\label{fig:app_ds_12}
\end{figure*}

\clearpage

\begin{figure*}
	\centering
	\subfloat[]{\includegraphics[trim=1.4cm 0.1cm 3.5cm 1.6cm, clip=true, height=0.224\textwidth]{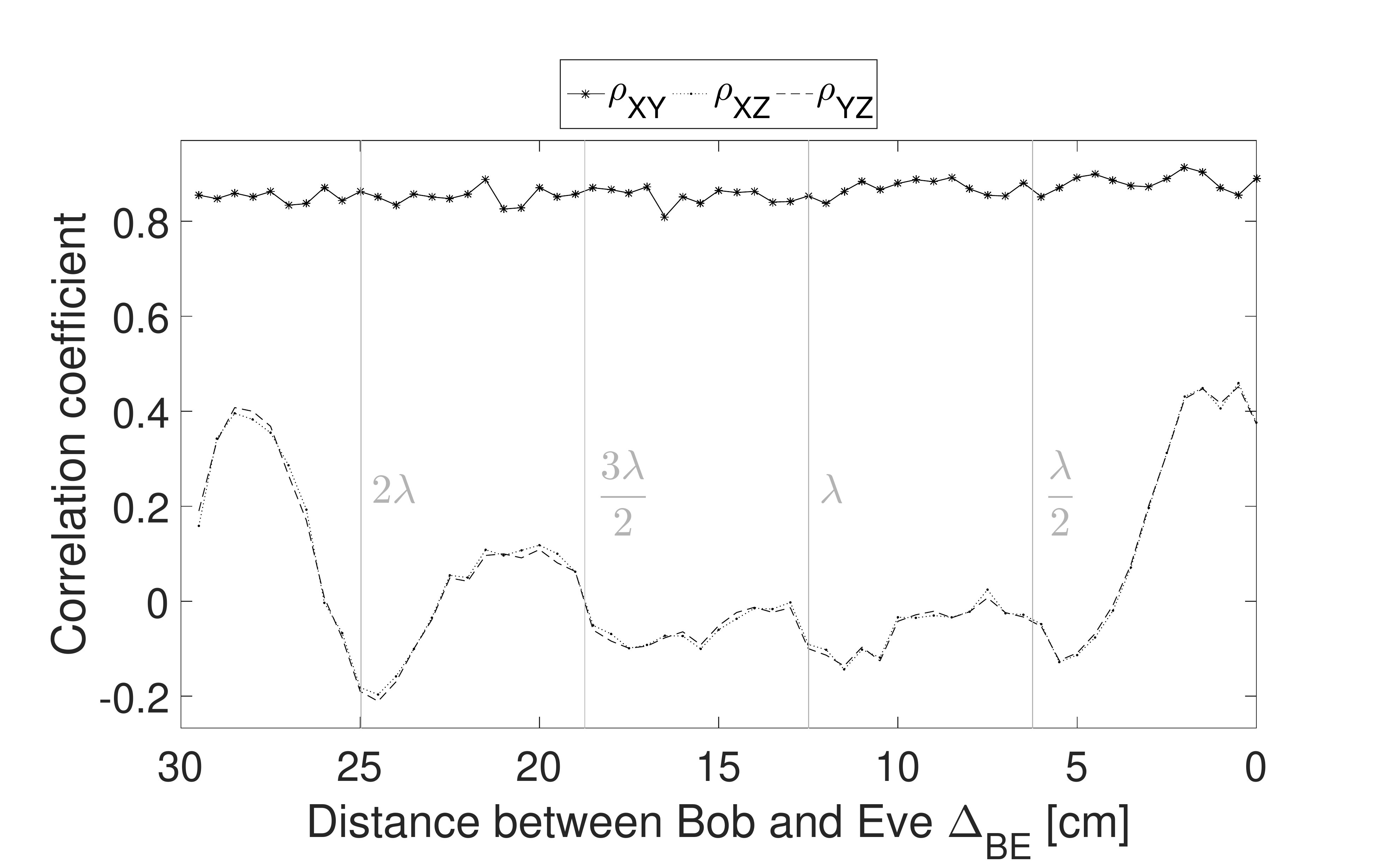}}
	\subfloat[]{\includegraphics[trim=1cm 0.1cm 3.5cm 1.6cm, clip=true, height=0.224\textwidth]{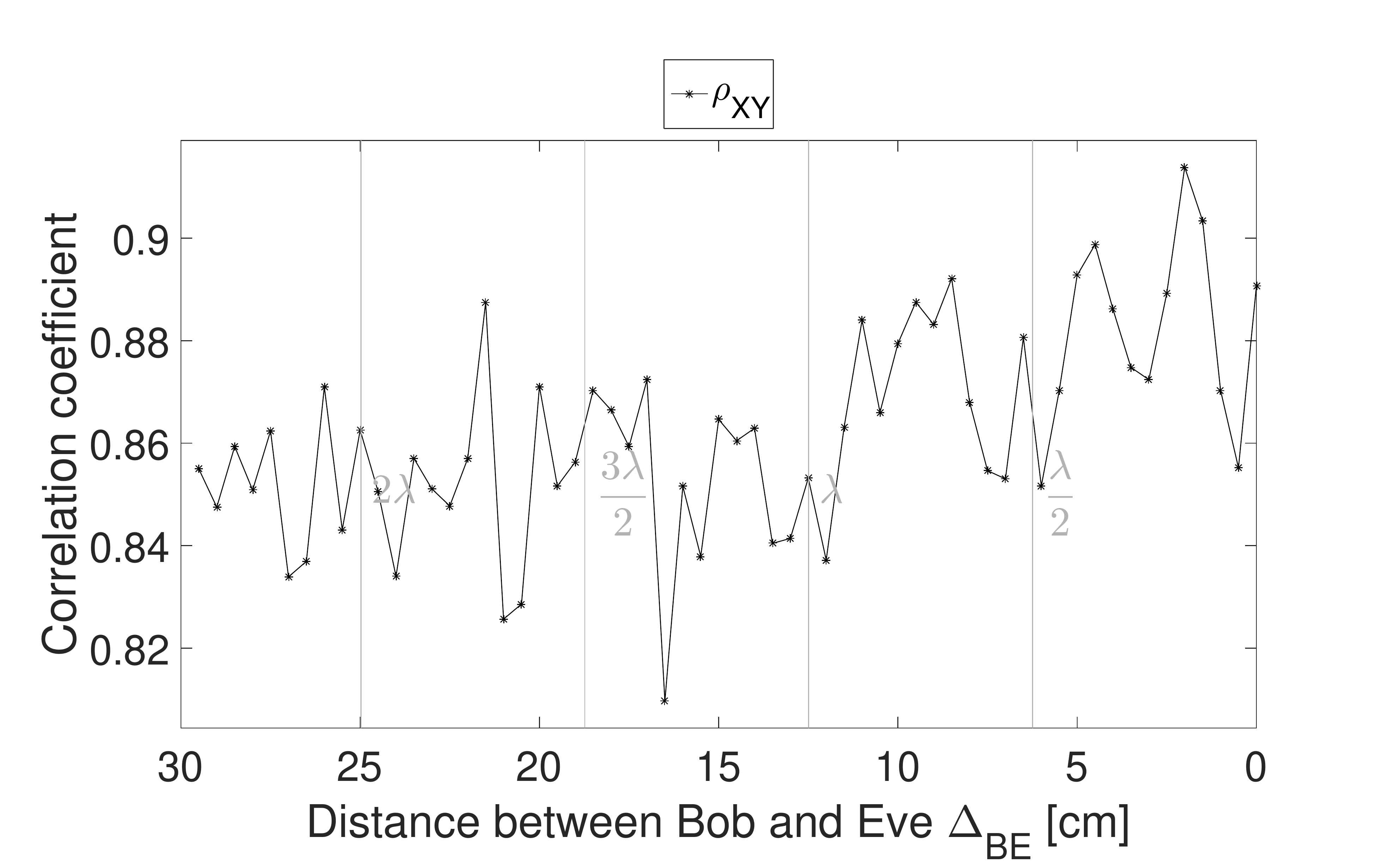}}
	\subfloat[]{\includegraphics[trim=1.8cm 0.1cm 3.5cm 1.6cm, clip=true, height=0.224\textwidth]{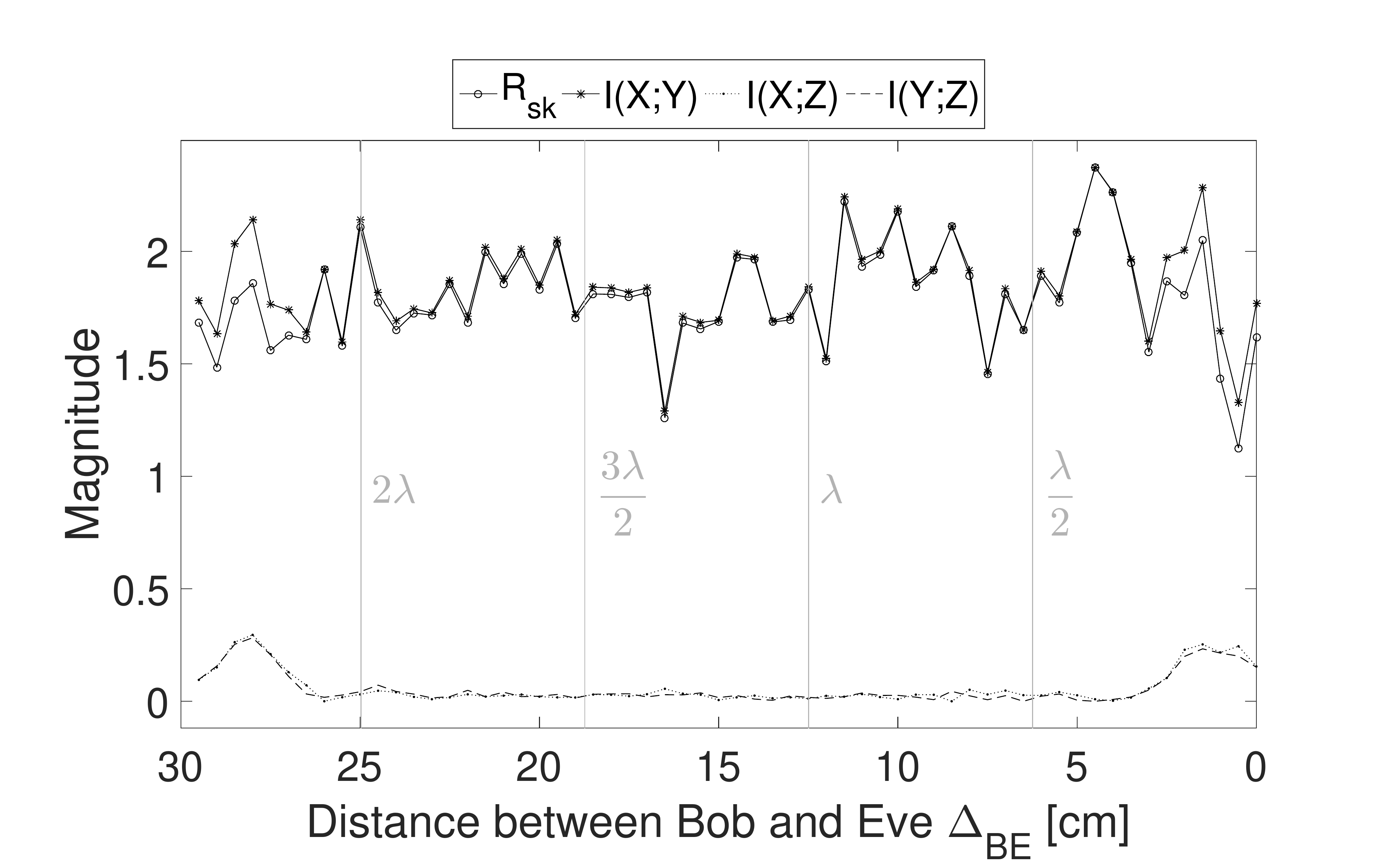}}
	\caption{Evaluation results of $\mybold{v}^{\text{de}}_k$. In (a) and (b) the cross-correlations is given; in (c) the mutual information as well as $\rsk$ is given. Position 12.}
	\label{fig:app_decorr_12}
\end{figure*}


\begin{figure*}
	\centering
	\subfloat[]{\includegraphics[trim=1.4cm 0.1cm 3.5cm 1.6cm, clip=true, height=0.224\textwidth]{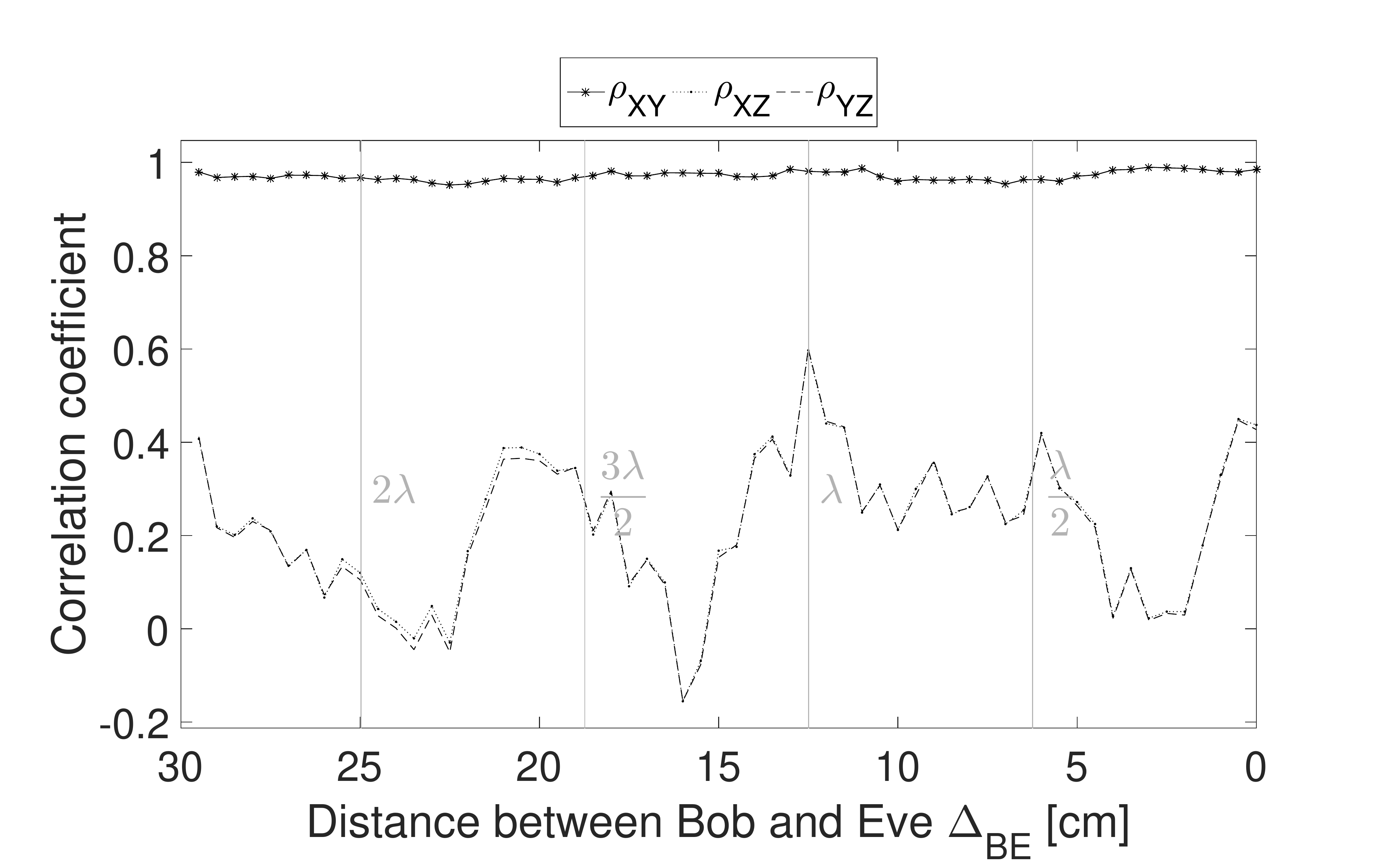}}
	\subfloat[]{\includegraphics[trim=0.5cm 0.1cm 3.5cm 1.6cm, clip=true, height=0.224\textwidth]{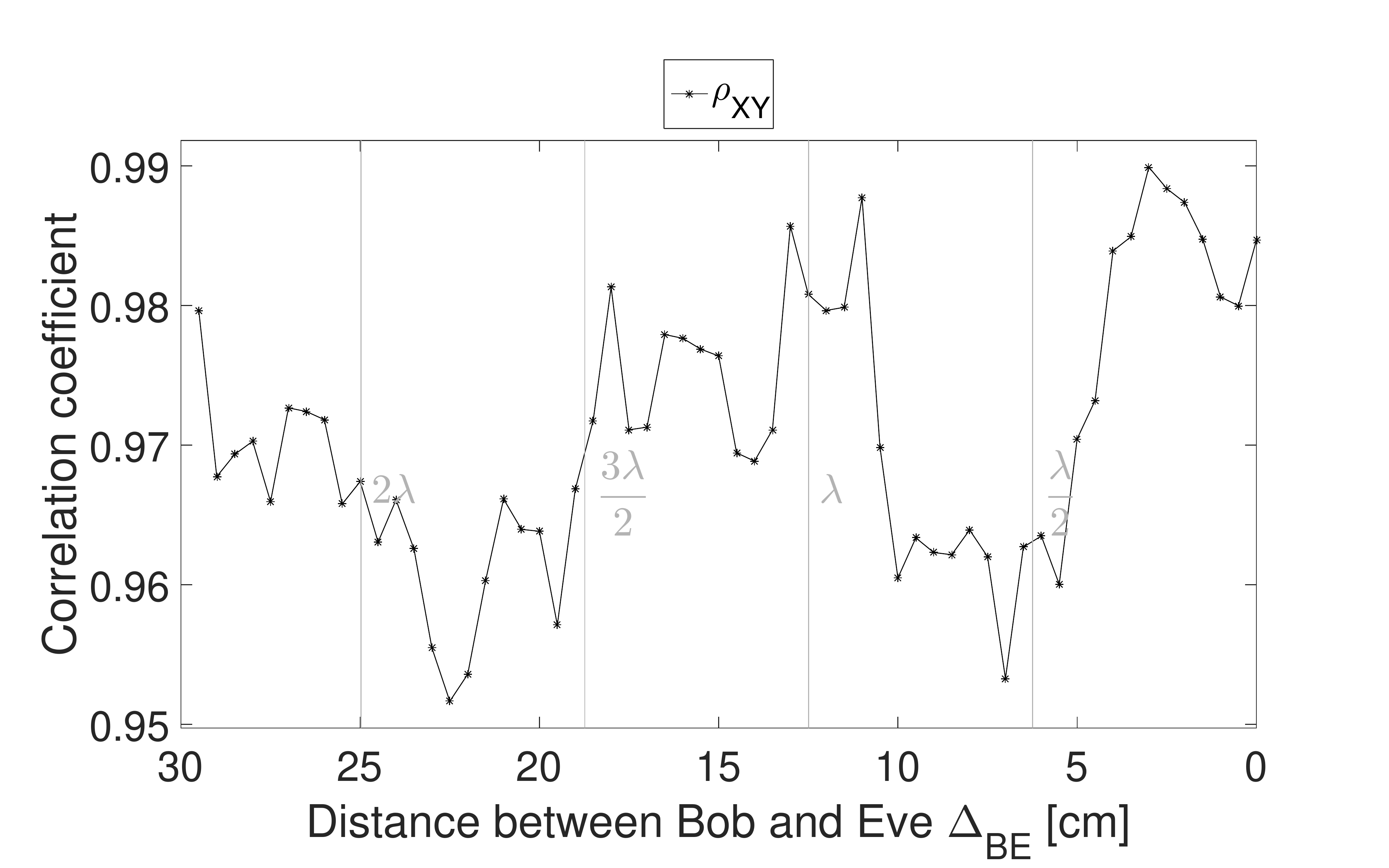}}
	\subfloat[]{\includegraphics[trim=2.2cm 0.1cm 3.5cm 1.6cm, clip=true, height=0.224\textwidth]{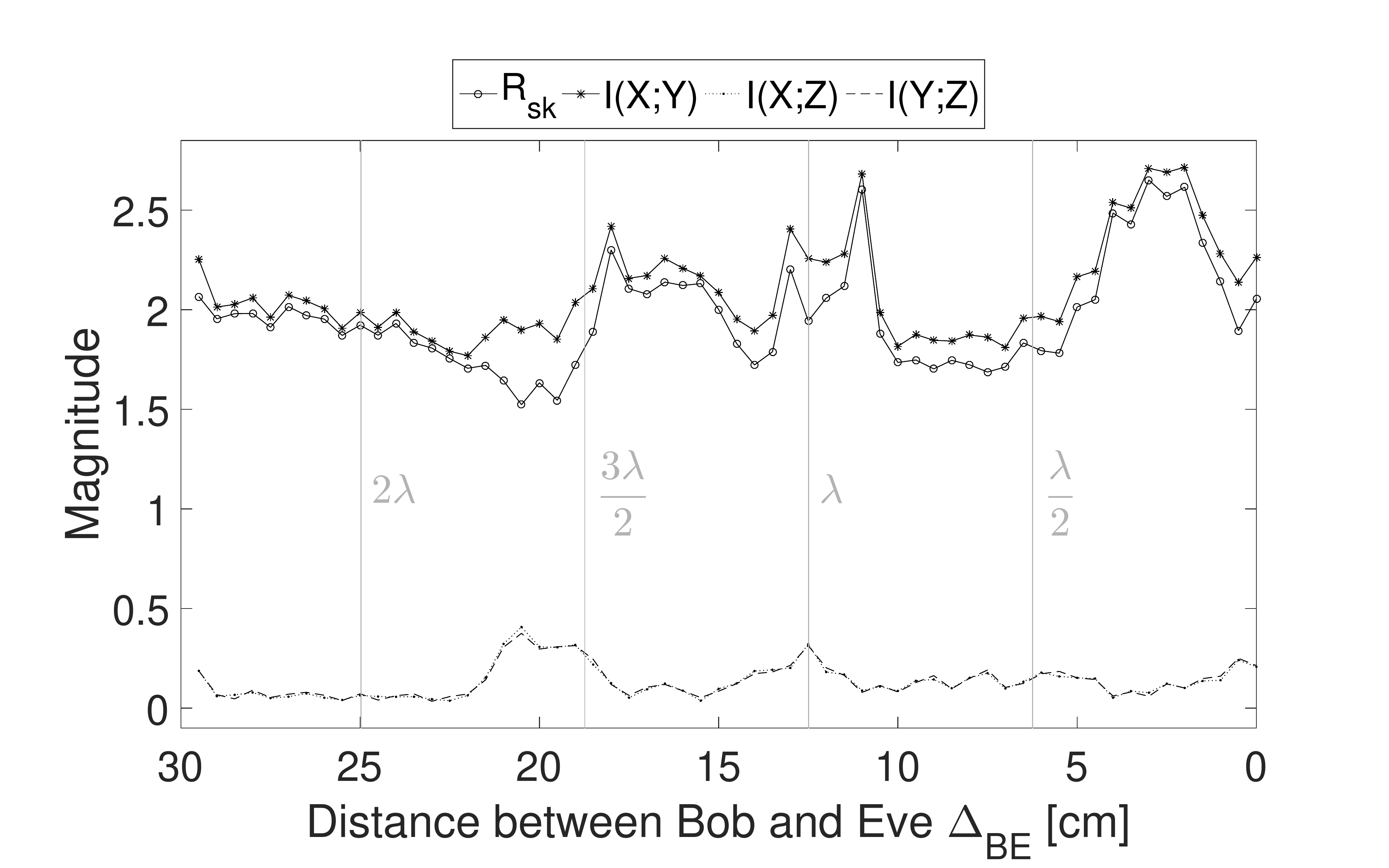}}
	\caption{Evaluation results of $\mybold{v}_k$. In (a) and (b) the cross-correlations is given; in (c) the mutual information as well as $\rsk$ is given. Position 13.}
	\label{fig:app_original_13}
\end{figure*}

\begin{figure*}
	\centering
	\subfloat[]{\includegraphics[trim=1.4cm 0.1cm 3.5cm 1.6cm, clip=true, height=0.224\textwidth]{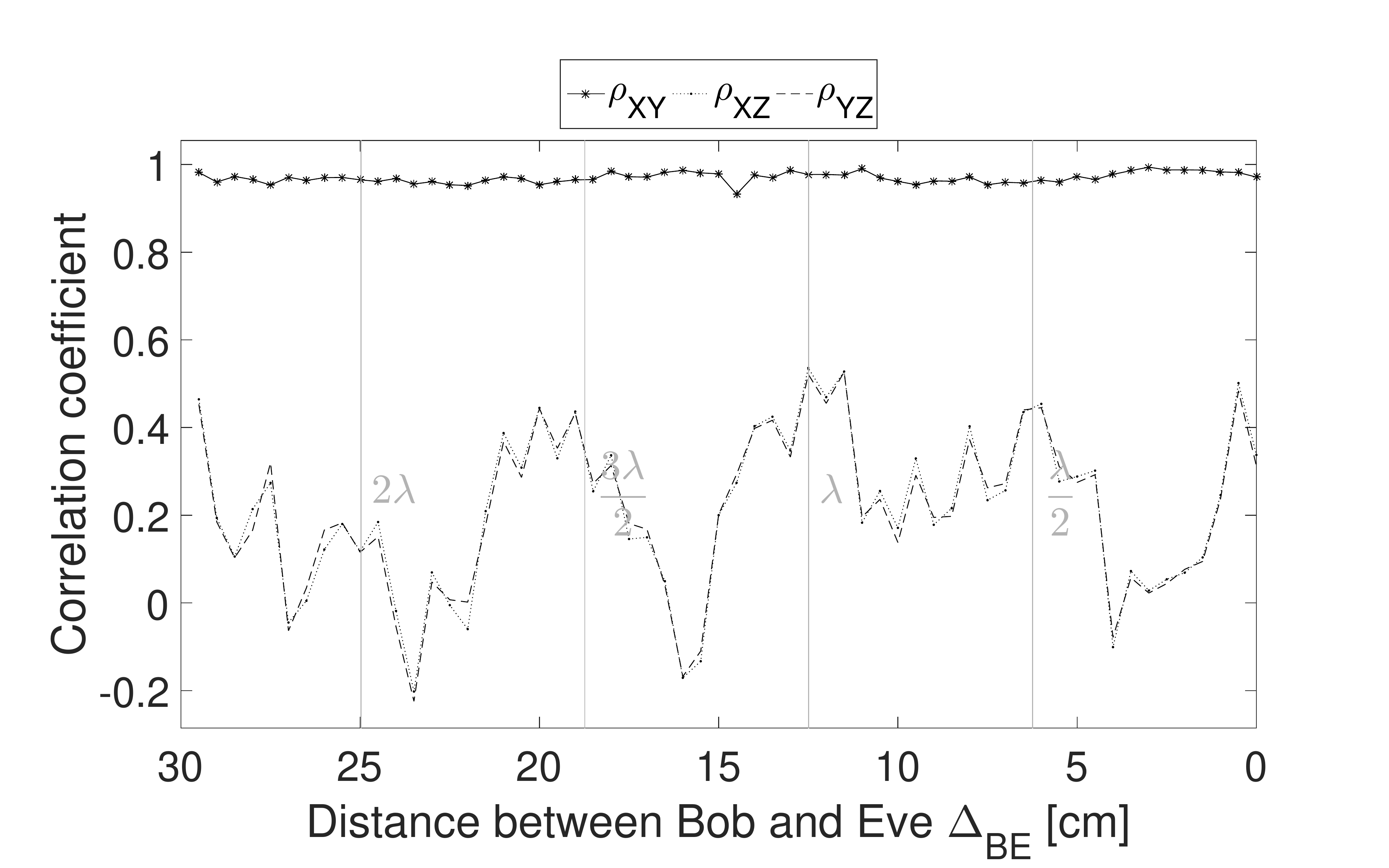}}
	\subfloat[]{\includegraphics[trim=0.5cm 0.1cm 3.5cm 1.6cm, clip=true, height=0.224\textwidth]{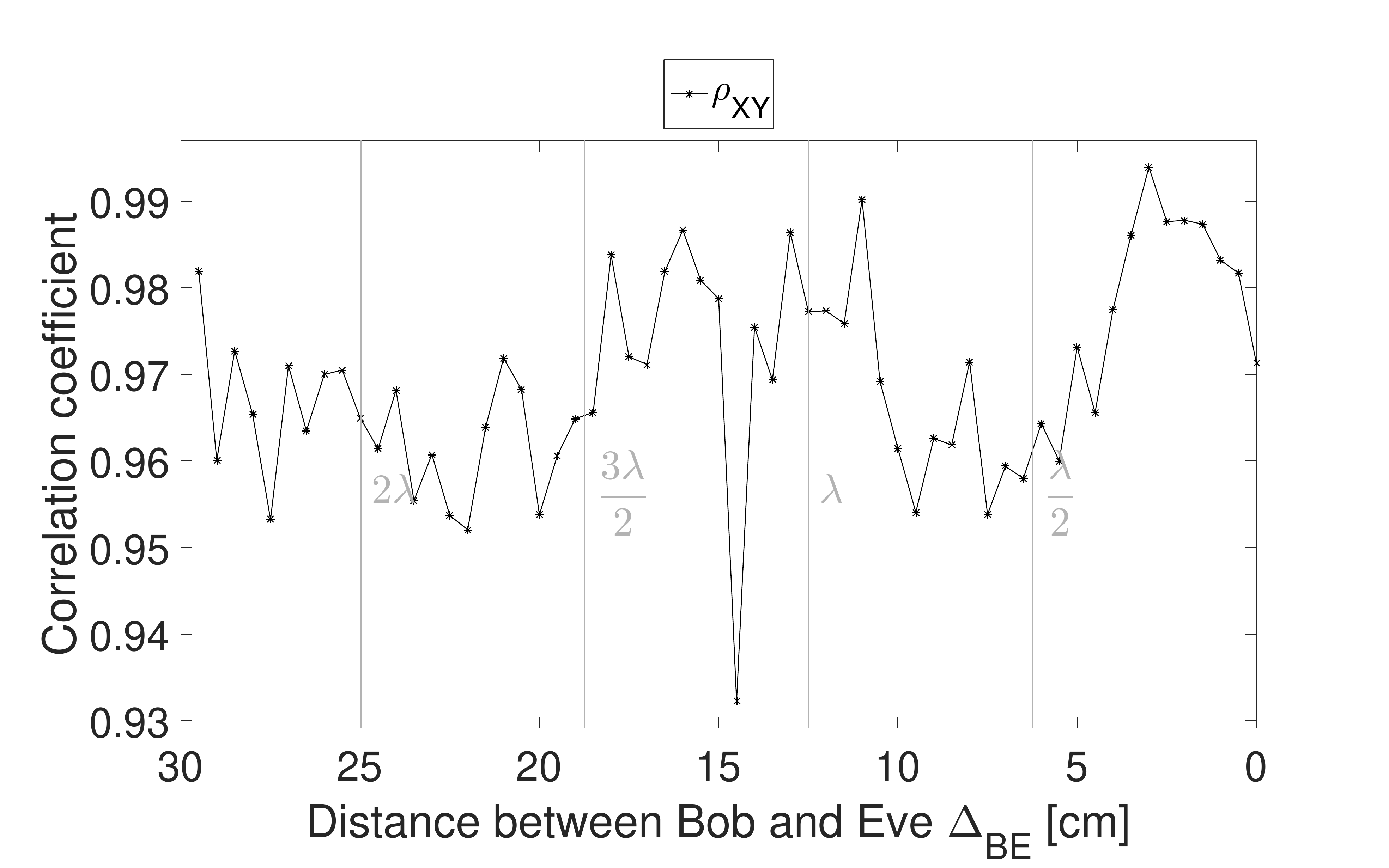}}
	\subfloat[]{\includegraphics[trim=2.2cm 0.1cm 3.5cm 1.6cm, clip=true, height=0.224\textwidth]{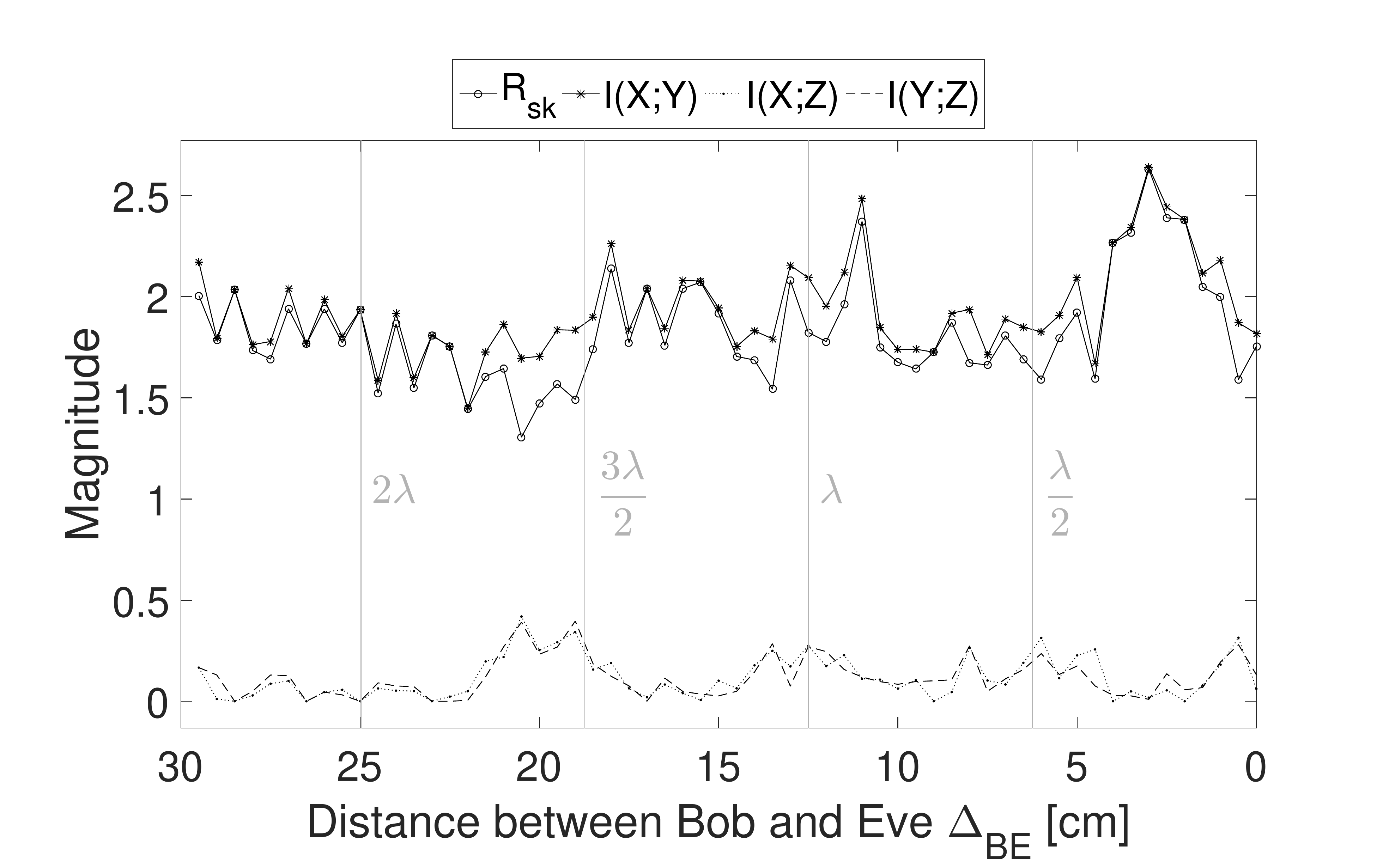}}
	\caption{Evaluation results of $\mybold{v}^{\text{ds}}_k$. In (a) and (b) the cross-correlations is given; in (c) the mutual information as well as $\rsk$ is given. Position 13.}
	\label{fig:app_ds_13}
\end{figure*}

\begin{figure*}
	\centering
	\subfloat[]{\includegraphics[trim=1.4cm 0.1cm 3.5cm 1.6cm, clip=true, height=0.224\textwidth]{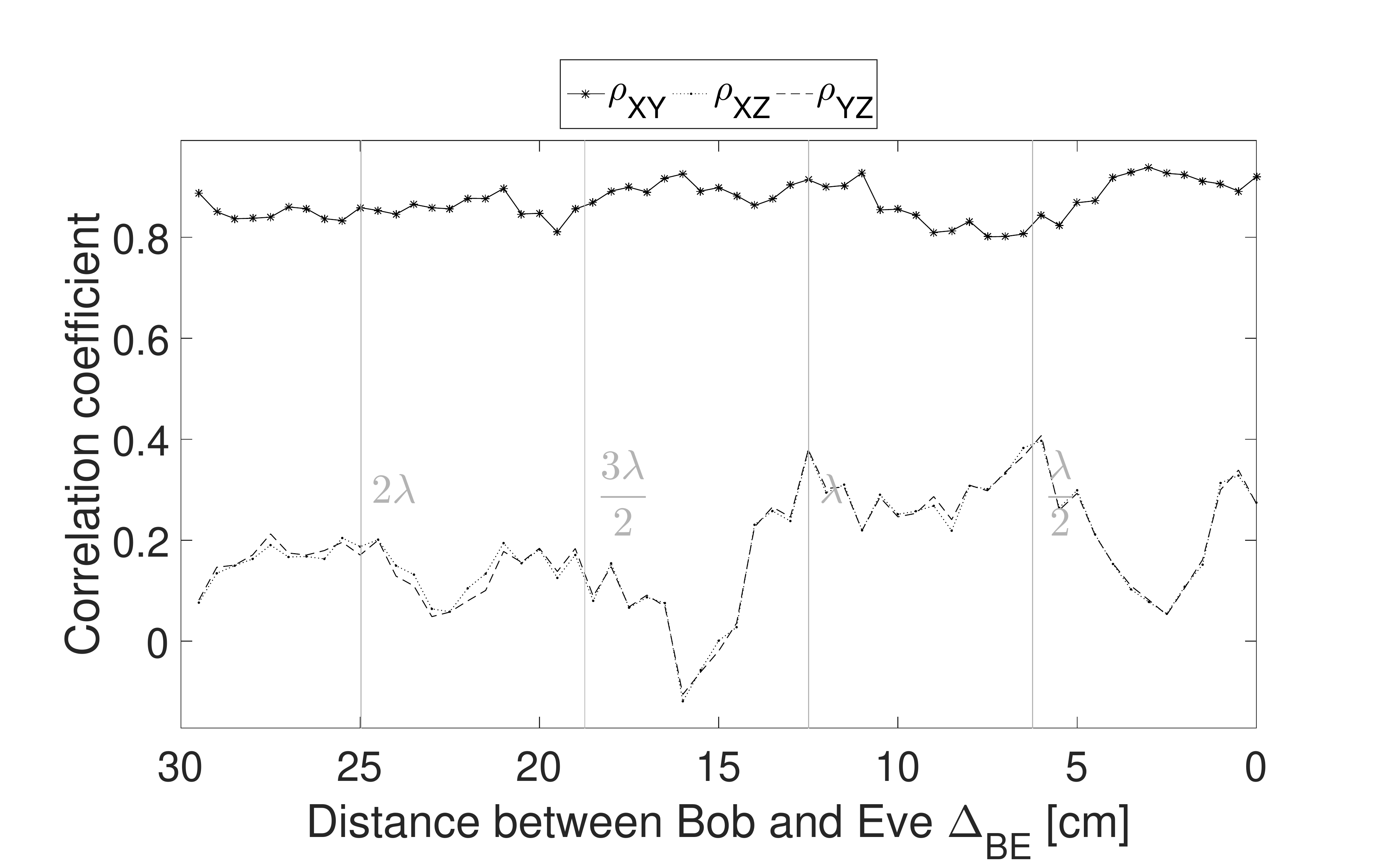}}
	\subfloat[]{\includegraphics[trim=1cm 0.1cm 3.5cm 1.6cm, clip=true, height=0.224\textwidth]{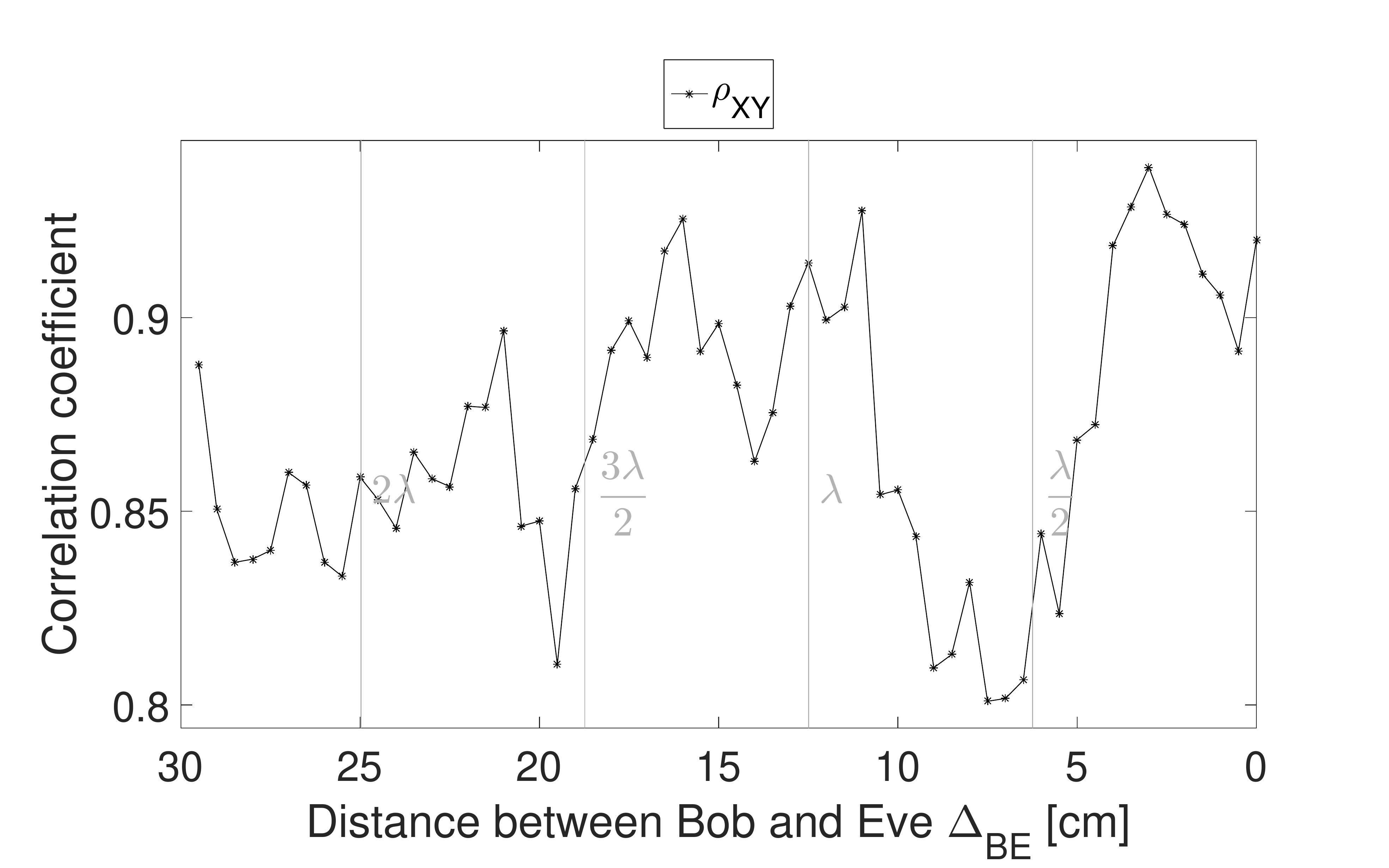}}
	\subfloat[]{\includegraphics[trim=1.8cm 0.1cm 3.5cm 1.6cm, clip=true, height=0.224\textwidth]{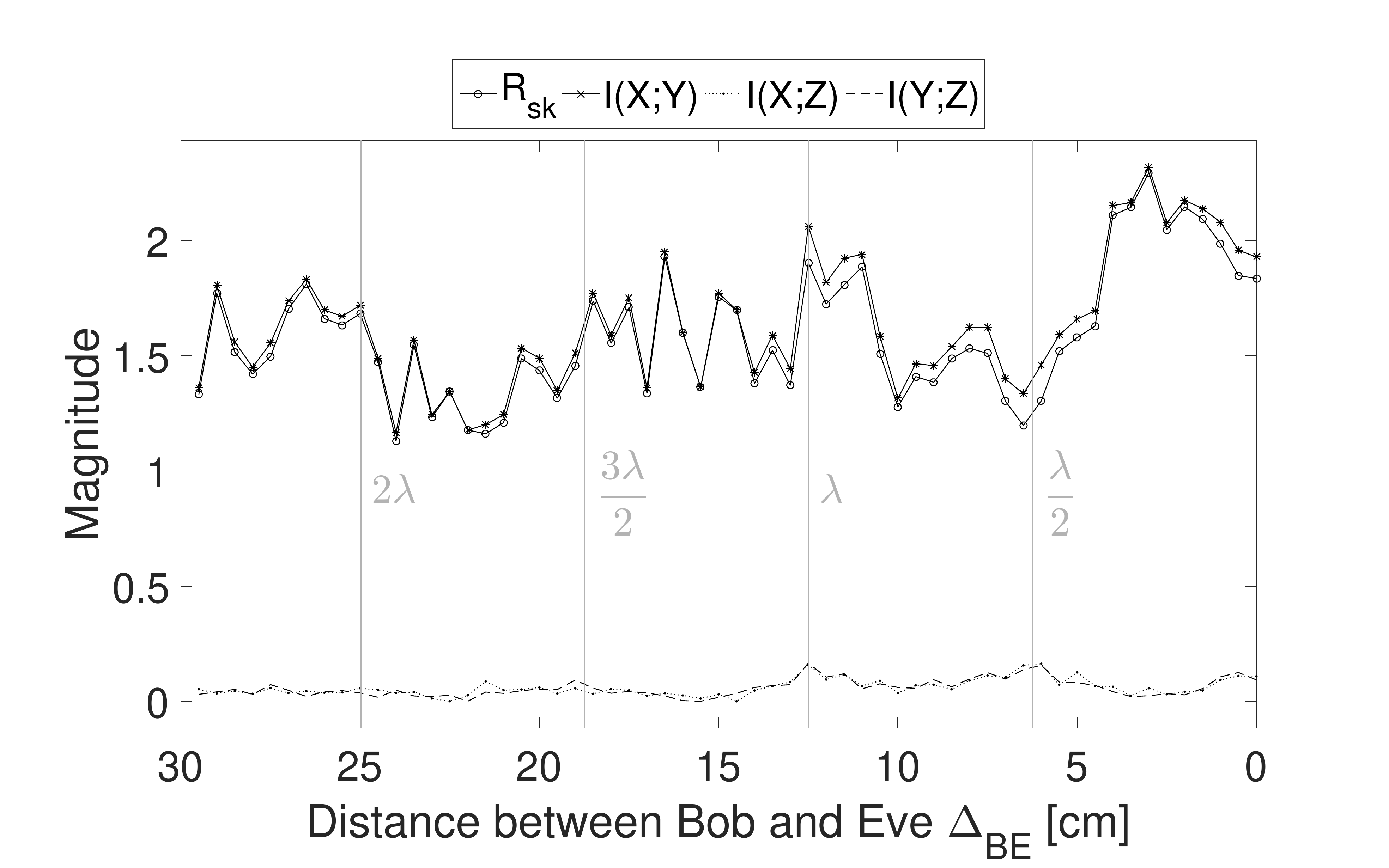}}
	\caption{Evaluation results of $\mybold{v}^{\text{de}}_k$. In (a) and (b) the cross-correlations is given; in (c) the mutual information as well as $\rsk$ is given. Position 13.}
	\label{fig:app_decorr_13}
\end{figure*}


\begin{figure*}
	\centering
	\subfloat[]{\includegraphics[trim=1.4cm 0.1cm 3.5cm 1.6cm, clip=true, height=0.224\textwidth]{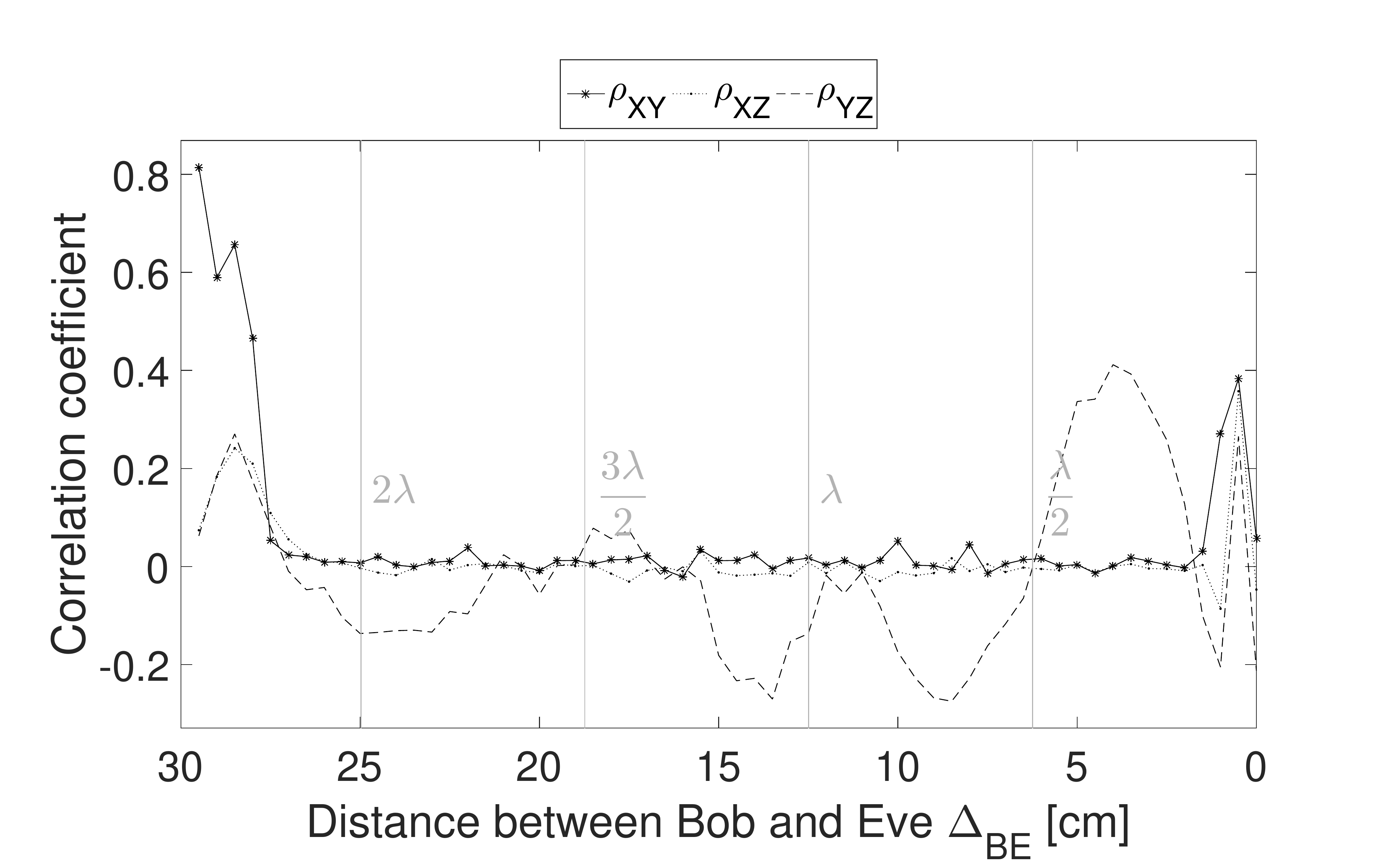}}
	\subfloat[]{\includegraphics[trim=0.5cm 0.1cm 3.5cm 1.6cm, clip=true, height=0.224\textwidth]{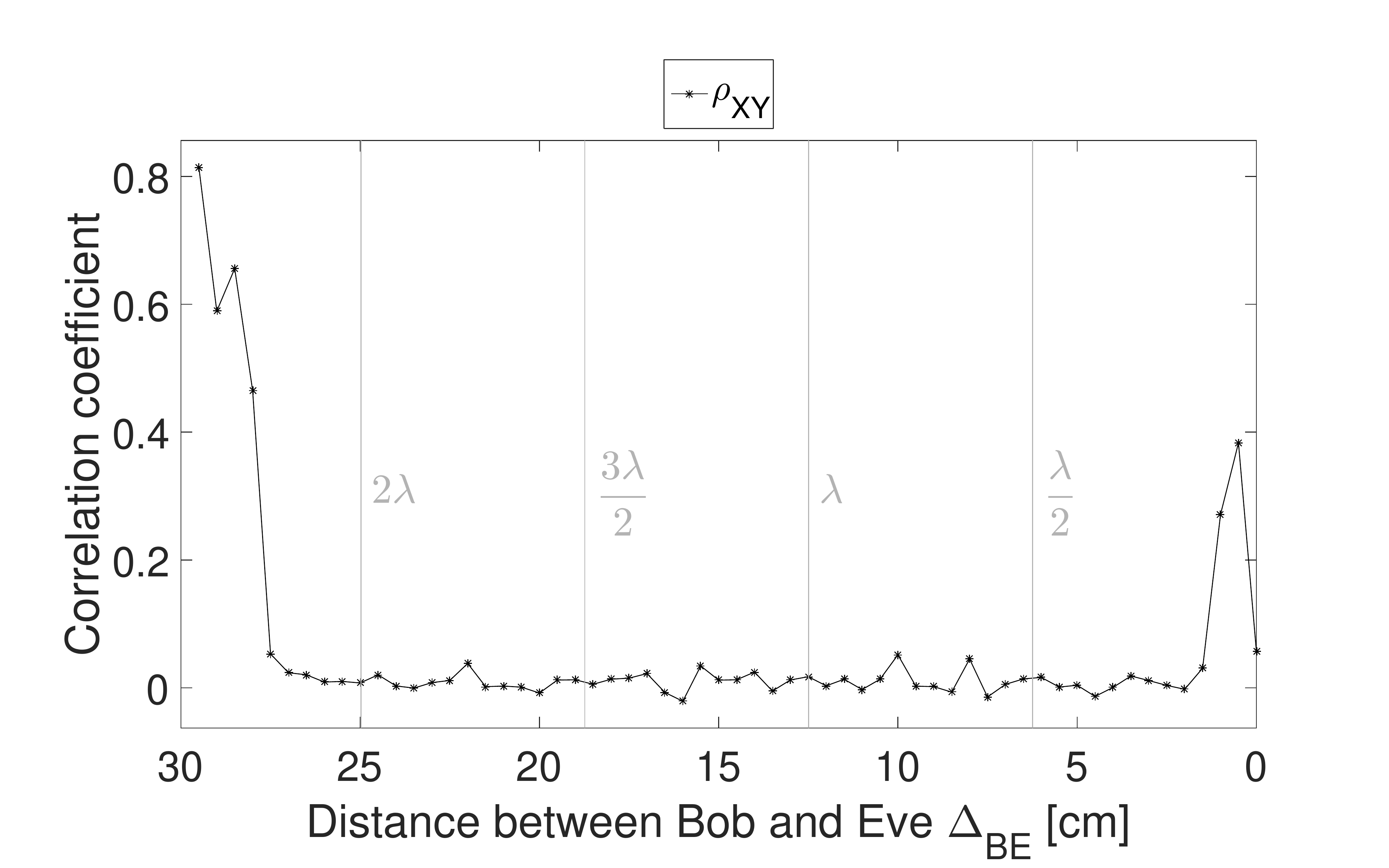}}
	\subfloat[]{\includegraphics[trim=2.2cm 0.1cm 3.5cm 1.6cm, clip=true, height=0.224\textwidth]{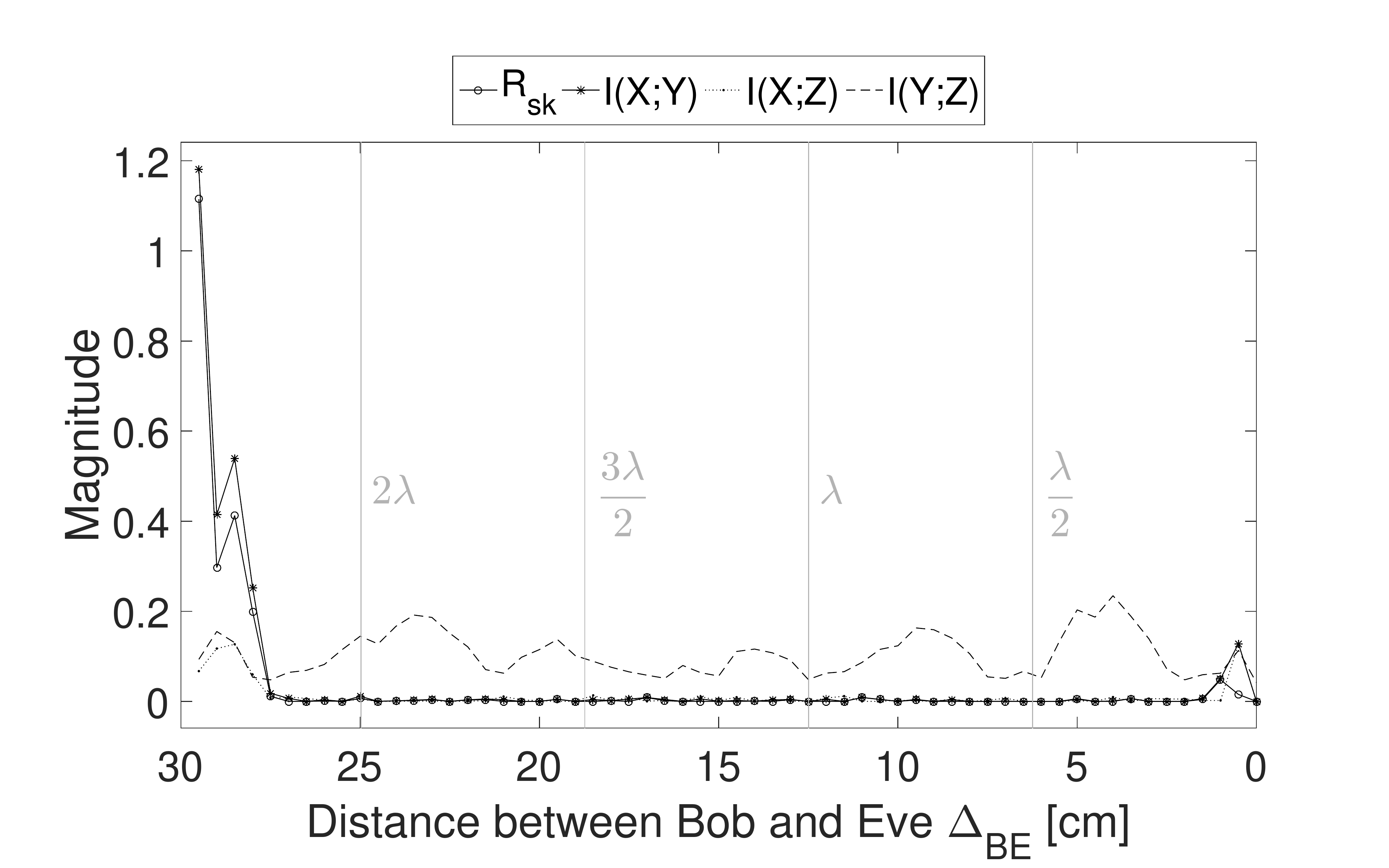}}
	\caption{Evaluation results of $\mybold{v}_k$. In (a) and (b) the cross-correlations is given; in (c) the mutual information as well as $\rsk$ is given. Position 14.}
	\label{fig:app_original_14}
\end{figure*}

\begin{figure*}
	\centering
	\subfloat[]{\includegraphics[trim=1.4cm 0.1cm 3.5cm 1.6cm, clip=true, height=0.224\textwidth]{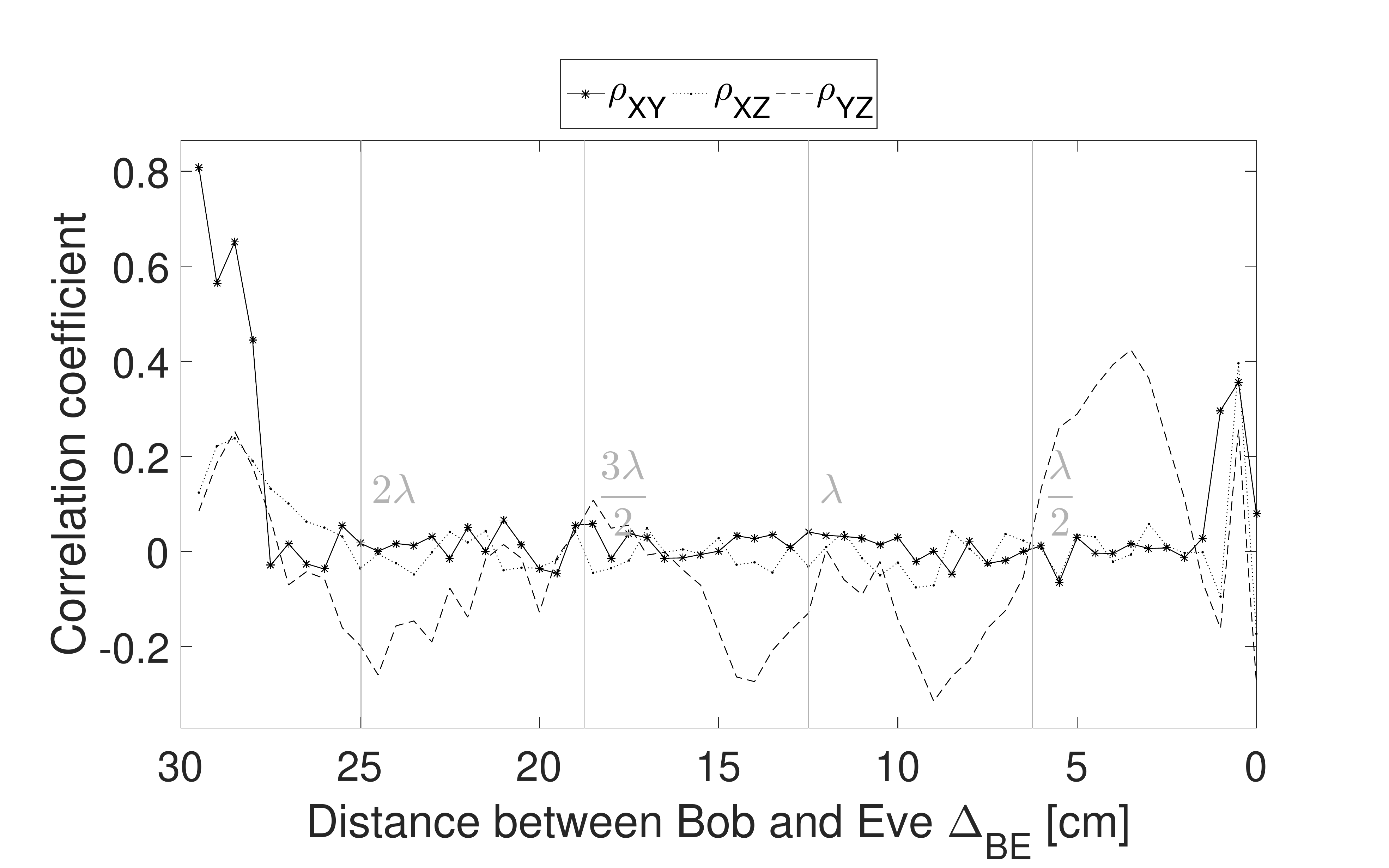}}
	\subfloat[]{\includegraphics[trim=0.5cm 0.1cm 3.5cm 1.6cm, clip=true, height=0.224\textwidth]{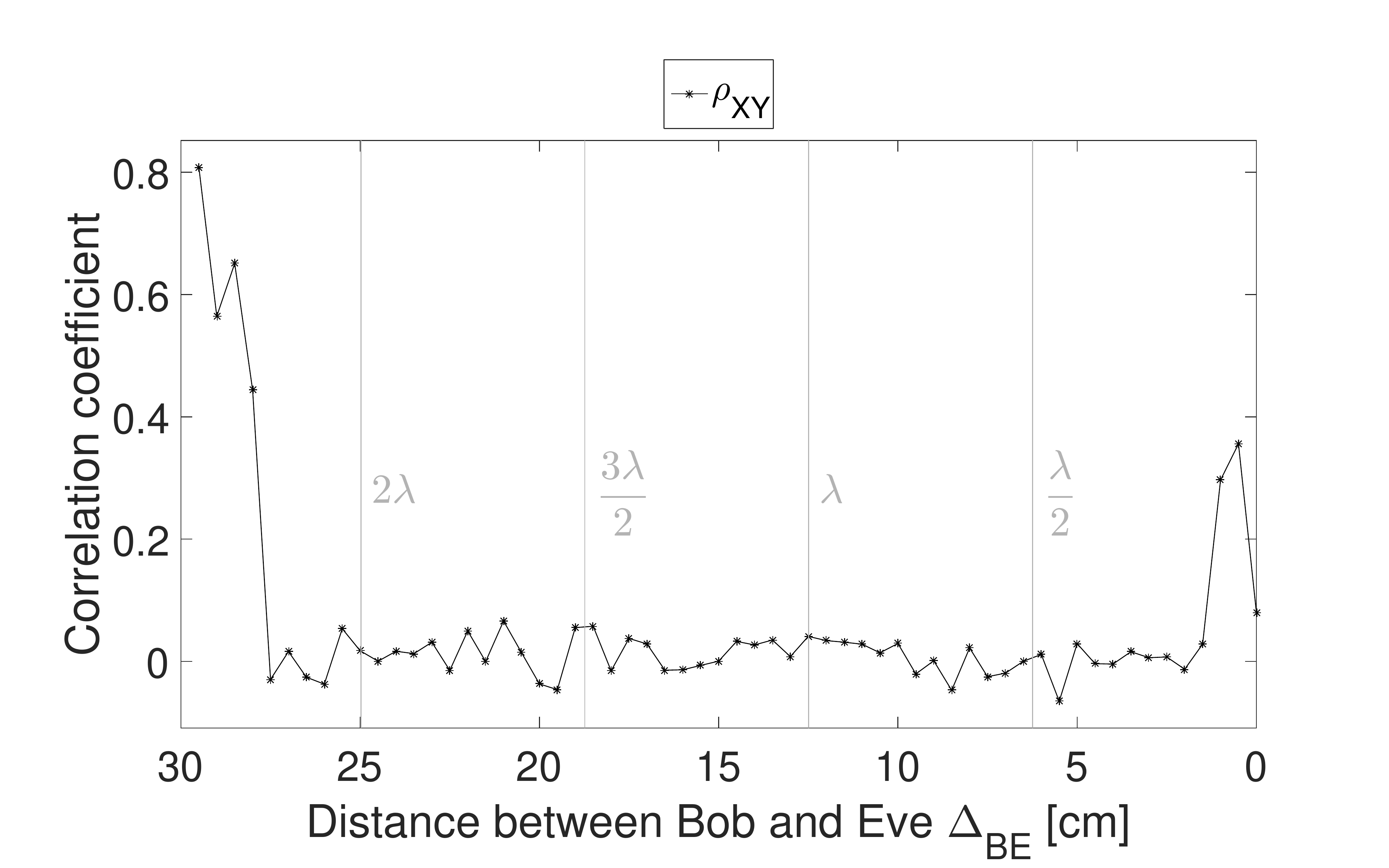}}
	\subfloat[]{\includegraphics[trim=2.2cm 0.1cm 3.5cm 1.6cm, clip=true, height=0.224\textwidth]{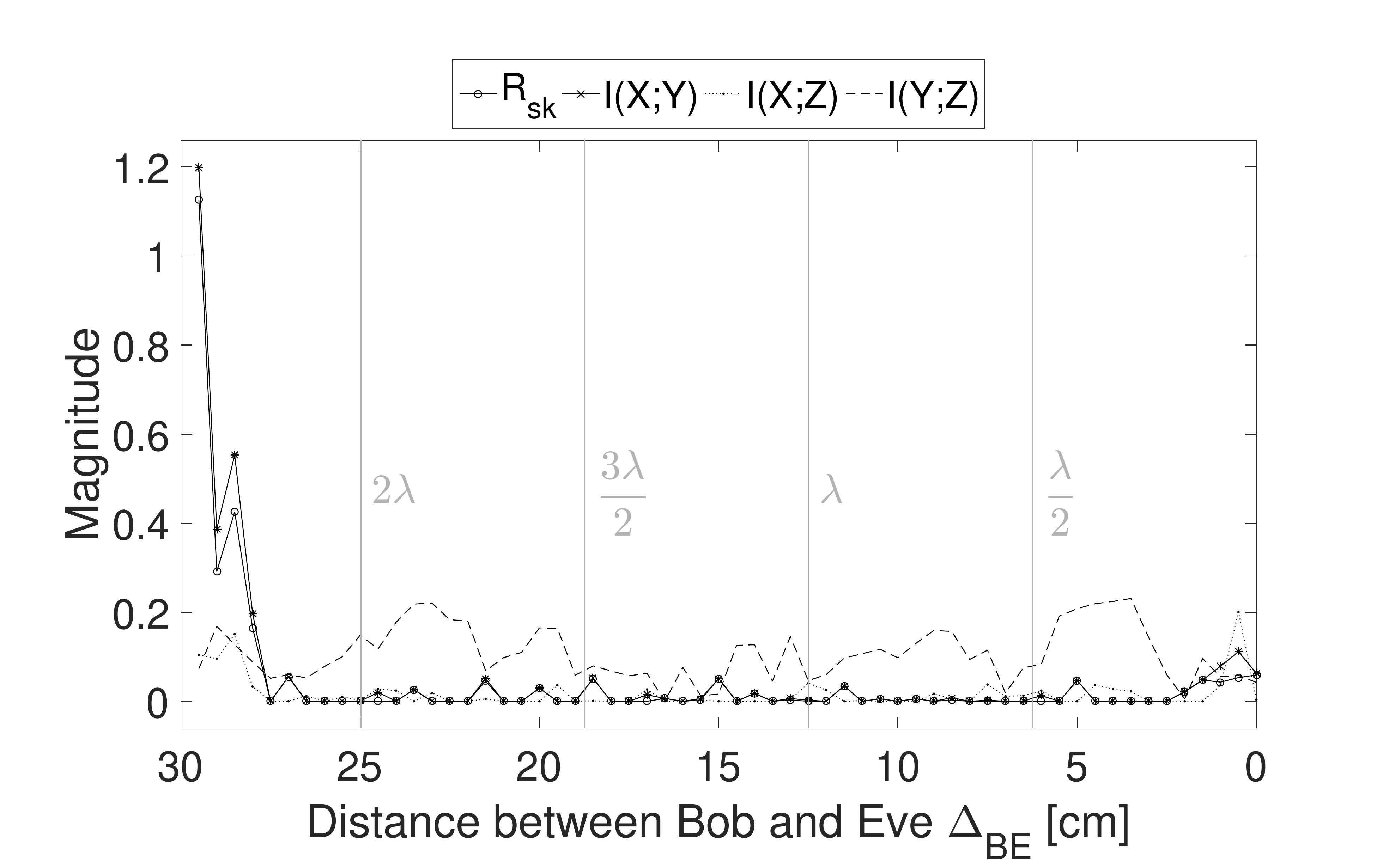}}
	\caption{Evaluation results of $\mybold{v}^{\text{ds}}_k$. In (a) and (b) the cross-correlations is given; in (c) the mutual information as well as $\rsk$ is given. Position 14.}
	\label{fig:app_ds_14}
\end{figure*}

\begin{figure*}
	\centering
	\subfloat[]{\includegraphics[trim=1.4cm 0.1cm 3.5cm 1.6cm, clip=true, height=0.224\textwidth]{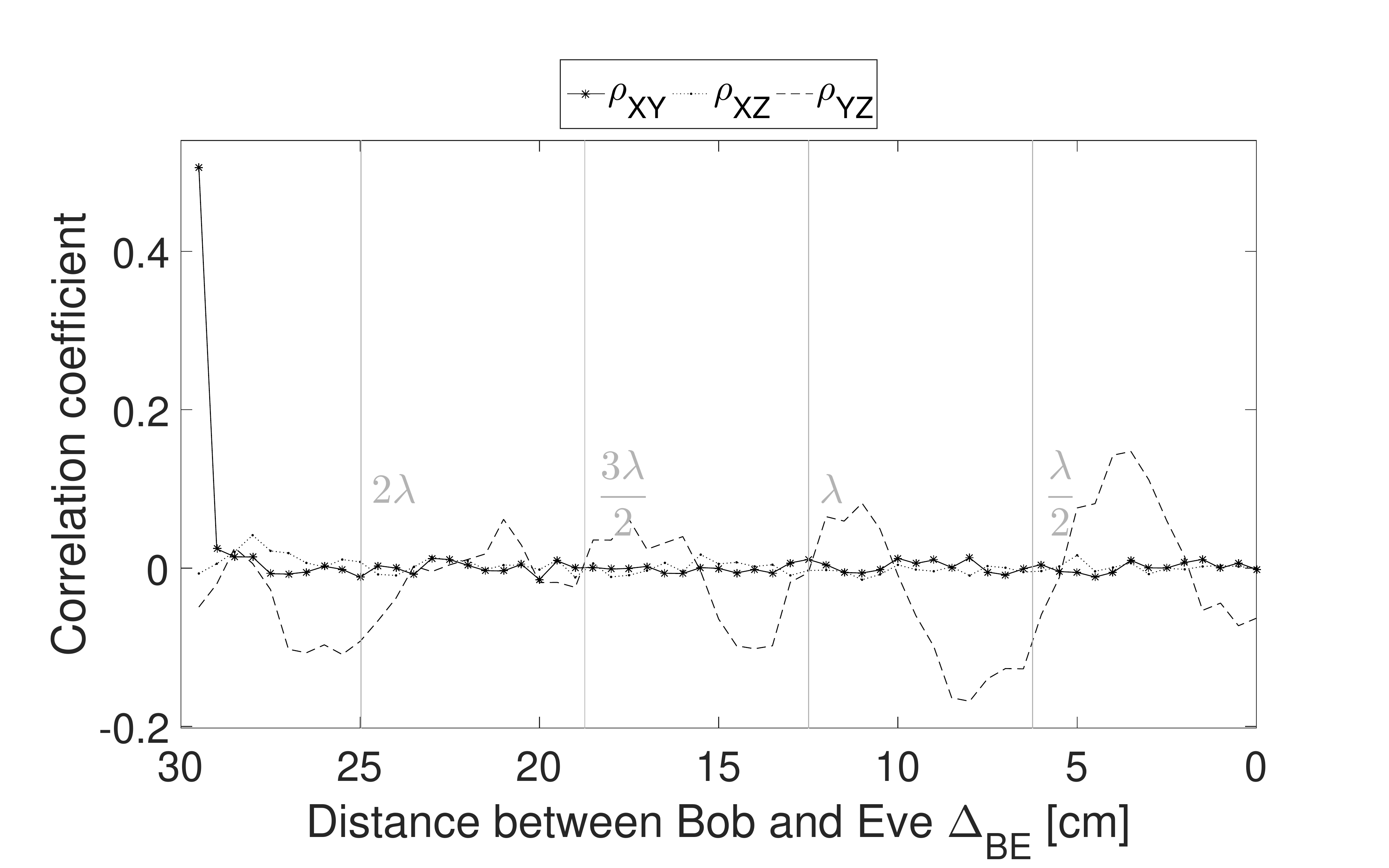}}
	\subfloat[]{\includegraphics[trim=1cm 0.1cm 3.5cm 1.6cm, clip=true, height=0.224\textwidth]{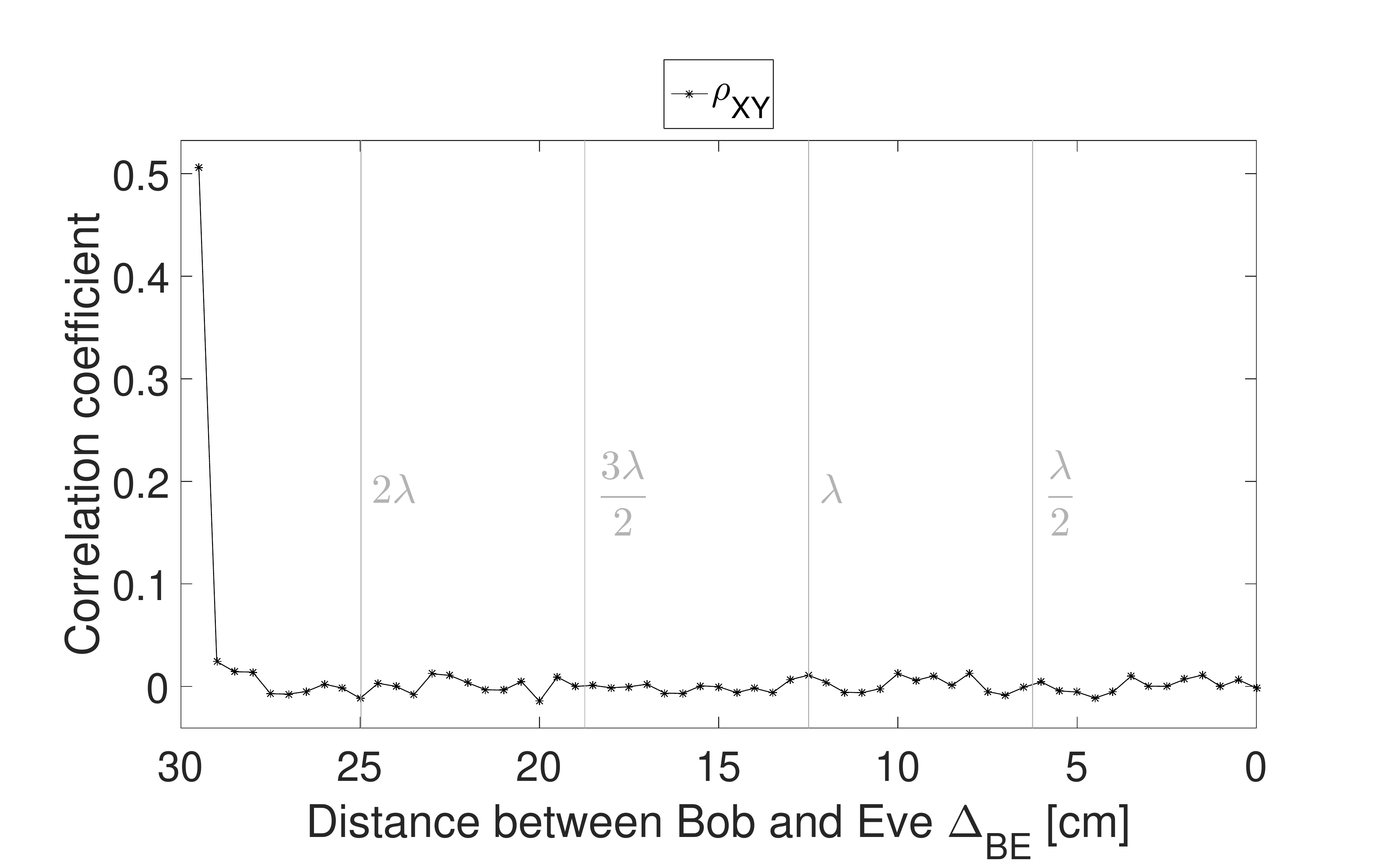}}
	\subfloat[]{\includegraphics[trim=1.8cm 0.1cm 3.5cm 1.6cm, clip=true, height=0.224\textwidth]{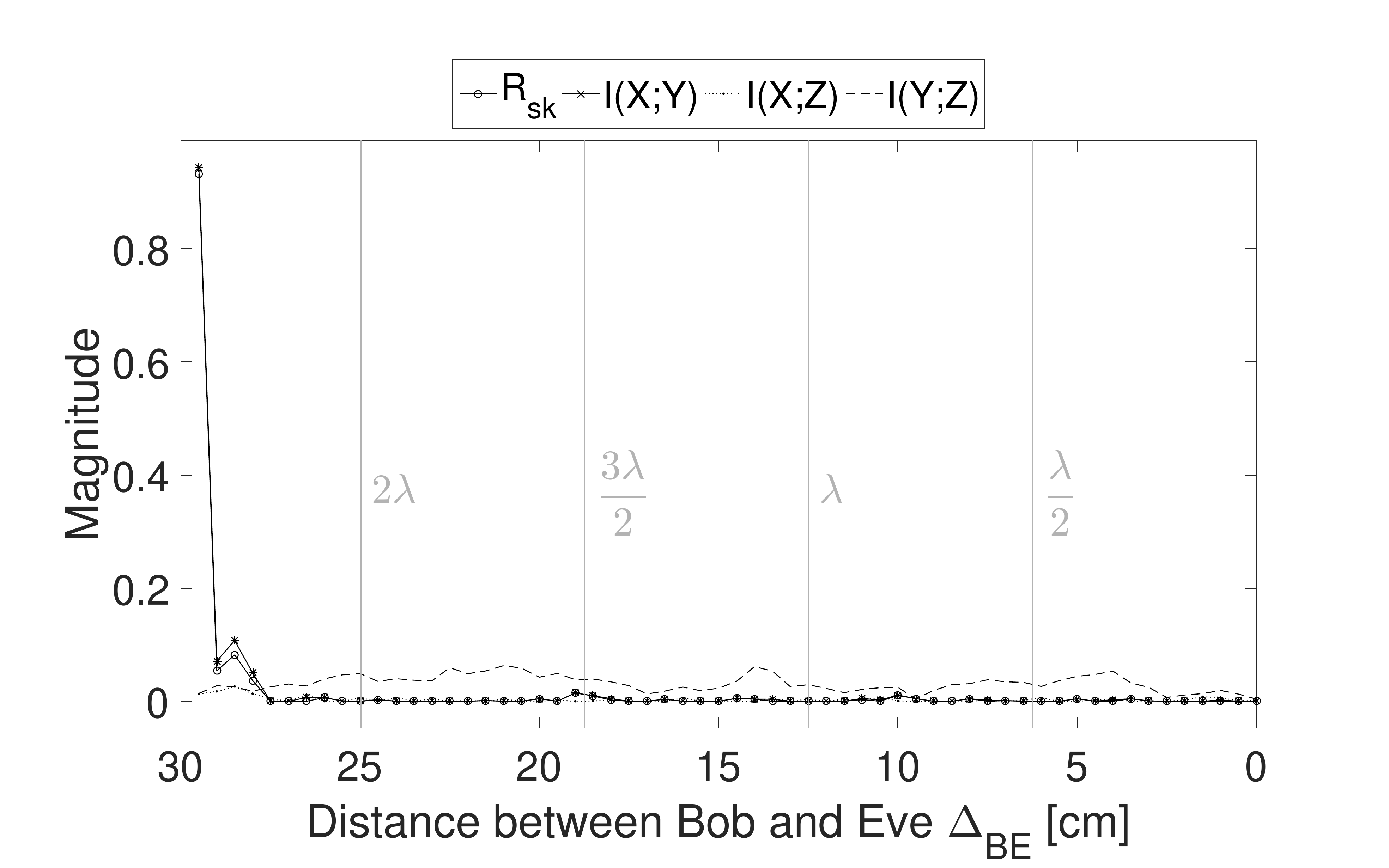}}
	\caption{Evaluation results of $\mybold{v}^{\text{de}}_k$. In (a) and (b) the cross-correlations is given; in (c) the mutual information as well as $\rsk$ is given. Position 14.}
	\label{fig:app_decorr_14}
\end{figure*}


\begin{figure*}
	\centering
	\subfloat[]{\includegraphics[trim=1.4cm 0.1cm 3.5cm 1.6cm, clip=true, height=0.224\textwidth]{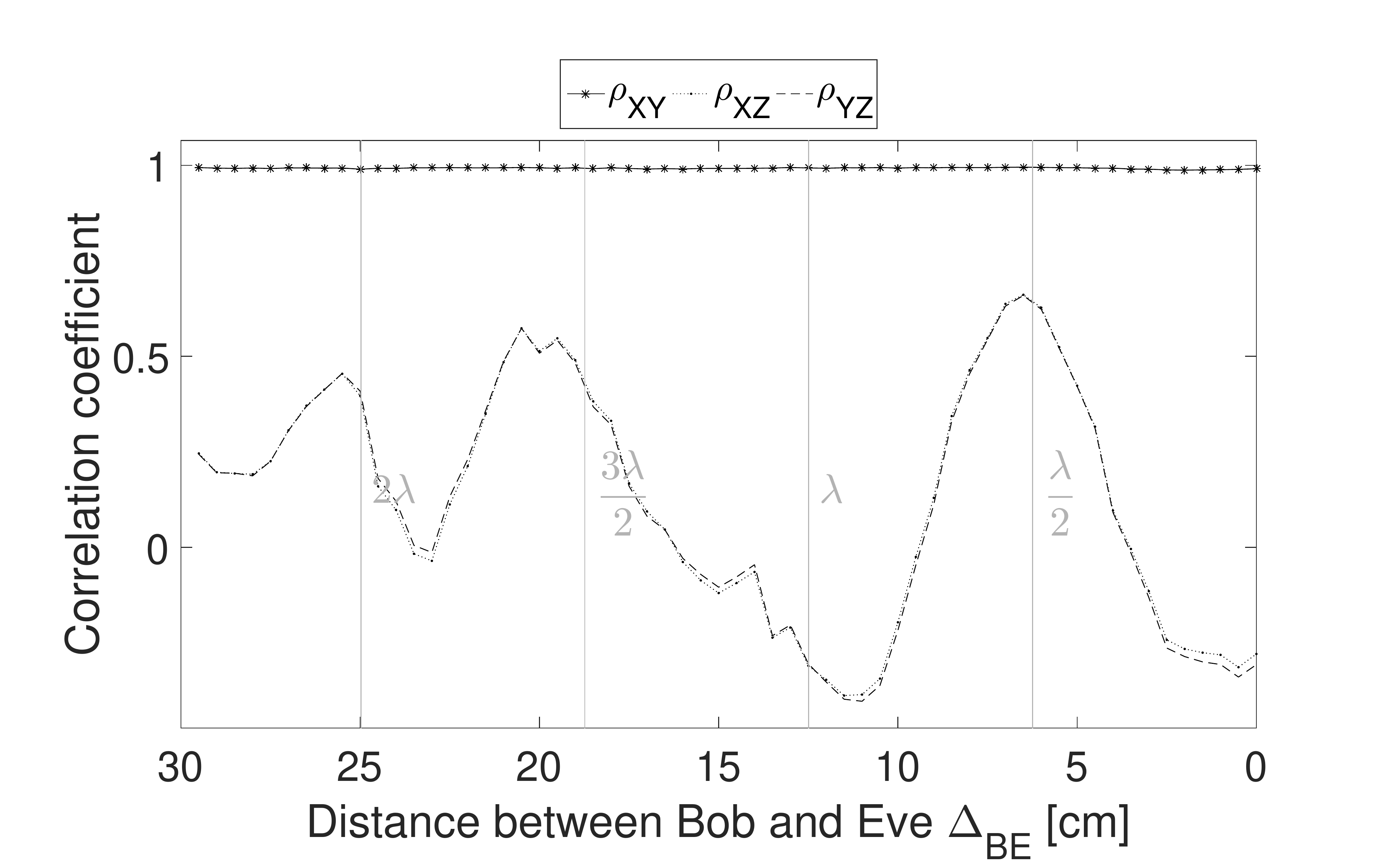}}
	\subfloat[]{\includegraphics[trim=0.5cm 0.1cm 3.5cm 1.6cm, clip=true, height=0.224\textwidth]{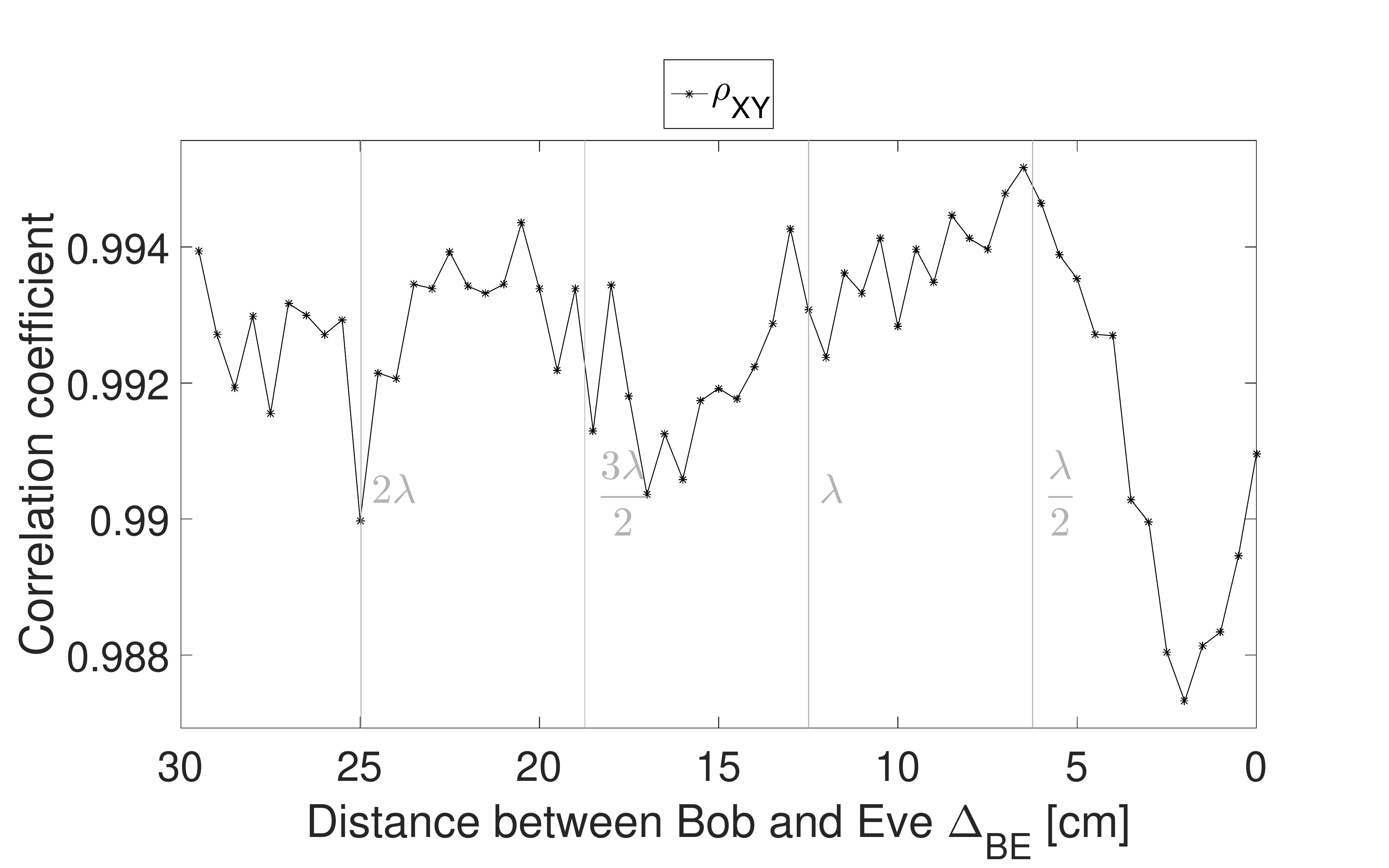}}
	\subfloat[]{\includegraphics[trim=2.2cm 0.1cm 3.5cm 1.6cm, clip=true, height=0.224\textwidth]{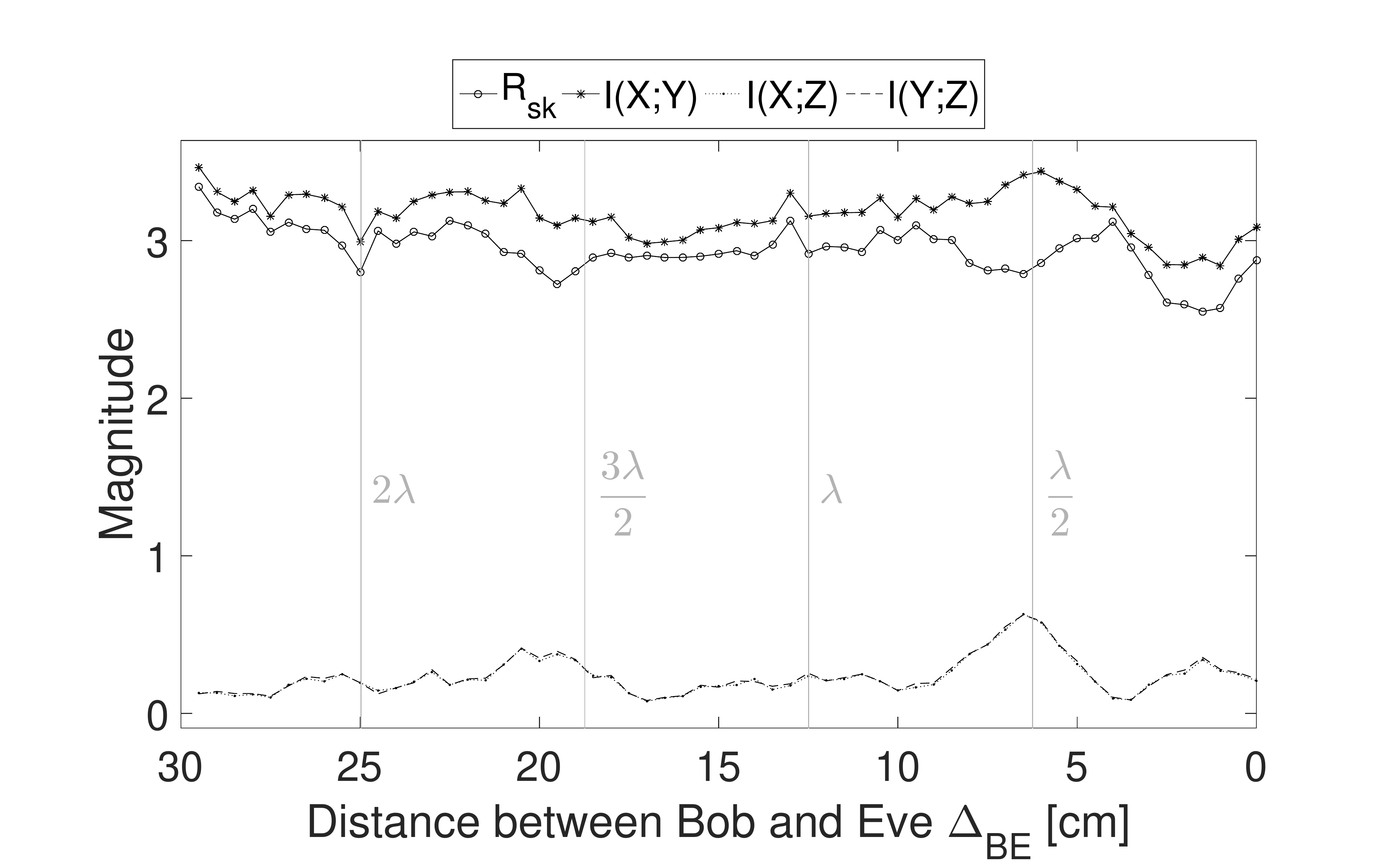}}
	\caption{Evaluation results of $\mybold{v}_k$. In (a) and (b) the cross-correlations is given; in (c) the mutual information as well as $\rsk$ is given. Position 15.}
	\label{fig:app_original_15}
\end{figure*}

\begin{figure*}
	\centering
	\subfloat[]{\includegraphics[trim=1.4cm 0.1cm 3.5cm 1.6cm, clip=true, height=0.224\textwidth]{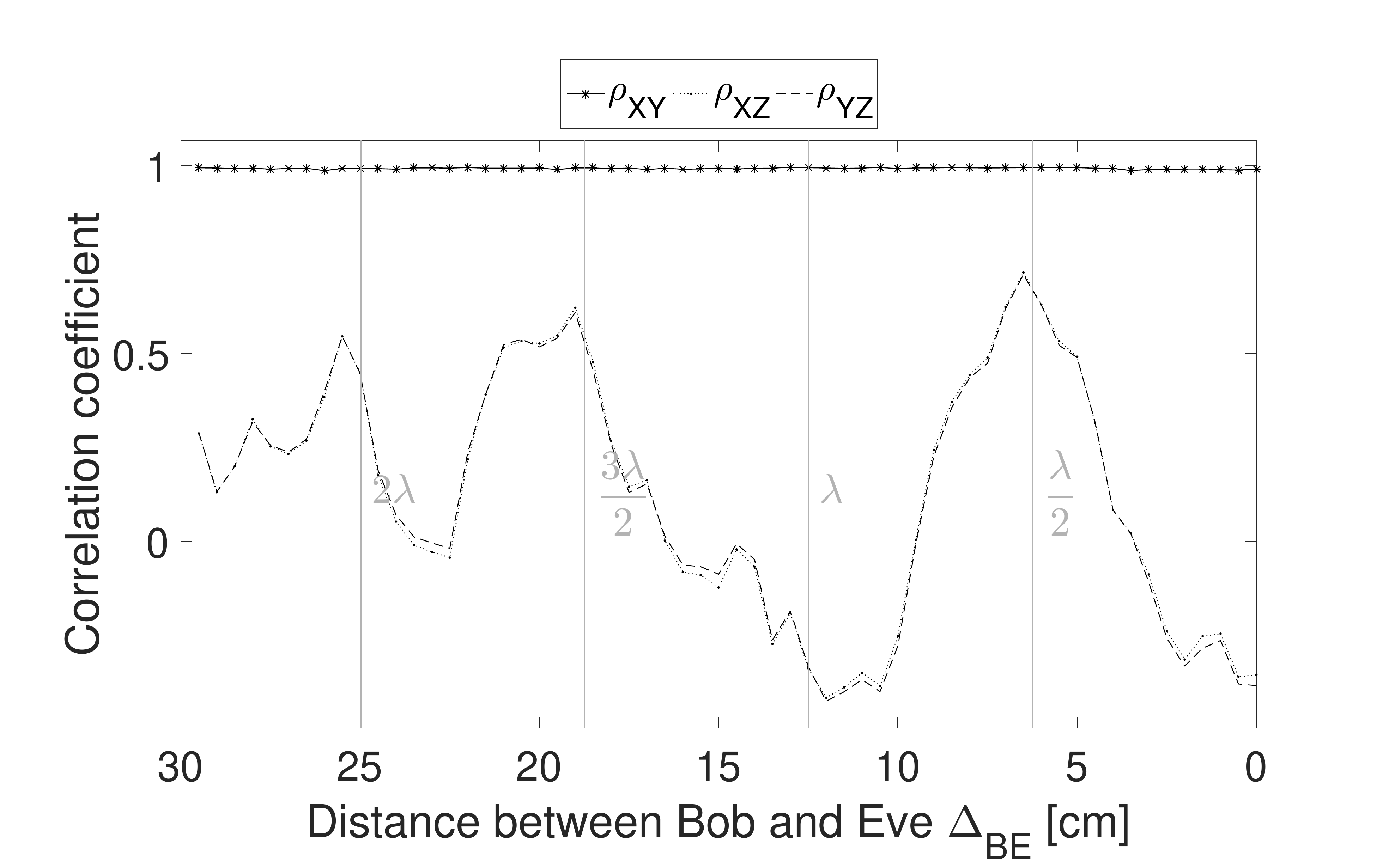}}
	\subfloat[]{\includegraphics[trim=0.5cm 0.1cm 3.5cm 1.6cm, clip=true, height=0.224\textwidth]{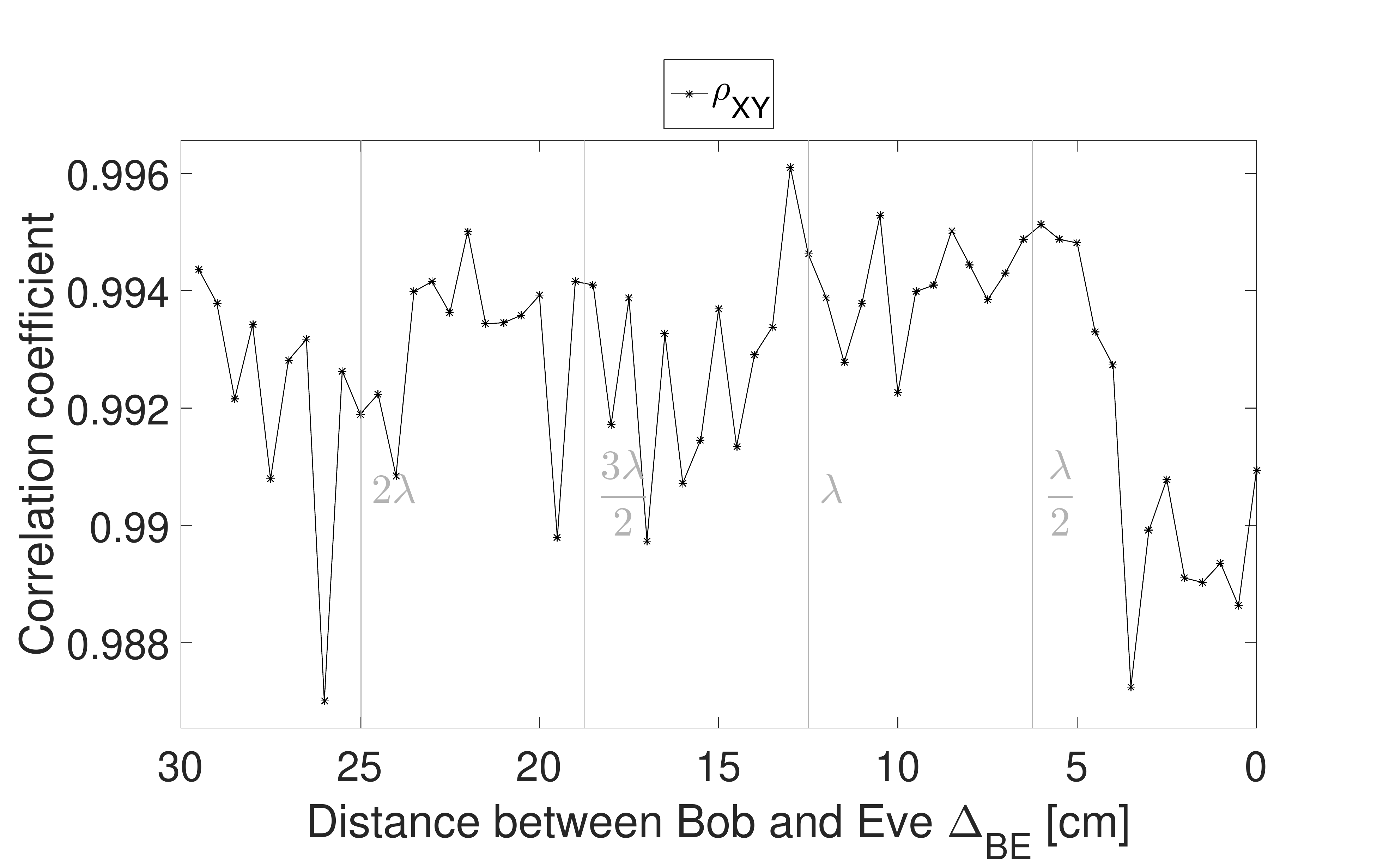}}
	\subfloat[]{\includegraphics[trim=2.2cm 0.1cm 3.5cm 1.6cm, clip=true, height=0.224\textwidth]{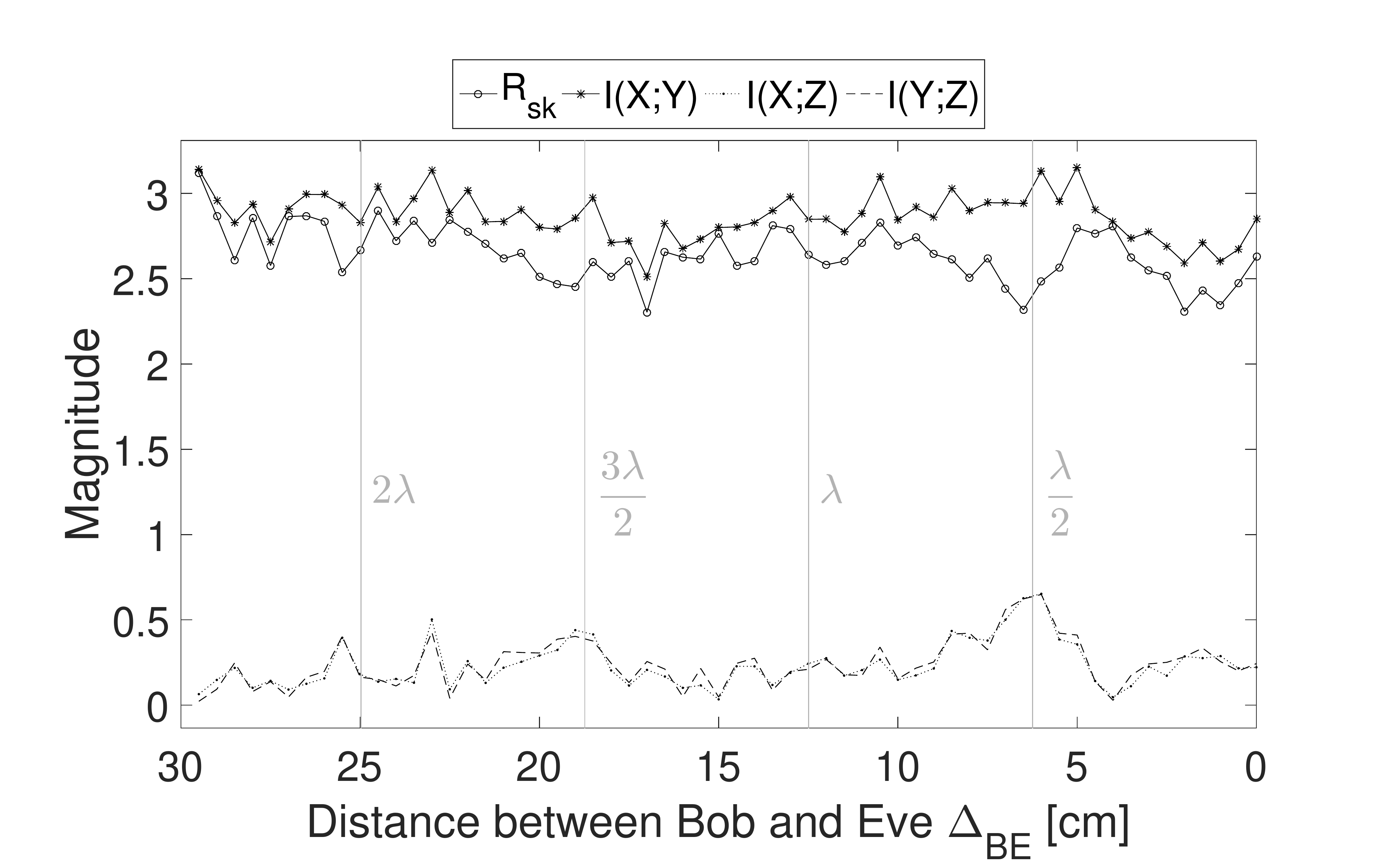}}
	\caption{Evaluation results of $\mybold{v}^{\text{ds}}_k$. In (a) and (b) the cross-correlations is given; in (c) the mutual information as well as $\rsk$ is given. Position 15.}
	\label{fig:app_ds_15}
\end{figure*}

\begin{figure*}
	\centering
	\subfloat[]{\includegraphics[trim=1.4cm 0.1cm 3.5cm 1.6cm, clip=true, height=0.224\textwidth]{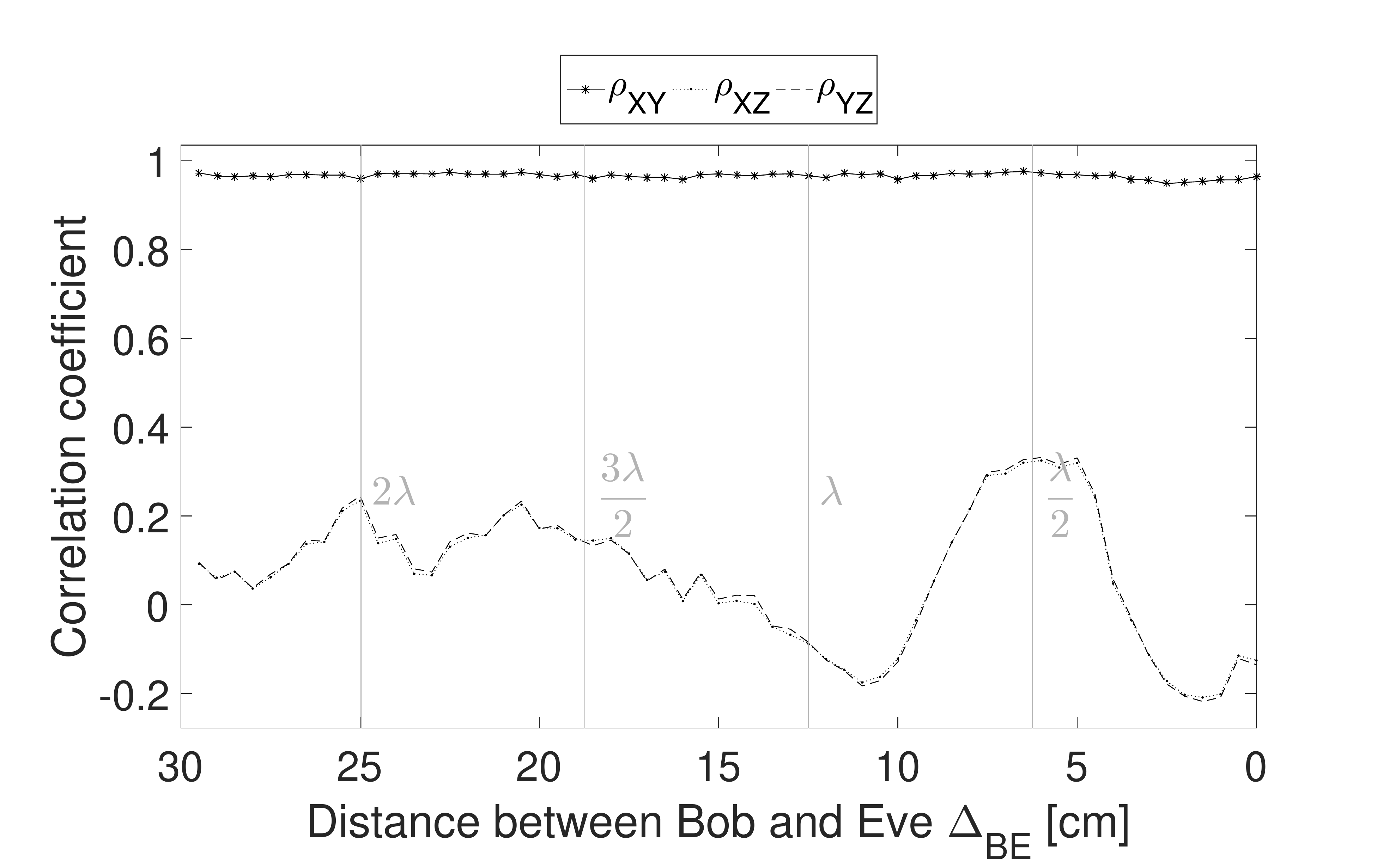}}
	\subfloat[]{\includegraphics[trim=1cm 0.1cm 3.5cm 1.6cm, clip=true, height=0.224\textwidth]{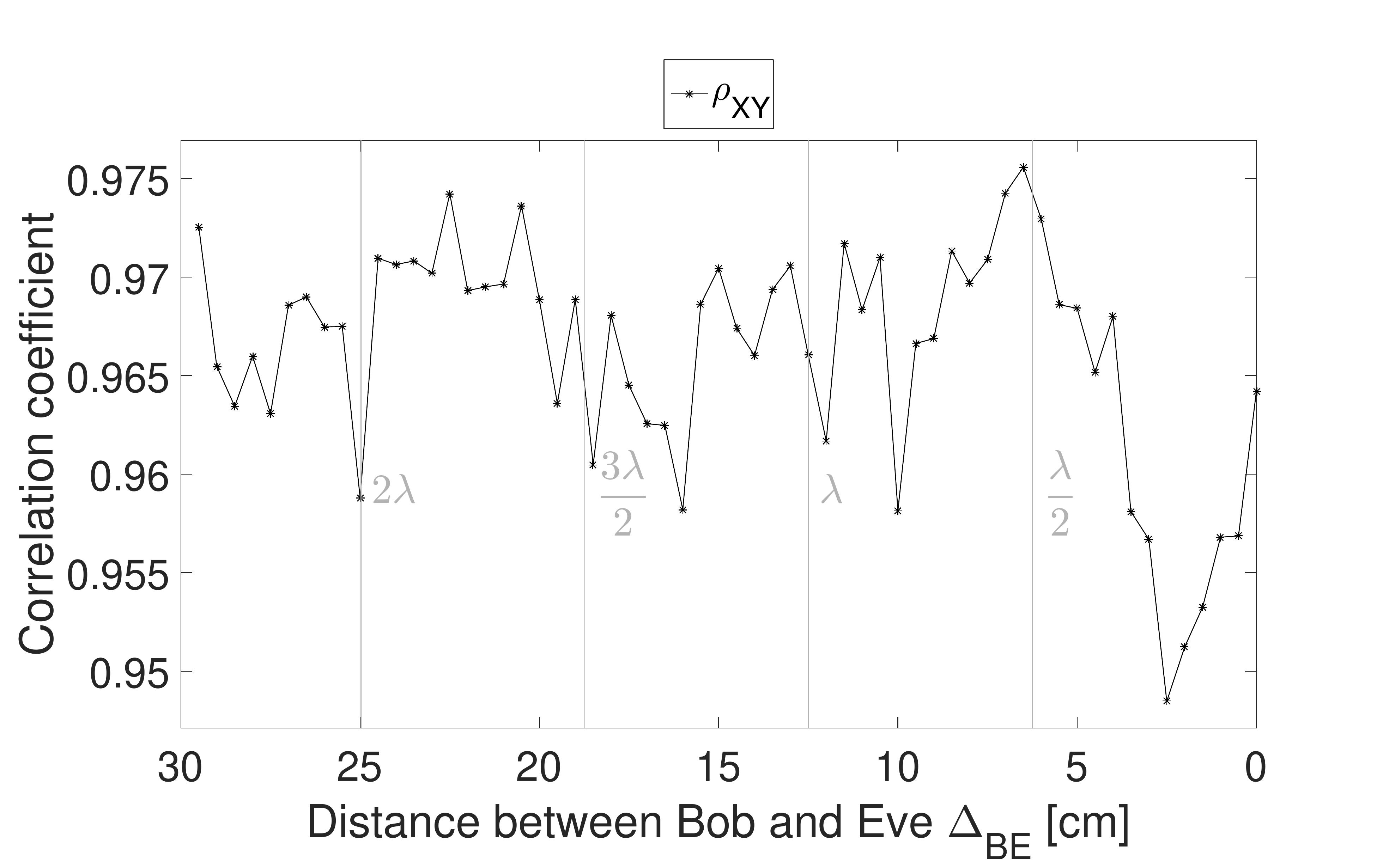}}
	\subfloat[]{\includegraphics[trim=1.8cm 0.1cm 3.5cm 1.6cm, clip=true, height=0.224\textwidth]{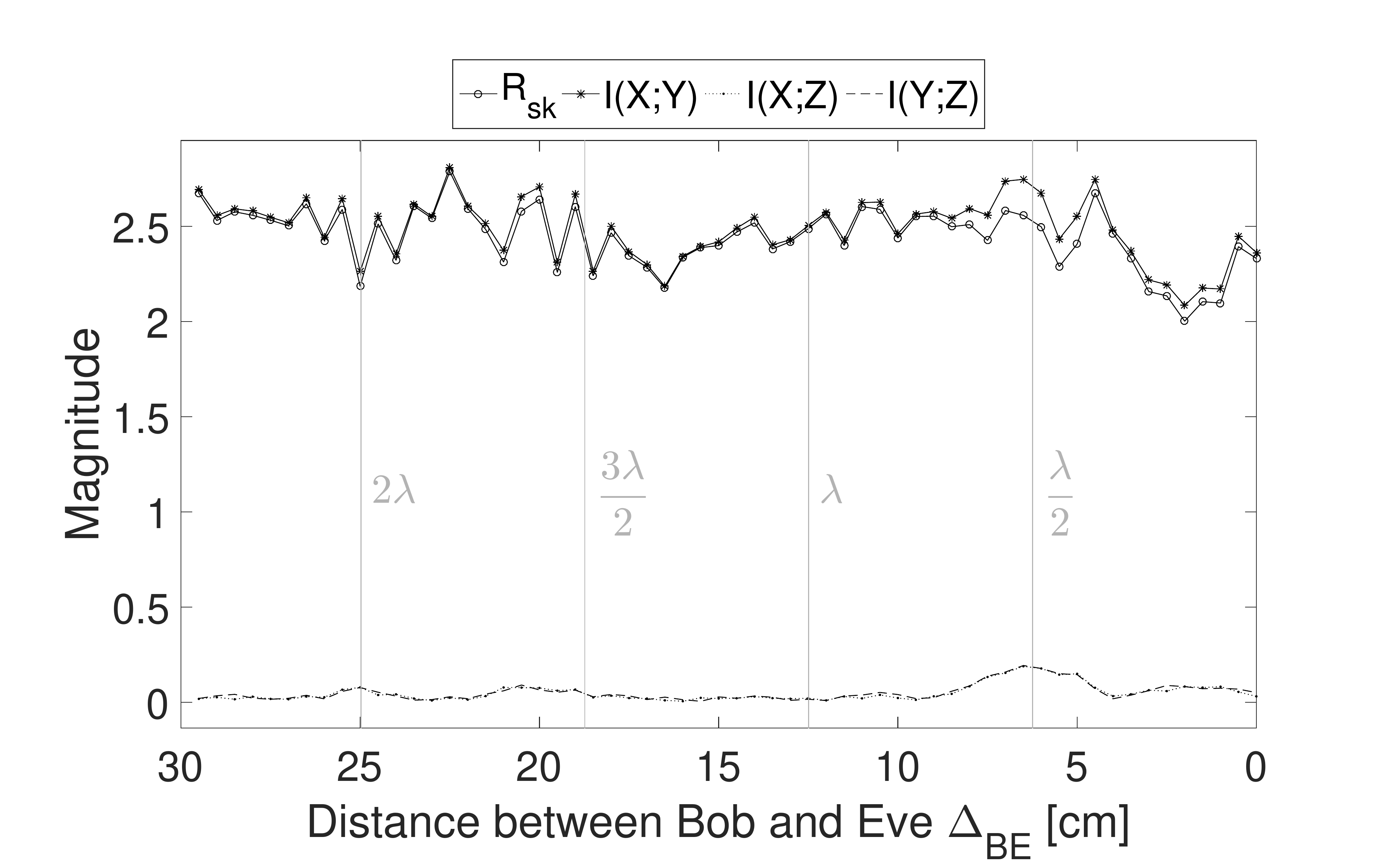}}
	\caption{Evaluation results of $\mybold{v}^{\text{de}}_k$. In (a) and (b) the cross-correlations is given; in (c) the mutual information as well as $\rsk$ is given. Position 15.}
	\label{fig:app_decorr_15}
\end{figure*}


\begin{figure*}
	\centering
	\subfloat[]{\includegraphics[trim=1.4cm 0.1cm 3.5cm 1.6cm, clip=true, height=0.224\textwidth]{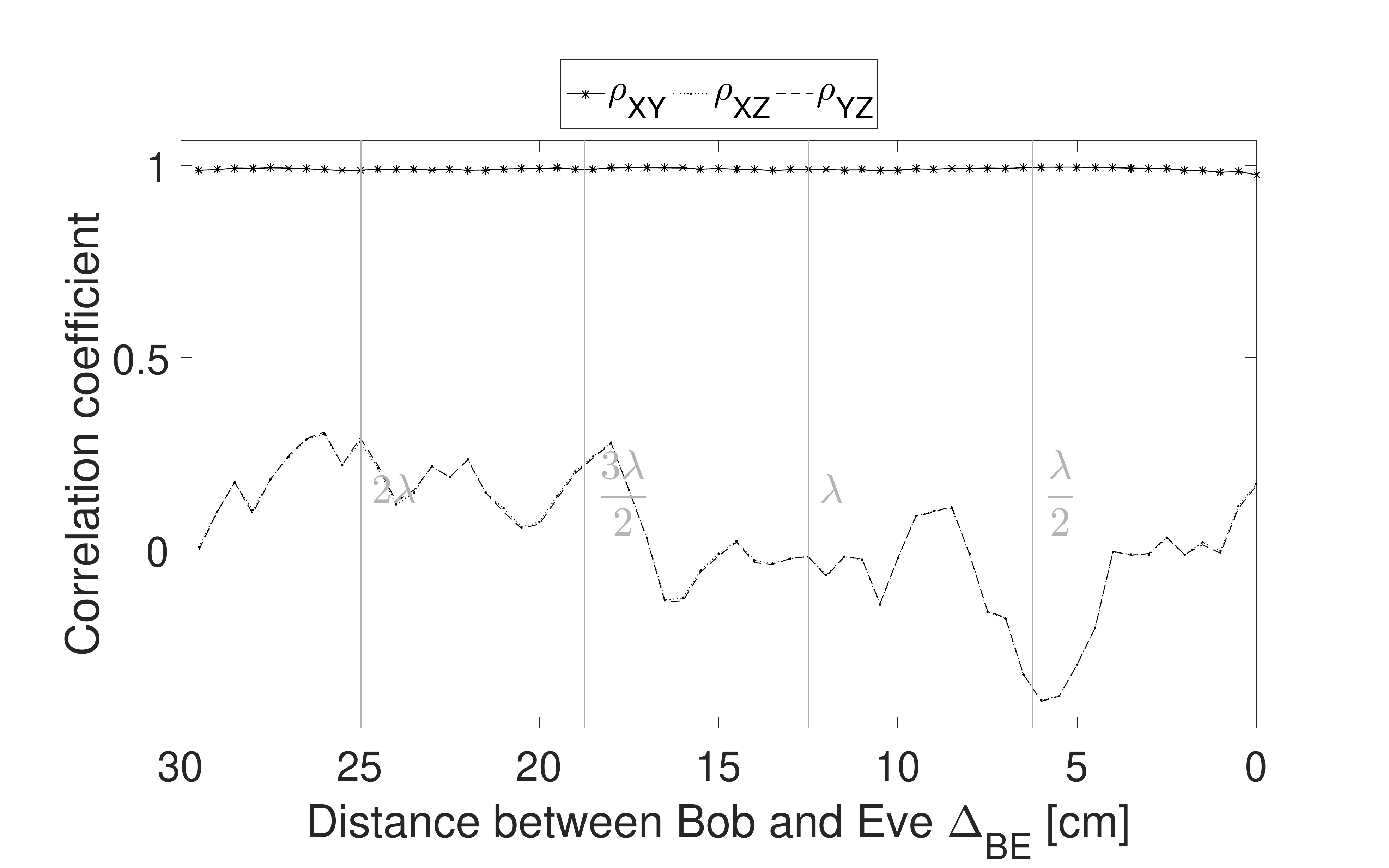}}
	\subfloat[]{\includegraphics[trim=0.5cm 0.1cm 3.5cm 1.6cm, clip=true, height=0.224\textwidth]{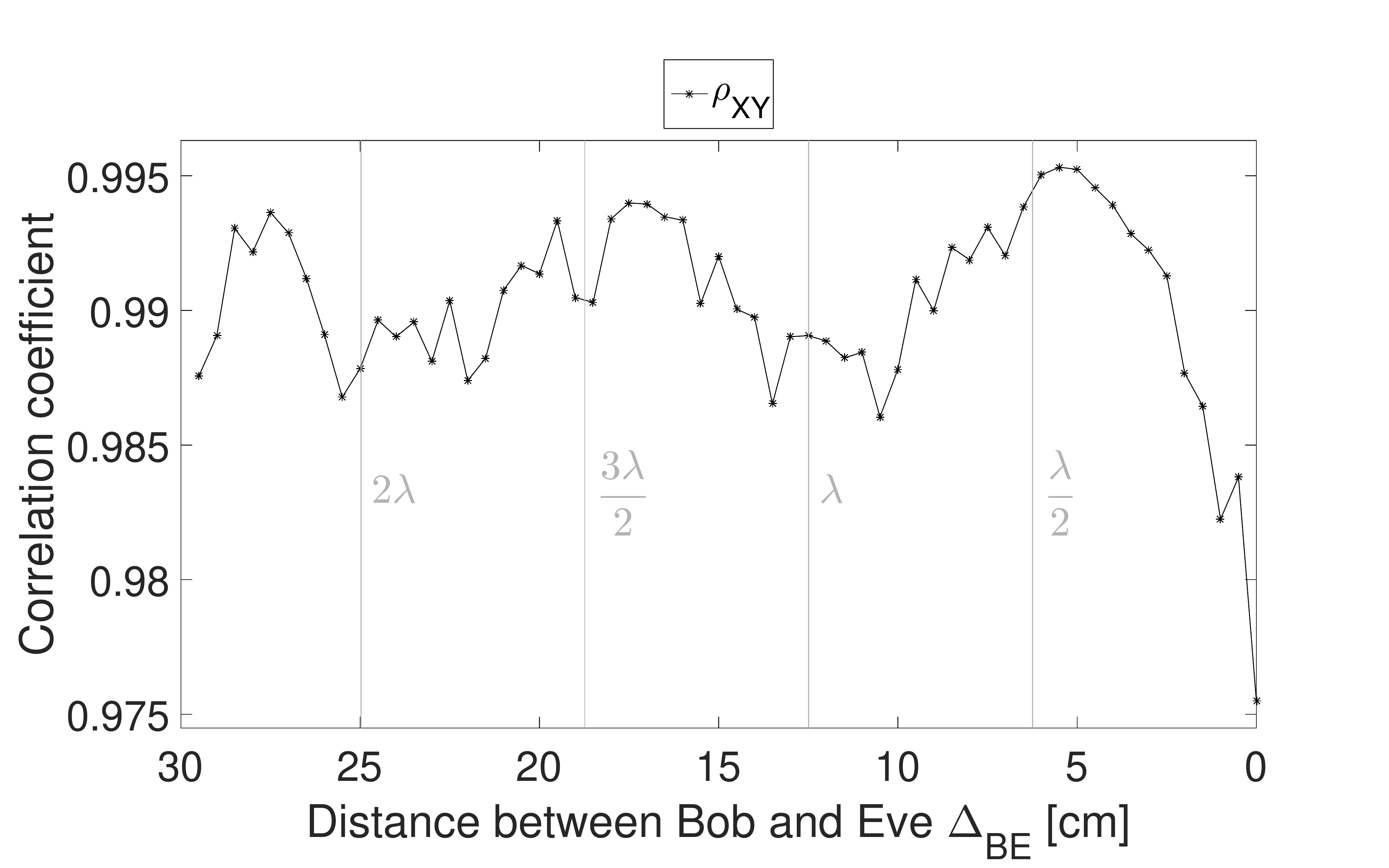}}
	\subfloat[]{\includegraphics[trim=2.2cm 0.1cm 3.5cm 1.6cm, clip=true, height=0.224\textwidth]{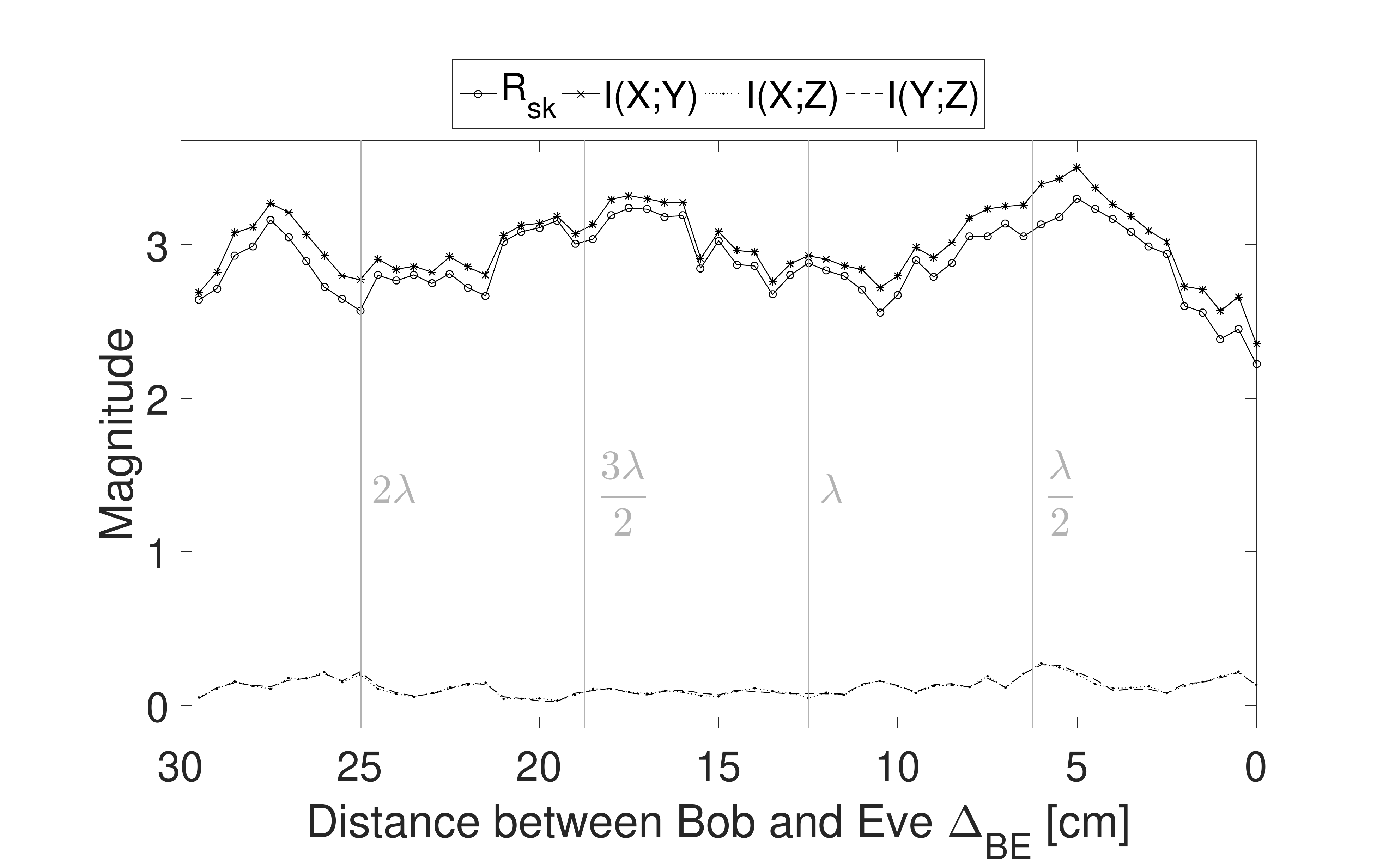}}
	\caption{Evaluation results of $\mybold{v}_k$. In (a) and (b) the cross-correlations is given; in (c) the mutual information as well as $\rsk$ is given. Position 16.}
	\label{fig:app_original_16}
\end{figure*}

\begin{figure*}
	\centering
	\subfloat[]{\includegraphics[trim=1.4cm 0.1cm 3.5cm 1.6cm, clip=true, height=0.224\textwidth]{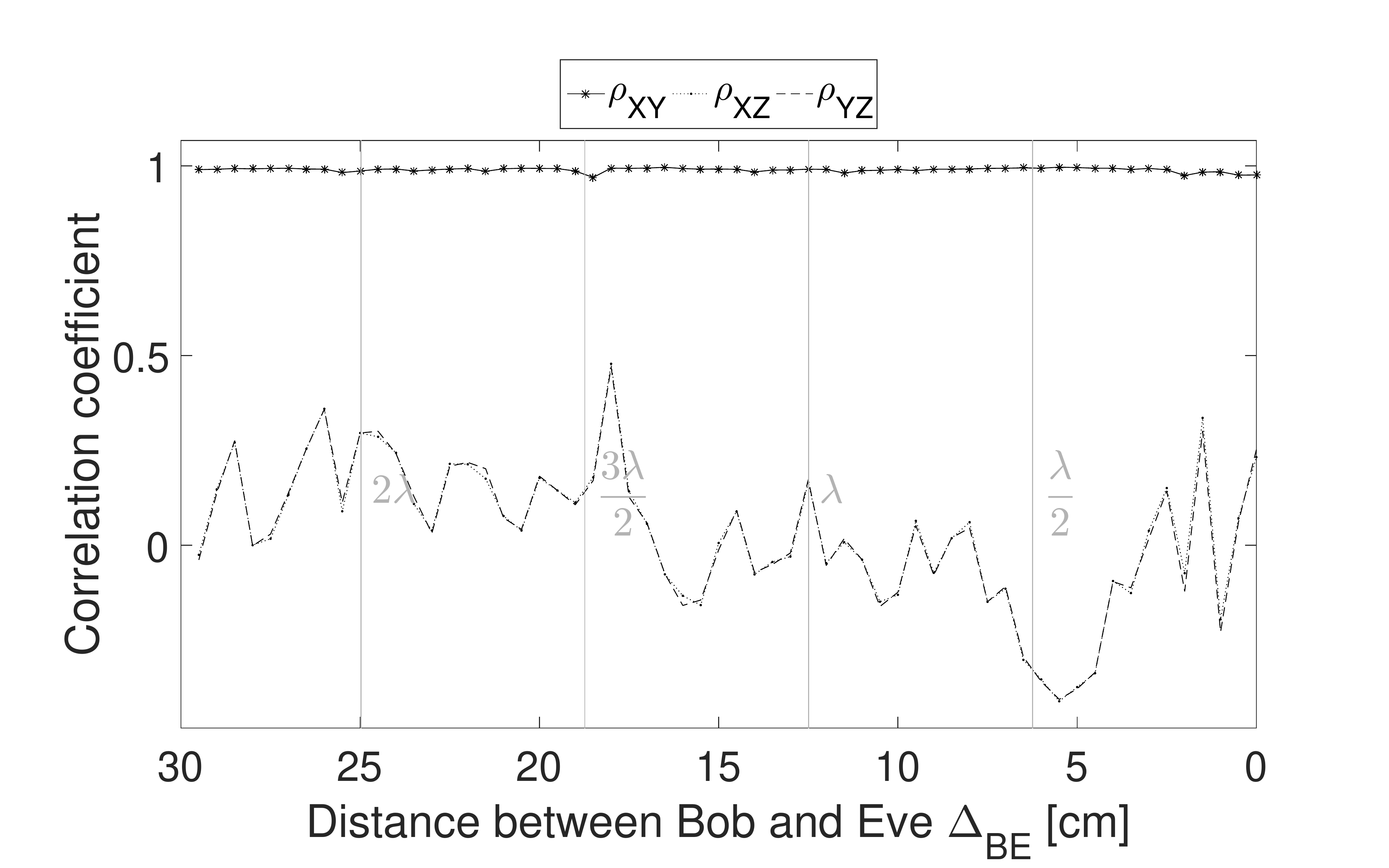}}
	\subfloat[]{\includegraphics[trim=0.5cm 0.1cm 3.5cm 1.6cm, clip=true, height=0.224\textwidth]{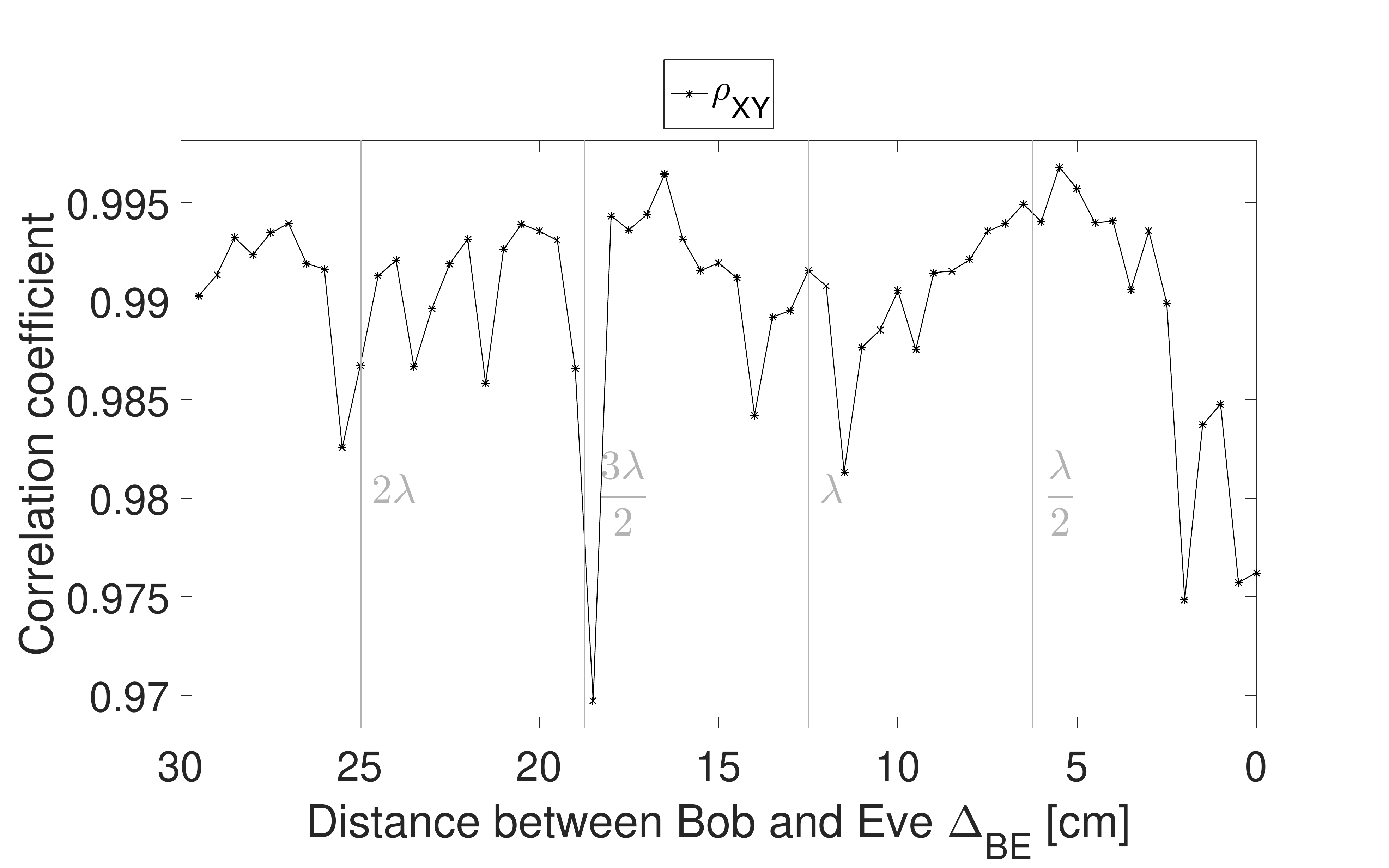}}
	\subfloat[]{\includegraphics[trim=2.2cm 0.1cm 3.5cm 1.6cm, clip=true, height=0.224\textwidth]{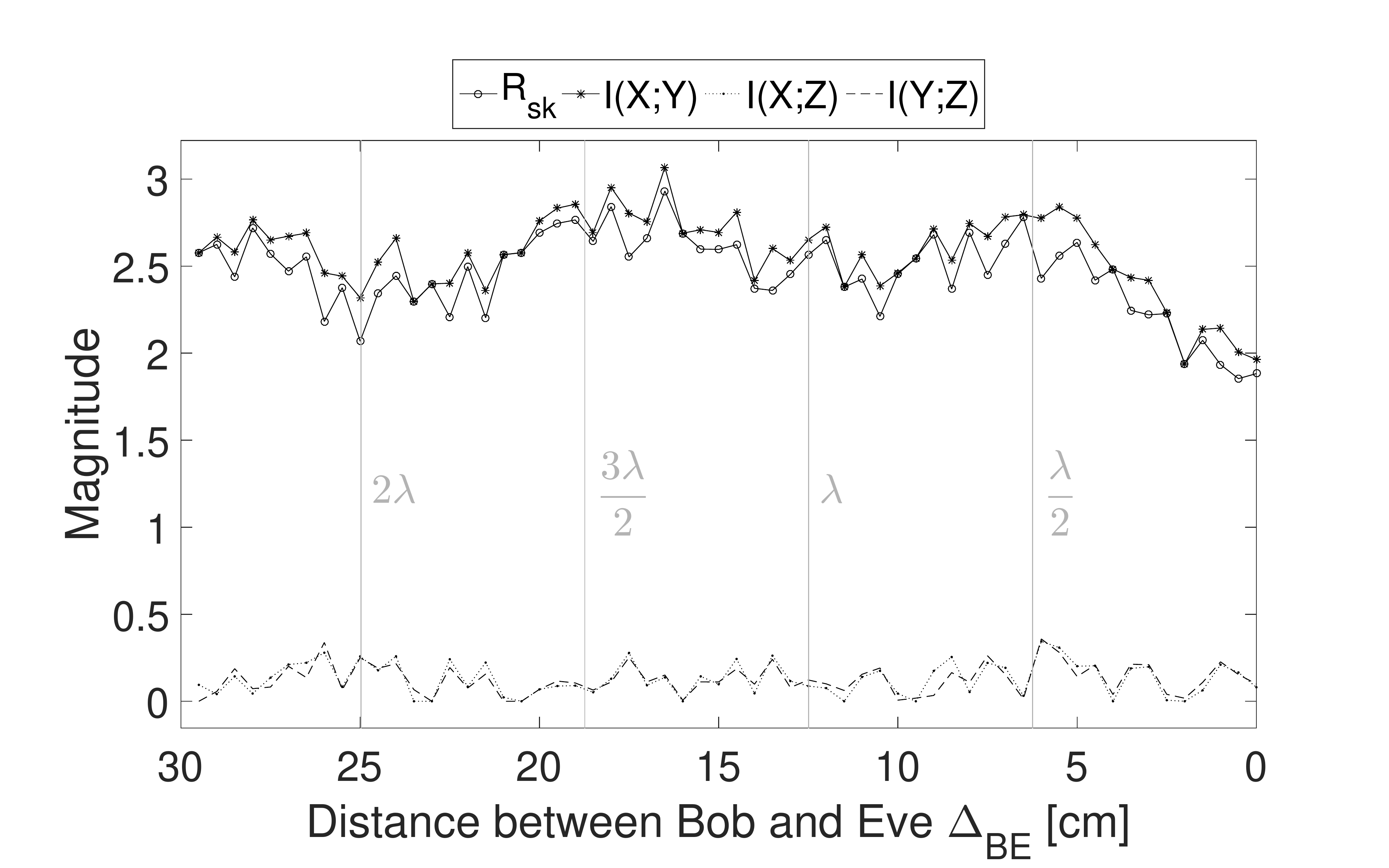}}
	\caption{Evaluation results of $\mybold{v}^{\text{ds}}_k$. In (a) and (b) the cross-correlations is given; in (c) the mutual information as well as $\rsk$ is given. Position 16.}
	\label{fig:app_ds_16}
\end{figure*}

\begin{figure*}
	\centering
	\subfloat[]{\includegraphics[trim=1.4cm 0.1cm 3.5cm 1.6cm, clip=true, height=0.224\textwidth]{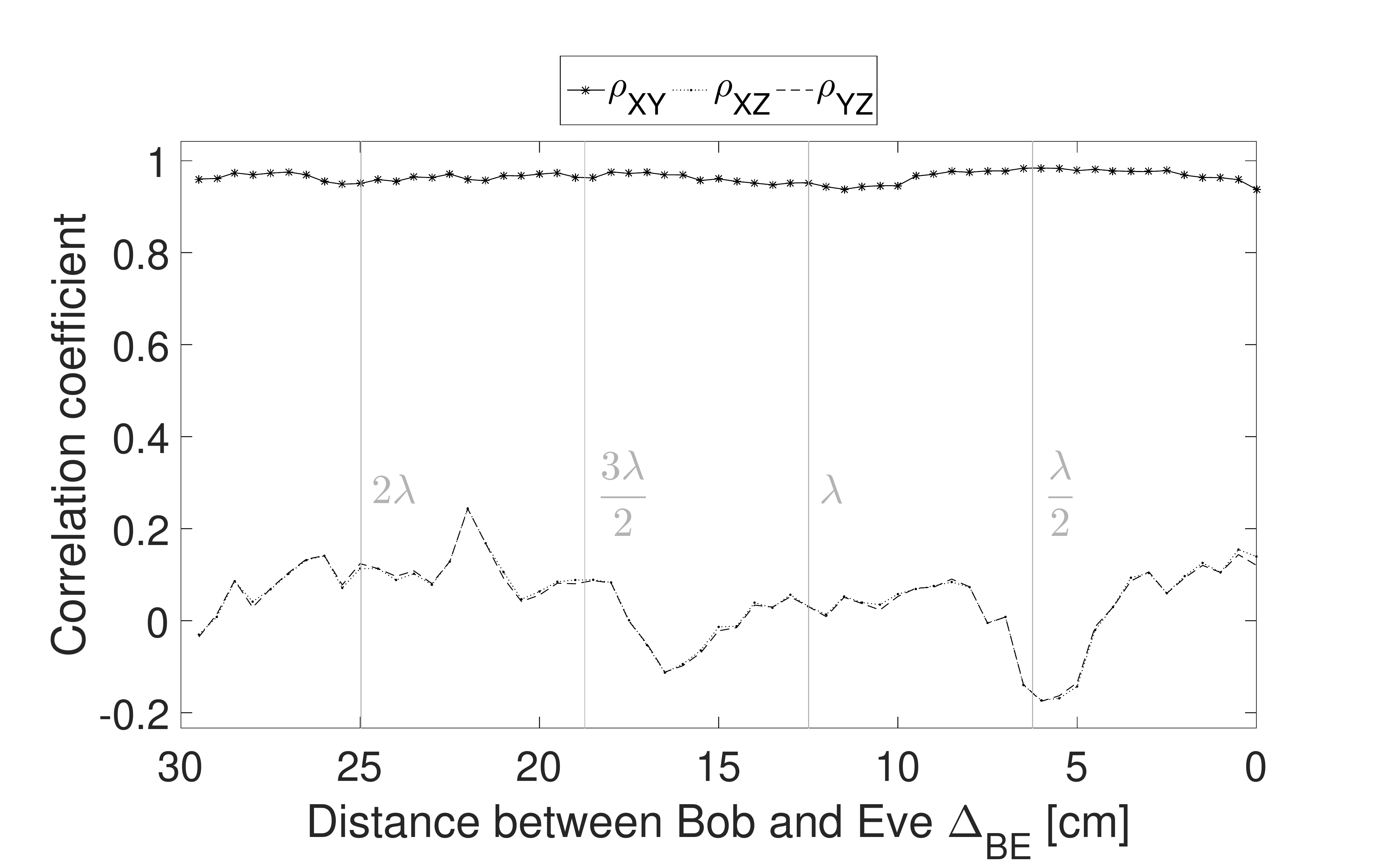}}
	\subfloat[]{\includegraphics[trim=1cm 0.1cm 3.5cm 1.6cm, clip=true, height=0.224\textwidth]{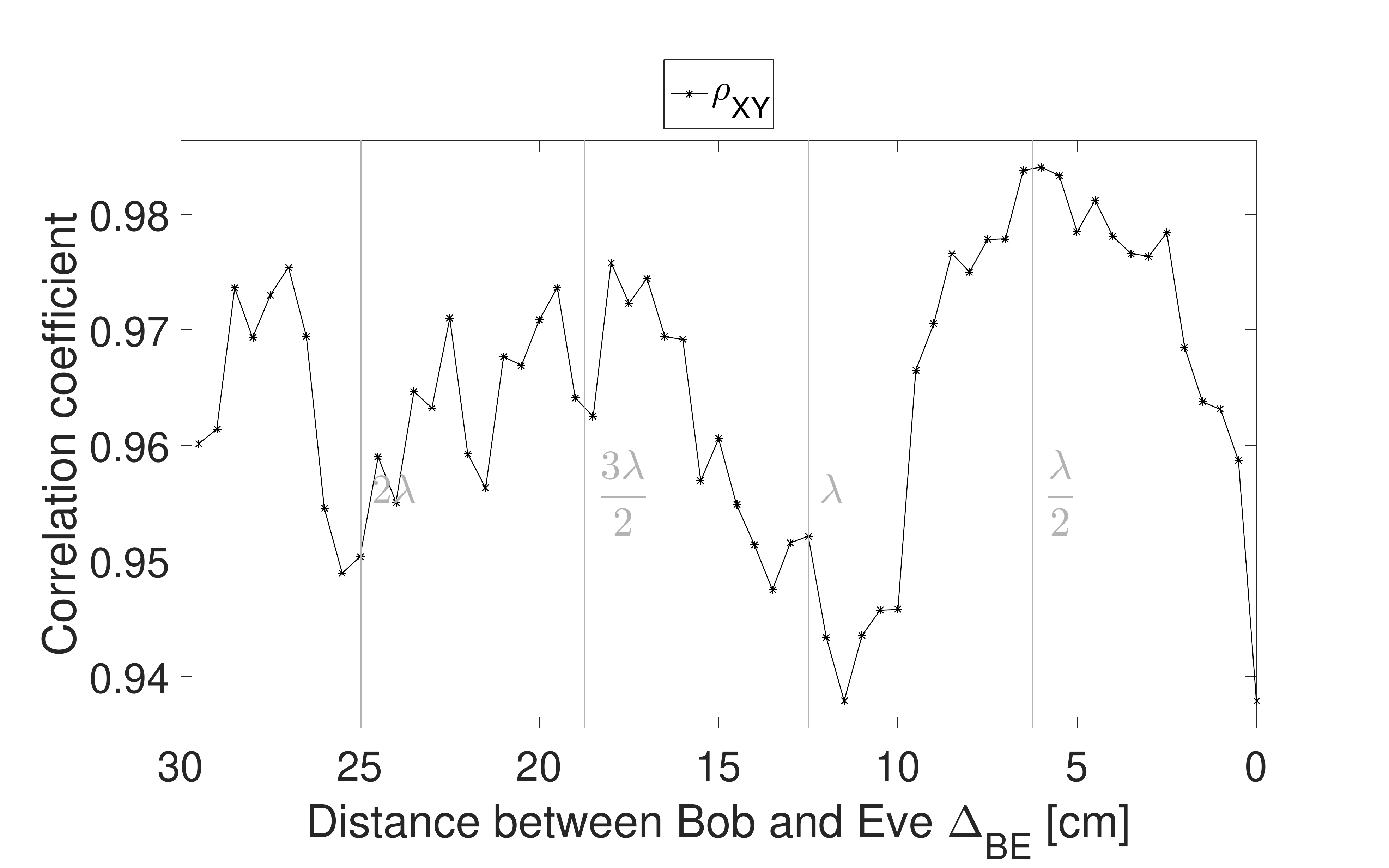}}
	\subfloat[]{\includegraphics[trim=1.8cm 0.1cm 3.5cm 1.6cm, clip=true, height=0.224\textwidth]{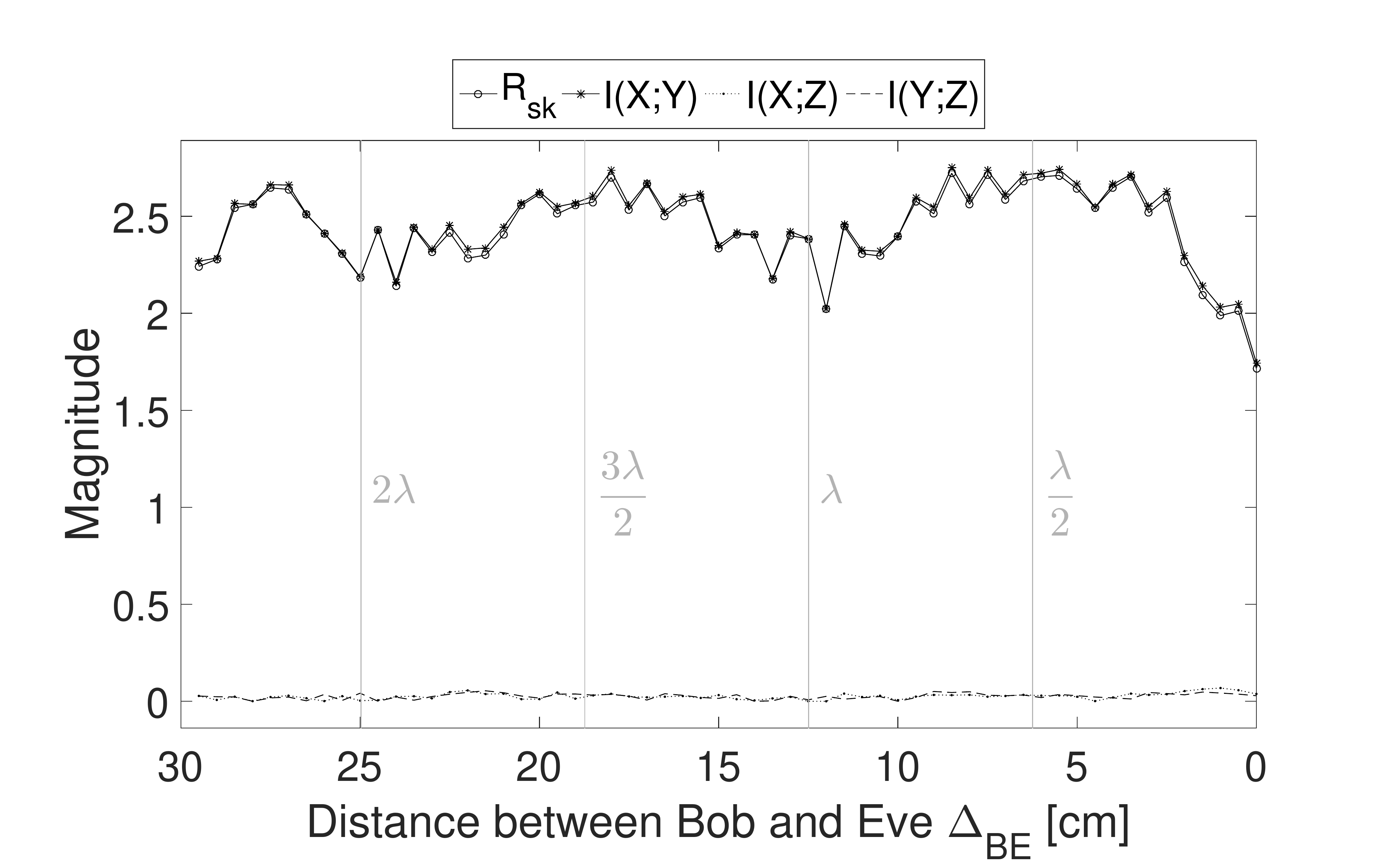}}
	\caption{Evaluation results of $\mybold{v}^{\text{de}}_k$. In (a) and (b) the cross-correlations is given; in (c) the mutual information as well as $\rsk$ is given. Position 16.}
	\label{fig:app_decorr_16}
\end{figure*}


\begin{figure*}
	\centering
	\subfloat[]{\includegraphics[trim=1.4cm 0.1cm 3.5cm 1.6cm, clip=true, height=0.224\textwidth]{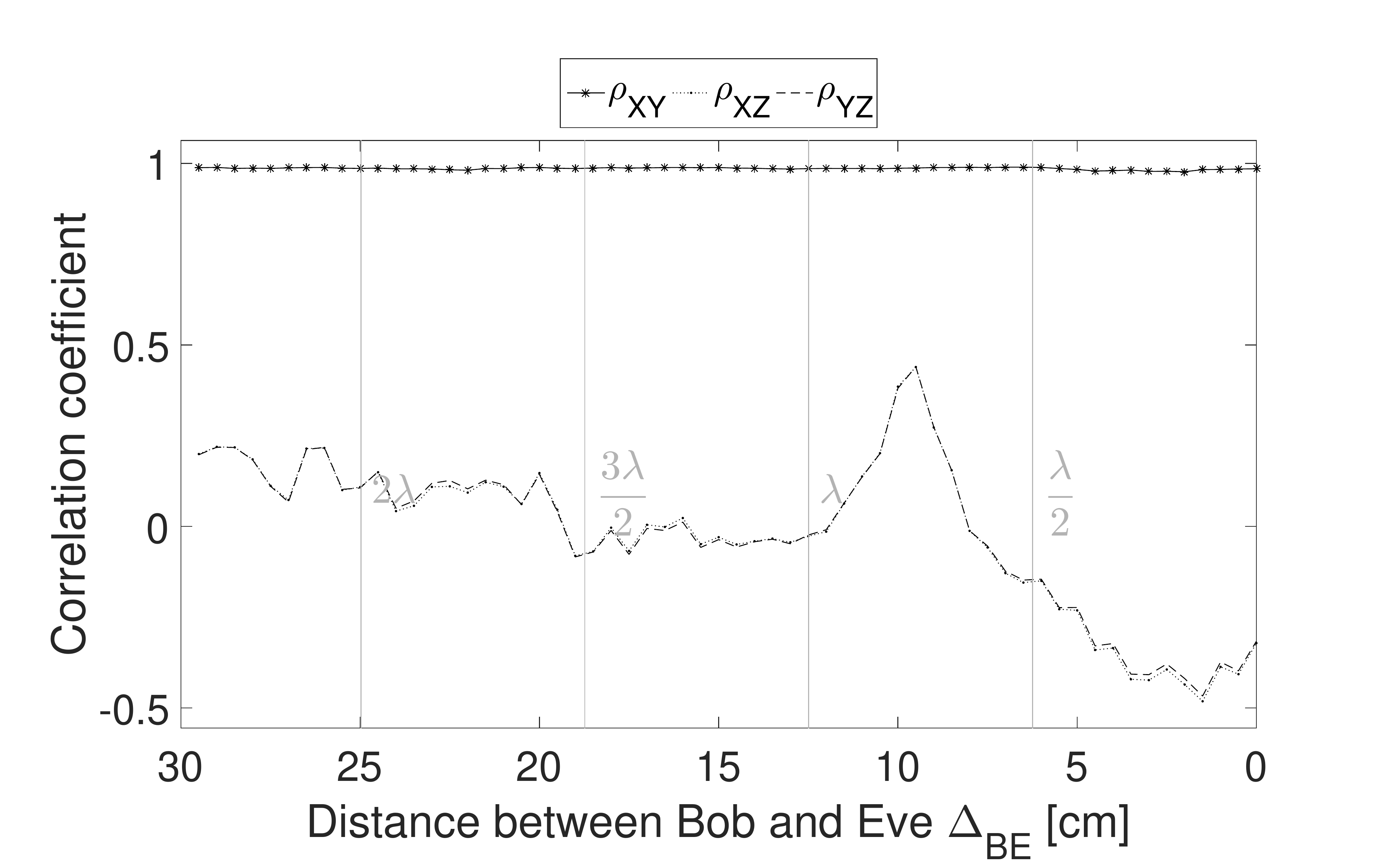}}
	\subfloat[]{\includegraphics[trim=0.5cm 0.1cm 3.5cm 1.6cm, clip=true, height=0.224\textwidth]{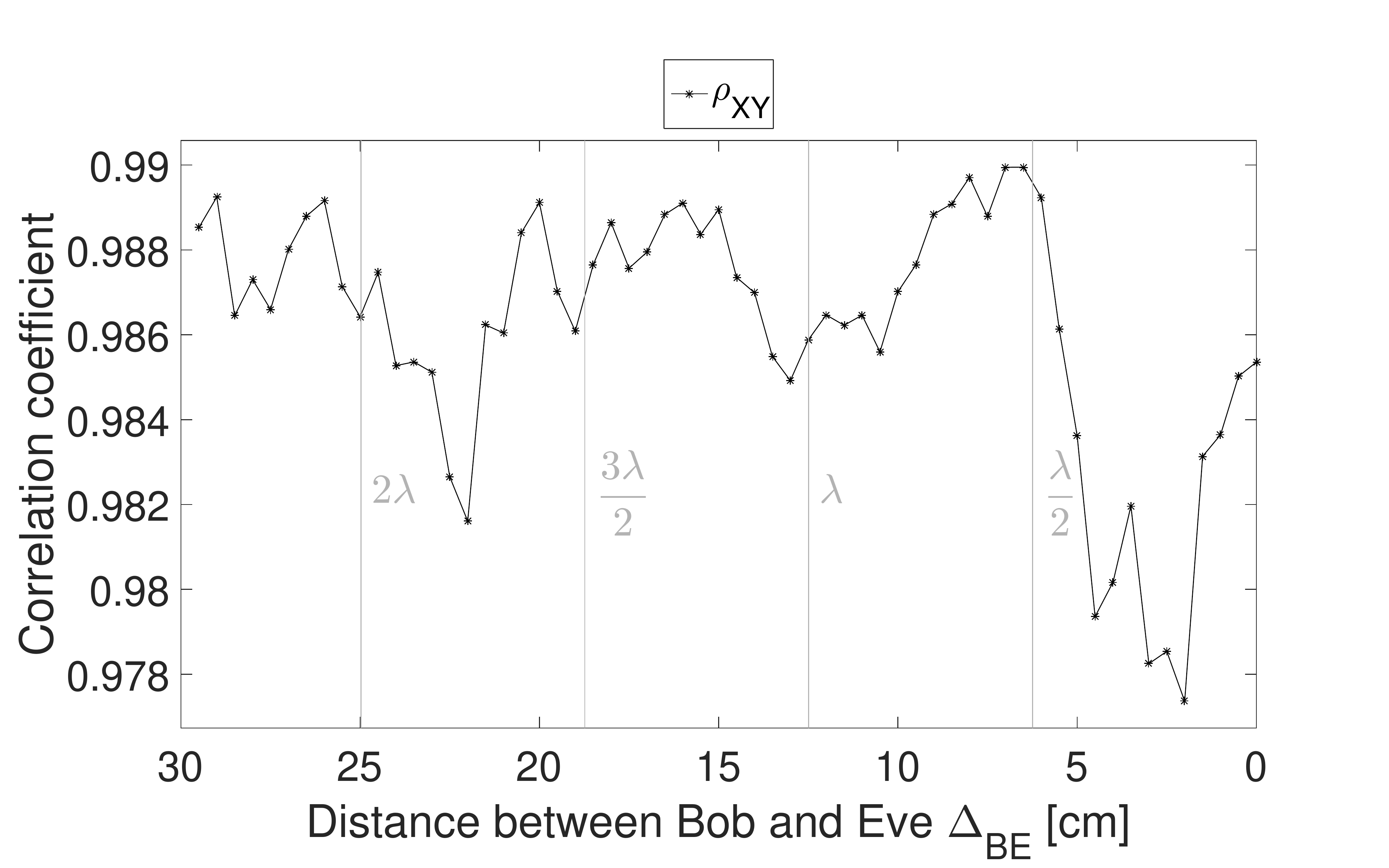}}
	\subfloat[]{\includegraphics[trim=2.2cm 0.1cm 3.5cm 1.6cm, clip=true, height=0.224\textwidth]{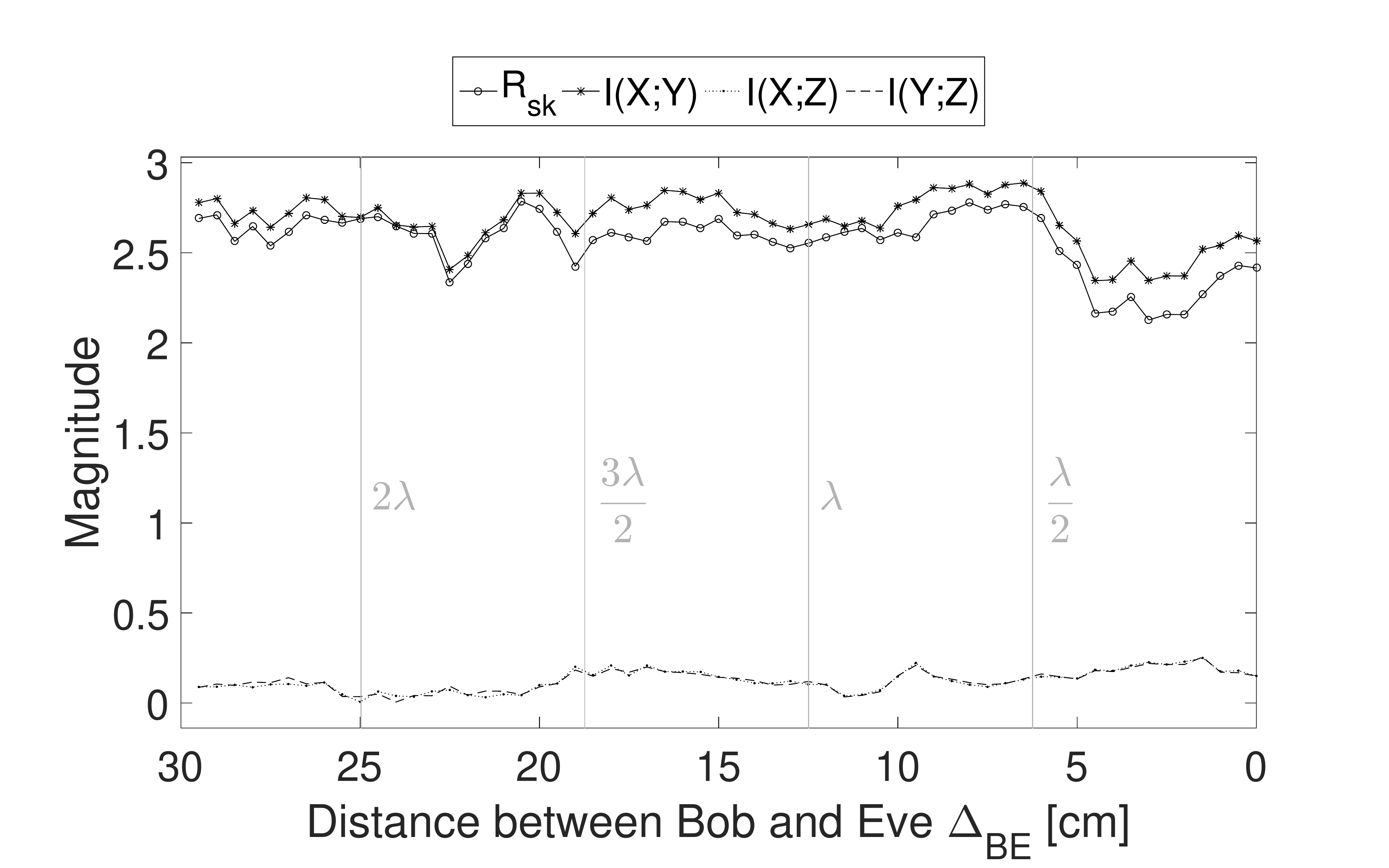}}
	\caption{Evaluation results of $\mybold{v}_k$. In (a) and (b) the cross-correlations is given; in (c) the mutual information as well as $\rsk$ is given. Position 17.}
	\label{fig:app_original_17}
\end{figure*}

\begin{figure*}
	\centering
	\subfloat[]{\includegraphics[trim=1.4cm 0.1cm 3.5cm 1.6cm, clip=true, height=0.224\textwidth]{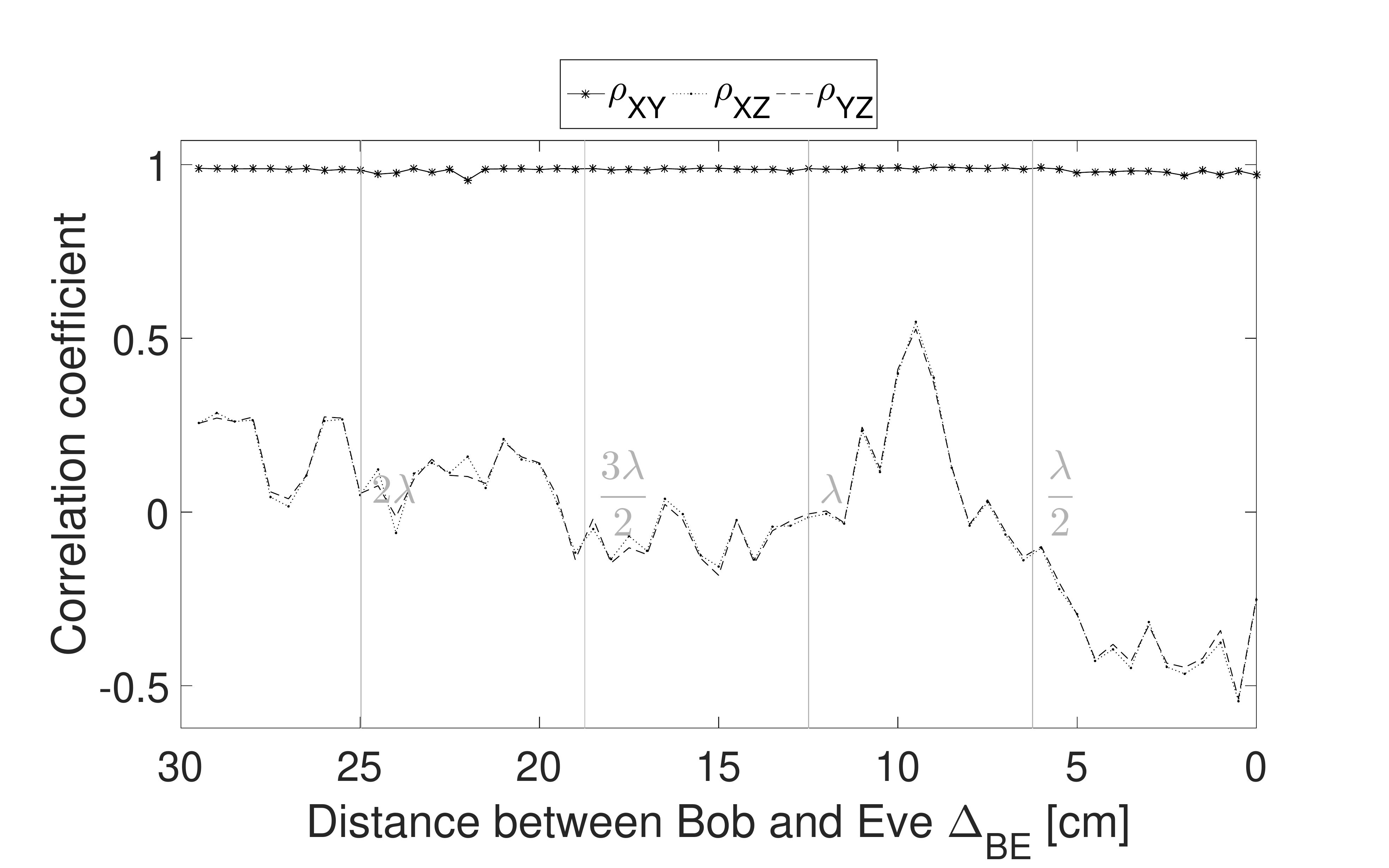}}
	\subfloat[]{\includegraphics[trim=0.5cm 0.1cm 3.5cm 1.6cm, clip=true, height=0.224\textwidth]{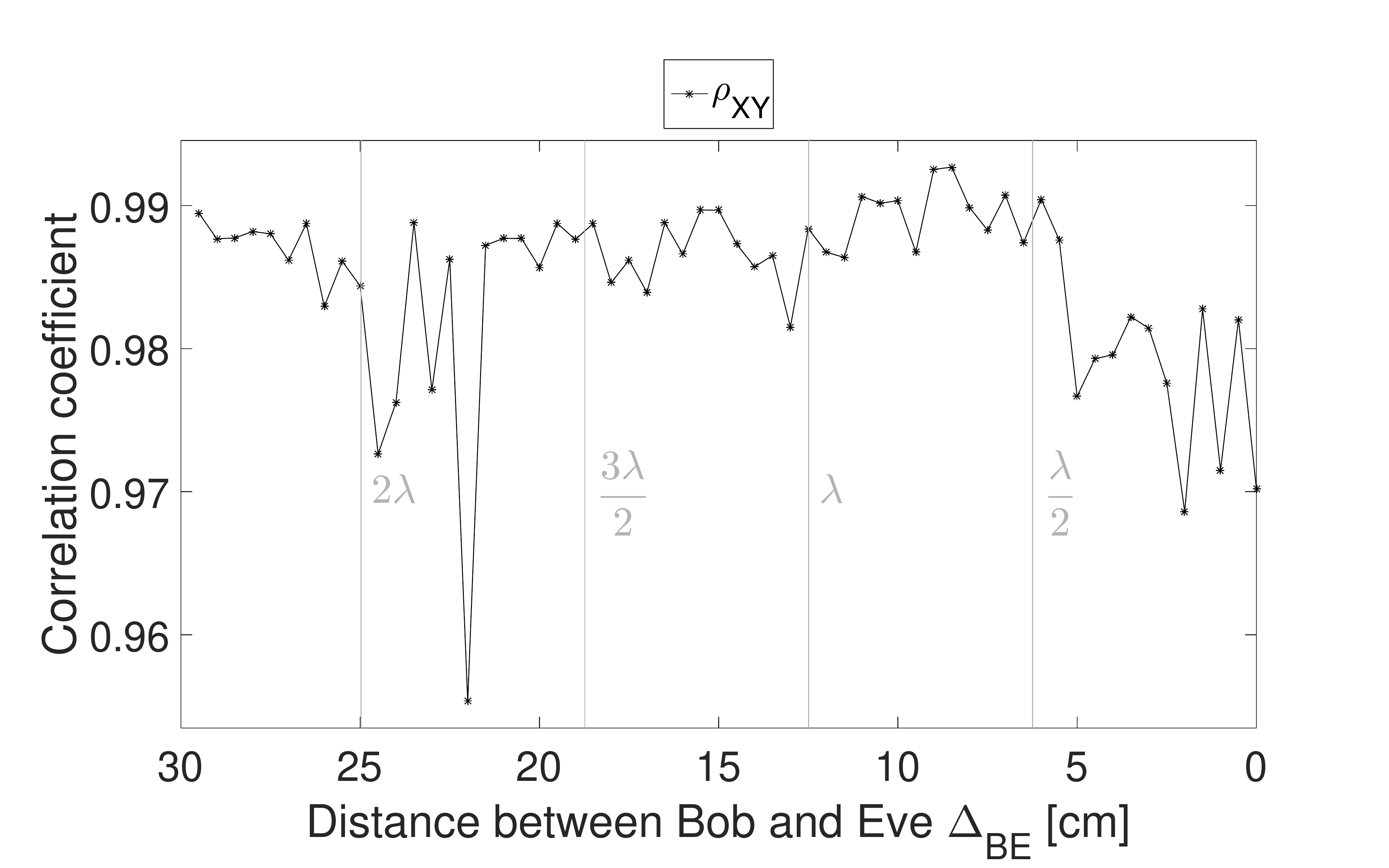}}
	\subfloat[]{\includegraphics[trim=2.2cm 0.1cm 3.5cm 1.6cm, clip=true, height=0.224\textwidth]{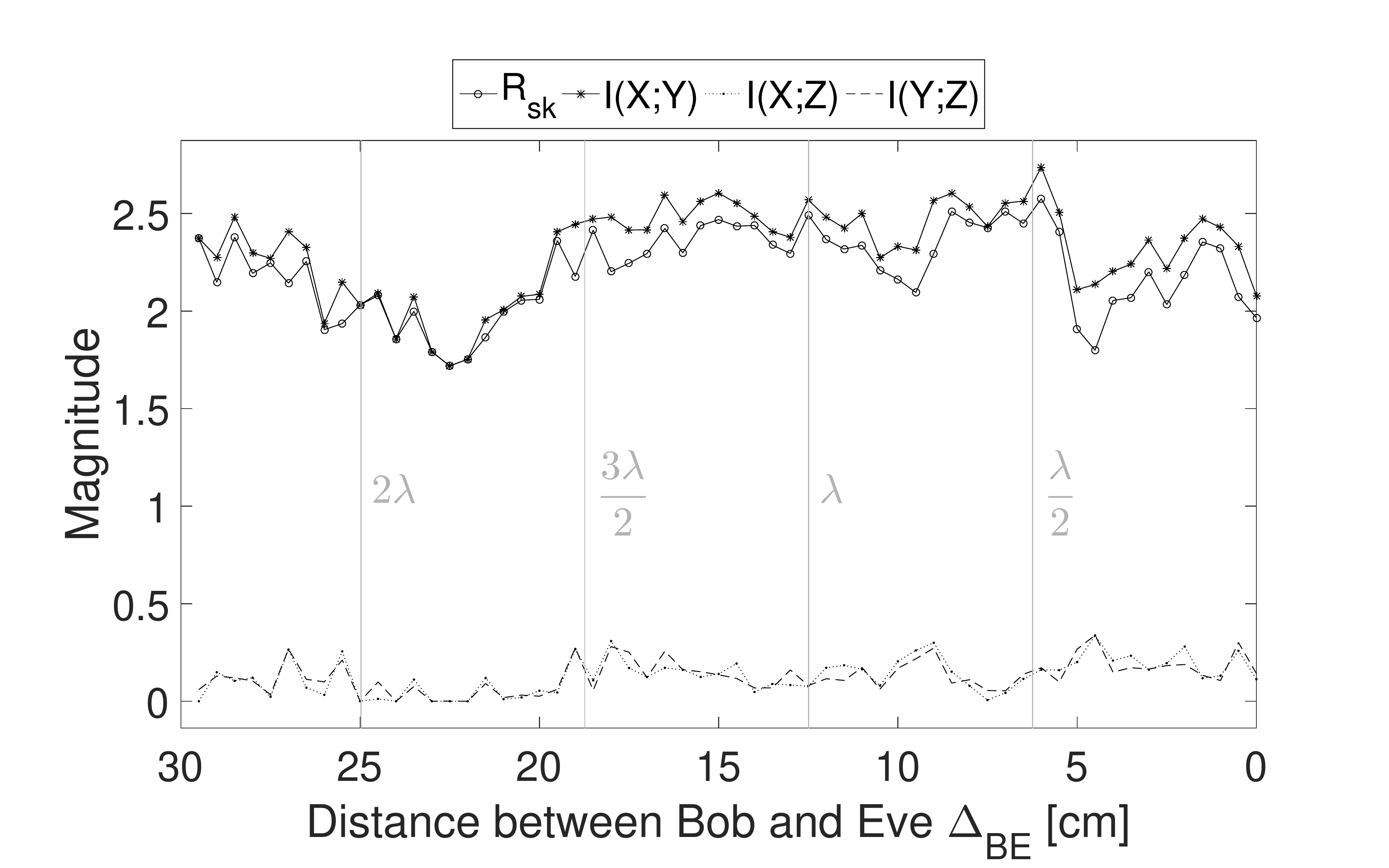}}
	\caption{Evaluation results of $\mybold{v}^{\text{ds}}_k$. In (a) and (b) the cross-correlations is given; in (c) the mutual information as well as $\rsk$ is given. Position 17.}
	\label{fig:app_ds_17}
\end{figure*}

\begin{figure*}
	\centering
	\subfloat[]{\includegraphics[trim=1.4cm 0.1cm 3.5cm 1.6cm, clip=true, height=0.224\textwidth]{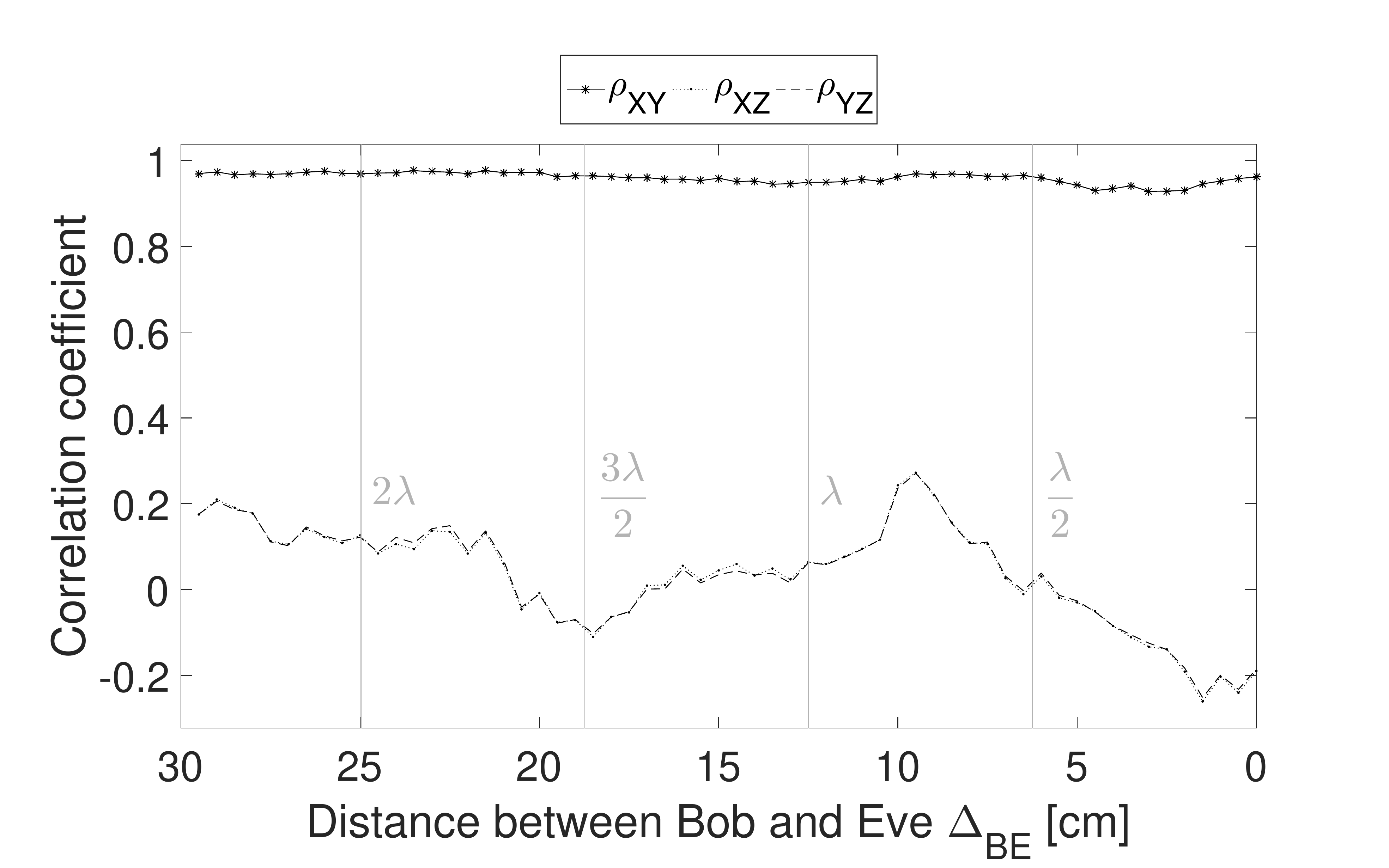}}
	\subfloat[]{\includegraphics[trim=1cm 0.1cm 3.5cm 1.6cm, clip=true, height=0.224\textwidth]{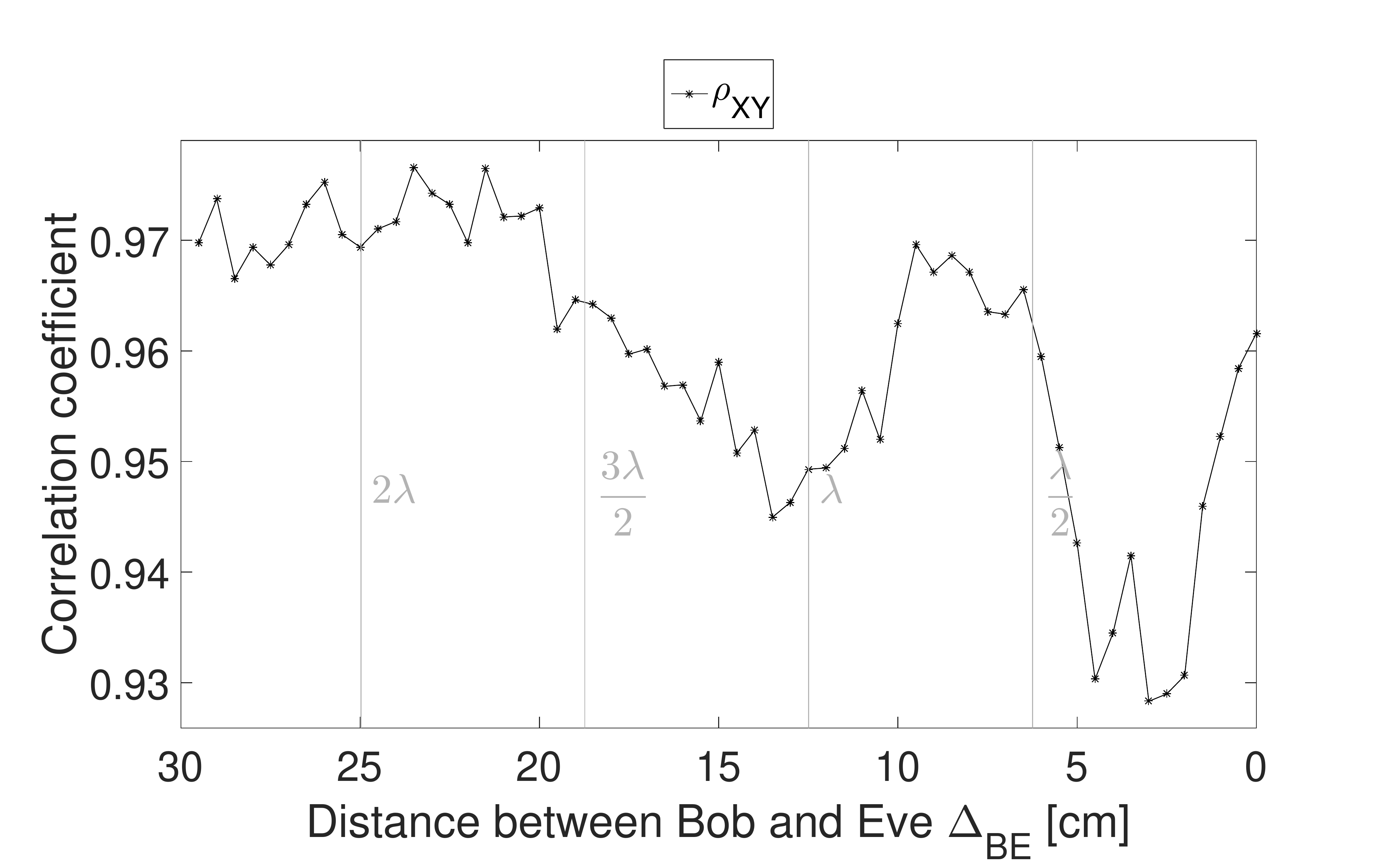}}
	\subfloat[]{\includegraphics[trim=1.8cm 0.1cm 3.5cm 1.6cm, clip=true, height=0.224\textwidth]{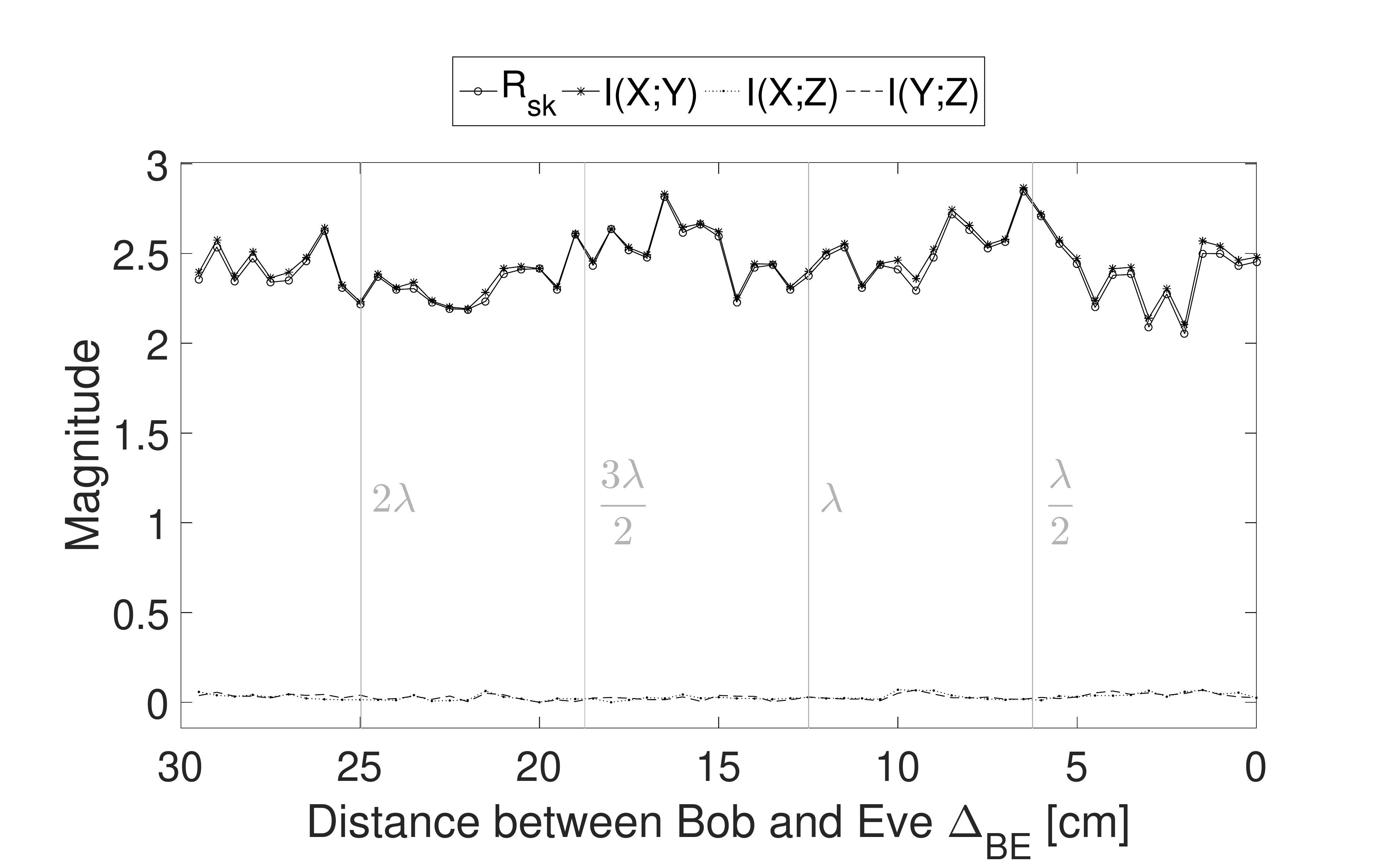}}
	\caption{Evaluation results of $\mybold{v}^{\text{de}}_k$. In (a) and (b) the cross-correlations is given; in (c) the mutual information as well as $\rsk$ is given. Position 17.}
	\label{fig:app_decorr_17}
\end{figure*}


\begin{figure*}
	\centering
	\subfloat[]{\includegraphics[trim=1.4cm 0.1cm 3.5cm 1.6cm, clip=true, height=0.224\textwidth]{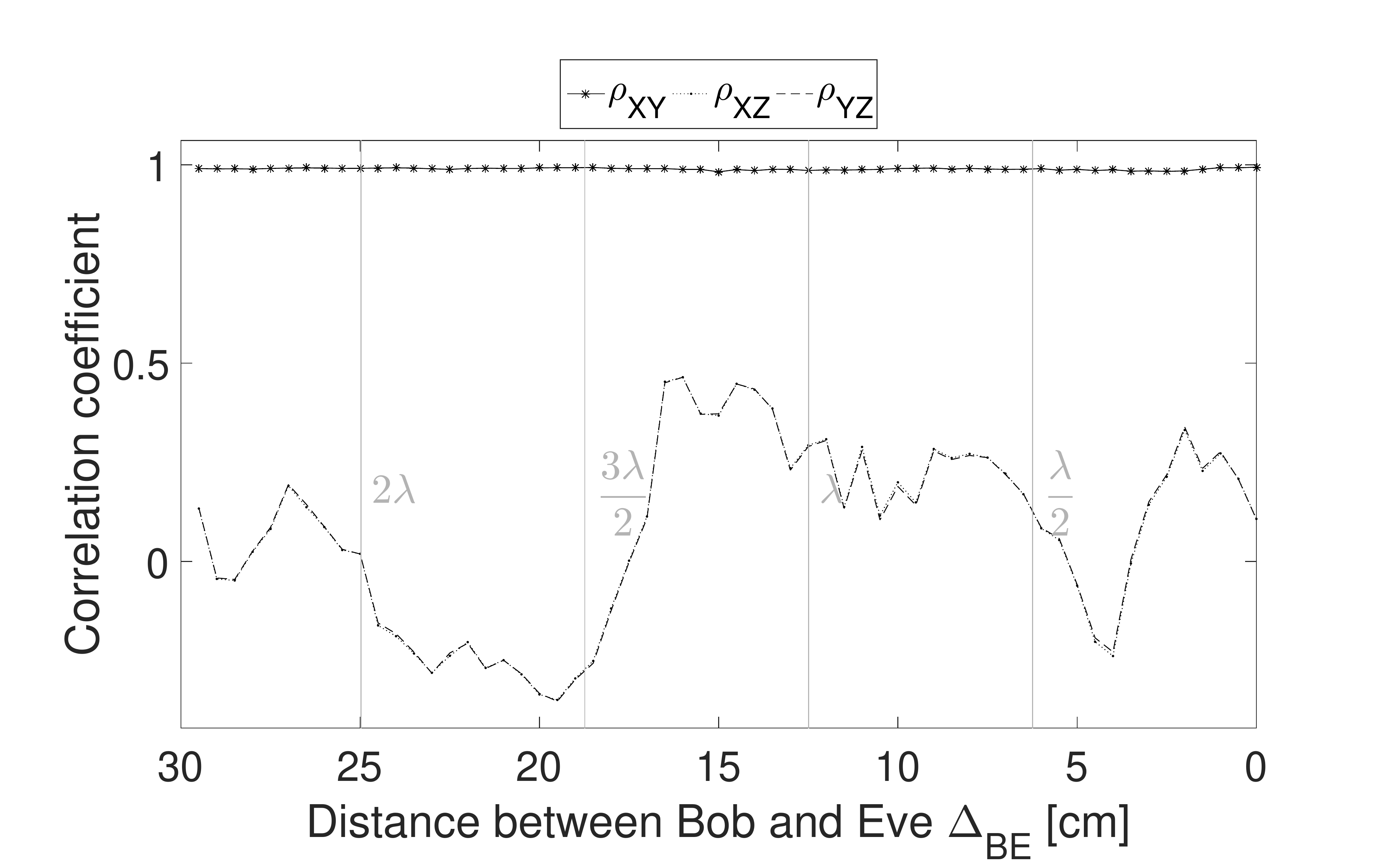}}
	\subfloat[]{\includegraphics[trim=0.5cm 0.1cm 3.5cm 1.6cm, clip=true, height=0.224\textwidth]{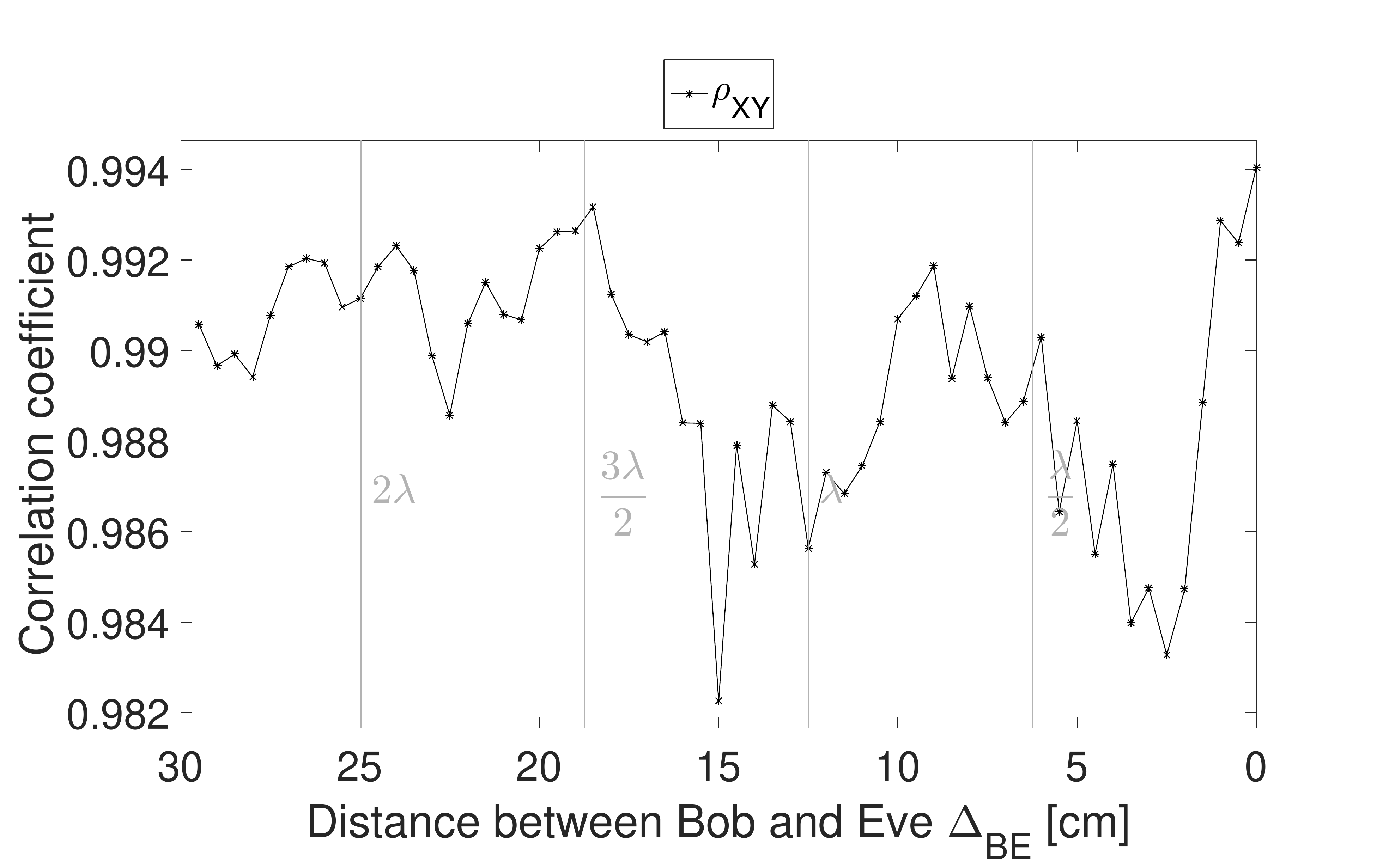}}
	\subfloat[]{\includegraphics[trim=2.2cm 0.1cm 3.5cm 1.6cm, clip=true, height=0.224\textwidth]{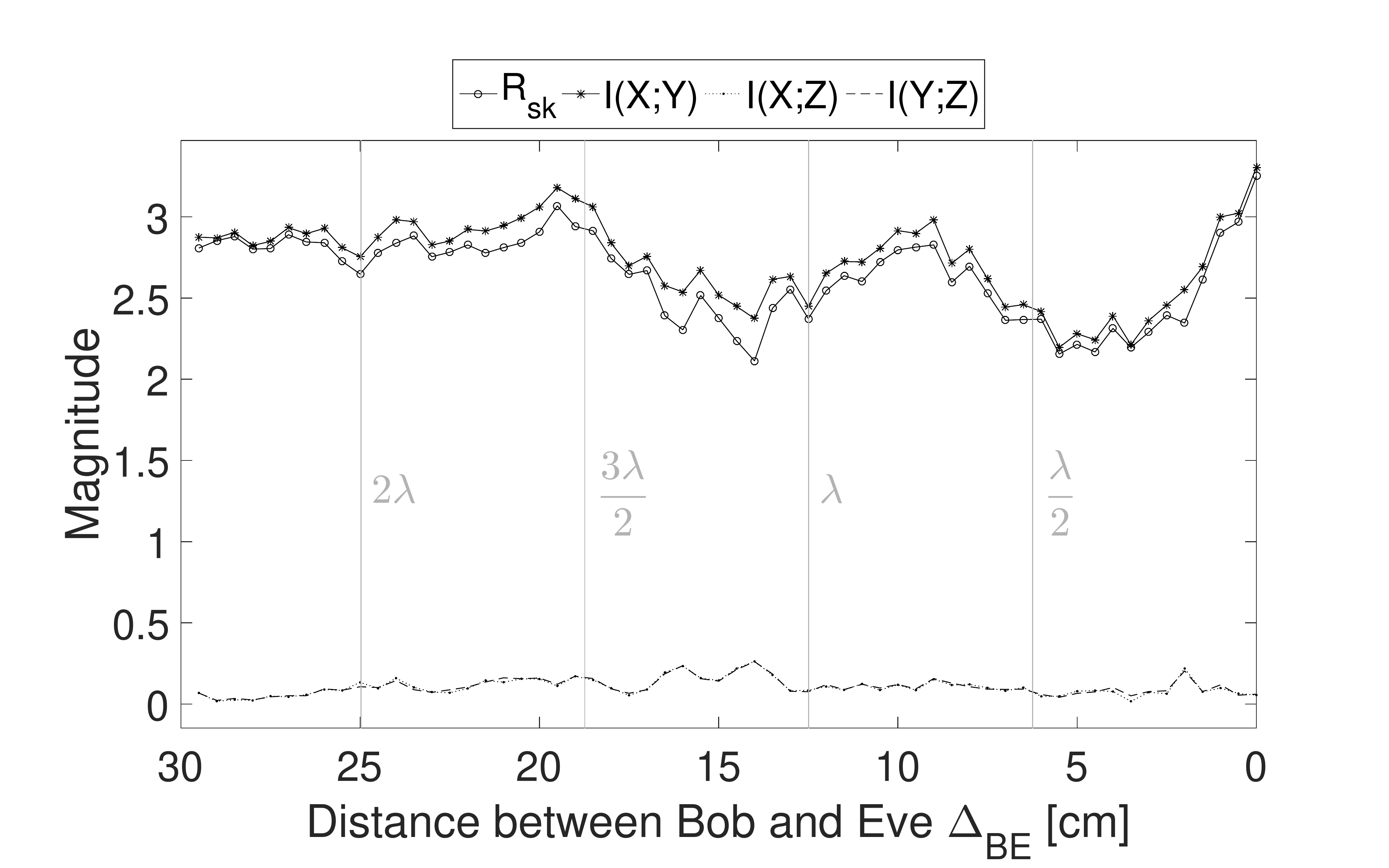}}
	\caption{Evaluation results of $\mybold{v}_k$. In (a) and (b) the cross-correlations is given; in (c) the mutual information as well as $\rsk$ is given. Position 18.}
	\label{fig:app_original_18}
\end{figure*}

\begin{figure*}
	\centering
	\subfloat[]{\includegraphics[trim=1.4cm 0.1cm 3.5cm 1.6cm, clip=true, height=0.224\textwidth]{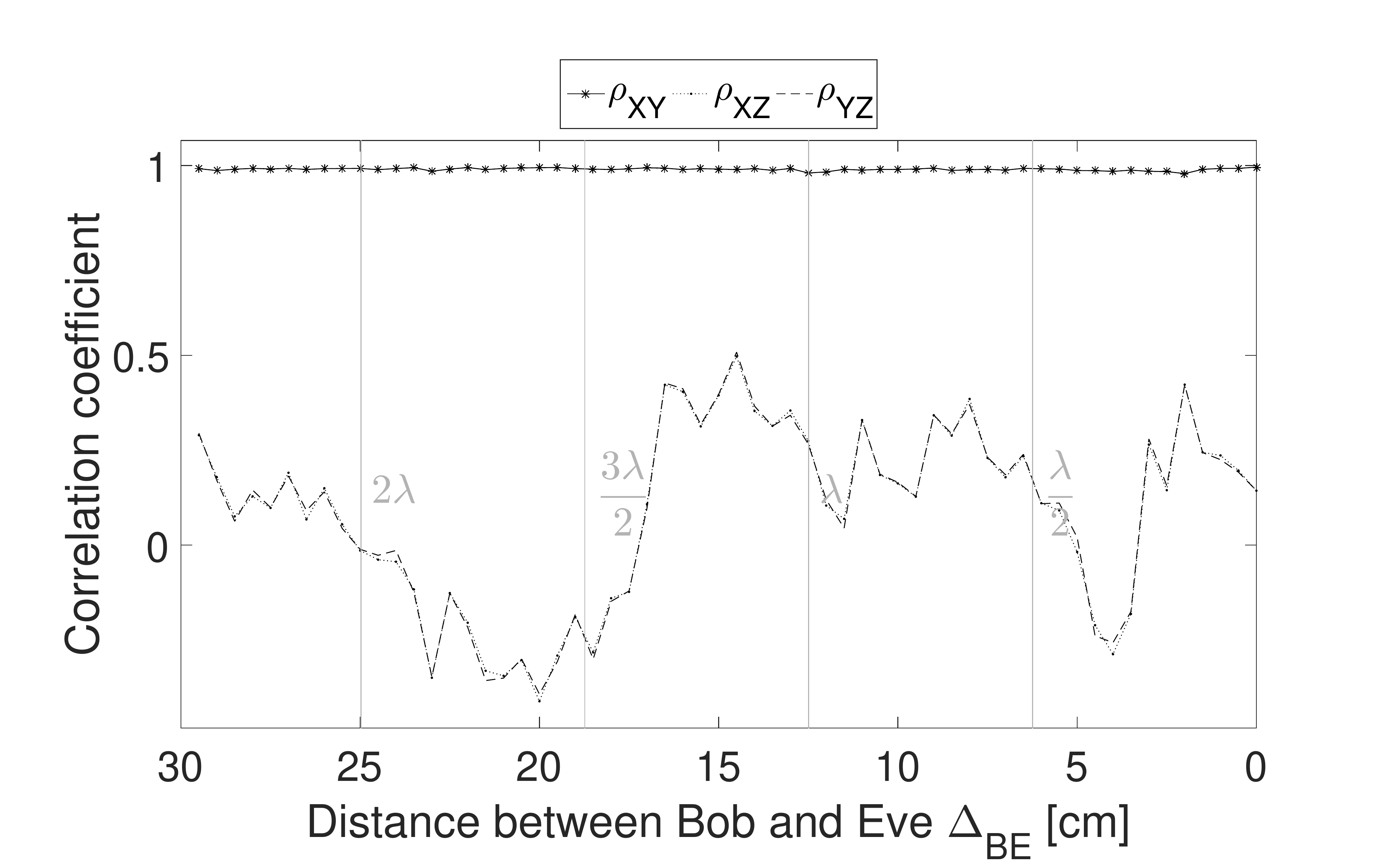}}
	\subfloat[]{\includegraphics[trim=0.5cm 0.1cm 3.5cm 1.6cm, clip=true, height=0.224\textwidth]{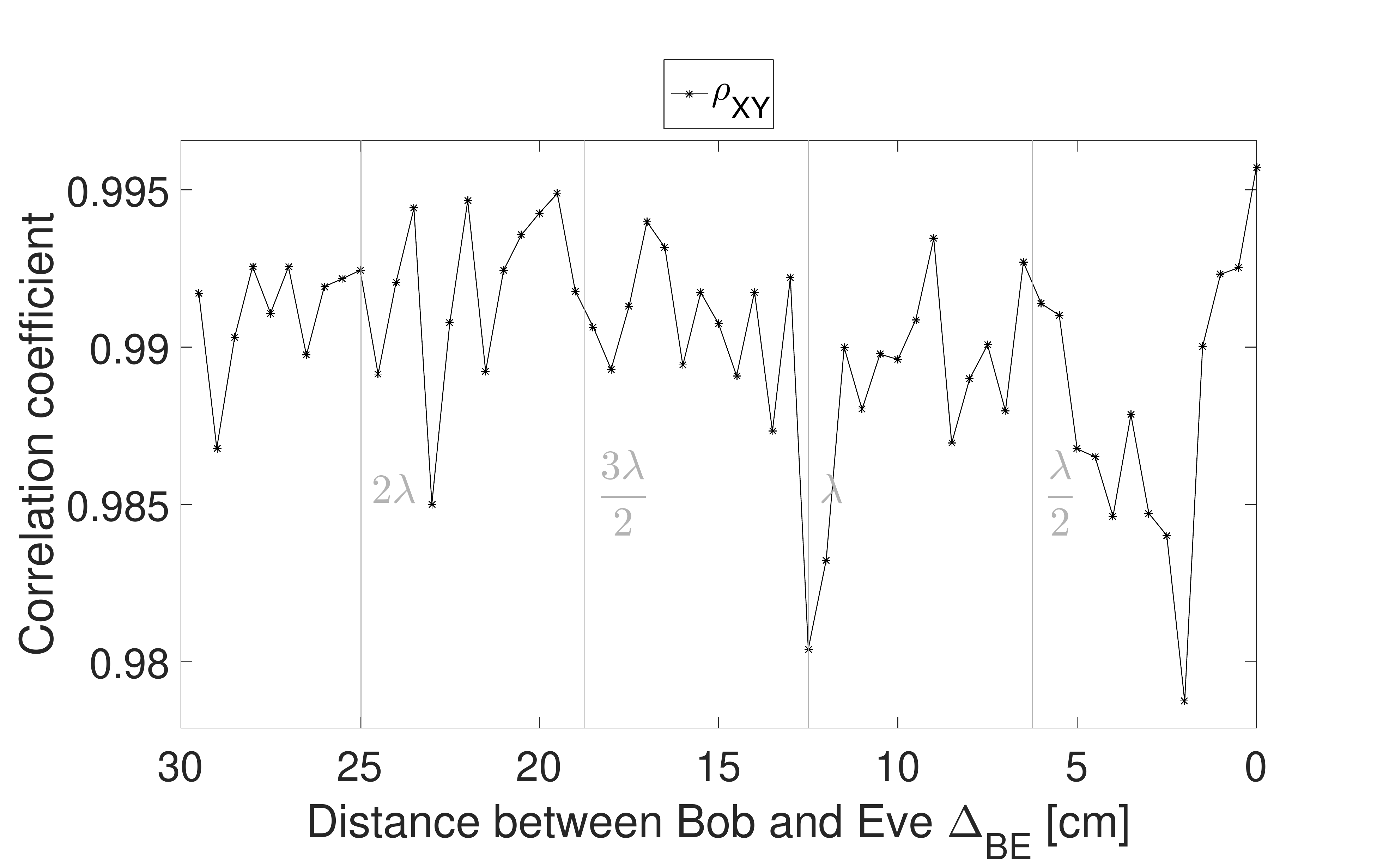}}
	\subfloat[]{\includegraphics[trim=2.2cm 0.1cm 3.5cm 1.6cm, clip=true, height=0.224\textwidth]{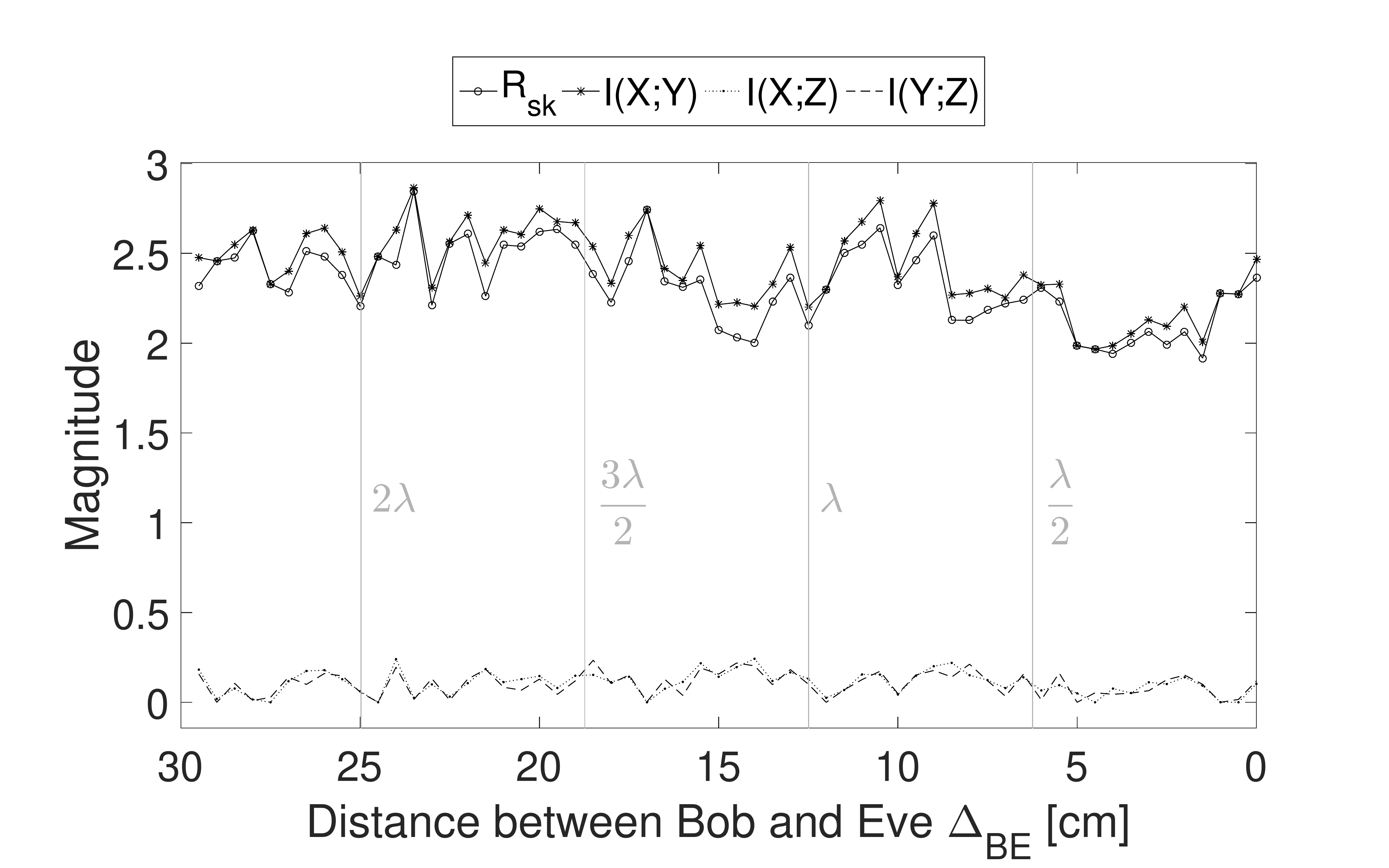}}
	\caption{Evaluation results of $\mybold{v}^{\text{ds}}_k$. In (a) and (b) the cross-correlations is given; in (c) the mutual information as well as $\rsk$ is given. Position 18.}
	\label{fig:app_ds_18}
\end{figure*}

\begin{figure*}
	\centering
	\subfloat[]{\includegraphics[trim=1.4cm 0.1cm 3.5cm 1.6cm, clip=true, height=0.224\textwidth]{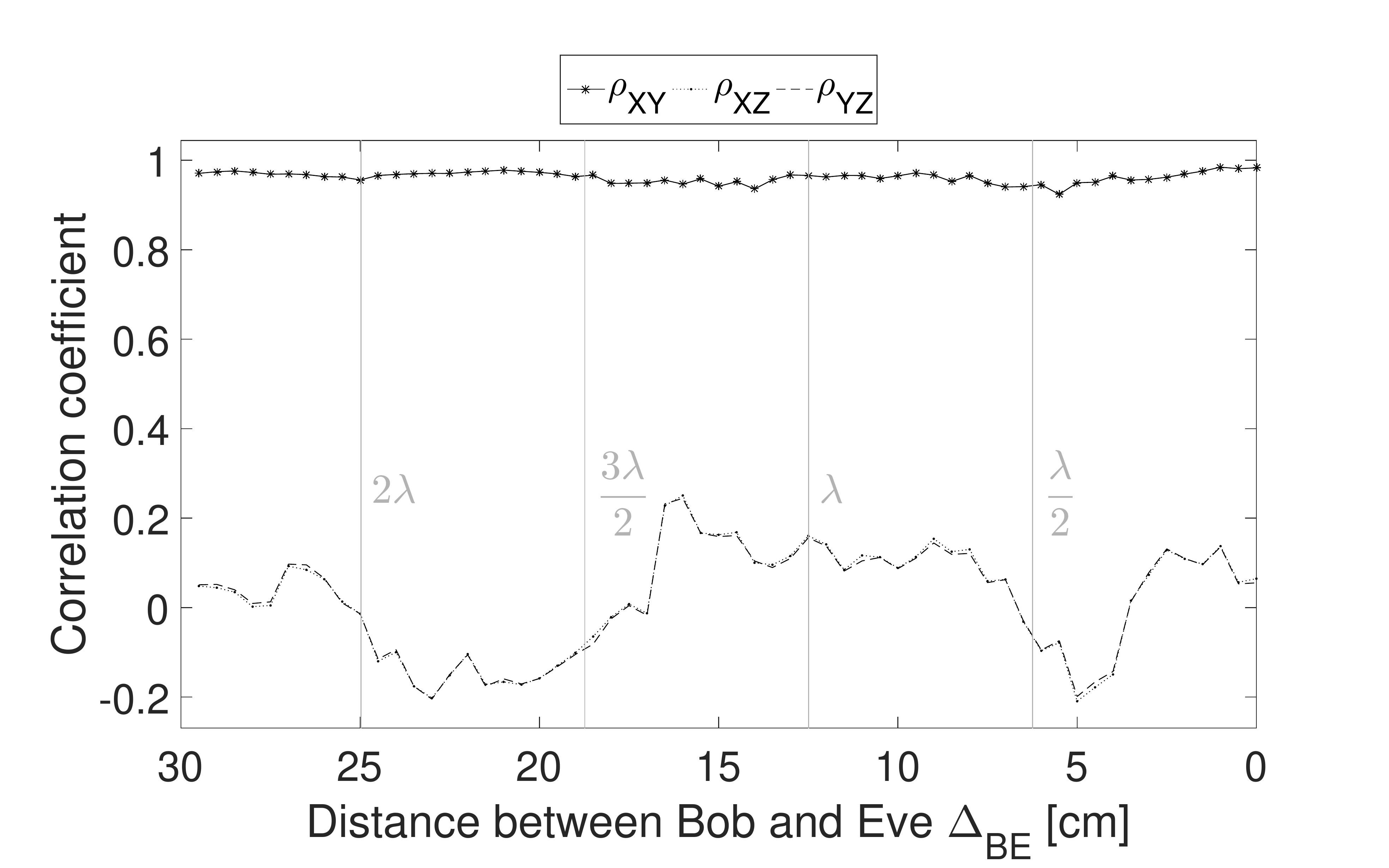}}
	\subfloat[]{\includegraphics[trim=1cm 0.1cm 3.5cm 1.6cm, clip=true, height=0.224\textwidth]{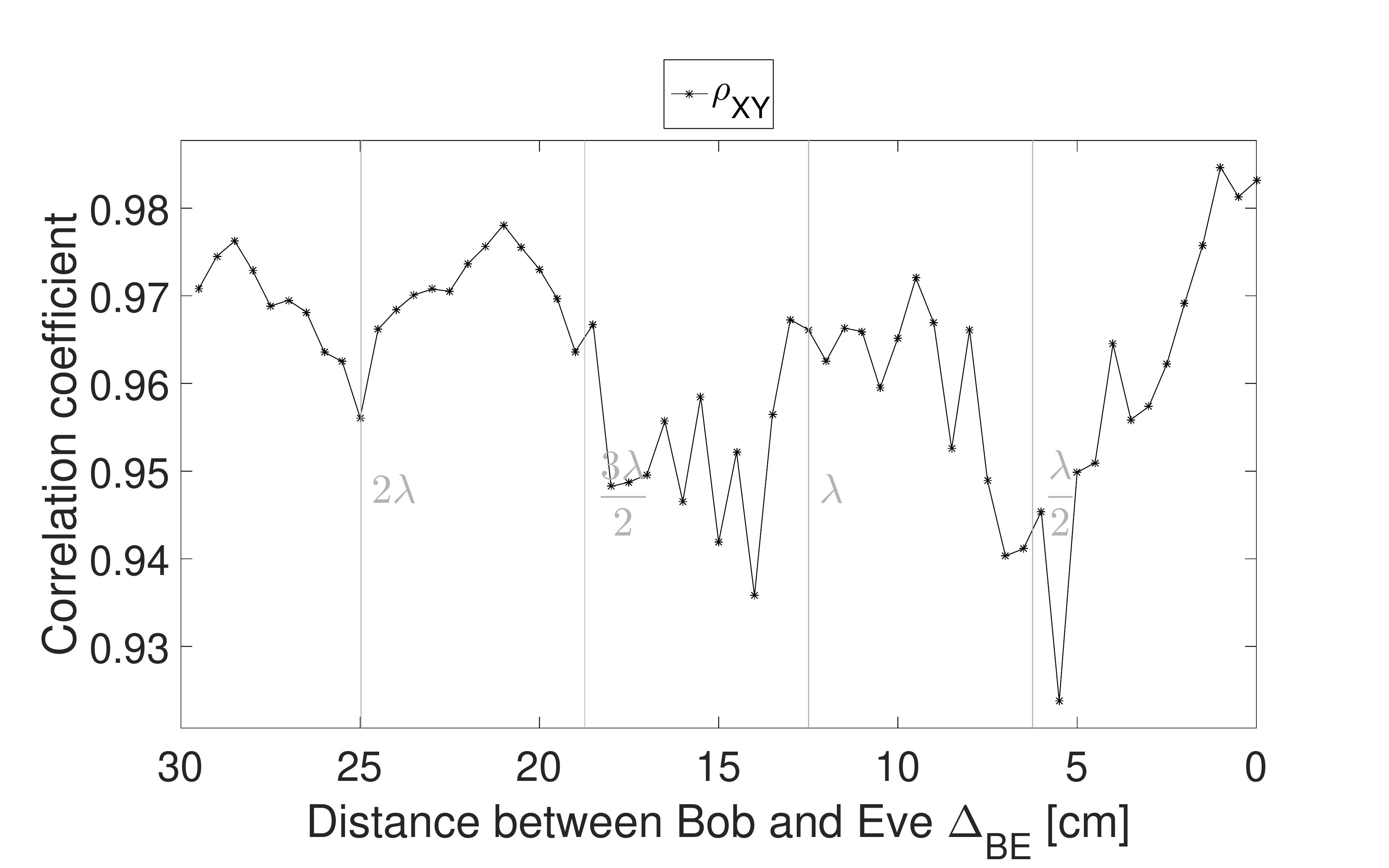}}
	\subfloat[]{\includegraphics[trim=1.8cm 0.1cm 3.5cm 1.6cm, clip=true, height=0.224\textwidth]{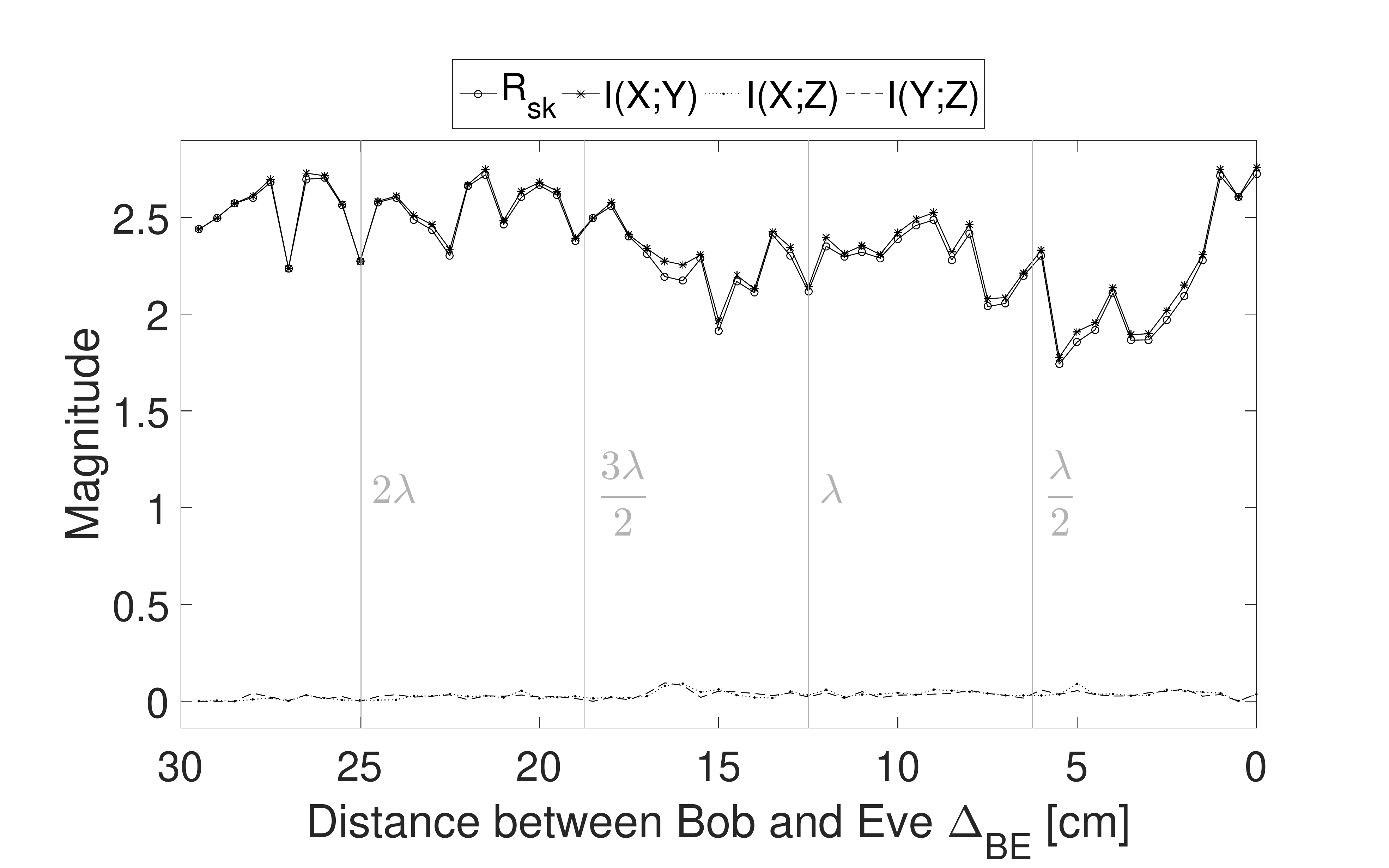}}
	\caption{Evaluation results of $\mybold{v}^{\text{de}}_k$. In (a) and (b) the cross-correlations is given; in (c) the mutual information as well as $\rsk$ is given. Position 18.}
	\label{fig:app_decorr_18}
\end{figure*}


\begin{figure*}
	\centering
	\subfloat[]{\includegraphics[trim=1.4cm 0.1cm 3.5cm 1.6cm, clip=true, height=0.224\textwidth]{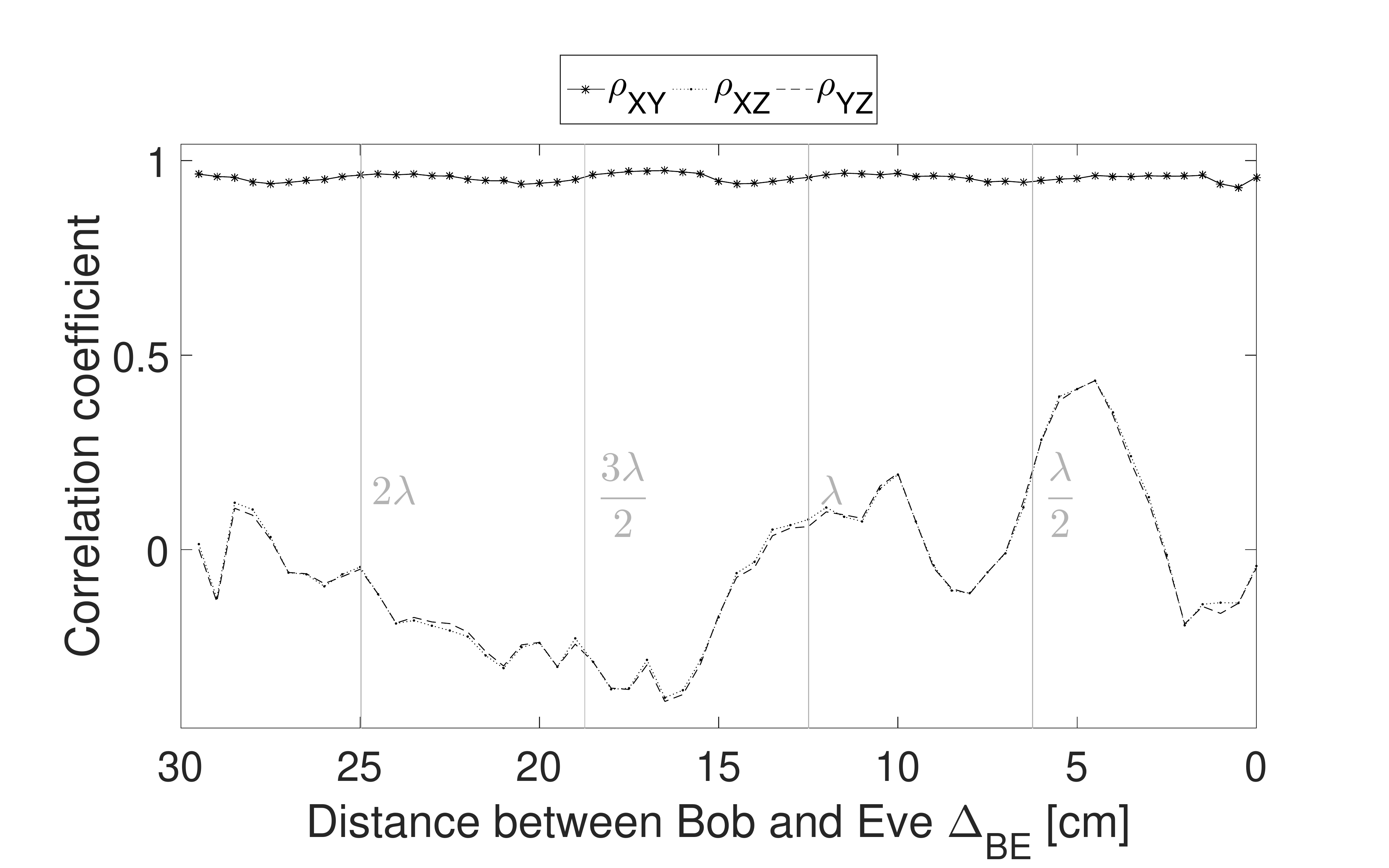}}
	\subfloat[]{\includegraphics[trim=0.5cm 0.1cm 3.5cm 1.6cm, clip=true, height=0.224\textwidth]{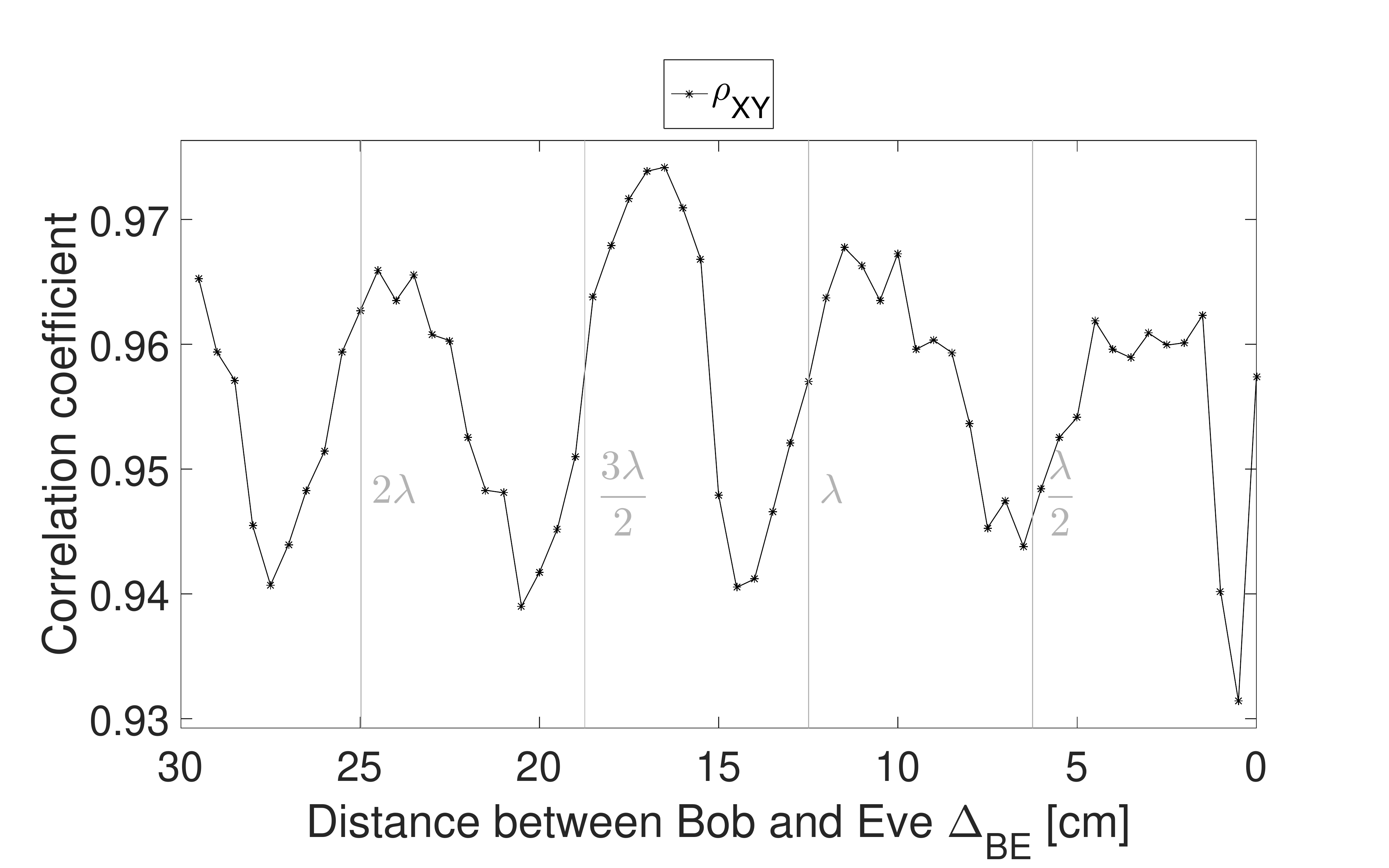}}
	\subfloat[]{\includegraphics[trim=2.2cm 0.1cm 3.5cm 1.6cm, clip=true, height=0.224\textwidth]{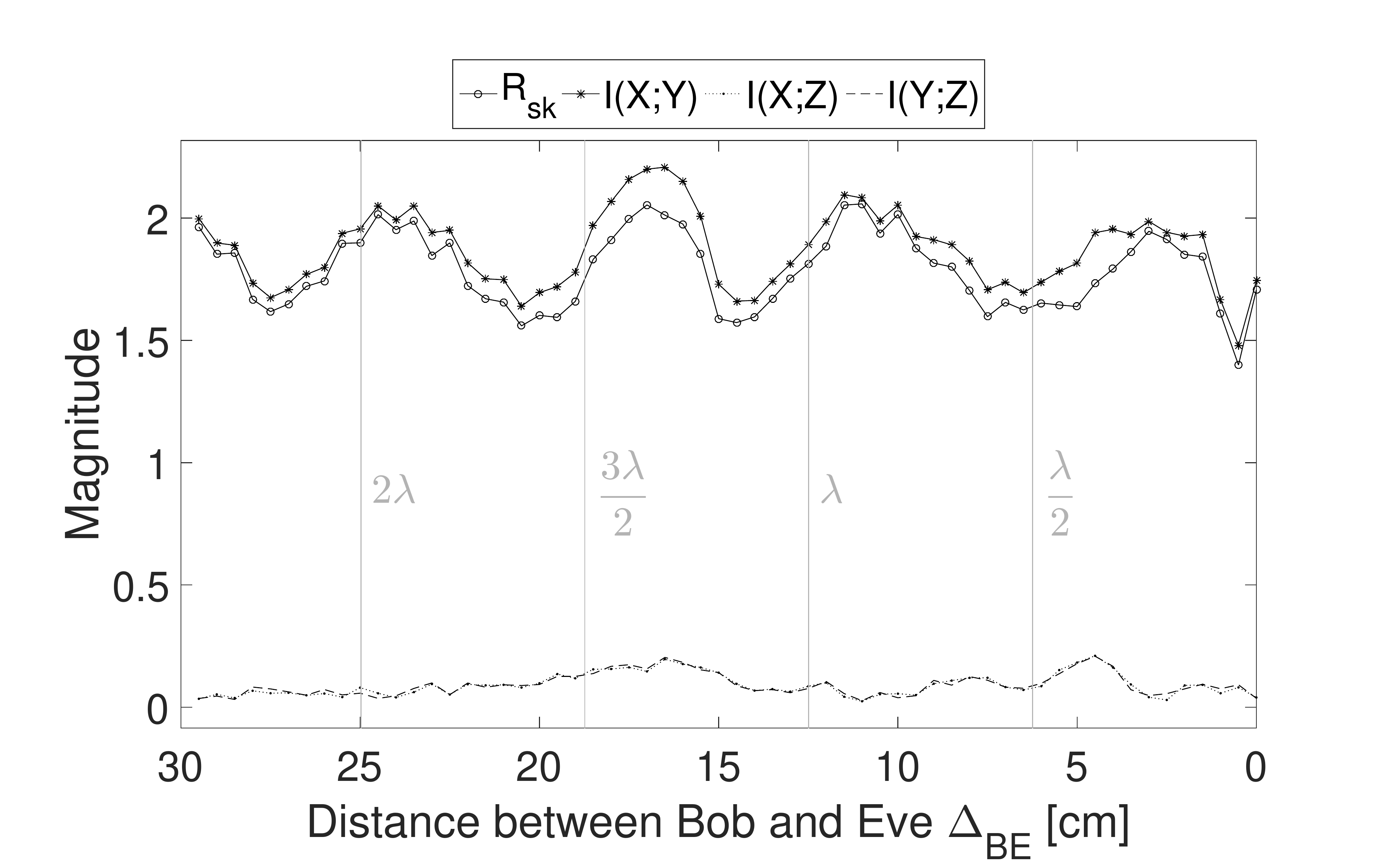}}
	\caption{Evaluation results of $\mybold{v}_k$. In (a) and (b) the cross-correlations is given; in (c) the mutual information as well as $\rsk$ is given. Position 19.}
	\label{fig:app_original_19}
\end{figure*}

\begin{figure*}
	\centering
	\subfloat[]{\includegraphics[trim=1.4cm 0.1cm 3.5cm 1.6cm, clip=true, height=0.224\textwidth]{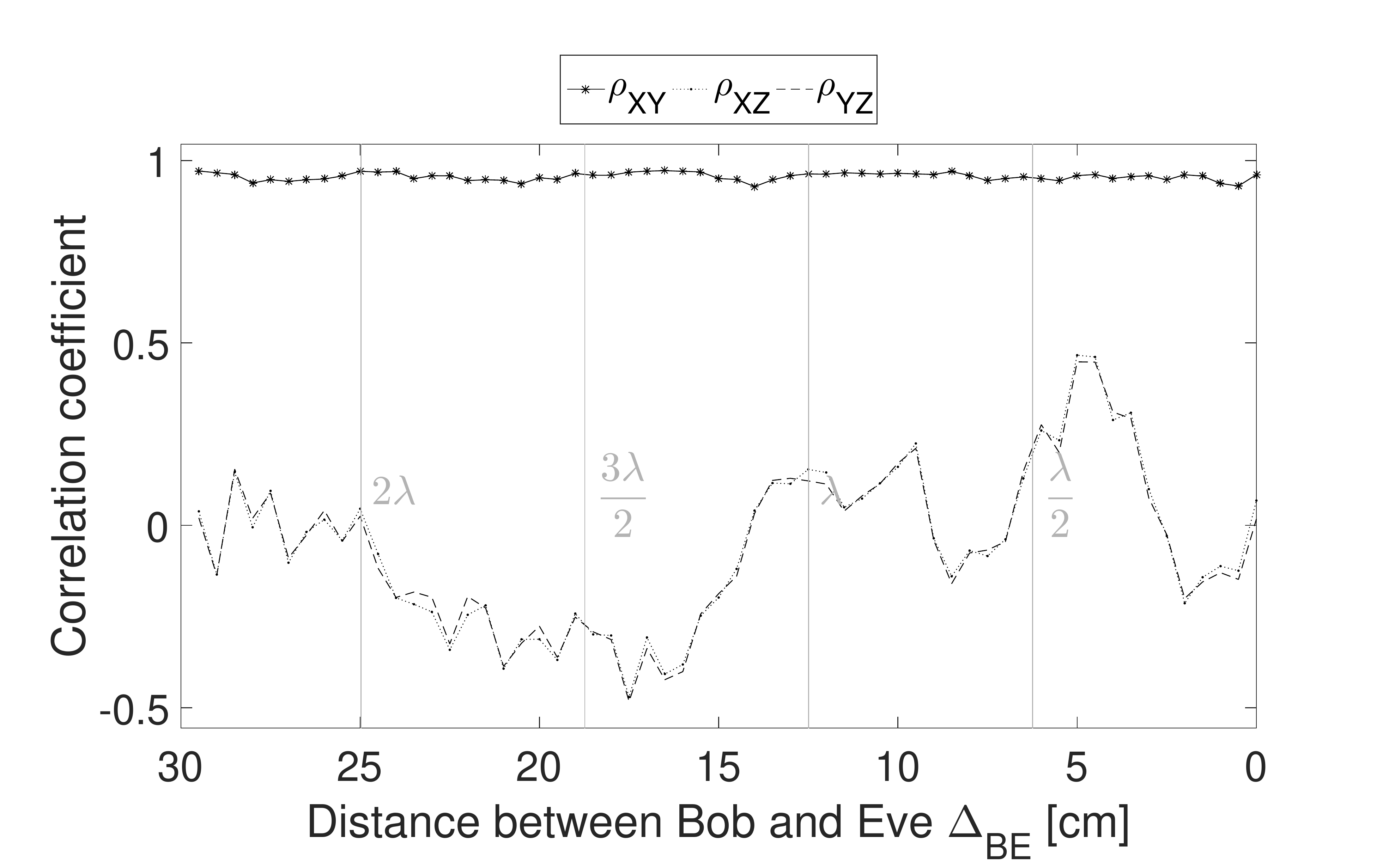}}
	\subfloat[]{\includegraphics[trim=0.5cm 0.1cm 3.5cm 1.6cm, clip=true, height=0.224\textwidth]{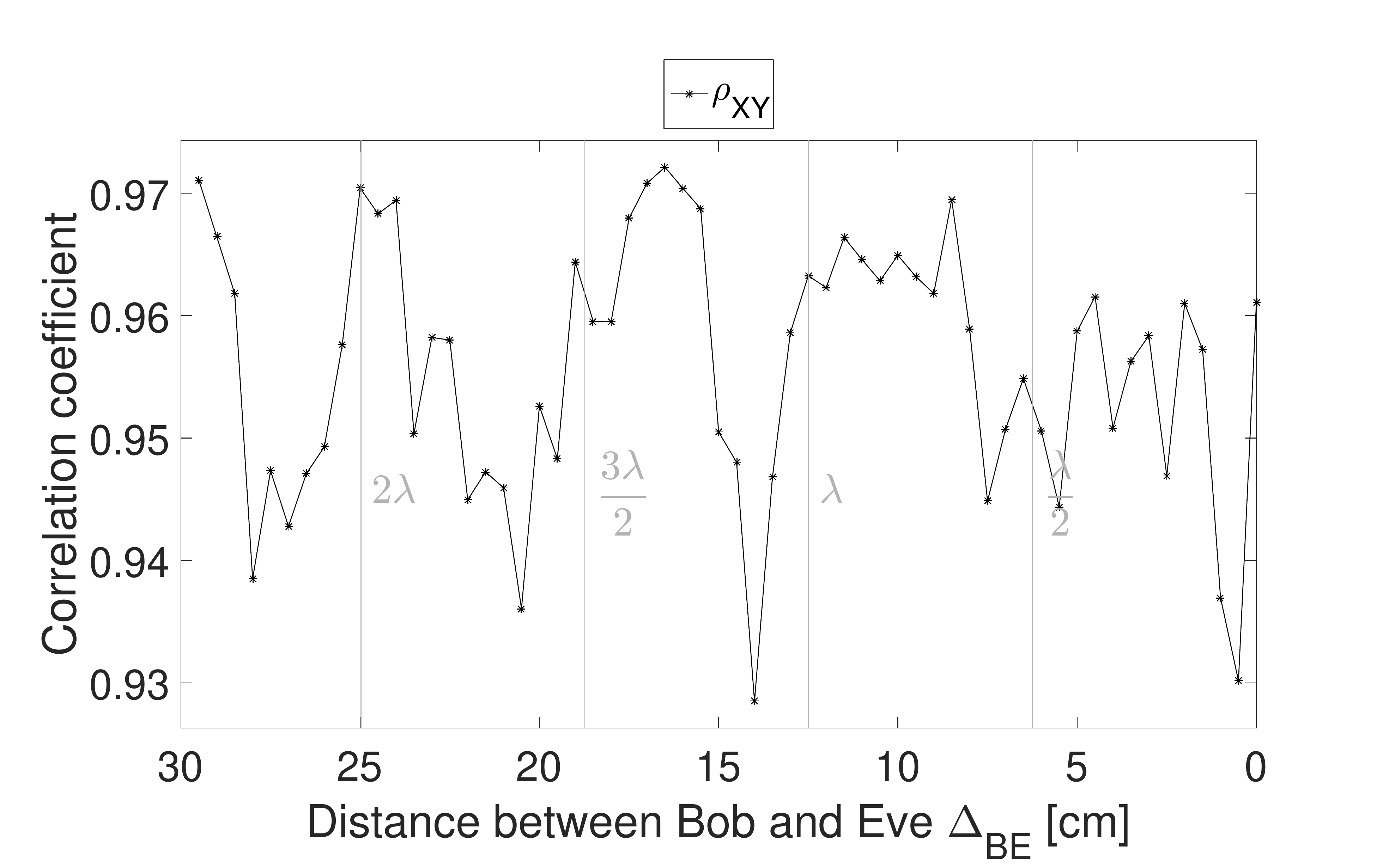}}
	\subfloat[]{\includegraphics[trim=2.2cm 0.1cm 3.5cm 1.6cm, clip=true, height=0.224\textwidth]{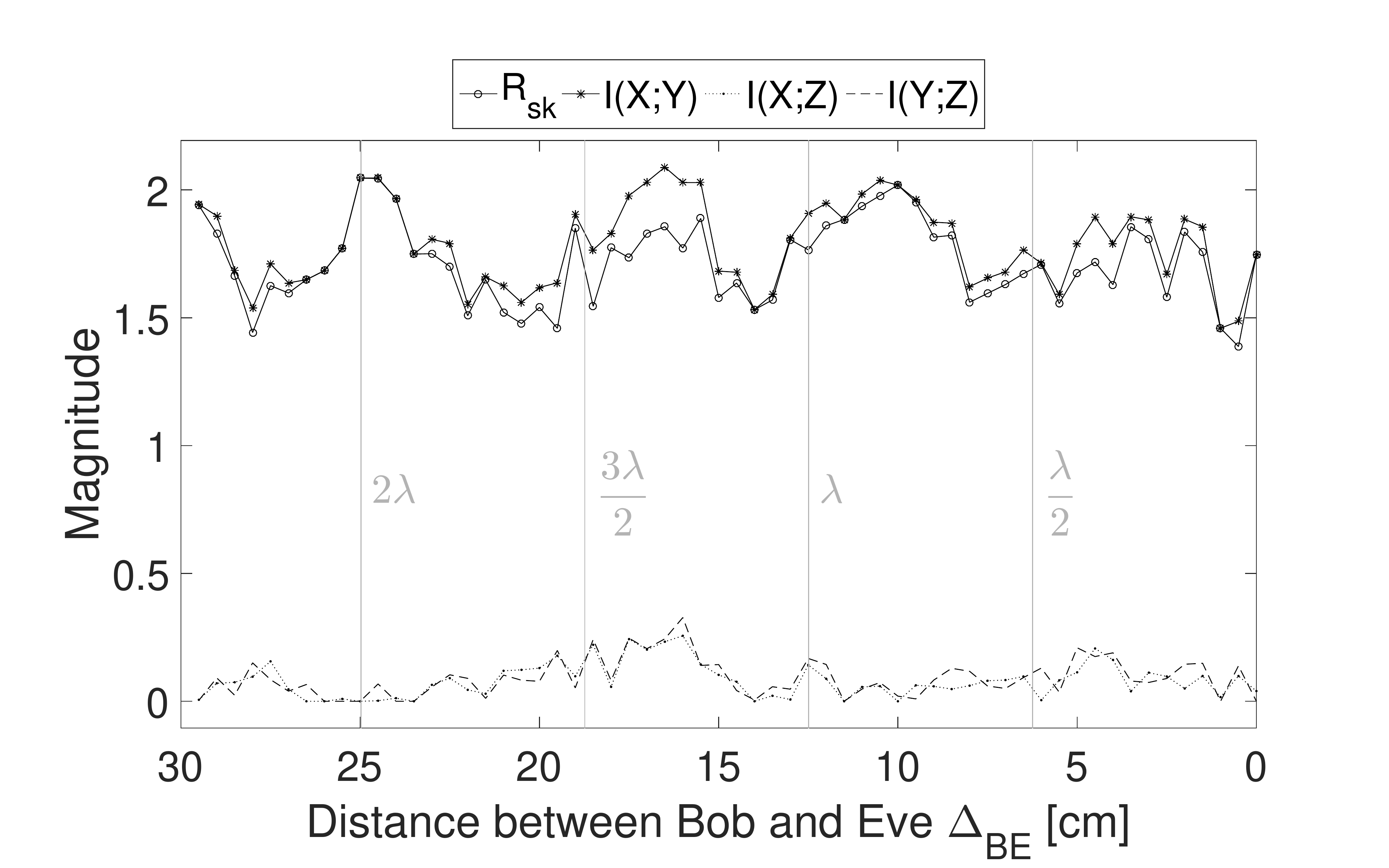}}
	\caption{Evaluation results of $\mybold{v}^{\text{ds}}_k$. In (a) and (b) the cross-correlations is given; in (c) the mutual information as well as $\rsk$ is given. Position 19.}
	\label{fig:app_ds_19}
\end{figure*}

\begin{figure*}
	\centering
	\subfloat[]{\includegraphics[trim=1.4cm 0.1cm 3.5cm 1.6cm, clip=true, height=0.224\textwidth]{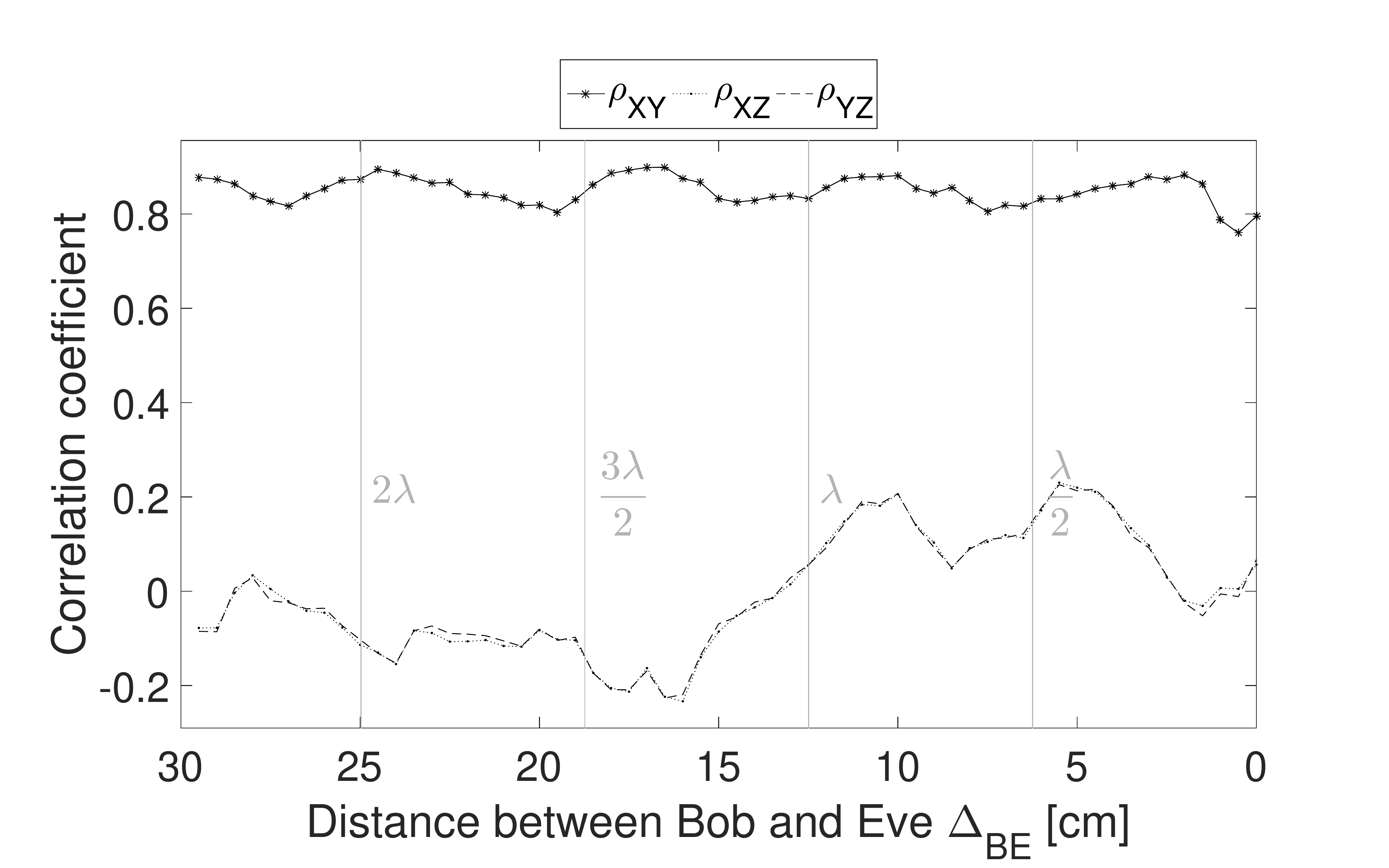}}
	\subfloat[]{\includegraphics[trim=1cm 0.1cm 3.5cm 1.6cm, clip=true, height=0.224\textwidth]{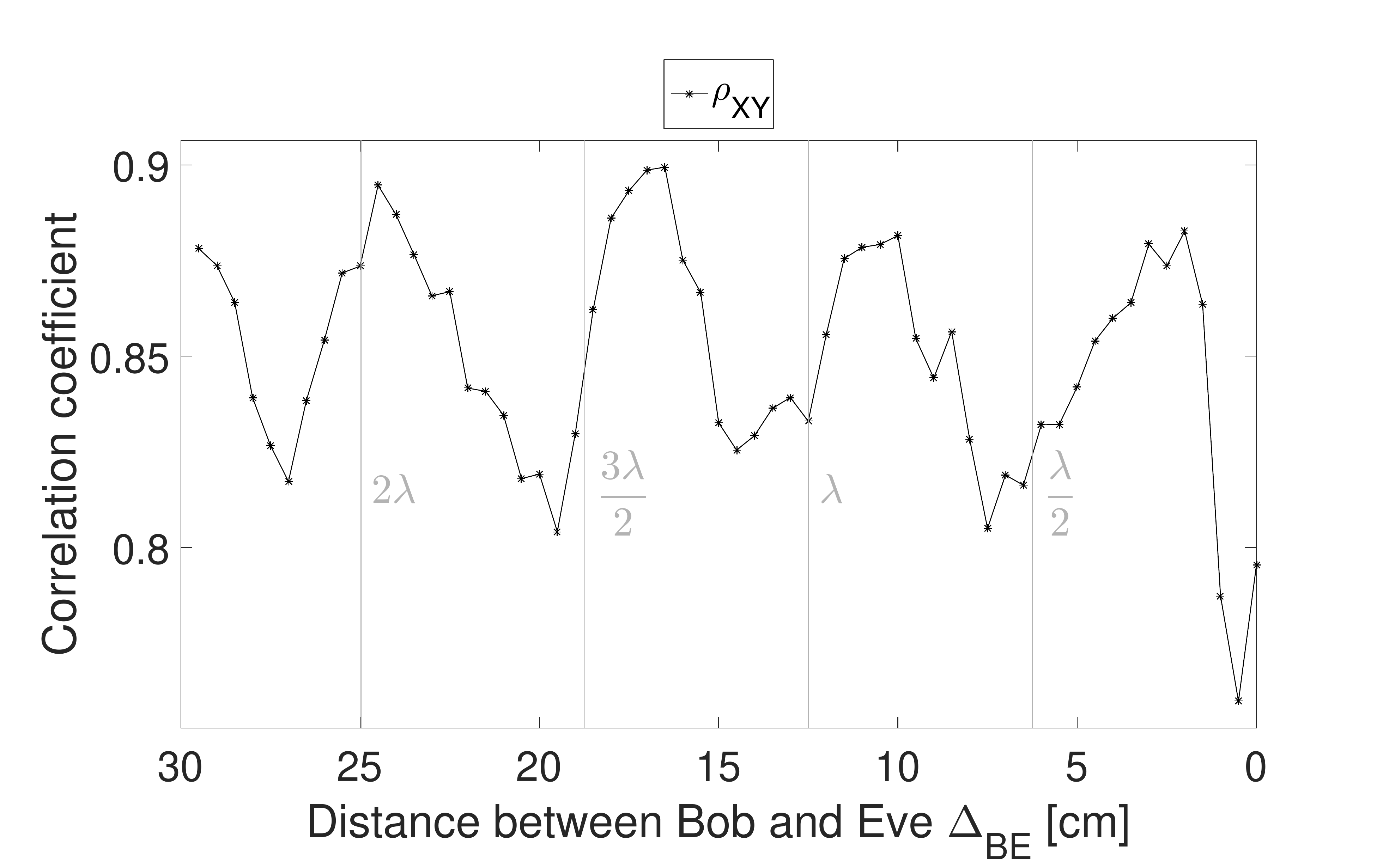}}
	\subfloat[]{\includegraphics[trim=1.8cm 0.1cm 3.5cm 1.6cm, clip=true, height=0.224\textwidth]{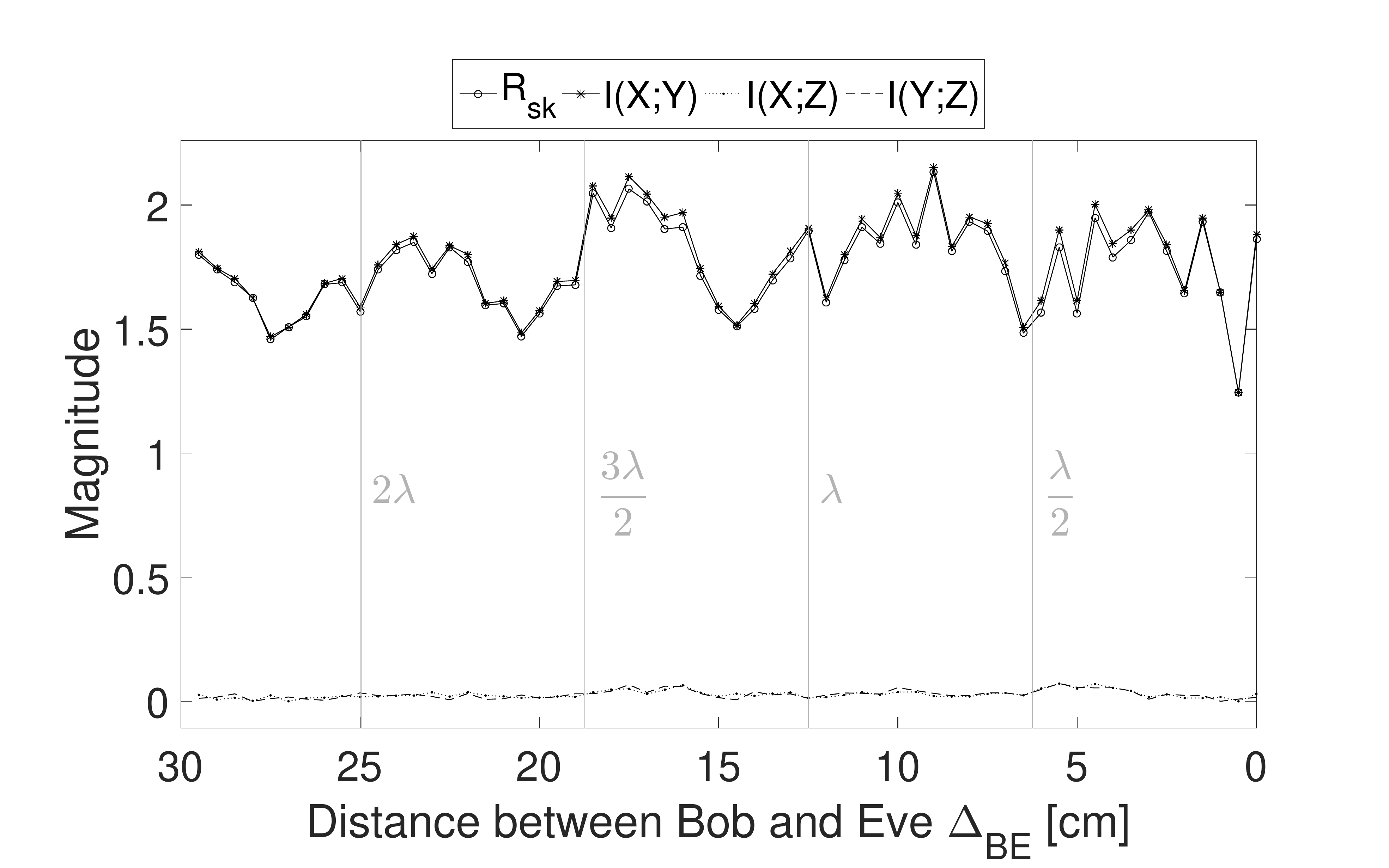}}
	\caption{Evaluation results of $\mybold{v}^{\text{de}}_k$. In (a) and (b) the cross-correlations is given; in (c) the mutual information as well as $\rsk$ is given. Position 19.}
	\label{fig:app_decorr_19}
\end{figure*}


\begin{figure*}
	\centering
	\subfloat[]{\includegraphics[trim=1.4cm 0.1cm 3.5cm 1.6cm, clip=true, height=0.224\textwidth]{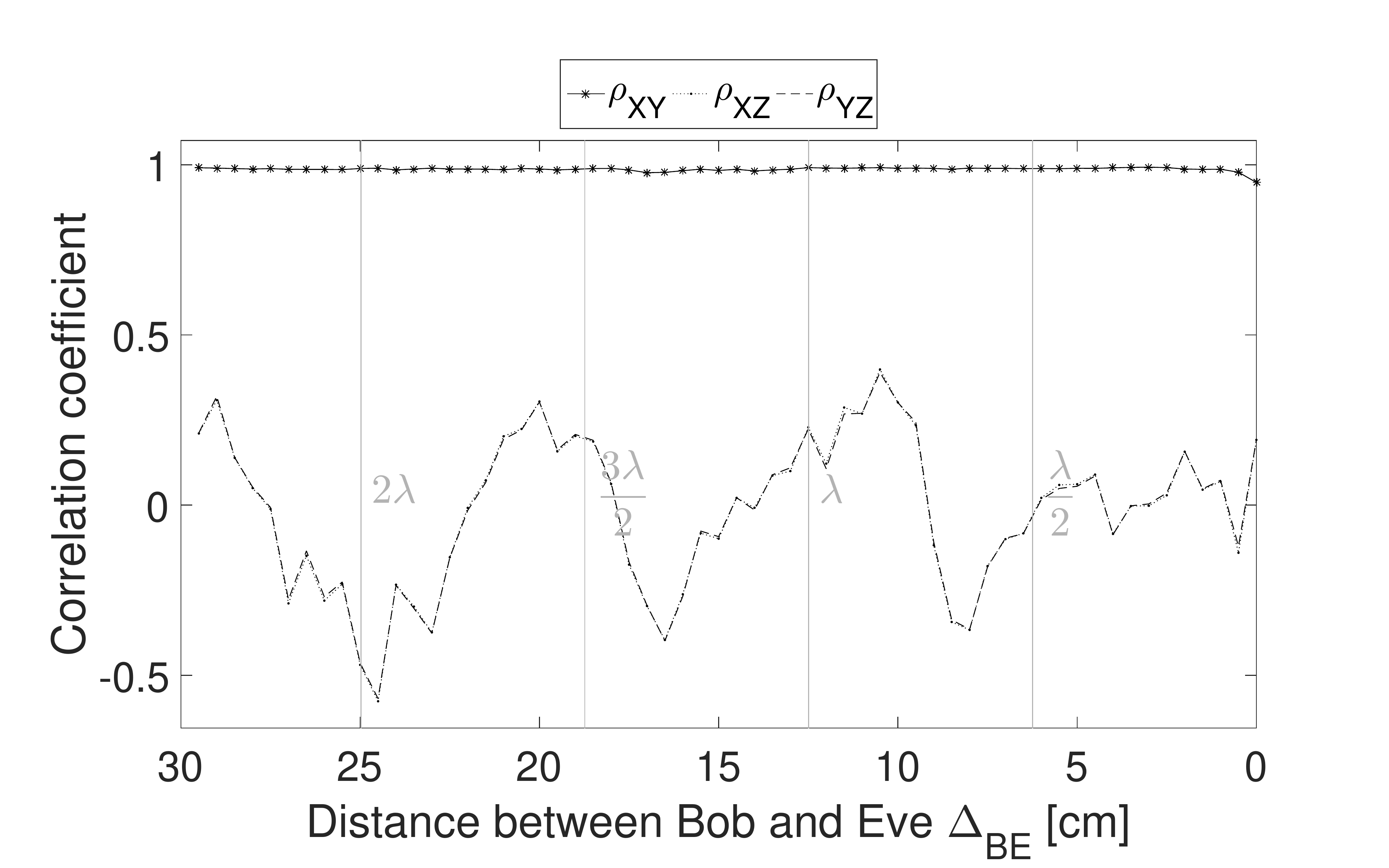}}
	\subfloat[]{\includegraphics[trim=0.5cm 0.1cm 3.5cm 1.6cm, clip=true, height=0.224\textwidth]{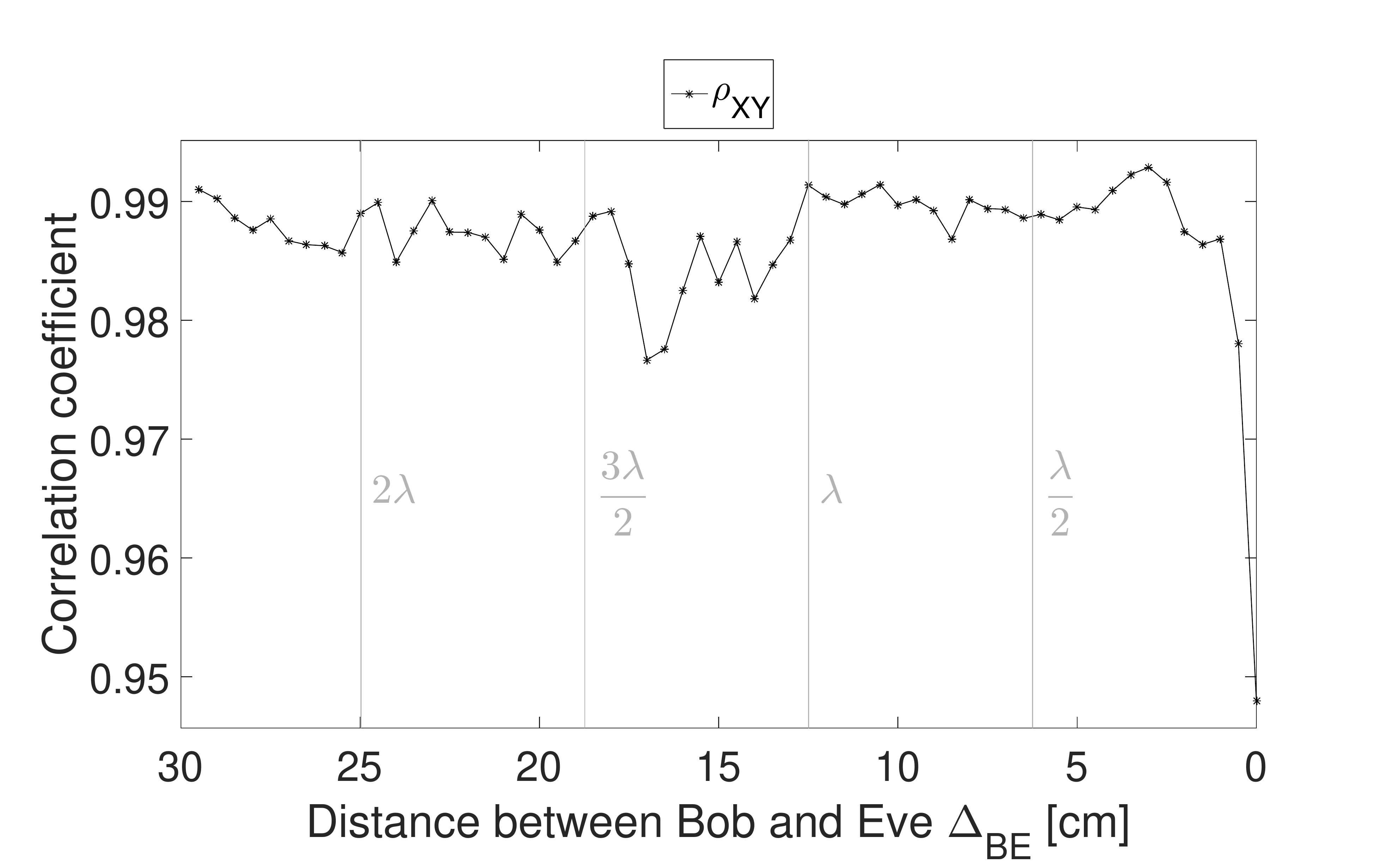}}
	\subfloat[]{\includegraphics[trim=2.2cm 0.1cm 3.5cm 1.6cm, clip=true, height=0.224\textwidth]{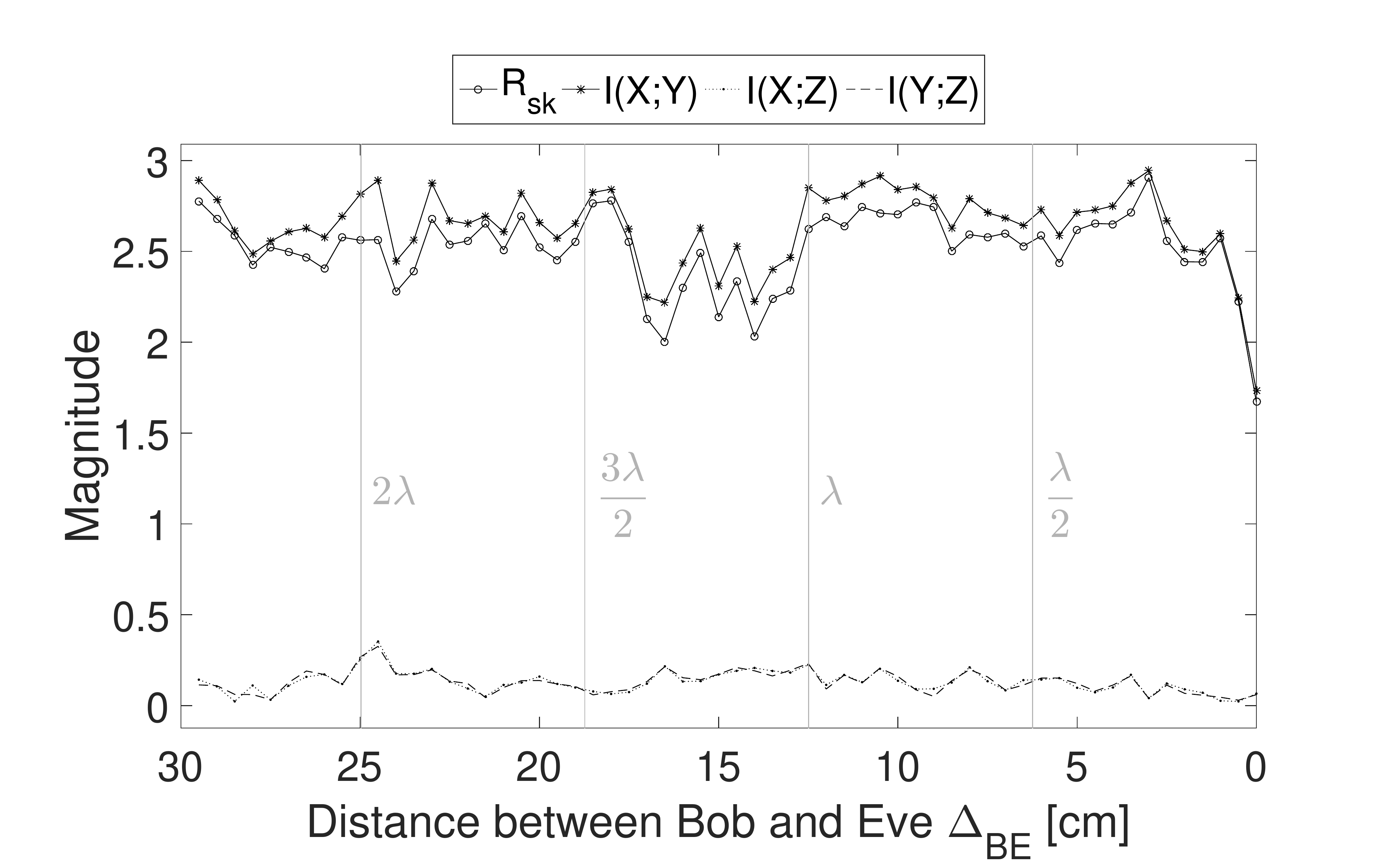}}
	\caption{Evaluation results of $\mybold{v}_k$. In (a) and (b) the cross-correlations is given; in (c) the mutual information as well as $\rsk$ is given. Position 20.}
	\label{fig:app_original_20}
\end{figure*}

\begin{figure*}
	\centering
	\subfloat[]{\includegraphics[trim=1.4cm 0.1cm 3.5cm 1.6cm, clip=true, height=0.224\textwidth]{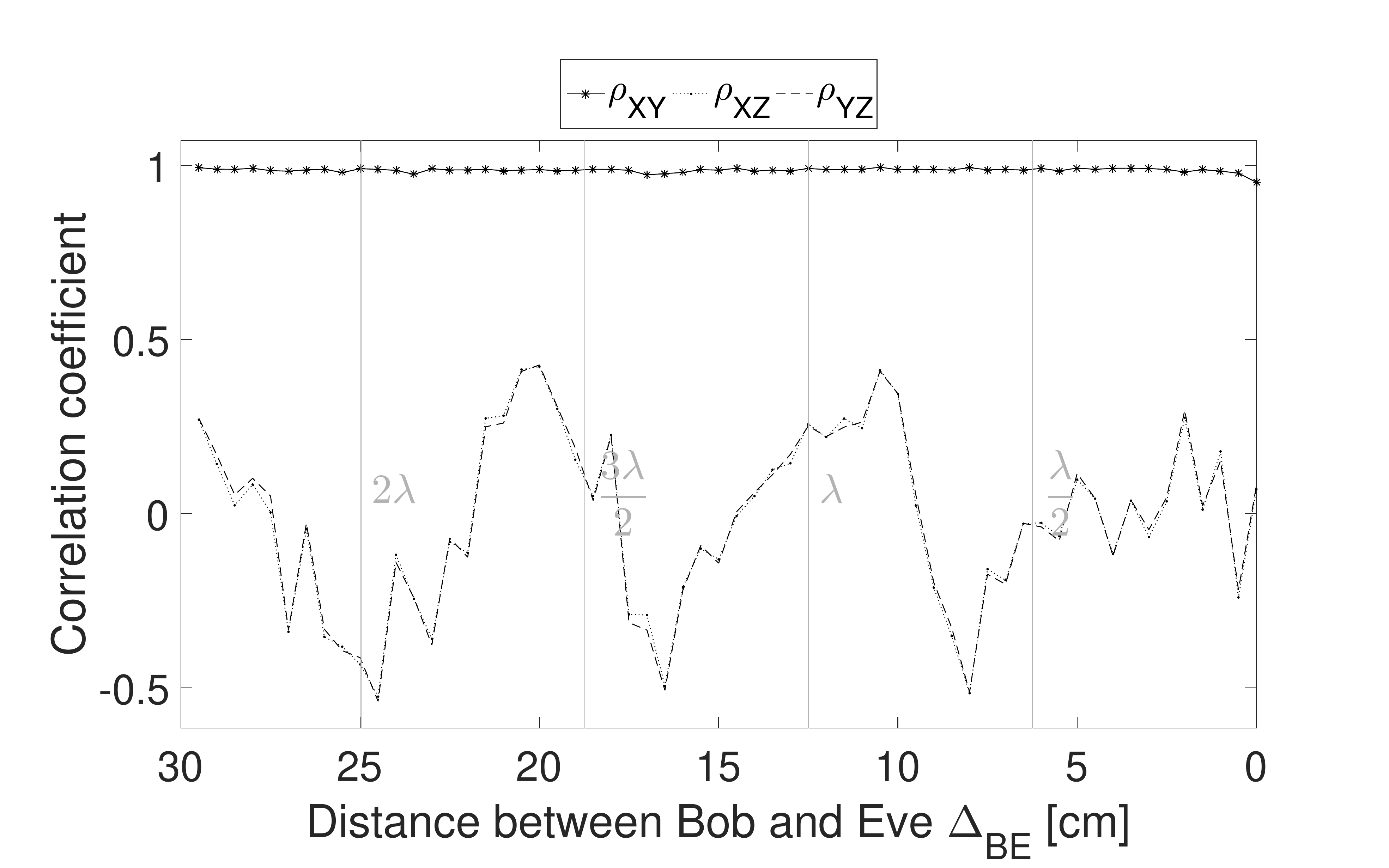}}
	\subfloat[]{\includegraphics[trim=0.5cm 0.1cm 3.5cm 1.6cm, clip=true, height=0.224\textwidth]{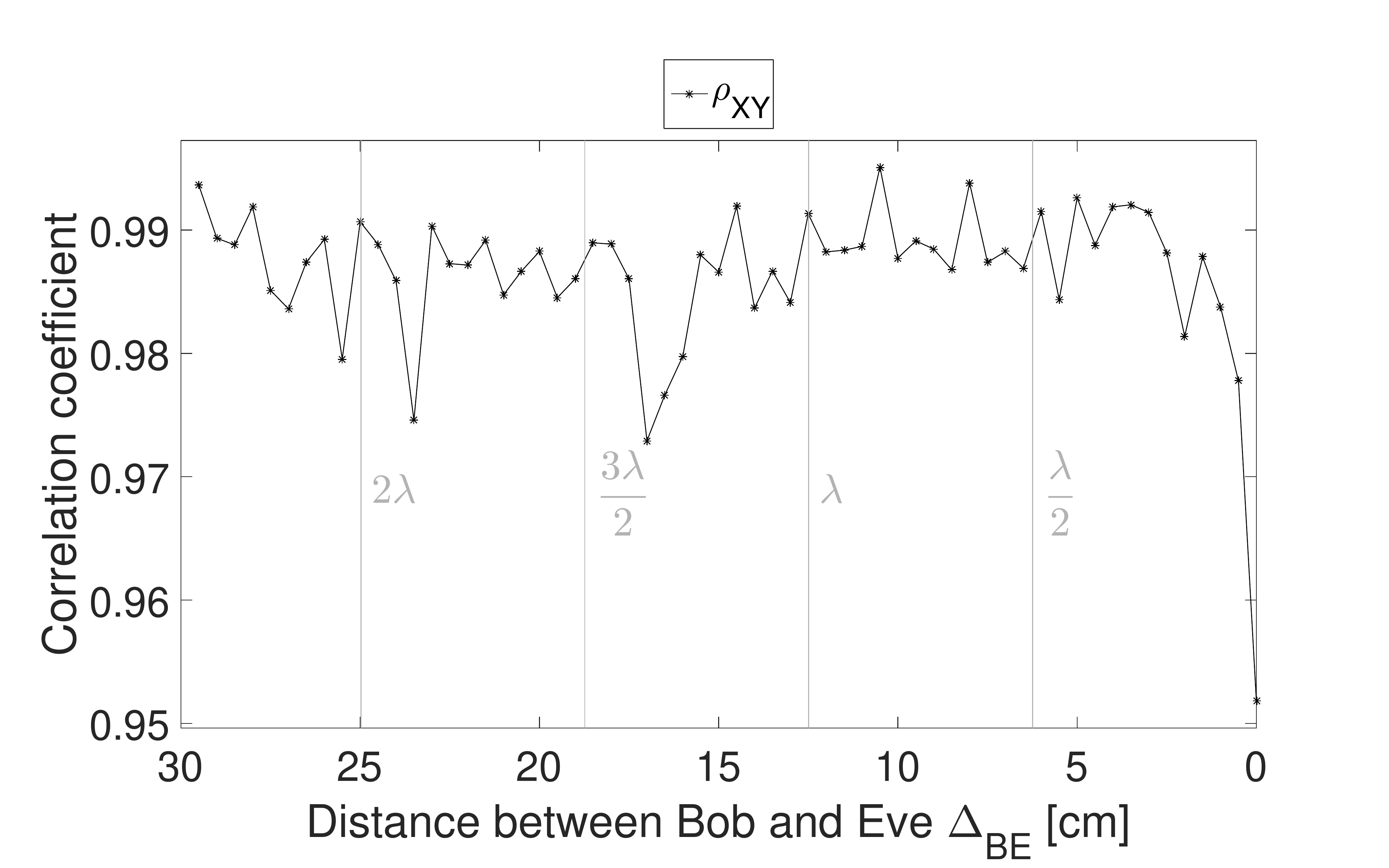}}
	\subfloat[]{\includegraphics[trim=2.2cm 0.1cm 3.5cm 1.6cm, clip=true, height=0.224\textwidth]{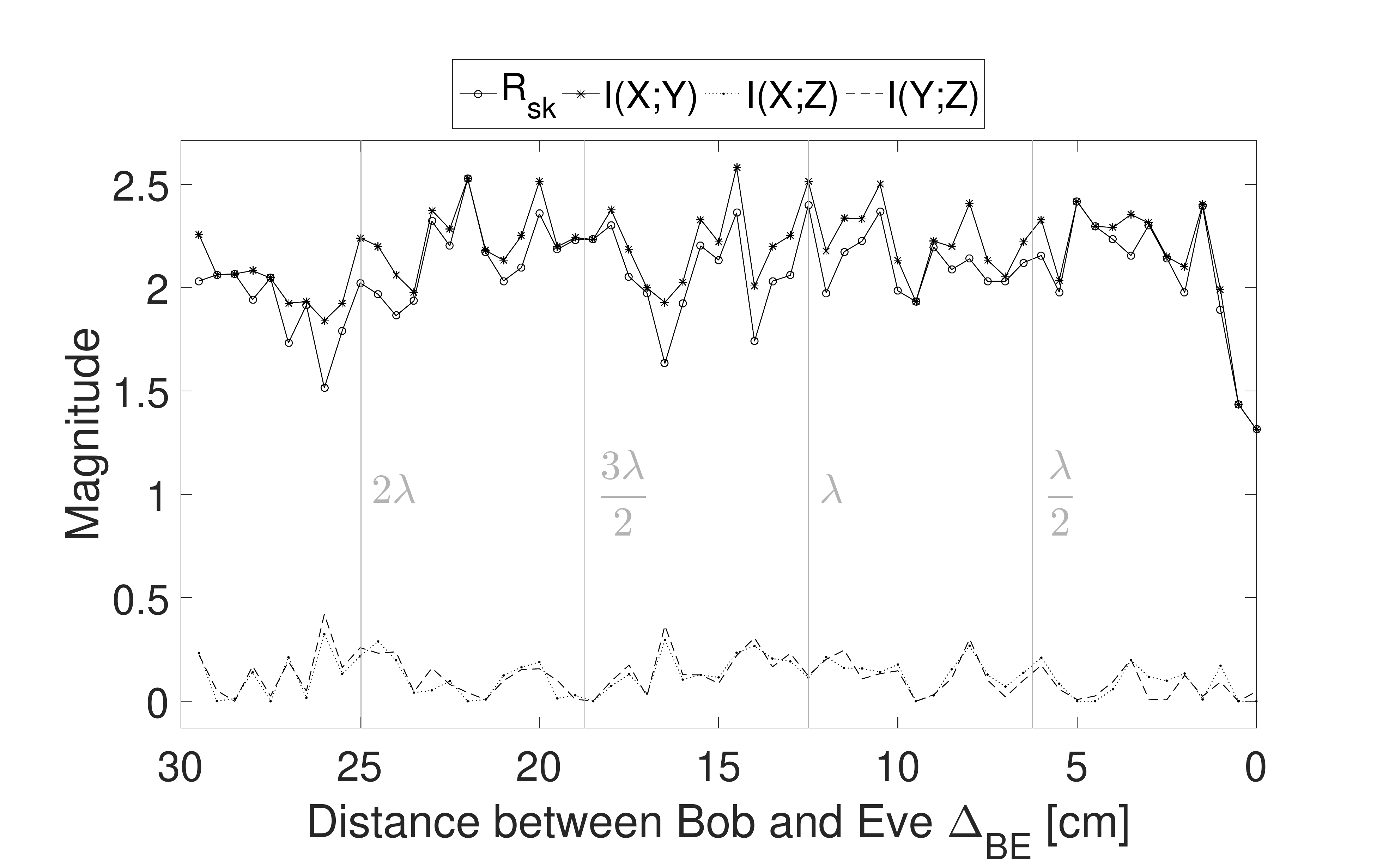}}
	\caption{Evaluation results of $\mybold{v}^{\text{ds}}_k$. In (a) and (b) the cross-correlations is given; in (c) the mutual information as well as $\rsk$ is given. Position 20.}
	\label{fig:app_ds_20}
\end{figure*}

\begin{figure*}
	\centering
	\subfloat[]{\includegraphics[trim=1.4cm 0.1cm 3.5cm 1.6cm, clip=true, height=0.224\textwidth]{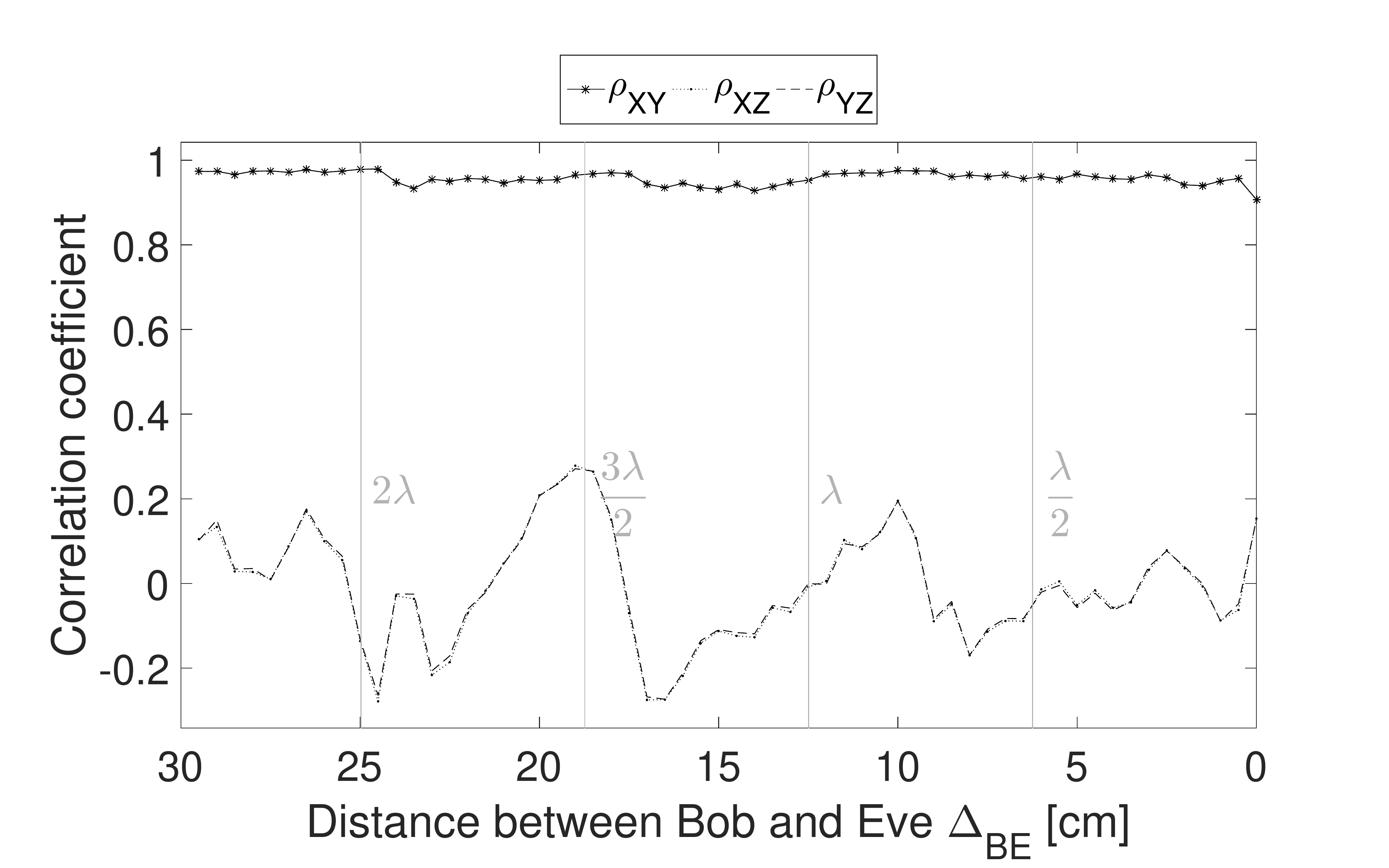}}
	\subfloat[]{\includegraphics[trim=1cm 0.1cm 3.5cm 1.6cm, clip=true, height=0.224\textwidth]{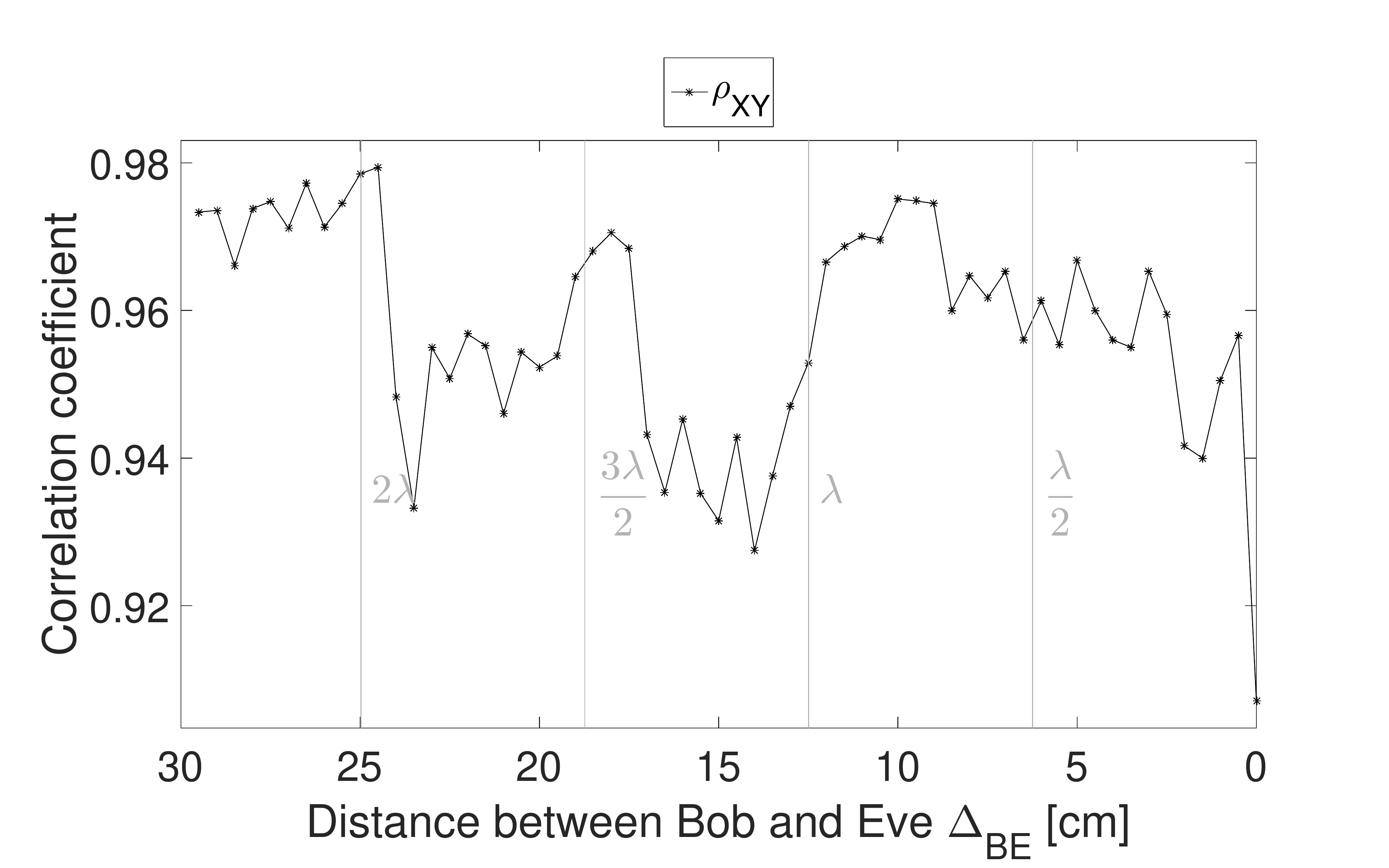}}
	\subfloat[]{\includegraphics[trim=1.8cm 0.1cm 3.5cm 1.6cm, clip=true, height=0.224\textwidth]{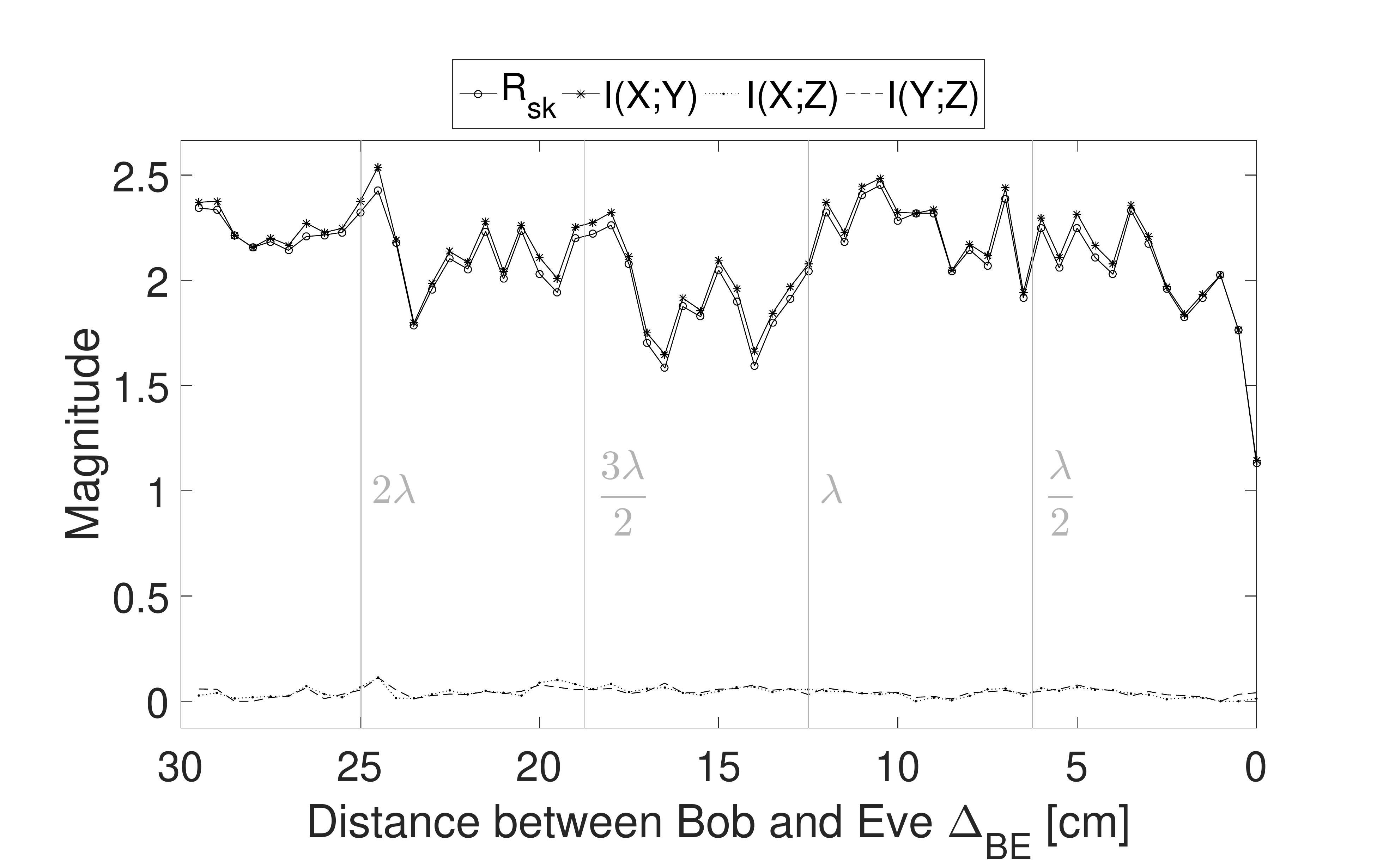}}
	\caption{Evaluation results of $\mybold{v}^{\text{de}}_k$. In (a) and (b) the cross-correlations is given; in (c) the mutual information as well as $\rsk$ is given. Position 20.}
	\label{fig:app_decorr_20}
\end{figure*}


\begin{figure*}
	\centering
	\subfloat[]{\includegraphics[trim=1.4cm 0.1cm 3.5cm 1.6cm, clip=true, height=0.224\textwidth]{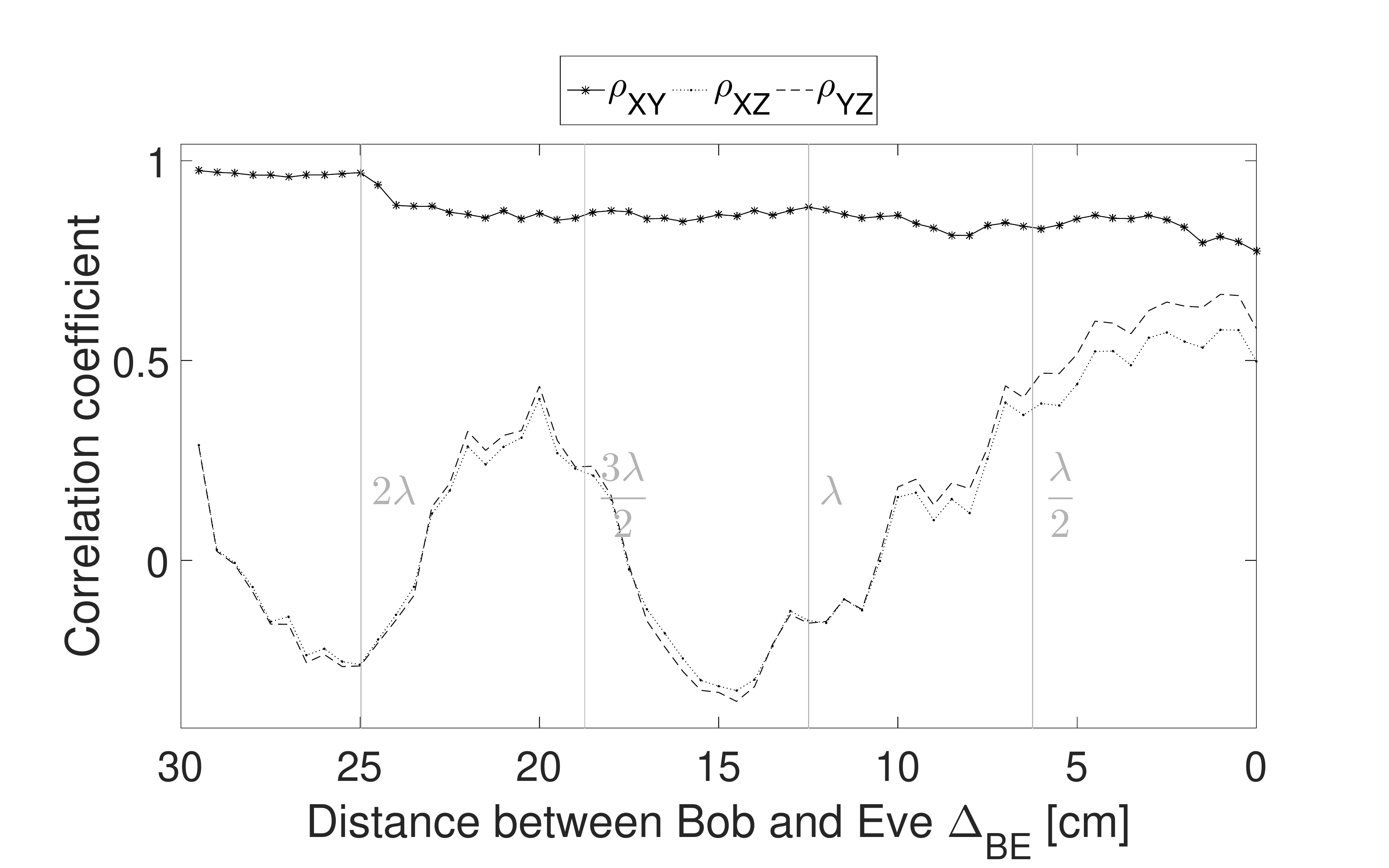}}
	\subfloat[]{\includegraphics[trim=0.5cm 0.1cm 3.5cm 1.6cm, clip=true, height=0.224\textwidth]{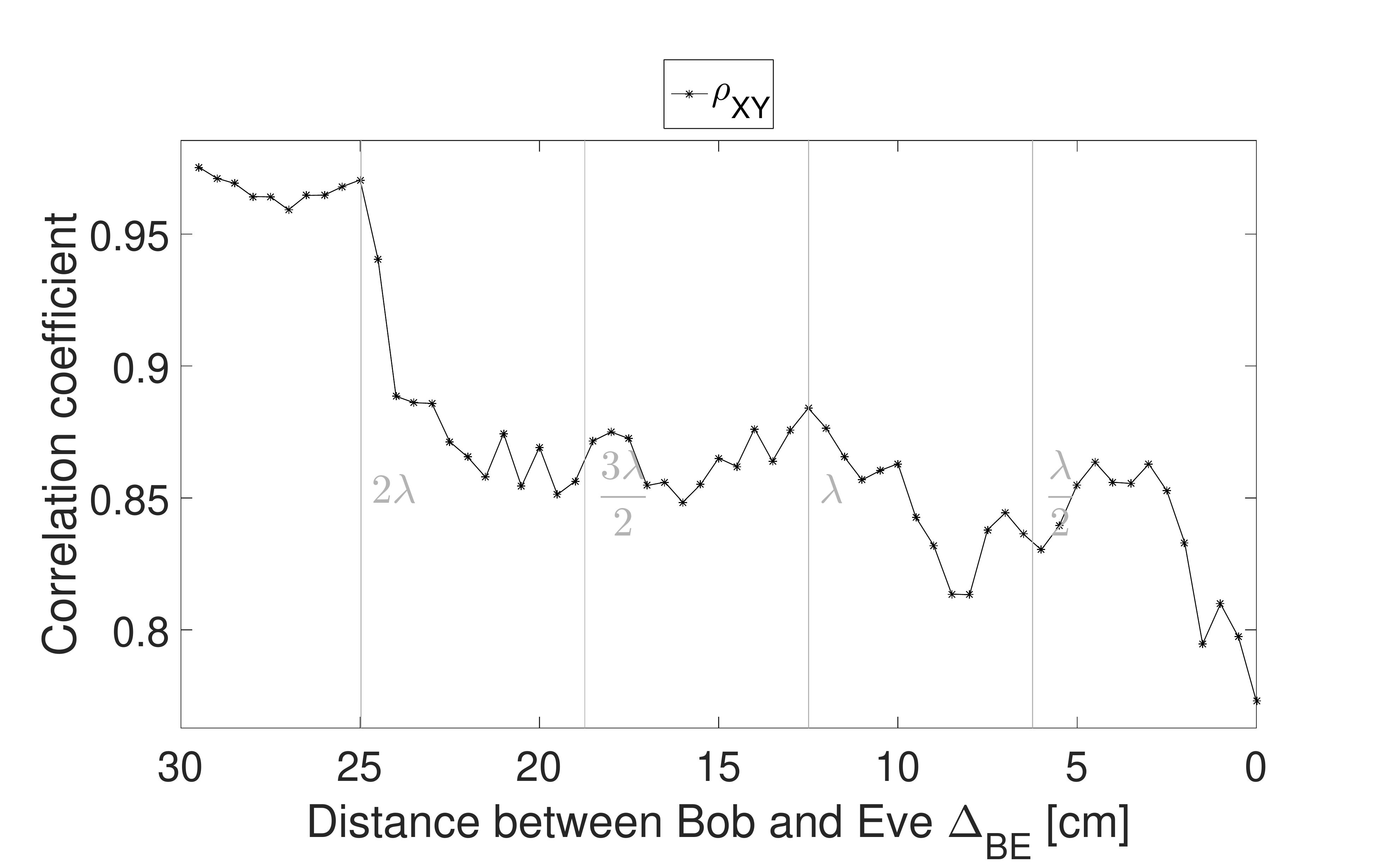}}
	\subfloat[]{\includegraphics[trim=2.2cm 0.1cm 3.5cm 1.6cm, clip=true, height=0.224\textwidth]{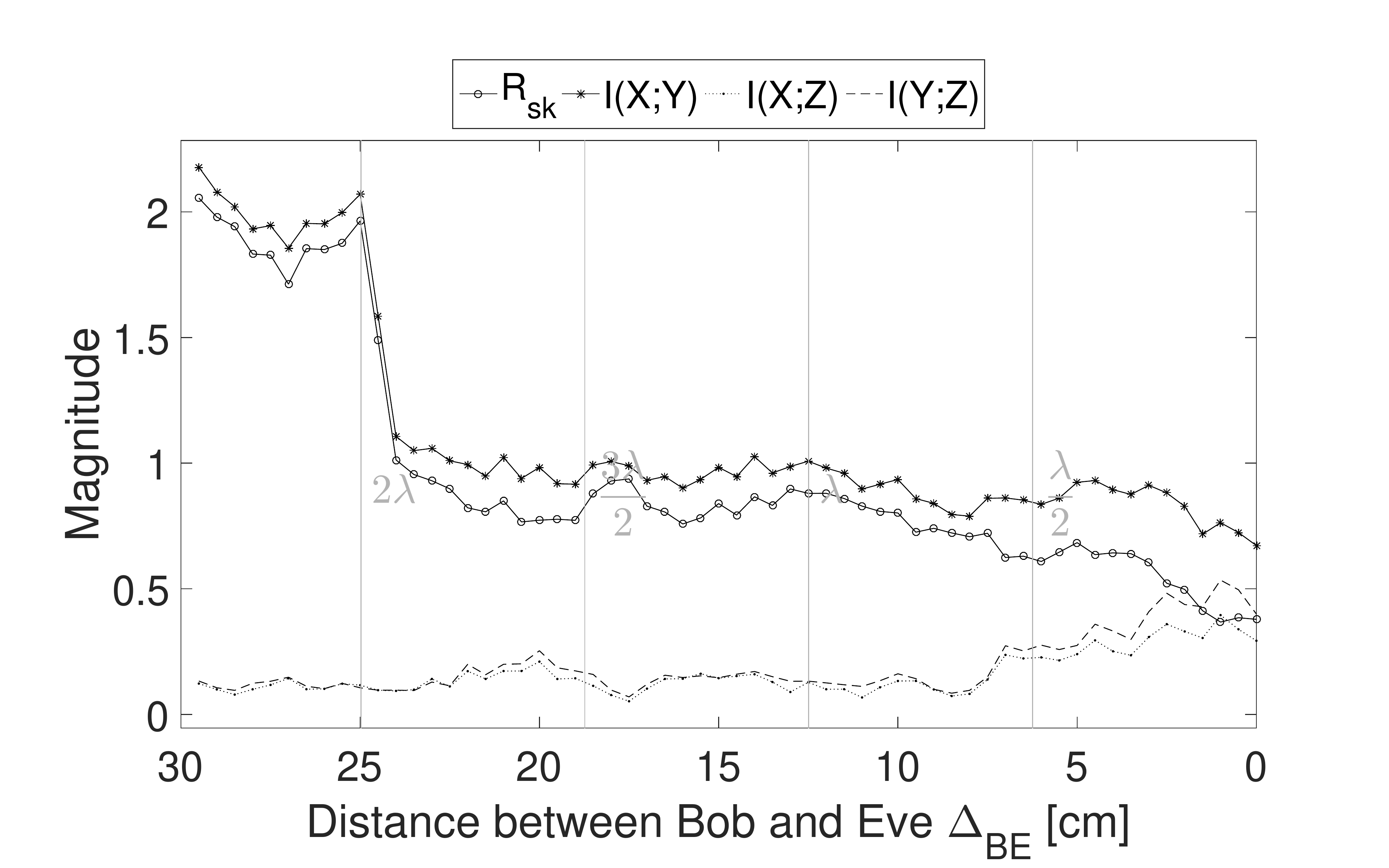}}
	\caption{Evaluation results of $\mybold{v}_k$. In (a) and (b) the cross-correlations is given; in (c) the mutual information as well as $\rsk$ is given. Position 21.}
	\label{fig:app_original_21}
\end{figure*}

\begin{figure*}
	\centering
	\subfloat[]{\includegraphics[trim=1.4cm 0.1cm 3.5cm 1.6cm, clip=true, height=0.224\textwidth]{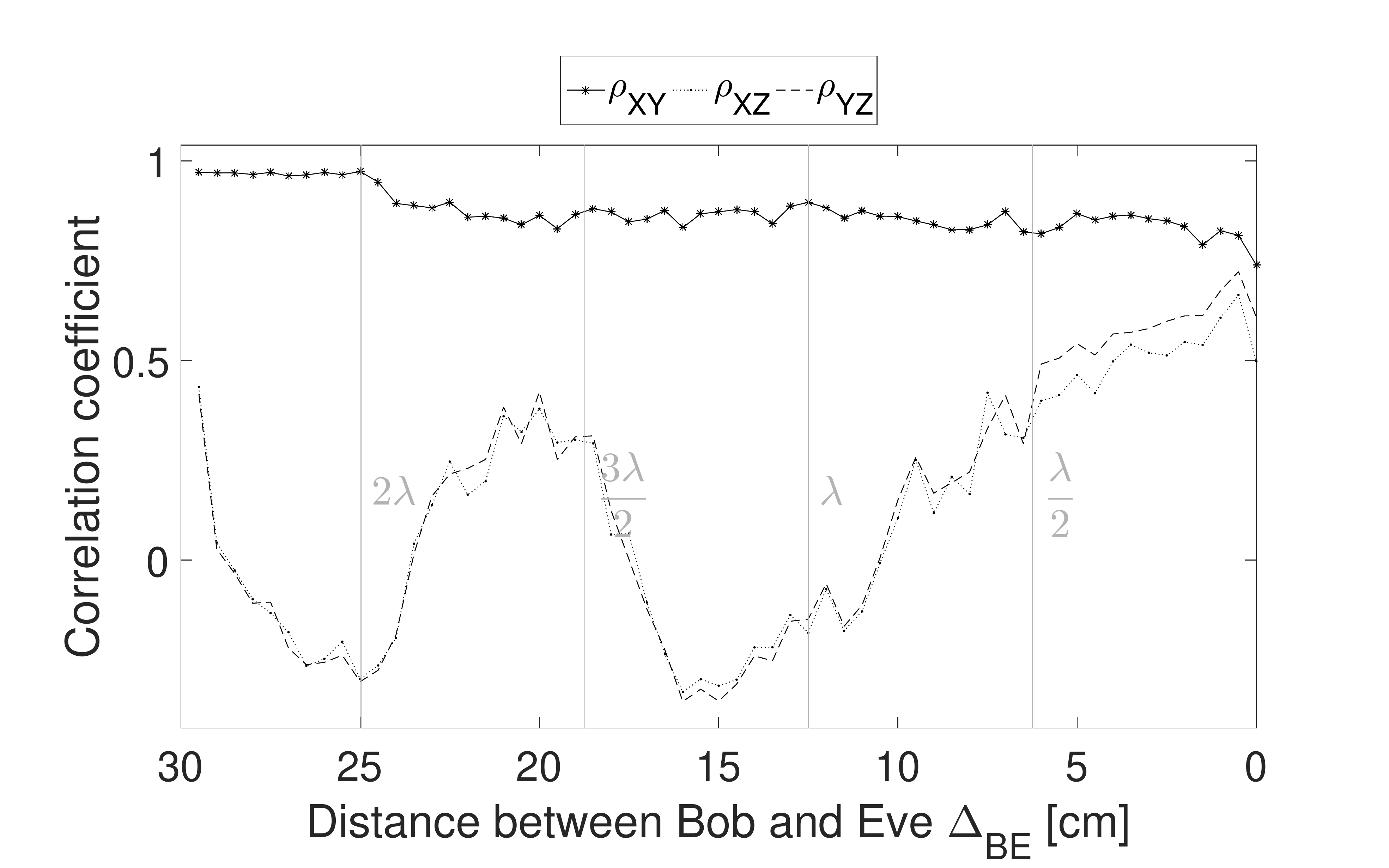}}
	\subfloat[]{\includegraphics[trim=0.5cm 0.1cm 3.5cm 1.6cm, clip=true, height=0.224\textwidth]{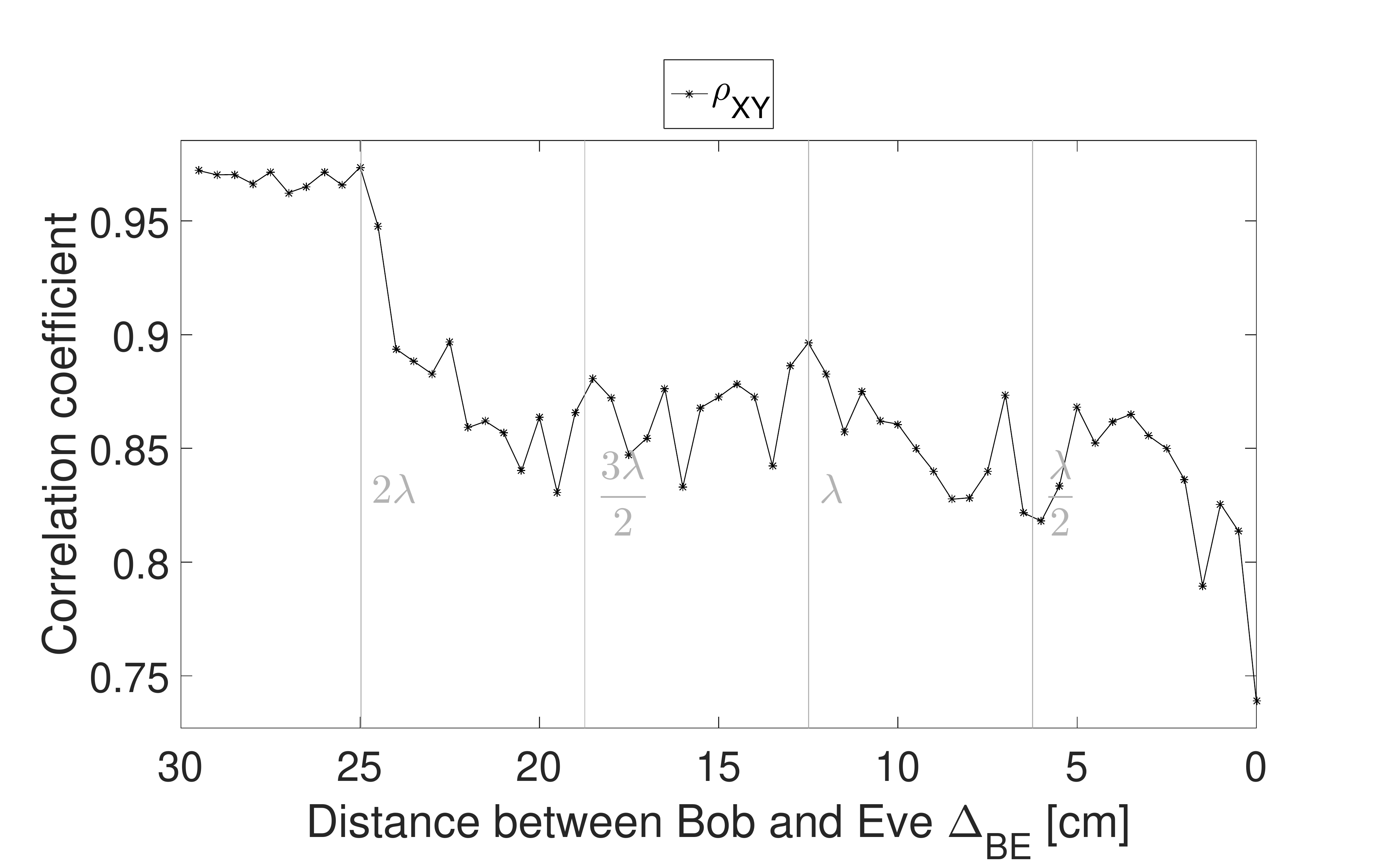}}
	\subfloat[]{\includegraphics[trim=2.2cm 0.1cm 3.5cm 1.6cm, clip=true, height=0.224\textwidth]{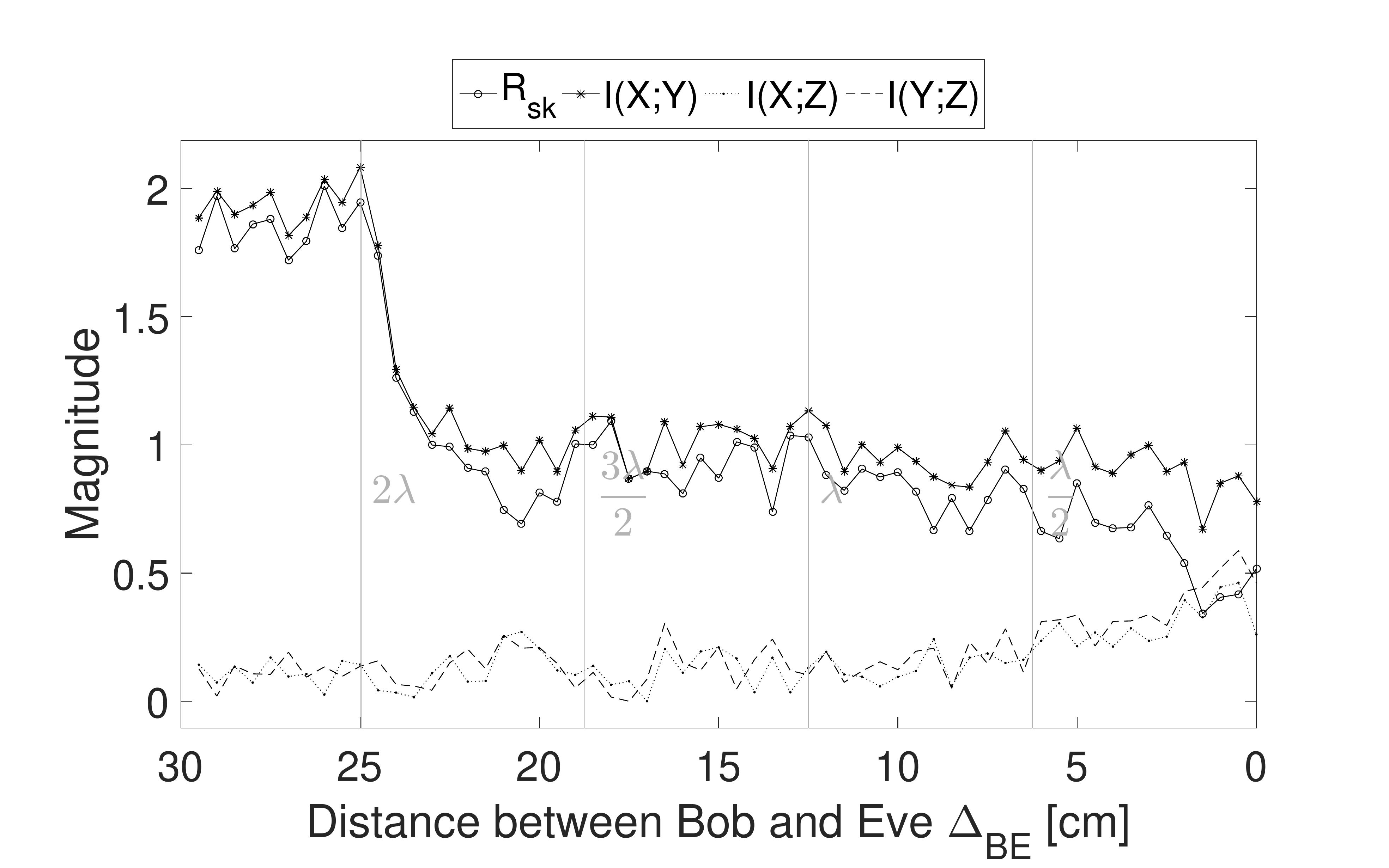}}
	\caption{Evaluation results of $\mybold{v}^{\text{ds}}_k$. In (a) and (b) the cross-correlations is given; in (c) the mutual information as well as $\rsk$ is given. Position 21.}
	\label{fig:app_ds_21}
\end{figure*}

\begin{figure*}
	\centering
	\subfloat[]{\includegraphics[trim=1.4cm 0.1cm 3.5cm 1.6cm, clip=true, height=0.224\textwidth]{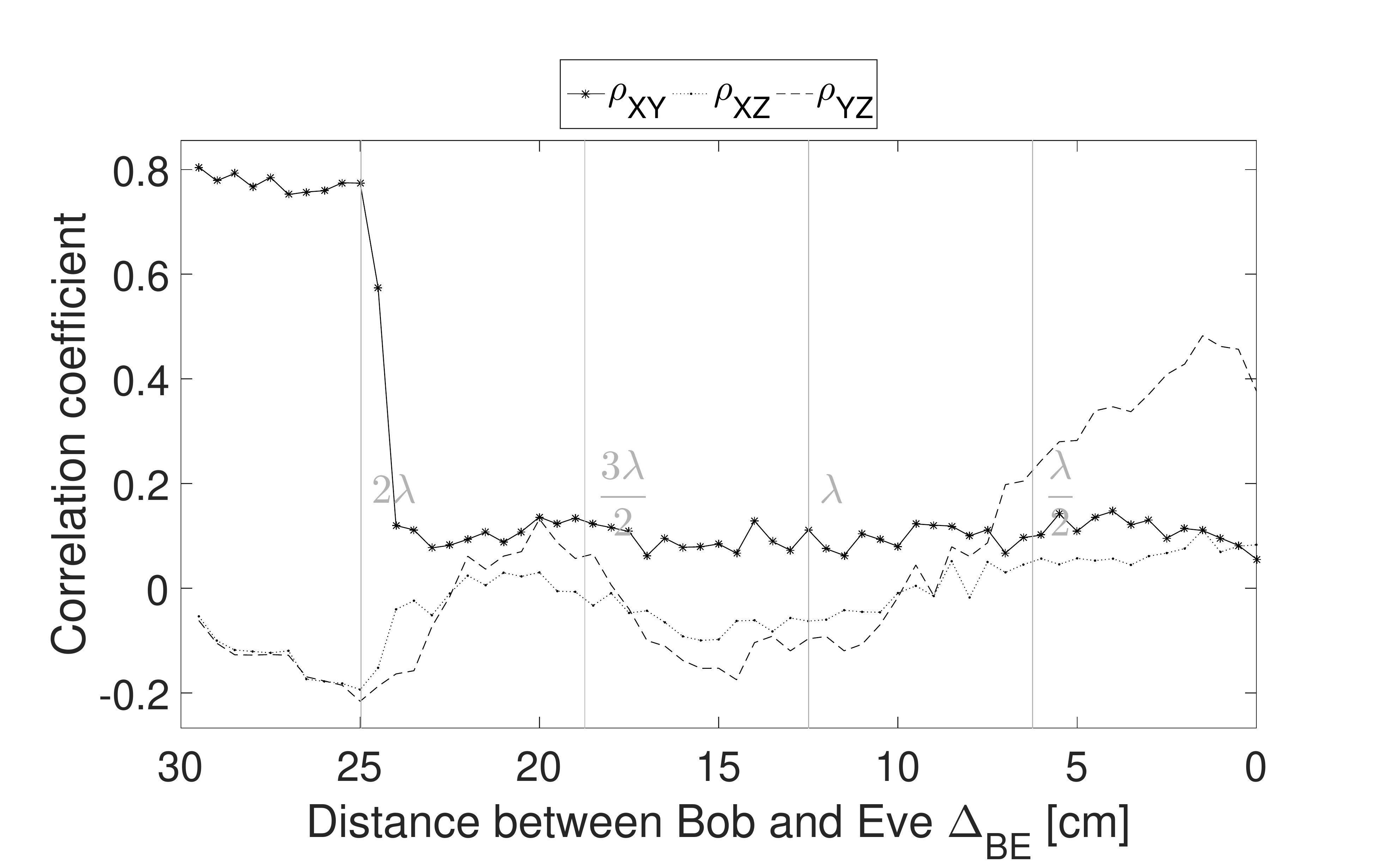}}
	\subfloat[]{\includegraphics[trim=1cm 0.1cm 3.5cm 1.6cm, clip=true, height=0.224\textwidth]{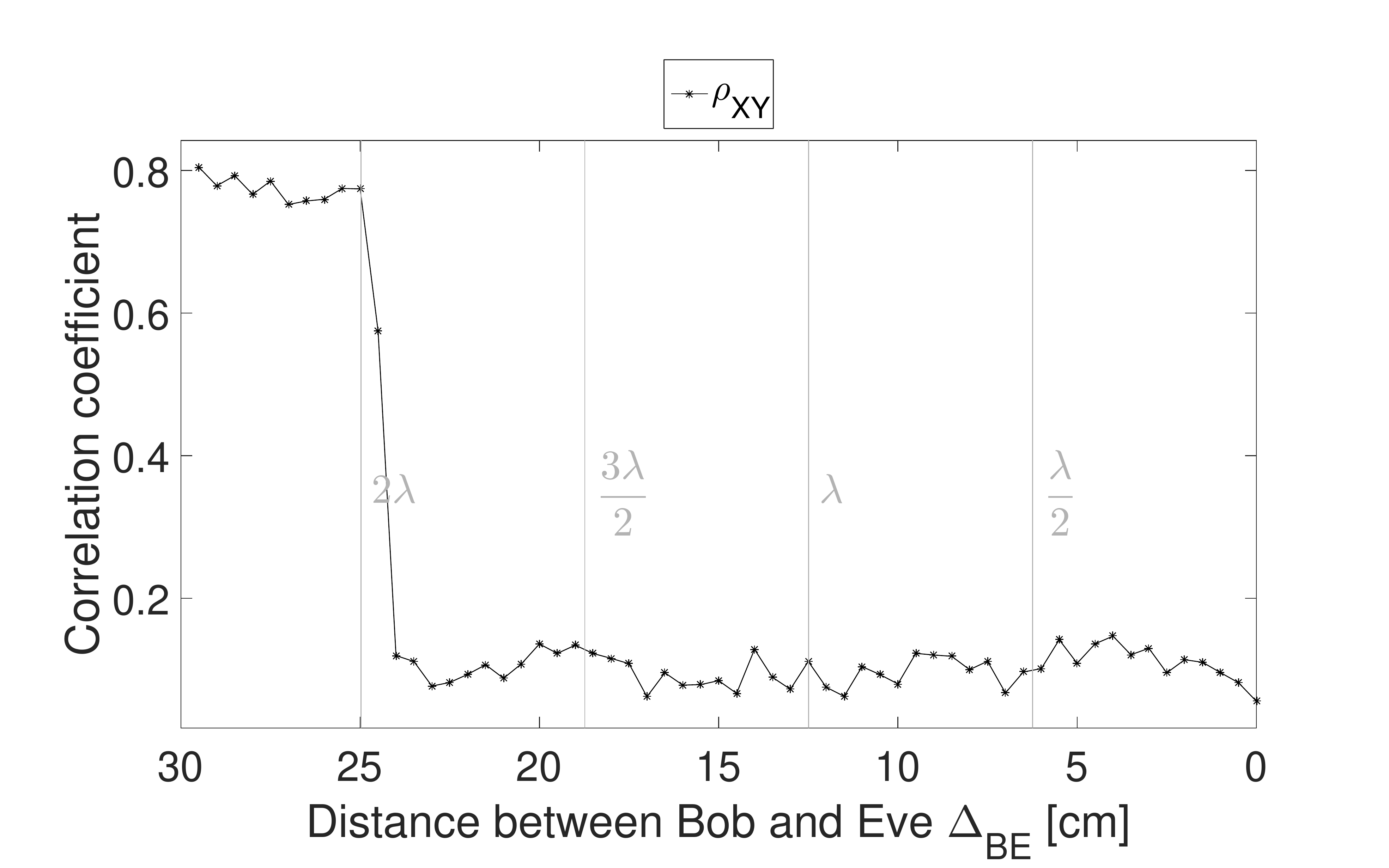}}
	\subfloat[]{\includegraphics[trim=1.8cm 0.1cm 3.5cm 1.6cm, clip=true, height=0.224\textwidth]{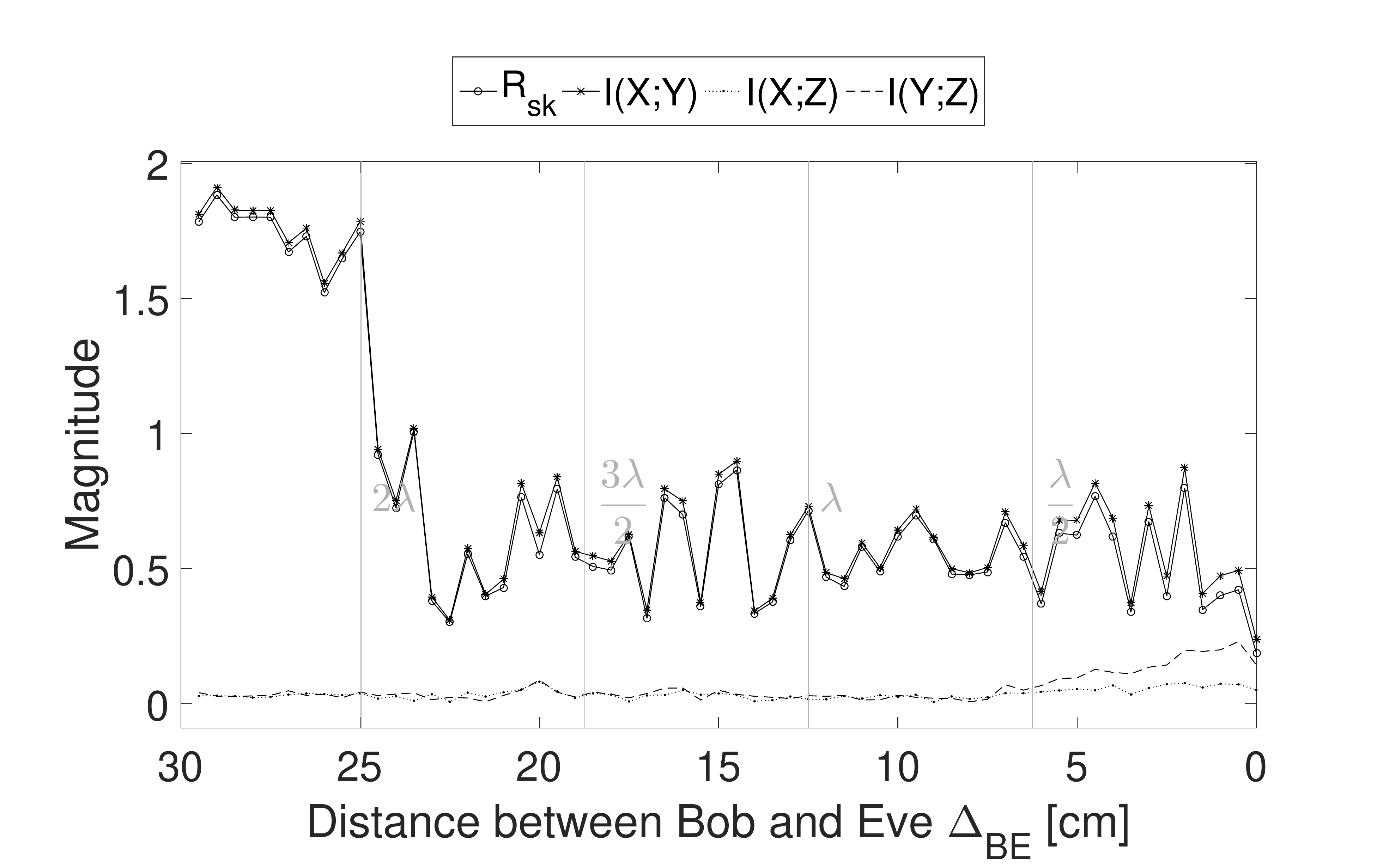}}
	\caption{Evaluation results of $\mybold{v}^{\text{de}}_k$. In (a) and (b) the cross-correlations is given; in (c) the mutual information as well as $\rsk$ is given. Position 21.}
	\label{fig:app_decorr_21}
\end{figure*}


\begin{figure*}
	\centering
	\subfloat[]{\includegraphics[trim=1.4cm 0.1cm 3.5cm 1.6cm, clip=true, height=0.224\textwidth]{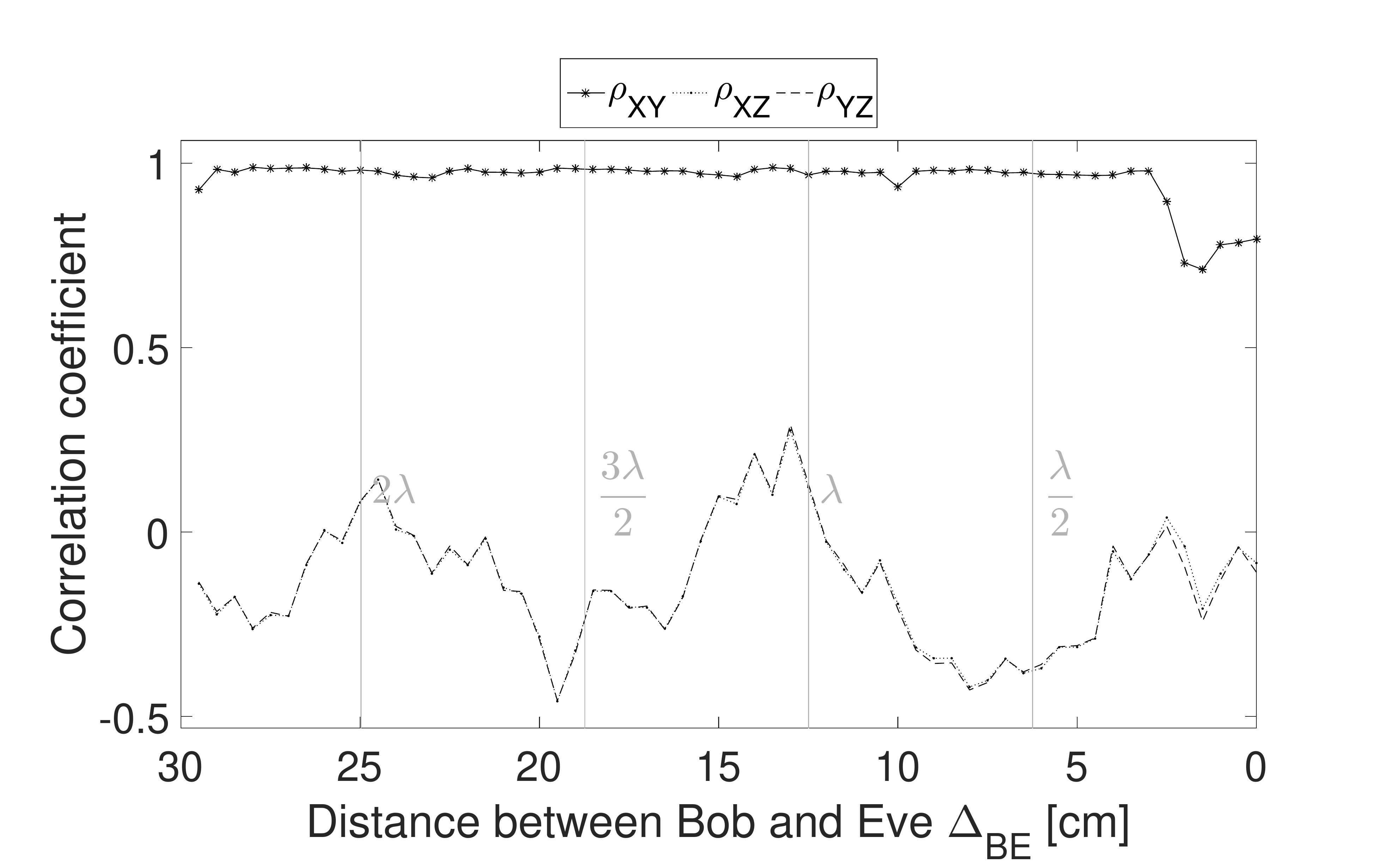}}
	\subfloat[]{\includegraphics[trim=0.5cm 0.1cm 3.5cm 1.6cm, clip=true, height=0.224\textwidth]{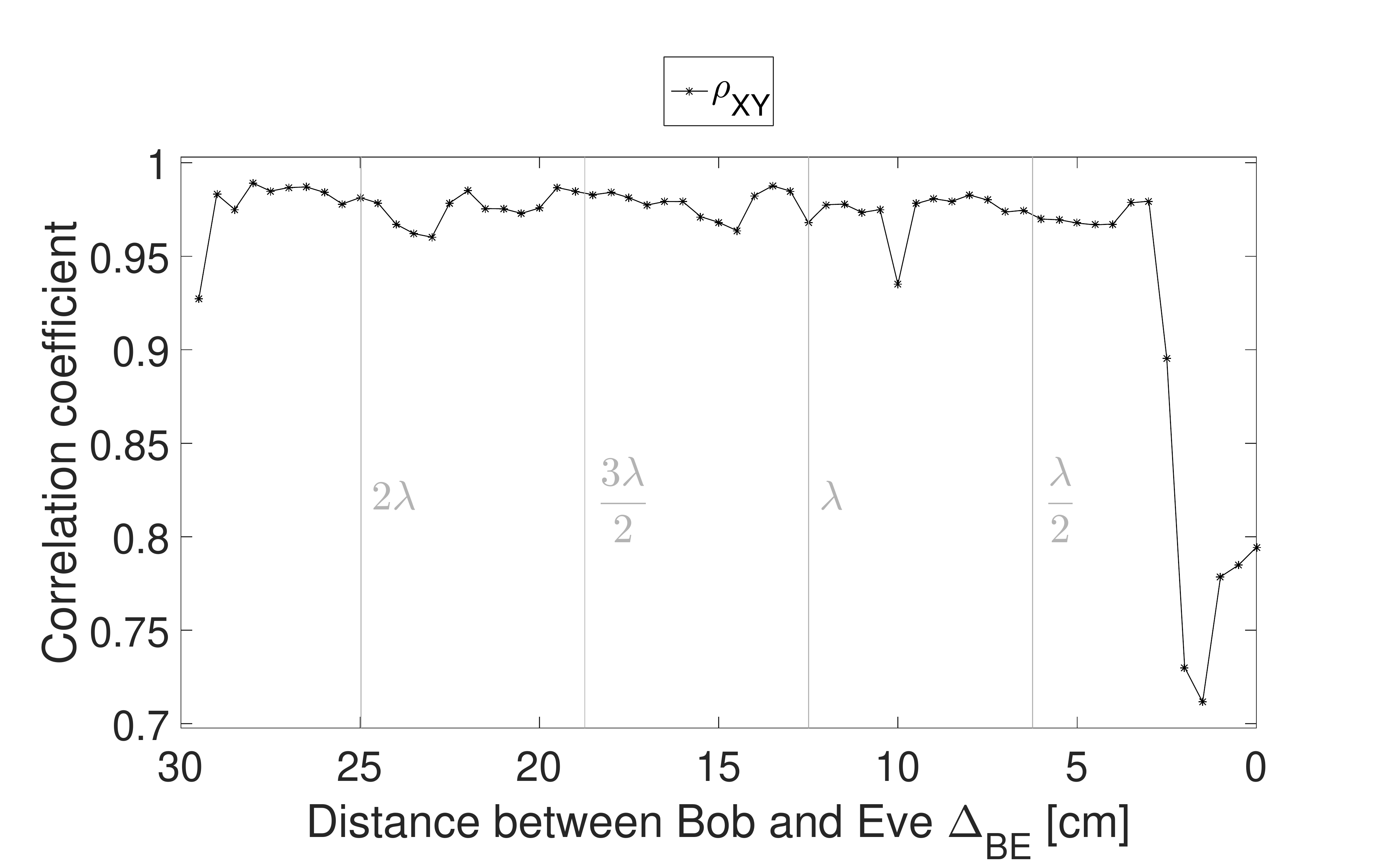}}
	\subfloat[]{\includegraphics[trim=2.2cm 0.1cm 3.5cm 1.6cm, clip=true, height=0.224\textwidth]{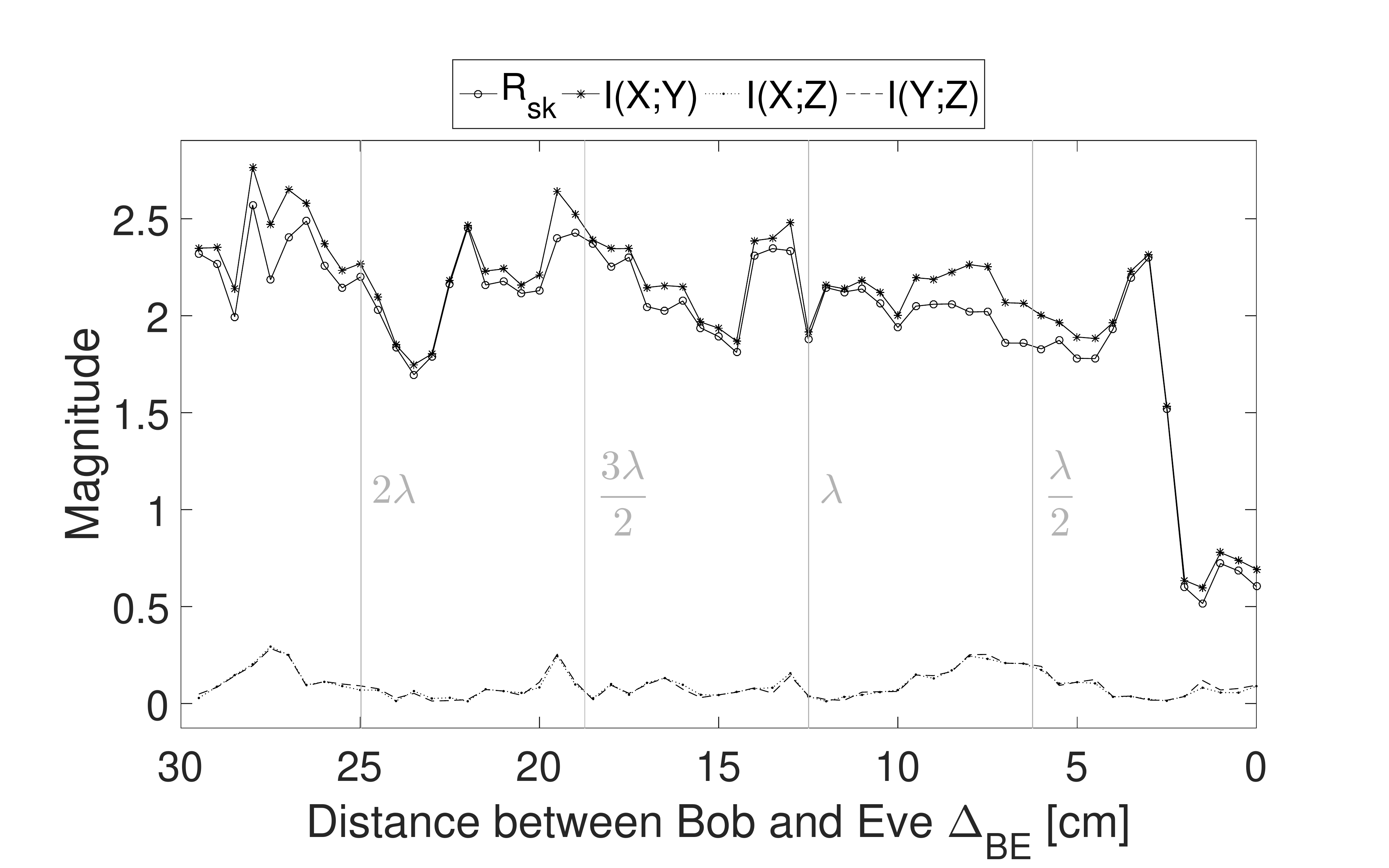}}
	\caption{Evaluation results of $\mybold{v}_k$. In (a) and (b) the cross-correlations is given; in (c) the mutual information as well as $\rsk$ is given. Position 22.}
	\label{fig:app_original_22}
\end{figure*}

\begin{figure*}
	\centering
	\subfloat[]{\includegraphics[trim=1.4cm 0.1cm 3.5cm 1.6cm, clip=true, height=0.224\textwidth]{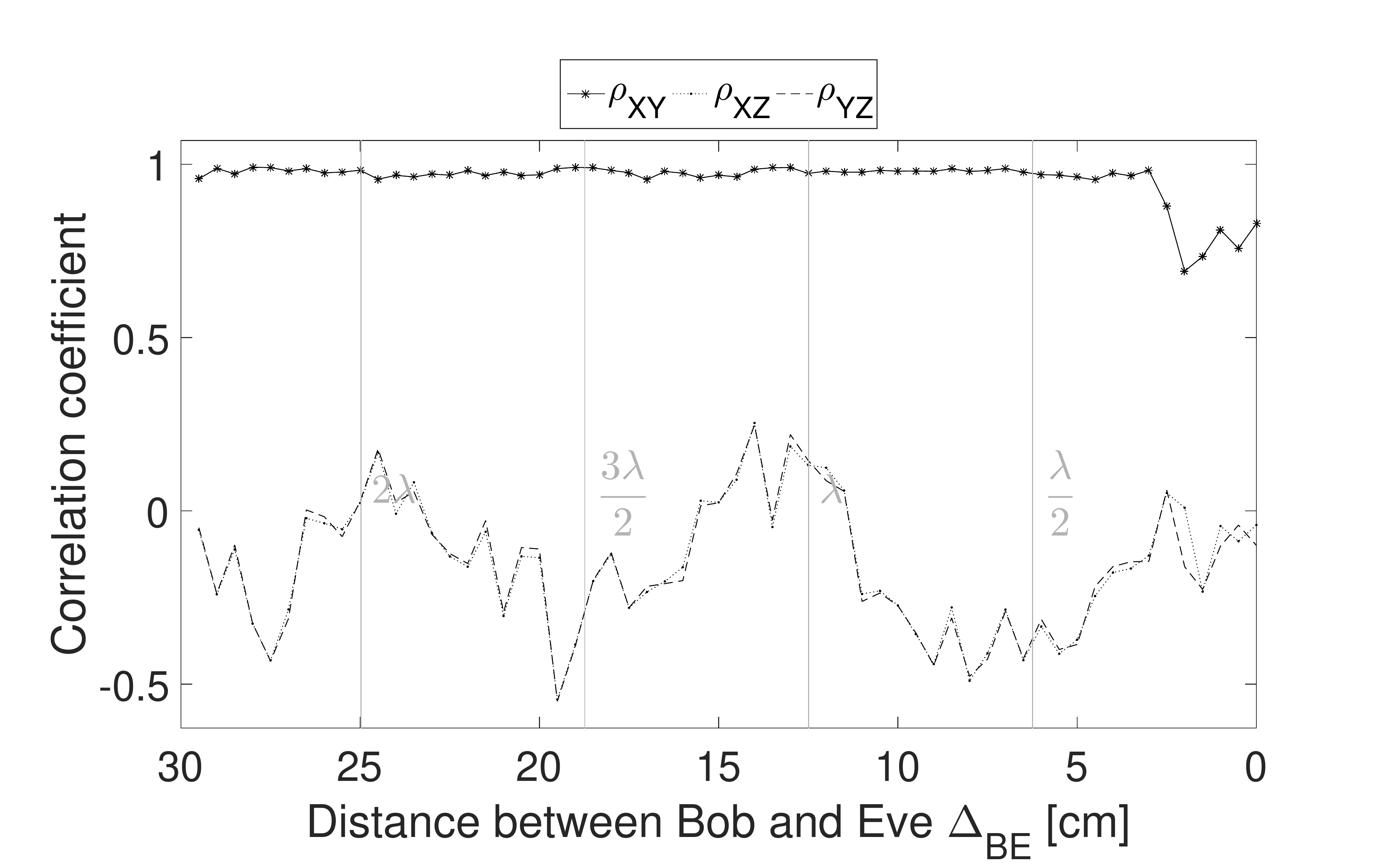}}
	\subfloat[]{\includegraphics[trim=0.5cm 0.1cm 3.5cm 1.6cm, clip=true, height=0.224\textwidth]{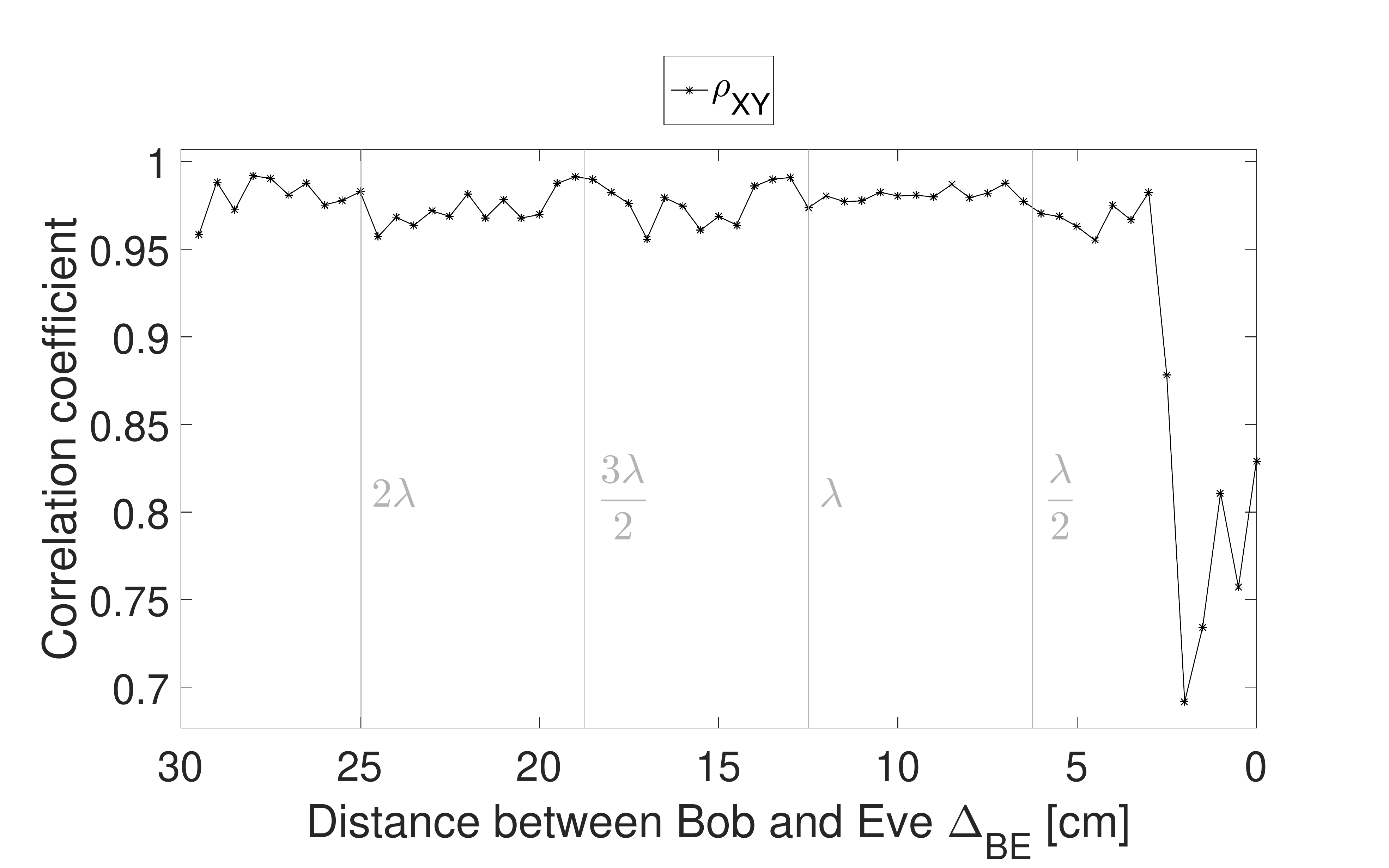}}
	\subfloat[]{\includegraphics[trim=2.2cm 0.1cm 3.5cm 1.6cm, clip=true, height=0.224\textwidth]{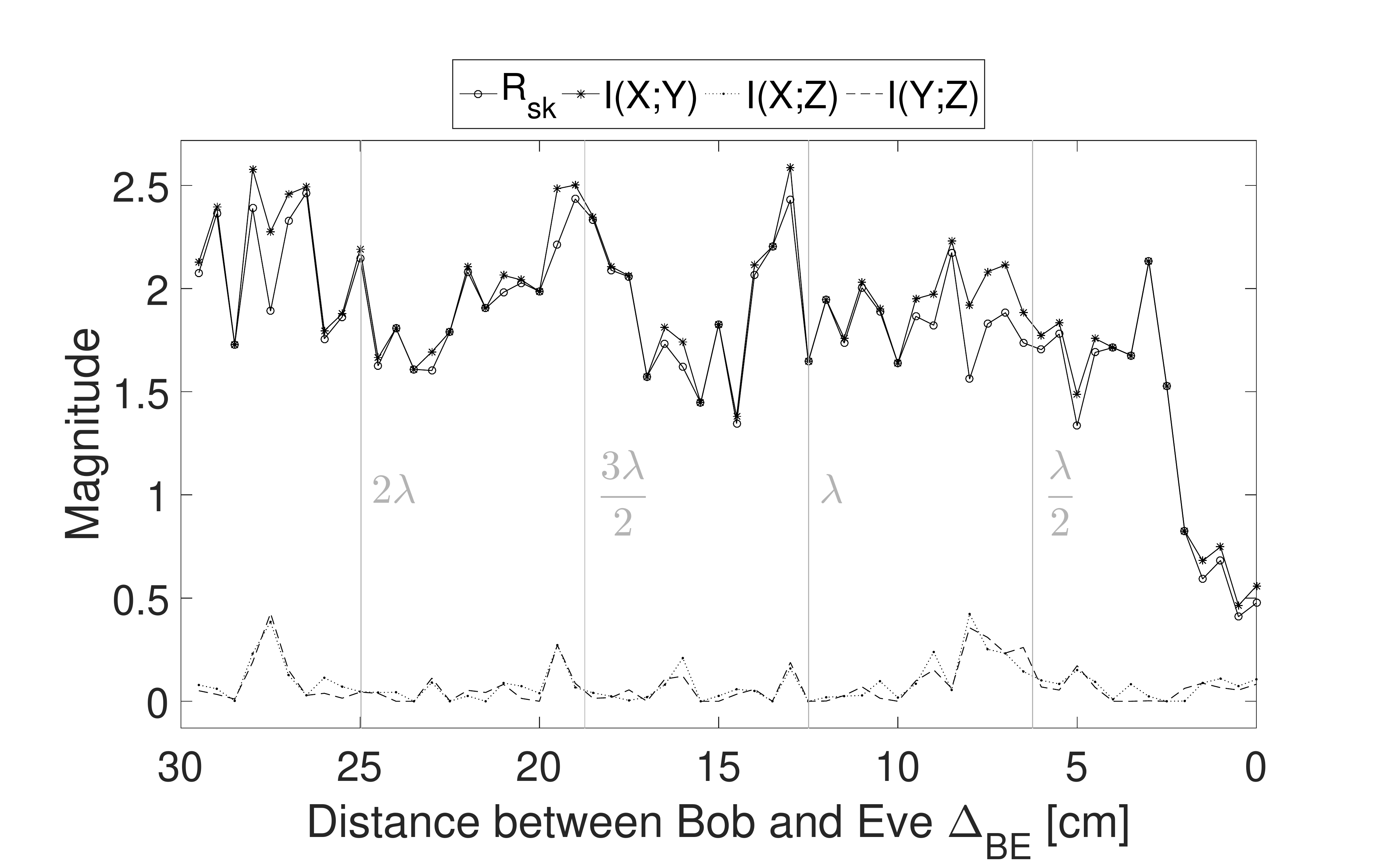}}
	\caption{Evaluation results of $\mybold{v}^{\text{ds}}_k$. In (a) and (b) the cross-correlations is given; in (c) the mutual information as well as $\rsk$ is given. Position 22.}
	\label{fig:app_ds_22}
\end{figure*}

\begin{figure*}
	\centering
	\subfloat[]{\includegraphics[trim=1.4cm 0.1cm 3.5cm 1.6cm, clip=true, height=0.224\textwidth]{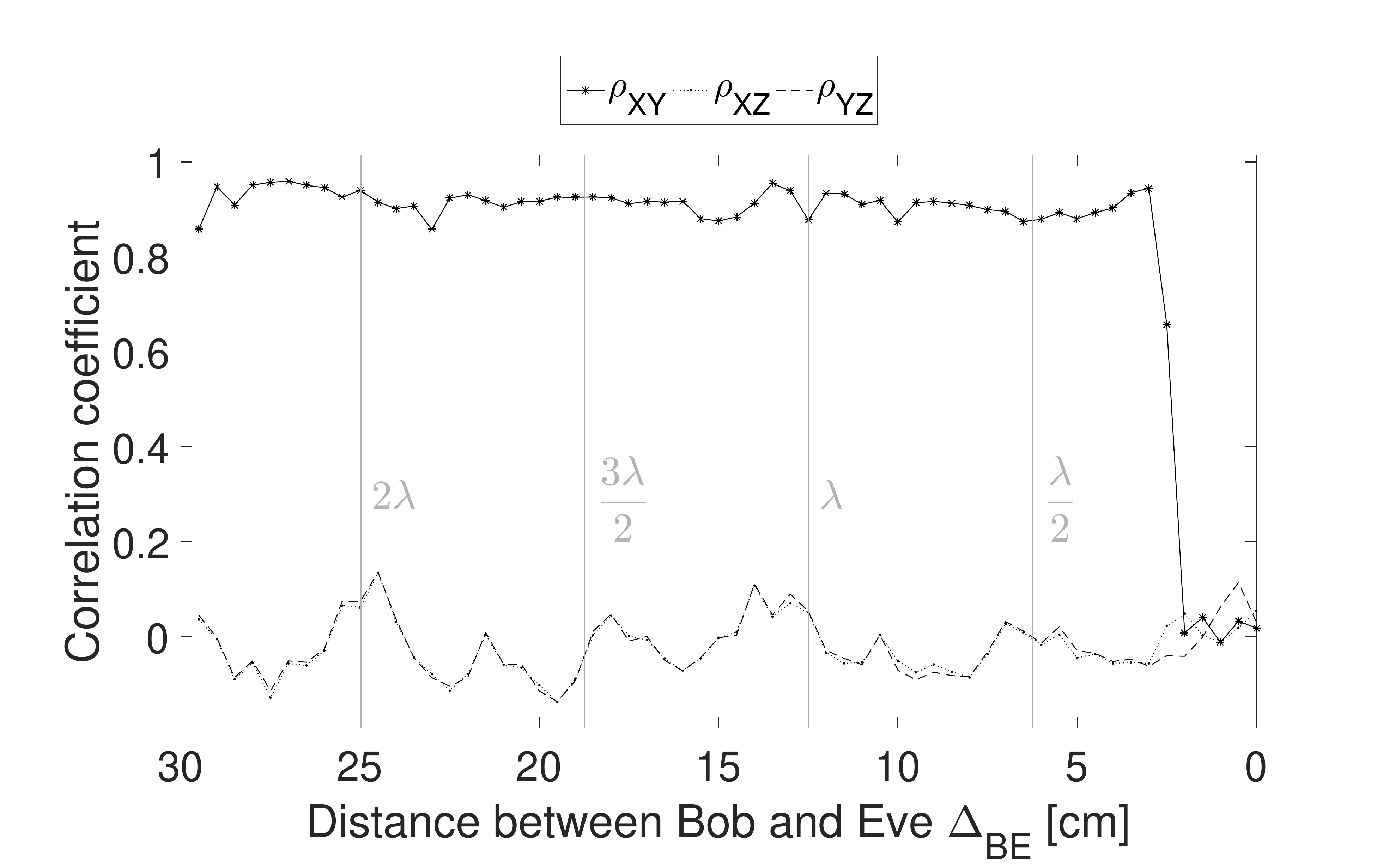}}
	\subfloat[]{\includegraphics[trim=1cm 0.1cm 3.5cm 1.6cm, clip=true, height=0.224\textwidth]{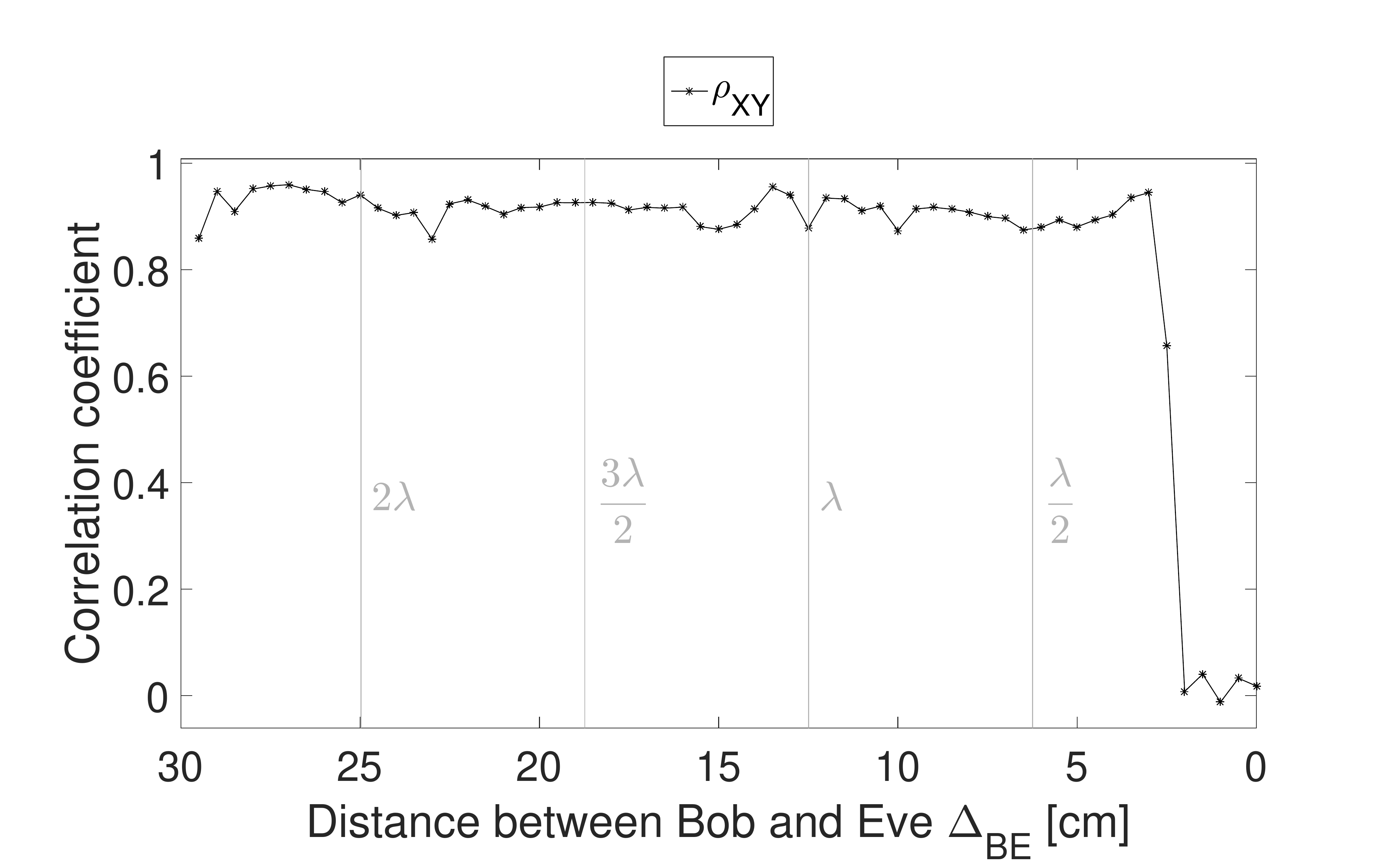}}
	\subfloat[]{\includegraphics[trim=1.8cm 0.1cm 3.5cm 1.6cm, clip=true, height=0.224\textwidth]{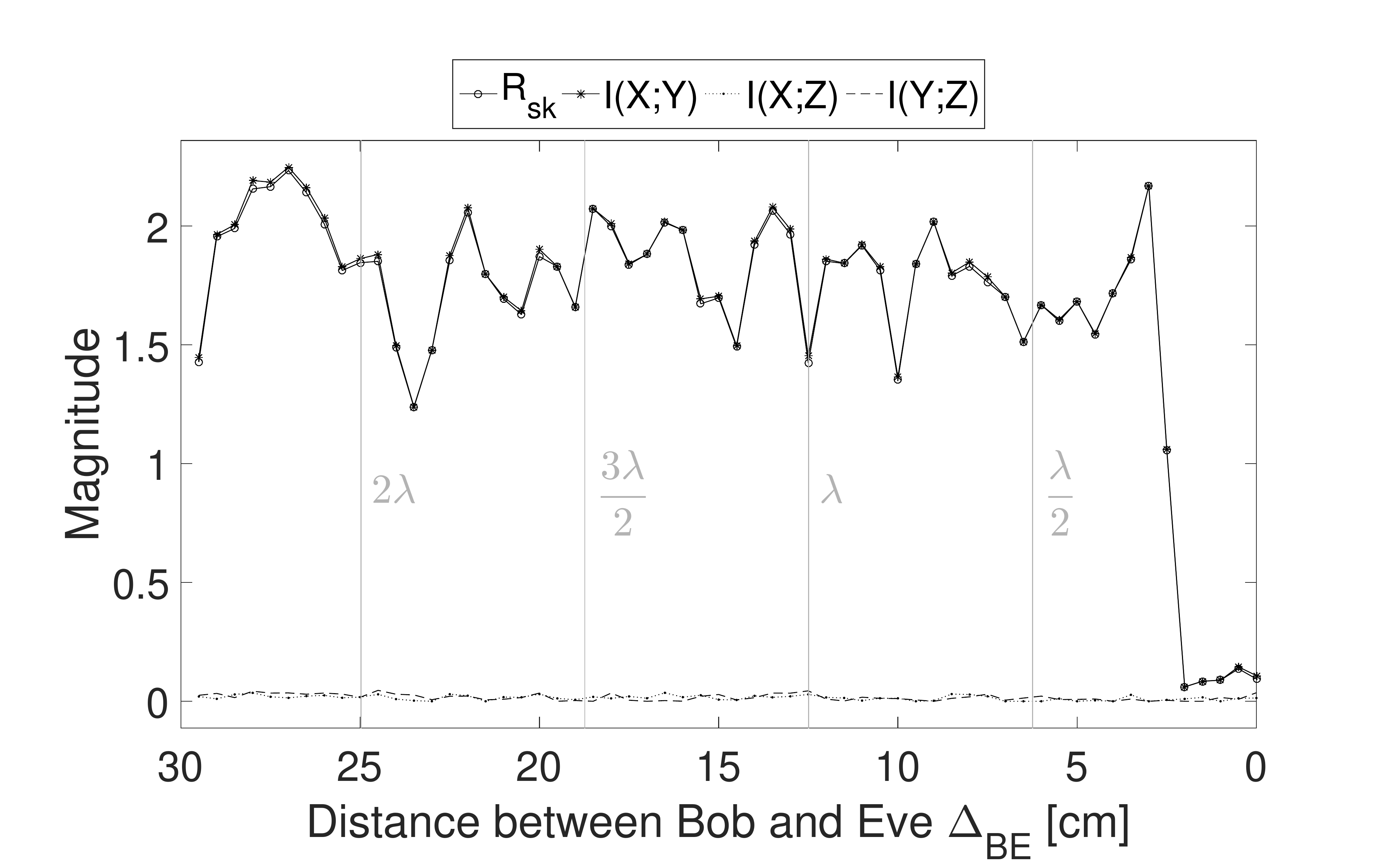}}
	\caption{Evaluation results of $\mybold{v}^{\text{de}}_k$. In (a) and (b) the cross-correlations is given; in (c) the mutual information as well as $\rsk$ is given. Position 22.}
	\label{fig:app_decorr_22}
\end{figure*}


\begin{figure*}
	\centering
	\subfloat[]{\includegraphics[trim=1.4cm 0.1cm 3.5cm 1.6cm, clip=true, height=0.224\textwidth]{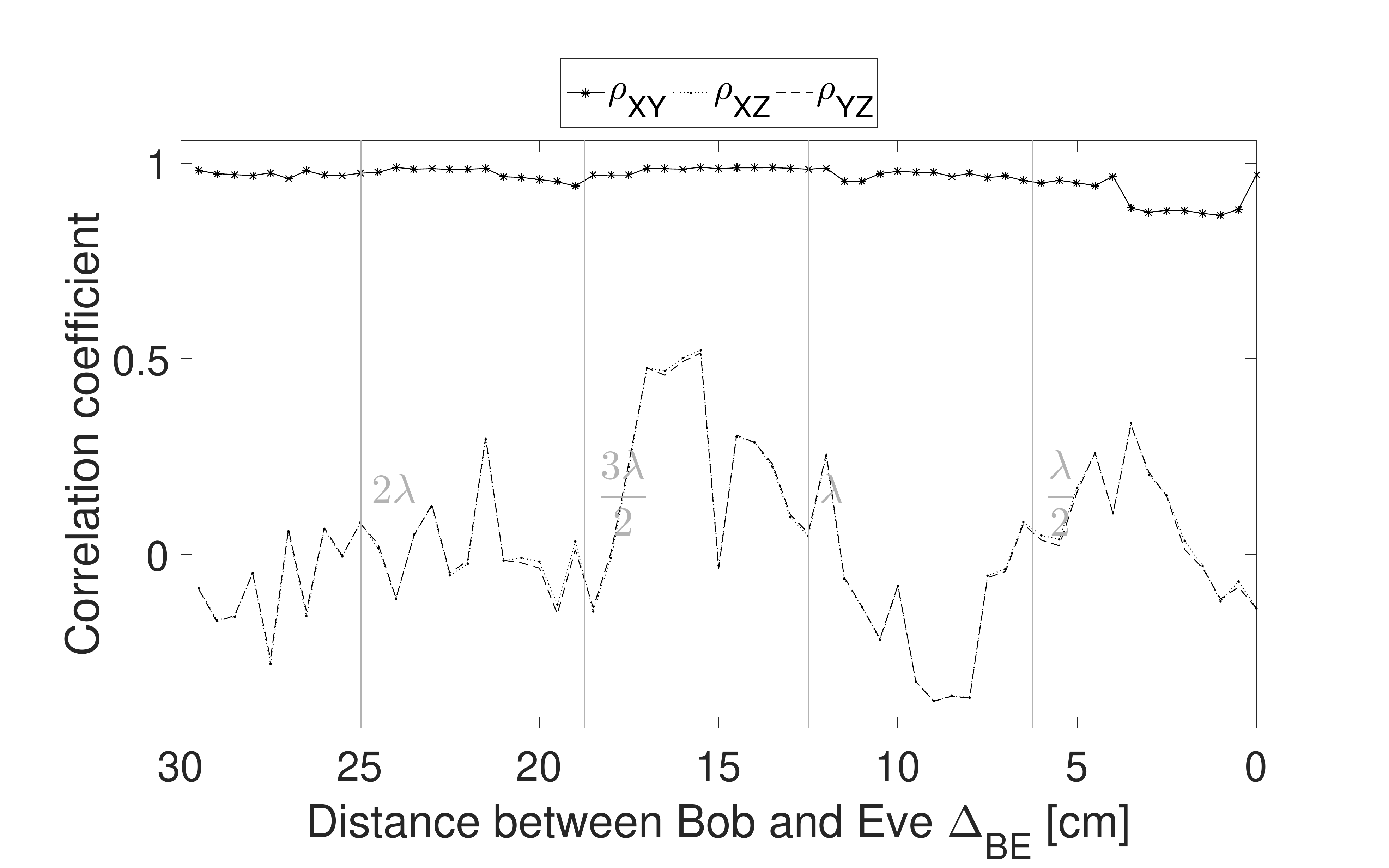}}
	\subfloat[]{\includegraphics[trim=0.5cm 0.1cm 3.5cm 1.6cm, clip=true, height=0.224\textwidth]{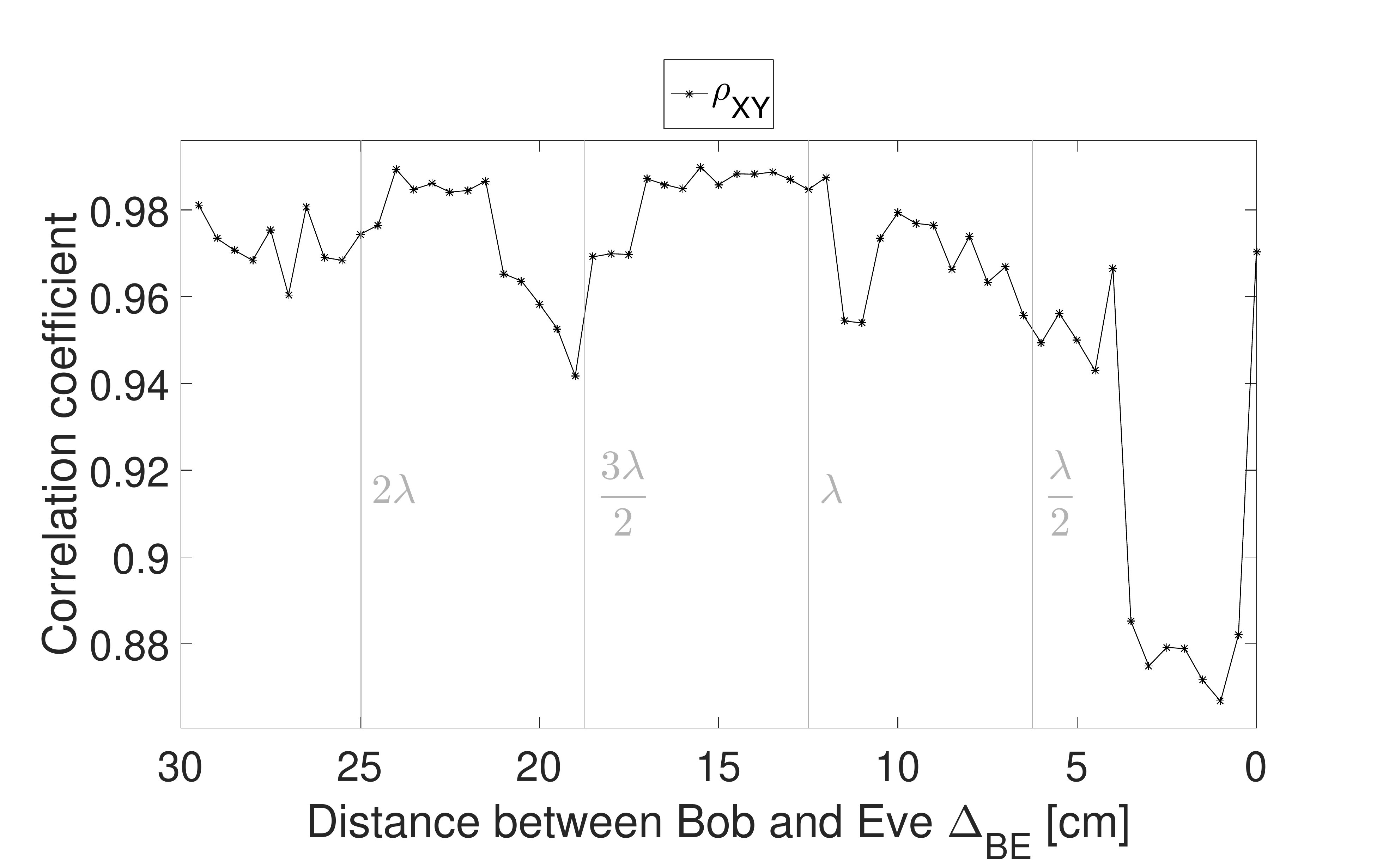}}
	\subfloat[]{\includegraphics[trim=2.2cm 0.1cm 3.5cm 1.6cm, clip=true, height=0.224\textwidth]{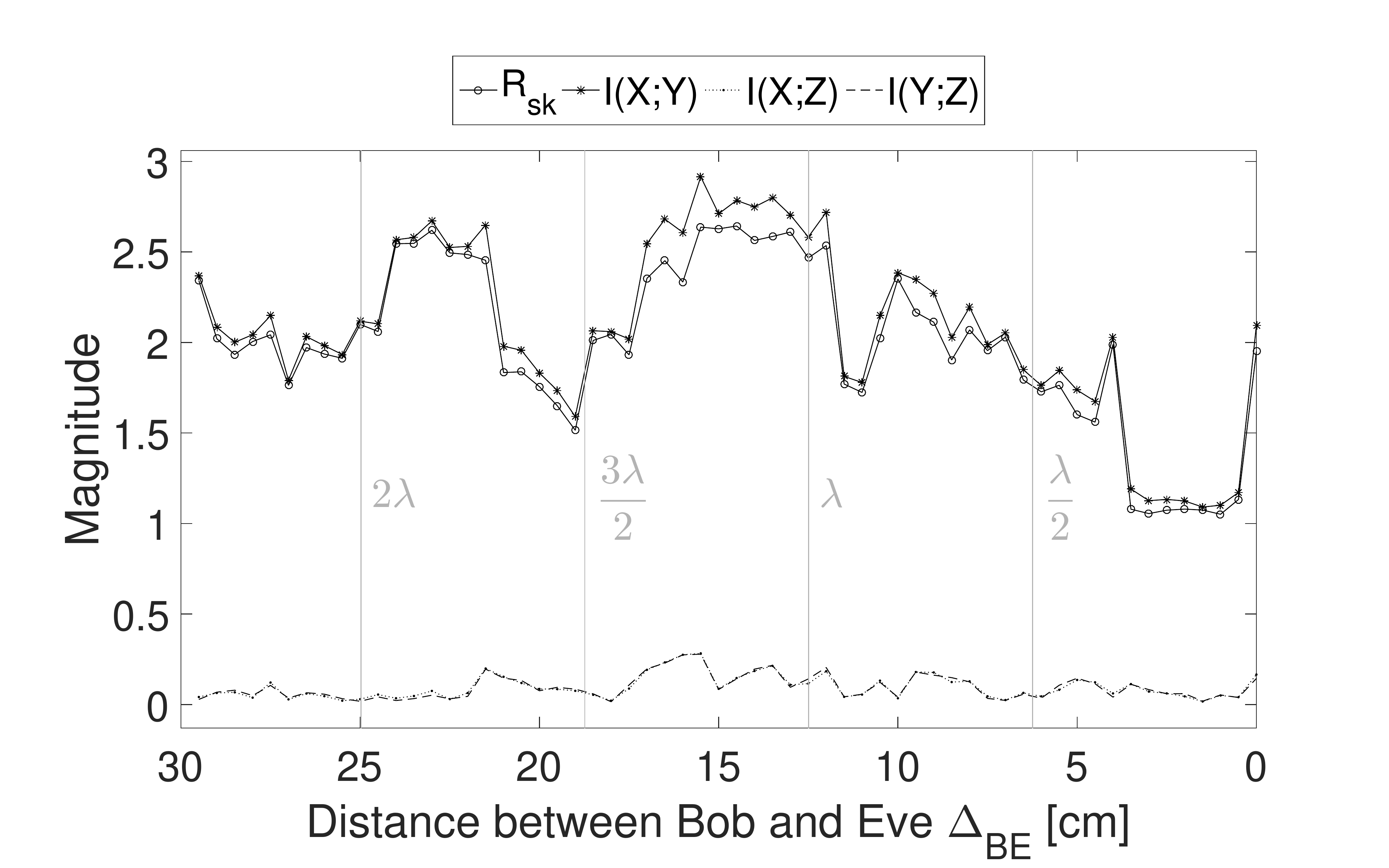}}
	\caption{Evaluation results of $\mybold{v}_k$. In (a) and (b) the cross-correlations is given; in (c) the mutual information as well as $\rsk$ is given. Position 23.}
	\label{fig:app_original_23}
\end{figure*}

\begin{figure*}
	\centering
	\subfloat[]{\includegraphics[trim=1.4cm 0.1cm 3.5cm 1.6cm, clip=true, height=0.224\textwidth]{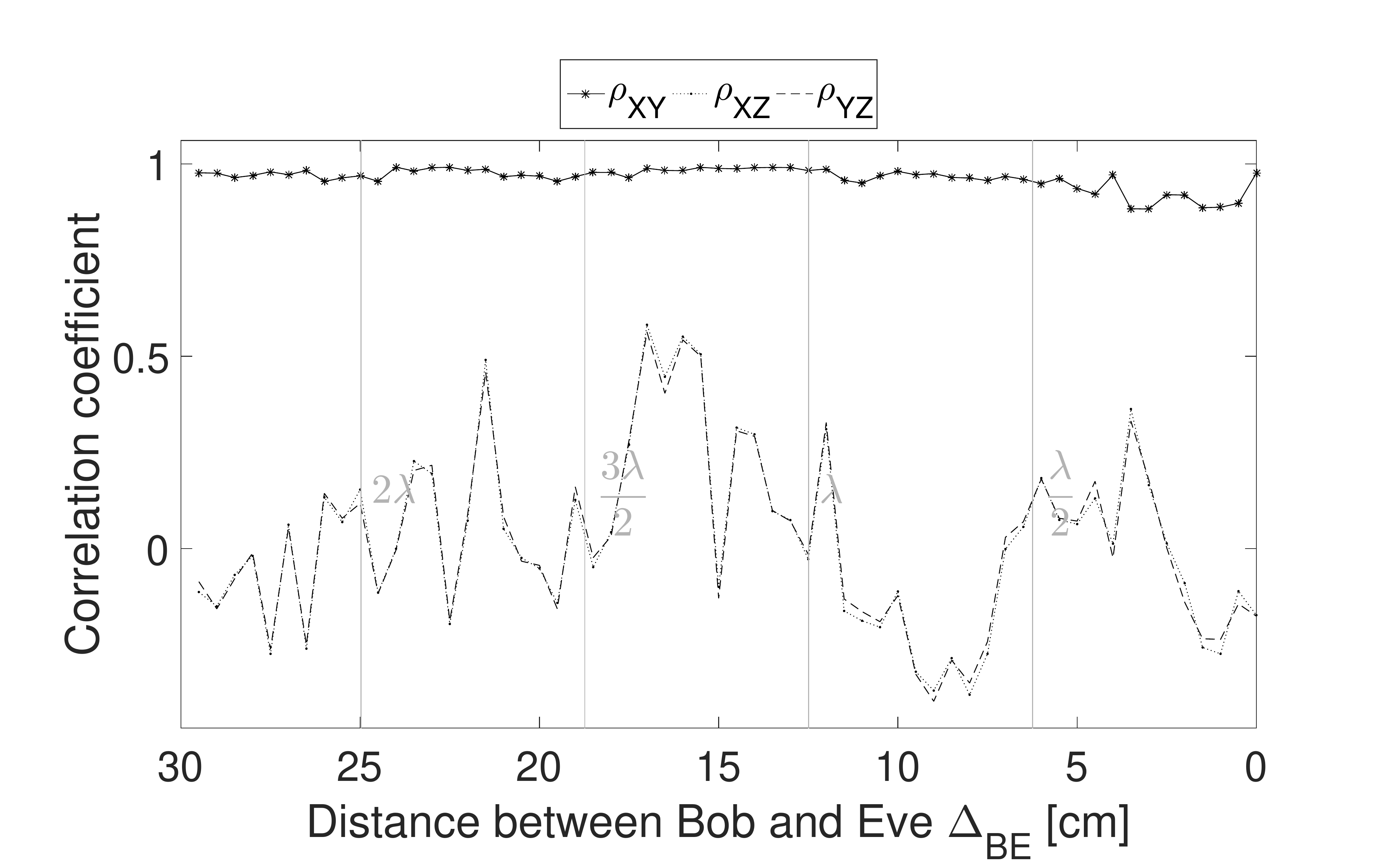}}
	\subfloat[]{\includegraphics[trim=0.5cm 0.1cm 3.5cm 1.6cm, clip=true, height=0.224\textwidth]{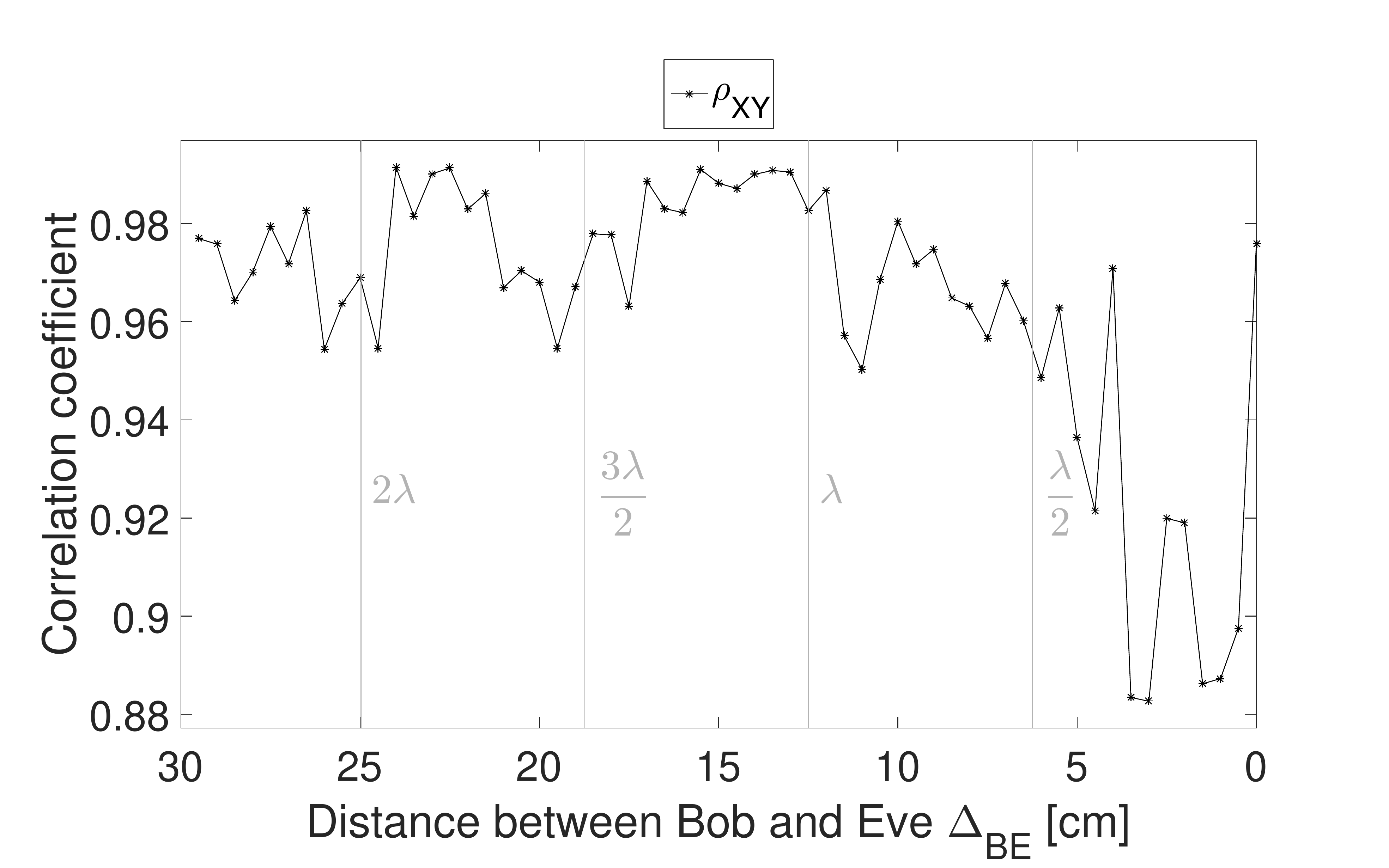}}
	\subfloat[]{\includegraphics[trim=2.2cm 0.1cm 3.5cm 1.6cm, clip=true, height=0.224\textwidth]{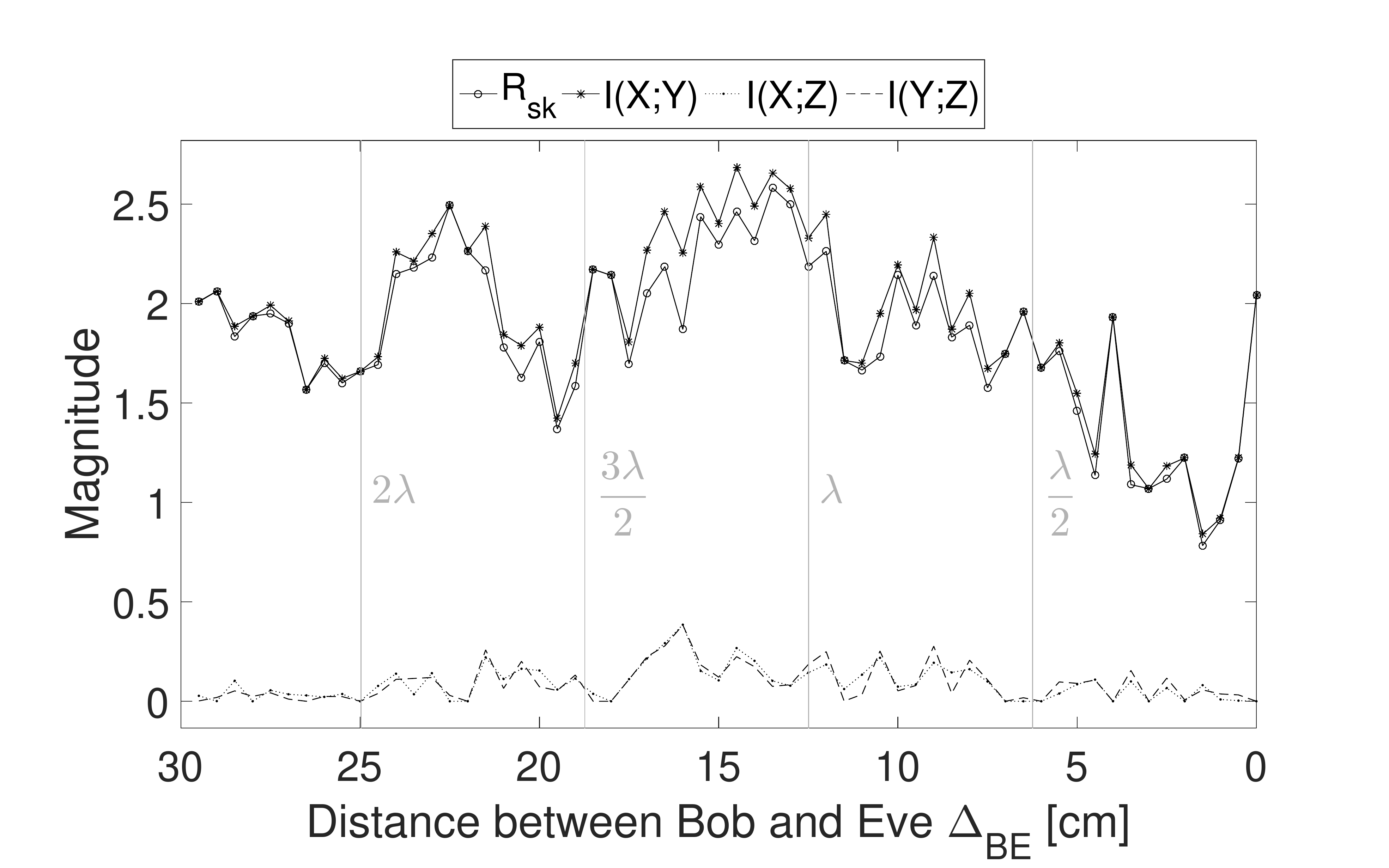}}
	\caption{Evaluation results of $\mybold{v}^{\text{ds}}_k$. In (a) and (b) the cross-correlations is given; in (c) the mutual information as well as $\rsk$ is given. Position 23.}
	\label{fig:app_ds_23}
\end{figure*}

\begin{figure*}
	\centering
	\subfloat[]{\includegraphics[trim=1.4cm 0.1cm 3.5cm 1.6cm, clip=true, height=0.224\textwidth]{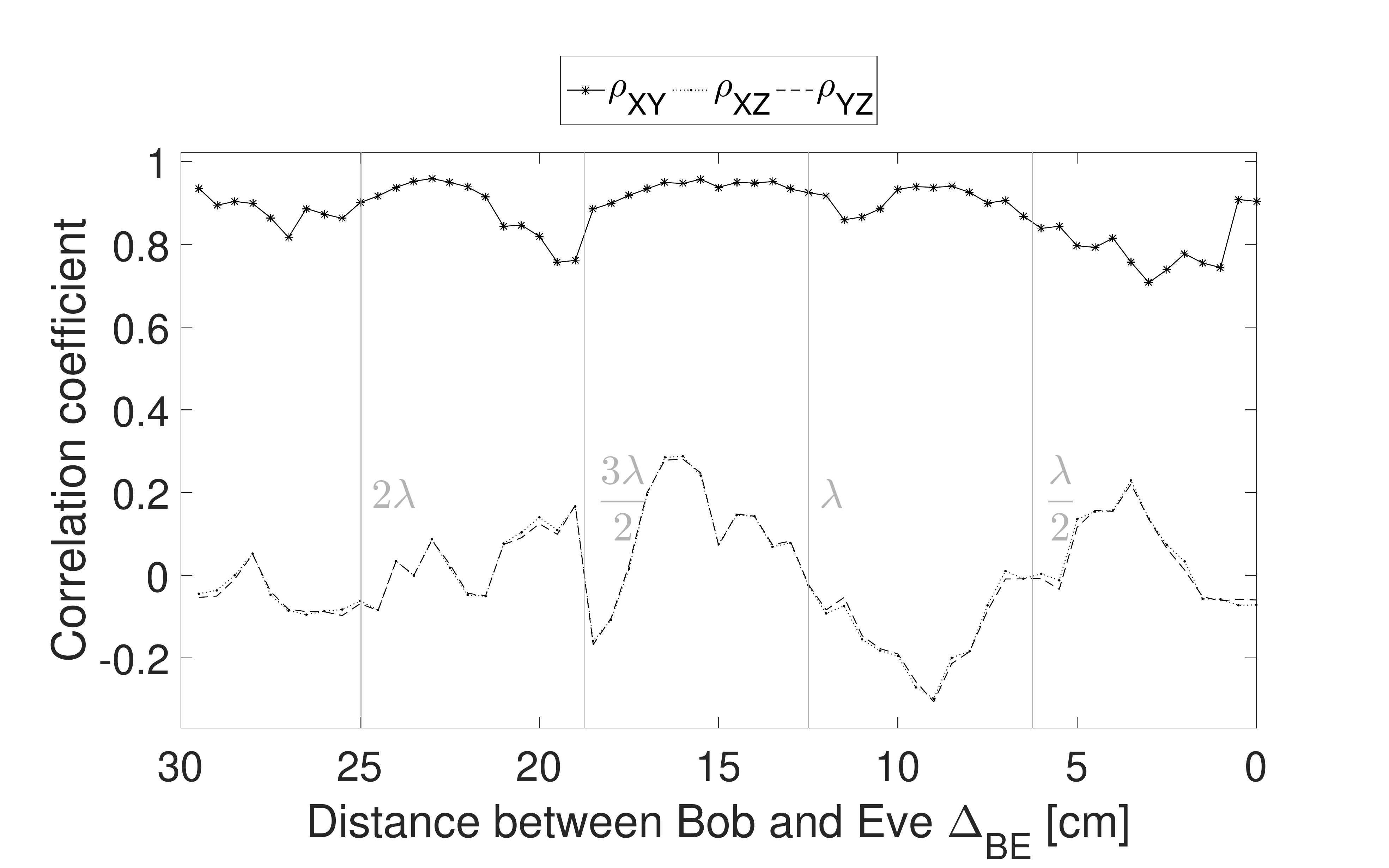}}
	\subfloat[]{\includegraphics[trim=1cm 0.1cm 3.5cm 1.6cm, clip=true, height=0.224\textwidth]{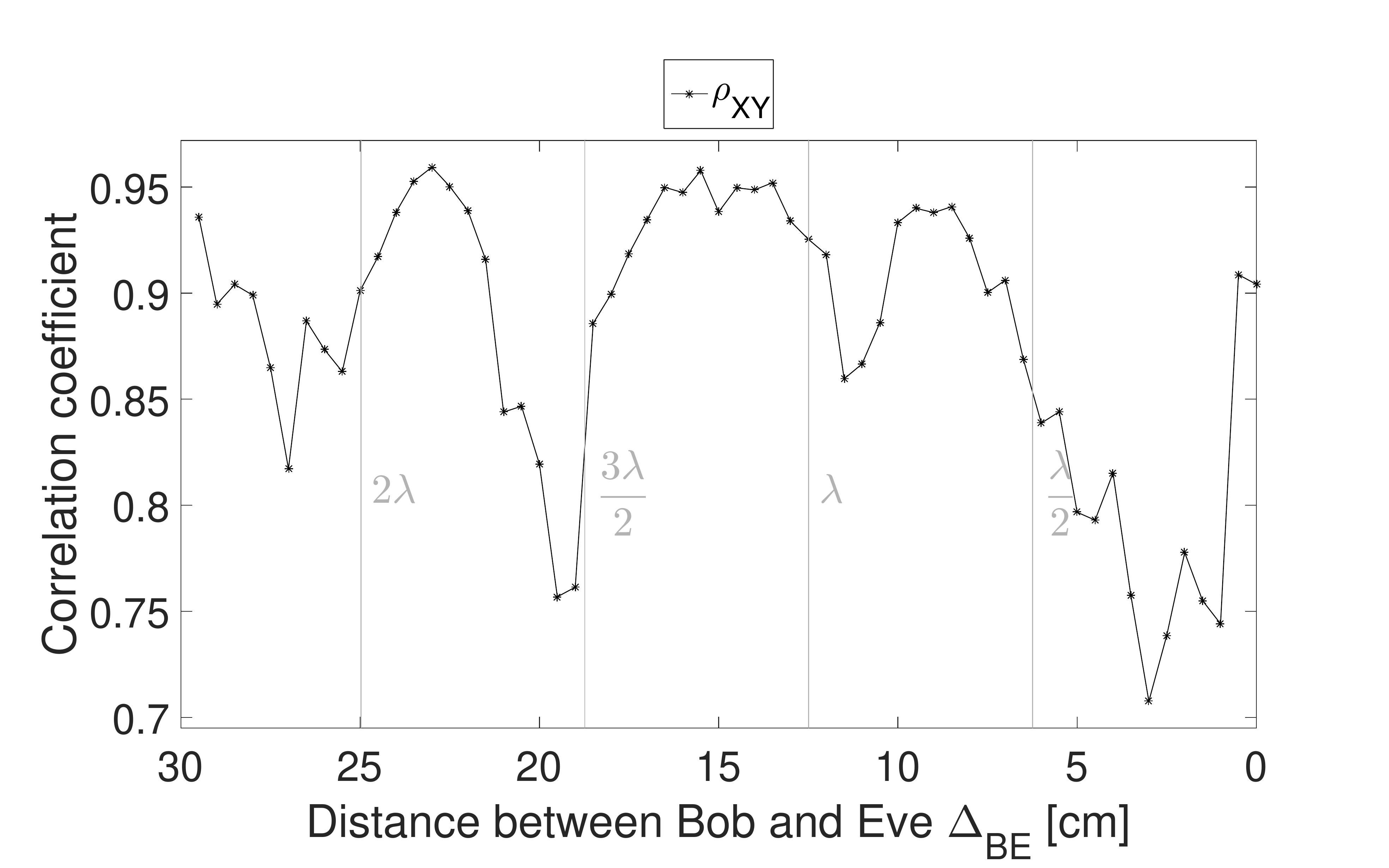}}
	\subfloat[]{\includegraphics[trim=1.8cm 0.1cm 3.5cm 1.6cm, clip=true, height=0.224\textwidth]{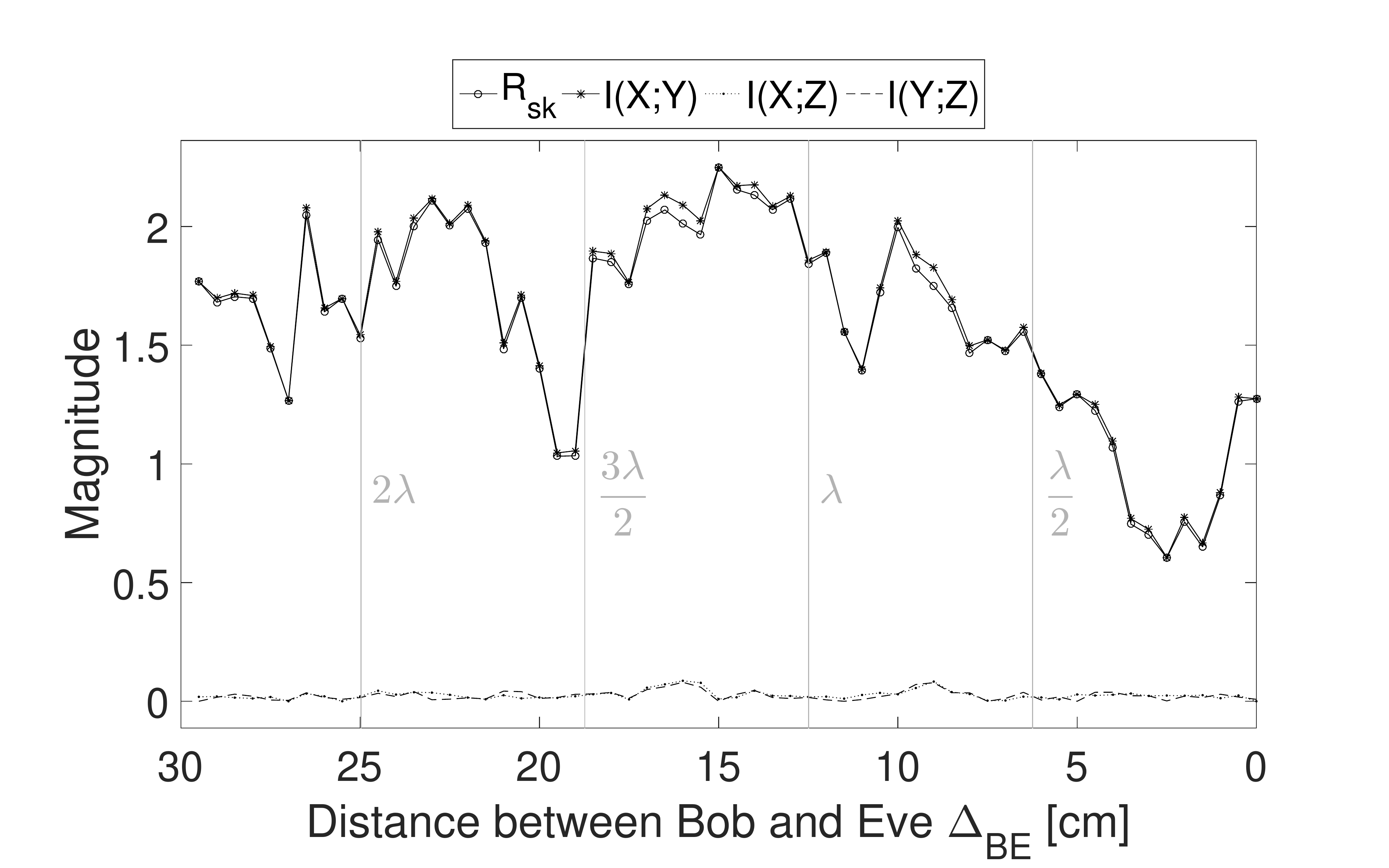}}
	\caption{Evaluation results of $\mybold{v}^{\text{de}}_k$. In (a) and (b) the cross-correlations is given; in (c) the mutual information as well as $\rsk$ is given. Position 23.}
	\label{fig:app_decorr_23}
\end{figure*}


\end{appendices}